\documentclass[12pt,a4paper,twoside,italian,english]{book}
\usepackage{uniudtesi}
\usepackage[nottoc]{tocbibind}
\graphicspath{{./img/}}
\usepackage{amsmath,amsfonts,amssymb,amsthm}
\usepackage{latexsym}

\usepackage{fancyhdr}
\renewcommand{\chaptermark}[1]{\markboth{#1}{}}

\titolo{NON QUESTO FRONTESPIZIO\\This is Just a Placeholder}

 \usepackage{hyperref} 
\usepackage[utf8]{inputenc}
\usepackage[english]{babel}
\usepackage{comment}
\usepackage[nolist,nohyperlinks]{acronym}
\usepackage{csquotes}
\usepackage{amsthm}
\usepackage{amsmath}
\theoremstyle{definition}

\usepackage{lmodern}
\usepackage{verbatim} 
\usepackage[square,sort,numbers]{natbib}
\usepackage{color, colortbl}

\usepackage{color}  
\usepackage{hyperref} 
\definecolor{dark-red}{rgb}{1,0.15,0.15}
\definecolor{dark-blue}{rgb}{0.15,0.15,1}

\hypersetup{%
   colorlinks   = true, 
	urlcolor     = black, 
	linkcolor    = black, 
	citecolor   = black, 
}
\usepackage[textsize=small]{todonotes}

\usepackage{xparse}

\newlength\longest
\usepackage{enumitem}
\usepackage{amssymb}

\setlist[itemize]{noitemsep, topsep=5pt}

\makeatletter
\newcommand{\Spvek}[2][c]{%
	\gdef\@VORNE{1}
	\left(\hskip-\arraycolsep%
	\begin{array}{#1}\vekSp@lten{#2}\end{array}%
	\hskip-\arraycolsep\right)}

\def\vekSp@lten#1{\xvekSp@lten#1;vekL@stLine;}
\def\vekL@stLine{vekL@stLine}
\def\xvekSp@lten#1;{\def\temp{#1}%
	\ifx\temp\vekL@stLine
	\else
	\ifnum\@VORNE=1\gdef\@VORNE{0}
	\else\@arraycr\fi%
	#1%
	\expandafter\xvekSp@lten
	\fi}
\makeatother

\usepackage[export]{adjustbox}
\usepackage{mathtools}

\usepackage{epigraph}
\usepackage{etoolbox}
\makeatletter
\newlength\epitextskip
\pretocmd{\@epitext}{\em}{}{}
\apptocmd{\@epitext}{\em}{}{}
\patchcmd{\epigraph}{\@epitext{#1}\\}{\@epitext{#1}\\[\epitextskip]}{}{}

\makeatother

\usepackage{pdfpages}
\usepackage{afterpage}

\usepackage{bibentry}
\nobibliography*

\usepackage[left,pagewise,displaymath, mathlines]{lineno}  
\linenumbers 
\nolinenumbers

\usepackage{xcolor}
\usepackage{soul}
\setstcolor{red}
\setul{}{.2ex}

\usepackage{adjustbox}

\usepackage{booktabs}
\usepackage{amsmath}
\usepackage{multirow}
\usepackage{mathtools}

\DeclarePairedDelimiter\floor{\lfloor}{\rfloor}
\DeclareMathOperator{\AP}{AP}
\DeclareMathOperator{\MAP}{MAP}
\DeclareMathOperator{\AAP}{AAP}
\DeclareMathOperator{\GMAP}{GMAP}
\DeclareMathOperator{\GAAP}{GAAP}
\usepackage[flushleft]{threeparttable}

\preto\tabular{\setcounter{mmagicrownumbers}{0}}
\newcounter{mmagicrownumbers}
\def\rrownumber{}
\usepackage{siunitx}

\usepackage{xfrac}
\DeclareMathOperator{\logitAP}{logitAP}
\usepackage[ruled]{algorithm2e} 

\usepackage{subfigure}

\newcommand{\cut}{cut\xspace}
\newcommand{\cuts}{cuts\xspace}
\newcommand{\hh}{{\tt H}\xspace}
\newcommand{\rr}{{\tt R}\xspace}
\newcommand{\mm}{{\tt M}\xspace}
\newcommand{\nn}{{\tt N}\xspace}

\definecolor{shadowcells}{HTML}{E7E4E3} 

\usepackage{epigraph}

\begin{document}
\frontmatter
\afterpage{\null\newpage}
\includepdf[pages=1]{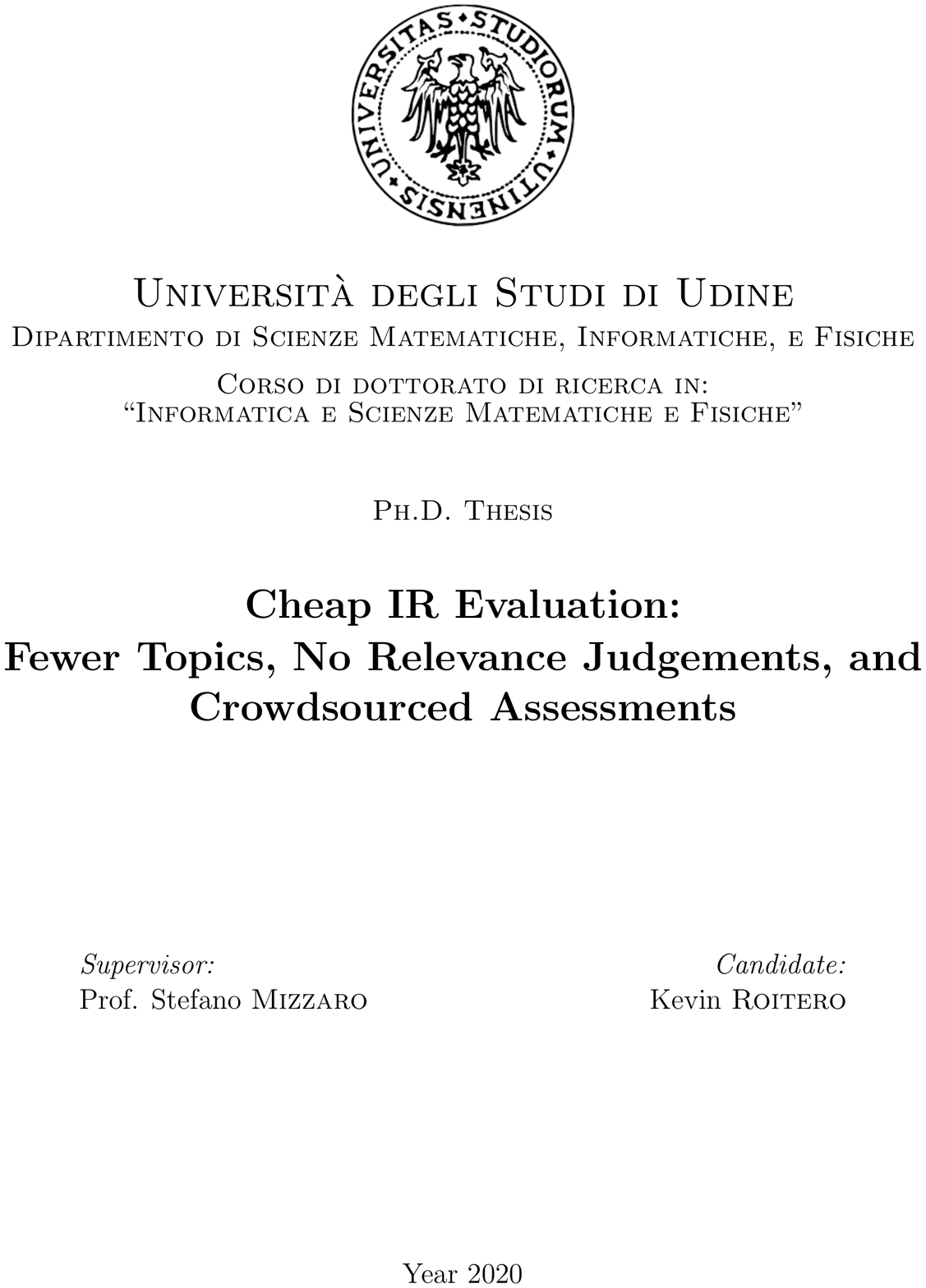}

\renewcommand{\theequation}{\arabic{equation}}
\renewcommand{\thesection}{\arabic{section}}

\vspace*{8em}
\begin{flushright}
%
%
%
%
%
%
%
%
%
%
%
%
%
\epigraph{``Things we lose have a way of coming back to us in the end, if not always in the way we expect.''}{J.K. Rowling (Luna Lovegood), Harry Potter and the Order of the Phoenix.}
%
%
\end{flushright}

\nolinenumbers
\chapter*{Abstract}
To evaluate Information Retrieval (IR) effectiveness, a possible approach is to use test collections, which are composed of a collection of documents, a set of description of information needs (called topics), and a set of relevant documents to each topic. 
Test collections are modelled in a competition scenario: for example, in the well known TREC initiative, participants run their own retrieval systems over a set of topics and they provide a ranked list of retrieved documents; some of the retrieved documents (usually the first ranked) constitute the so called pool, and their relevance is evaluated by human assessors; the document list is then used to compute effectiveness metrics and rank the participant systems. 
Private Web Search companies also run their in-house evaluation exercises; although the details are mostly unknown, and the aims are somehow different, the overall approach shares several issues with the test collection approach.

The aim of this work is to: 
(i) develop and improve some state-of-the-art work on the evaluation of IR effectiveness while saving resources,
and 
(ii) propose a novel, more principled and engineered, overall approach to test collection based effectiveness evaluation.

In this thesis we focus on three main directions: 
the first part details the usage of few topics (i.e., information needs) in retrieval evaluation and shows an extensive study detailing the effect of using fewer topics for retrieval evaluation in terms of number of topics, topics subsets, and statistical power.
The second part of this thesis discusses the evaluation without relevance judgements, reproducing, extending, and generalizing state-of-the-art methods and investigating their combinations by means of data fusion techniques and machine learning.
Finally, the third part uses crowdsourcing to gather relevance labels, and in particular shows the effect of using fine grained judgement scales; furthermore, explores methods to transform judgements between different relevance scales. 

\nolinenumbers

\tableofcontents

\mainmatter

\renewcommand{\theequation}{\thechapter.\arabic{equation}}
\renewcommand{\thesection}{\thechapter.\arabic{section}}

\nolinenumbers
\chapter{Introduction}

\section{Information Retrieval Evaluation}

\subsection{Historical Context}
The meaning of Information Retrieval (IR) can be very broad. \citet[p. 26]{DBLP:books/daglib/0021593} defines the IR as  following:
\begin{displayquote}
	Information retrieval (IR) is finding material  of an unstructured nature that satisfies an information need from within large collections.
\end{displayquote}
The word ``material'' used in the definition refers to items that are usually documents; the usage of ``unstructured nature'' means that the structure is not fundamental in the task of retrieving a document, that is, IR techniques must not rely on it; most of the documents are indeed in plain text format, but web pages and other formats are possible. The term ``collections'' refers to a set of document with certain properties which are often stored on a computer. The ``Information need" is a very important concept in IR and it will be detailed in the following.

The birth of information retrieval, as stated by \citet{DBLP:journals/debu/Singhal01}, can be traced back to around 3000 BC, when the Sumerian community designated areas of storage  to keep clay tablets with  inscriptions  in order to be able to efficiently retrieve and identify each tablet. Furthermore, they developed a classification method to identify every tablet content.

The need to retrieve documents  efficiently together with their content became more and more important over time, but has an outstanding grown after the discovery of the printing press performed by Joannes Gutenberg around 1440. In 1945 \citet{DBLP:journals/theatlantic/Bush45} published an article  that gave birth to the idea of automatic access to large amounts of stored knowledge by the construction of a large index that can be accessed automatically by a workstation he called ``Memex''. 

Driven by the technological progress, in the mid 1950s several works elaborated upon the basic idea of searching text with the aim of a computer. \citet{DBLP:journals/ibmrd/Luhn57}  proposed ``the compilation of a thesaurus-type dictionary and index'' used in order to ``searching pattern for finding pertinent information'' by an automated machine.
In the 1950s the idea of measure in a repeatable way the effectiveness of the systems of information retrieval born. To this aim, in the late 1950s test collections were introduced. The first test collection was the Cranfield collection \cite{misc:cranfield,cleverdon1967cranfield}, which allowed precise measurements of information retrieval effectiveness. The creation of the Cranfield collection has started the development of various test collections: this project hallowed, among others, the born of the Text REtrieval Initiative (TREC) in 1992.

In the 1960s, \citet{DBLP:conf/sigir/Salton91}  proposed the Smart Environment, the computer programs that he and his colleagues created to facilitate  the research. It has been used for implementing and evaluating a large number of different automatic search and retrieval processes. The  Smart Environment was a theoretical and experimental program to explore and evaluate various indexing and retrieval techniques. 
In 1967, the National Library of Medicine (NLM) contracted with Systems Development Corporation (SDC), which had developed a  IR system in order to install a system that would allow medical institutions across US to search NLM’s electronic database of medical journal citations. This system was the Medical Literature Analysis and Retrieval System On-line (MEDLINE).

In the 1970s and 1980s  many developments followed the advances of the 1960s. In those years new techniques where proved to work on the collections, which were made of thousands of articles, all of them with a good level of quality information.
In 1977, a major project began at Syracuse University to design an on-line bibliographic retrieval system that used techniques pioneered by Salton’s SMART experiments. The project was called the Syracuse Information Retrieval Experiment (SIRE).

In the 1990s the situation changed when in 1989 \citet{misc:lee:proposal} proposed a new way to handle management of general information about accelerators and experiments at CERN with a distributed hypertext system. After that event, there has been an exponential growth in the number of hypertext documents published. In 1992 was created, co-sponsored by the National Institute of Standards and Technology (NIST) and U.S. Department of Defence, the Text REtrieval Conference (TREC) as an extension of research conducted during the early 1990s as part of the TIPSTER Program. The purpose of this initiative is  to support research within the information retrieval community by providing the infrastructure necessary for large-scale evaluation of text retrieval methodologies \cite{DBLP:conf/trec/Harman92}.

After the birth of TREC many new IR techniques have been developed, with the large diffusion of collections and the development of technology and research in the IR field.

\subsection{The Relevance Concept}
The ``Information need'' concept  refers to the necessity that  the users  have to obtain some information of a certain kind. In particular, as stated by \citet{Belew:2000:FOC:368213}, we must find out about a topic of interest, looking for those things that are relevant to our search. 
As stated by \citet{DBLP:books/daglib/0022709}, it can be difficult for people to identify their Information Need (IN), because there is the so called Anomalous State of Knowledge (ASK) \cite{belkin:anomalous}; that is, the user might not know what he really wants, or needs, to know. 
This state of ``not knowing'' has  important consequences: the user may require different search techniques,  and furthermore may require different search iterations and attempts in order to satisfy his/her information need. This leads  the user to establish a dialogue between he/she and the Information Retrieval System.
Furthermore, the concept or relevance is difficult to formalize, since even the user does not always know precisely what it needs to know.
The definition of IN lead to consider a definition of relevance, in order to state how much a document is relevant to the information need of the user.	

Some attempts to formalize the relevance concept have been developed \citet{mizzarorelevances},
 \citet{DBLP:journals/jasis/Saracevic07},
  \citet{DBLP:journals/ipm/CosijnI00}, 
  \citet{DBLP:journals/ipm/SpinkGB98}, and even more.
An exhaustive overview of the relevance history can be found in works by \citet{mizzarorelevances}.
According to \citet{mizzarorelevances}, we can formalize the relevance concept along a four dimensional space definition: 
the formalization of the resources,
the formalization of the user problem,
time, and 
the so called ``components'',  which are topic, task, and context.

\subsection{Metrics in IR}
Then we have a set of documents and a query, we can divide the result of the query to the system along two axes, the \emph{relevance axis} and the \emph{retrieval axis}. 
The relevance axis leads to consider the subset of documents which are relevant, or not, to a given query. The retrieval axis division, instead, considers the subset of documents that are returned, or not, by the information retrieval system to the user.
If we combine this two axes we obtain then four categories of documents:
(i) The Relevant-Retrieved set, which represents the relevant documents that are returned by the system, thus represents the part that is both useful to the informational need and retrieved, so the part that a good system should return (and maximize);
(ii) The Relevant-Not Retrieved set represents the part of relevant documents that are not returned by the IR system to the user; this represents a lack of the system performance;
(iii) The Not Relevant-Retrieved set represents the returned documents which are not useful to the user  informational need. This part represent as well a lack of the performance of the system, and a good IR system should therefore minimize this set of documents;
(iv) The Not Relevant-Not Retrieved represents the set of documents that are not shown to the user, and contemporary are not useful to the user.

A good IRS, that is an optimal IR system,  should maximize the Relevant-Retrieved and the Not Relevant-Not Retrieved documents, minimizing the other two sets of documents.
Using this division of documents, and considering the task to measure the system performance in an objective way, we can define some metrics that are useful to compare the performance of different IRS, detailed in the following.
Precision is the fraction of the documents retrieved that are relevant to the user's information need. 
Recall is the fraction of relevant documents that are retrieved by a certain query.
After computing both precision and recall, in order to visualize the performance of a IRS, we can plot precision viewed as a function of recall.  
This curve is then interpolated and  the precision is measured at eleven fixed recall levels ($0, .1, .2, .3, .4, .5, .6, .7, .8, .9, 1$). This curve is useful to compare a single system performance along different queries, and also the performance of different IRS.
Average precision computes the average value of precision viewed as a function of recall ($p(r)$)  in the interval from $r=0$ to $r=1$. Thus, it is represented by  the area under the precision-recall interpolated curve. 
This metric measures the average of precision values at the rank levels corresponding to relevant and  retrieved documents. 

Then the performances of a system over a set of queries can be averaged to compute a single score for such system. 
Mean average precision (MAP) represents the mean precision score for each query
This metric represents the average precision of the AP measure over a set Q of queries.
In order to avoid the behaviour of the arithmetic mean (i.e., MAP) to overestimate systems that are good on easy queries, and to underestimate systems which are good on hard queries, alternatives such as GMAP~\citet{Robertson:2006:GOT:1183614.1183630} are possible.
The Geometric Mean Average Precision (GMAP) considers the geometric mean instead of the arithmetic mean.

\subsection{Evaluation by Means of User Studies}
The effectiveness of information retrieval systems when estimating the user relevance can be measured in different ways. One way of doing so experimentally consists in perform  user studies.

The main advantages of the user studies are that the evaluation is performed on real human being, using measures like: the satisfaction of the user in using a system, eye tracking, physiological conditions like the skin conductance response, the perceived level of stress, and even more.
On the contrary the repeatability of the studies can be difficult and is not always possible. Furthermore, it is difficult to recruit a large set of users, which are usually required for user studies  in order to make the studies valid.
In order to obtain results that are repeatable, thus to provide a benchmark for measure effectiveness, an approach  that does not include users developed: it is the case of the test collections. This methodology to measure effectiveness is described in the following section. Since in our experiments we use the evaluation of effectiveness computed using test collections, we will focus more on this part, without going into details of the user studies.

\subsection{Evaluation by Means of Test Collections}
In this section we present the test collection setting, with a focus on the TREC initiative, which data has been used for  all the experiments described in this thesis.

Each test collection is composed by  different parts: it contains a document repository or document collection, that is a repository of documents that will be used to perform the retrieval tasks; the collection contains a set of descriptions of information needs, that is a set of queries that  represents the information need that have to be submitted by the systems to the collection of documents; finally, a test collection must include, for each interrogation (i.e., query), a set of relevant documents; in fact this is the set that each system should return in order to satisfy the user information need. 

The terminology used in test collections is peculiar: the descriptors of information needs are called ``topics''; each evaluation of a system over a set of topics is called ``run''; the set of relevant documents  used in the evaluation of the effectiveness of IR systems is made by a pooling of the full document set: this set is usually referred as ``pooled set of documents'' or just as ``pool''. 

The test collections are used in order to test the effectiveness of different IR systems, which can be brand new systems or adapted versions of the implementation of a particular retrieval model.
Test collections were introduced in the late 1950s by the Cranfield collection, which allowed precise measurements of information retrieval effectiveness. It contains 1398 abstracts of aerodynamics journal articles, a set of 225 queries (topics), and exhaustive relevance judgements of all the (query, document) pairs \cite{misc:cranfield}.
In 1992 Text Retrieval Conference (TREC)  test collection was build by the joint effort of the U.S. National Institute of Standards and Technology (NIST) in cooperation with the U.S. Defence Department. Its aim is to provide a benchmark for researchers and stakeholders,  in order to allow repeatable experiments over a test collection.
The TREC collection started as a small collection and in more recent years, NIST has done evaluations on larger document collections, including the 25 million pages of the ``GOV2 web page collection''.
In recent years, other test collections have been made available. 
In 1999 the NII Testbeds and Community for Information access Research (NTCIR) project has built various test collections of similar sizes to the TREC collections, focusing on East Asian language and cross-language information retrieval \cite{w:NTCIR}. 
In 2010, the Conference and Labs of the Evaluation Forum (CLEF) initiative, promotes multilingual and multi-modal system testing, unstructured, semi-structured, highly-structured, and semantically enriched data, creation of reusable test collections, exploration of new evaluation methodologies  \cite{w:CLEF}.
Some  other initiatives, like the Initiative for the Evaluation of XML retrieval (INEX) \cite{w:INEX}, focus more on retrieval of structured data.

The main advantages of use test collections are: the repeatably of the experiments, the availability of the data, the use of the collections for different research purposes: test new algorithms on a large set of documents, test the implementation of a new retrieval model, and even more.

The disadvantages are that the collections are artificial, thus the information needs used are not real ones (i.e., from real users that have real information needs); furthermore, the quality  of the collection is biased by the choice of the documents that are included, as well as the information needs, which are usually artificial as well, and they are biased by the curators of the collection.  Finally, as stated in from the report from \citet{tassey2010economic}, the cost of the whole process is very high, both in terms of money and in terms of human effort that is required in order to arrange the whole TREC competition.

\section{Motivations of the Thesis}
Effectiveness evaluation by means of test collections is not the only possible approach (user studies and log analysis, particularly in the case of companies, are also widely used), but its importance is indisputable and, perhaps, it is even what differentiates IR from related areas. 
However, some limitations of such an approach can be identified, from both the practical/engineering and the theoretical/scientific viewpoints.

From a pragmatical viewpoint, it can be observed that the whole evaluation process is rather expensive, in terms of both human time and money: TREC cost from 1999 to 2009 has been estimated to be about \$30M \cite{tassey2010economic}. 
For a commercial company that needs to evaluate its own IR system, the cost will of course be different, but it is still an important concern. 
A significant component of this cost is due to human relevance assessments: this can be reduced by using smaller document collections, relying on shallower pools, resorting to crowd-sourced assessors (which are probably of lower quality), or using fewer topics.

From a more general standpoint, when analysing the vast literature on IR evaluation, one can have a twofold attitude. 
On the one hand, one can notice that the literature is, indeed, enormous, and a large amount of work has been carried out in the last 50 years. 
On the other hand, one can clearly feel that a lot of work seems more ``artisan-ship'' than engineering. 
To cite a few examples, the topics used in test collection initiatives are often chosen ``manually'', by a few assessors, with no guarantee that they are indeed an unbiased sample of real life needs; when using query logs (a luxury that, although available to search companies, is usually unavailable to the research community), the need behind the query is often unknown; relevance assessment are often carried out without a rigorous quality control; and so on. Nevertheless, It should be remarked that when building a test collection many  decisions must be made~\cite{tague1992pragmatics,Voorhees:2005:TEE:1121636}, as for example considering many trade-offs between different types of experimental validity, as well as overall evaluation cost. Such trade-offs require deep expertise and trial and error phases  over a long period of time to fully understand and improve. Furthermore, many (if not all) of the approaches adopted to better engineer the experimental environment (including the ones detailed in this thesis), might include some ``artisanal'' step, as for example the value selected for the parameters of an algorithm. Thus, it is necessary to remark that all the work done in the last 50 years to better improve the evaluation paradigm include careful and considerate thoughts, and the last paragraph is intended as a consideration that some evaluation steps can be improved, and it should not be interpreted as a critique to such evaluation paradigm.

Having said so, an attempt to describe more precisely the situation is by studying what is happening in the non-ideal scenario. 
An ideal test collection should have a perfect sample of topics, an adequate document collection, etc. 
In general the evaluation setting is not ideal, along different ``dimensions'' as, for example:
\begin{itemize} 
\item Topics and documents: these might be too many or too few, selected using a not optimal sampling strategy, etc.
\item Quality of assessments:  it is well known that the relevance of a document to a topic is to some extent subjective, 
and the increasingly common practice of using low quality crowdsourced assessments might further compound this issue.
\item Effectiveness metrics: more than 100 metrics exist, 
and  choosing a wrong metric 
can harm the evaluation result.
\item Pooling: although the pooling approach, combined with the competition scenario, is a practical solution, it introduces some bias in the evaluation process, which is not neglectable anymore with today's large document collections.
\end{itemize}

We do not yet have an overall and complete understanding of what happens when the theoretical ideal evaluation setting is somehow degraded, as it is always the case in practice. 
From a more general viewpoint, this scenario makes one wonder if there is a more principled approach to address the evaluation problem. 
For example, it is particularly striking that in both test collection based initiatives and in-company private evaluation exercises, enormous amount of data are produced and call for a deeper relationship with the disciplines of data science, big data, and machine learning, that have much recently increased their importance --- but such a relationship is nowhere in sight.

\section{Aims of the Thesis}
This thesis sets in the Information Retrieval field, precisely in the branch of research which investigates how to reduce the cost and the effort in the evaluation of Information Retrieval systems, in particular using test collections. 
Specifically, this thesis investigates about the reduction of the cost and the effort in the evaluation of Information Retrieval systems by means of three different approaches:
the reduction of the topic set currently used (Part~\ref{part:fewtopics}),
the evaluation performed with no human intervention (Part~\ref{part:eewrj}), and 
the evaluation performed collecting crowdsourced relevance judgements (Part~\ref{part:cs}). 

It has been estimated that, on average, according to the  IR researchers who responded to the survey conducted by \citet{tassey2010economic}, end users of web search products would be able to satisfy an information need 215\% faster in 2009 than in 1999 as a result of improvements in web search engine performance.  It has been estimated, by \citet{tassey2010economic},  that 32\% of this improvement was enabled by TREC Program activities.

This thesis aims to reduce the effort of this whole process evaluation, preserving the benefits.
In the first part of this thesis we aim to reduce the number of information needs (i.e., queries) which are used in the evaluation of collections of IR. In fact, when we evaluate the effectiveness of IR systems using test collections, each system is required to run over a set of queries. If we reduce the number of queries used in the evaluation, preserving the ability to evaluate the effectiveness of the systems of IR, we would have obtained an actual saving of resources in the whole evaluation process.

Furthermore, in the second and third parts of this thesis, we aim in providing reliable relevance judgements, focusing on a totally automatic approach (Part~\ref{part:eewrj}), and a semi-automatic one which relies on cheap crowdsourced relevance labels (Part~\ref{part:cs}).
Also in this case, if we reduce the cost of gathering reliable relevance judgements preserving the ability to reliably estimate system effectiveness, we would have reduced the total cost of the whole evaluation process, by reducing the actual cost needed to build a test collection.

\section{Structure of the Thesis}
This work is structured as follows.

Part~\ref{part:fewtopics} details the usage of few topics in the effectiveness evaluation: Chapter~\ref{chapt:few:evolutionary} describes a novel approach based on a multi-objective Evolutionary Algorithm which allows to run a battery of new experiments to select topic subsets in test collections.
Chapter~\ref{chapt:few:topicsubsets} explores what happens to measurement accuracy when the number of topics in a test collection is reduced, using the Million Query 2007,
TeraByte 2006, and Robust 2004 TREC collections.

Part~\ref{part:eewrj} discusses the idea of evaluation without relevance judgements: 
Chapter~\ref{chapt:eewrj:reproduce} reproduces notable work on the evaluation without relevance judgements and generalize some of the obtained results to other collections (including a recent one), evaluation metrics, and a shallow pool.
Chapter~\ref{chapt:eewrj:combinations} compares such methods when they are used under the same conditions, using different collections and different measures, and investigates combinations of the various methods.

Part~\ref{part:cs} discusses the usage of crowdsourcing to gather relevance judgements, and in particular the effect of different judgement scales.
Chapter~\ref{chapt:cs:S100} discusses and experimentally evaluates by means of a large scale crowdsourced relevance judgements the use of a fine-grained scale on 100 levels.
Using the proposed scale, the human assessor judges the relevance of a document with respect to a query by means of a number in the $[0..100]$ range (extremes included, thus the levels are actually 101; we name it S100 anyway). 
Using the data from such a scale crowdsourcing experiment (we collect more than 50 thousand labels on such a scale), we discuss its advantages and disadvantages with respect to the already proposed alternatives.
Chapter~\ref{chapt:cs:trans} discusses the transformation between relevance scales. 
That chapter looks at the effect of scale transformations in a systematic way. We perform extensive experiments to study the transformation of judgements from fine-grained to coarse-grained. We use different relevance judgements expressed on different relevance scales and either expressed by expert annotators or collected by means of crowdsourcing.  The objective is to understand the impact of relevance scale transformations on IR evaluation outcomes and to draw conclusions on how to best transform judgements into a different scale, when necessary.

Finally,  Chapter~\ref{chapt:thesisconclusion} provides the conclusions and directions for future work.

\section{Publications}
This work is based on the following peer-reviewed publications:
\begin{enumerate}
\item \bibentry{DBLP:conf/sigir/Roitero18}\\
conference rank: A*\\
This work details the idea of the cheap evaluation of information retrieval systems.
\item \bibentry{Roitero:2018:RIE:3282439.3239573}\\
This work sets the basis for Chapter~\ref{chapt:few:evolutionary}.
\item \bibentry{Roitero2019ffew}\\
This work sets the basis for Chapter~\ref{chapt:few:topicsubsets}.
\item \bibentry{DBLP:conf/iir/RoiteroM16}\\
This work contains some of the clustering experiments detailed in Section~\ref{FewTopics:sec:clustering}.
\item \bibentry{Roitero:2018:EES:3209978.3210108}\\
conference rank: A*\\
This work contains material used in Chapters~\ref{chapt:few:topicsubsets} and \ref{chapt:eewrj:combinations}.
\item \bibentry{Roitero:2018:RGE:3282439.3241064}\\
This work sets the basis for Chapter~\ref{chapt:eewrj:reproduce}.
\item \bibentry{toapp:ipm}\\
This work sets the basis for Chapter~\ref{chapt:eewrj:combinations}.
\item \bibentry{Roitero:2018:FRS:3209978.3210052}\\
conference rank: A*\\
This work sets the basis for Chapter~\ref{chapt:cs:S100}.
\item \bibentry{toapp:transf}\\
conference rank: A\\
This work sets the basis for Chapter~\ref{chapt:cs:trans}.
\end{enumerate}

Furthermore, during my Ph.D. and up to the thesis submission date (i.e., 2019-10-31), I produced the following peer-reviewed publications:

\begin{enumerate}

\item \bibentry{toapp:ipm}

\item \bibentry{toapp:impact}

\item \bibentry{toapp:strat}\\
conference rank: A*

\item \bibentry{DBLP:conf/iir/RoiteroMS19}

\item \bibentry{toapp:stoc}\\
conference rank: A

\item \bibentry{toapp:transf}\\
conference rank: A

\item \bibentry{DBLP:conf/sigir/SopranoRM19}

\item \bibentry{Roitero2019ffew}

\item \bibentry{DBLP:conf/sigir/ZampieriRCKM19}\\
conference rank: A*

\item \bibentry{DBLP:conf/wsdm/HanRGSCMD19}\\
conference rank: A*

\item \bibentry{toapp:howmany}

\item \bibentry{Roitero:2018:RGE:3282439.3241064}

\item \bibentry{Roitero:2018:RIE:3282439.3239573}

\item \bibentry{Roitero:2018:FRS:3209978.3210052}\\
conference rank: A*

\item \bibentry{Mizzaro:2018:QPP:3209978.3210146}\\
conference rank: A*

\item \bibentry{RoiteroMPM18}\\
conference rank: A*

\item \bibentry{Roitero:2018:EES:3209978.3210108}\\
conference rank: A*

\item \bibentry{DBLP:conf/sigir/Roitero18}\\
conference rank: A*

\item \bibentry{DBLP:conf/hcomp/CheccoRMMD17}

\item \bibentry{Maddalena:2017:CAA:3121050.3121060}

\item \bibentry{DBLP:conf/iir/RoiteroMS17}

\item \bibentry{Roitero2017}\\
conference rank: A

\item \bibentry{DBLP:conf/iir/RoiteroM16}

\end{enumerate}



\part{On Using Fewer Topics in Information Retrieval Evaluation}\label{part:fewtopics}

\chapter{Introduction and Background}

This chapter is structured as follows:
Section~\ref{part:few:intro} introduces the work,
Section~\ref{part:few:numberoftopics} discusses the usage of fewer topics in retrieval evaluation,
Section~\ref{part:few:topicsubsets} details the studies which investigated topic subsets,
Sections~\ref{part:few:guiver}, \ref{part:few:ecir11}, and \ref{part:few:berto} explain in detail notable work on topic subsets, and
Section~\ref{part:few:bestsubsoftware} describes the BestSub software and its limitations.

\section{Introduction}\label{part:few:intro}
When evaluating the effectiveness of Information Retrieval (IR) systems,
the design of the measurement process has been examined by
researchers from many `angles': e.g. the consistency of relevance
judgements; the means of minimizing judgements while maintaining
measurement accuracy; and the best formula for measuring effectiveness.
One aspect -- the number and type of queries (\emph{topics} in TREC
terminology) needed in order to measure reliably -- has been discussed
less often. In general, there has been a trend in test collection construction
of increasing the number of topics, but without much consideration of
the benefits of such an approach.
In many areas of measurement via sampling, it is generally accepted
that there are diminishing returns from increasing the sample size \citep{kotrlik2001organizational}. Beyond
a certain point, improvements in measurement accuracy are small and
the cost of creating the sample becomes prohibitive. We are not
aware of work in IR that establishes if such an optimal sample size exists.

\begin{figure}[t]
  \centering
   \includegraphics[width=.6\linewidth]{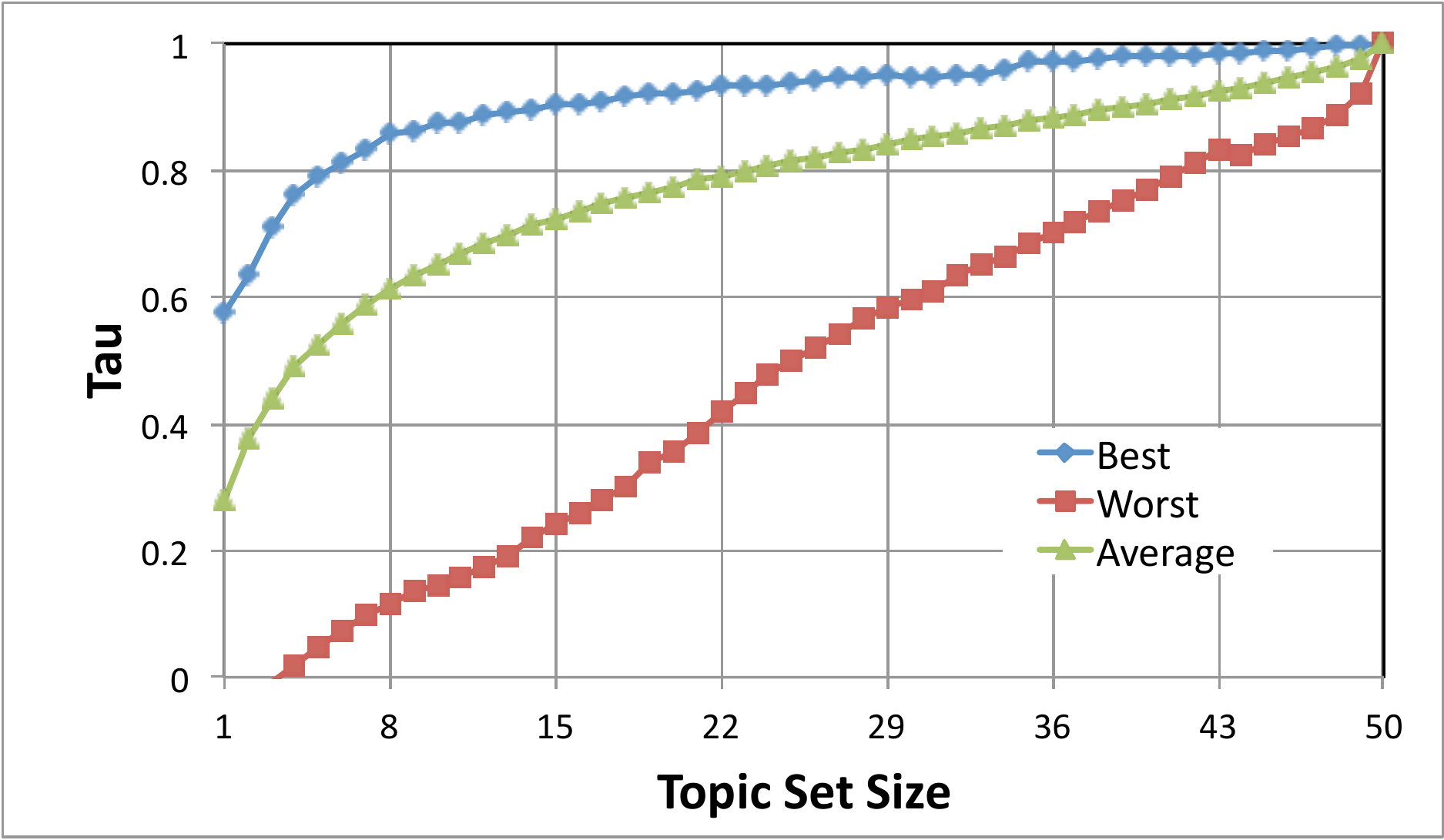}
\caption{Kendall's $\tau$ correlation curves for AH99, 
adapted 
from \citet[Figure 2]{Guiver:2009:FGT:1629096.1629099}.
}
  \label{FewTopics:fig:tois1}
\end{figure}

Other work has been conducted on whether smaller topic sets
(\emph{subsets}) could be used in a test collection, examining early
TREC ad hoc collections \citep{Guiver:2009:FGT:1629096.1629099,ecir11,ictir13}, and the 2009
Million Query (MQ) Track \citep{MQ2009,Carterette:2009}.
These approaches, in general, ask how similarly a set of
retrieval runs are ranked when using such a subset versus a full set
of topics. Note that in these experiments, the full set of topics is taken to be the
\emph{ground truth}. The similarity of the two rankings is measured
using Kendall's Tau (henceforth, $\tau$). 
Figure~\ref{FewTopics:fig:tois1} shows an example result
from this work, taken from \citet{Guiver:2009:FGT:1629096.1629099}. On the x-axis are topic
subsets of increasing cardinality, the y-axis measures $\tau$.
Three types of subset are shown for each cardinality:
\begin{itemize}
\item Best -- the subset of a given cardinality that results in a ranking that is closes to
  the ranking of runs using the full topic set;
\item Average -- the average $\tau$ of all the topic subsets
  examined; 
\item Worst -- the topic subset that results in a ranking that is furthest from the ranking of
  runs from the full topic set.\footnote{\label{FewTopics:fn:2}\citet{Guiver:2009:FGT:1629096.1629099}
    use the terminology Best/Average/Worst, and we adopt it in this
    paper in order to be consistent with past work.}
\end{itemize}

The best correlation curve shows that even when using a topic subset
of cardinality $6$, a relatively high $\tau$ ($>0.8$) can be
found. 
The curve for the average topic set reaches a $\tau$ of $0.8$ at
cardinality $22$.
The generality of this basic result was questioned by \citet{ecir11},
and revisited again by \citet{ictir13} with results that confirmed the
original conclusions.\footnote{\label{FewTopics:fn:after}It is important to remark that this line of research focuses on an \emph{a posteriori}, i.e., after-evaluation setting: it is not aimed at predicting in advance a good topic subset, but only at determining if such a subset exists.}
\citet{MQ2009} conducted similar experiments though only measuring the
average. 
However, they also examined different topic types, which will be
discussed later.

In the following we describe the state-of-the art in the evaluation of IR systems using fewer topics.

\section{Number of Topics}
\label{part:few:numberoftopics}
\citet{BuckleyVoorhees} examined the accuracy of common evaluation
measures relative to the number of topics used.
They suggested using at least 25 topics,
though stated having more was better. The authors concluded that
50 topics produce reliable evaluations. The conclusion on the number
of topics was broadly confirmed by \citet{carterette-etc06} who
considered a larger number of topics (200). 

While the methods used in earlier work to determine the appropriate number
of topics for a test collection involved a range of empirical approaches,
\citet{webber08CIKM} proposed the use of statistical power analysis
when comparing the effectiveness of runs.
The authors argued that a set of nearly 150 topics was necessary to
distinguish runs. Building on suggestions by \citet{SandersonZobel},
they also argued that using more topics with a shallow assessment
pool was more reliable than using few topics with a deep assessment
pool.
\citet{Carterette:2007:HTI:1321440.1321530} used power analysis statistics to study both topic set size and judgement set size.

Using the approach of Test Theory, introduced by \citet{Bodoff_Li_2007},
\citet{urbano13} examined test collection reliability considering
all aspects of the collection. The authors tabulated their measures
of reliability across a large number of TREC collections, and suggested
that the number of topics used in most current test collections is
insufficient.

More recently, \citet{Sak14,Sakai2016} used power analysis to argue that
more topics than are currently found in most test collections are
required. 
He showed that many
significant results may be missed due to the relatively small
number of topics in current test collections. He concludes that
potentially, hundreds of topics are required to achieve reasonable
power in current test collections.

While the works here seem to draw contradictory conclusions
of different minimum numbers, a common theme to the work is
that the minimum number needed to separate the effectiveness
of two runs depends on how similar the runs are. The earlier
work examined runs more widely separated than more recent work.

\section{Topic Subsets} \label{part:few:topicsubsets}
Separate to the question of how many topics are required, researchers have
asked if some form of targeted topic sample could achieve the same measurement
effect.

Subsequent to the work of \citet{hits-trec07} on topic subsets, 
\citet{hauff_subsets_CIKM} presented three approaches to measure effectiveness
estimation using topic subsets:
greedy, median Average Precision (AP), and estimation accuracy.
\citet{hauff_subsets_ECIR} then presented evidence showing
that the accuracy of ranking the best runs depended on the degree of 
human intervention in any manual runs submitted,
and went on to show that this problem can be somewhat alleviated by
using a subset of ``best'' topics.
\citet{cattelan_ICTIR} also studied whether it is possible to evaluate 
different runs with different topics. \citet{Roitero2017} generalized the approach to other collections and metrics, further investigating the correlations between topic ease and its capability of predicting system effectiveness.

In contrast to the work conducted by \citeauthor{hits-trec07} -- which
looked for best and worst subsets in a ``bottom up'' approach, finding
any topics that would fit into each subset -- \citet{MQ2009} took a
``top down'' approach. They manually split the topics of the MQ collections
into subsets based on groups of categories from \citet{Rose_Levinson_2004}.
They found little difference examining the groups. They also looked at 
different combinations of hard, medium, and easy topics (determined by
the average score that runs obtained on the topics) and found similar
conclusions to earlier topic subset work.

In related work, \citet{hosseini_ICTIR11} presented an approach to 
expand relevance judgements when new runs are evaluated.
The cost of gathering additional judgements was offset by selecting
a subset of topics that discriminated the runs best, 
determined using Least Angle Regression (LARS) and
convex optimization, up to a maximum topic set cardinality of 70.
Later, \citet{hosseini_CIKM11} used convex optimization to select
topics that needed further relevance judgements when evaluating
new runs.
The algorithm estimates the number of unjudged documents for a topic
and identifies a set of query-document pairs that should be judged
given a fixed budget.

\citet{Hosseini:SIGIR:2012} proposed a mathematical framework to select topic subsets based on modeling the evaluation metric's uncertainty obtained when dealing with incomplete or missing relevance judgements for a set of topics. This work is particularly relevant as we will be able to compare some of our results with theirs.

\citet{KUTLU201837} developed a method for topic selection based on learning-to-rank; they took into account the effect of pool depth and focused on deep vs.\ shallow judging.

We now detail the approaches of topic set reduction based on the  theoretical best possible choice, which three main contributions are by \citet{Guiver:2009:FGT:1629096.1629099}, \citet{ecir11}, and \citet{Berto:2013:UFT:2499178.2499184}. Since in Chapter~\ref{chapt:few:evolutionary} we focus on reproducing those results, we describe such papers more in detail.

\section{The Study by \texorpdfstring{\citeauthor*{Guiver:2009:FGT:1629096.1629099}}{Guiver:2009:FGT:1629096.1629099}} \label{part:few:guiver}


\begin{table}[tb]
\centering
  \begin{tabular}[t]{|c|ccc|c|}
  \hline	
  & $t_1$ &  $\cdots$ & $t_n$ &	$\MAP$\\\hline
  $s_1$ & $\AP(s_1,t_1)$& $\cdots$ &$\AP(s_1,t_n)$&$\MAP(s_1)$\\
  \vdots&\vdots &$\ddots$&\vdots&$\vdots$\\
  $s_m$& $\AP(s_m,t_1)$& $\cdots$ &$\AP(s_m,t_n)$&$\MAP(s_m)$\\\hline
  \end{tabular}
  \caption{AP and MAP, for $n$ Topics and $m$ Systems (adapted from \cite{Guiver:2009:FGT:1629096.1629099})
    \label{JDIQGenetic:tab:AP}
}
\end{table}

\citet{Guiver:2009:FGT:1629096.1629099} propose a theoretical analysis
on topic set reduction.
Their analysis starts from  TREC evaluation results as represented in Table~\ref{JDIQGenetic:tab:AP}.
The process is described as follows \cite[page~21:4]{Guiver:2009:FGT:1629096.1629099}:
\blockquote{
The basic method is as follows. We start from a set of $n$ topics ($n$ = 50 or
25 in the experiments that follow). We now consider, for any $c \in \{1,\ldots, n\}$ and
for any subset of topics of cardinality $c$, the corresponding values of MAP for
each system calculated on just this subset of topics: that is, we average only a
selected set of $c$ of the $n$ columns in Table I [our Table~\ref{JDIQGenetic:tab:AP}].
For each such subset, we calculate the correlation of these MAP values with the MAP values for the whole set of
topics. This correlation measures how well the subset predicts the performance of different systems on the whole set.
Now for each cardinality $c$, we select the best subset of topics, that is the one
with the highest correlation. We also select the worst, and finally we calculate an
average correlation over all subsets of size $c$.
}

\begin{figure}[tbp]
  \centering
  \begin{tabular}{cc}
     \includegraphics[width=.45\linewidth]{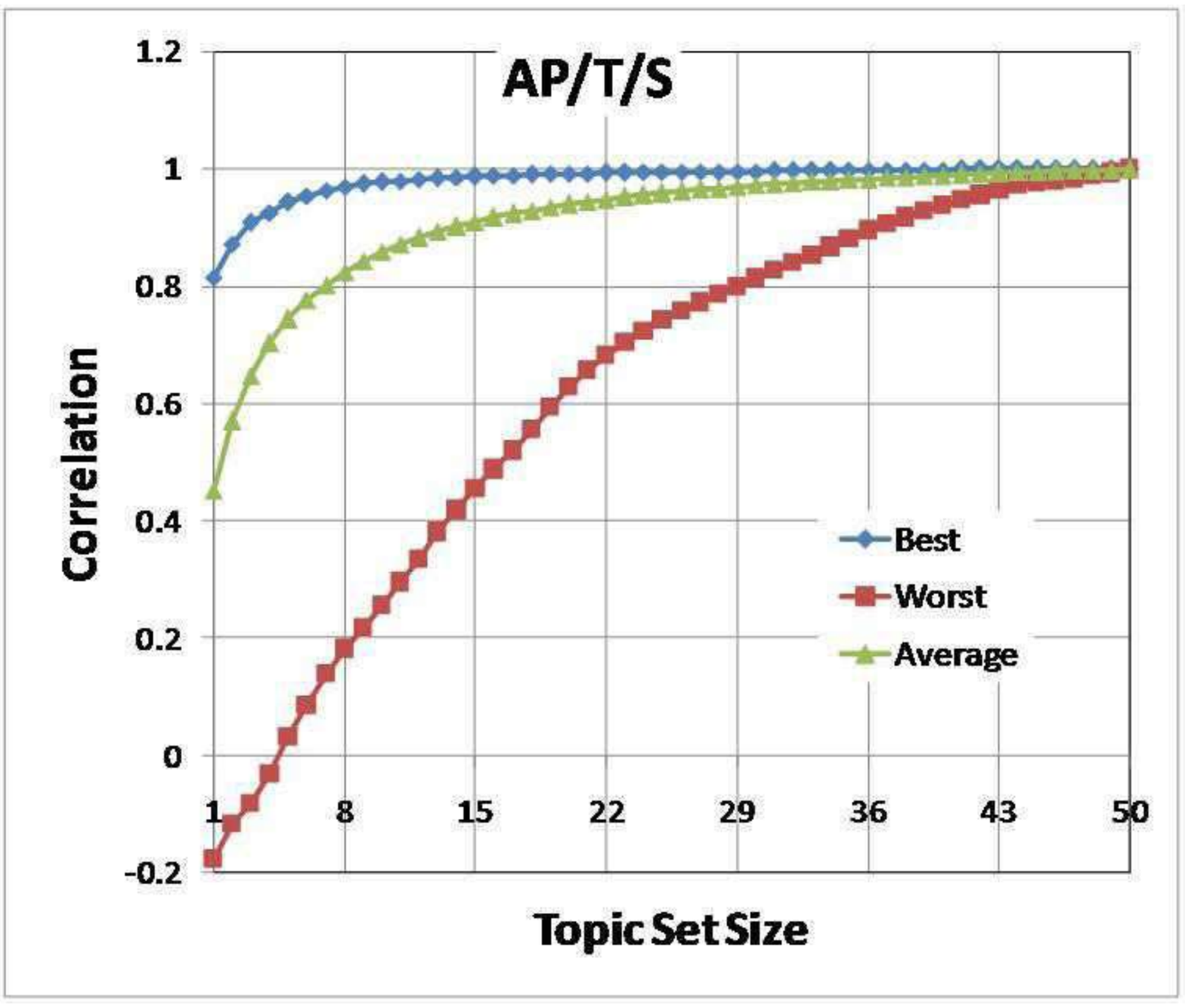}&
     \includegraphics[width=.45\linewidth]{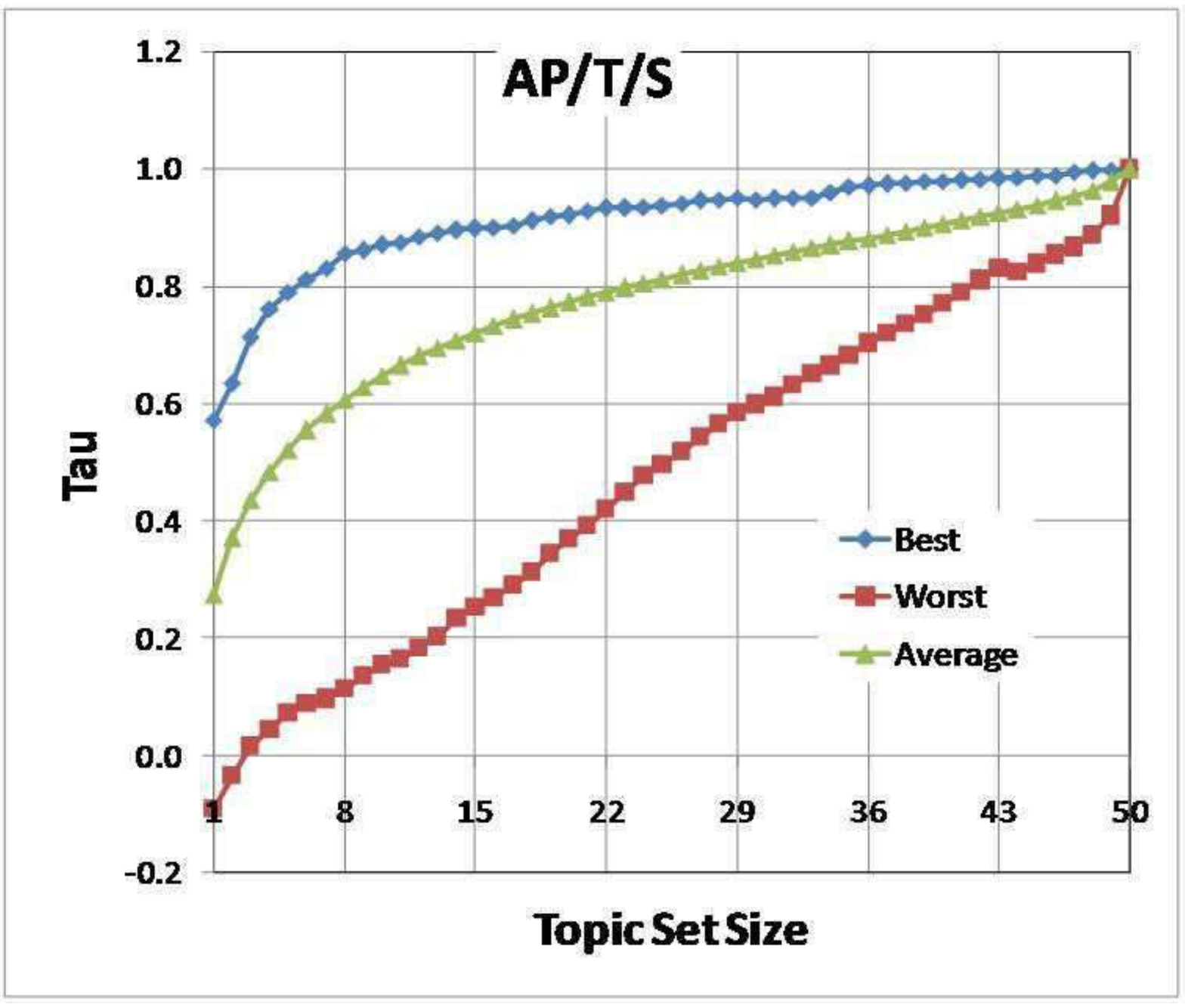}
    \\
 (a)&(b)
  \end{tabular}
  \caption{Correlation values from Pearson's $\rho$ (a) and Kendall's $\tau$ (b), obtained with BestSub on TREC-8 
  (from \cite[Figs.~2 and 3]{Guiver:2009:FGT:1629096.1629099}).
  }
  \label{JDIQGenetic:fig:guiver_correlation}
\end{figure}

The outcome of the process is shown in Figure~\ref{JDIQGenetic:fig:guiver_correlation}, that shows  for each cardinality (x-axis) the correlation values (y-axis): the best possible correlations are the curves in blue, the average correlation (i.e., that obtained by a random topic selection) is in green, and the worst is in red.
Results show that the best topic subset is much better than the average one in predicting the system performance on the full set of topics.
Furthermore, the worst topic subset is much worse than the average one, and the gap between the two correlations is high.
To make an example for the Pearson Correlation, with just 8 best topics we obtain correlations higher than 0.95; to achieve the same result we would need 23 topics for the average series, and more than 40 for the worst one. 
Furthermore, results appear stable across measures: \citeauthor{Guiver:2009:FGT:1629096.1629099} include in the study  R-prec, P\@10, and logAP (or GMAP).
Whereas the usefulness of the Best series is intuitive, the interestingness of Worst series is perhaps less straightforward and deserves a brief justification, besides simply stating that it was studied by the previous authors and we are reproducing it. Indeed, knowing how a topic subset can rank the systems in a so different way from the official ranking (i.e., the one provided by TREC) is useful to understand how ``wrong'' one can be when using a topic set of a given cardinality. The Best series is an optimum to aim at; the Worst series is something that needs to be avoided.

It has to be noted that the computational complexity is high and finding an exact solution becomes intractable even for rather small $n$, since the number of subsets to be analyzed increases exponentially. For this reason,  \citeauthor{Guiver:2009:FGT:1629096.1629099} rely on a heuristic search in their analysis, that works as follows \cite[page~21:10]{Guiver:2009:FGT:1629096.1629099}:
\blockquote{%
a heuristic is to search recursively:
having identified the best set for cardinality $c$, to seek the best for cardinality
$c+1$ among sets which differ from the best $c$ set by not more than 3 topics (the
number 3 was chosen primarily because 4 is intractable).%
}

 \citeauthor{Guiver:2009:FGT:1629096.1629099} also addressed two questions which are strongly related to the heuristic algorithm used:
\begin{enumerate}[label=(\roman*)] 
\item how much difference there is considering the topics of the Best/Worst topic subset at a given cardinality ($c$), and the topics of the subset at the next one ($c+1$)? and
\item what happens when performing a neighborhood analysis, i.e., when selecting not only the single Best/Worst subset, but also the 2nd Best/Worst subset, and the subsequent Best/Worst ones? In particular, they analyzed the 10 Best/Worst topic subsets for each cardinality.
\end{enumerate}
%
\begin{figure}[tbp]
  \centering
  \begin{tabular}{@{}cc@{}}
      \includegraphics[width=.45\linewidth, valign=c]{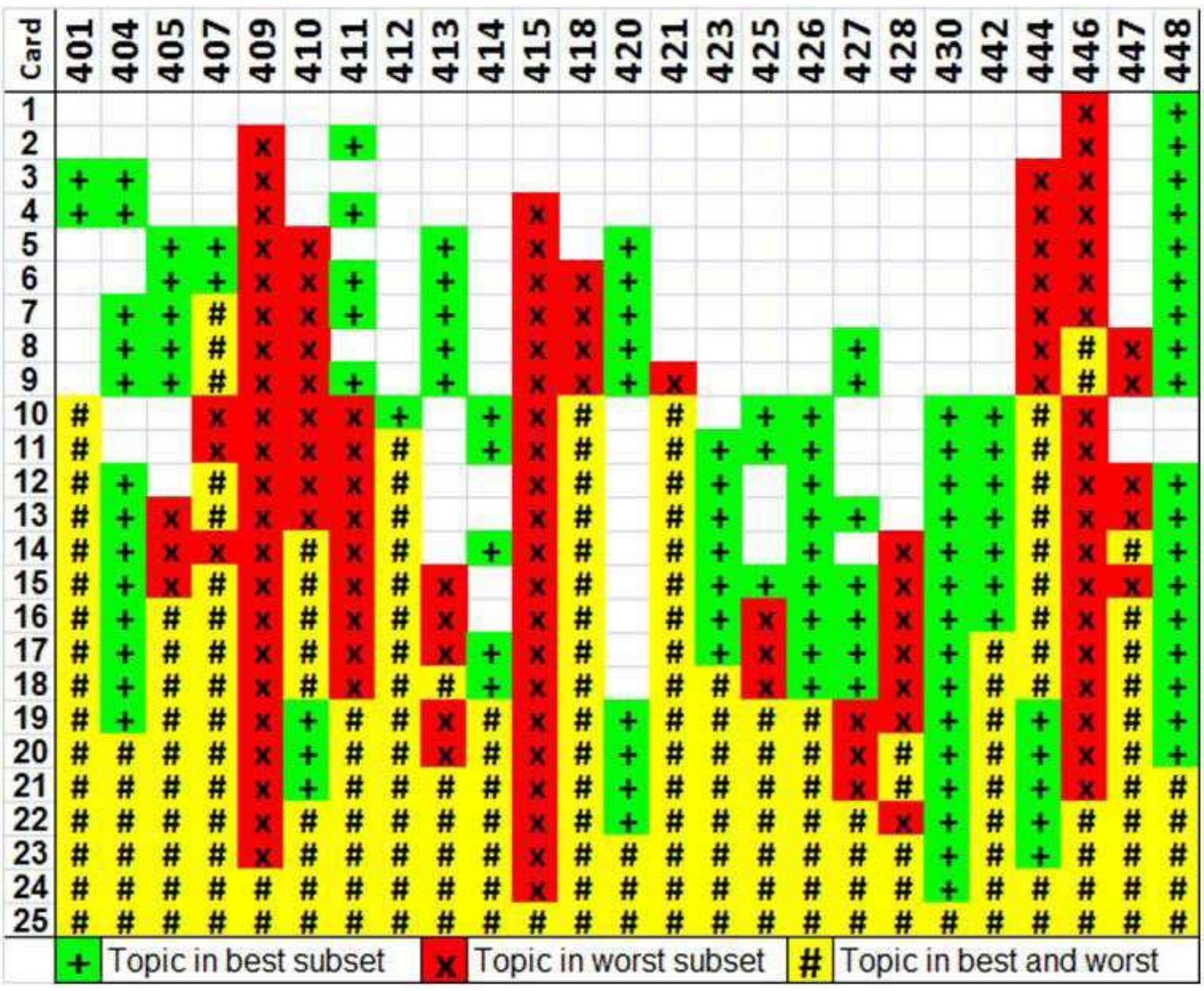}&
     \includegraphics[width=.45\linewidth, valign=c]{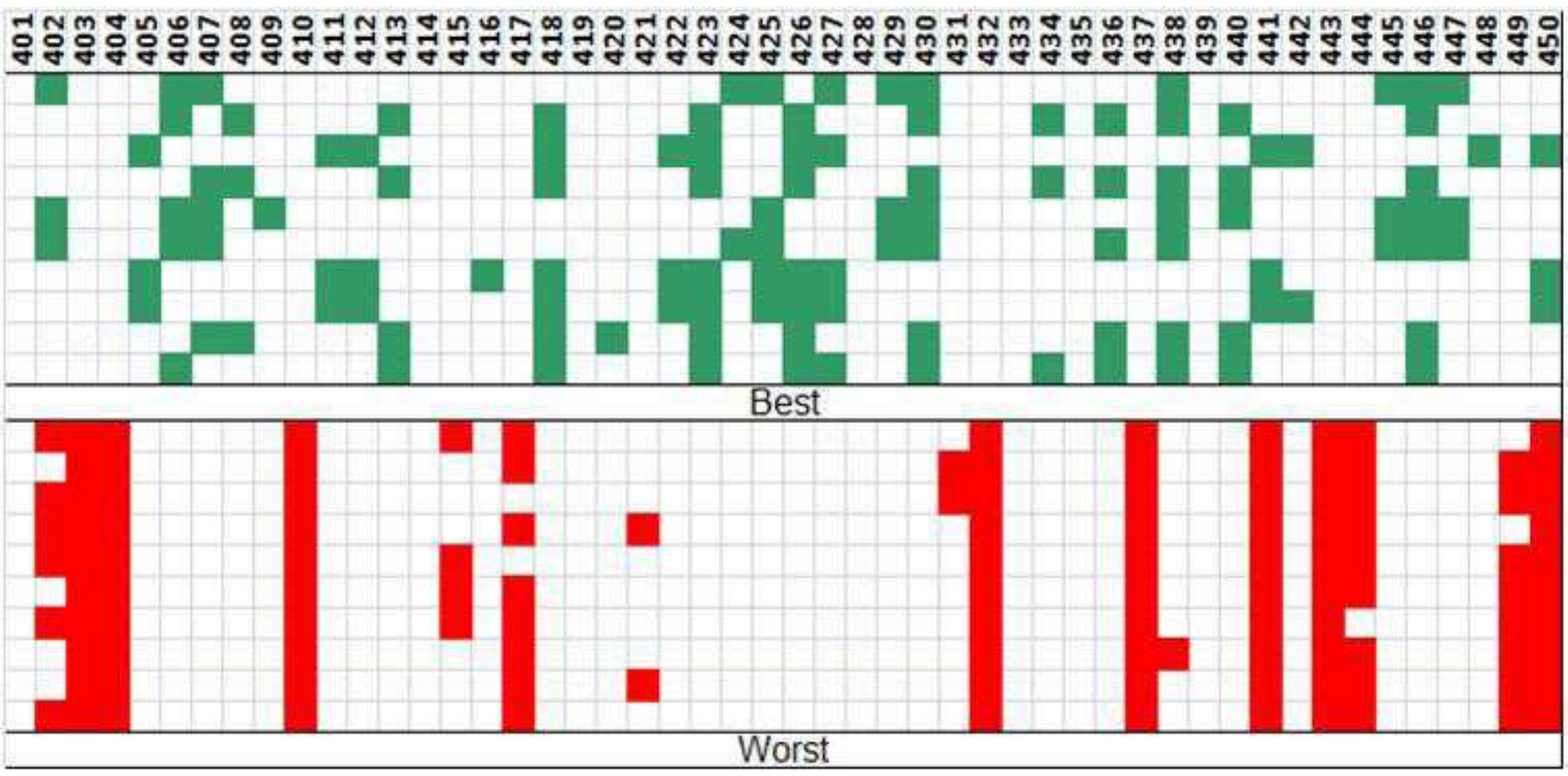}
    \\
 (a)&(b)
  \end{tabular}
  \caption{Stability pixel-maps for Pearson's $\rho$ for the whole dataset (a) and for Pearsons's $\rho$ for the top 10 sets at cardinality 12 (b)
  (from \cite[Figs.~5 and 6]{Guiver:2009:FGT:1629096.1629099}).
  }
  \label{JDIQGenetic:fig:guiver_pixel}
\end{figure}

Results are shown in Figure~\ref{JDIQGenetic:fig:guiver_pixel}. 
Figure~\ref{JDIQGenetic:fig:guiver_pixel}(a) is a  topic by cardinality pixel-map, for $\rho$ correlation; in each cell the value is ``+'' if for that cardinality the topic is part of the Best subset, ``x'' if the topic is in the Worst subset, and ``\#'' if the topic is present in both sets.
Figure~\ref{JDIQGenetic:fig:guiver_pixel}(a) shows that in general single topics appear to be either good or bad; this statement is more true for the bad topic subset: once a topic enters in the worst set at a given cardinality it tends to be in the worst set also for the next cardinality, while for the best topic subset some variation arises in this pattern.

Figure~\ref{JDIQGenetic:fig:guiver_pixel}(b) is a  topic by ``goodness'' (i.e., 1st Best/worst set, 2nd, etc.) pixel-map, for cardinality 12 for $\rho$ correlation; in the upper part (i.e., first 10 rows) of the map the Best 10 subsets are shown, while in the lower part (i.e., last 10 rows) of the map the Worst 10 subsets are shown; in each cell the value is colored if for that set the topic is part of the Best subset (upper part of the map), or in the Worst subset (lower part of the map).
This figure confirms that while the quality of being a bad topic is true for individual topics, a good topic set is not formed by individual good topics only, but by a set of topics that somehow contribute in predicting systems effectiveness.

\section{The Study by \texorpdfstring{\citeauthor*{ecir11}}{ecir11}} \label{part:few:ecir11}
The second contribution is by \citet{ecir11}, who extended the work by \citeauthor{Guiver:2009:FGT:1629096.1629099} in four ways:
\begin{itemize}
\item He used two more collections: TREC87 and another one, named Terrier, in which he used a set of different  configurations of the Terrier system\footnote{\url{http://terrier.org}} to obtain another system population. 
\item He used a different evaluation measure (i.e., logitAP).
\item He studied a particular method to build good topic subsets. He built a matrix, as proposed by \citet{Mizzaro:2007:HHT:1277741.1277824},  to represent interactions between systems and topics, and in particular correlations between topic ease and the ability to predict system effectiveness; on that matrix, using the HITS algorithm \cite{Kleinberg:1999:ASH:324133.324140}, he computed topic hubness, and he analyzed if such a feature can be used to find a few good topics.
\item He performed three generalization experiments \cite[page 138]{ecir11}:
	\begin{enumerate}
	\item a HITS analysis: he compared hubness vectors of topics from the three collections, for a given metric; 
    \item a Best/Worst subset analysis: he used the Best/Worst subsets computed on the TREC data on the Terrier collection;
    \item a topic selection strategy: he used the HITS analysis of TREC data to predict good topic sets on Terrier.
	\end{enumerate}
\end{itemize}
Conclusions of the analysis confirm that choosing a good topic subset is not just a matter of selecting good individual topics.

\section{The Study by \texorpdfstring{\citeauthor*{Berto:2013:UFT:2499178.2499184}}{Berto:2013:UFT:2499178.2499184}} \label{part:few:berto}
\citet{Berto:2013:UFT:2499178.2499184} generalize the work by \citet{Guiver:2009:FGT:1629096.1629099} and \citet{ecir11} by:
(i) investigating how many good topic subsets exist;
(ii) extending the generalization analysis of \citeauthor{ecir11}, considering various evaluation metrics; and
(iii) extending the results of \citeauthor{Guiver:2009:FGT:1629096.1629099} on the best 10 topic subsets by including a generalization experiment (i.e., what happens when considering the rank of the top topic subsets and a different topic subset).
Results show that:
(i) many good topic subsets exist, so there is hope that some of them are general;
(ii) even if the single best topic subset is not able to generalize to a new system population, some of the subsequent ones might be adequate under certain conditions; and
(iii) the metric has a major role when dealing with best topic subsets.

Note that results by \citet{Guiver:2009:FGT:1629096.1629099}, \citet{ecir11}, and \citet{Berto:2013:UFT:2499178.2499184} are \emph{a-posteriori}: their analysis is conducted after the whole evaluation process is finished; thus, 
their results are not immediately applicable to obtain a practical topic selection strategy, which should be able to identify the few good topics \emph{a priori} (or, at least, during the relevance assessment process).
We agree with the authors of the previous studies that, even if a-posteriori, such a theoretical approach is interesting and useful since the theoretical maximum, minimum, and average correlation values for the different topic subsets constitute useful baselines to compare with when testing and proposing a novel topic selection strategy. 

In addition to the previously mentioned work examining topics
\citep{Guiver:2009:FGT:1629096.1629099,ecir11,ictir13}, a wide variety of studies
analyze the components of test collections. Here, we focus
on those that consider the number of topics needed and topic subsetting.

\section{The BestSub Software}  \label{part:few:bestsubsoftware}
The above results (i.e., \citep{Guiver:2009:FGT:1629096.1629099,ecir11,ictir13}) have been obtained by using the BestSub software, specifically implemented in 2006 (and revised some years later) to study the fewer topics approach. BestSub is written in C\#, and receives in input the topic-system table (see Table~\ref{JDIQGenetic:tab:AP}) and a parameter $k$, used for the heuristic search.
BestSub searches the Best/Worst topic subsets (i.e., the one which correlates more/less with the ground truth) at each cardinality $c$ between 1 and $n$ (i.e., the number of topics).
BestSub computes the Best/Worst sets for each cardinality using the Pearson's $\rho$, and Kendall's $\tau$ correlations, plus the Error-Rate measure \cite{Guiver:2009:FGT:1629096.1629099}.
The output of BestSub can be represented as in Figure~\ref{JDIQGenetic:fig:guiver_correlation}.

BestSub has been a valuable and useful tool that allowed to obtain the previous results, but it is not free from limitations, as we now discuss.

The heuristic is quite rough. In the BestSub implementation, the side effect of the  heuristic is that the topic subset at cardinality $c$ and the one at cardinality $c+1$ differ  for at most $k$ elements. The more $k$ is close to 1, the more the search process becomes a greedy algorithm.
\citeauthor{Guiver:2009:FGT:1629096.1629099} set this parameters to a maximum number of 3.
\citeauthor{ecir11} investigated the outcome of the experiments with a parameter $k$ close to 1.
Concerning exhaustive search, \citet[page~21:10]{Guiver:2009:FGT:1629096.1629099} write:
\blockquote{%
when using the Kendall's $\tau$ , searching exhaustively
takes around 7 days for $c = 11$, and around 20 days for $c = 12$, even
with efficient $O(n \log{n})$ calculation of $\tau$ using Knight's algorithm [Boldi et al.
2005]. Exhaustive search for correlation runs is much faster due to the simpler
calculation, and wider scope for optimization of the algorithm; however, even
there, computation becomes a real issue beyond $c = 15$.%
}

It is particularly worrying that the heuristic can distort the stability results. Although the correlation values obtained are probably not heavily affected by using the heuristics, some effect on stability (see Figure~\ref{JDIQGenetic:fig:guiver_pixel}) cannot be excluded. 

Finally,  BestSub efficiency is not ideal. The search algorithm of BestSub, even if optimized, results in an extremely slow search even for a small topic set. With $k=3$ on a top class PC (2013 Mac Pro) for 50 topic the algorithm takes approximately 10.5 hours to finish, and for 250 topics with $k=2$ (i.e., almost a greedy search) the algorithm takes more than 1 month to finish the computation.

\chapter{An Evolutionary Approach to Identify Topics Subsets}\label{chapt:few:evolutionary}
This chapter deals with the design and implementation of an evolutionary algorithm used to identify topic subsets.
Section~\ref{JDIQGenetic:sec:intro} introduces and frames the research questions,
Section~\ref{JDIQGenetic:sect:reimplement} discusses the reimplementation of BestSub software, detailing the Evolutionary Algorithm approach we used,
Section~\ref{JDIQGenetic:sect:experiments} presents the experiments, and 
Section~\ref{JDIQGenetic:sect:conlusions_future} concludes the chapter.

\section{Introduction and Research Questions}\label{JDIQGenetic:sec:intro}

In this chapter we focus on one particular approach to reduce the cost of the evaluation process~\cite{DBLP:conf/sigir/Roitero18}, which consists of limiting the number of topics used in the evaluation. This approach has been studied by \citet{Guiver:2009:FGT:1629096.1629099}, \citet{ecir11}, and \citet{Berto:2013:UFT:2499178.2499184}. Their results have been obtained by using the BestSub software, that presents several limitations, discussed in the following.

Our contribution in this chapter is threefold: 
\begin{enumerate}[label=(\arabic*),ref=(\arabic*)]
\item \label{JDIQGenetic:i:aim1} We re-implement the BestSub software using a novel approach based on a multi-objective Evolutionary Algorithm (EA). We also release the software making it freely available to the research community.
\item We then reproduce the main results by \citet{Guiver:2009:FGT:1629096.1629099}, \citet{Berto:2013:UFT:2499178.2499184}, and \citet{ecir11} using the novel implementation, as well as discuss its advantages w.r.t.\ the original approach. 

\item The novel and more efficient implementation allows us to run a battery of new experiments. We therefore generalize the previous  results to other datasets and collections.

\end{enumerate}

We remark that reproduction seems particularly important in this case, since the previous results have been obtained by a single research group, that used a specific, ad hoc, software that has never been released widely and officially, and they have been published in potentially high impact venues: \cite{Guiver:2009:FGT:1629096.1629099} has been published in an important journal (ACM TOIS, 1.070 impact factor);
\citet{Berto:2013:UFT:2499178.2499184} in ICTIR, and 
\cite{ecir11} in ECIR, two important conferences for the IR community. 


\section{NewBestSub} \label{JDIQGenetic:sect:reimplement}

We now turn to our first aim, see item~\ref{JDIQGenetic:i:aim1} in Section~\ref{JDIQGenetic:sec:intro}. 
In this section we first briefly present some background on Evolutionary Algorithms (EAs); we then detail our overall approach and the NewBestSub software, the reimplementation of BestSub by means of EAs. 

\subsection{Evolutionary Algorithms}
In the current work we make use of an approach based on \emph{Evolutionary Algorithms} (EAs), i.e., population-based metaheuristics which rely on mechanisms inspired by the process of biological evolution and genetics in order to solve optimization problems \cite{Eiben:2003:IEC:954563}.  
Unlike blind random search algorithms, EAs are capable of exploiting historical information to direct the search into the most promising regions of the search space, relying on methods designed to imitate the processes that in natural systems lead to adaptive evolution. 

In nature, a population of individuals tends to evolve, in order to adapt to the environment in which they live; in the same way, EAs are characterized by a population, where each individual represents a possible solution to the optimization problem. Every solution is evaluated with regard to its degree of \lq\lq adaptation\rq\rq{} to the problem through a single- or multi-objective \emph{fitness} function.

During the computation of the algorithm, the population iteratively goes through a series of \emph{generations}. At each generation step, some of the individuals are picked by a {\em selection strategy}, and go through a process of reproduction, by the application of suitable \emph{crossover and mutation operators}. The selection strategy is one of the main distinguishing factors between meta-heuristics, although typically individuals with high degree of adaptation are more likely to be chosen. 

NSGA-II~\cite{Deb:2002:FEM:2221359.2221582}, on which our method is based, uses a Pareto-based multi-objective strategy with a {\em binary tournament selection} and a {\em rank crowding better} function. To the selected individuals, operations such as crossover and mutation are applied with a certain degree of probability, with the goal of generating new offspring, creating a new generation of solutions. Crossover is the EA equivalent of natural reproduction, by which the characteristics of two individuals are combined. Mutation is used to maintain the genetic diversity in the elements of the population, through applying random changes in the encoding of the selected solution. Typically, a high crossover probability tends to pull the population towards a local minimum or maximum, while a high degree of mutation allows to explore the search space more broadly.

The algorithm terminates when a predefined criteria is satisfied, which can be a bound on the number of generations, or a minimum fitness increment that must be achieved between subsequent evolution steps of the population. 

\emph{Multi-objective} EAs are designed to solve a set of minimization/maximization problems for a tuple of $n$ functions $f_1(\overrightarrow{x}),\ldots,f_n(\overrightarrow{x})$, where $\overrightarrow{x}$ is a vector of parameters belonging to a given domain. A set $\mathcal S$ of solutions for a multi-objective problem is said to be {\em non-dominated} (or {\em Pareto optimal}) if and only if for each $\vec{x}\in\mathcal S$, there exists no $\vec{y}\in\mathcal S$ such that: (i) $f_i\left(\vec{y}\right)$ improves $f_i\left(\vec{x}\right)$ for some $i$, with $1\leq i\leq n$, and (ii)  for all $j$, with $1\leq j\leq n$ and $j\neq i$, $f_j\left(\vec{x}\right)$ does not improve $f_j\left(\vec{y}\right)$. The set of non-dominated solutions from $\mathcal S$ is called \emph{Pareto front}. 

Note that, although we are indeed interested in finding a set of Pareto optimal individuals (the best or worst topic set for each cardinality), it would also be possible, in principle, to choose a single one of them as the final solution. This presupposes the existence of a suitable \emph{a-posteriori} selection strategy, such as, for example \lq\lq keep the subset characterized by the highest correlation value, among the ones having less than 10 topics\rq\rq.

\subsection{EAs and Fewer Topics Subsets}

Multi-objective approaches are particularly suitable for solving multi-objective optimization problems, as the one treated in this chapter, because they are capable of searching for multiple optimal solutions in parallel. 
Indeed, given the wide search space of topic sets, and the two antithetical objectives which characterize the topic reduction process (i.e., reducing the topic subset size while keeping a high correlation value), EAs seem to be a natural solution to solve such a problem. 
We propose a method capable of optimizing, together, the size of a subset of topics and its accuracy in evaluating information systems, with respect to the whole set of topics. For each subset cardinality, we look for the best and the worst topics to be included for such an extent.

\begin{figure}[tbp]
  \centering
  \begin{tabular}{@{}cc@{}}
    \includegraphics[width=.6\linewidth, valign=c]{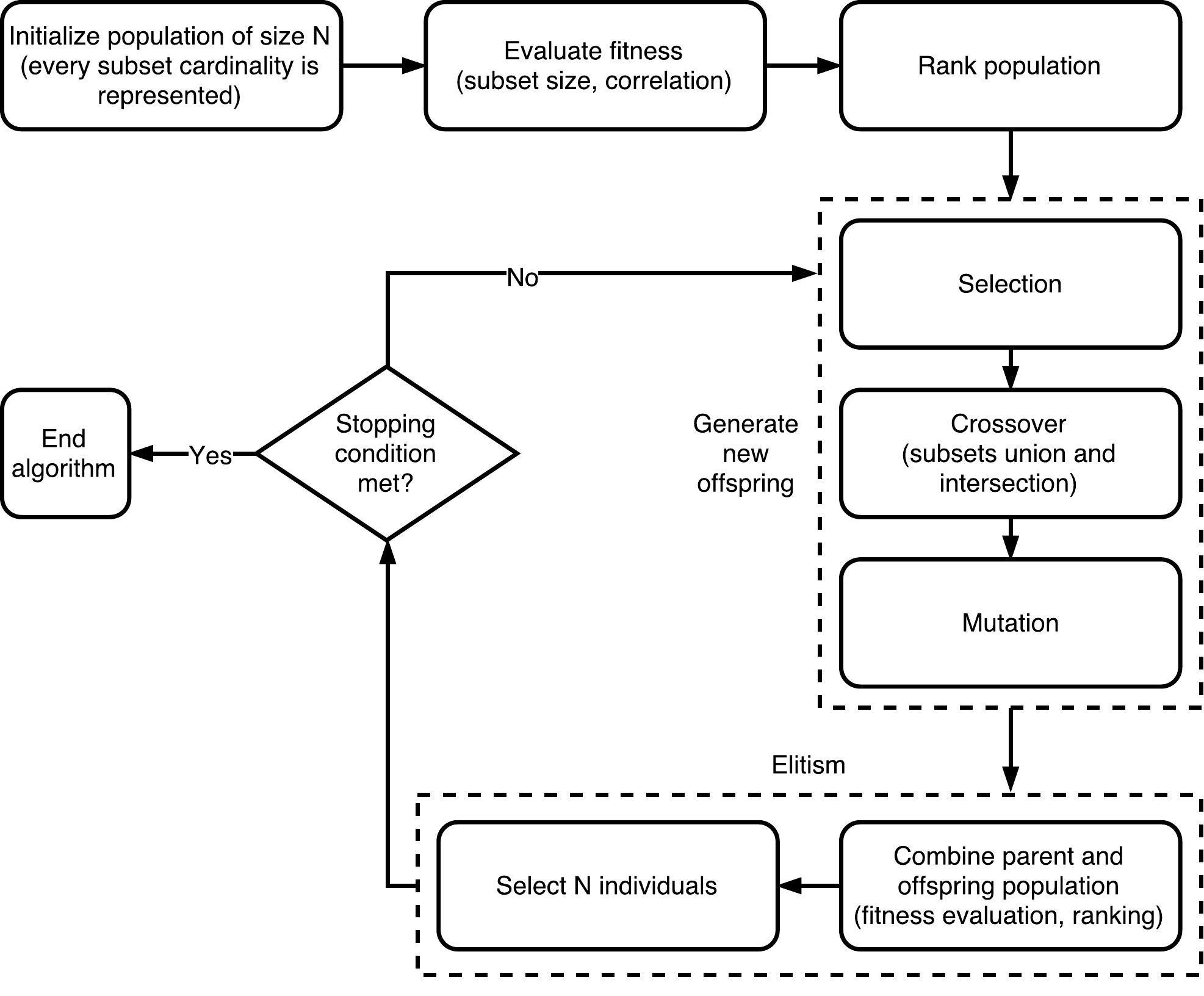}&
    \includegraphics[width=.33\linewidth, valign=c]{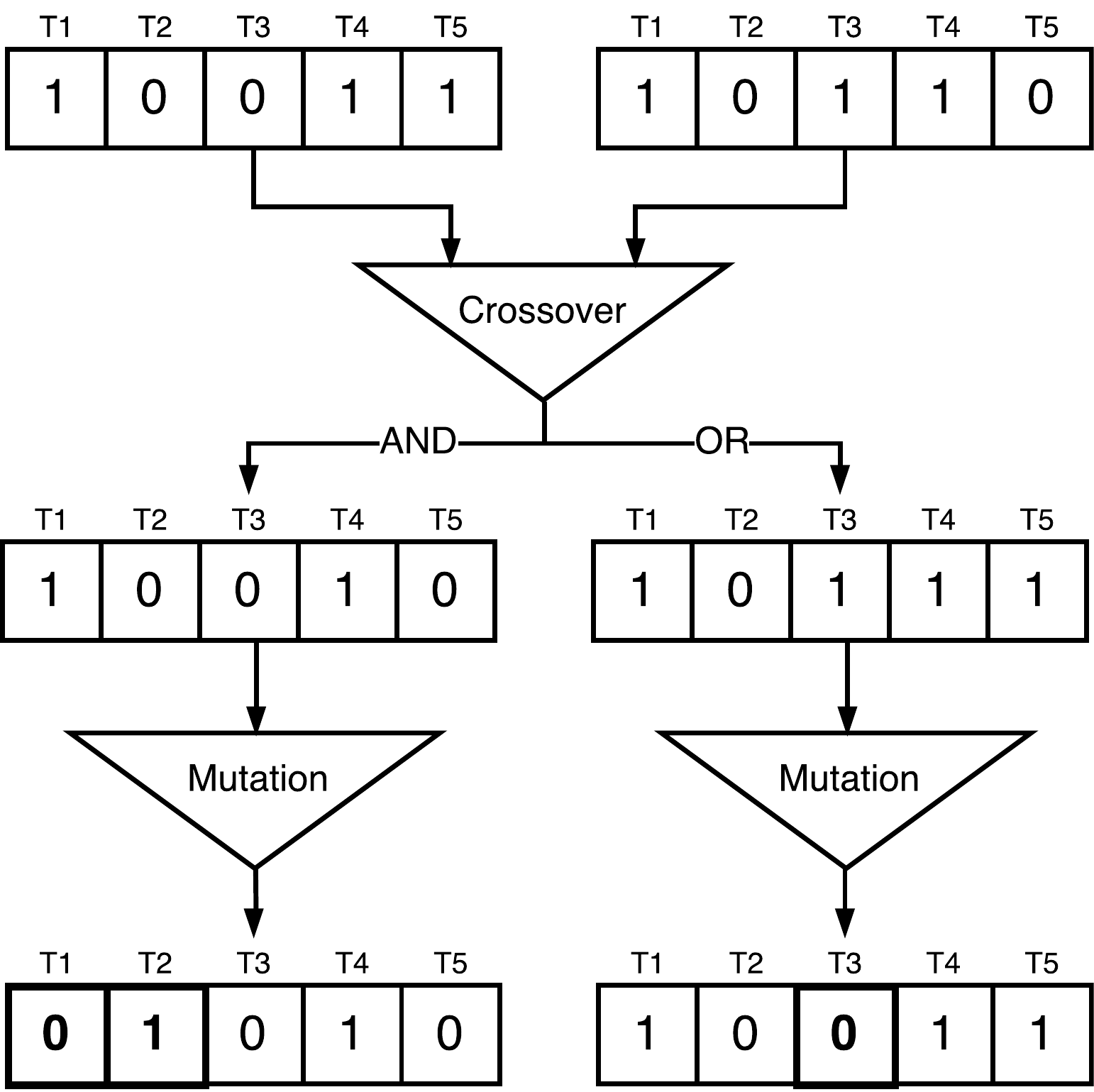}
    \\
 (a)&(b)
  \end{tabular}
  \caption{
  Overview of the algorithm NSGA-II applied to the few topics selection problem (left), and detail of the crossover and mutation process (right); the bold squares represent the mutated topics.
  }
  \label{JDIQGenetic:fig:evo:scheme}
\end{figure}

An overview of the operation of the NSGA-II Evolutionary Algorithm, applied to the topic selection problem, is presented in Figure~\ref{JDIQGenetic:fig:evo:scheme}; further details are provided in the following.

\subsection{Initial Population} 

Each individual of the population is represented by a binary array, having length equal to the number of topics in the full topic set ($n$). The $i$-th cell of the array tracks whether the corresponding topic is included in the encoded solution or not. Note that in the population, topic subsets of the different cardinalities $1 \leq c \leq n$ co-exist.

As for the size of the initial population, the minimum allowed value is equal to the number of topics, to ensure that every subset cardinality is initially represented; also, to guarantee population heterogeneity, we simply randomly generate a set of individuals for each subset cardinality.

\subsection{Operators}

The crossover operator merges information from two (or more) randomly selected parents, generating one or more offspring. The underlying idea is that of combining the features of two different but desirable individuals. In our implementation, given two parent solutions, two children are generated by simply performing a pairwise logic AND and a pairwise logic OR of the two corresponding binary arrays. Although many other options for the operators are possible \cite{Eiben:2003:IEC:954563}, we chose to rely on those two since they have an intuitive, clear meaning for the problem at hand: one child will contain only the topics which are in common between the two parents, and the other one will contain the union of the topics of the two parents.
A graphical representation of the crossover operator is shown in Figure~\ref{JDIQGenetic:fig:evo:scheme}(b).

A proper mutation operator should not try to improve a solution on purpose, since this would bias the evolution process of the population. Rather, it should cause random, unbiased changes  in the solution components.
In our implementation, mutation is performed by scanning the array left to right and deciding, for each cell, whether to flip its value or not according to a given probability. That is, for every topic in the vector representing the solution, we randomly generate a number between zero and 1: if the draw number is less than the given probability we perform the mutation (in doing so, if the topic is included in the solution then we remove it, and vice versa).
This is a classic strategy, which has been described, for example, in \cite{Eiben:2003:IEC:954563}.
A graphical representation of the mutation operator is shown in Figure~\ref{JDIQGenetic:fig:evo:scheme}(b).

For the selection of the parent candidates, i.e., the sets of topics to be used for the application of crossover and mutation operators, we rely on the classic strategy implemented in NSGA-II, based on the concepts of ranking and crowding distance: the entire population is sorted into fronts, according to non-domination. A first front is made by the individuals which are non-dominated. A second one is composed of the individuals which are dominated by the elements in the first front only, and so on for the remaining fronts. Then, pairs
of individuals are randomly selected from the population; finally, a \emph{Binary Tournament} selection is carried out for each  pair, considering a better-function based on the concepts of \emph{rank} (which considers the front which the instance belongs to) and \emph{crowding distance} (intuitively, it measures how close an individual is to its neighbors). For further details see \cite{Deb:2002:FEM:2221359.2221582}.

Observe that, after the offspring generation phase, the population has doubled in size. In order to select the individuals to pass to the next generation, the entire population is sorted again based on non-domination, and the best ones, according to the same function as before, are selected (elitist criterion).

\subsection{Fitness Function}
We use two fitness functions in order to optimize two antithetical objectives: the cardinality of the topic subset, and its correlation value with respect to the full topic set. In particular, we investigate two instances of the problem, forcing the constraint to find a best/worst topic set for each cardinality:
\begin{itemize}
\item  Finding the \emph{Best} topic set: the cardinality has to be minimized, while the correlation has to be maximized.
\item Finding the \emph{Worst} topic set: the cardinality has to be maximized, while the correlation has to be minimized.
\end{itemize}

\subsection{Choice of the Final Solution}
Given the aim of the work, we are not interested in finding a single solution, but rather we collect the best (or worst, depending on the instance of the problem to be solved) solution for each possible topic subset cardinality. In order to do that, we store the Pareto-front of the initial population. Then, at the end of each iteration of the  algorithm, we merge the current population with the stored ones, keeping only the non-dominated solutions (if two solutions for the same cardinality have also the same correlation value, then one of the two is randomly taken).
We choose to consider the best 10 subsets for each cardinality, but a deeper analysis is possible.

\subsection{Implementation}

We implemented our algorithm named NewBestSub extending the jMetal framework\footnote{\url{https://github.com/jMetal/jMetal}.} (version 5.0); jMetal is an Object Oriented framework based on Java, used for multi-objective optimization problems through meta-heuristics.
More in detail, we extended the NSGA-II algorithm.

Concerning the parameters used for the collections (which mostly depend on the number of topics of the collection), we use:
2,000 as \emph{population number}, i.e., the number of individuals in the population which, in our case, correspond to individual topic sets;
10 million as \emph{max evals}, i.e., an upper bound on the number of evaluations carried out during the computation, and used as a stopping condition for the algorithm;
0.3 as \emph{mutation probability} and 0.7 as \emph{crossover probability}, which represent respectively the probability of applying the mutation and crossover operators to the selected individuals (i.e., the individual topic sets);
finally, we choose 5,000 as \emph{average repetitions}.

\begin{sloppypar}
We implemented the software using the Kotlin\footnote{\url{https://kotlinlang.org/}.} programming language, a multi-platform programming language 
developed by JetBrains, 
fully interoperable with Java.\footnote{See \url{https://kotlinlang.org/docs/reference/} and \url{https://kotlinlang.org/docs/reference/comparison-to-java.html}.}
The software is about 2,000 lines of code (plus comments); the full project code is available at \url{https://github.com/Miccighel/newbestsub}.
\end{sloppypar}

\subsection{Discussion} \label{JDIQGenetic:sect:discussion3}
As we will discuss later in more detail, compared to BestSub, our implementation is capable of achieving higher/lower correlation values; to do so, several runs of the evolutionary algorithm are carried out starting from different initial populations. The final result is obtained by merging the several intermediate outcomes; this is in fact a commonly used (and trivial) practice to improve the results of EAs algorithms, and simultaneously to avoid overfitting.  
We run some experiments with 10 executions on various datasets: when comparing the correlation values obtained by running NewBestSub 10 times with the ones obtained by a single run of BestSub we found a small but not significant improvement in the correlation values of the Best/Worst topic in datasets with 50 topics, while major improvements (i.e., higher/lower correlation values for the Best/Worst subset) are observed for datasets with more than 50 topics.

The software presents also some limitations:
in the current version it is not granted that, for each cardinality $c$, we can obtain the Best/Worst $x$ (let us say 10) sets.
However, experimentally we found that we obtain at least 10 solutions for each cardinality when using a ground truth of 50 topics, and at least 10 solutions for most of the cardinalities using a larger ground truth (i.e., 1000 topics);
we leave for future work to analyze the relation between the number of Best/Worst solutions we can obtain for a given cardinality and the selected algorithm parameters, as well as to study an alternative approach able to avoid this problem completely.

We remark that we decided to use a state-of-the-art setting, without any fine tuning of the parameters. By doing so, we aim to (i) avoid overfitting, (ii) keep the implementation simple, and (iii) obtain a fast algorithm,  rather than focusing on finding the absolute Best and Worst topic subsets, which are extreme results per-se. The latter remark is also motivated by the result by \citet{Berto:2013:UFT:2499178.2499184}, that  shows that  a high number of good topic sets exist; in detail, \cite[Figure 3]{Berto:2013:UFT:2499178.2499184} shows that for the TREC96 collection (our AH99\_top96 dataset, see below), at a cardinality of half of the full topic set (i.e., 25 topics out of a ground truth of 50) more than 50\% of the topics are ``good'' (i.e., subsets with $\tau > 0.85$ and $\rho > 0.96$), and 99\% of the topics are good  after cardinality 35. 

\section{Experiments} \label{JDIQGenetic:sect:experiments}

\subsection{Aims, Motivations}

With the novel implementation of NewBestSub, we now turn to:
(i) reproduce previous work, to see if the past results hold;
(ii) compare the efficiency of NewBestSub with BestSub; and
(iii) generalize and extend some results, performing some novel experiments that would not be feasible with the old BestSub.

\subsection{Data}
In our experiments we use the following nine datasets derived from TREC and summarized in Table~\ref{JDIQGenetic:tab:coll}:
\begin{enumerate}
\item AH99: the dataset obtained using the full TREC-8 Ad Hoc collection, with all the runs.
\item AH99\_top96: the dataset obtained  selecting from AH99 only the top96 runs, i.e., circa the top 75\%  of the most effective systems (this is the choice done by \citet{Guiver:2009:FGT:1629096.1629099}).
\item AH99\_logAP: the dataset obtained using the logAP metric (this is equivalent to using the logarithm of AP values in Table~\ref{JDIQGenetic:tab:AP}).
\item AH99\_logAP\_top96: the dataset obtained using the logAP metric on the top runs, to further study the effect of considering the top 75\% of the most effective systems.
\item AH99@20: the dataset obtained using a shallow pool (AP@20 is used in place of AP: values are computed considering the first 20 retrieved documents only).
\item  AH99@20\_top96: the dataset obtained combining shallow pool and the top 75\% of the most effective systems.
\item WEB14:  the dataset obtained from the TREC Web Track of 2014. This allows us to  include in our analysis a more recent collection, and to compare  it with TREC-8. This seems important to us since previous results were obtained on rather old collections. 
\item WEB14\_top25: the dataset obtained  when selecting the circa top 75\% of the most effective systems for WEB14.

\item WEB14B: WEB14 also allows us to use the official NDCG metric as well as two binarized AP metrics: 
in the former we consider as not relevant the qrels values -2 and 0, and as relevant the qrels values 1, 2, and 3;
in the latter we consider as not relevant the qrels values -2, 0, and 1, and as relevant the qrels values  2 and 3;
in the following we report results for the former binarization since it leads to better results.
\end{enumerate}

\begin{table}[tb]
  \centering
\begin{adjustbox}{max width=\textwidth}
  \begin{tabular}{llll rr}
  \toprule
    & \textbf{Acronym} & \textbf{TREC Official Name}  &\textbf{Year} & \textbf{Topics} & \textbf{Runs} \\
    \midrule
	1.&  AH99 &	Ad Hoc &	1999 & 50 & 	129 \\
   	2.&  AH99\_top96 & 	Ad Hoc &	1999 & 50 & 	96 \\
    3.&  AH99\_logAP & 	Ad Hoc &	1999 & 50 & 	129 \\
    4.&  AH99\_logAP\_top96 & 	Ad Hoc &	1999 & 50 & 	96 \\
   	5.&  AH99@20 & 	Ad Hoc & 	1999 & 50 & 	129 \\
    6.&  AH99@20\_top96 & 	Ad Hoc & 	1999 & 50 & 	96 \\
    7.& WEB14 & 	Web Track (ad Hoc Task) & 	2014 & 50 & 	30 \\
    8.&  WEB14\_top25 & 	Web Track (ad Hoc Task) & 	2014 & 50 & 	25 \\
    9.&  WEB14B & 	Web Track (ad Hoc Task) & 	2014 & 50 & 	30 \\
 \bottomrule
  \end{tabular}
  \end{adjustbox}
   \caption{The test collections used in the experiments}
  \label{JDIQGenetic:tab:coll}
\end{table}

We decided to focus our analysis on the effects of different evaluation metrics, pool depth, number of systems, and collections; to do so, we choose all our datasets to have a fixed number of 50 topics.
We leave as future work the study of the effects of varying the number of topics.

\subsection{Reproduce Previous Work}\label{JDIQGenetic:sec:repro}
\begin{figure}[tbp]
  \centering
  \begin{tabular}{cc}
    \includegraphics[width=.45\linewidth]{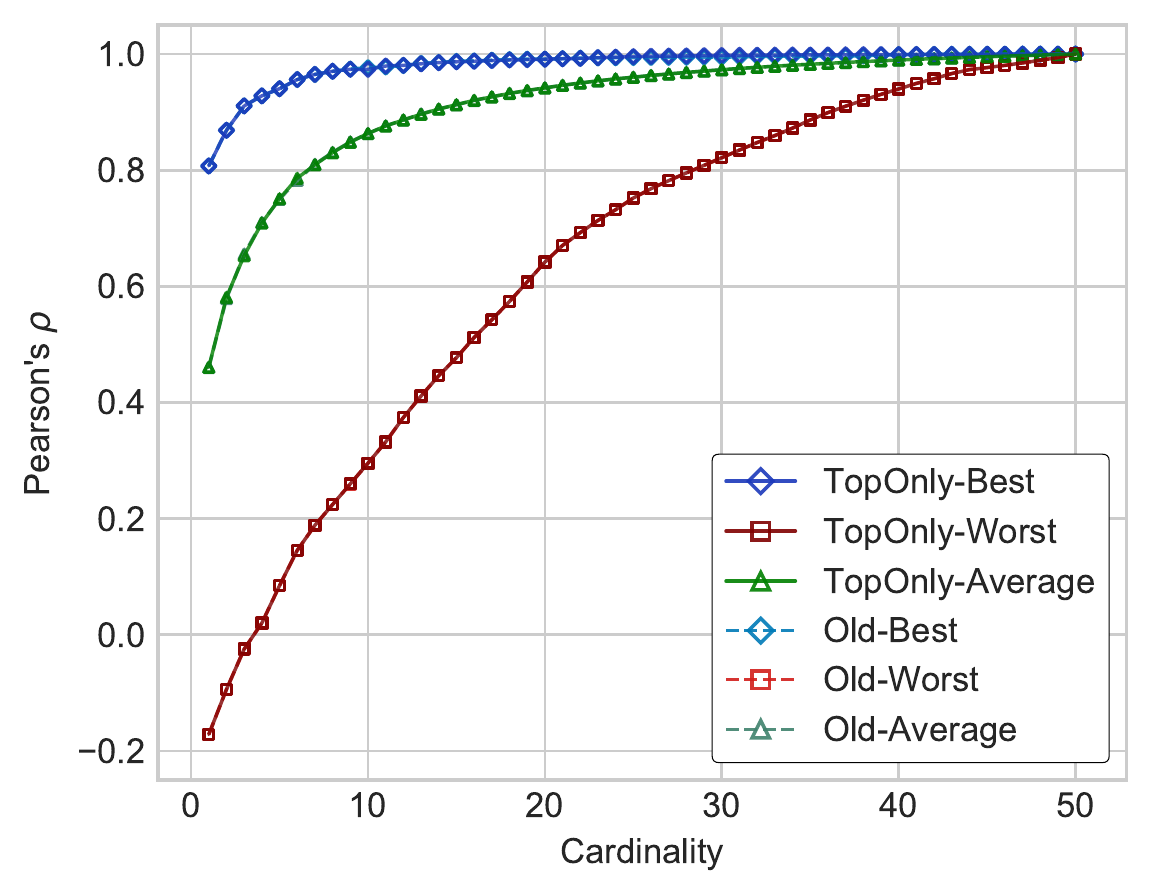}&
    \includegraphics[width=.45\linewidth]{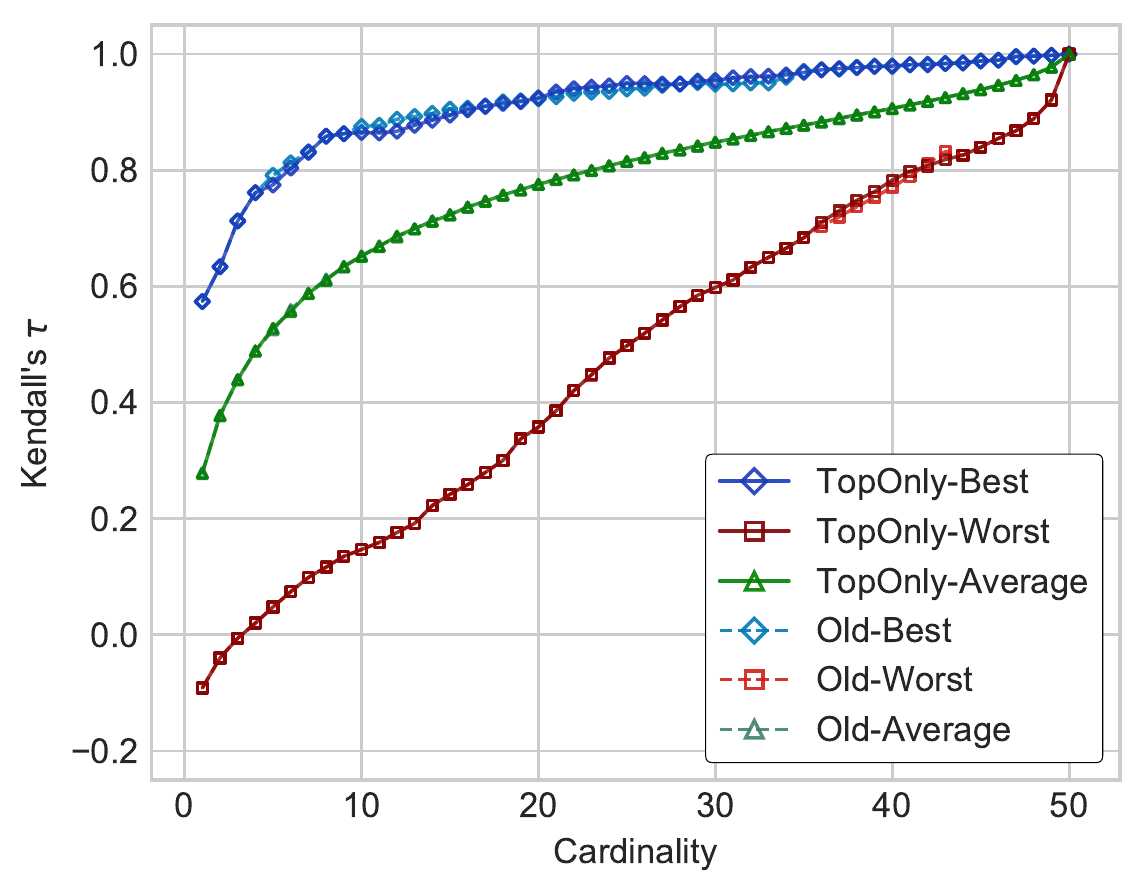}\\
    (a)&(b)
    \end{tabular}
  \caption{Correlation curves, Pearson's $\rho$ (a) and Kendall's $\tau$ (b) for BestSub and NewBestSub on AH99\_top96. For most of the cardinalities the series are indistinguishable.
}
  \label{JDIQGenetic:fig:repro}
\end{figure}

The first experiment is to reproduce the same results as \citet{Guiver:2009:FGT:1629096.1629099}.
Figure~\ref{JDIQGenetic:fig:repro} shows the comparison between BestSub and NewBestSub  for the Best, Worst, and Average series.
NewBestSub obtains almost the same results  (i.e., correlation values) as BestSub. For some cardinalities, NewBestSub provides slightly different correlation values from the original BestSub: for example see, in the $\tau$ chart, cardinalities around 12 and 31 for the Best series and cardinalities around 40 for the Worst series.
The Average series appears to be stable with  $10,000$ repetitions, and overlaps perfectly with the BestSub Average series.

\subsection{Efficiency}

\begin{table}[tb]
  \centering
  \begin{small}
  \begin{tabular}{cc}
\begin{tabular}{rrr}
  \toprule
    \textbf{Topics}  &\textbf{Runs} &  \multicolumn{1}{c}{\textbf{Time}} \\
    && \multicolumn{1}{c}{\textbf{NewBestSub}} \\
    \midrule
    50 & 5  & 2 min, 00 sec\\ 
    50 & 10 & 2 min, 04 sec\\ 
    50 & 25 & 2 min, 08 sec\\ 
    50 & 40 & 2 min, 15 sec\\ 
    50 & 50 & 2 min, 18 sec\\ 
    50 & 75 & 2 min, 30 sec\\ 
    50 & 90 & 2 min, 38 sec\\ 
    50 & 96 & 2 min, 41 sec\\ 
 \bottomrule
 \\
  \end{tabular}
\\ (a) \\ \vspace*{1em} 
\begin{threeparttable}
  \begin{tabular}{rrr c rrr}
  \\
  \toprule
    \textbf{Topics}  &\textbf{Runs} & \multicolumn{1}{c}{\textbf{Time}} &&\textbf{$k$} & \multicolumn{1}{c}{\textbf{Time}}  &  \textbf{Speedup}   \\
    && \multicolumn{1}{c}{\textbf{NewBestSub}} && & \multicolumn{1}{c}{\textbf{BestSub}}  &   \\

    \midrule
    50&  96   & 3 min&& 2 & 30 min     					&   10x \\
    50&  96   & 3 min&& 3 & 12 hours  				 	&   240x \\
    250& 96   & 10 min&& 2 & 1 month   			  &   4380x  \\
    250& 96    & 10 min&& 3 & $>$4 months\tnote{*}  &   $>$ 17,520x \\
    500& 96   & 20 min&& 2 & $>$1 year\tnote{*}    &   $>$ 26,280x \\
    750& 96   & 35 min&& 2 & $>$1 year\tnote{*}    &   $>$ 15,017x\\
    1000& 96  & 60 min&& 2 & $\gg$1 year\tnote{*}    &   $\gg$ 8760x\\
    1100& 96  & 80 min&& 2 & $\gg$1 year\tnote{*}    &   $\gg$ 6570x\\
 \bottomrule
  \end{tabular}
  \begin{tablenotes}
  \item[*] The execution was stopped before terminating.
  \end{tablenotes}
  \end{threeparttable}
  \\
  (b)
  \end{tabular}
  \end{small}
    \caption{Time comparison between BestSub and NewBestSub, on varying the number of runs (a) and the number of topics (b)
  }
    \label{JDIQGenetic:tab:efficiency}
\end{table}


Having shown that we are able to reproduce previous results, we now focus on the efficiency of NewBestSub.
To test efficiency we run two kinds of experiments:
(i) we keep constant the number of topics (we use 50 topics, as in AH99\_top96) and we vary the number of systems / runs, from 5 to 96 (the number of runs in AH99\_top96); and (ii) we keep constant the number of systems (we use 96 runs, as in AH99) and we vary the number of topics, from 50 to 1100 (the latter being an approximation of the number of topics in the largest collection available, i.e., the Million Query 2007 collection \cite{MQ2007}).

The effect of the number of runs can be seen by analyzing the results of experiment (i), in Table~\ref{JDIQGenetic:tab:efficiency}(a).
Of course, the more the number of runs $m$ increases, the longer will be the vectors used to compute the correlation values; in NewBestSub, as in BestSub, the complexity is $O(m)$ for $\rho$ and $O(m\log{m})$  for $\tau$. 
However, in practice  the effect is rather small. The time complexity appears to be more than linear, but with a quite small growth.

The number of topics affects computation time to a much greater extent, as it can be seen in Table~\ref{JDIQGenetic:tab:efficiency}(b). 
The efficiency of NewBestSub is mainly influenced by the number of topics: the time complexity appears to be slightly more than linear, but with a much higher growth.

The rightmost  three columns in Table~\ref{JDIQGenetic:tab:efficiency}(b) compare the efficiency of NewBestSub with the old BestSub. Speedup is  defined as
\begin{equation*}
\mathrm{Speedup} = \frac{\mathrm{Time}(\mathrm{BestSub})}{\mathrm{Time}(\mathrm{NewBestSub})}.
\end{equation*}
The speedup obtained by NewBestSub, even when the old BestSub is used with a smaller $k$ for a faster heuristics, is clear and very large. Even though the  computation time is  drastically reduced, the results are not affected: the correlation values obtained are similar when the comparison between the two softwares is possible, i.e., up to 250 topics.

Even better, NewBestSub is much more effective also when focusing on extreme value results (which might not be necessary, as detailed in Section~\ref{JDIQGenetic:sect:discussion3}): we run  10 executions of NewBestSub in order to maximize/minimize the correlation of Best/Worst subset as much as possible; we use 96 runs and 250 topics (the third row in  Table~\ref{JDIQGenetic:tab:efficiency}(b)). 
The total time for NewBestSub is 10 times 10 minutes = 100 minutes. Correlation appears to be much lower for the Worst set (i.e., with a $\tau$ correlation difference up to 0.2) when compared with results of BestSub, with $k=2$ as heuristic parameter; we also run BestSub with $k=3$ only for correlation up to cardinality 50 using a ground truth of 250 (results obtained in a 1.5 months circa); results of NewBestSub with 10 executions are still significantly more optimal, especially  for the  Worst series. 

The much higher efficiency allows us to conveniently perform many novel experiments. In the remaining part of this chapter  we discuss some of them.

\subsection{Generalize the Results}

In this section we generalize previous work, considering 
the inclusion of the all runs or just the most effective one (Section~\ref{JDIQGenetic:subsect:top_nontop}),
the stability of the Best/worst sets (Section~\ref{JDIQGenetic:subsect:stability}),
and  of the top 10 Best/Worst sets (Section~\ref{JDIQGenetic:subsect:stability_top10}),
as well as a more recent collection,
new metrics, and shallow pool effects (Section~\ref{JDIQGenetic:subsect:metric_pool}).

\subsubsection{Top-only vs.\ All} \label{JDIQGenetic:subsect:top_nontop}
\begin{figure}[tbp]
  \centering
  \begin{tabular}{cc}
    \includegraphics[width=.45\linewidth]{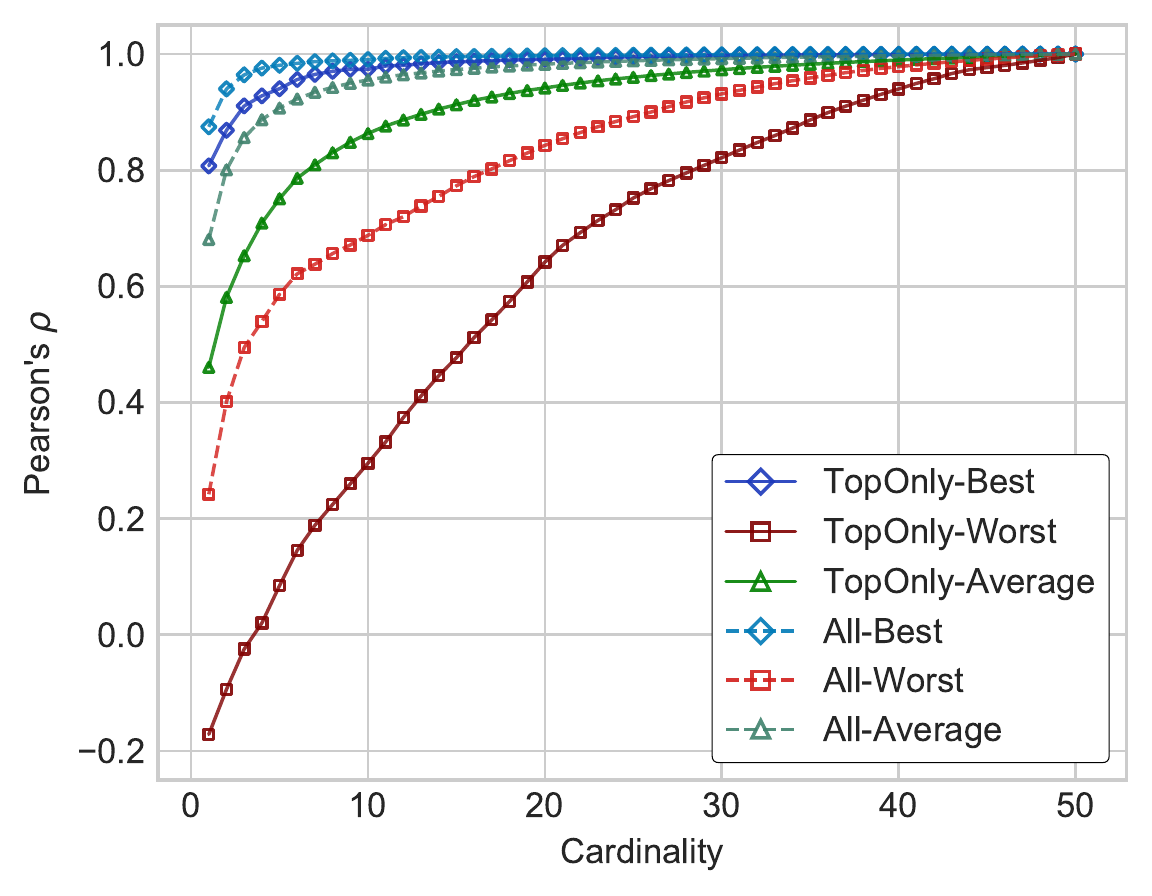}&
    \includegraphics[width=.45\linewidth]{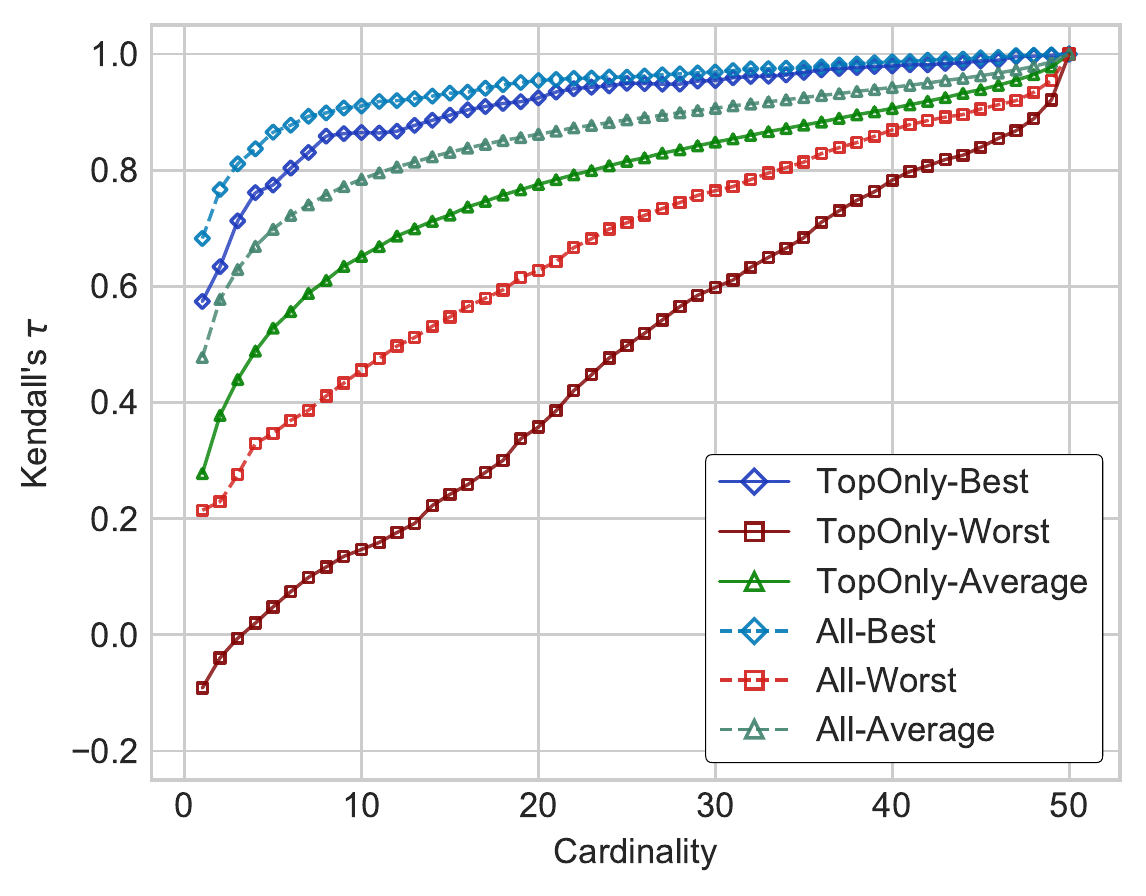}\\
    (a)&(b)
    \end{tabular}
  \caption{Comparison of Correlation curves when including top 75\% or all runs, for Pearson's $\rho$ (a) and Kendall's $\tau$ (b) correlation measures.
  }
  \label{JDIQGenetic:fig:top-all}
\end{figure}
One issue that, in general, hinders reproducibility  is not using a whole dataset \cite{Ferro:2017:RCI:3035914.3020206,ferro2016increasing}. In our case this corresponds to not using the whole set of runs of a TREC track. This  is a common practice in TREC data analysis \cite{VoorheesBuckley02}, and it is usually justified by the need of removing buggy and not informative systems from the analysis.
Figure~\ref{JDIQGenetic:fig:top-all} compares Correlation curves (both $\rho$ and $\tau$) obtained when using  the whole AH99 dataset, to those obtained when including only the top 75\% best runs, as it is done in the original AH99\_top96 dataset.
Considering only the top  75\% of all systems leads to rather different results: the Best (and, especially, the Worst) series appear to achieve  much higher (respectively lower) correlation values. For example, when considering $\tau$ correlation and the Worst series, a correlation  of $0.4$ is achieved with 8 topics in AH99\_top96, while 21 topics are needed in the case of considering the full AH99 dataset.

\subsubsection{Stability of the Best and Worst Sets} \label{JDIQGenetic:subsect:stability}
%
The heuristic adopted in BestSub might seriously affect the stability results. Conversely, also NewBestSub is based on a, yet more sophisticate, heuristic, and its effect on the stability of the Best/Worst sets needs to be studied.

\begin{figure}[tbp]
  \centering
  \begin{tabular}{cc}
    \includegraphics[width=.45\linewidth, valign=c]{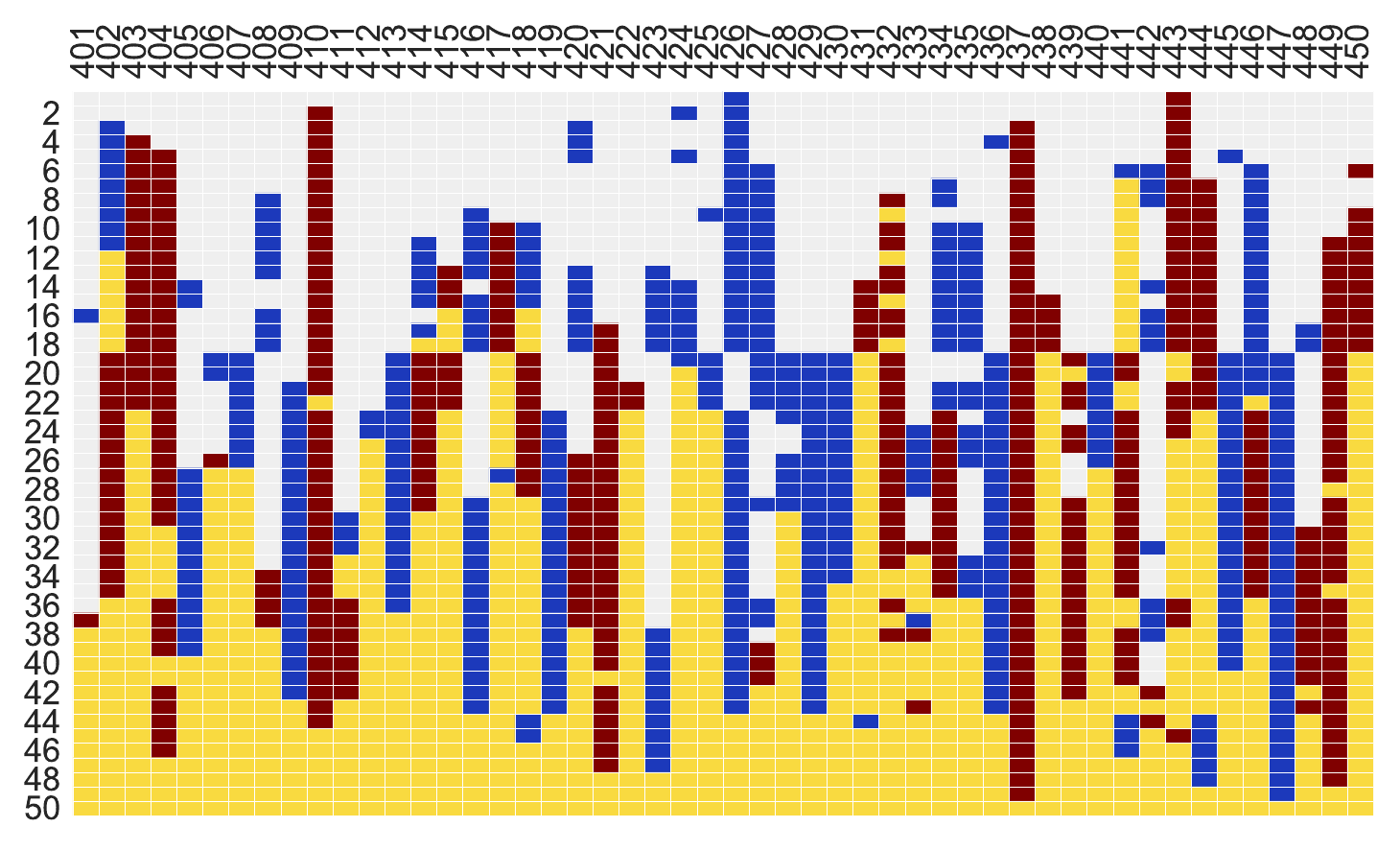}&
    \includegraphics[width=.45\linewidth, valign=c]{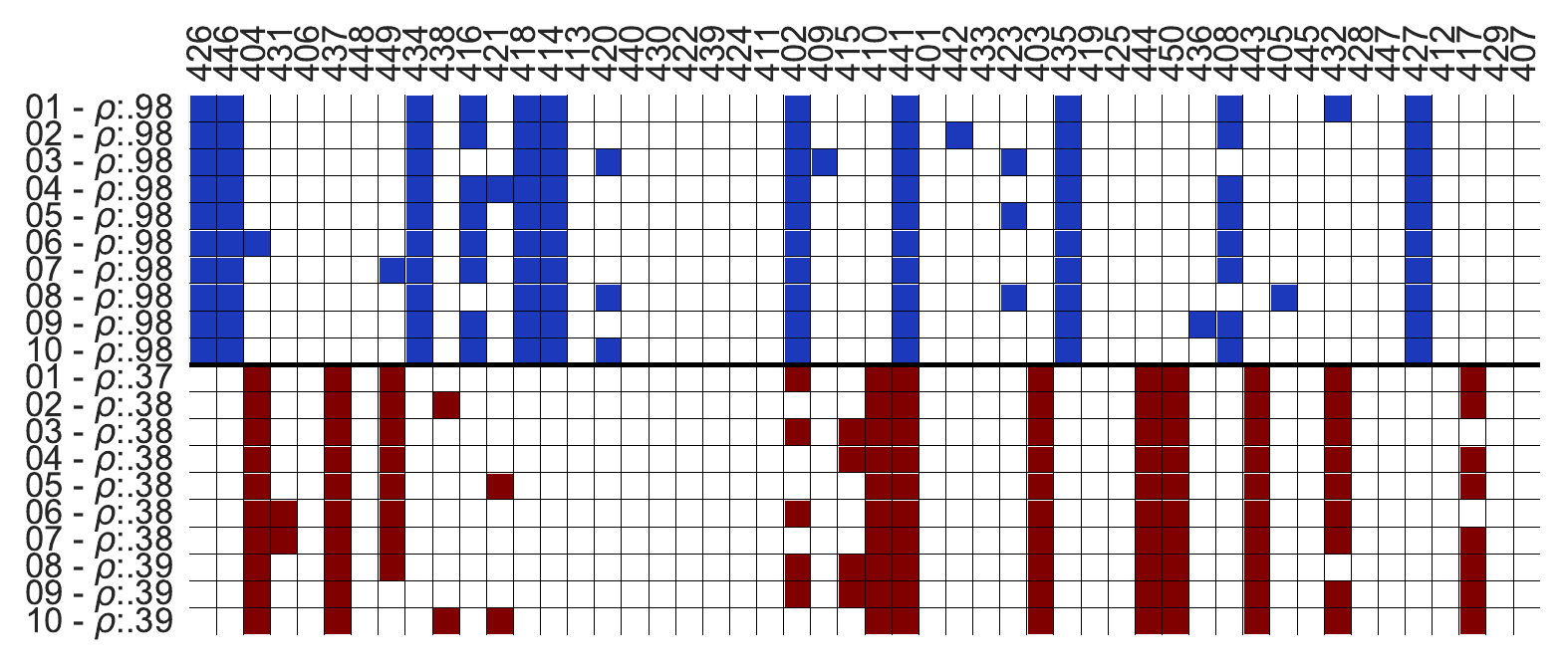}\\
    (a)&(b)
    \end{tabular}
  \caption{
  Stability pixel-maps for Pearson's $\rho$ for the whole AH99\_top96 dataset (a) and for Pearsons's $\rho$ for the top 10 sets at cardinality 12 (b).
  The color blue is used to represent topics in the Best set (differently from Figure~\ref{JDIQGenetic:fig:guiver_pixel}), red for the topics in the Worst set, and yellow for the topics in the Best and Worst set.}
  \label{JDIQGenetic:fig:pixelmaps}
\end{figure}

Figure~\ref{JDIQGenetic:fig:pixelmaps} shows two pixel-maps created on NewBestSub output. When comparing them to the old BestSub pixel-maps in Figure~\ref{JDIQGenetic:fig:guiver_pixel}, it might appear that the stability of the Best topic subset, and of the 10 Best topics subsets, is higher with NewBestSub (Worst topic subsets seems instead similar). To quantify this sensation, we use the  measure of stability of the Best/Worst set defined as follows by \citet[page~21:14]{Guiver:2009:FGT:1629096.1629099}:
\blockquote{
For each topic in a set of a given size, a counter is incremented if the topic is in the set next size up. For the exhaustive search, this counter has a minimum and maximum possible value, where the minimum is greater than 0 due to the fact that as topic sets get larger there is some inevitable overlap from one topic set to the next. The stability value (expressed as a percentage) is given by the ratio
\begin{equation*}
\frac{\mbox{counter actual} - \mbox{counter min}}{\mbox{counter max} - \mbox{counter min}}.
\end{equation*}
For worst subsets the average stability for Average Precision across the three goodness measures is 93\%. Best subsets are somewhat less consistent, with an average of 86\%.
}
Note that \citet{Guiver:2009:FGT:1629096.1629099} do not specify how to compute the counter max/min values. We derive that they can be computed as follows (where $n$ is the number of topics, as usual): 
\begin{align*}
\mbox{counter min} &= 
\begin{cases}
    0, & \mbox{if } c \leq \floor*{\frac{n}{2}}-1 \\
    2c+1-n,  & \mbox{otherwise}
\end{cases}
\\
\mbox{counter max} &= c.
\end{align*}

\begin{table}[tb]
\centering
\begin{adjustbox}{max width=\textwidth}
\begin{tabular}{ll c cc c cc}
\toprule
                         &&& \multicolumn{2}{c}{\textbf{Pearson's $\rho$}}       && \multicolumn{2}{c}{\textbf{Kendall's $\tau$}}     \\
\cmidrule{4-5} \cmidrule{7-8}
             &&& Best set & Worst set  && Best set & Worst set         \\
\midrule
0.& BestSub on AH99\_top96  && .88 & .98 && .84 & .95   \\
\midrule
1.& AH99                    && .85 & .97 && .86 & .97   \\
2.& AH99\_top96             && .88 & .98 && .88 & .95   \\
3.& AH99\_logAP             && .83 & .96 && .85 & .92   \\
4.& AH99\_logAP\_top96      && .89 & .95 && .82 & .93   \\
5.& AH99@20                 && .83 & .96 && .85 & .92   \\
6.& AH99@20\_top96          && .89 & .94 && .89 & .90   \\
7.& Web14                   && .88 & .92 && .84 & .89   \\
8.& Web14\_top25            && .87 & .95 && .82 & .88   \\
9.& Web14B                  && .85 & .97 && .61 & .91   \\
\addlinespace
&Average  && .86 & .95 && .82 & .92 \\  
\bottomrule
\end{tabular}
\end{adjustbox}
\caption{Stability values for the Best and Worst sets using \citet{Guiver:2009:FGT:1629096.1629099} measure, for $\rho$ and $\tau$ correlation}
\label{JDIQGenetic:tab:stability}
\end{table}

Table~\ref{JDIQGenetic:tab:stability} shows the stability values for the Best/Worst set for all the dataset used, plus the values for BestSub on AH99\_top96.  When comparing BestSub and NewBestSub (i.e., rows 0.\ and 2.\ of the table),
differently from the first sensation from Figure~\ref{JDIQGenetic:fig:pixelmaps}, the stability values are almost identical.
This is also clear when comparing the averages of stability for NewBestSub across all the nine datasets (last row in the table) with the old BestSub.
Also, in general, correlation values are similar across all datasets,
with the stability values for the Worst set always higher than the value for the Best set; the only peculiar dataset is WEB14B, but the lower value for the $\tau$ Best set stability can be caused by the binarization process.\footnote{\label{JDIQGenetic:fn:WEB14}As well as by the particular nature of that dataset that is known to be rather incomplete due to shallow pools and low number of participants \cite{Lu:2016:EPE:2975219.2975241}.}
Therefore, these results confirm previous findings:
\begin{itemize}
\item  once a topic set enters in the Worst set at a certain cardinality, it tends to remain in the Worst set also for the consequent cardinalities;
\item the previous finding is less true for Best topics;
\item  a Worst topic set is formed by individual Worst topics, while Best topic sets are not necessarily formed by individual Best topics. In other words, a set formed by Worst individual topics in general is Worst, while this is not true for Best topics.
\end{itemize}
Thus, we can conclude that the stability results of \citet{Guiver:2009:FGT:1629096.1629099} still hold with the completely different heuristic that we used, and therefore it seems unlikely that they depend on the (quite rough) heuristic they used.

\subsubsection{Stability of the Top 10 Best and Worst Sets} \label{JDIQGenetic:subsect:stability_top10}
We turn now to study the stability of the top 10 Best/Worst sets, as done by \citet{Guiver:2009:FGT:1629096.1629099} and \citet{Berto:2013:UFT:2499178.2499184}.
We can generalize the stability measure  used in Section~\ref{JDIQGenetic:subsect:stability} to include the top $p$ sets (in this case $p=10$) at a given cardinality $c$.
Thus, as in Section~\ref{JDIQGenetic:subsect:stability}, we can compute a stability value counter actual as done by \citet{Guiver:2009:FGT:1629096.1629099}
where, in this case,  
\begin{align*}
\mbox{counter min} &= 
\begin{cases}
    (2c-n)(p-1), & \mbox{if } c > \floor*{\frac{n}{2}} \\
    0,  & \mbox{otherwise}
\end{cases}\\
\mbox{counter max} &= c(p-1).
\end{align*}
\begin{table}[tb]
\centering
\begin{adjustbox}{max width=\textwidth}
\begin{tabular}{ll c cccccc c cccccc}
\toprule
                         &&& \multicolumn{6}{c}{\textbf{Best 10 sets}}       && \multicolumn{6}{c}{\textbf{Worst 10 sets}}     \\
\cmidrule{4-16}
&\textbf{Collection}               && \multicolumn{6}{c}{cardinality}        && \multicolumn{6}{c}{cardinality}        \\
\cmidrule{4-9} \cmidrule{11-16}
                         &&& 5 & 10 & 20 & 30 & 40 &45 && 5  & 10 & 20 & 30 & 40 & 45 \\
\midrule
1.& AH99                    && .60 & .74 & .89 & .92 & .80 & .65 && .74 & .84 & .94 & .93 & .84 & .71 \\
2.& AH99\_top96             && .49 & .82 & .92 & .90 & .84 & .64 && .71 & .84 & .93 & .91 & .72 & .67 \\
3.& AH99\_logAP             && .58 & .78 & .88 & .88 & .85 & .73 && .73 & .83 & .91 & .92 & .88 & .73 \\
4.& AH99\_logAP\_top96      && .60 & .81 & .80 & .86 & .62 & .62 && .60 & .86 & .91 & .88 & .86 & .71 \\
5.& AH99@20                 && .54 & .74 & .90 & .83 & .62 & .62 && .67 & .82 & .90 & .92 & .78 & .56 \\
6.& AH99@20\_top96          && .56 & .77 & .94 & .89 & .70 & .56 && .67 & .82 & .92 & .89 & .71 & .51 \\
7.& WEB14                   && .67 & .72 & .91 & .61 & .53 & .22 && .73 & .83 & .92 & .93 & .86 & .58 \\
8.& WEB14\_top25            && .47 & .64 & .89 & .46 & .21 & .20 && .78 & .82 & .90 & .89 & .84 & .62 \\
9.& WEB14B                  && .29 & .20 & .47 & .43 & .25 & .18 && .69 & .85 & .75 & .89 & .82 & .60 \\
\addlinespace
& Average					&& .53 & .69 & .84 & .75 & .60 & .49 && .70 & .83 & .90 & .91 & .81 & .63 \\
\bottomrule
\end{tabular}
\end{adjustbox}
\caption{Stability values for the Best/Worst 10 sets for $\tau$ correlation
}
\label{JDIQGenetic:tab:stability_top10}
\end{table}


Table~\ref{JDIQGenetic:tab:stability_top10} shows the stability values for the Best/Worst 10 sets at some selected cardinality. We report the results for $\tau$ correlation values only; the outcome for $\rho$ correlation is similar.
As we can see reading Table~\ref{JDIQGenetic:tab:stability_top10} column-wise (i.e., for each cardinality),
results are quite similar across collections, with the exception of (again) WEB14 and, to a lesser extent, WEB14\_top25 (see again Footnote~\ref{JDIQGenetic:fn:WEB14}). 
As we can see reading the table
row-wise (i.e., for each collection),
the stability values are different at different cardinality values: the maximum stability value is reached around cardinality 20 and 30 for all the collections, whereas the value is minimal at cardinalities of 5 and 45. But the main remark is that, generalizing the result of Section~\ref{JDIQGenetic:subsect:stability}, it is clear that the 10 Worst sets are much more stable than the 10 Best ones.
Including the top 75\% of the runs or the whole datasets changes the stability values: for example, considering cardinality 5, the stability  of the Best 10 sets for AH99 is 0.60, while for AH99\_top96 is 0.49.

\subsubsection{A Recent Collection, other Evaluation Metrics, and a Shallow Pool} \label{JDIQGenetic:subsect:metric_pool} We use NewBestSub on a more recent collection with a new evaluation metric: the NDCG metric of the WEB14 collection.
We do not show the corresponding charts as the overall results look very similar to those of the other collections.
Some of the more specific results are quite similar as well. Including or not the top 75\% of the runs still makes a difference with WEB14 and NDCG, as results are comparable with Figure~\ref{JDIQGenetic:fig:top-all}; and
stability values for the Best/Worst single sets are similar  to the other collections.
On the contrary, stability values  for the Best/Worst 10 sets are lower for Web14 than for the other collections, especially at cardinalities 5 and 45.

Concerning using a shallow pool (i.e., AP@20), we see that when comparing AH99 with AH99@20, results are similar both for the stability of single and Best/Worst 10 sets.

As a final remark, we can state that the reproduced results (Section~\ref{JDIQGenetic:sec:repro}) seem quite stable also on the more recent collection, when using NDCG, and/or with a shallower pool, with very few exceptions.

\subsubsection{Larger Ground Truth} \label{JDIQGenetic:subsect:topic_size}
We now briefly discuss the effect of the size of the initial topic set on the Best and Worst series. 
To this aim, we design an experiment as follows: we make use of the Robust track from 2004, which includes 249 topic and 110 runs \cite{voorhees2004overview}. 
We simulate five larger ground truth of different size by randomly sampling $t \in \{ 50, 100, 150, 200, 249 \}$ topics. We then run NewBestSub on the selected topic set. 
To limit noise and give stability to the results, we repeat the process 20 times for each sampling. 
Finally, we compute, for each cardinality $c\in \{1,\ldots,t\}$, the average $\tau$ correlation over the 20 repetitions.  

\begin{figure}[tbp]
  \centering
  \begin{tabular}{cc}
    \includegraphics[width=.45\linewidth]{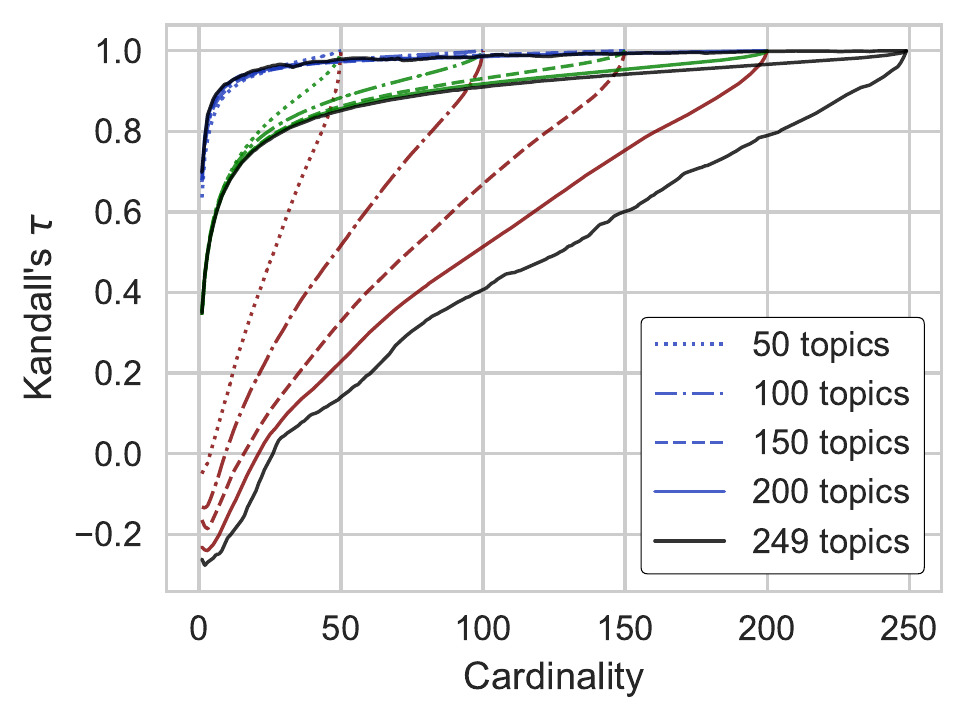} &
    \includegraphics[width=.45\linewidth]{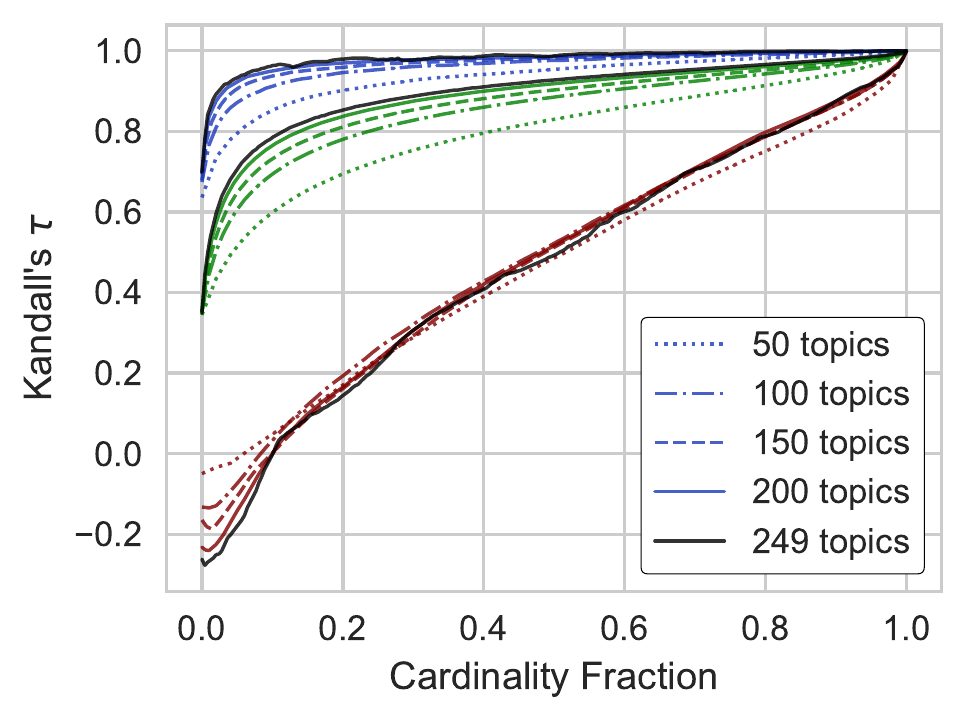}\\
    (a)&(b)
  \end{tabular}
  \caption{
  $\tau$ correlation series when starting from a topic population of various sizes taken from Robust 2004 dataset, as a function of the absolute cardinalities (a), and of the cardinality fractions (b).
  }
  \label{JDIQGenetic:fig:topic_size}
\end{figure}

Figure~\ref{JDIQGenetic:fig:topic_size} shows the five Best, five Worst, and five Average series for the different five initial topic set size. 
Besides the usual correlation chart with the cardinality on the horizontal axis (Figure~\ref{JDIQGenetic:fig:topic_size}(a)), we also show the  chart with the correlation as a function of the cardinality fraction of the full topic set (Figure~\ref{JDIQGenetic:fig:topic_size}(b)), which is perhaps more informative in this case.
From this second chart we see that the initial size of the ground truth has an impact on both the Best and Worst series; the larger the initial topic set, the higher (respectively, lower) is the correlation for the Best (respectively, Worst) series. The behavior of the average series is similar to that of the Best ones.

So far, the effects of using fewer topics have been studied on collections having a ground truth of 50 topics. These results hint that these effects still hold on larger collections, and they even become more extreme. However, it has to be noted that there are many variables that need to be taken into consideration, such as the topic set (i.e., different collections have different topics), the collection task and track, the participating runs, the effectiveness metric, etc. Therefore, we leave for future work a more systematic study of the effect of the initial ground truth size, which is now made possible by NewBestSub.

\section{Conclusions} \label{JDIQGenetic:sect:conlusions_future}

In this chapter we show that our approach based on evolutionary algorithms is able to both reproduce the main state of the art results and to allow us to analyze the effect of the collection, metric, and pool depth used for the evaluation.
Differently from previous studies which have been mainly theoretical, we are also able to discuss some practical topic selection strategies, integrating results of automatic evaluation approaches.

More in detail, our contribution is threefold:
(i) we re-implement BestSub using a novel approach based on evolutionary algorithms;
(ii) we successfully reproduce the results by \citet{Guiver:2009:FGT:1629096.1629099}, \citet{ecir11}, and \citet{Berto:2013:UFT:2499178.2499184}; and
(iii) we generalize such results to other metrics and collections.
%

NewBestSub, the novel implementation of BestSub, besides allowing us to obtain the previous results, also leaves plenty of space for future work.

The consistent speedup allows for many new interesting  research possibilities. For example, 
we aim at reproducing and extending the generalization experiments of \citet{Guiver:2009:FGT:1629096.1629099}, and we aim also at developing an effective topic selection strategy.
 
In the next chapter we will use the novel implementation of the NewBestSub algorithm to perform an extensive set of experiments on the effect of the reduction of the topic set size in retrieval evaluation.


%
\chapter{On Topic Subsets in Test Collections}\label{chapt:few:topicsubsets}

This chapter deals with the effect of using topic subsets for IR evaluation.
Section~\ref{FewTopics:sec:introduction} introduces and frames research questions,
Section~\ref{FewTopics:sec:experiments} describes the experimental setting. 
Section~\ref{FewTopics:sec:rq1:-larger-ground} discusses the results related to the first research question RQ\ref{FewTopics:RQ:1}, highlighting the existence of even more extreme results when the number of topics increases.
Section~\ref{FewTopics:sec:rq2:stat} focuses on RQ\ref{FewTopics:RQ:2} and addresses statistical significance, specifically discussing what kind of errors are more likely when using fewer topics. 
Section~\ref{FewTopics:sec:clustering} examines RQ\ref{FewTopics:RQ:3}, about clustering, and highlights how a rather natural approach turns out to be only slightly more effective than randomly chosen topics.
Section~\ref{FewTopics:sec:concl-future-work} summarizes the contribution of this chapter and sketches future developments.

\section{Introduction and Research Questions}
\label{FewTopics:sec:introduction}
In the previous chapters, we discussed notable results on topic subsets for retrieval evaluation.
There are a number of limitations with these past studies:
\begin{enumerate}
\item Researchers have examined relatively small ground truth topic
  sets: $n=50$~\citep{Guiver:2009:FGT:1629096.1629099,ecir11,ictir13} and $n=87$~\citep{MQ2009}. 
  However, little is known about the generality of these results for larger $n$.
  Because the existing studies sampled from topic sets that are
  relatively small, as the cardinality of the subset becomes a
  substantial fraction of the ground truth set, the properties of the
  sample and the full set are guaranteed to become similar and the
  correlations between the rankings of runs will tend to $1$. 
  The observation in Figure~\ref{FewTopics:fig:tois1} that a topic subset of
  cardinality $22$ has similar properties to the full set of $50$
  topics may not hold with a larger ground truth.
  This limitation is striking in the light of recent work
  by \citet{Sakai2016}, who showed that for test collections to have
  reasonable statistical power, ground truth topic sets
  size should be at least around 200, if not higher.
  Therefore the results obtained on the basis of a ground truth of far
  fewer than 100 topics calls for further confirmation on
  higher cardinalities.
\item A limitation of past work \citep{Guiver:2009:FGT:1629096.1629099,ecir11,ictir13} is that
  the statistical significance of the differences between runs was not
  taken into account: $\tau$ values do not explain if a
  different run ranking is due to minor fluctuations or to statistically significant differences in measurement
  values. 
  This is a notable omission, in the light of recent work
  from \citet{Sak14} that emphasizes the link between topic set size and
  statistical power.
\item Almost no characterization of the best topic sets has been
    attempted (apart some results on stability of such sets, see
    e.g. Figures~5 and~6 in \cite{Guiver:2009:FGT:1629096.1629099}). 
    However, it seems intuitive that smaller topic sets should be
    obtained by removing redundancy, for example by clustering topics
    and selecting representatives from each cluster.

\end{enumerate}

In this chapter we address three research questions:
\begin{itemize}
\item \label{FewTopics:RQ:1} \textbf{RQ1} What effect does a larger ground truth topic set have on correlation curves?
Are the results obtained in past work \citep{Guiver:2009:FGT:1629096.1629099,ecir11,ictir13} confirmed when using a larger ground truth?
How does the minimum cardinality of a topic subset, needed in order to achieve
a high correlation, depend on the cardinality of the ground truth, when using data from test collections?
\item \label{FewTopics:RQ:2} \textbf{RQ2} Are the results on topic subset size, obtained in past work \citep{Guiver:2009:FGT:1629096.1629099,ecir11,ictir13}, still valid when statistical significance is considered?
\item \label{FewTopics:RQ:3} \textbf{RQ3} Is clustering an effective strategy to potentially\footnote{Consistently with this line of research (see Footnote~\ref{FewTopics:fn:after}), we investigate clustering of topics using an \textit{a posteriori} setting; thus, we study an after-evaluation characterization of Best topic subsets, but do not aim at providing a methodology to find such subsets in practice.} find and characterize the best topic sets? 
Does the choice of a specific clustering setting (features, algorithms, distance functions, etc.) make important differences? 
If so, what clustering settings are most effective in finding topic sets featuring high correlations? 
\end{itemize}


\section{Experimental Setting and Data}\label{FewTopics:sec:experiments}

We describe the test collections, methods, and means of evaluation used in our experiments.

\subsection{Data and Collections}
\label{FewTopics:sec:data}

Our experiments require test collections with more than $50$ topics,
and for which a sufficient number of runs are available to be analyzed.
The three instantiations of the Million Query track collections feature more than $1,000$ topics
each year that are sampled from a query log. 
We use the data from the 2007 track.
However, the Million Query datasets are not free from disadvantages:
runs are evaluated using the statMAP and E[MAP] metrics, which are slightly
different from classical Mean AP (MAP),\footnote{The effect of
  statMAP, on which we focus in this chapter, is discussed in more detail in Section~\ref{FewTopics:sec:metrics}}.
In addition, not as many runs are available ($25$-$35$). 
We also employ the TREC 2004 Robust and 2006 TeraByte track collections, using automatic runs only.
To enable a comparison with the results obtained in previous studies
\citep{Guiver:2009:FGT:1629096.1629099,ecir11,ictir13}, we also use the TREC 8 ad hoc (AH) track
(1999).
Table~\ref{FewTopics:tab:coll} summarizes the four test collections.
For the analyses in this work, when not otherwise noted, we work on a subset of the runs. As is usual for the analysis of TREC run data (see e.g. \citet{VoorheesBuckley02}), we remove the least
effective runs, obtaining the number of runs in the last column.
For AH99 we removed the 25\% least
effective runs to have the same situation as in prior work
\citep{Guiver:2009:FGT:1629096.1629099,ecir11,ictir13}; for R04 we did the same; for TB06 and
especially MQ07, which feature a smaller number of runs, we removed
fewer (20\% and 10\%, respectively). The number removed was determined
by manually examining
the distribution of run effectiveness values, and pruning runs with a 
clear drop in effectiveness compared to others that are ranked higher.

\begin{table}[tb]
  \centering
  \caption{Test collections used for all experiments.}
  \begin{tabular}{lll rrr}
  \toprule
    \multicolumn{1}{l}{\textbf{Acronym}} & \multicolumn{1}{l}{\textbf{TREC}}  & \multicolumn{1}{l}{\textbf{Year}} & \multicolumn{1}{l}{\textbf{Topics}} & \multicolumn{1}{l}{\textbf{Total}} & \multicolumn{1}{l}{\textbf{Used}}\\
     &  \multicolumn{1}{l}{\textbf{Collection}}  &  & & \multicolumn{1}{l}{\textbf{Runs}} & \multicolumn{1}{l}{\textbf{Runs}}\\
    \midrule
 AH99 & Ad Hoc & 1999  & $50$ & $129$ & $96$\\
    R04&  Robust & 2004  & $249$& $110$&  $82$\\
   TB06 &  TeraByte & 2006& $149$ & $61$ & $49$\\
 MQ07 & Million Query & 2007 & $1153$ & $29$ & $26$ \\
 \bottomrule
  \end{tabular}
  \label{FewTopics:tab:coll}
\end{table}

\subsection{Software}
\label{FewTopics:sec:software}

For our analysis, we employed both the \textsf{BestSub} software that was
used in previous studies \citep{Guiver:2009:FGT:1629096.1629099,ecir11,ictir13}, and its genetic implementation \citep{Roitero:2018:RIE:3282439.3239573}, detailed in the previous Chapter.
The number of all subsets of a topic set of cardinality $n$ is $2^n$. The number of all possible topic subsets of cardinality $c$ drawn from the larger set is $\binom{n}{c} = \frac{n!}{c!(n-c)!}$. Therefore, the \textsf{BestSub} software uses a heuristic to cope with the combinatorial explosion. The heuristic builds the best set of cardinality $c+1$ on the basis of the best set of cardinality $c$ by looking at those subsets of cardinality $c+1$ that differ from the best set of cardinality $c$ by at most $k$ topics. In the previous studies, $k=3$.

Since in our case $n>50$ (i.e. $149$, $249$,
and $1,153$), the complexity is higher. This would mean that using \textsf{BestSub} was impractical, with months if not years of computation time required, even by resorting to lower $k$ values. We therefore used also its genetic counterpart \citep{Roitero:2018:RIE:3282439.3239573}. 
This change has no effect when tested on small
cardinalities: both versions of \textsf{BestSub} produce almost completely overlapping and graphically indistinguishable correlation curves. For higher cardinalities, the curves obtained are not distant from  interpolating the curves from \textsf{BestSub}. 
We also needed stable results to work on the percentiles (as we discuss below). For this reason, the average correlation curves are obtained by averaging one million samples in place of $50,000$ that was used in past work. Again, this larger sample did not substantially affect the average curves.

Using such heuristic searches means that the best and worst curves are not
optimal: there could be topic sets that are even better or
worse. However,  correlation values should not change significantly, as shown by \citet[Section~5.1]{Guiver:2009:FGT:1629096.1629099}.

\subsection{Effectiveness Metrics}
\label{FewTopics:sec:metrics}

The MQ07 collection differs from the other collections in that it uses statAP and
statMAP (together with E[MAP], that we do not use in this chapter), rather than AP and MAP, as its primary
evaluation measure.
The measure \citep{MQ2007,MQ2009,pavlu2007practical} is a
version of MAP that is used to create a pool with a sampling strategy: each document is associated with an \emph{inclusion
  probability}, used both to decide whether a document is in the pool,
and to weight the importance of the document when computing the
metric.
Since the differences between statMAP and MAP may have implications for
our analysis, we consider two approaches for comparing them.

The first is to produce scatter plots showing how the run ranks
change when using the two metrics. 
This has been explored several times, and on different datasets, in
previous work, e.g. 
over AH99 data by \citet[Figure~7]{pavlu2007practical}, and over TB06
data by \citet[Figure~5]{MQ2007}; both analyses showed that while
variations exist, they are limited.

\begin{figure}[t]
  \centering
   \includegraphics[width=.66\linewidth]{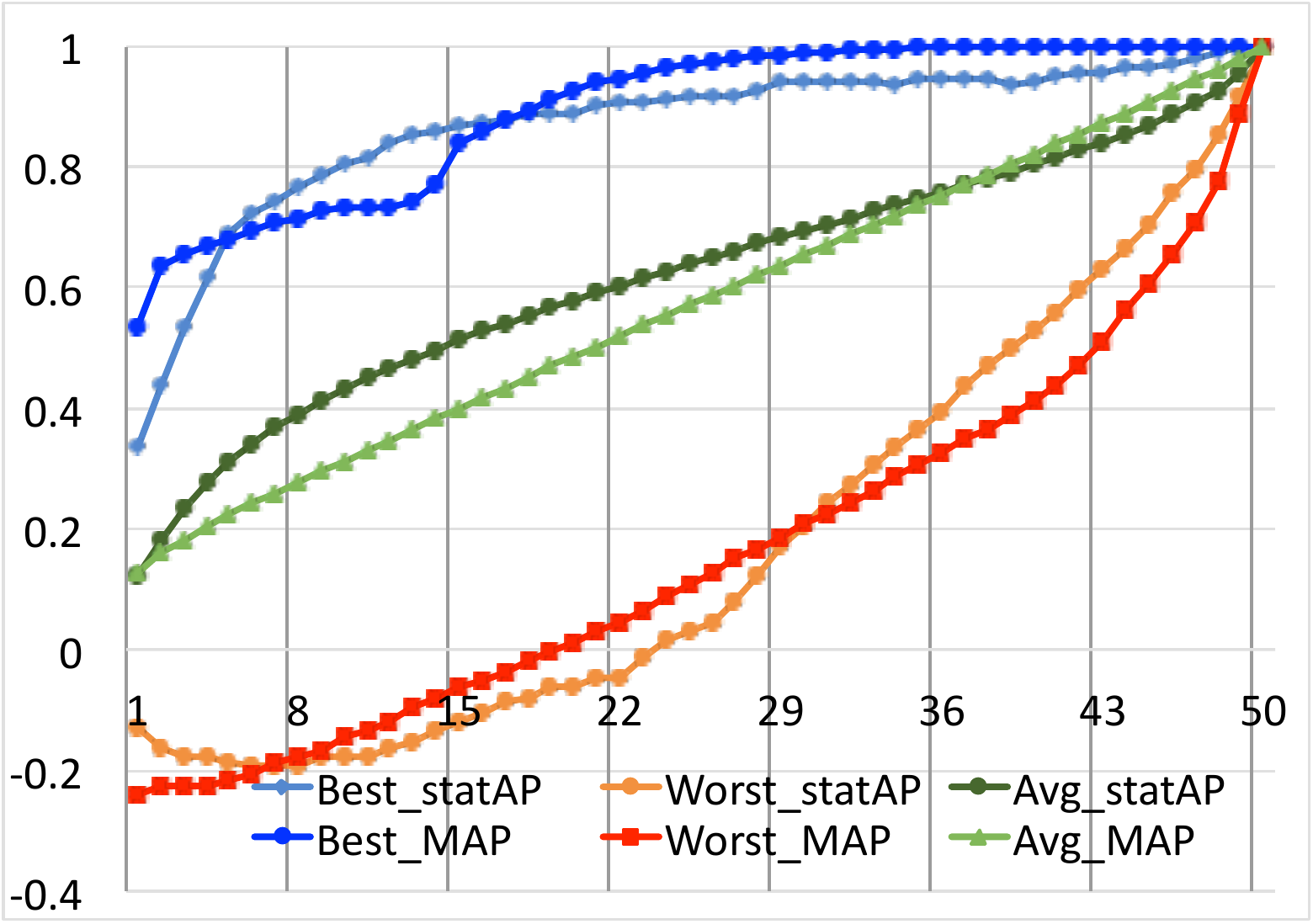}
\caption{Kendall's $\tau$ correlation curves for AH99, on a subset of runs, for both MAP and statMAP.}
  \label{FewTopics:fig:statMAP}
\end{figure}

A second approach is to compare the correlation curves produced by
\textsf{BestSub} when using statMAP and MAP. 
To do so, we re-evaluated AH99 using statMAP. We selected the
$57$ runs in AH99 for which the statMAP sampling algorithm does not
select any unjudged documents, and used statMAP software from
MQ07,\footnote{Note, several versions of statMAP exist, we used
\textsf{statAP\_MQ\_eval\_v3.pl}: \url{http://trec.nist.gov/data/million.query07.html}}
thereby implementing ``stratified sampling''
\cite[Section~2.4]{pavlu2007practical}, where each document has a
probability of being sampled that is proportional to its rank in the
run outputs.
We ran \textsf{BestSub} using both statMAP and MAP.
The (best, average, and worst) correlation curves that we obtained for
statMAP and for MAP are shown in Figure~\ref{FewTopics:fig:statMAP}. 
The lines are similar, and often overlap or cross each other. 
In fact the differences are much larger when comparing them with the
full AH99 dataset, such as in Figure~\ref{FewTopics:fig:tois1}; this is likely
due to the different (smaller) number of runs, and the range of metric
values, which have a larger impact than using statMAP in place of MAP. 
We therefore conclude overall that, although statMAP does create some
differences, these appear to be smaller than the differences
introduced by other variables, and that using statMAP in place of MAP
should not introduce any strong bias into our analysis. 
This confirms the results obtained by previous studies
\citep{Guiver:2009:FGT:1629096.1629099,ecir11,ictir13}, where the evaluation metric usually did
not make any noticeable difference.

\section{RQ\ref{FewTopics:RQ:1}: Larger Ground Truth}
\label{FewTopics:sec:rq1:-larger-ground}

To address RQ\ref{FewTopics:RQ:1}, we first present a simulation experiment on synthetic data in Section~\ref{FewTopics:sec:simulation}. We then focus on real data starting with an overview of the results in Section~\ref{FewTopics:sec:general-results}, followed by descriptions of best, average, and worst curves in Section~\ref{FewTopics:sec:best-average-worst}. In Section~\ref{FewTopics:sec:hosseini} we compare our results with those by \citet{Hosseini:SIGIR:2012}
and, finally, the worst sets are analyzed in more detail in Section~\ref{FewTopics:sec:worstanalysis}. 

\subsection{A Simulation Experiment}\label{FewTopics:sec:simulation}


Intuitively, given a larger initial topic set, it will be easier to find good (and bad) subsets, as the degrees of freedom increase.
Analogously, when the number of runs in a test collection decreases, it should be easier to find good (and bad) topic subsets, as it is simpler to reorder fewer items in a given way since the size of the gaps between the runs becomes larger and the number of constraints is smaller.
To have a first, less qualitative and more concrete, insight on what might happen when varying the number of topics and runs, we perform the following experiments. We randomly generate synthetic AP values for datasets having different sizes of topics (20, 50, 100, 1000) and runs (25, 50, and 100), using two strategies: 
(i) we generate random AP values normally distributed ($\mathcal{N}(\mu,\,\sigma^{2})$), setting the $\mu$ and $\sigma^2$ parameters equal to the real $\mu$ and $\sigma^2$ values of AH99; and (ii) we randomly sample with replacement real AP values from AH99 thus obtaining the same distribution of AH99.
We then run \textsf{BestSub} on the synthetic datasets to obtain the best, worst, and average 
correlation values at each topic subset cardinality. 

\begin{figure*}[tb]
  \centering 
  \begin{tabular}[b]{c}
    \includegraphics[width=.95\linewidth]{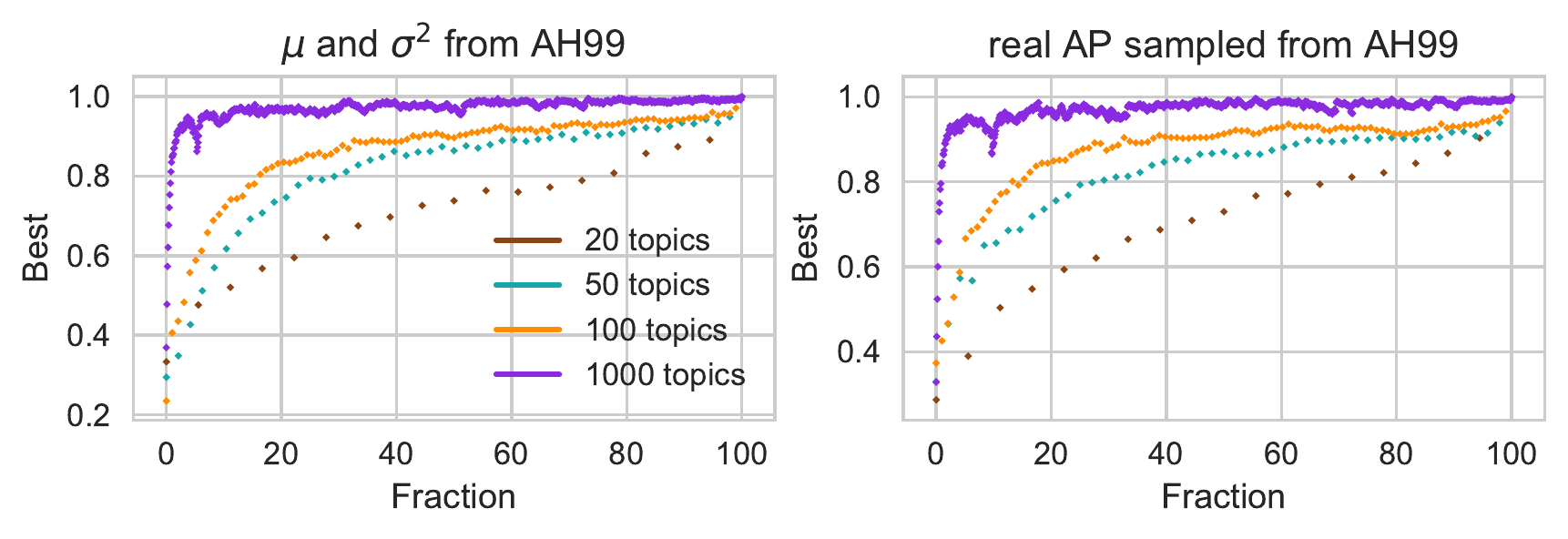}\\
    \includegraphics[width=.95\linewidth]{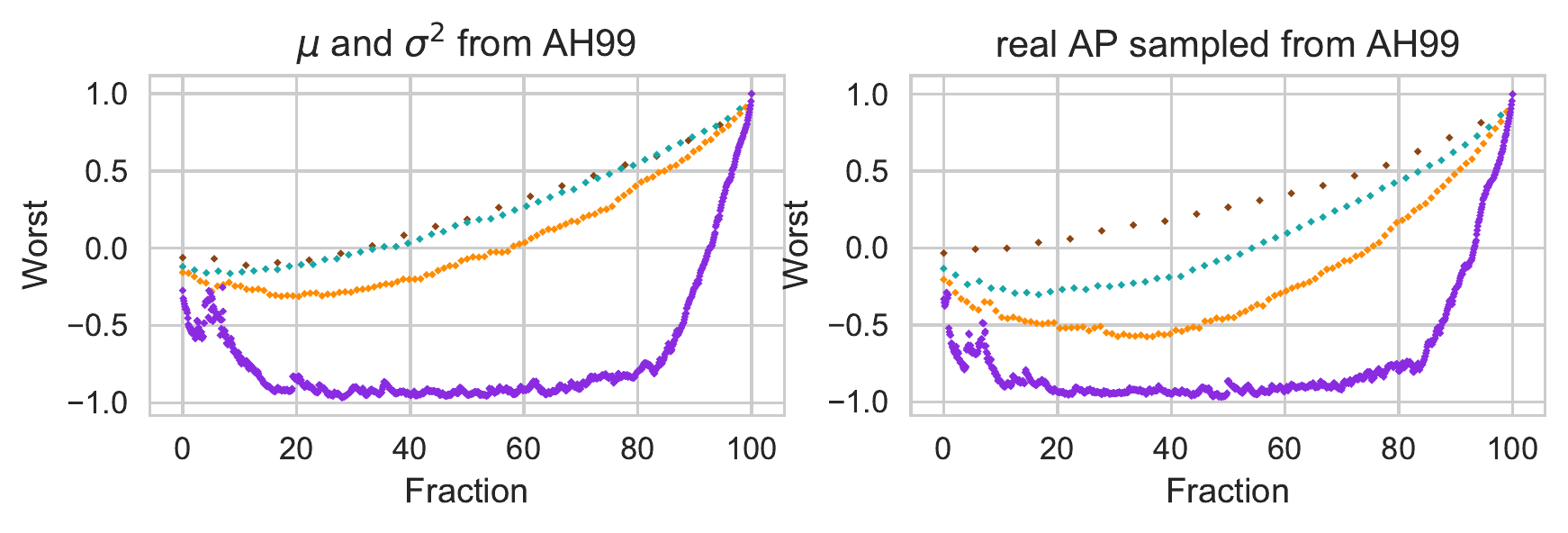}\\
    \includegraphics[width=.95\linewidth]{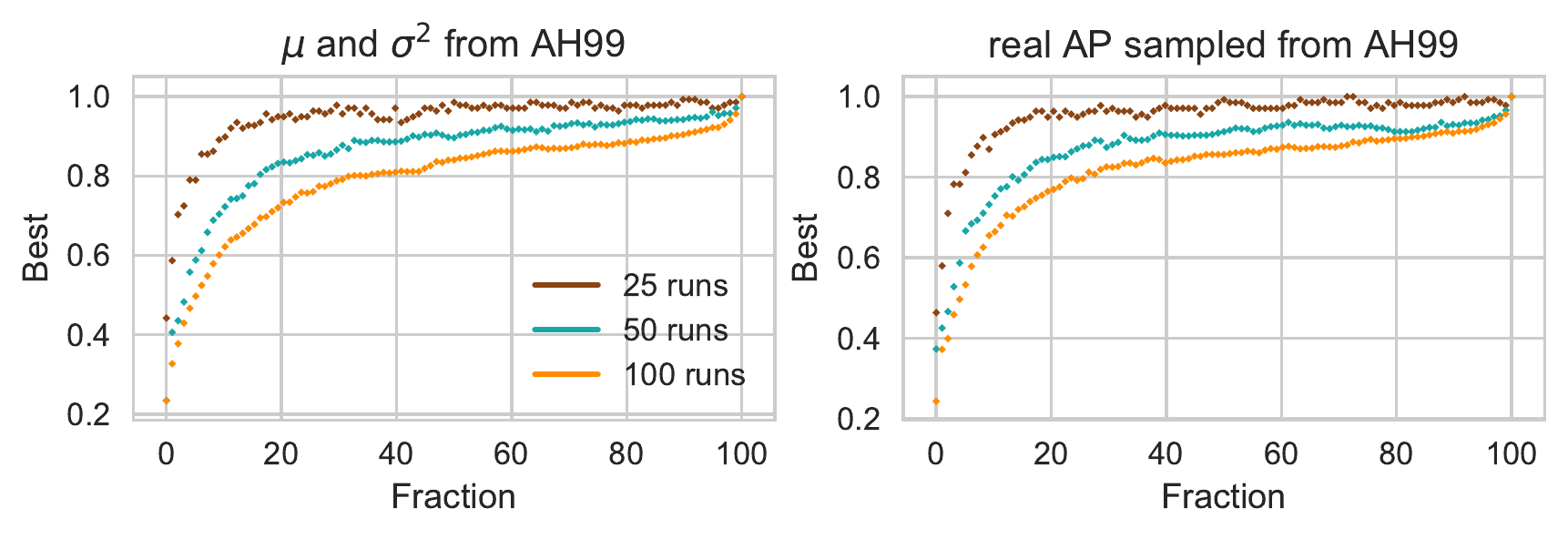}\\
    \includegraphics[width=.95\linewidth]{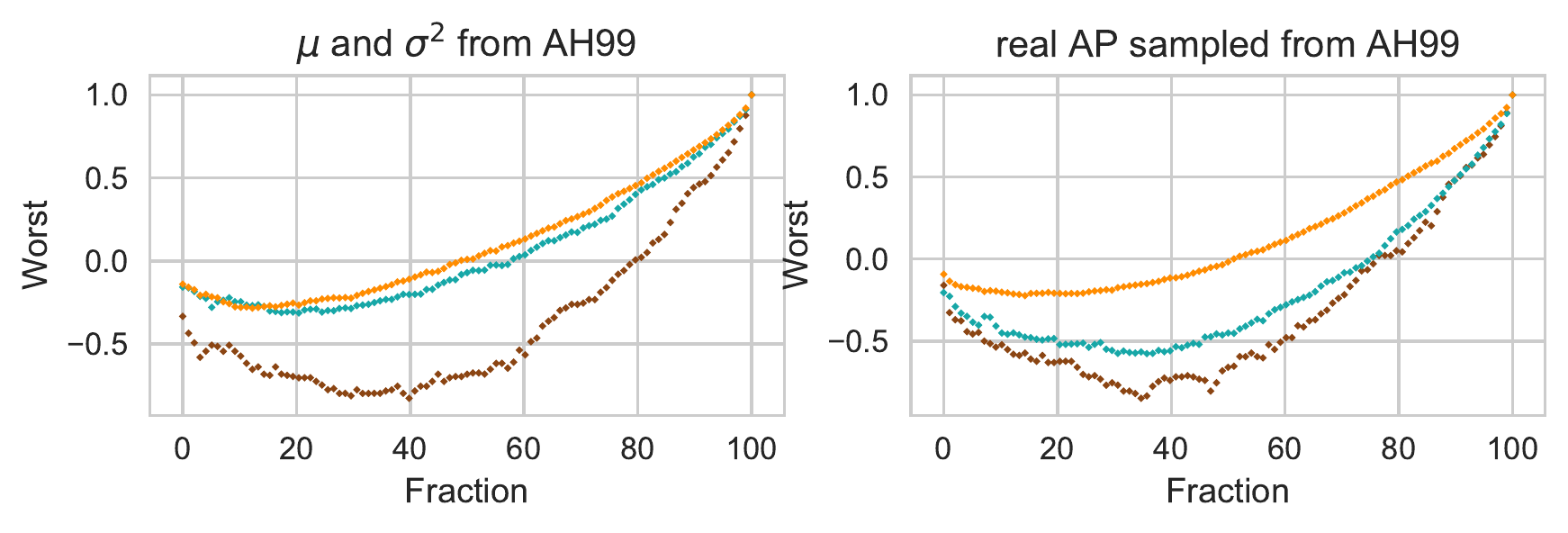}\\
\vspace*{-2mm}
  \end{tabular}
\caption{
Kendall's $\tau$ correlation curves for fractions of the full topic set, on synthetic data: best and worst curves on random data generated using $\mu$ and $\sigma$ values of AH99 (left column), and on real AP scores sampled from AH99 (right column).
 In the first four plots, 50 runs are used; in the last four, 100 topics are used.
  \label{FewTopics:fig:simulation}
}
\vspace{-35pt}
\end{figure*}

Figure~\ref{FewTopics:fig:simulation}  shows the results as correlation charts having the fraction of the full set of topics cardinality on the x-axis and $\tau$ on the y-axis. 
The  four charts of the first two rows are obtained by using 50 runs and varying the number of topics. They clearly show that correlation curves become more extreme as the number of topics in the ground truth increases. 
The effect on the average curves (not shown) is less clear but are much smaller as they are quite similar to each other.

When using a fixed ground truth of 100 topics and varying the number of runs, the results are similar. 
The four charts on the last two rows of Figure~\ref{FewTopics:fig:simulation} show the correlation curves when varying the number of runs and using 100 topics; 
the best and worst correlation curves become more extreme as the number of runs decreases, as expected (and the correlation for the average series does not vary much).
This is perhaps a less interesting result than the previous one, since the number of topics is related to test collection design and can be decided when building a test collection, whereas the number of runs depends on factors that are more difficult to control. Therefore in this chapter we focus on the number of topics.
Regardless, this confirms that the number of runs in a test collection can have an effect. 
Overall, comparing the two sampling strategies (i.e., the left and right columns in Figure~\ref{FewTopics:fig:simulation})  we see that their behaviour is similar, although not identical, when considering a fixed number of both runs and topics.

The results of this simulation experiment hint that the extreme nature of the curves found in previous studies \citep{Guiver:2009:FGT:1629096.1629099,ecir11,ictir13}  not only is confirmed on datasets with a larger topic set ground truth, but it can even become more striking in some cases. For example, in the worst curve for 1000 topics, even 75\% of the topics (i.e., 750) would rank the 50 runs in almost the opposite way to the full topic set. Note that this is a setting similar to MQ07 (see Table~\ref{FewTopics:tab:coll}): if these results were confirmed in the real datasets, they might have important practical implications.

However, the simulation experiment has some limitations: it relies on assumptions that might be not true in a real-world scenario, as different collections have different distributions and parameters, and complex systems topics interactions exist, as shown for example by \citet{Urbano2016} and \citet{Urbano:2018:SST:3209978.3210043}. 
For example, the charts on the left in Figure~\ref{FewTopics:fig:simulation} need to be interpreted with care, as real AP values are usually not normally distributed in practice. When running the Anderson-Darling normality test on each of the four test collections that we use in this chapter (see Table~\ref{FewTopics:tab:coll}), the set of all AP (or statAP) values for all topic/run pairs is not normally distributed  (neither with $p<0.05$ nor with $p<0.01$). When considering the AP values for each single run, the distribution of values is not normal, accordingly to the same test, for 186 ($p<0.01$) and 219 ($p<0.05$) cases out of the total 253.
%
By using random AP values we are assuming that the AP values of one run across different topics and the AP values of one topic across different runs are independent, both of which are false as usually the performance of a system across topics is relatively stable, and each research group submits usually many runs, which are somehow related.

Summarising, real test collections include many more variables and interactions than what our simulations can capture: the number of runs, dependencies between runs, the similarity and documents overlap of run variants, the topic system interactions \citep{Urbano2016, Urbano:2018:SST:3209978.3210043}, etc.
Moreover, it is of course interesting to see what happens in a real dataset, and in particular if there are particular ``pathological'' cases that might have occurred. For these reasons we turn to experiments on the real datasets. 



\subsection{General Results}
\label{FewTopics:sec:general-results}

\begin{figure*}[pt]
  \centering 
  \begin{tabular}[b]{cc}
    \includegraphics[width=.45\linewidth]{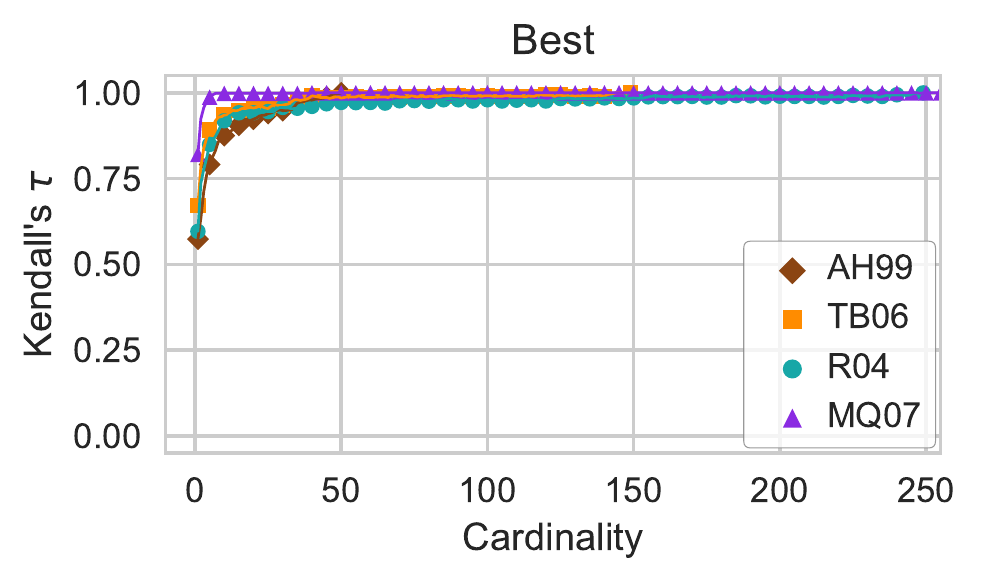}
    \includegraphics[width=.45\linewidth]{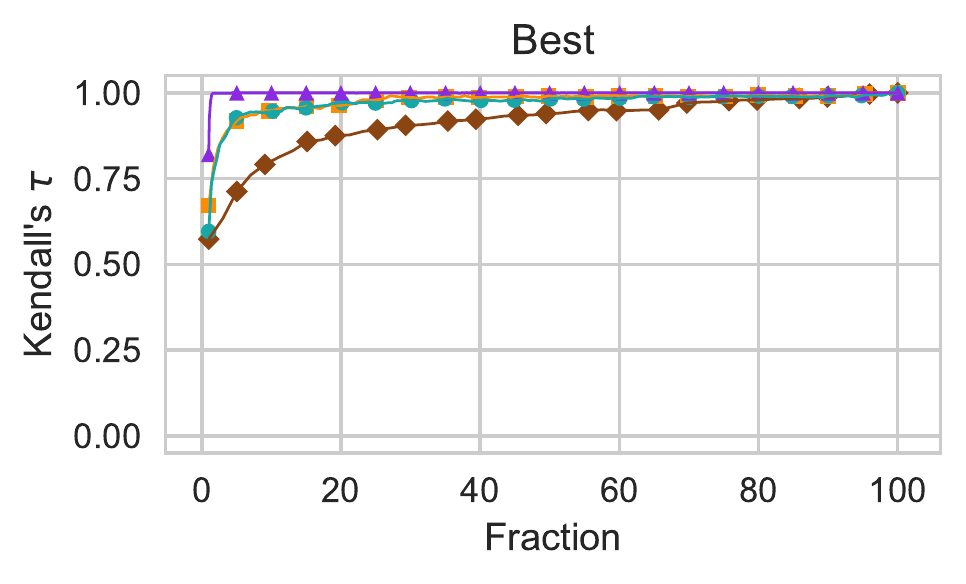}
\vspace*{-2mm}
  \end{tabular}
  \begin{tabular}[b]{cc}
    \includegraphics[width=.45\linewidth]{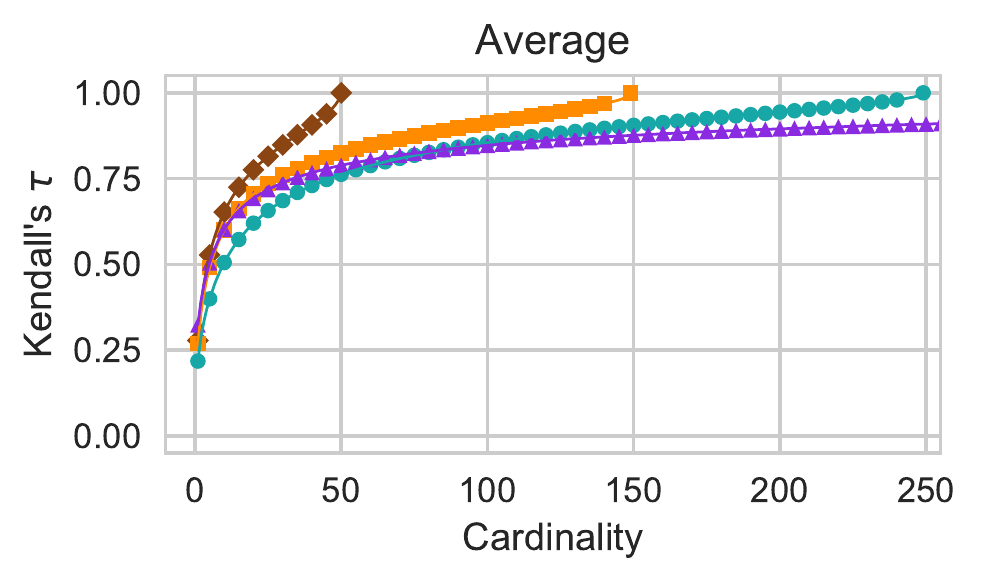}
    \includegraphics[width=.45\linewidth]{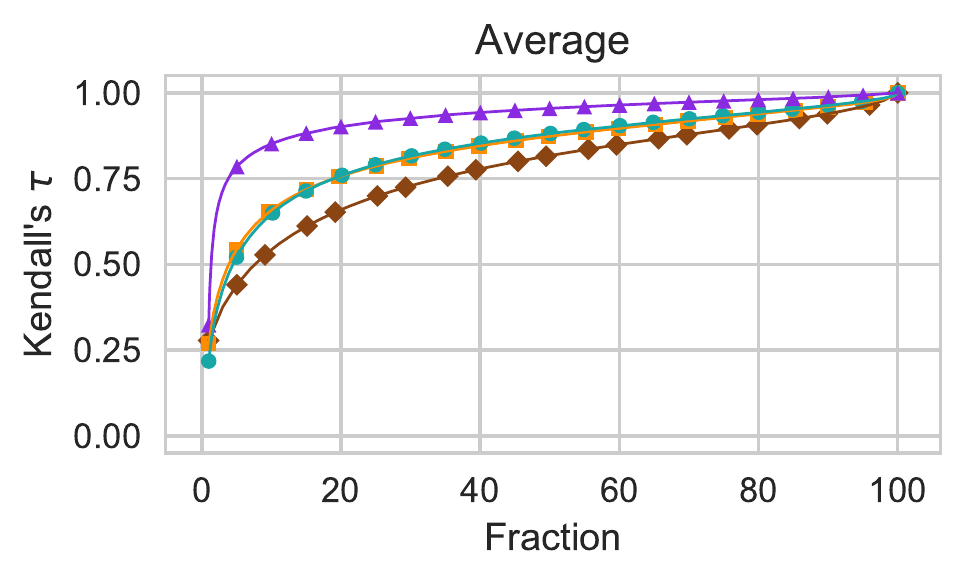}
\vspace*{-2mm}
  \end{tabular}
  \begin{tabular}[b]{cc}
    \includegraphics[width=.45\linewidth]{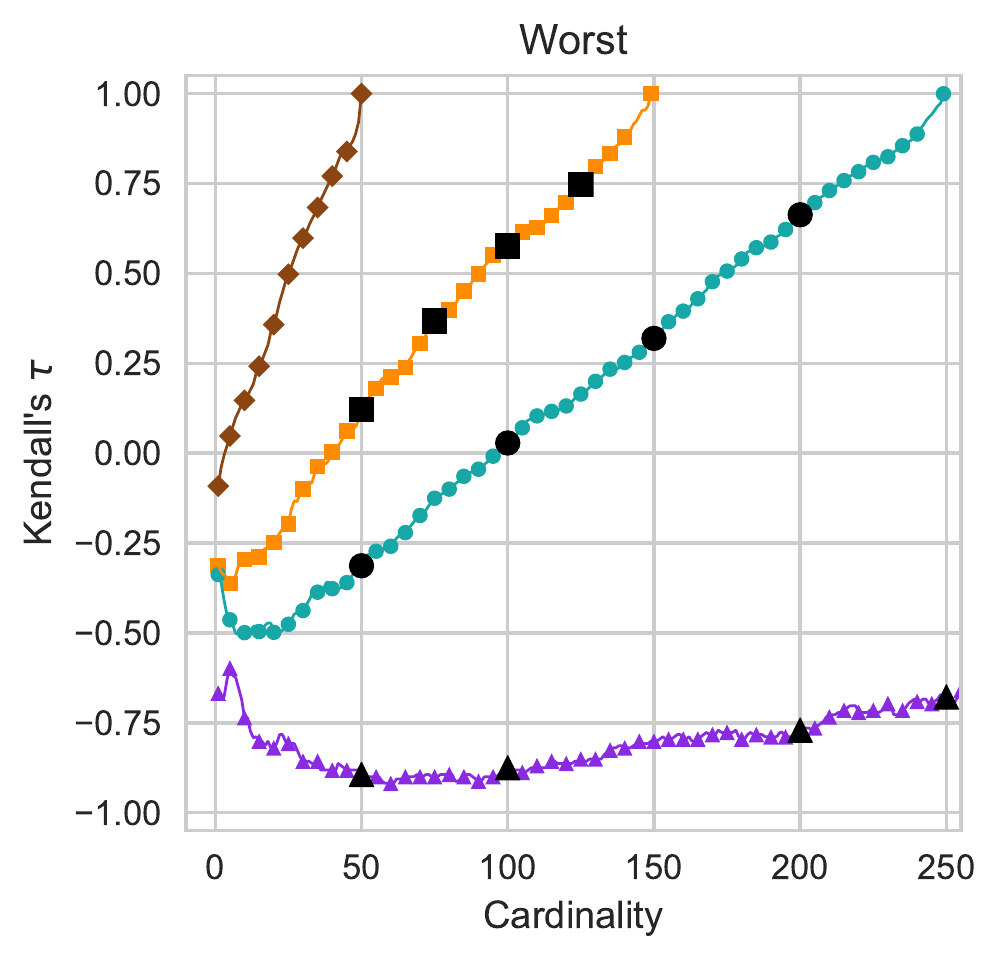}
    \includegraphics[width=.45\linewidth]{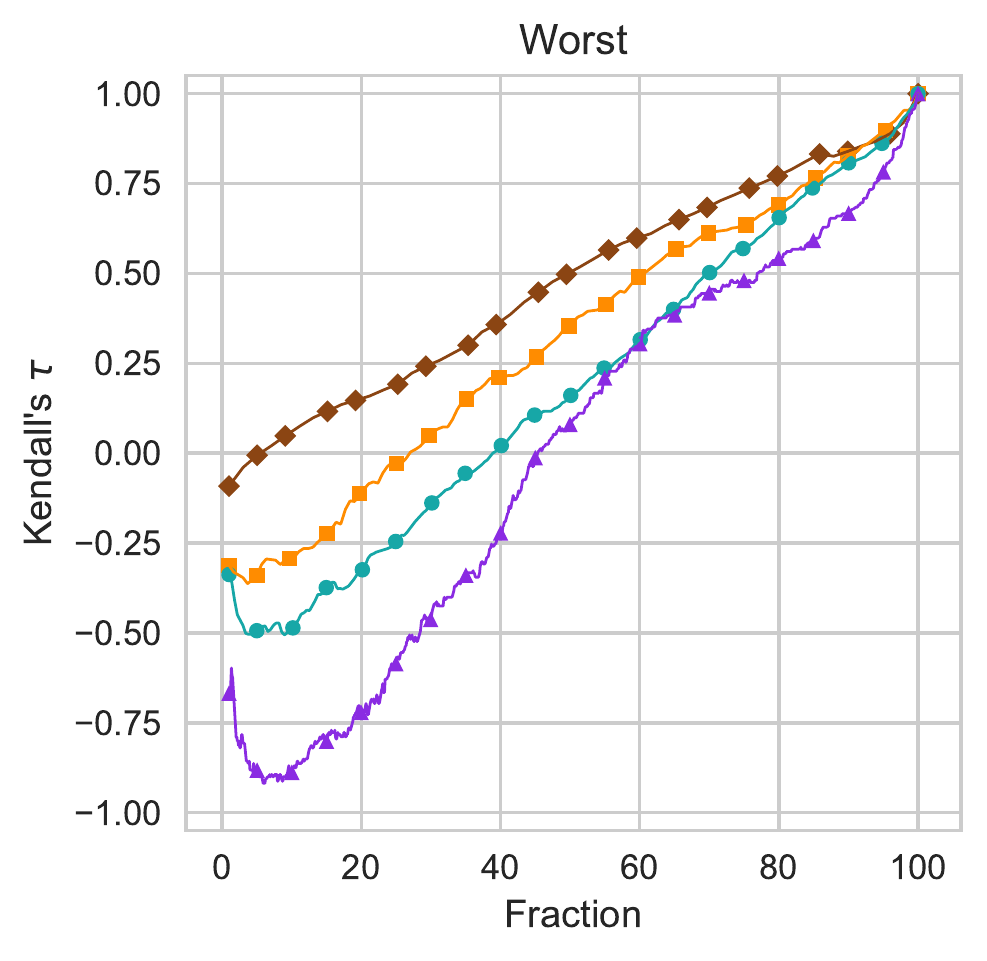}
    \vspace*{-2mm}
  \end{tabular}
\caption{Kendall's $\tau$ correlation curves for absolute cardinalities (left-side, cardinalities up to 250) and fraction of full set (right-side). Black markers in the Worst curves are further analyzed in Figure~\ref{FewTopics:fig:scatterplots}.
  \label{FewTopics:fig:corr_BWA}}
\end{figure*}

\begin{figure*}[tb]
  \centering 
  \begin{tabular}[b]{cc}
    \includegraphics[width=.45\linewidth]{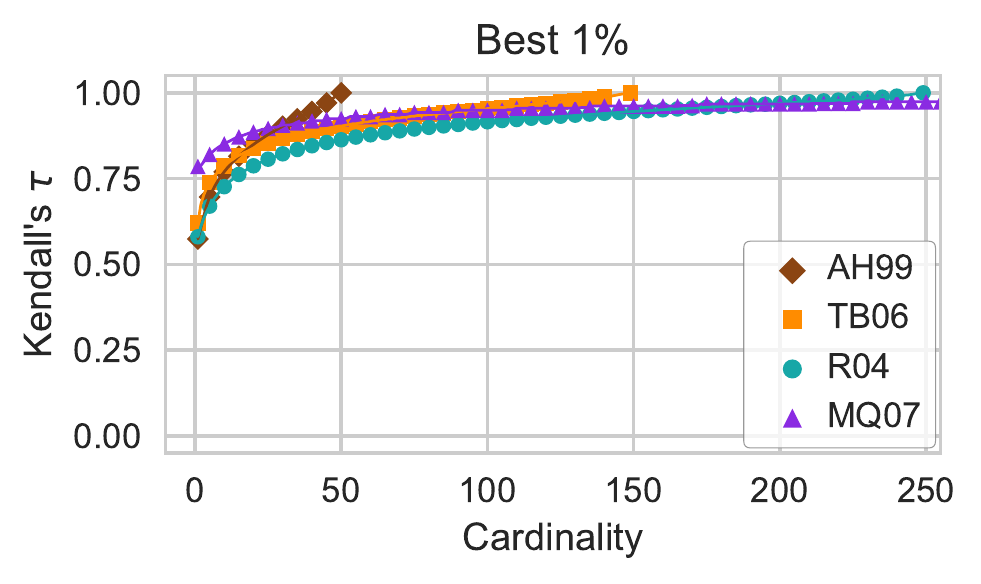}
    \includegraphics[width=.45\linewidth]{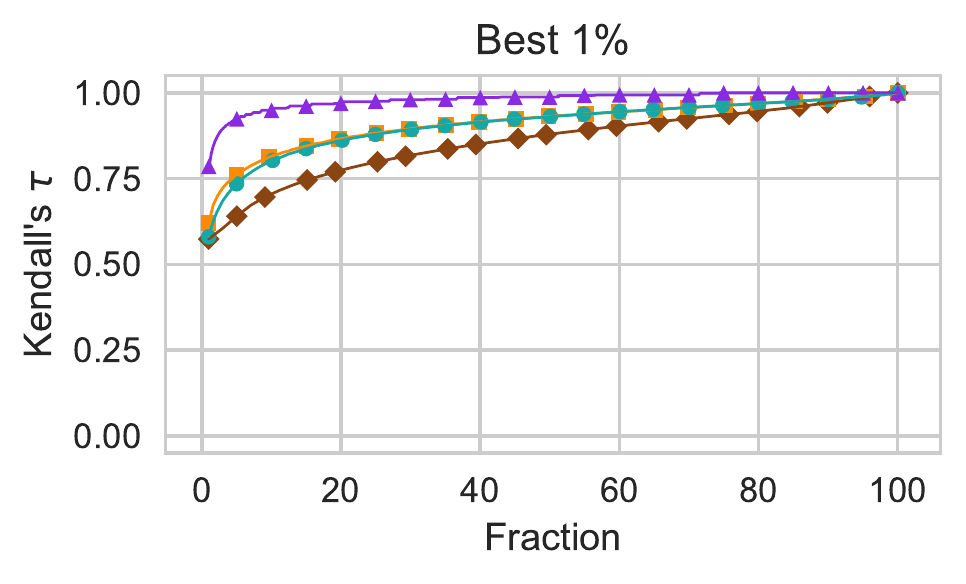}
\vspace*{-2mm}
  \end{tabular}
  \begin{tabular}[b]{cc}
    \includegraphics[width=.45\linewidth]{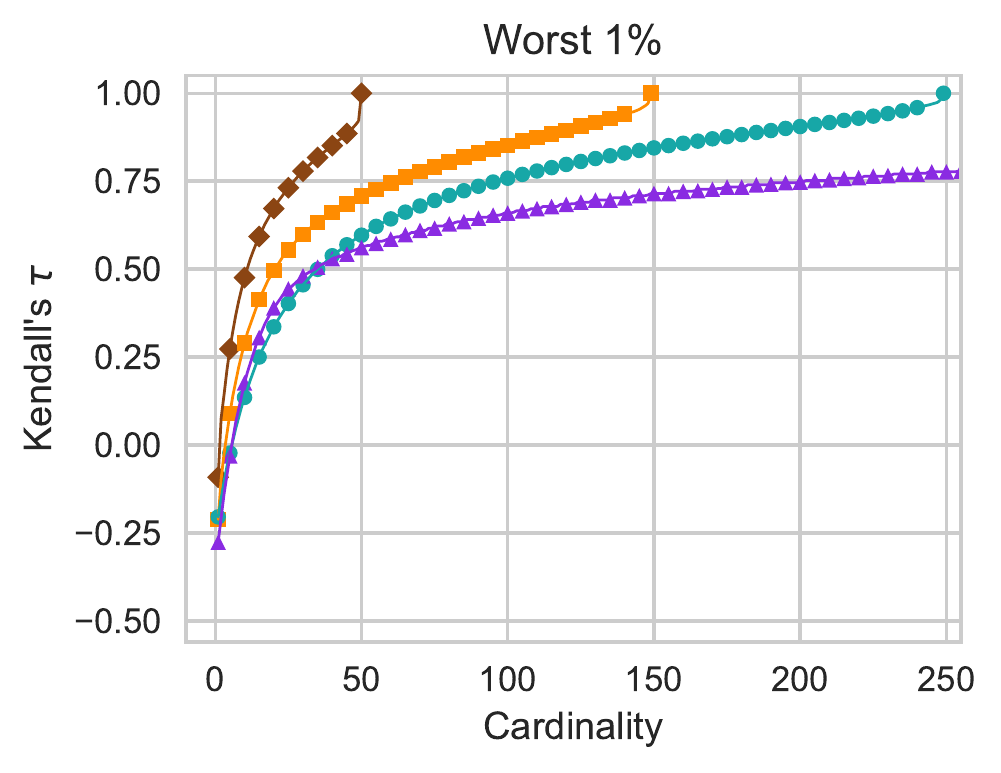}
    \includegraphics[width=.45\linewidth]{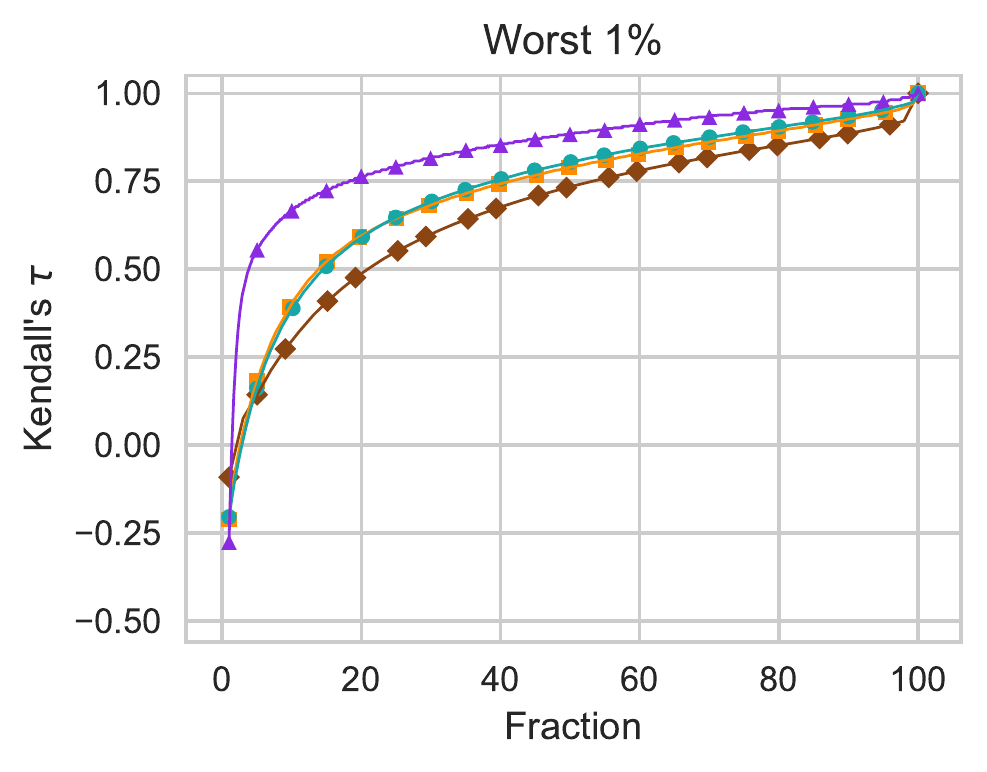}
    \vspace*{-2mm}
  \end{tabular}
\caption{Kendall's $\tau$ correlation curves for absolute cardinalities (left-side, cardinalities up to 250) and fraction of full set (right-side). 
  \label{FewTopics:fig:corr_BW1Perc}}
\end{figure*}

Figures~\ref{FewTopics:fig:corr_BWA} and \ref{FewTopics:fig:corr_BW1Perc} show correlation charts for the three new datasets TB06, R04, MQ07, as well as AH99 (correlation values are obtained using statMAP for MQ07 and MAP for the other datasets).
Correlation is measured using $\tau$. We plot best,
average, and worst in separate charts. We also plot the
best and worst 1\% topic subsets found.
In the graphs on the left side of the figures, the x-axis shows subset size measured by cardinality; the graphs on the right, subset size is measured as the cardinal fraction of the ground truth set.
The graphs on the left have $250$ as maximum cardinality so that we can fully represent the curves for AH99, TB06, and R04. As a consequence, the MQ07 curves are truncated, but their complete trend can be seen in the graphs on the right.
To avoid clutter, we do not plot the markers for all cardinalities: on the left hand side markers are shown at multiples of $5$, plus cardinality $1$ and full set. On the right hand side, a marker is plotted at each multiple of $5\%$  (or, when not available because of rounding, the closest value), plus the $1\%$ marker. The lines in the charts are not interpolations, they follow the real values at each cardinality.

While there are similarities between the current charts and those previously published, the
best and average curves seem higher when the ground truth cardinality increases (as predicted by the simulation experiment in Section~\ref{FewTopics:sec:simulation}). The
worst curves are lower, particularly for the MQ07 dataset. 
For example, for the MQ07\_W\footnote{We use the suffix B/A/W
to indicate the correlation curve for the best/average/worst topic set.} curve, a $\tau$ of $0$ is reached at around $0.45$ of the
full cardinality set (around $500$ topics) and a $\tau$ of $0.5$ is 
reached at $0.8$ (around $900$ topics).
In other words, it would appear that one can build a subset of around $500$
MQ topics that ranks the runs randomly, compared to the ground truth. A subset
of $900$ topics can be found that ranks the runs in a still different way to the
ground truth set. We analyze these curves in more detail in the following.



\subsection{Best, Average, and Worst Curves}
\label{FewTopics:sec:best-average-worst}

\subsubsection{Best Correlation Curves}
\label{FewTopics:sec:best-corr-curv}
From the best correlation curves 
we see that fewer topics can potentially be used on ground truth
cardinalities of $n\gg50$: the MQ07\_B curve is highest, followed by R04\_B and TB06\_B, which are
in turn both consistently higher than AH99\_B.
This answers the research question RQ1 by supporting, together with the experiment on synthetic data described in Section~\ref{FewTopics:sec:simulation}, the hypothesis that
having a larger topic set as ground truth increases the possibility of finding a subset
of good topics, thereby leading to higher correlation curves.

A further confirmation of that hypothesis comes from the fraction curves (right-side).
Here, the two best curves R04\_B and TB06\_B are almost exactly
overlapping, and they both stay well above the best curve AH99\_B.
The MQ07\_B is even higher.
Compared with the previous three studies 
\citep{Guiver:2009:FGT:1629096.1629099,ecir11,ictir13} we see that when using a higher
cardinality ground truth ($149$, $249$, or $1,153$ topics in place of
only $50$), run effectiveness can be predicted by using even fewer
topics.

When comparing across the four test collections, it is prudent to examine
other properties of the collections that might impact on the trend observed.
One can see from Table~\ref{FewTopics:tab:coll} that as well as a change in ground
truth topic cardinality, there is also a change in the number of runs associated
with each of the test collections and that this might impact on the $\tau$
values. 

\citet{sanderson} illustrated that the range of scores that a set of runs have has the greatest impact on $\tau$ and
other correlation measures.
As will be seen
in Figure~\ref{FewTopics:fig:scatterplots}, the range of scores of the runs is similar
across the four test collections. 
However, as discussed in Section~\ref{FewTopics:sec:simulation}, a decreasing number
of runs is another factor leading to more extreme curves. 
In fact, if the goal is to find extreme topic sets, as the number of run increases, there are more runs that need to be re ordered, and the chances of finding extreme topic subsets is lower.
Although we leave to future work a complete study of the interplay between the number of topics and of runs, we observe that 
    the effect of the number of topics seems to dominate that of the number of runs, as it can be seen by comparing the worst curves of R04 and TB06 and observing that R04 is clearly the most extreme. The reason is that R04 has more topics than TB06; even if R04 also has more runs (which leads to less extreme curves), this is less important.

\subsubsection{Average Correlation Curves}
\label{FewTopics:sec:aver-corr-curv}

When examining the average $\tau$ 
across topic subsets,
we see that $\tau$ for AH99\_A is higher than R04\_A, TB06\_A, and MQ07\_A:
on average, by selecting a random subset of topics of a given cardinality, this 
appears to be a better predictor of 
run rankings in the AH99
dataset than in R04, TB06, and MQ07. Returning to the example in Figure~\ref{FewTopics:fig:tois1}
an average topic subset of cardinality 22 drawn from the collections with larger
ground truth has a lower $\tau$ than on AH99.

The corresponding fraction curves tell a different story
however: on average, by selecting a given fraction of the ground truth,
the topic subset of AH99  turns out to be a worse predictor of
run rankings than that of R04, TB06, and MQ07. Collections with
larger ground truths appear to need a smaller fraction of topics to
achieve high values of $\tau$.

A particular feature of the MQ07\_A curve is that its trend seems
more similar to the best than to the average curves of the other
datasets. For this dataset, on average,
a good prediction of run ranks can be obtained with a small
fraction of topics (around $5$-$10\%$) and a very good prediction
of run ranks can be obtained with $20\%$.
This result needs to be examined on other test collections with
similarly large topic sets.


The curves for R04 and TB06 on the fraction charts are almost
exactly overlapping. This might be an indication
that a ground truth of cardinality $50$ is somewhat different from a larger
ground truth. There might be some numerical/statistical effect that does
not appear when using only $50$ topics.



\subsubsection{Worst Correlation Curves}
\label{FewTopics:sec:worst-corr-curv}
The most noticeable difference between AH99 and the larger datasets is in
the worst curves: 
whereas best and average are broadly similar to past work,
the worst curves are quite different.

The correlation values for the worst curves are strongly negative.
This is a novel situation, not observed in the previous three studies
\citep{Guiver:2009:FGT:1629096.1629099,ecir11,ictir13} where $\tau$ values were at worst negative
with a low absolute value (around $-0.2$).
Negative correlations show topic subsets that evaluate runs in
broadly opposite ways.
Also, the negative correlation values in R04\_W persist for
cardinalities much larger than $50$, the usual number of topics used in
evaluation exercises.
The MQ07\_W curve is even lower and stays below
$-0.5$ up to $250$ (and, as can be seen from the fraction curve on the right, even up to $300$).

Although this is something expected after the simulation experiment in Section~\ref{FewTopics:sec:simulation}, it is somehow
striking that on MQ07, a subset of more than $250$ topics can be found that
negatively correlates with the ground truth topic set. As mentioned
above,  a set of around $45\%$ of the MQ07 topics
(around $500$ topics)  results in a
$\tau$ of zero.

Note, the reason three of the curves drop as the cardinality of the topic
sets increase from $1$ is due to the degrees of freedom there are when searching
for topic subsets that are the worst: the value of $\binom{n}{c}$ initially increases
as $c$ gets larger. Therefore, the range of possible topic sets that are searched
to find the worst also gets larger.

\subsubsection{Best and Worst First Percentile Curves}
\label{FewTopics:sec:worst-1st}
Given the extreme nature of the best and worst curves, we also
computed the average $\tau$ of the best and worst $1^{st}$
percentile of topic subsets.
Figure~\ref{FewTopics:fig:corr_BW1Perc} shows the resulting charts.

The Best 1\% curves emphasize that although the quest of finding
  the best topic subsets is rather difficult since they are extremely
  rare, reasonably good results that can more easily be obtained in
  practical cases do exist. 
  The Worst 1\% curves are less worrying than the
  Worst ones, since they do not feature the same extremely low, if not
negative correlations.
Although these curves look more like those from the Average, it is worth
noting that when trying to find subsets of topics for an effective
test collection, a low positive correlation is not satisfying either.
For example, the R04 $1^{st}$ percentile curve has low $\tau$ ($<0.6$)
even for cardinality $45$, and the MQ07 $1^{st}$ percentile curve has
a $\tau$ of about $0.75$ at cardinality $250$.
These are not extremely unlikely topic sets, and it is possible that
some test collections have been created with topics that rank runs
quite differently from what might be expected.



\subsection{Comparison with \texorpdfstring{\citet{Hosseini:SIGIR:2012}}{Hosseini et al.}'s Results}\label{FewTopics:sec:hosseini}


\citet{Hosseini:SIGIR:2012} report in their paper some numeric correlation values for the AH99 and R04 collections to which we can compare. Since \citeauthor{Hosseini:SIGIR:2012} use all the runs in a collection, for this comparison we performed again our experiments using all the runs instead of the top 75\% (thus, 129 for AH99 and 29 for MQ07, see Table~\ref{FewTopics:tab:coll}), and all the values reported in this section concern such a setting.

\begin{table}[tbp]
 \centering
    \begin{tabular}{l c rrr c rrr}
    \toprule
                 &&\multicolumn{3}{c}{AH99} && \multicolumn{3}{c}{R04}\\
    \cmidrule{3-5}\cmidrule{7-9}
                 && 20\% &   40\%    & 60\% && 20\% &   40\%    & 60\% \\
\midrule
    Best        && 0.92 & 0.96 & 0.97 && 0.97 & 0.98 & 0.99 \\
    Best 1\%    && 0.86 & 0.91 & 0.94 && 0.91 & 0.95 & 0.97 \\
    Average     && 0.80 & 0.87 & 0.91 && 0.85 & 0.91 & 0.94 \\
    Worst 1\%   && 0.70 & 0.81 & 0.87 && 0.76 & 0.86 & 0.90 \\
    Worst       && 0.48 & 0.64 & 0.77 && 0.17 & 0.43 & 0.63 \\
    \addlinespace
    Oracle      && 0.88 & 0.93 & 0.95 && 0.90 & 0.92 & 0.94 \\
    Adaptive    && 0.83 & 0.90 & 0.93 && 0.77 & 0.87 & 0.91 \\
    Random      && 0.72 & 0.77 & 0.87 && 0.68 & 0.80 & 0.85 \\  
    \addlinespace
    Clustering  && 0.79 & 0.88 & 0.93 && 0.84 & 0.92 & 0.95 \\
    \bottomrule
    \end{tabular}
    \caption{Kendall's $\tau$ values for comparison with \citet{Hosseini:SIGIR:2012}'s results. All runs used.}
    \label{FewTopics:tab:hosseini}
\end{table}

Table~\ref{FewTopics:tab:hosseini} shows the results of the comparison. The first five rows in the table report Kendall's $\tau$ correlation values obtained for best, best 1st percentile, average, worst first percentile, and worst subsets at the specified fractions (20\%, 40\%, and 60\%) of the full set cardinality, for the two collections AH99 and R04. Since we are using all the runs in this computation, the results do not exactly match with those presented in previous Figures~\ref{FewTopics:fig:corr_BWA} and~\ref{FewTopics:fig:corr_BW1Perc}. The next three rows in the table are the values reported in \citet[Table~1]{Hosseini:SIGIR:2012}: ``Oracle'' is their attempt to find the highest possible correlation, and so it somehow corresponds to our Best topic subsets; ``Random'' is a random selection of topics, so it should correspond to our Average; and ``Adaptive'' values are those obtained by their topic selection algorithm. The values in the last row of the table will be discussed when focusing on RQ\ref{FewTopics:RQ:3} on clustering in the following. 

We can draw several remarks.
\begin{itemize}
\item When comparing  the correlation values in the first five rows of the table with those obtained on the top 75\% runs (Figures~\ref{FewTopics:fig:corr_BWA} and~\ref{FewTopics:fig:corr_BW1Perc}), it is clear that the correlation values obtained using all systems are higher. This is expected, as the bottom runs are usually consistently ineffective on all topics. In other terms, focusing on the top systems only as we are doing in this chapter is a more difficult setting for our task than using all systems.
\item When comparing our Best with \citeauthor{Hosseini:SIGIR:2012}'s Oracle, we note that Best values are always higher than Oracle. Indeed, Oracle is always closer to Best 1\% than to Best, and for R04 it is even closer to Average than to Best. 
\item When comparing Average with  Random, we expected no differences, but it turns out that some clear differences exist. Our Average values are clearly higher than their Random. Indeed, Random is always closer to, and often lower than, Worst 1\%. 
 We have not been able to obtain the original code used by \citeauthor{Hosseini:SIGIR:2012} to replicate their experiment and although we tried, we could not obtain their values.
We double checked and we are quite confident that Average values are correct, for several reasons: they correspond to the values in previous work \citep{Guiver:2009:FGT:1629096.1629099,ecir11,ictir13} obtained with the different implementation of \textsf{BestSub} (see Section~\ref{FewTopics:sec:software}); the new implementation of \textsf{BestSub} is publicly available (\url{https://github.com/Miccighel/NewBestSub}) and it has been flawlessly used in other experiments \citep{Roitero:2018:EES:3209978.3210108,Roitero:2018:RIE:3282439.3239573}, including some specifically aimed at reproducing previous results  \citep[Section~4.3]{Roitero:2018:RIE:3282439.3239573}. As a further check, we also reproduced the random series of the plots in \citep[Figure~1]{KUTLU201837} for two datasets (Robust 2003 and 2004 reduced to 149 topics): also in this case our average values correspond to \citeauthor{KUTLU201837} random ones. 
\item When looking at the Adaptive values (that will be further analyzed in the last part of this chapter), one can notice that Adaptive is clearly higher than Random (the baseline used by \citeauthor{Hosseini:SIGIR:2012}) but, as a consequence of the previous remark, it is very similar to Average for AH99 and even always lower than Average for R04. Therefore, it turns out that Adaptive is not effective when compared to our, higher, baseline.
\end{itemize}


\subsection{Worst Subset Analysis}
\label{FewTopics:sec:worstanalysis}

Although exceptionally rare, the very worst topic subsets feature extremely
low correlations.
In this section we try to better understand how the subsets produce
such low $\tau$ correlations.

\subsubsection{Overlap}
\label{FewTopics:sec:overlap}

Examining intersections between the best and worst topic subsets,
we find that there is a quite large overlap between them: at
cardinality $100$, R04 and MQ07 have a topic overlap of around $40\%$.
This means that it is possible to select a set of $40$ topics, then to
add to it either a first or a second set of $60$ (different) topics, and
obtain completely different, even almost opposite, rankings of
runs.\footnote{Note, the
overlap that we find might be an effect of the heuristic used; we
can say no more than it is possible to build a best and a worst
set of topics with a high overlap.}

A possible explanation for this overlap could be that there are two
small subsets of topics, one good and one bad, that are used to
build the low cardinality best and worst sets; then a set of common
``neutral'' topics are added to both to obtain the higher
cardinality sets.
However, this needs further study, as this possibility is not consistent with the data, 
since the $40\%$ overlap can be found from cardinality $50$ up to
$200$.




\subsubsection{Comparing Worst with Best}
\label{FewTopics:sec:extr-value-stat}
\begin{figure}[tb]
  \centering
  \includegraphics[width=\linewidth]{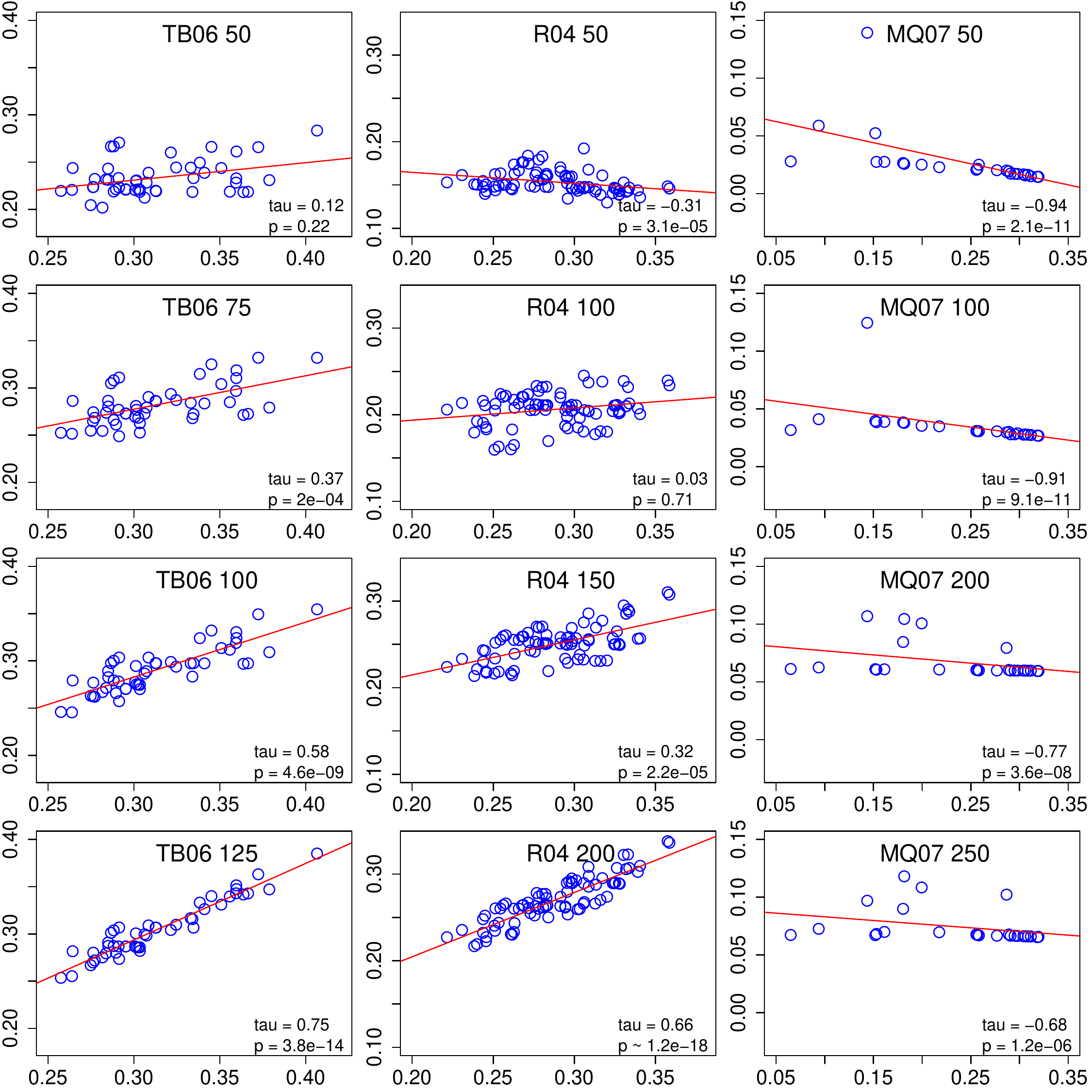}
  \caption{Scatter plots for some selected notable worst topic subsets.
Each dot is a run, the x-axis shows MAP (statMAP for MQ07) computed
on the ground truth full topic set, the y-axis shows effectiveness
computed on the worst subset indicated. The $\tau$ of
the correlation along with the significance of the correlation
(indicated by a $p$ value) is detailed on each plot.
}
  \label{FewTopics:fig:scatterplots}
\end{figure}

It is also possible that some conceptual features of the topic 
subsets exist that could explain the low correlations.
Therefore, some of the worst topic subsets are characterized here
for analysis.
We manually selected illustrative topic subsets that have low
$\tau$ correlations and high cardinalities: 

\begin{itemize} 
\item  TB06: cardinalities $50$, $75$, $100$, and $125$.
\item  R04: cardinalities $50$, $100$, $150$, and $200$.
\item  MQ07: cardinalities $50$, $100$, $200$, and $250$.
\end{itemize}
These are the subsets represented by black markers in Figure~\ref{FewTopics:fig:corr_BWA}.
Figure~\ref{FewTopics:fig:scatterplots} shows scatter plots for these
subsets. We see that the effectiveness measure
computed on the worst subset (y-axis) usually has both a smaller range and lower
values when compared to the measure computed on the ground truth set (x-axis).
This is especially true for MQ07, but the same effect can also be found
on the other datasets.
To better understand this observation, the mean effectiveness over all topic
subset cardinalities was computed for the best and worst topic subsets.
The results are shown in the left part of Table~\ref{FewTopics:tab:MAPs}.
It can be seen that the best subsets contain topics that lead to higher
effectiveness values than the worst subsets.
The right part of the table shows the $\Delta$ between the
average subset effectiveness and the ground truth effectiveness.
As might be expected, in all cases the best subsets contain topics that lead to values more 
similar to ground truth effectiveness.

\begin{table}[tb]
  \centering
\caption{Effectiveness measures (MAP, except statMAP for MQ07) over the best and worst subsets and between ground truth.}
  \begin{tabular}{@{}l   @{ }c@{ } c@{ } c@{ } c@{ }   c   c@{  } c@{    }    c@{  } c@{ }}
\toprule
  & \multicolumn{4}{@{ }c@{ }}{Av. Effectiveness of Subset} && \multicolumn{4}{@{ }c@{ }}{Subset $\Delta$ from Ground Truth}  \\
\cmidrule{2-5} \cmidrule{7-10}
 & \multicolumn{1}{@{ }c@{ }}{AH99} & \multicolumn{1}{@{ }c@{ }}{TB06} & \multicolumn{1}{@{ }c@{ }}{R04} & \multicolumn{1}{@{ }c@{ }}{MQ07} &&
    \multicolumn{1}{@{ }c@{ }}{AH99} & \multicolumn{1}{@{ }c@{ }}{TB06} & \multicolumn{1}{@{ }c@{ }}{R04} & \multicolumn{1}{@{ }c@{ }}{MQ07}  \\
\midrule
Best   & 0.298  & 0.277 & 0.264 & 0.148 &&   0.017 & 0.036 & 0.025 & 0.092 \\
Worst  & 0.201 & 0.263 & 0.224 & 0.049 &&   0.080 & 0.050 & 0.066 & 0.191 \\
\bottomrule
\end{tabular}
\label{FewTopics:tab:MAPs}
\end{table}


\section{RQ\ref{FewTopics:RQ:2}: Statistical Significance}
\label{FewTopics:sec:rq2:stat}

We now turn to RQ\ref{FewTopics:RQ:2}. 
While the previous results demonstrate that it is possible to find
topic subsets that lead to run rankings that are highly correlated
with the rankings obtained when using a full (ground truth) set of topics,
in order for one run to be considered more effective
than another, a statistical significance test is usually carried out.
The number of topics that are used to evaluate effectiveness
has a direct impact on significance calculations.
For example, for a paired t-test,
the test
statistic includes the sample size~\citep{She07}, and the larger the
sample, the lower the $p$-value. In IR experiments the sample size
is the number of topics. 
Some analysis of statistical significance  is therefore due in the fewer topics scenario. We present two different and somehow dual approaches to do such an analysis in the next two subsections: the first approach is based on the work by \citet{Sakai2016} that determines the number of topics needed when aiming at a given statistical power; the second is aimed at determining the amount of error that is introduced when using topic subsets, as well as at understanding what kind of errors are made.

\subsection{Power Analysis}
\label{FewTopics:sec:sakai}

\citet{Sakai2016} recently proposed three methods to compute the cardinality of a topic set size to ensure that a test collection  has enough statistical power to distinguish effectiveness of the systems/runs. 
The  methods compute the estimated topic set size on the basis of three different tests:
\begin{itemize}
\item Method 1, based on t-test, and used when one wants to compare two system scores, or the score of one system against all the other systems.
\item Method 2, based on one way ANOVA, and used when one wants to compare the scores of more than two systems, or to compare all systems against each other.
\item Method 3, similar to Method 1, but it allows one to specify a confidence interval $\delta$ to ensure that the outcome of this test is bounded with precision $\delta$.  As \citeauthor{Sakai2016} points out: ``a wide confidence interval that includes zero implies that we are very unsure as to whether systems X and Y actually differ''.
\end{itemize}

We computed and/or estimated the parameters required by \citeauthor{Sakai2016}'s methods and ran them on our four collections, using the software (Excel spreadsheets) provided by \citeauthor{Sakai2016}. 
Tables~\ref{FewTopics:tab:sakai:t-test}, \ref{FewTopics:tab:sakai:ANOVA}, and~\ref{FewTopics:tab:sakai:CI} 
show the results. 

\begin{table}[tb]
\centering
\caption{Number of estimated topics using the first method, based on t-test. The required $\sigma^2_t$ parameter has values $0.096$ (for AH99),  $0.071$ (TB06), $0.100$ (R04), and $ 0.118$ (MQ07).
The values in bold represent the maximum and minimum estimated number of topics for the given parameters, for each collection.
}
\label{FewTopics:tab:sakai:t-test}
\begin{tabular}{c@{\hskip3pt}c@{\hskip3pt} l@{\hskip3pt} r@{\hskip3pt}r@{\hskip3pt}r@{\hskip3pt} l@{\hskip3pt} r@{\hskip3pt}r@{\hskip3pt}r@{\hskip3pt} l@{\hskip3pt} r@{\hskip3pt}r@{\hskip3pt}r@{\hskip3pt} l@{\hskip3pt} r@{\hskip3pt}r@{\hskip3pt}r}
\toprule
\multicolumn{1}{c}{\textbf{}} & \multicolumn{1}{c}{\textbf{}} && \multicolumn{3}{c}{\textbf{AH99}} && \multicolumn{3}{c}{\textbf{TB06}} && \multicolumn{3}{c}{\textbf{R04}} && \multicolumn{3}{c}{\textbf{MQ07}} \\ \cmidrule{4-18} 
\multicolumn{1}{c}{\multirow{2}{*}{\textbf{$\alpha$}}} & \multicolumn{1}{c}{\multirow{2}{*}{\textbf{$\beta$}}} && \multicolumn{3}{c}{\textbf{$minD_t$}} && \multicolumn{3}{c}{\textbf{$minD_t$}} && \multicolumn{3}{c}{\textbf{$minD_t$}} && \multicolumn{3}{c}{\textbf{$minD_t$}} \\ 
\cmidrule{4-6} \cmidrule{8-10} \cmidrule{12-14} \cmidrule{16-18} 
 &  && .05 & .1 &.2 && .05 & .1 & .2 && .05 & .1 &.2&& .05 & .1 & .2 \\ 
\midrule
\multirow{2}{*}{.01} & .1 && \textbf{575} & 147 & 40 && \textbf{426} & 109 & 30 && \textbf{599} & 153 & 41 && \textbf{706} & 179 & 48 \\ 
 & .2 && 452 & 116 & 32 && 336 & 87 & 25 && 471 & 121 & 33 && 554 & 142 & 38 \\ 
\addlinespace
\multirow{2}{*}{.05} & .1 && 406 & 103 & 28 && 301 & 77 & 21 && 422 & 106 & 29 && 498 & 126 & 33 \\ 
 & .2 && 304 & 78 & \textbf{21} && 225 & 58 & \textbf{16} && 315 & 81 & \textbf{22} && 373 & 95 & \textbf{26} \\ 
 \bottomrule
\end{tabular}
\end{table}

\begin{table}[tb]
\centering
\caption{Number of estimated topics using the second method, based on ANOVA. The required $\sigma^2$ parameter has values $0.048$ (for AH99), $0.036$ (TB06), $0.050$ (R04), and  $0.59$ (MQ07). The values in bold represent the maximum and minimum estimated number of topics for the given parameters, for each collection.
}
\label{FewTopics:tab:sakai:ANOVA}
\begin{tabular}{c@{\hskip3pt}c@{\hskip3pt} l@{\hskip3pt} r@{\hskip3pt}r@{\hskip3pt}r@{\hskip3pt} l@{\hskip3pt} r@{\hskip3pt}r@{\hskip3pt}r@{\hskip3pt} l@{\hskip3pt} r@{\hskip3pt}r@{\hskip3pt}r@{\hskip3pt} l@{\hskip3pt} r@{\hskip3pt}r@{\hskip3pt}r}
\toprule
\multicolumn{1}{c}{\textbf{}} & \multicolumn{1}{c}{\textbf{}} && \multicolumn{3}{c}{\textbf{AH99}} && \multicolumn{3}{c}{\textbf{TB06}} && \multicolumn{3}{c}{\textbf{R04}} && \multicolumn{3}{c}{\textbf{MQ07}} \\ \cmidrule{4-18} 
\multicolumn{1}{c}{\multirow{2}{*}{\textbf{$\alpha$}}} & \multicolumn{1}{c}{\multirow{2}{*}{\textbf{$\beta$}}} && \multicolumn{3}{c}{\textbf{$minD$}} && \multicolumn{3}{c}{\textbf{$minD$}} && \multicolumn{3}{c}{\textbf{$minD$}} && \multicolumn{3}{c}{\textbf{$minD$}} \\ 
\cmidrule{4-6} \cmidrule{8-10} \cmidrule{12-14} \cmidrule{16-18} 
 &  && .05 & .1 &.2 && .05 & .1 & .2 && .05 & .1 &.2&& .05 & .1 & .2 \\ 
\midrule
\multirow{2}{*}{.01}& .1 && \textbf{2352} & 588 & 147 &&  \textbf{1341} & 336 & 84 && \textbf{2295} & 574 & 144 && \textbf{3446} & 862 & 216 \\ 
& .2 && 2001 & 501 & 126 &&  1131 & 283 & 71 &&1948 & 487 & 122 && 2879 & 720 & 180 \\
\addlinespace
\multirow{2}{*}{.05} & .1 && 1860 & 465 & 117 &&  1050 & 263 & 66 && 1810 & 453 & 114 && 2669 & 668 & 167 \\ 
& .2 && 1529 & 383 & \textbf{96} && 855 & 124 & \textbf{54} &&  1485 & 372 & \textbf{93} && 2151 & 538 & \textbf{135} \\ 
\bottomrule
\end{tabular}
\end{table}

\begin{table}[tb]
\centering
\caption{Number of estimated topics using the third method, based on confidence intervals. The required $\sigma^2_t$ parameter has values $0.096$ (for AH99), $0.071$ (TB06), $0.100$ (R04), and $0.118$ (MQ07).
}
\label{FewTopics:tab:sakai:CI}
\begin{tabular}{rl c rrrr}
\toprule
\multicolumn{1}{c}{\textbf{$\alpha$}} & \multicolumn{1}{c}{\textbf{$\delta$}} && \textbf{AH99} & \textbf{TB06} & \textbf{R04} & \textbf{MQ07} \\ 
\midrule
\multirow{3}{*}{.01} & .05 && 1019 & 754 & 1061  & 1253 \\ 
 & .1 && 255 & 189 & 266  & 314 \\ 
 
 & .2 && 64 & 48& 67  & 79 \\ 
 \addlinespace
\multirow{3}{*}{.05} & .05 && 591 & 437& 615  & 726 \\ 
 & .1 && 148 & 110 & 154  & 182 \\ 
 & .2 && 37  & 28& 39 & 46 \\ 
 \bottomrule
\end{tabular}
\end{table}

The topic set size cardinalities in Table~\ref{FewTopics:tab:sakai:t-test} are those required to find statistical significance when comparing two systems, or a system against a set of other systems (e.g. when trying to understand if  system $s_1$ is better than both systems $s_2$ and $s_3$).
The three parameters are:
$\alpha$, which is the probability of Type I error (to find a difference that does not exist; that is, one concludes that $s_1$ is more/less effective than $s_2$ but this is not true); 
$\beta$, which is the probability of Type II error (not to find a difference that does exist; that is, 
one does not conclude that $s_1$ is more effective than $s_2$ when it is in fact better); and 
$minD_t$, which is the minimum detectable difference in MAP values. 
We use the same values for these three parameters as adopted by \citeauthor{Sakai2016} in the examples in his paper.
$minD_t$ is computed considering the estimated within-system variance from past collections, $\sigma^2$. To compute $\sigma^2$ we used, as \citeauthor{Sakai2016} suggests, Formula (36) of \citet{Sakai2016}, that is the residual variance from one-way ANOVA computed considering all the AP values for a given collection (i.e., all the systems and all the topics):
we applied Formula (36) to our collections when using the AP (statMAP for MQ07) metric. As discussed by \citet[Section~3.2]{Sakai2016}, $\sigma^2$ represents the common system variance computed under the so called homoscedasticity assumption, which means that $\sigma^2$ is considered to be common for all the systems. \citet{carterette2012multiple} shows that this assumption does not hold for IR evaluation, and discusses how this fact is not important; indeed, as remarked
by  \citet[Footnote~16]{Sakai2016}, ANOVA is widely used in the IR field. 

The values in the table (the estimated required number of topics) range from $16$ to $706$. Besides the considerations that could be made on the values of the three parameters $\alpha$, $\beta$, and $minD_t$ (probabilities of Type I and II errors, and the minimum detectable difference), what is important to note for our purposes is that quite often the required number of topics is even higher than the cardinality of the full topic set size for the corresponding collection. 

This is even more true when using the second method (based on ANOVA), see Table~\ref{FewTopics:tab:sakai:ANOVA}: in this case values range from $54$ to $3446$. The parameters for this method have a similar meaning to the previous method based on the t-test, with some technical differences. It is important to notice that the estimates obtained with the second method are probably more related to the approach in this chapter, since we generally compare all the systems together, rather than a single system to the other ones.
%
%
The third method returns intermediate results (see Table~\ref{FewTopics:tab:sakai:CI}).

This analysis led to reappraising the results on the best correlation curves: whereas it is true that small good topic sets exist, using them would, unsurprisingly, lead to less statistical power (which is defined 
according to \citeauthor{Sakai2016} 
as $1-\beta$, and represents the capability of finding a difference between system scores which is statistically significant), or in other words it is a move away from the number of topics required to have such statistical power.

We note that this approach \citep{Sakai2016} does not directly quantify how much statistical power we are losing when using the smaller good topic sets. In future work we intend to further explore the relationship between the factors of (sub-)topic set size and quality, and statistical power. 
Moreover, this method does not consider what kind of errors are made: when using fewer topics, there are different possible specific outcomes besides the result of a statistical test: one might find significance for a sub-set while according to the full set of topics there is not, or vice versa one might not find significance for a sub-set while for the full there is; one might even find statistically significant disagreement; and so on.
For these reasons, we conducted another, more general, experiment, described in the following subsection.

\subsection{Statistically Significant Agreement and Disagreement}
\label{FewTopics:sec:stat-sign-agre}

We  conduct an empirical investigation into the relationship
between the number of topics considered in an IR experiment and the
observed outcomes of statistically significant differences between
runs.
We first discuss some methodological issues and then describe our
experimental results.

\subsubsection{Methodology}

Consider a typical IR effectiveness experiment, where a
researcher is seeking to demonstrate that one retrieval approach is superior to another.
The researcher chooses a test collection consisting of (say) 50 topics,
and generates two sets of 50 effectiveness scores (two runs).
If the mean score for one run is higher than that for the other, it
is standard to carry out a significance test such as a paired t-test.
This will indicate whether the two scores are indeed likely to come
from populations with different means, at some specified level of
confidence.

We are interested in investigating the question: if the researcher had
carried out the same experiment but with a subset of topics, would the
same results have been observed?
This is somehow related to a similar question that has been investigated by \cite{Sakai:2007:AB:1277741.1277756}, who studied the effect of collection incompleteness on the discriminative power using Sakai's bootstrap sensitivity method; however, we focus on subsets of topics rather than subset of documents.
More concretely, let us consider a test collection with a ground truth set, $G$,
of topics of cardinality $b$. Let there also be a subset of topics, $S$, with
cardinality $a$, where $a<b$.
For a pair of runs X and Y, calculate their MAP using topic set
$S$, and carry out a paired 2-tailed $t$-test to determine whether the
difference is statistically significant.
Repeat the process for the same pair of runs, but using the topic set
of full cardinality, $G$. There are five possible outcomes~\citep{MofSch12}: 

\begin{itemize}
\itemsep 0mm
\item
    SSA: run X is significantly better than run Y on both topic sets, $S$ and $G$.
\item
    SSD: run X is significantly better than run Y on one topic set, but Y is significantly better than X on the other topic set.
\item
    SN: one run is significantly better than the other on topic set $S$, but there is no significant difference on topic set $G$.
\item
    NS: one run is significantly better than the other on topic set $G$, but there is no significant difference on topic set $S$.
\item
    NN: there is no significant difference between the runs on either topic set.
\end{itemize}

\noindent The first two letters of each label indicate the outcome of
the experiment (Significant or Not significant) on topic set $S$ and $G$,
respectively, while A and D stand for Agreement and Disagreement, respectively.
Note that only two of the five outcomes, SSA and NN, are cases where
consistent conclusions would be drawn from the experiments regardless
of which topic set is used.
For the other three, a researcher who happened to use a topic subset
$S$ would reach a different conclusion about relative run
effectiveness, than if they had used the ground truth $G$.

When considering topic subsets, it is desirable to maximize the number of SSA and NN cases (SSA if the researcher is looking for a publishable result), 
and to avoid SN and NS cases (where significant differences are found with one topic set but not with the other) and in particular SSD (where significant differences are found with both topic sets, but with different runs being indicated as being better).

\subsubsection{Results}
\label{FewTopics:sec:results-1}


\begin{figure*}[tb]
  \centering
  \rotatebox[origin=c]{-90}{%
  \begin{tabular}{@{}l@{}l@{}l@{}l@{}}
\includegraphics[width=4.5cm]{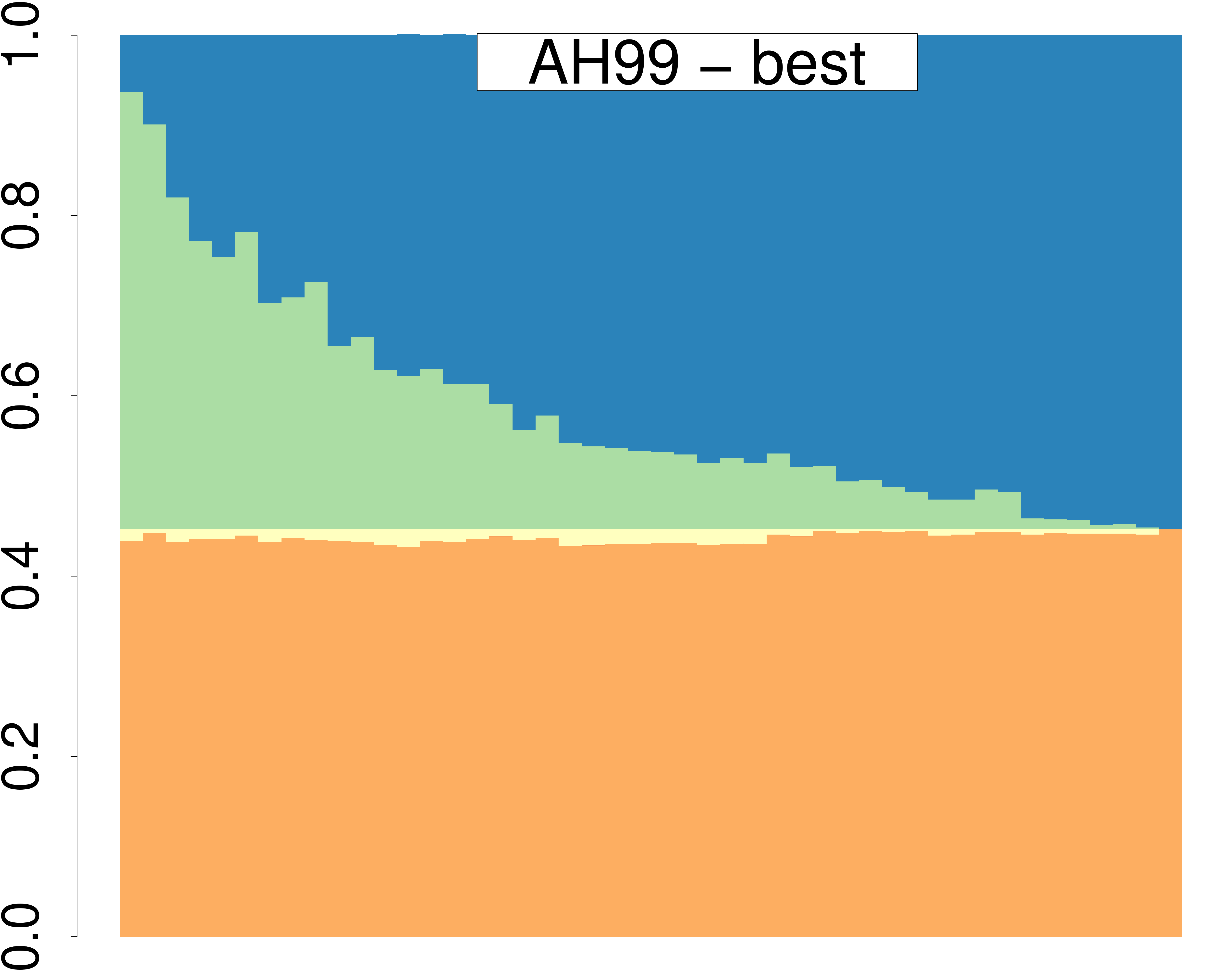}&
\includegraphics[width=4.5cm]{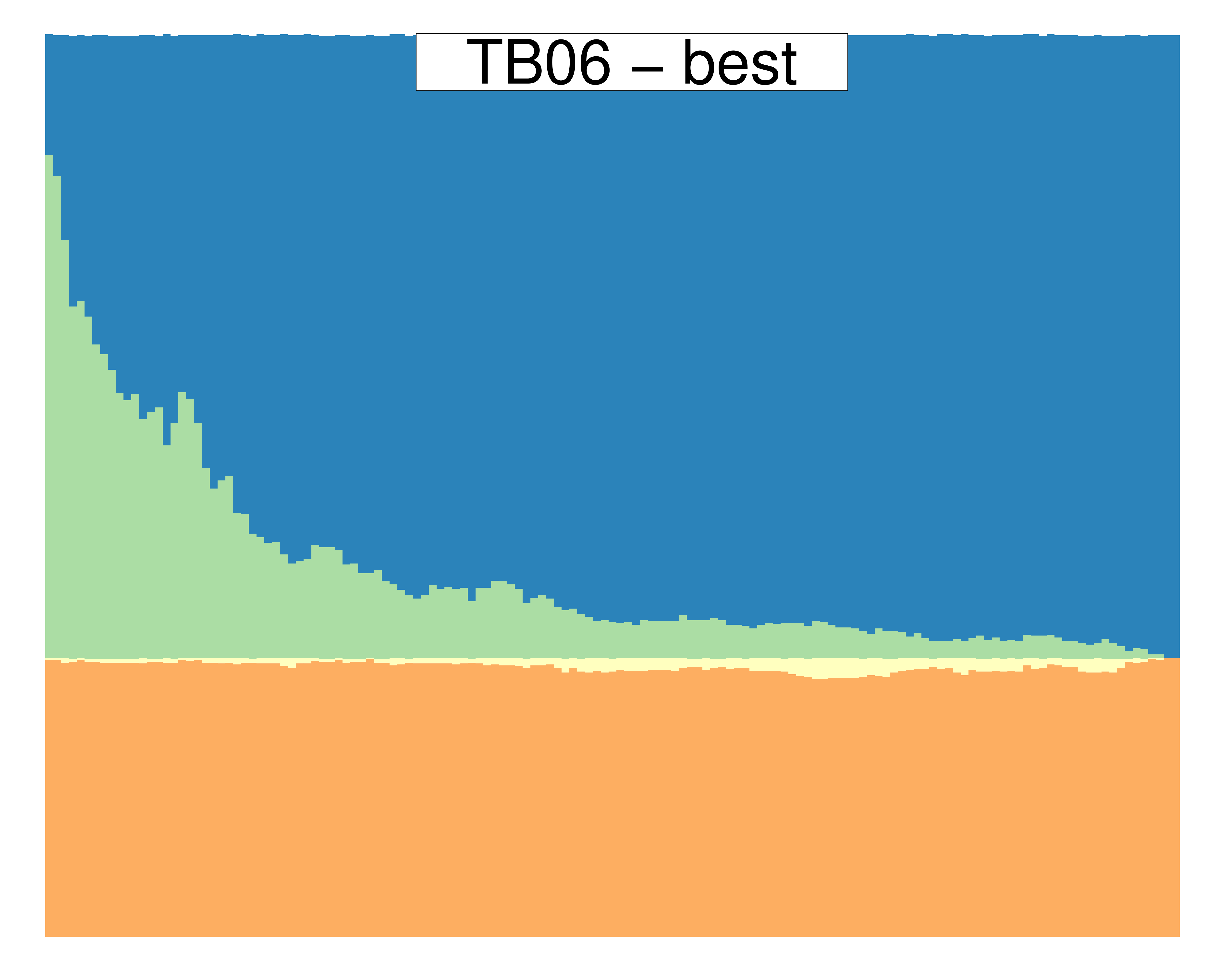}&
\includegraphics[width=4.5cm]{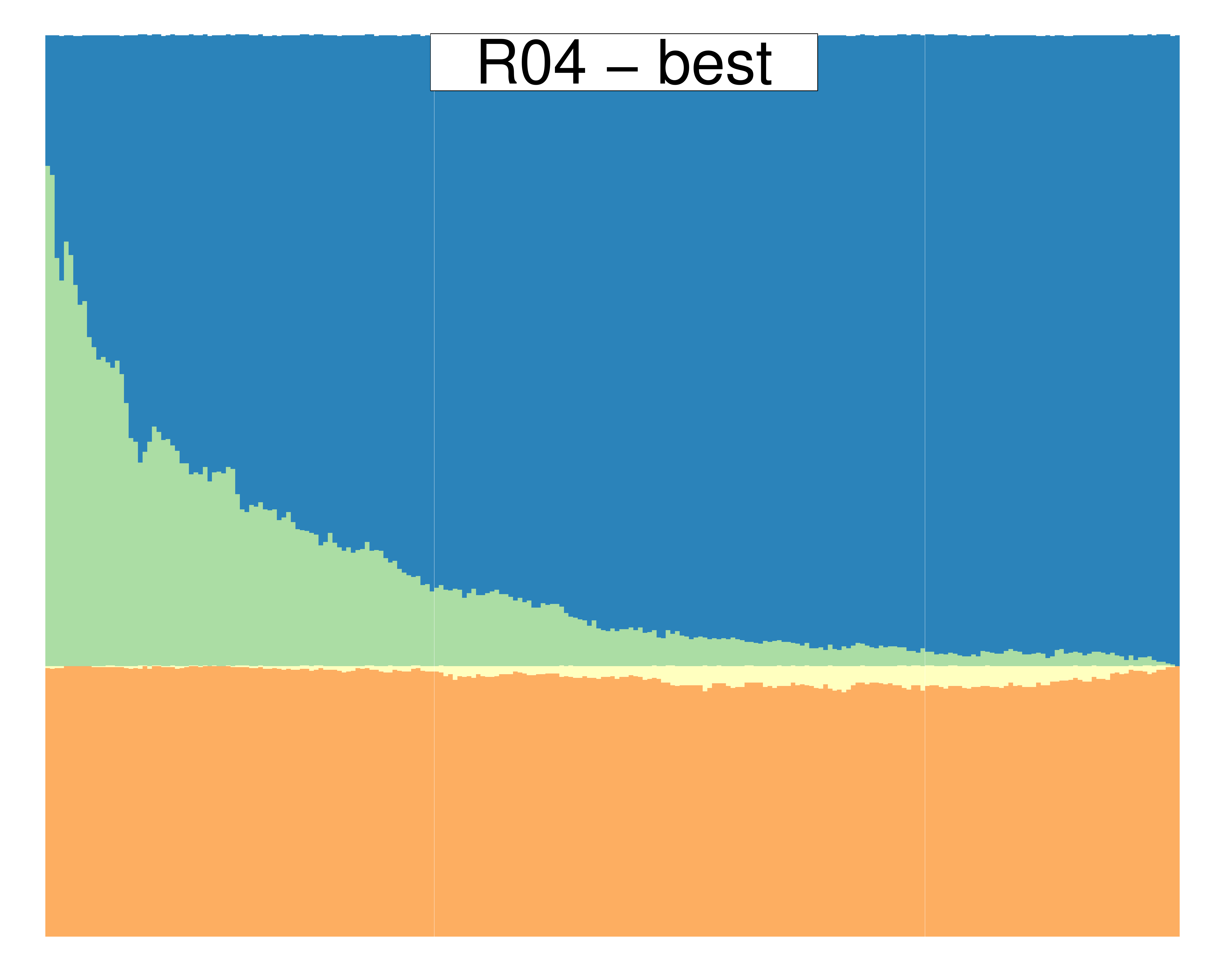}&
\includegraphics[width=4.5cm]{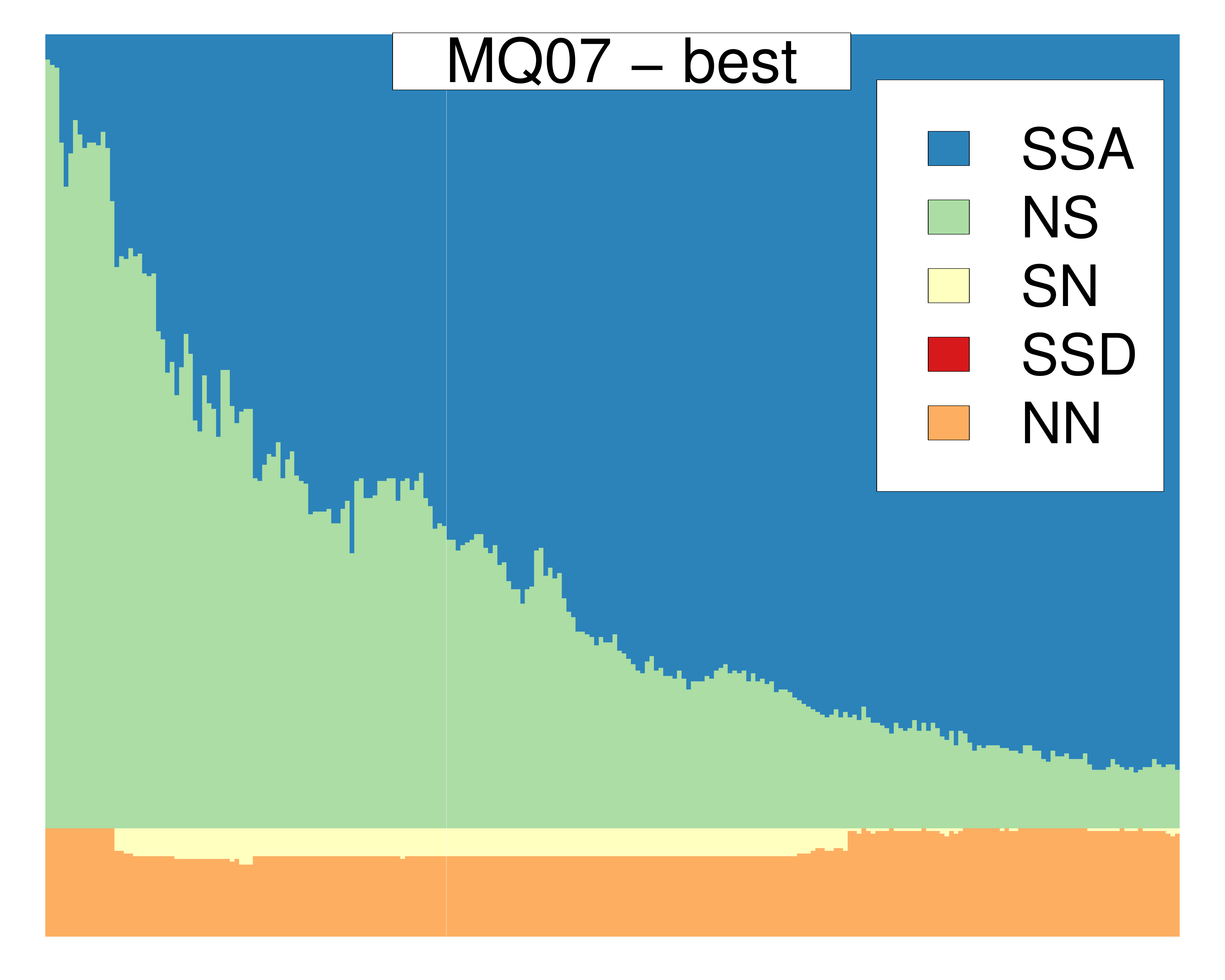}\\

\includegraphics[width=4.5cm]{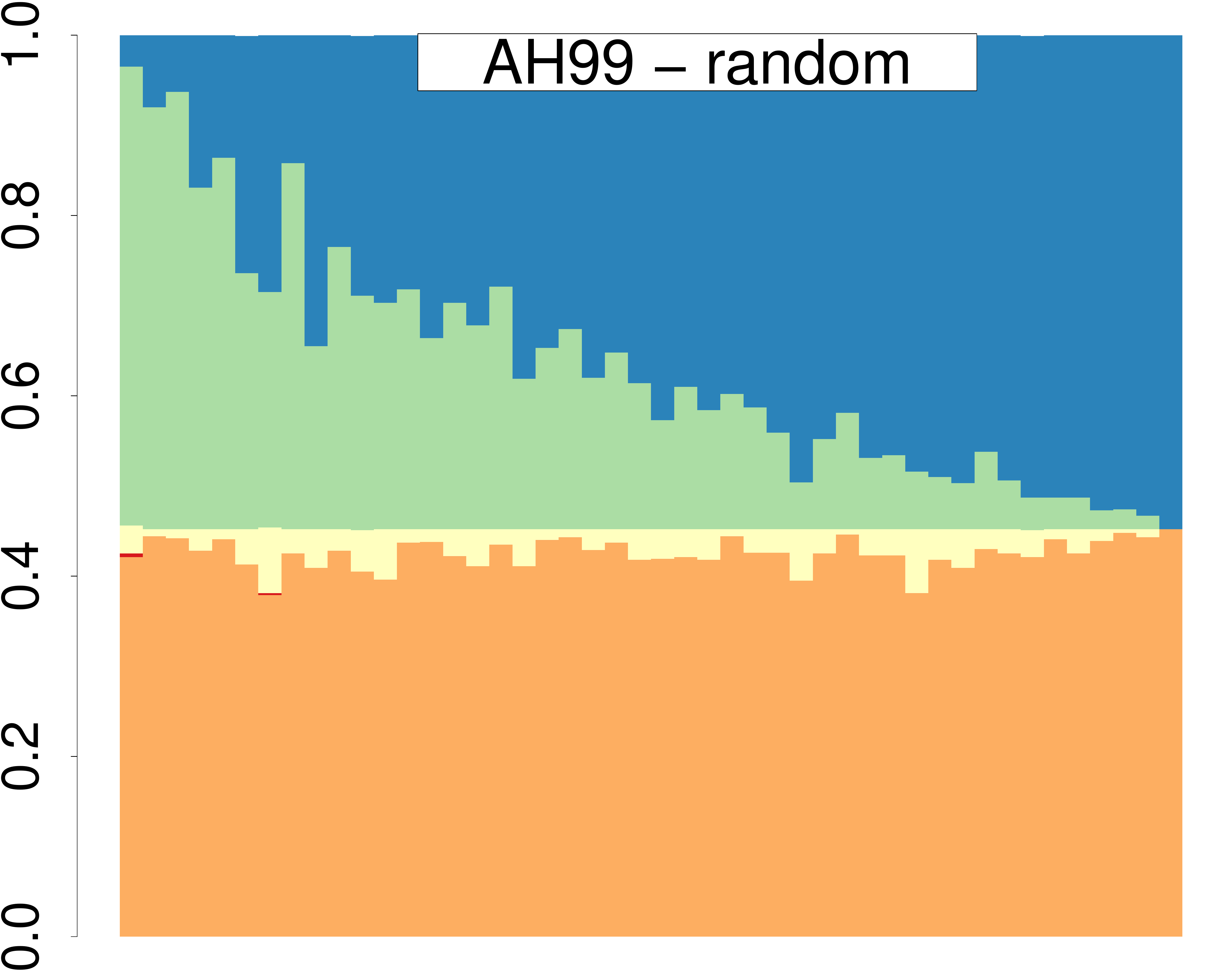}&
\includegraphics[width=4.5cm]{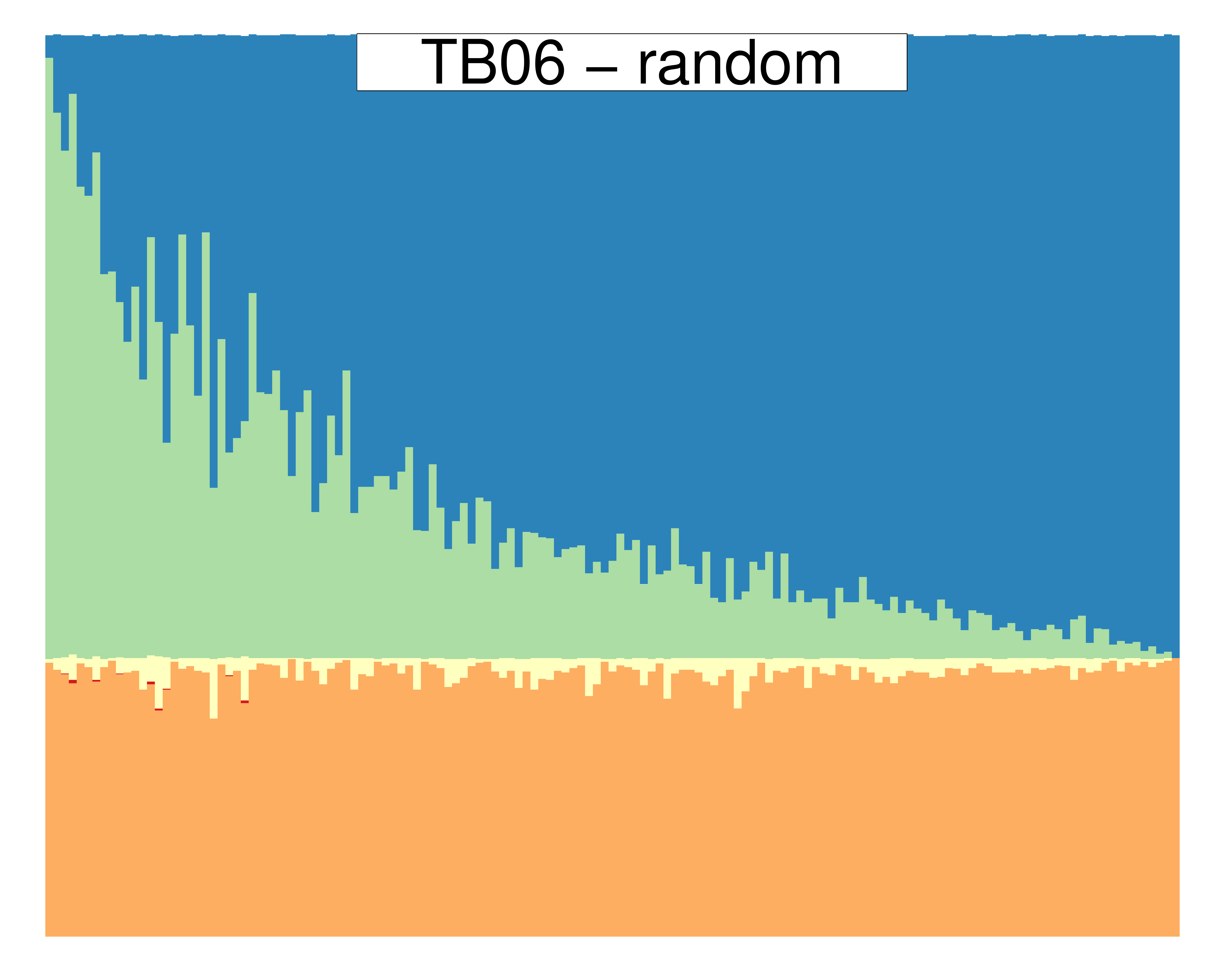}&
\includegraphics[width=4.5cm]{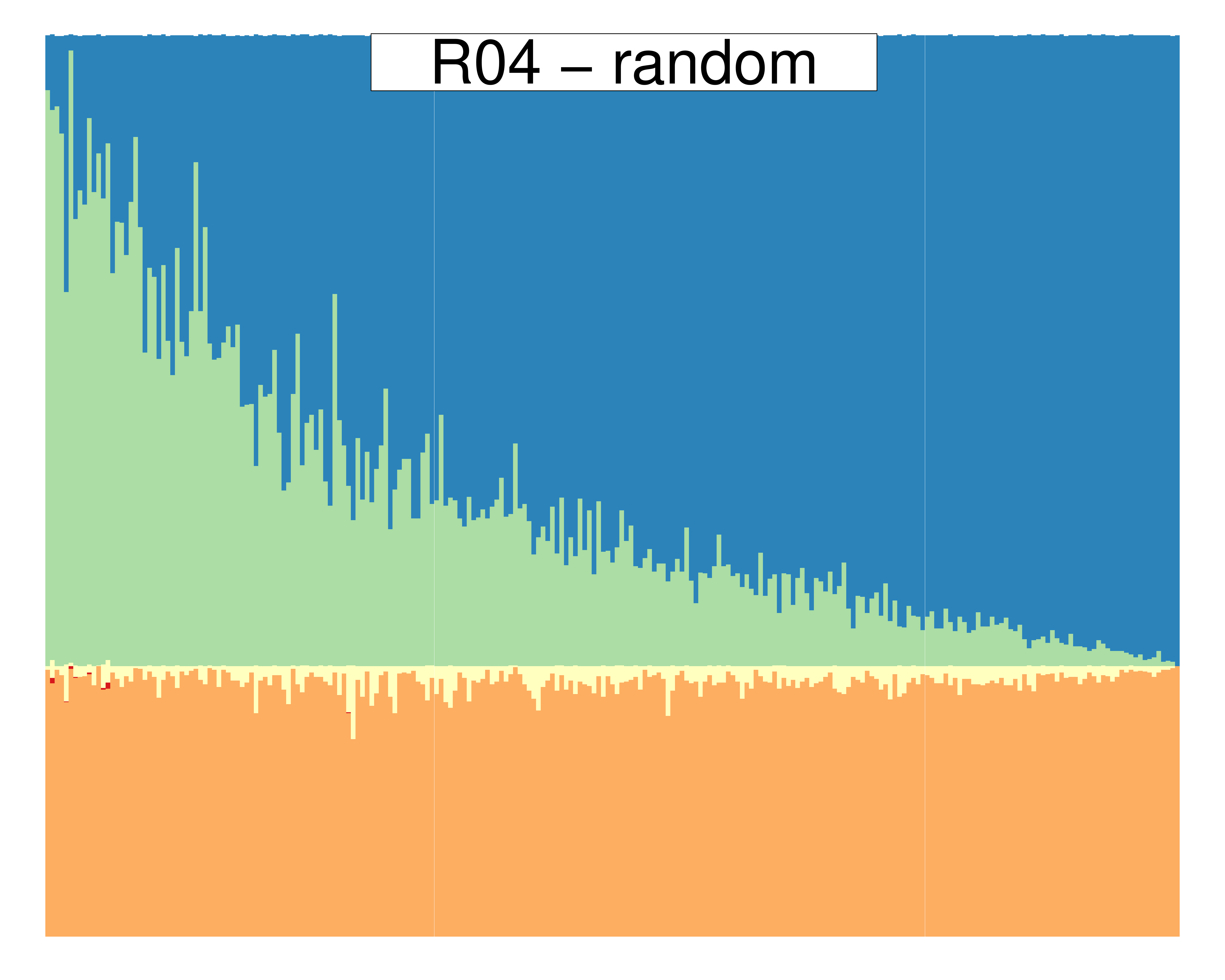}&
\includegraphics[width=4.5cm]{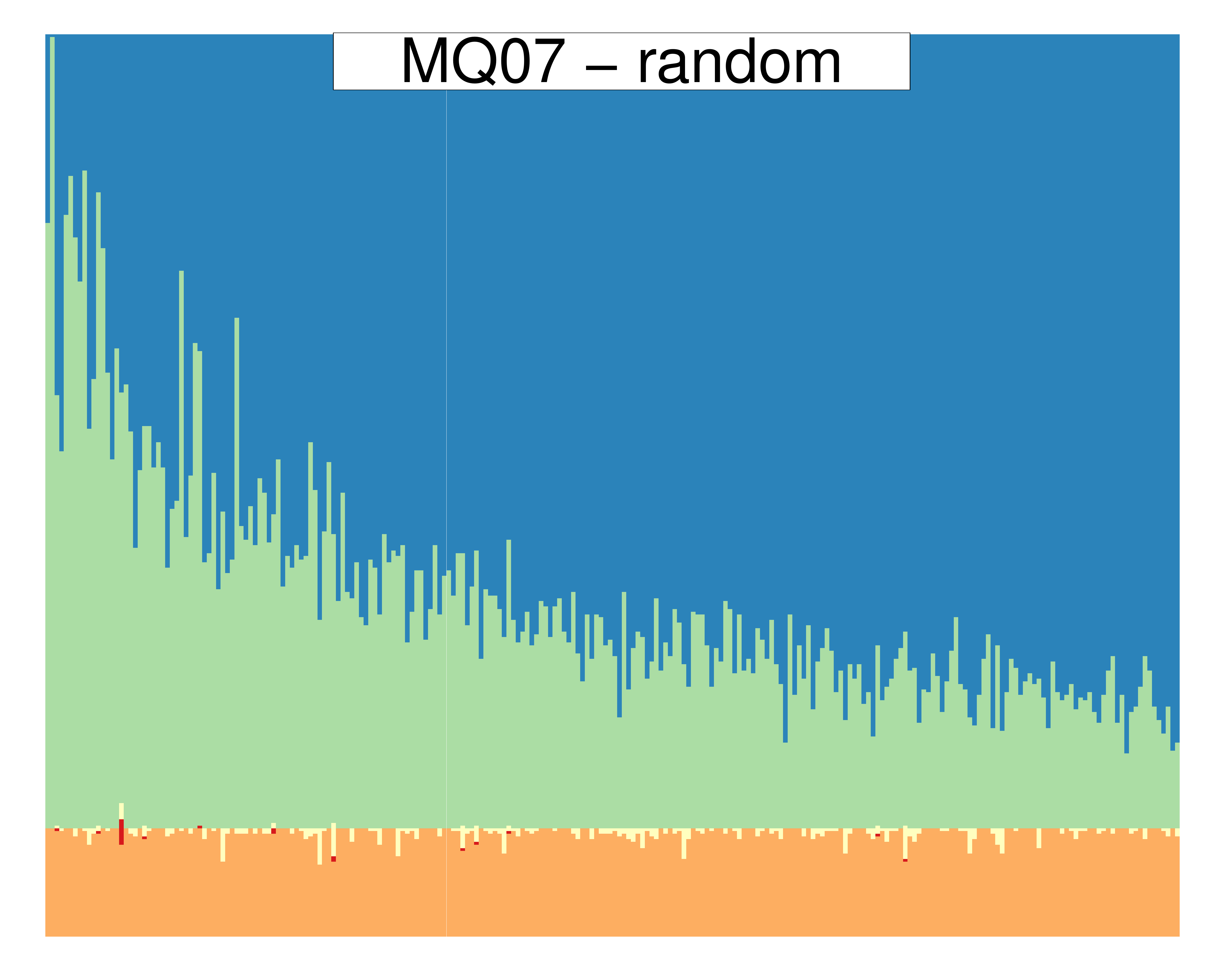}\\

\includegraphics[width=4.5cm]{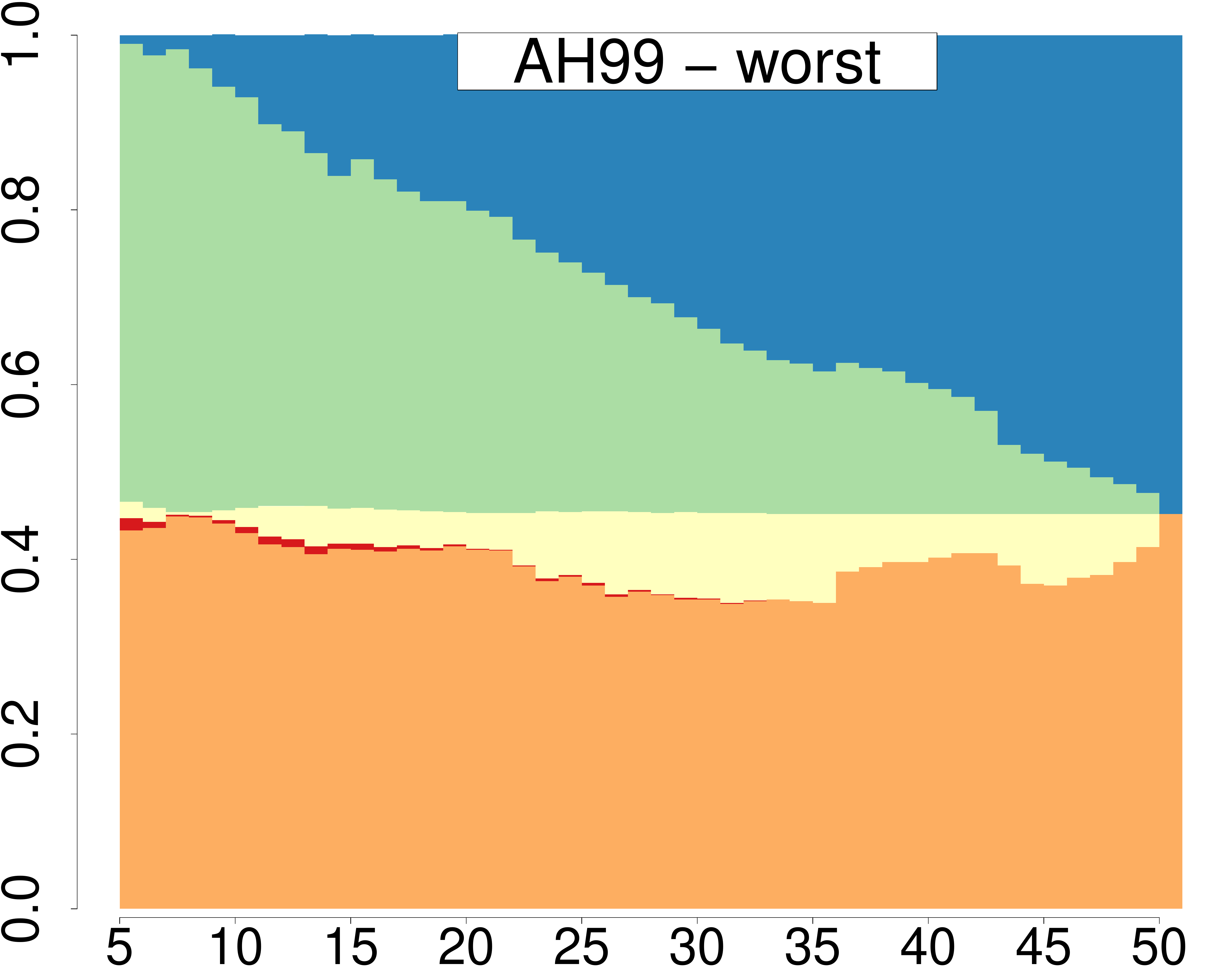}&
\includegraphics[width=4.5cm]{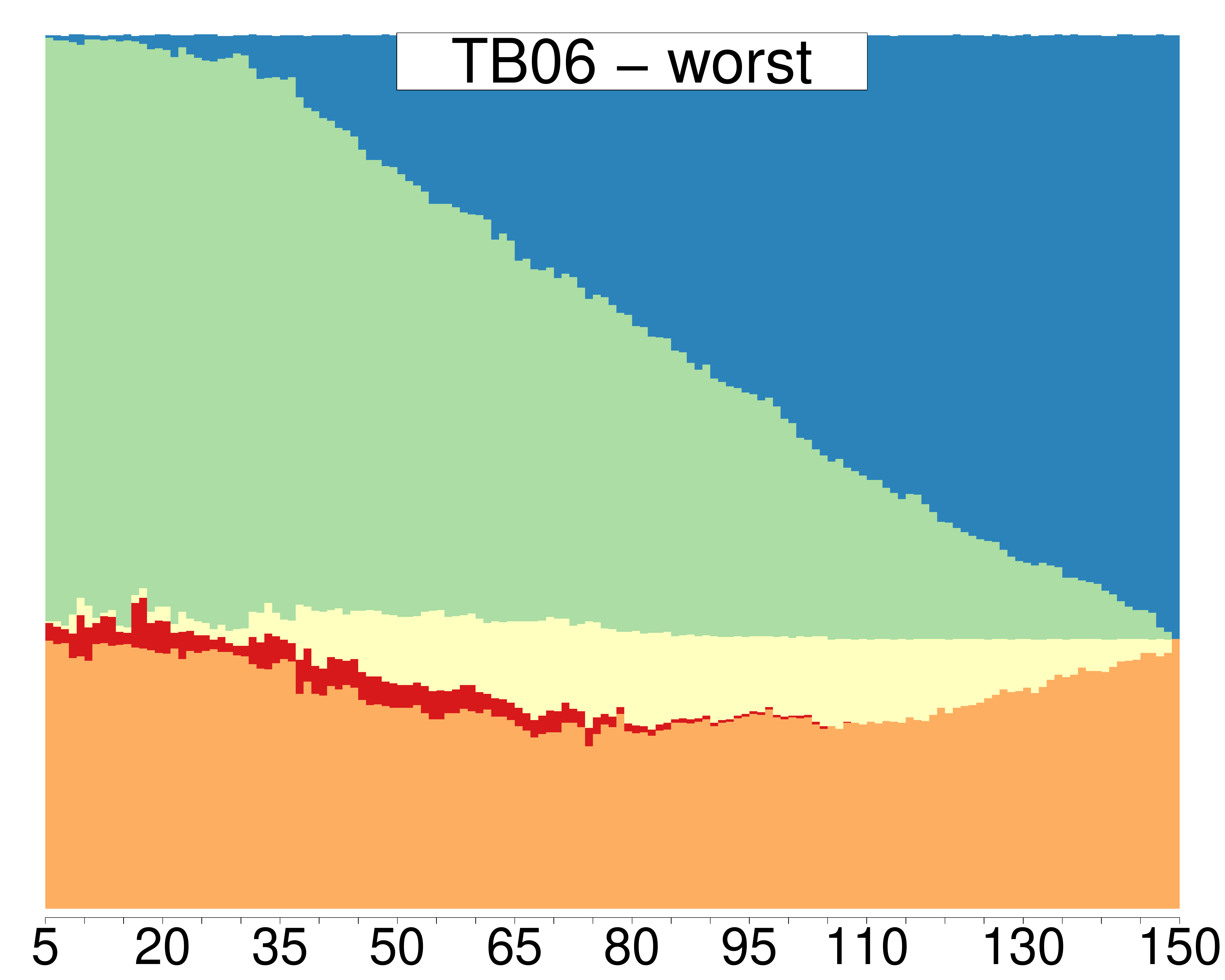}&
\includegraphics[width=4.5cm]{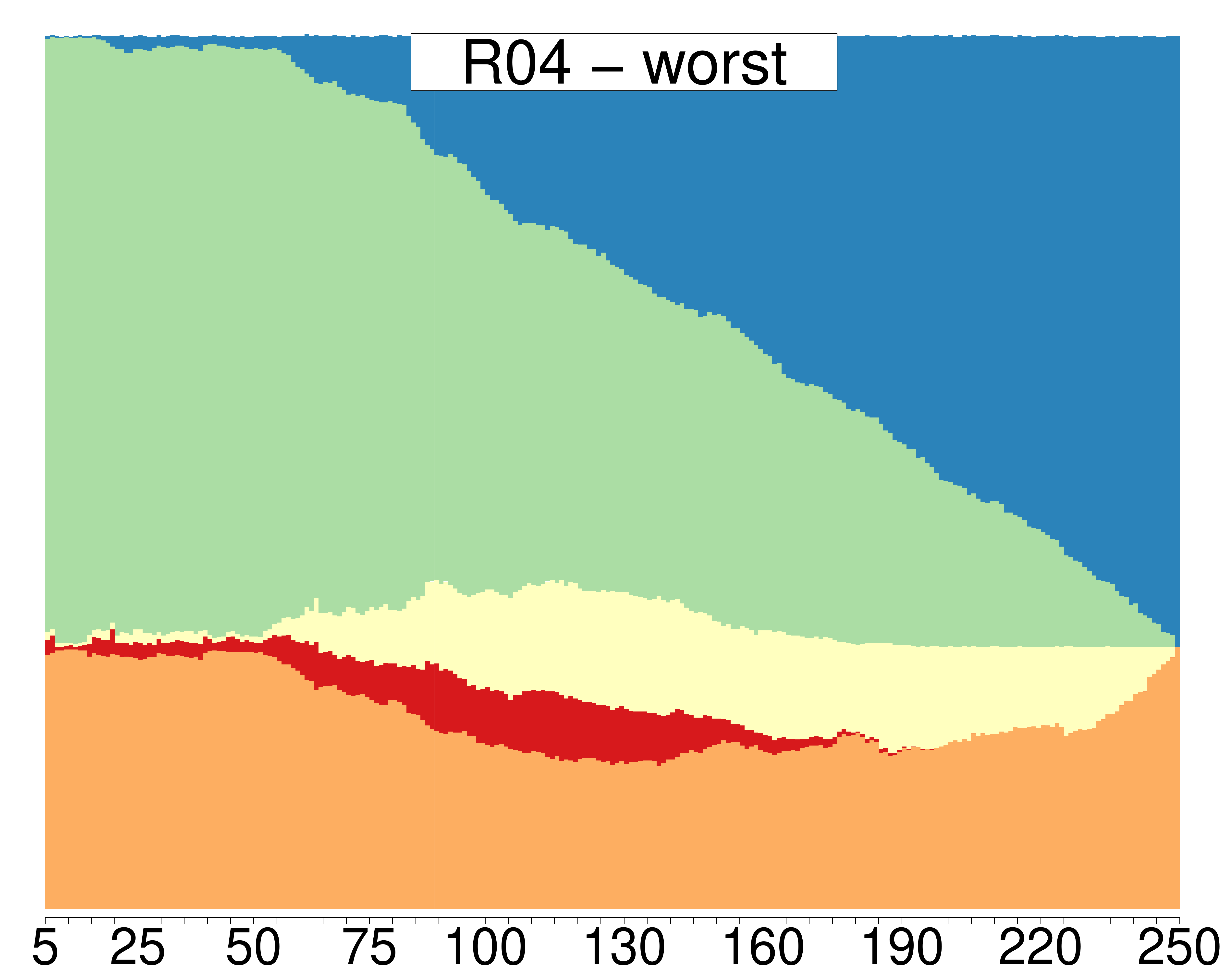}&
\includegraphics[width=4.5cm]{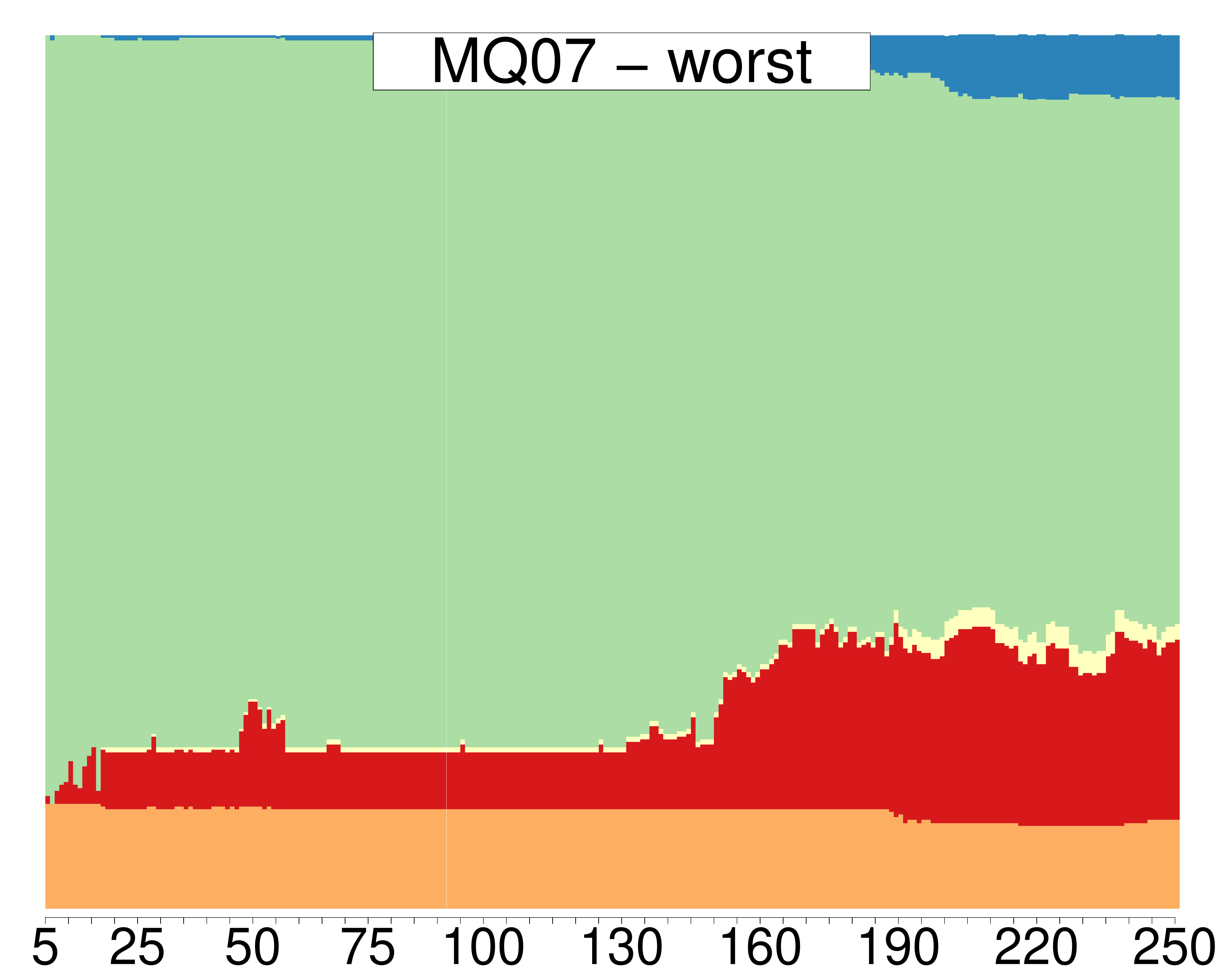}
  \end{tabular}
\vspace*{-2mm}
}
\caption{The results of typical IR effectiveness experiments, showing the proportion of cases
where statistically significant differences (two-tailed paired t-test, $p<0.05$) are observed
between two runs when comparing them using a reduced cardinality (shown on the x-axis) compared to
the full topic set for a collection.
  \label{FewTopics:fig:signif}
}
\vspace{-8pt}
\end{figure*}

The results of the simulated experiments are shown in
Figure~\ref{FewTopics:fig:signif} for the four collections (columns), and for the
best, random,\footnote{Here, for speed of calculation reasons only
a single random topic subset is drawn from the
set of all topic subsets of a given cardinality. The histograms of
random are consequently more ``spiky'' than if we averaged several
random subsets. However, the broad signal of the result is still visible
in the plots.} 
and worst subsets (rows). 
For each sub-figure, the x-axis shows the cardinality of topic set $S$,
which is being compared to the full cardinality ground truth, topic set $G$. The
y-axis shows the proportion of occurrences for each of the five
experimental outcomes: SSA (blue), NS (green), SN (yellow), SSD (red)
and NN (orange).
It can be observed that when the subsets reach their maximum
cardinality (on the right of the plots), only two outcomes are possible,
SSA and NN.
This must be the case, since at full (ground truth) cardinality $S$ and
$G$ are the same, and so the outcomes of the two experiments are
identical.
(Recall that in the MQ07 collection, subsets do not reach full
cardinality by 250.)
When the full topic sets are used, SSA is dominant, accounting for
around 55--70\% of cases.
This is reassuring, since it shows that using the full collections, it
is possible to statistically distinguish between the runs more often
than not.

The figures also clearly confirm that the larger the cardinality, the
higher the likelihood that the same significant evaluation results will
be observed as when using the ground truth topic set.
The NS class shows the cases where a significant difference would be
found between two runs using $G$, but no corresponding difference is
found when using $S$.
For these cases, the reduction in topic set cardinality has compromised
the ability of the significance test to identify significant effects, a
false negative.
Moreover, when comparing the charts ``horizontally'', the NN orange
areas decrease with the full cardinality of the dataset, in both best
and random, while the SSA blue areas increase.
As expected, results tend to be statistically significant more often
when the full cardinality of the ground truth is higher.

Considering the best, random, and worst topic subsets, over all four
collections, as cardinality increases, the best subsets lead to the
most rapid maximizing of the SSA class (and quickest reduction of the
NS class), though on the MQ07 collection, the best subset is only
somewhat better than random.
The best subsets, besides allowing to use fewer topics in evaluation,
also lead to finding SSA results (both in agreement with the
ground truth and statistically significant) more often than random
topic subsets.

The worst subsets lead to experimental significance results that
have the least correspondence with the ground truth topic set.
Perhaps unsurprisingly, the heuristic that selected the best and
worst subsets, which was optimized to maximize and minimize
$\tau$ respectively, also maximized and minimized on significance.


The size of the SN category (false positive) is generally very small --
there are few cases where significant differences are detected on $S$
while no significance is found on $G$.

The most problematic case, SSD, where one run is significantly better
than another on topic set $S$, while the other run is found to be
significantly better on topic set $G$, is fortunately rare, although it
should be noted that it is possible, for all collections except AH99,
to find a (worst) subset of topics of rather high cardinality that
would lead to such contradicting results.
In particular, for the MQ07 collection, even by cardinality 250, for
the worst subset, the proportion of SSD cases is substantially higher
than SSA cases.
In other words, the worst chart for MQ07 shows that we can not only
generate a topic subset of cardinality 250 with strong, and
statistically significant, negative correlation with the full set (as
already shown by the MQ07 series in the worst plots in
Figure~\ref{FewTopics:fig:corr_BWA} and, more in detail, by the last plot of
Figure~\ref{FewTopics:fig:scatterplots}), but moreoever that the ``aberrant''
subset of MQ07 topics of cardinality 250 would also feature a very
small amount of SSA and NN, and many NS and even SSD.
Experiments using that subset would lead a researcher to derive a
statistically significant result that is very different from the full
set.
Whether this is a temporary manifestation, or is maintained into higher
cardinalities, needs to be investigated in future work.
However, it must be noted that best, random, and worst charts for MQ07
are consistent with the other datasets (once the fact that we do not
reach the full cardinality for MQ07 is taken into account).

By and large the NN cases stay constant over the best and random topic
subsets and there is only some variation of these on the worst subset.

\subsubsection{Conclusions}
Overall, this analysis demonstrates that while it is possible to find a
subset of topics that lead to run effectiveness rankings that
are highly correlated with rankings from a ground truth set,
a side effect of doing so is that a researcher is
sacrificing the ability to identify statistically significant
differences between runs.

An experimenter using a topic subset in general does not risk
having to deal with false positive significant results, however, they
do risk having a number of false negatives in their experiments. As
seen in the ratio of SSA to NS in the plots, the magnitude of the
problem reduces as the subset cardinality increases. Indeed,
many experimenters might view a small amount of NS acceptable
if it means they can build their test collection more quickly using
fewer resources.

Perhaps more worryingly,  it is difficult to state that the topics used in IR test collections
are  sampled randomly and independently from the
population of all topics: TREC topics are created by analyzing the document collection and by selecting those topics that, for example, guarantee a minimal number of relevant documents. The bias introduced by such a process is far from being clearly measured. Therefore, one might question the general applicability, in IR evaluation,
of statistical tests which usually require specific conditions, and how much confidence one
should attach to such results in terms of estimating the
generalizability of experiments to larger topic sets.

\subsubsection{Caveat} \label{FewTopics:sect:caveat}
It should be noted that this simulation of typical IR experiments
includes a large number of pairwise significance tests.
One might therefore argue that corrections for multiple testing, such
as the Bonferroni correction~\citep{Fei02}, should be applied.
However, while individual researchers might use such corrections from
time to time, the fact is that IR test collections are used again and
again, often to compare against standard baselines, and there is no way
of knowing what corrections should be made to account for all (reported
as well as unreported) tests that are ever carried out by the
population of IR researchers as a whole. 
Not applying multiple comparison corrections is therefore a more
accurate simulation of the typical IR experimental environment.
This choice is also supported by \citet{carterette2012multiple}, who argues that it is not clear how to properly correct values in a TREC-like setting, or whether it should be done at all.
In this respect, it has to be remarked that in the second method of Section~\ref{FewTopics:sec:sakai} the variance estimates are indeed computed applying the correction method for multiple comparisons. 

Finally, we note that the $t$-test is the most widely used statistical test in IR
experiments~\citep{Sakai16}; however, we also repeated the experiment
using the 
Wilcoxon signed rank
test instead of the $t$-test, and the trends were consistent. 
\section{RQ3: Topic Clustering}
\label{FewTopics:sec:clustering}

We now turn to RQ\ref{FewTopics:RQ:3}.
As already stated in Section~\ref{FewTopics:sec:introduction}, it seems
intuitive that by (i) clustering the topics, and (ii) selecting
representatives from each cluster, the topic set obtained should be
more representative of the full ground truth than an average or random
topic subset of the same cardinality~\cite{DBLP:conf/iir/RoiteroM16}. Furthermore, the selection of a subset of representative queries has been proven to be effective in a Learning to Rank scenario \citep{Mehrotra:2015:RIQ:2766462.2767753}, and indeed the clustering of topics approach follows the same principles as it is clearly based on the representativeness notion.
Therefore, a topic clustering process should be an effective strategy to find good topic subsets.
However, such a process could involve many different settings.
We present several approaches, their results, and a discussion on clustering effectiveness.

We start by presenting in Section~\ref{FewTopics:sec:cl:setting} the overall experimental setting. We then discuss two possible approaches: the first in Section~\ref{FewTopics:sec:card_driver_clust}, that will is not effective despite attempting many variants, and the second, in Section~\ref{FewTopics:sec:card_independent_clust}, which is slightly more effective than the first. Section~\ref{FewTopics:sec:cl:disc} discusses the clustering approach.

\subsection{Experimental Setting}
\label{FewTopics:sec:cl:setting}

We start by defining the experimental settings and notation that are common to the experiments described below. 

\subsubsection{The Clustering Process}

We denote with $n$ the number of topics and with  $c \in \{1,\ldots, n\}$ the cardinality of the topic subset; also, $m \in \{2, \ldots, n\}$ is the number of clusters obtained when performing a clustering process. Our method is composed of the following three steps. 

\begin{enumerate}
\item For each cardinality $c$, we build a set of $m$ clusters.  
\item Then a topic subset of cardinality $c$ is formed by selecting  random representatives from each cluster. In the following we refer to this selection method as \emph{one-for-cluster} (note that one might devise different methods, e.g., selecting a number of topics proportional to cluster size, selecting from some clusters only, etc.)
\item Finally, we build the usual correlation curves, and we compare the one-for-cluster series with random topic selection, which is the average series (such as the ones represented in Figure~\ref{FewTopics:fig:corr_BWA}, left-side, second row).
\end{enumerate}

We use a standard, effective clustering algorithm, \emph{hierarchical} clustering with a \emph{complete linkage} method, and the \emph{cosine similarity} as the distance function. We also try variants, as specified below.
We conduct 10,000 repetitions to compute the one-for-cluster series, to avoid noise.\footnote{\label{FewTopics:fn:1M}We tried with up to 1 million repetitions, but the series are already stable with 1,000 repetitions.}
Note that we are only considering clustering as the main analysis technique. We leave as future work more complex machine learning approaches, that could make use of multiple features such as for example the $\mu$ and $\sigma^2$ parameters from Section~\ref{FewTopics:sec:simulation}.

\subsubsection{Feature Space}
We take as topic features the AP (or statAP) values over the run population, by clustering topics in a multi-dimensional space, where each dimension is the effectiveness on a specific run,
and each topic is a vector of AP (statAP) values.
The idea is that topics that have similar AP values for all runs are redundant: one topic should be as effective as all of the ``similar'' ones. 
Clustering should group together those topics that have similar scores, and by picking representatives from each cluster we should select a good topic subset.
For each dataset, the number of dimensions is therefore the number of
runs (the last column in Table~\ref{FewTopics:tab:coll}). We also experiment with a variant of this approach, as detailed below.

\subsubsection{Number of Clusters and Topic Cardinality}

We can think of two possible overall settings, that affect Steps 1 and 2 above. For each cardinality $c$ and number of clusters $m$: 
\begin{itemize}
\item  We can perform clustering with the constraint $c=m$; we refer to this setting with the term \emph{cardinality-driven clustering};
\item  We can determine the number of clusters a priori, independently from $c$, and subsequently select the topic subset;  we refer to this setting with the term \emph{cardinality-independent clustering}.
\end{itemize}

Both settings have pros and cons. 
The first approach forces the clustering algorithm to produce a clustering of exactly $m=c$ clusters, which might be unnatural for certain $c$ values: for example, if the topics are naturally form two clusters, forcing them into three will produce clusters that are less complete and more heterogeneous, thus potentially of lower quality. However, once the clusters are formed, the selection of topics is straightforward, since there is the guarantee that when $c$ topics are to be represented, there are exactly $c$ clusters.  Furthermore, even if the $c=m$ constraint might lead to unnatural clusters for certain $c$ values, in general just decent clusters, even if not perfect, might be of a sufficient quality to guarantee higher correlation values for the one-for-cluster series than for random topic selection.
	
Conversely, with the second setting, the topics can be clustered in a more natural way, but then the selection process is slightly more complicated: there is no equivalence between the number of clusters and the number of topics to select, thus there is not a unique selection method, and the selection process has to take into account the empty clusters that might occur during the process. Finally, whereas with the first setting the choice of the number of clusters $m$ is straightforward and determined, with the second setting $m$ is a parameter to be chosen, and it is not clear which criteria should be used.
In the following Sections~\ref{FewTopics:sec:card_driver_clust} and~\ref{FewTopics:sec:card_independent_clust} we analyze both settings, starting with the first one.

\subsection{Cardinality-driven Clustering} \label{FewTopics:sec:card_driver_clust}

\subsubsection{A First Attempt}

We compute the clustering as described above, with the constraint $c=m$; then, we compare the one-for-cluster with the average series.
It is found, however, that this clustering of topics approach does not result in any
topic subset having a $\tau$ correlation higher than the average; indeed usually $\tau$ is lower.
There are multiple possible explanations for this behavior.
First is the choice of clustering algorithm. Therefore, we tried
different variations of the clustering, for example,
using a non hierarchical algorithm such as K-means (with the algorithm variations Hartigan-Wong, Lloyd, and MacQueen\footnote{See the R function ``kmeans'' in the ``stats'' package (\url{https://stat.ethz.ch/R-manual/R-devel/library/stats/html/kmeans.html}), and ``k-means'' of ``scikit-learn'' for Python 3 (\url{http://scikit-learn.org/stable/modules/generated/sklearn.cluster.KMeans.html}).
      }), and/or using different distance
functions (including as different kinds of proximity
  measures\footnote{For an exhaustive list see the R package ``proxy'' (\url{https://cran.r-project.org/web/packages/proxy/proxy.pdf}), and the ``Distance computations'' section of Python 3 (\url{https://docs.scipy.org/doc/scipy/reference/spatial.distance.html}).
    }
  both linear metrics, e.g., Euclidean, Manhattan, Divergence, etc.,
  and similarity-angular distances, e.g., Cosine, Correlation,
  Jaccard, Phi, etc.), or using
different methods to join clusters (thus different linkage techniques
including single, average, mean, median, Ward).
However, $\tau$ was never found to be higher than average for any of these clustering methods, and 
we can be confident that these negative results are not affected by a particular clustering setting.

A second possible explanation is related to the feature vector:
our feature vectors are in a high-dimensional space, and therefore most of the distances tend to be similar, and vectors tend to be orthogonal.
To be more precise, as soon as the number of dimensions grows, the number of possible distance values drops.  This is a well known phenomenon, referred to as ``the curse of dimensionality''
\cite[Chapter~7]{ullman:mining}, and it occurs for both linear and angular distance values (Cosine, Euclidean, and Manhattan).
This could of course  harm the clustering process.
To address this limitation we tried to combine clustering with dimensionality reduction, as described in Section~\ref{FewTopics:sec:pca}.
Finally, in this setting we have the constraint that the number of clusters $m$ must be equal to the topic subset cardinality $c$, and that could lead to forming unnatural clusters, as already mentioned; we discuss this third possible explanation in Section~\ref{FewTopics:sec:clust:artificial:exp}.

\subsubsection{Dimensionality Reduction} \label{FewTopics:sec:pca}
To deal with the curse of the dimensionality effect,  a second attempt makes use of Principal Components Analysis (PCA).
To express around 85-90\% of the total variance of the data, three components/dimensions are needed for AH99, R04, and TB06, and five for MQ07. 
Each topic vector is then heavily reduced, to very few components:
from the values in the last column of Table~\ref{FewTopics:tab:coll} to 3, 3, 3, and 5, respectively.
We then repeat the clustering process with the same primary settings as above
($c=m$, hierarchical algorithm, cosine distance, and complete linkage).

\begin{figure}[tb]
  \centering
  \begin{tabular}{@{}c@{}c@{}}
    \includegraphics[width=.5\linewidth]{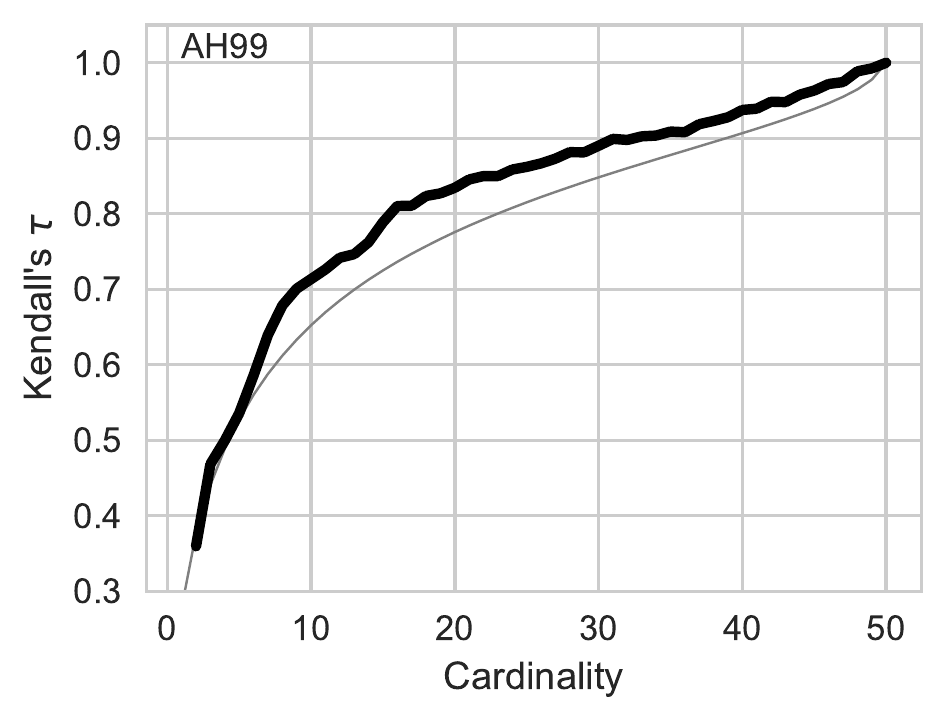}&
        \includegraphics[width=.5\linewidth]{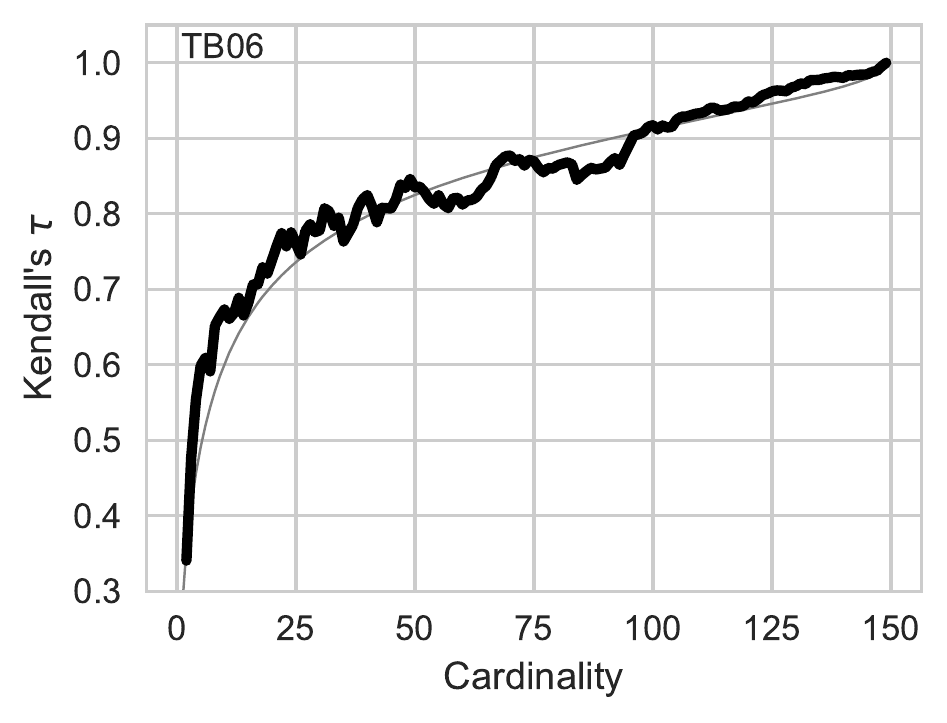}\\
        \includegraphics[width=.5\linewidth]{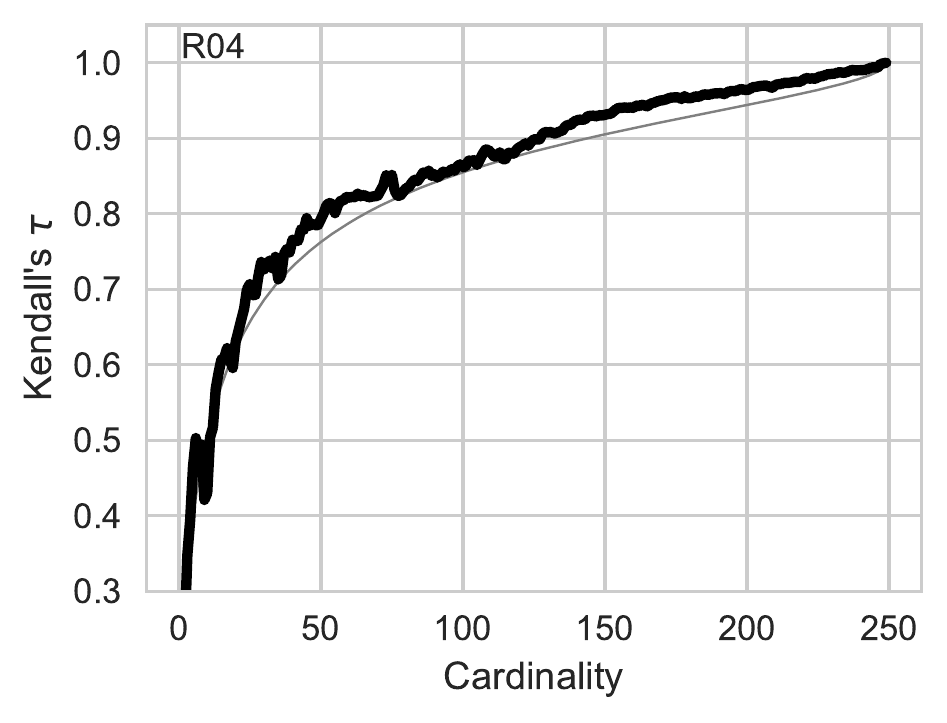}&
        \includegraphics[width=.5\linewidth]{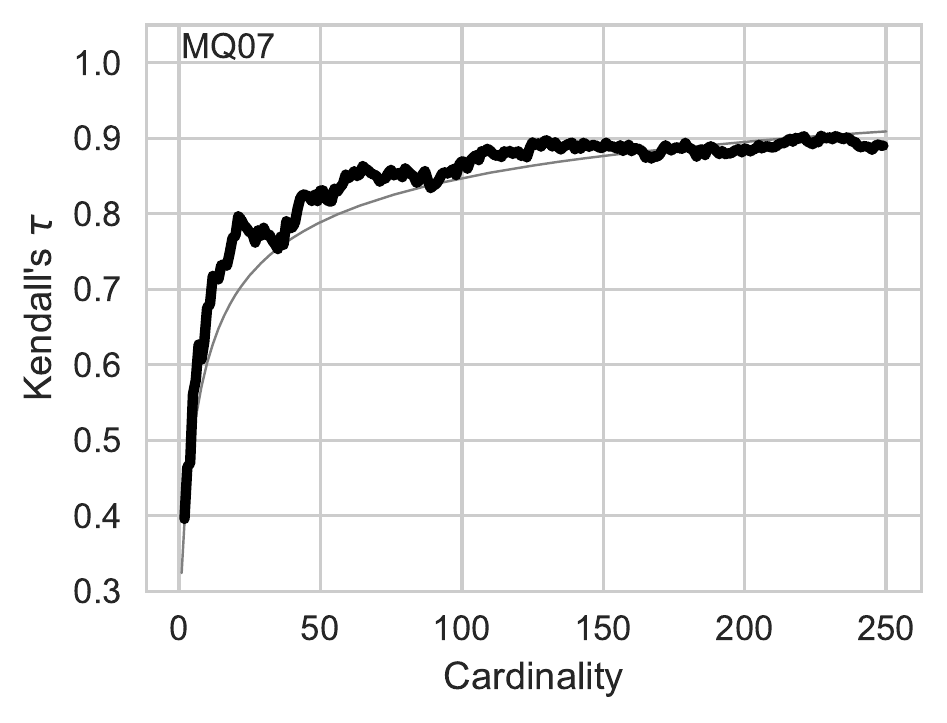}
  \end{tabular}
\caption{Kendall's $\tau$ correlation curves for the four datasets: 
averaged (thinner gray lines), and obtained using clustering (thicker darker lines).
}
  \label{FewTopics:fig:clustering_PCA}
\end{figure}

With PCA, the results are different to clustering.
Figure~\ref{FewTopics:fig:clustering_PCA} compares the correlation curves for average
subsets, which are gray and thin in the figure, with the
correlation curves obtained with the one-for-cluster method:
the latter are usually above the former.
Moreover, the differences between one-for-cluster and average correlation
values are statistically significant for most of the cardinalities: in
around $90\%$ of the $50+249+149+250=698$ total cases for the four
collections, the difference is statistically significant according to
the Wilcoxon signed rank test, $p<.01$,  and there are no noticeable
differences across datasets (the number of statistically significant
cases varies between $86\%$ and $92\%$). 

In summary, topic subsets found by clustering combined with
dimensionality reduction show correlations with the ground truth that are statistically significantly higher than average / random subsets.
However, the difference is rather small: although clustering helps, it helps just a little. Indeed, considering the results of Figure~\ref{FewTopics:fig:clustering_PCA}, one might be tempted to conclude that clustering of topics is not an effective technique, at least with the constraint $c=m$. 
Also, the oscillations of the one-for-cluster correlation curves that can be seen in Figure~\ref{FewTopics:fig:clustering_PCA} call for an explanation.
To address these issues, and to present a detailed analysis of clustering of topics with the constrain $c=m$, we perform another experiment, described in the next section.

\subsubsection{A Simulation Experiment} 
\label{FewTopics:sec:clust:artificial:exp}

To further understand what is happening during the clustering process, and 
to further investigate the capabilities of the clustering process with the constraint $c=m$, as well as the limitations, we design the following simulation experiment.
The aim of the simulation is to show what happens with clustering of topics in an ideal situation, where the data is distributed with a minimum and controlled amount of noise, and the topics are artificially clustered in a neat way. This represents the most favorable scenario for the topic clustering process.
We will discuss the same experiment for cardinality-independent clustering in Section~\ref{FewTopics:sec:card_independent_clust}.

The experiment is as follows.
We select $s$ topics, called \emph{seeds}. We experiment with choosing as seeds the topics from a collection in two ways: either randomly, or choosing a set of well separated topics after projecting the multidimensional topic space onto two dimensions.
In the following, we report the results of the random selection only, as the other one provides a comparable result.

Given the seed topics, we form a set of new topics, placing in the neighborhood of each seed $r$ fictitious topics in a hyper-sphere of radius $\epsilon$; we call these topics the \emph{surrounding} topics of the seed topics.
Thus, we simulate an ideal scenario for clustering of topics where we have $s$ ideal clusters of $r$ 
topics each; $2\epsilon$ is the maximum distance, in terms of AP (statAP), between two topics in the same ideal cluster.
Note that, the higher $\epsilon$, the higher the probability that 
the ideal clusters overlap, and therefore that
a topic, during an automatic clustering process, is placed in a cluster 
different from that of its seed, and of the other topics in the same ideal cluster.

We now perform clustering as we did in Section~\ref{FewTopics:sec:pca}; we use the constraint $c=m$, PCA, hierarchical clustering with a complete linkage method, and the cosine similarity as the distance function.
We vary the three parameters as follows:
$s\cdot r=150$,
with $s \in \{15,30,50\}$, and thus 
$r \in \{10, 5, 3\}$, and
$\epsilon \in \{0.01, 0.02, 0.05\}$.

\begin{figure}[tb]
  \centering
  \begin{tabular}{@{}c@{}c@{}}
    \includegraphics[width=.49\linewidth]{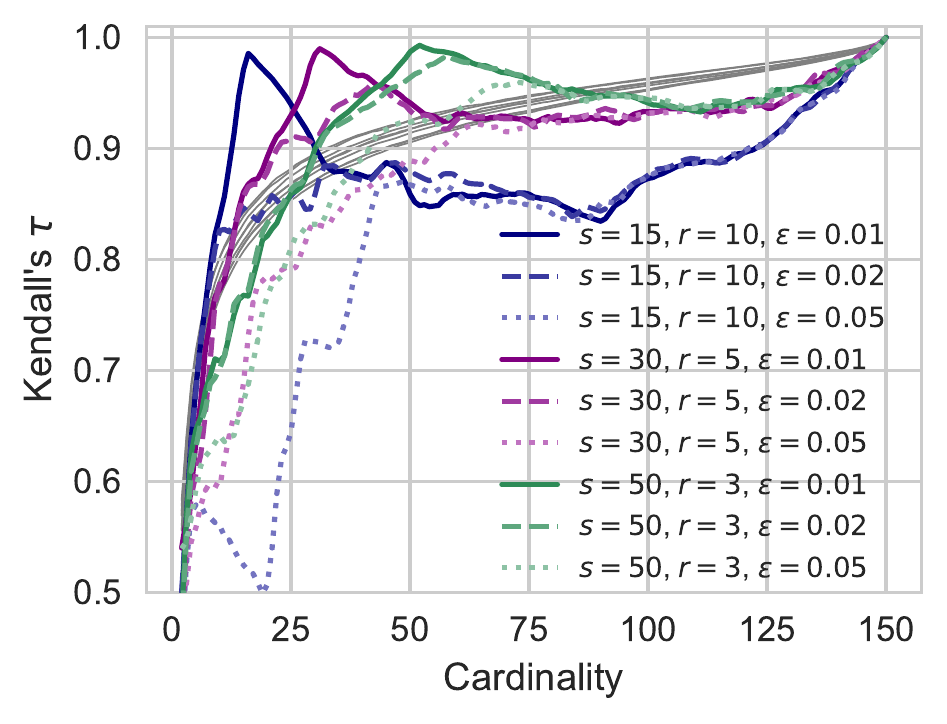}&
    \includegraphics[width=.49\linewidth]{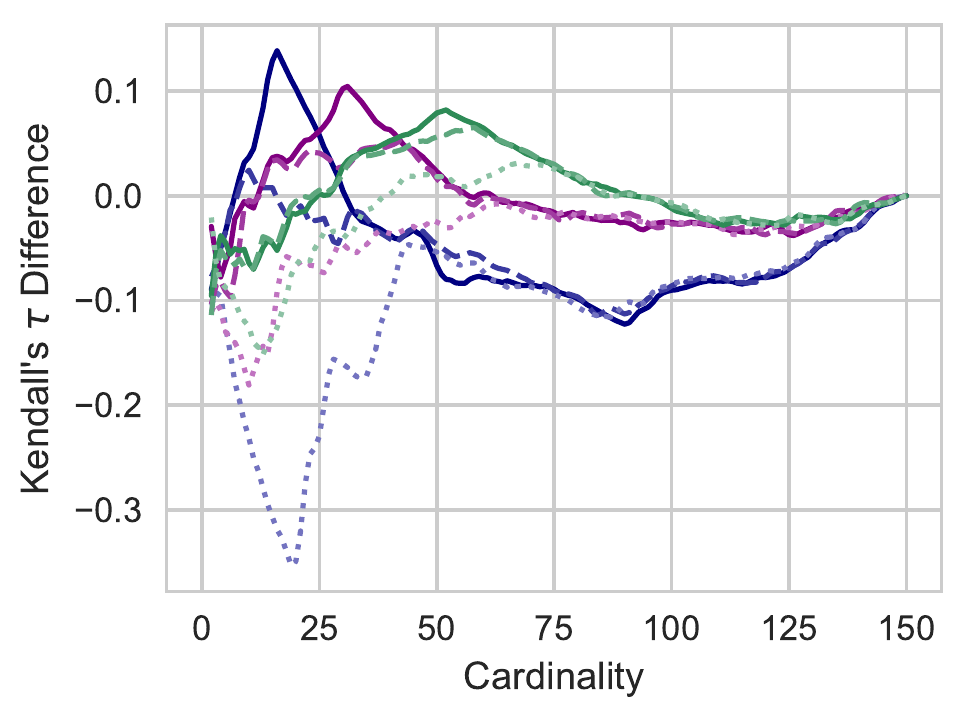}
    \\
 (a)&(b)
  \end{tabular}
  \caption{
  On the left, Kendall's $\tau$ correlation values for the average and the one-for-cluster series.
On the right, the series obtained subtracting the average series to the one-for-cluster one.
  The three series represent the three number of topic seeds: 15, 30, and 50, represented with different colors.
  The different colors and line types of the series represents different epsilon values. 
  The series are smoothed using a mobile mean with a window of three elements.
  The  gray  lines represent the average series.
  }
  \label{FewTopics:fig:clustering_SEED_PCA_AllCollections}
\end{figure}

Results of the experiment are shown in Figure~\ref{FewTopics:fig:clustering_SEED_PCA_AllCollections}. In panel (a), 
the  one-for-cluster series for the three topic seeds (15, 30, and 50) are represented with different colors, and the different line types (continuous, dashed, and dotted) identify  the different $\epsilon$ values (0.01, 0.02, and 0.05). The figure also shows the average series as gray thin lines.
Figure~\ref{FewTopics:fig:clustering_SEED_PCA_AllCollections}(b) shows the same data with a different representation. Each series is obtained subtracting the corresponding average series from the one-for-cluster one. The horizontal gray line highlights the value of zero: if the series in the plot is above zero it means the one-for-cluster series has higher correlation values  with the ground truth than the average one, if the series is below zero vice-versa the average series has higher correlation values.

We can draw several conclusions from these results. Looking at the highest peaks, one can see that they occur at cardinalities corresponding with the number of ideal clusters (equal to the number of seeds $s$): the clustering approach works well if the topics can be ``naturally clustered'' in a number of clusters corresponding to the cardinality of the subset of a few good topics; this is true when all the surrounding topics are placed in the same cluster as the corresponding seed topic.
However, the further the cardinality is from this ideal number of clusters (the number of seeds), the more the correlation of the one-for-cluster series decreases, and becomes comparable with the random selection of topics (the average series), or even worse.

Focusing on the ``negative'' peaks (e.g. for the series with 15 seeds, for $\epsilon= 0.05$ at the cardinalities around 20, and for all three $\epsilon$ values at cardinalities around 90) we note that the negative peaks achieve lower values of correlation as $\epsilon$ increases, as expected.
These negative peaks confirm that, if a natural clustering of topics is not possible, clustering of topics worsens the selection of a few good topics with respect to random selection.
This effect can be explained by looking at the composition of the clusters produced during the cluster process, where we notice that
surrounding topics of different seeds indeed tend to be clustered together even when $\epsilon$ is small. This is likely caused by the constraint $c=m$, that forces the number of clusters.
Furthermore, the desired behavior would be that when increasing cardinalities, the clusters split into balanced sub-clusters; for example, with $s=15$, at cardinality 30 each cluster containing the seed should split into 2 balanced clusters, at cardinality 45 into 3 balanced clusters, and so on. However, in practice this is not the case: on the contrary, there are always few clusters split into smaller clusters, while other larger clusters remain intact.
This results in a ``bad'' clustering of topics: in the one-for-cluster series the majority of topics come from more fragmented clusters. We can say that the more fragmented weight more than the other in the evaluation; on the contrary, the average series chooses topics uniformly.

Finally, we note that there are some lower positive peaks in the series. 
For example, see in the chart on the right the series with 30 seeds with $\epsilon= 0.02$, for the cardinalities around the values of 18, 22, 39, and 41.
These lower peaks suggest that it is not always the case that the data can be explained with only one number of clusters, but multiple numbers of clusters are possible to obtain a natural clustering of topics.

Summarizing, it seems reasonable to conclude that the $c=m$ constraint makes clustering ineffective for most of the cardinalities, even in the most favorable scenario. 
Moreover, considering real data, $\epsilon$ will be quite high, since in general it is unlikely that our vectors (topics) have similar values, with just a small $\epsilon$ difference. Thus cardinality-driven clustering does not seem to be a feasible technique to be applied on real data. For this reason, in the following we study cardinality-independent clustering, starting by repeating the simulation experiment of this section.

\subsection{Cardinality-independent Clustering} \label{FewTopics:sec:card_independent_clust}

In our previous experiments, the number of clusters is equal to the number of selected topics.
Now, we perform clustering of topics with a  number of clusters $m$ independent from the topic subset cardinality $c$ and hopefully matching the number of clusters in a natural clustering.
%

\subsubsection{The Clustering Process}\label{FewTopics:sec:indep-process}

In the case of cardinality-independent clustering, differently from cardinality-driven clustering, $m$ is a parameter to be chosen.
There are several ways of selecting such a parameter. 
The first alternative is to try all possible values from 2 to the number of topics.
A second approach could be to rely on some index of goodness of the obtained clusters.
Another possibility is to look at the results of cardinality-driven clustering: in cardinality-driven clustering, due to the constraint $c=m$, the positive peaks in the one-for-cluster series (see Figures~\ref{FewTopics:fig:clustering_PCA} and~\ref{FewTopics:fig:clustering_SEED_PCA_AllCollections}) correspond to $m$ values leading to an effective clustering of topics; this fact can be exploited to choose the value of $m$ for the cardinality-independent clustering: we can focus on the cardinalities corresponding to  the positive peaks of the one-for-cluster series in cardinality-driven clustering.
In the following we investigate the latter approach;
we also tried various indexes on clustering goodness 
(e.g Connectivity, Dunn, and Silhouette)
with no positive result, and we leave for future work the study of other feasible a priori approaches to find $m$.

Once the $m$ clusters are formed, the probably most natural algorithm for selecting the topics from the clusters is as follows.
Considering the one-for-cluster series, there exist three possibilities for each cardinality $c \in \{1,\ldots,n\}$:
 
\begin{itemize}
\item Case $c<m$: we select randomly $c$ clusters, and then we select $c$ elements, one for each cluster.
\item Case $c=m$: we select one topic per cluster, as we did in the case of cardinality-driven clustering (Section~\ref{FewTopics:sec:card_driver_clust}). 
\item Case $c>m$: 
we select $m$ topics as in the previous $c=m$ case; we then repeat for the remaining $c-m$, until we fall in the first $c<m$ case. When a cluster becomes empty during the process, we skip it in the following iterations.
\end{itemize}
Note that cardinality-driven and cardinality-independent clustering coincide only when $c=m$.

\subsubsection{Cardinality-independent Clustering on the Simulated Example} \label{FewTopics:sec:toy-fixed}

\begin{figure}[tb]
  \centering
  \begin{tabular}{@{}c@{}c@{}}
    \includegraphics[width=.49\linewidth]{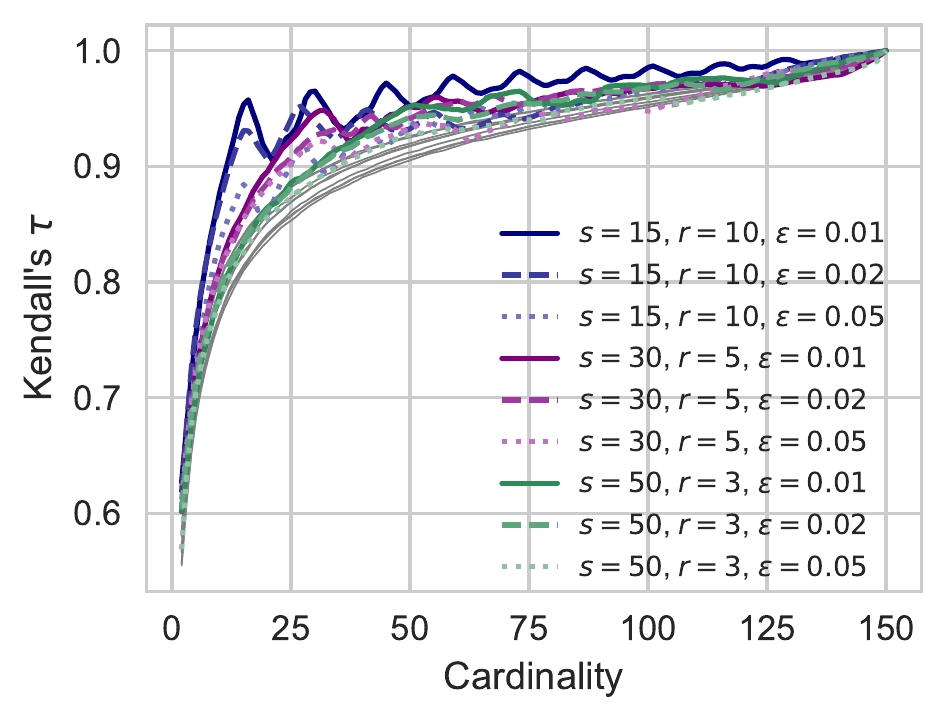}&
    \includegraphics[width=.49\linewidth]{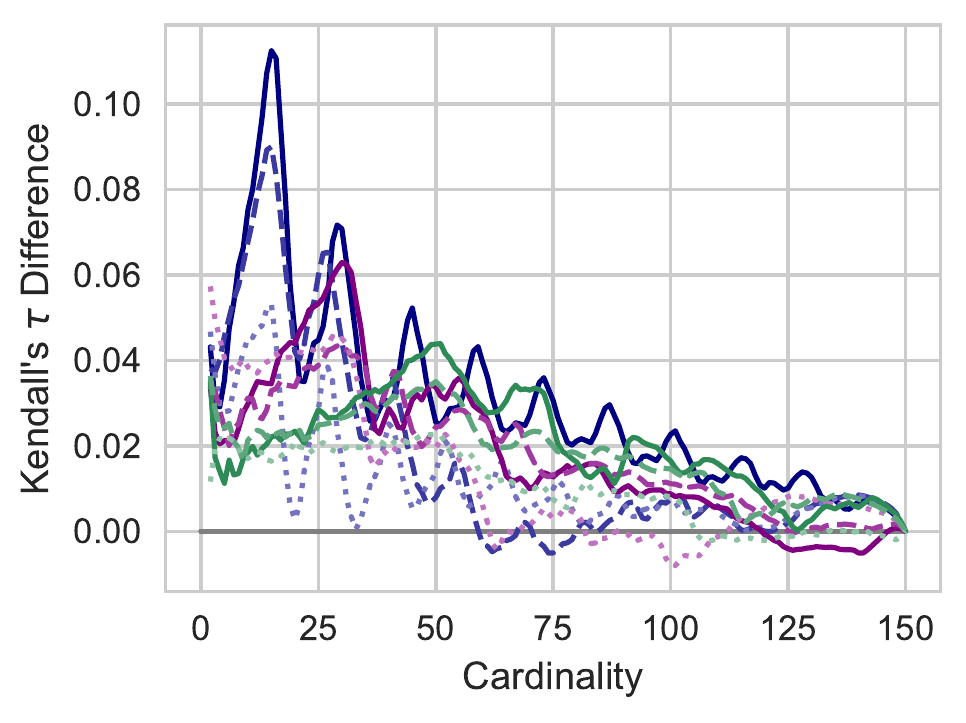}
    \\
 (a)&(b)
  \end{tabular}
  \caption{
    Results of the cardinality-independent clustering for the artificial experiment.
  Compare with Figure~\ref{FewTopics:fig:clustering_SEED_PCA_AllCollections}.
  }
  \label{FewTopics:fig:clustering_SEED_PCA_FixedCard_AllCollections}
\end{figure}
Figure~\ref{FewTopics:fig:clustering_SEED_PCA_FixedCard_AllCollections} shows the results for cardinality-independent clustering for the same simulated  experiment. The figure shows that in general, we obtain topic subsets that always have higher $\tau$ values than the average; this holds for almost all the $s$, $r$, and $\epsilon$ values.

Also, there are several positive peaks in the series. These occur at cardinalities corresponding to multiples of the number of topic seeds $s$; e.g. considering $s=15$, the positive peaks are around cardinalities 15, 30, 45, and so on. This is an indication that multiple effective $m$ values exist. 
Indeed, clustering is effective not only for $m$ corresponding exactly to the cardinalities of the peaks, but also for near values, and this fact can be exploited for $m$ selections. 

Finally, the lower negative peaks of Figure~\ref{FewTopics:fig:clustering_SEED_PCA_AllCollections} almost disappear, even for the largest $\epsilon$ value of $0.05$: even if the topics are difficult to cluster, the clustering process is still effective.

\subsubsection{Cardinality-independent Clustering on Real Data} \label{FewTopics:sec:real_data_fixed}

\begin{figure}[tb]
  \centering
  \begin{tabular}{@{}c@{}c@{}}
    \includegraphics[width=.5\linewidth]{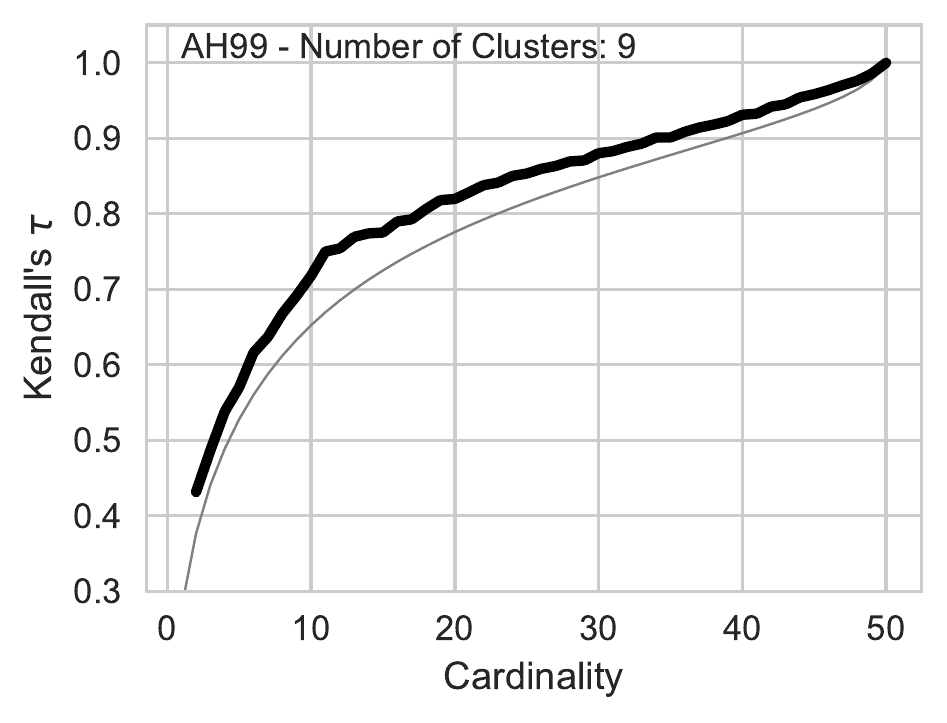}&
        \includegraphics[width=.5\linewidth]{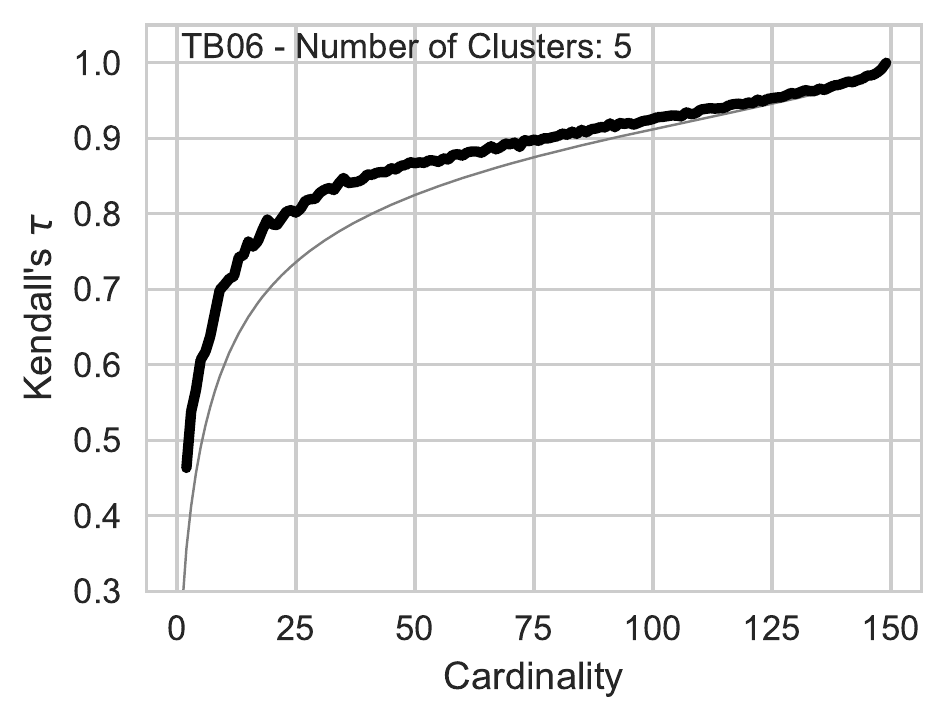}\\
        \includegraphics[width=.5\linewidth]{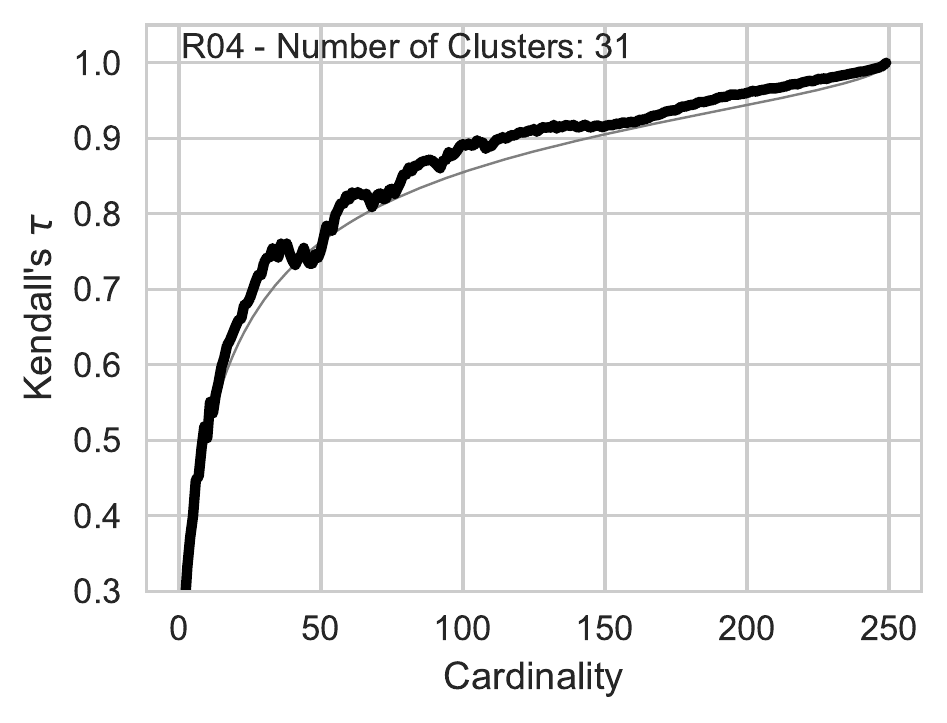}&
        \includegraphics[width=.5\linewidth]{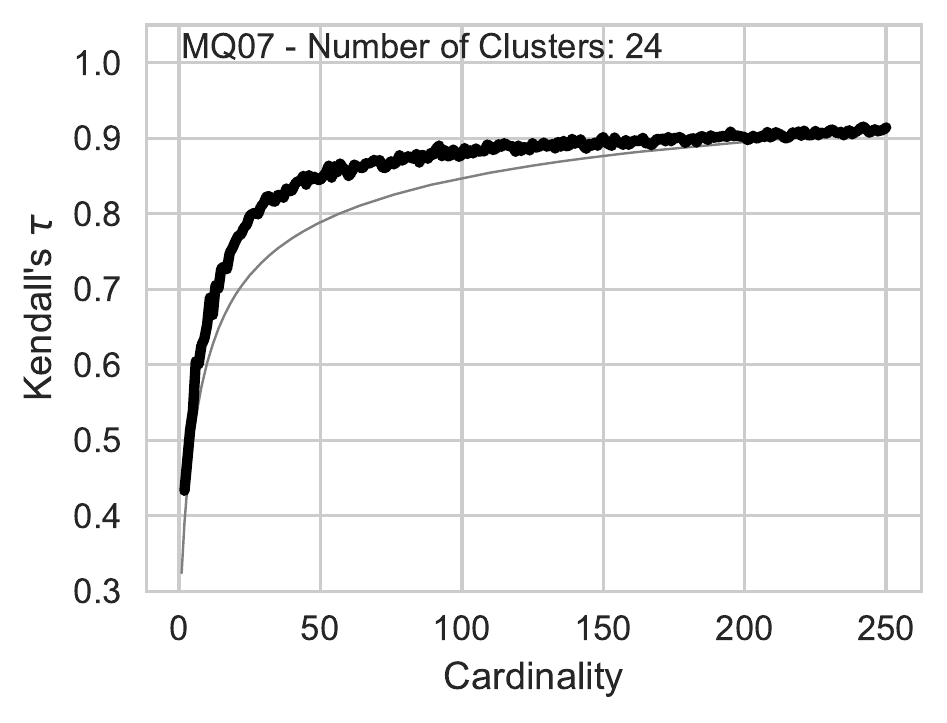}
  \end{tabular}
\caption{Kendall's $\tau$ correlation curves for the four datasets: 
averaged (thinner gray lines), and obtained using clustering (thicker darker lines).
}
  \label{FewTopics:fig:clustering_PCA_FixedCard}
\end{figure}
Figure~\ref{FewTopics:fig:clustering_PCA_FixedCard} shows the results of cardinality-independent clustering for the real-data experiment, for some selected $m$ values, corresponding to the cardinalities of the positive peaks of the series of Figure~\ref{FewTopics:fig:clustering_PCA}: 9 clusters for AH99, 
5 clusters for TB06, 
31 clusters for R04, and 
24 clusters for MQ07.
We choose to report the results corresponding to the highest peak at the lowest possible cardinality: 
for AH99 a similar behavior is found for cardinalities  16, and 31,
for TB06 for cardinalities 22, 43, 45, and 75, 
for R04 for cardinalities 16, 25, 31, 45, 60, and 75, 
and finally for MQ07 for cardinality 64.

The figure
shows that the one-for-cluster series always has higher $\tau$ values than the average series, for all the collections, with a single exception for R04. Cardinality-independent clustering is effective. 
As we have seen in Section~\ref{FewTopics:sec:card_driver_clust}, in the cardinality-driven clustering, even with PCA, the one-for-cluster series is 
often significantly lower than the average in the simulation experiment (see Figure~\ref{FewTopics:fig:clustering_SEED_PCA_AllCollections}) and sometimes lower in the real datasets (see Figure~\ref{FewTopics:fig:clustering_PCA}). In cardinality-independent clustering, this never happens: in the least favorable case, the one-for-cluster and average series are equivalent (the series overlap).

We also verified that the series oscillations do not depend on noise, as they still occur with 1M repetitions, as noted previously (see Footnote~\ref{FewTopics:fn:1M}): the one-for-cluster series always fluctuates a little, but these oscillations are small and do not affect the results.

Another final result is that, in cardinality-independent clustering, the choice of the number of clusters $m$ can be critical.
A detailed analysis of our data shows that good $m$ values can be found by looking at the positive peaks on the one-for-cluster series of the cardinality-driven clustering. For the $m$ values corresponding to the cardinalities of such peaks, as well as  the nearest cardinalities, the one-for-cluster series of the cardinality-independent clustering tend to have higher $\tau$ values than the average series, for almost any cardinality. 
It has to be noted that these are not all the good $m$ values, as there exist other $m$ values such that for the cardinality-independent clustering the one-for-cluster series is always above the average, but there is not a corresponding peak in the cardinality-driven clustering. However, this provides a general criterion for the choice of $m$.
For example, considering our datasets, to obtains one-for-cluster series that are better than random topic selection:
for AH99 any $m$ value can be used 
(but cardinalities around 8, 15, and 30 are better),
for TB06 the best values are around 10,
for R04 the best  values are 25, 75, and 110; and, finally,
for MQ07 the best values are around 25, 45, and 60.

 \subsection{Discussion}\label{FewTopics:sec:cl:disc}
 
The above results show that cardinality-independent clustering of topics is an a posteriori  topic selection strategy that is more effective than the random selection of topics. The effectiveness increase is still not large but it is consistent across all cardinalities and collections. As all the other results of this line of research, this is an a posteriori strategy that is only potentially useful and cannot be applied in practice. However, it can give useful insights for a priori strategies, like suggesting the  number of clusters to be used.

Note that although the setting is still a posteriori, clustering of topics shows only a limited effectiveness as a strategy to find good topic subsets. 
That is, even if we focused on a context where we expected clustering to be clearly effective, this was not the case.
This is perhaps surprising and might even cast some doubts on the effectiveness of clustering also for an a priori approach; however, in that case the features used would be very different, and therefore this claim needs to be verified with further experiments, that we leave as future work.

Also, note that comparing the clustering curves with the average series, as we have done, might even be unfair, since the clustering approach needs the whole topic set to produce the topic subset at a given cardinality $c$, whereas the average series are produced using just $c$ topics at cardinality $c$. In this respect, the clustering is even less effective. For instance, focusing on cardinality $50$ for MQ07 (fourth chart in Figure~\ref{FewTopics:fig:clustering_PCA_FixedCard}), we can indeed say that clustering has a higher correlation than average ($0.85$ vs. $0.79$), but that a clustering-based topic subset is generated using all $1153$ topics, whereas by using around $100$ random topics, one would get the same correlation.

We can also compare clustering correlations with \citet{Hosseini:SIGIR:2012}'s ``Adaptive'' ones. 
As in Section~\ref{FewTopics:sec:hosseini}, we need to change again our setting to perform clustering on the dataset with all the runs (instead of the top 75\% only); the obtained correlation values are shown in the last row of Table~\ref{FewTopics:tab:hosseini}. 
Incidentally, by doing so, we are not able to obtain correlation values higher than the average as those on the top 75\% runs; indeed, as it can be seen in Table~\ref{FewTopics:tab:hosseini}, when using all the runs the correlation values obtained by our clustering are hardy distinguishable from Average values. This is consistent with the remark in Section~\ref{FewTopics:sec:hosseini}: when using all the runs in a collection, the Average curves achieve higher values of $\tau$, and therefore it is more difficult to do better than the Average baseline in such a case.

Focusing on the comparison between Adaptive and clustering, we see from Table~\ref{FewTopics:tab:hosseini} that Adaptive is more effective than clustering on the smaller (having a lower number of topics) AH99 dataset, and conversely clustering is more effective than Adaptive on the larger R04 dataset. This result will need to be confirmed by further experiments, but it suggests that the two approaches could be fruitfully combined.

As a final remark, we conjecture that one general reason for the less than satisfactory effectiveness obtained with a posteriori clustering could simply be a ``tyranny of majority'' effect.
If there is a large subset of topics that can be ``naturally clustered'' together, and that cluster is indeed recognized by the clustering algorithm (as is quite likely), then the one-for-cluster selection method will be forced to pick up just one topic from that largest cluster. However, the topics in that large natural cluster are driving the evaluation in a specific direction -- these topics ``weigh more'' than the other topics. This will result in penalizing the one-for-cluster selection method, that is forced to not recognize this majority. 
%
%
%
This conjecture is true at least to some extent in our datasets: in our experiments the largest cluster usually contained around $75$--$90\%$ of the topics. 

To analyse this conjecture, we performed a last experiment. Given a clustering of topics, and the topic subset obtained from it, we computed not only the MAP by averaging the AP values, but also a Weighted MAP (WMAP) in which the AP values are averaged with a weight corresponding to the size of the cluster the topic belongs to. Note that both MAP and WMAP make sense: the WMAP approach somehow assumes that the full topic set is a representative sample of the whole topic population, and therefore if some topics are clustered together, that happens because the whole topic population contains many topics like those; conversely, the MAP approach is based on the assumption that since some topics are very similar, picking just one of them avoids a biased sampling, in which the topics of larger clusters are over represented.  Therefore, the two approaches differ on the weight given to each sampled topic; MAP assumes all topics to be of the same importance, conversely WMAP assumes topics that are sampled from a larger population are more representative, and thus more important. 
We also remark that by using WMAP we are not guaranteed that by using the full topic population we reach correlation 1 with MAP.
Results are clearly negative: all correlation values obtained when using WMAP are not only lower than those obtained when using MAP, but also always lower than Average.

\section{Conclusions}
\label{FewTopics:sec:concl-future-work}

Compared to previous work on using fewer topics in the evaluation of IR systems, our contributions are threefold.  
Addressing RQ\ref{FewTopics:RQ:1}, we show that examining subsets of a larger ground
truth topic set results in average and best subsets that are more highly correlated
with the ground truth topic set than found in previous work 
\citep{Guiver:2009:FGT:1629096.1629099,ecir11,ictir13}. It would appear that as the cardinality of the ground
truth increases, the size of the subset (relative to ground truth) required to
obtain a high correlation also decreases.

We also find that under large cardinalities, worst topic subsets are notably worse than shown in past work. 
Although finding a few bad topics was perhaps to be expected, when a larger pool of topics could be drawn from the large size of worst topic subsets that still had very low correlations was striking. 
Examination of the effectiveness of worst subsets shows that they were mainly composed of topics with poor effectiveness scores.

Addressing RQ\ref{FewTopics:RQ:2}, we analyze the role of statistically significant
differences between runs for different topics subsets. The ability to
distinguish statistically between the effectiveness of two runs is
impaired when topic cardinality is lowered. The main problem is an
increase in false negatives (type II errors) when making comparisons.
This issue has not been shown before in this area of topic subsetting
research, although it has been addressed in conjunction with incomplete relevance judgements: see for example \citet[Table~2]{Carterette:2007:HTI:1321440.1321530}, which agree with our findings. Some subsets were shown to be better than others at
minimizing type I and II errors. The analysis showed that the level
of error reduced relatively quickly as subset cardinality increased.
Nevertheless, because all of our experiments still use relatively small populations
of topics when compared to ``the set of all topics in the world'', it is not clear if the level of type II error will reduce sufficiently.
The collections still don't give us a sense of what the ``true'' population of
possible topics is like, and we have no way to be sure that the full cardinality is the truth. In a way, the results in this chapter suggest that all test collections are
suspect, since their very small subset of topics might be completely
un-correlated with the ``true'' population of all possible topics.
 
Our findings on the overlap of best and worst topics sets confirm that
being a good topic largely depends on the other topics in the subset.
In general, the previously established terminology of 
best/worst topic sets is perhaps misleading since it can be argued
that the worst topics are actually the most interesting ones (they
rank runs in ways contrary to the majority of topics), whereas
the best topics feature a high degree of redundancy that might lead to
a waste of resources.
Indeed, the high degree of redundancy is manifested in the best
correlation curves, that have high correlation values also for low cardinalities.


Addressing RQ\ref{FewTopics:RQ:3}, our analysis showed that clustering is effective in
finding topic subsets that are more representative than simply taking average or
random subsets, as long as the clustering is combined with dimensionality reduction.
However, the topic subsets obtained by clustering are only slightly more effective than random topic subsets, and are far from featuring
correlations that are as high as the best topic sets. A comparison with, and an analysis of, related work shows that we are in good company, though: good topics subsets exists but finding them seems a rather daunting task. While the work here
is a first step in finding representative and effective topic subsets, there
is still much work to be done to improve topic subset selection.

\part{On Effectiveness Evaluation Without Relevance judgements}\label{part:eewrj}

\chapter{Introduction and Background}
This chapter is structured as follows:
Section~\ref{part:eewrj:intro}  provides the introduction,
Section~\ref{part:eewrj:eewrj} discusses background on evaluation without relevance judgements, and
Sections~\ref{part:eewrj:soboroff}, \ref{part:eewrj:wu}, and \ref{part:eewrj:spoerri} explain in detail notable work on topic subsets.

\section{Introduction}\label{part:eewrj:intro}
The evaluation of Information Retrieval (IR) systems by means of test collections allows researchers to evaluate, develop, and compare different retrieval systems or algorithms in a well-defined experimental setting.
%
%

Probably the most expensive part of building a test collection is to produce, for every topic, the relevance assessment for the documents retrieved by the retrieval systems participating in the competition. 
To reduce the effort of this process, it is common practice to pool a subset of the top 1000 documents retrieved by each system; the relevance assessment is then performed only for the pooled documents \cite{Voorhees00overviewof}. 
The pooling method leads to reliable results in evaluating the effectiveness of retrieval systems \cite{zobel98}; but even with the pooling strategy the cost required to produce the relevance judgements is still high.

Many researchers tried to reduce the effort of producing relevance assessment, in several different ways. For example,  \citet{Lu:2016:EPE:2975219.2975241} and \citet{alonso2009can} proposed to crowd-source relevance judgements,  \citet{Lipani:2016:IFP:2970398.2970429} and \citet{LOSADA20171005} developed novel pooling strategies to build test collections with a reduced number of judgements, \citet{Guiver:2009:FGT:1629096.1629099}, \citet{ecir11}, and \citet{Berto:2013:UFT:2499178.2499184} studied the evaluation of IR systems using fewer topics, and many others tried to propose more sensitive and reliable evaluation metrics  \cite{Yilmaz_cikm06},  \citet{phdthesisBen,10.1145/1390334.1390437,10.1145/1571941.1572022,10.1145/1390334.1390445} discussed sampling of collection components.

A perhaps more extreme approach is to produce automatic relevance assessment, i.e., to evaluate the systems participating in a test collection initiative without any relevance judgements, in a completely automatic way \cite{Soboroff:2001:RRS:383952.383961,Wu:2003:MRI:952532.952693,spoerri:2007}. In this part we focus on this approach.

\section{Evaluation Without Relevance judgements}\label{part:eewrj:eewrj}

\begin{table}[tbp]
  \caption{The 17 Prediction Methods Used in this chapter 
    \label{IPM:tab:methods}}
  \centering
\begin{tabular}{@{}l @{  }r    @{  }l @{  }l @{  }l@{  }l@{}}
    \toprule
    &\#&Acronym &   Accuracy  & Datasets & Effectiveness\\
    && (version)&   Measures & \ & Metrics\\

    \midrule
    \addlinespace
    \multicolumn{3}{@{}l@{}}{\citet{Soboroff:2001:RRS:383952.383961}}& $\tau$, charts & TREC 3,5,6,7,8 & MAP\\
    & 1 & SNC \\
    \addlinespace
    \multicolumn{3}{@{}l@{}}{\citet*{Wu:2003:MRI:952532.952693}}&$r_s$&TREC 3,5,6,7,&R-Precision,\\
    \multicolumn{3}{@{}l@{}}{}&&2001& P@10,30,50,100\\
    & 2 & \multicolumn{3}{@{}l}{WUCv0 (Basic)}\\
    & 3 & \multicolumn{3}{@{}l}{WUCv1 (Variation 1)}\\
    & 4 & \multicolumn{3}{@{}l}{WUCv2 (Variation 2)}\\
    & 5 & \multicolumn{3}{@{}l}{WUCv3 (Variation 3)}\\
    & 6 & \multicolumn{3}{@{}l}{WUCv4 (Variation 4)}\\
    \addlinespace
    \multicolumn{3}{@{}l@{}}{\citet*{Aslam:Savell:2003}}&$\tau$, $\rho$,  & TREC 3,5,6,7,8 & MAP\\
    \multicolumn{3}{@{}l@{}}{} & scatterplots\\
    & 7 &AS&\\
    \addlinespace
    \multicolumn{3}{@{}l@{}}{\citet*{nuray:can:2006}}&$r_s$&TREC 3,5,6,7 & MAP\\
    &8& \multicolumn{3}{@{}l}{NC-NRP (Normal Rank Position)} \\
    &9&\multicolumn{3}{@{}l}{NC-NB  (Normal Borda)} \\
    &10&\multicolumn{3}{@{}l}{NC-NC  (Normal Condorcet)} \\
   &11& \multicolumn{3}{@{}l}{NC-BRP (Bias Rank Position)} \\
    & 12 & \multicolumn{3}{@{}l}{NC-BB   (Bias Borda)} \\
    & 13& \multicolumn{3}{@{}l}{NC-BC  (Bias Condorcet)} \\
    \addlinespace
    \multicolumn{3}{@{}l@{}}{\citet*{spoerri:2007}}&$\rho$, scatterplots & TREC 3,6,7,8 & MAP, P@1000\\
    &14& \multicolumn{3}{@{}l}{SPO-S (Single)} \\
    &15& \multicolumn{3}{@{}l}{SPO-A (AllFive)} \\
    &16& \multicolumn{3}{@{}l}{SPO-SA (Single - AllFive)}\\
    \addlinespace
    \multicolumn{3}{@{}l@{}}{\citet*{sakai-lin2010}}& $\tau$, $\tau_{ap}$,  & R03, R04, CLIR6-JA, & MAP,  \\
    \multicolumn{3}{l}{}&charts,&  CLIR6-CT,  & nDCG,\\
    \multicolumn{3}{l}{}&scatterplots&   IR4QA-CS & Q-measure\\
    &17&SL \\
    \bottomrule
  \end{tabular}
\end{table}

Table~\ref{IPM:tab:methods} summarizes the proposals to use  no human relevance assessments when evaluating IR effectiveness. 
The first proposal is by \citet{Soboroff:2001:RRS:383952.383961}: 
their method performs a random sample from the pool of documents (i.e., the documents retrieved by at least one system); the sampled documents are deemed to be relevant, while the remaining ones are non relevant, and the evaluation is performed accordingly.
The underlying assumption is that relevant documents tend to be retrieved by many systems, and thus pooled.

Another method, proposed by \citet*{Wu:2003:MRI:952532.952693}, is based on data fusion, and consists in merging the ranked lists of documents retrieved by each retrieval system querying the same test collection for a certain topic. The idea is to assign a weight to each retrieved document and to use such weights to rank retrieval systems. Thus, good systems are those that retrieve ``popular'' documents.
In the simplest version of the algorithm (WUCv0), the weight, called reference count, sums up the occurrences of each document retrieved by a system which is present in the ranked lists of other systems.
The four variants 
assign a weight to the reference count differentiating the position in which each document appears in the ranked list. 

\citeauthor*{Aslam:Savell:2003}'s method \cite{Aslam:Savell:2003} measures the similarity of each system to the others (by computing the ratio between the cardinality of the intersection of the documents of the ranked lists and their union) and uses this similarity to evaluate them. 
This evaluation is 
highly correlated to \citeauthor{Soboroff:2001:RRS:383952.383961}'s 
method.
One issue is that the average similarity is computed by means of  ``the grossest possible measure'' \cite[p.~362]{Aslam:Savell:2003}.
This work also presents one of the main criticisms to this approach: the observation that runs are ranked by popularity rather  than effectiveness.
Such ``tyranny of the masses'' effect is  
penalizing for best runs, that are underestimated.
We use a slightly modified version of this method, 
keeping the raw topic scores instead of computing their mean value over the topic set.

The method by \citet*{Nuray:2003,nuray:can:2006}
 consists of three phases: (i) select the runs, (ii) compute the popularity of each document according to various methods, and (iii) the top 30\% of the most popular documents are said to be relevant.
The run selection can be done in two ways: either ``normal'', where each run is selected, or ``bias'', where the runs selected are the top 50\% of runs which have a list of retrieved document that is farther from the ``norm''.
The document ranking can be performed according to three strategies taken from theory of voting: ``Rank Position'', ``Borda'' \cite{Emerson2013}, and ``Condorcet'' \cite{fishburn1977condorcet}.

The method by \citet*{SpoerriMEET:MEET14504201175} selects one run for each participating organization, and forms a set of trials containing five runs (we borrow this terminology from \citet{sakai-lin2010}) in a way that each run occurs exactly five times (in different trials); then, it computes the percentage of the set of documents either retrieved by the run exclusively (called ``Single''), the set of documents retrieved by all the five runs in the trial (``AllFive''), and the ``Single minus AllFive'' measure. Finally, to obtain a trial-independent behavior, the three computed measures  for each run are averaged over the five trials in which the run occurs.


The method by \citet{sakai-lin2010}
is very similar to Condorcet method, even if statistically different and more efficient.


All the above methods have been experimentally evaluated using as datasets some TREC test collections as detailed in Table~\ref{IPM:tab:methods} (third column), with the only exception of \citet{sakai-lin2010} who used, to run their experiments, also 
 NTCIR collections. The table also shows in the last column the IR effectiveness measure(s) used in each experimental evaluation. The accuracy of the methods\footnote{In an attempt of avoiding confusion, and consistently with other authors \cite{Wu:2003:MRI:952532.952693,sakai-lin2010,nuray:can:2006}, we reserve the term ``effectiveness'' for retrieval effectiveness and ``accuracy'' for the accuracy in predicting system effectiveness by a method.} has been measured as correlations between the predicted and actual MAP values, again as detailed in the table. Overall (but we will see a more detailed analysis in the following sections) the accuracy of the methods is rather limited and they often do not significantly outperform the original proposal by \citet{Soboroff:2001:RRS:383952.383961}.
\citet{hauff:dejong:2010} noted that the low accuracy might depend on having human intervention (the ``manual runs'') in the best systems: in the datasets where the best systems are completely automatic, human relevance assessments are less needed. Later, \citet*{hauff:dejong:2010} compared the no assessment and the fewer assessment approaches, finding a rather good correlation and 
claiming that it is still unclear whether manual assessments are really needed. Moreover, as noted by \citet{sakai-lin2010} if the organizers of a test collection initiative can release a so called ``system ranking forecast'', this can be useful when no ``true'' assessments are available.

\citet{Roitero:2018:RGE:3282439.3241064} provide a full re-implementation of such algorithms and discuss their reproducibility.
Recent work \cite{Roitero:2018:EES:3209978.3210108,Roitero:2018:RIE:3282439.3239573} proposes to use the described methods in a practical way: 
 reproduce some of the previous result and use the discussed methods to identify a subset of few good topics for retrieval evaluation; \citet{Mizzaro:2018:QPP:3209978.3210146} use the methods in the setting of query and topic performance prediction. When compared to their work, in this work we use more datasets, more methods, and we also analyze several fusion strategies including those based on machine learning techniques. Moreover, we do not simply aim at reproducibility but we also focus on comparisons across methods and collections, as we detail in the following.

Since in Chapter~\ref{chapt:eewrj:reproduce} we will reproduce some notable results on evaluation without relevance judgements, we now discuss such works more in detail.
We first discuss the three main contributions that can be found in the literature and that will be the focus of Chapter~\ref{chapt:eewrj:reproduce}.

\section{The Method by \texorpdfstring{\citeauthor*{Soboroff:2001:RRS:383952.383961}}{Soboroff et al.}}
\label{part:eewrj:soboroff}

The approach proposed by \citet*{Soboroff:2001:RRS:383952.383961} is the first work investigating the ranking of retrieval systems without human assessments. With almost 100 citations%
\footnote{Source:  \url{http://dl.acm.org/citation.cfm?id=383961}; date: 
21 April 2018.}
this work is considered by the research community a strong baseline in this context. 

\citeauthor{Soboroff:2001:RRS:383952.383961} start questioning what happens if relevant documents are chosen randomly from the pool considering the hypothesis that relevant documents occur in the pool according to a defined probability distribution.
To address this question, they design an experiment in which they estimate a probabilistic model which describes the occurrence of the relevant documents in the pool. Specifically, they choose to model relevant document occurrence with a Normal Distribution $\mathcal{N}(\mu,\sigma)$
that requires only two parameters to be estimated from queries, namely
the mean percentage of relevant documents occurring in the pool $\mu$ and the 
standard deviation $\sigma$:
\begin{align*}
\mu &= \frac{1}{n} \sum_{i=1}^{n} \mu_{i} \\
\sigma &= \sqrt{ \frac{ \sum_{i=1}^{n} \left( \mu_{i} - \mu \right)^2}{n-1} },
\end{align*}
where $n$ is the number of topics of the test collection and $\mu_{i}$ is the percentage of relevant documents occurring in the pool for the $i$-th topic.

%
%
Using this model, \citeauthor{Soboroff:2001:RRS:383952.383961} randomly sample a set of documents and labeled them as ``relevant'', to form a set of pseudo relevance judgements called \emph{pseudo-qrels}, in three ways:
\begin{itemize}
\item sampling the documents from the official qrels using $\mu$ and $\sigma$;
\item sampling the documents from the official qrels considering each topic separately, i.e., using $\mu_{1}, \ldots, \mu_{n}$ and $\sigma$. \citeauthor{Soboroff:2001:RRS:383952.383961} named this strategy ``Exact--fraction Sampling'';
\item sampling the documents from the pool at depth 100 including duplicate documents, and using $\mu$ and $\sigma$. The rationale of this experiment is that the higher the number of systems that retrieve a  document, the more it is likely to be relevant.
\end{itemize}
We focus on the first approach since it is the more realistic and because it does not include a-posteriori knowledge (i.e., knowledge that can be obtained only after the human relevance assessments have been gathered), except for the mean and standard deviation parameters; furthermore, we provide an experiment in which we estimate the $\mu$ and $\sigma$ parameters to present a realistic ``without relevance judgements'' scenario.

Using the \emph{pseudo-qrels}, \citeauthor{Soboroff:2001:RRS:383952.383961} present experimental evaluation on all the runs of the TREC-3, TREC-5, TREC-6, TREC-7, and TREC-8 collections.
The effectiveness of the method is measured by (i) computing, on each topic, the Kendall's $\tau$ correlation between the ground truth of the official rank of the systems and the rank obtained on the basis of the pseudo-qrels, and (ii) taking the mean $\tau$ over all topics.
The highest mean $\tau$ value
is of $0.487$, obtained for the TREC-5 collection (more details on these results will be shown in Table~\ref{JDIQNorel:tab:SNC_reproduced_cors} in Section~\ref{JDIQNorel:sec:Reproduce}).
Although this approach achieves a reasonable  performance in terms of correlation, the method fails mostly on top-ranked systems (i.e., the most interesting ones for the evaluation process), whose effectiveness is usually heavily underestimated.

Based on the work by \citeauthor{Soboroff:2001:RRS:383952.383961}, \citet{Aslam:Savell:2003} 
propose a strategy to infer the similarity of retrieval systems by assessing the similarity of their retrieved documents. Considering two retrieval systems and their ranked lists, the measure is simply  the ratio between the number of documents that they have in common divided by the total number of retrieved documents. 
Although the proposed measure is trivial and easy to compute, the authors show that it is able to achieve correlation values
with the ground truth similar to those by \citeauthor{Soboroff:2001:RRS:383952.383961}.
They also observe that  both methods are 
affected by a ``tyranny of the masses''  phenomenon: top ranked systems (i.e., the systems that lower the correlation in \citeauthor{Soboroff:2001:RRS:383952.383961}'s experiments), are being punished for  retrieving documents significantly different from the average systems in a competition.
Therefore, \citeauthor{Aslam:Savell:2003}  observe that \citeauthor{Soboroff:2001:RRS:383952.383961}'s method assesses the retrieval systems more in terms of ``popularity'' than actual ``performance''.

This effect has been also investigated by \citet{SpoerriMEET:MEET14504201175}, who
remarks that ``the potential relevance of a document increases exponentially as the number of systems retrieving it increases'', calling it the ``Authority effect''.
\citeauthor{SpoerriMEET:MEET14504201175} also suggests that  selecting only a single run per participant group\footnote{Initiatives like TREC allow participant groups to submit the results obtained with different variations of their system. One of them is called ``run''.} 
``would help to sharpen the signal and make the Authority Effect more dominant'' \cite[page~1061]{spoerri:2007}.

\section{The Method by \texorpdfstring{\citeauthor{Wu:2003:MRI:952532.952693}}{Wu and Crestani}}\label{part:eewrj:wu}

The work proposed by \citet{Wu:2003:MRI:952532.952693} presents another approach for ranking retrieval systems without relevance judgement. This technique uses a measure called  ``reference count'', which is developed by  the same authors within the data fusion context \cite{Wu:2002}. 

Specifically, suppose we have a topic and a set of retrieval system results on the same data collection, the reference count of a retrieval system result can be obtained as follows.
Given the set
of retrieved documents by a system (called \textit{original documents}), the reference count is the sum of the occurrences of these documents in the results of all the other retrieval systems (called \textit{reference documents}) up to number of retrieved document per topic, which is usually 1000. 

This approach is called by the authors Basic reference count. However, this technique does not take into account the different position of \textit{reference documents} and the position of the \textit{original document}. To overcome these limits, \citeauthor{Wu:2003:MRI:952532.952693} present four different variations by changing either or both of the aspects. 
The first variation (V1) assigns different weighs to \textit{reference documents} based on their ranking positions. 
The second variation (V2) assigns different weights to \textit{original documents} based on their ranking positions. 
The third variation (V3) consists of assigning different weights to both the \textit{reference documents} and the \textit{original document}. 
Finally, the fourth variation (V4) assigns different weights based on the \textit{reference documents} ranking positions and the \textit{original document}'s normalized scores (instead of their ranking positions).  

\citeauthor{Wu:2003:MRI:952532.952693} present their results using Spearman's average $r_S$ correlation values over the topics (the detail of their values over different TREC collections will be shown in Table~\ref{JDIQNorel:tab:WUC_reproduced_cors}, discussed in the following). 
The values are not directly comparable to \citeauthor{Soboroff:2001:RRS:383952.383961}'s ones, who use $\tau$; however, \citeauthor{Wu:2003:MRI:952532.952693}  compute  $r_S$ also for \citeauthor{Soboroff:2001:RRS:383952.383961}'s  method and find that 
they obtain average $r_S$ correlation values 
that are lower than, or comparable with, \citeauthor{Soboroff:2001:RRS:383952.383961}'s ones. 
\citeauthor{Wu:2003:MRI:952532.952693} observe that the results are mostly based on two effects: (i) the overlap of the relevant and non relevant documents retrieved by the systems is quite different, as also shown by \citet{Lee:1997:AME:258525.258587}; (ii) there is a connection between reference count and the percentage of relevant documents in a ranked list of a system. Furthermore, they observe that their results are affected by the same problem of the top ranked systems as in \citeauthor{Soboroff:2001:RRS:383952.383961} Again, this phenomenon is probably due to the fact that top ranked systems are quite peculiar, in fact they retrieve documents that not many other systems retrieve.

\section{The Method by \texorpdfstring{\citeauthor{spoerri:2007}}{Spoerri}} \label{part:eewrj:spoerri}

\citet{spoerri:2007} proposes another method to rank retrieval systems without human relevance judgements. The proposed approach starts from the fact  that retrieval systems tend to retrieve similar sets of relevant documents and dissimilar sets of non-relevant documents \cite{Lee:1997:AME:258525.258587}. 
\citeauthor{spoerri:2007}'s approach estimates the relative performance of multiple retrieval systems computing the \textit{structure of overlap} between their retrieved documents. 
Specifically, for each system he counts the number of documents retrieved by it, which are also retrieved by a specific number of other systems; in \citeauthor{spoerri:2007}'s method implementation this number is five. This process is called \textit{random grouping}.
More in detail, the ``structure of overlap''
is used to extrapolate two measures: Single\% (S\% in the following), the percentage of a system's documents not found by other systems, and AllFive\% (A\%), the percentage of a system's documents found by all five 
systems. A third measure, $\text{Single\%} - \text{AllFive\%}$ (S-A\%), is also computed as the arithmetic difference between the two previously found percentages.
In his experiments \citeauthor{spoerri:2007} builds the structure of overlap both when considering the top 1000 documents retrieved by each system and a shallow pool, to study the effect of pool depth on the effectiveness of his method.

Analyzing these evaluation measures, the author shows that the percentage of a system's documents that are only found by it and not by other systems (S\%) increases as the system retrieval quality decreases. 
For the structure of overlap to be computed it is critical that a single run for each system participating in the track is included. 
In fact, although runs of a system participating in a track are different, they usually share the same technique and system architecture. Therefore, the structure of overlap could be affected by this dependence.
Furthermore, the author in the paper demonstrates that there exists an optimal number of retrieval systems and topics needed for building the structure of overlap.

In the experimental evaluation \citeauthor{spoerri:2007} selects a subset of the runs submitted to  the selected TREC tracks. 
\citeauthor{spoerri:2007} does not use multiple runs of a system, because these would have very strong structure of overlap compromising the comparison with the other selected systems.  Therefore, for each participant group he selects one run, called \emph{short-run}, according to the following criterion:
the short-run is selected considering the one with the highest AP value among the runs which use ``automatic'' as ``Query Method'' and ``Title+Description'' as ``Topic Length'' (i.e., runs that did not use the Narrative field of the topic). 
%
%
%
Although the selection of the most effective short-run for each group allowed to run some experiments to analyze the effectiveness of his approach, it must be noticed that this kind of selection would be impossible in a real scenario without a pre-evaluation process.

\citeauthor{spoerri:2007}'s approach achieves Spearman $r_S$ correlation values up to 
0.96. However, this value is not comparable to \citeauthor{Wu:2003:MRI:952532.952693}'s ones, since it is not computed by taking the mean $r_S$ over topics, but as the $r_S$ of predicted and real MAP values. 
The first and third variant (i.e., S\% and S-A\%) show the best effectiveness (details will be shown in Table~\ref{JDIQNorel:tab:SPO_reproduced_cors}). The results are also confirmed by using other metrics (i.e. P@100 and R@100).

\chapter{Reproduce and Generalize Notable Results on Evaluation Without Relevance judgements}
\chaptermark{Notable Results on Evaluation Without Relevance judgements}
 \label{chapt:eewrj:reproduce}
This chapter deals with the reproduction and generalization of notable results on evaluation without relevance judgements.
Section~\ref{JDIQNorel:sec:Intro} introduces and frames research questions.
In Section~\ref{JDIQNorel:sec:Experiments} we describe data, methods, and measures used in our experiments.
In Section~\ref{JDIQNorel:sec:Reproduce} we focus on \ref{JDIQNorel:I:R} and reproduce some of the most important work on evaluating IR systems without relevance judgements.
In Section~\ref{JDIQNorel:sec:Generalize} we turn to \ref{JDIQNorel:I:G} and generalize some of the obtained results to other collections, including a more recent one, evaluation metrics, and a shallow pool.
Finally, in Section~\ref{JDIQNorel:sec:con} we conclude.

\section{Introduction and Research Questions}\label{JDIQNorel:sec:Intro}
In this chapter we focus on an automatic approach to produce relevance judgements, and we pursue the threefold aim of reproducing the main previous results (Aim~\ref{JDIQNorel:I:R}), generalize them to other collections, metrics, and pool depth (\ref{JDIQNorel:I:G}), and expand the approach to derive some insights on related problems not studied yet (\ref{JDIQNorel:I:E}).
More in detail, our aims can be stated as follows: 
\begin{enumerate}[label=\textbf{A\arabic*.},ref=A\arabic*] 
\item \label{JDIQNorel:I:R} To reproduce the main results of the notable works on automatic evaluation of retrieval systems, as well as present such results in a uniform way.
\item \label{JDIQNorel:I:G}  To generalize such previous work, in particular:
\begin{enumerate}[label=\textbf{A2\alph*.},ref=A2\alph*,leftmargin=8.5mm] 
\item  \label{JDIQNorel:I:G:C} To analyze the effect of using further test collections, featuring different properties from those used in the original experiments;
\item \label{JDIQNorel:I:G:M} To study the effect of different evaluation metrics; and 
\item \label{JDIQNorel:I:G:S} To study the effect of a shallow pool.
\end{enumerate}

\item \label{JDIQNorel:I:E} To expand the idea; in detail:
\begin{enumerate}[label=\textbf{A3\alph*.},ref=A3\alph*,leftmargin=8.5mm] 
\item \label{JDIQNorel:I:E:MIX} To experiment with a mixed approach, in which a part of the evaluation is automatic and a part of it is manual; and
\item \label{JDIQNorel:I:E:AAP} To apply the same approach to a dual problem, the estimation of topic difficulty.\footnote{The duality of topic difficulty and system effectiveness will be discussed in detail Section~\ref{JDIQNorel:sec:E:AAP}.}
\end{enumerate}
Both these ideas have not yet been explored, although they do seem quite natural in this context.
\end{enumerate}

Apart from \citet*{Soboroff:2001:RRS:383952.383961},  \citet*{Wu:2003:MRI:952532.952693}, and \citet{spoerri:2007}, other proposals of studies about effectiveness evaluation without relevance judgements exist: for example, \citet{Aslam:Savell:2003}'s work can be exploited, and  \citet{Nuray:2003} is another viable alternative. However, in the rest of this chapter we will focus on the three above described methods, that are somehow the most classical and well known approaches.


\section{Experimental Setting: Data, Methods, and Measures }\label{JDIQNorel:sec:Experiments}


To pursue the aims of this chapter, namely to reproduce and then to generalize as well as expand the above presented results, we run  
a battery of experiments. In this section we describe the common features of those experiments, and in the rest of the paper we will provide further details when needed, and the results. 

\subsection{Data}\label{JDIQNorel:sec:data}

\begin{table}[tb]
  \centering
  \caption{Test collections used for the reproducibility experiments in the upper part of the table, and for the other experiments in the lower part.
  }
  \begin{threeparttable}
\begin{adjustbox}{max width=\textwidth}
  \begin{tabular}{lll rrl r}
  \toprule
    \textbf{Acronym} & \textbf{Track}  &\textbf{Year} & \textbf{Topics} & \textbf{Runs}  & \textbf{Used Topics} & \textbf{Manual}\\
     &  & & & & & \textbf{Runs}\\
     \midrule
    TREC-3  & Ad Hoc    & 1994 & 50  & 40   & 151-200 & 11 \\
    TREC-5  & Ad Hoc    & 1996 & 50  & 61    & 251-300 & 31 \\
    TREC-6  & Ad Hoc    & 1997 & 50  & 74   & 301-350 & 17\\
    TREC-7  & Ad Hoc    & 1998 & 50  & 103 & 351-400 & 17 \\
    TREC-8  & Ad Hoc    & 1999 & 50  & 129 & 401-450 & 13 \\
    TREC-01 & Ad Hoc    & 2001 & 50  & 97    & 501-550 & 2 \\
    \midrule
	TREC-8  & Ad Hoc    & 1999 & 50  & 129  & 401-450 & 13 \\
   	TB06    & TeraByte  & 2006 & 149 & 61   & 701-850\tnote{$\dagger$} & 0 \\
    TB06M   & TeraByte  & 2006 & 50  & 80   & 801-850 & 19 \\
 	R04     & Robust    & 2004 & 249 & 110  & 301-450, 601-700\tnote{$\dagger$} &  0 \\
    WEB14   & Web       & 2014 & 50  & 30   & 251-300 & 4\\
 \bottomrule
  \end{tabular}
  \end{adjustbox}
  \begin{tablenotes}
  \item[$\dagger$] Not all, see text. 
  \end{tablenotes}
  \end{threeparttable}
  \label{JDIQNorel:tab:coll}
\end{table}

For reproducibility purposes we use the  datasets from the TREC\footnote{See \url{http://trec.nist.gov/}} editions that have been used by at least one of the previous studies: the
Ad Hoc tracks of TREC-3, TREC-5, TREC-6, TREC-7, TREC-8, and TREC-2001.
Furthermore, to extend and generalize such results, we use some more datasets: besides  again TREC-8,
also, 
from other TREC editions,
the Robust track of 2004  (R04),  
the TeraByte track of 2006 (TB06), and  
the Ad Hoc Web track of 2014 (WEB14).
All the test collections used in this chapter are detailed in Table~\ref{JDIQNorel:tab:coll}.

Concerning R04, we use 249 topics, removing topic 672, as described by \citet[Section 1]{voorhees2004overview}: ``the TREC 2004 track used a set of 250 topics (one of which was subsequently dropped due to having no relevant documents)''.

Concerning the TeraByte collection, we choose to investigate two versions of the original dataset (see \citet[Section 3.1]{buttcher2006trec}: ``Manual runs used only the 50 new topics; automatic runs used all 149 topics from 2004-2006''):
\begin{itemize}
\item We select the subset of 61 runs that run over all the 149 topics, as done by \citet{Roitero2017}; this collection has no manual runs; we denote this dataset as TB06. Specifically, the dataset uses the 150 track topics and removes topic 703.
\item We consider all the participating runs (i.e., 80 runs including the manual ones) that run over a common subset of 50 topics; we call this collection TB06M (M denotes the inclusion of the manual runs).
\end{itemize}

Concerning WEB14, we choose two different versions of the track, to investigate the effect of the evaluation measures: this approach is detailed in Section~\ref{JDIQNorel:subsect:evaluation_measures}.

For all the other datasets we use the standard settings proposed in the track.

\subsection{Evaluation Measures} \label{JDIQNorel:subsect:evaluation_measures}
\begin{table}[tb]
\centering
\caption{AP, MAP, and AAP for $n$ Topics and $m$ Systems (adapted from \citet{Mizzaro:2007:HHT:1277741.1277824})
\label{JDIQNorel:tab:AP}}
  \begin{tabular}[t]{|c|ccc|c|}
  \hline	
  	& $t_1$ &  $\cdots$ & $t_n$ &	$\MAP$\\\hline
  $s_1$ & $\AP(s_1,t_1)$& $\cdots$ &$\AP(s_1,t_n)$&$\MAP(s_1)$\\
  \vdots& \vdots &$\ddots$& \vdots &$\vdots$\\
  $s_m$& $\AP(s_m,t_1)$& $\cdots$ &$\AP(s_m,t_n)$&$\MAP(s_m)$\\\hline
  AAP & $\AAP(t_1)$ & $\cdots$ & $\AAP(t_n)$ & $ $ \\\hline
  \end{tabular}
\end{table}

The outcome of the TREC evaluation process can be represented as in Table~\ref{JDIQNorel:tab:AP}:
$s_i$ represents a system/run,
$t_j$ represents a topic,
$\AP(s_i, t_j)$ represents the effectiveness of the system $s_i$ on the topic $t_j$ according to an evaluation measure. 
Average Precision (AP) is perhaps the most widely used metric, however system effectiveness can be expressed by means of many other alternative metrics, like  logAP, logitAP, NDCG, etc.
Since systems in TREC usually are required to retrieve $1000$ documents for each topic, we use the truncated versions of these metrics, e.g., we use $\AP@1000$.
To rank retrieval systems, a common approach is to average the performance over the set of topics according to a measure (e.g., MAP = Mean AP); thus, 
\begin{equation}\label{JDIQNorel:eq:MAP}
\MAP(s_i) = \frac{1}{n} \sum_{j=1}^{n} \AP(s_i, t_j). 
\end{equation}

More in detail, for TREC-8, R04, TB06, and TB06M we use as evaluation measure AP@1000; the official track measure for TB06 is AP@10.000 (which is very close to AP). We also present results in terms of GMAP, and logitAP (metrics detailed in Section~\ref{JDIQNorel:sub:generalize_other_metrics}). 
For WEB14 we use the official track measure NDCG. 
To present AP values for this dataset as well, we binarize WEB14 \emph{qrels}. As it is usually done, we attempt two slightly different binarizations:
in the first version we map the original relevance values -2 and 0 into not relevant and the values 1, 2, 3 into  relevant;
in the second version we map the values -2, 0, and 1 into not relevant and the values 2 and 3 into  relevant. 
In the following we focus on   the first binarization only, which incidentally provides better results in terms of the final correlation obtained.
Note that selecting the binarization which leads to higher correlation values has no consequence in the experiment results since we are not competing against any baseline.
The meaning of AAP (see the last row in Table~\ref{JDIQNorel:tab:AP}) will be discussed in Section~\ref{JDIQNorel:sec:E:AAP}.

\subsection{Methods Configuration} \label{JDIQNorel:sec:methods_config}

For brevity, in the following we denote the methods by \citet*{Soboroff:2001:RRS:383952.383961},  \citet*{Wu:2003:MRI:952532.952693}, and \citet{spoerri:2007}  with SNC, WUC, and SPO, respectively.

\begin{table}[tbp]
\centering
\caption{$\mu$ and $\sigma$ values: comparison for reproducibility (leftmost four columns) and estimation (two rightmost columns).
}
\label{JDIQNorel:tab:SNC_reproduced}
\begin{threeparttable}
\begin{tabular}{l cc c cc c cc}
\toprule
       & \multicolumn{2}{c}{SNC original} &  & \multicolumn{2}{c}{Our obtained values}  & & \multicolumn{2}{c}{Estimated} \\
       \cmidrule{2-3} \cmidrule{5-6} \cmidrule{8-9}
       & $\mu$       & $\sigma$    && $\mu$       & $\sigma$ && $\mu$       & $\sigma$       \\
       \cmidrule{2-6} \cmidrule{8-9} 
TREC-3 & 14.90 & .123  && 10.414 (14.902)\tnote{$\dagger$} & .097 (.123)\tnote{$\dagger$} && 23.15 & .110 \\
TREC-5  & 3.90  & .043  && 3.956                      & .043   				  && 13.39 & .074 \\
TREC-6  & 6.32  & .067  && 6.351                      & .067   				  && 10.13 & .062 \\
TREC-7  & 5.78  & .047  && 5.834                      & .047   				  && 5.82 & .046 \\
TREC-8  & 5.35  & .048  && 5.497                      & .048   				  && 3.60 & .038 \\

\bottomrule
\end{tabular}
\begin{tablenotes}
  \item[$\dagger$] Pool built with all the participating runs at depth 100.
  \end{tablenotes}
  \end{threeparttable}
\end{table}

Considering SNC, we start estimating a probability distribution by randomly selecting relevant documents from \emph{qrels} for building \emph{pseudo-qrels}.
First, based on the official \emph{qrels} we compute the mean ($\mu$) and the standard deviation ($\sigma$) values of the percentage of relevant documents in the pool.
Table~\ref{JDIQNorel:tab:SNC_reproduced} shows in the left side  the comparison between SNC and our computed values of $\mu$ and $\sigma$ (the rightmost column is discussed in the following).
Comparing these values with \citeauthor{Soboroff:2001:RRS:383952.383961}'s ones, we observe that we are able to reproduce the same $\mu$ and $\sigma$ values, 
apart from TREC-3. Our hypotheses is that, in this case, all participating runs have been used, in place of the official runs only (i.e., the ones which are selected to form the pool);\footnote{Note that a detailed list of the official runs is not provided by NIST.} 
and indeed when using this approach we obtain the values in parentheses in the first row of the table, which perfectly match the original ones.
For completeness and for (future) reproducibility, we also report the  $\mu$ (and $\sigma$) values that we obtain for R04, TB06, TB06M, and WEB14: 5.12 (0.043), 13.39 (0.074), 8.98 (0.057), and 32.59 (0.145). 

Once we have the estimated Normal Distribution (based on the mean and the standard deviation values computed before) we can build the \emph{pseudo-qrels} by simply performing a random sampling on \emph{qrels} based on this distribution;
then using both the official ``trec\_eval'' software (version 9.0)\footnote{\url{http://trec.nist.gov/trec_eval/}} and IRevalOO\footnote{\url{https://github.com/KevinRoitero/IRevalOO}} \cite{RoiteroMPM18} we compute an approximated average precision (i.e., obtained with the sampled \emph{qrels}) value for each run and each topic. 
Based on these values, we compared the approximated 
AP and MAP values (i.e., the ones originated from the \emph{pseudo-qrels}) with the real 
AP and MAP values.
In order to provide a realistic (i.e., without any post-evaluation knowledge) setting for \citeauthor{Soboroff:2001:RRS:383952.383961}'s work, 
we also estimate $\mu$ and $\sigma$ values using a best-fit interpolation with an order one
polynomial trend-line,
obtaining:
\begin{equation}
\begin{aligned}\label{JDIQNorel:eq:estimate}
\mu &= \frac{1133.3}{\mbox{ no. runs }} - 5.1841 \\ 
\sigma &= 0.0037 \mu + 0.0242 .
\end{aligned}
\end{equation}
The two rightmost columns in Table~\ref{JDIQNorel:tab:SNC_reproduced} show the estimated values.

For WUC, we start by computing the so called document ``reference count'', for each topic, run, and position of the rank; then, we sum and normalize the reference count to compute the Basic, V1, V2, V3, and V4 measures according to \citeauthor{Wu:2003:MRI:952532.952693}'s definition. We consider all the runs submitted to the TREC tracks.

For SPO, we start by selecting the systems according to the selection method described in Section~\ref{part:eewrj:spoerri}.
Having the subset of systems, we compute the structure of overlap for all the runs; we compare then the percentages of overlap given by the structure of overlap with the real MAP values. 
As stated before, the structure of overlap is built forming random groupings of five retrieval systems and this structure is used to extrapolate S\%, A\%, and S-A\% measures.
Based on our experiments, we observe 
that by following the proposed selection method (see also Section~\ref{part:eewrj:spoerri}) the  ``Title+Description'' runs are often not enough to reach the number of runs selected by \citeauthor{spoerri:2007}.
Most likely, \citeauthor{spoerri:2007} then included in his experiments some runs which have only ``Description'' as ``Topic Length''; we follow this approach.\footnote{\label{JDIQNorel:fn:URL}To make our run selection process reproducible, we report the runs that we selected in the spreadsheet available at \url{https://users.dimi.uniud.it/~kevin.roitero/OUTSIDE/Reproducibility_SI_EvalNoJudg} and at \url{https://github.com/KevinRoitero/Reproducibility_SI_EvalNoJudg} (where we also include all the code used to carry out our experiments and some additional tables).} 

To avoid noise and give stability to our results, we performed 20 repetitions for SNC and SPO, which have a non-deterministic part. In the following we report the results obtained when averaging the AP and correlation values over the 20 repetitions.


\subsection{Correlation Measures}
In this chapter we focus on reporting correlation values between the official system rank provided by TREC, and the system rank obtained by the automatic evaluation methods.
In the three methods that we reproduce there is no homogeneity concerning the correlation coefficient used to compare the computed scores (representing the automatic evaluation of systems) with the real evaluation of systems (e.g., the real MAP values).
Thus, to present the results in a homogeneous way, we report the correlation values using value- and rank-based correlations, as well as top-heavy correlation measures.
More in detail, we use:
\begin{itemize}
\item Pearson's $\rho$, which measures linear correlation; 
\item Kendall's $\tau$, which measures rank correlation;
\item Spearman's $r_S$,  which measures rank correlation;
\item Rank Biased Overlap (RBO) \cite{Webber:2010:SMI:1852102.1852106}, a parametric rank correlation measure that is top-heavy, i.e., weights more the first positions of the rank. The rationale is that usually it is more important to correctly estimate the effectiveness of the top-ranked systems, i.e, the most effective ones.
We choose to give the top 10\% of the systems 
the 75\% of the weight evaluation; thus we estimate the parameter $p$ for RBO as detailed by \citet[Eq.~21]{Webber:2010:SMI:1852102.1852106}; 

\item AP \emph{correlation} ($\tau_{AP}$), proposed by \citet{Yilmaz:2008:NRC:1390334.1390435}, a top-heavy rank correlation coefficient based on the AP measure.
\end{itemize}

We also present the results by means of scatterplots, that allow to compare the real effectiveness with the effectiveness obtained by the three methods.

It has to be noted that while  the  SNC method actually  predicts an effectiveness (e.g., MAP) value,  the other two methods provide values that have  a different meaning: 
SPO returns a percentage (a number between 0 and 100) which represents the structure of overlap between  systems and
WUC produces a (normalized) reference count value (a value between 0 and $+\infty$). 
Thus, we normalize SPO and WUC scores in 3 ways, as discussed in Section~\ref{JDIQNorel:sub:generalize_gmap}.
Furthermore, we change the sign of the value returned by SPO method, to obtain positive correlations and to easily compare the results with SNC and WUC.

\section{\ref{JDIQNorel:I:R}: Reproduce} \label{JDIQNorel:sec:Reproduce}



We now turn to our first (and maybe most important) aim~\ref{JDIQNorel:I:R}, namely to reproduce the results previously published in the literature.

\subsection{Results} \label{JDIQNorel:sec:repro_result}

We first study the reproducibility of each of the three methods, and then address their comparison.

\subsubsection{SNC}

\begin{table}[tbp]
\centering
\caption{
Mean  and standard deviation of $\tau$ values: comparison for reproducibility. Original values from \citet[Tables~2 and~3]{Soboroff:2001:RRS:383952.383961} (first and third pair of columns) and our obtained values (second and fourth pair of columns). 
}
\label{JDIQNorel:tab:SNC_reproduced_cors}
\small
 %
 \begin{threeparttable}
\begin{tabular}{l cc c cc}
\toprule
 & \multicolumn{2}{c}{Original SNC} && \multicolumn{2}{c}{Our obtained values\tnote{$\dagger$}} \\
       \cmidrule{2-3} \cmidrule{5-6}
       & avg $\tau$ & std $\tau$  && avg $\tau$ & std $\tau$  \\
       \cmidrule{2-6} 
TREC-3 & .430 & .0312  && .401 (.411) & .0259 (.0276) \\
TREC-5 & .487 & .0462  && .359 (.382) & .0766 (.0279) \\
TREC-6 & .408 & .0354  && .391 (.387) & .0370 (.0299) \\
TREC-7 & .369 & .0363  && .377 (.379) & .0474 (.0470) \\
TREC-8 & .459 & .0340  && .460 (.444) & .0402 (.0567) \\
\bottomrule
\end{tabular}
\begin{tabular}{l cc c cc }
 &  \multicolumn{2}{c}{Orig.\ SNC dups} && \multicolumn{2}{c}{Our obtained dups\tnote{$\dagger$}\ \tnote{$\ddagger$}} \\
       \cmidrule{2-3} \cmidrule{5-6}
       & avg $\tau$ & std $\tau$  && avg $\tau$  & std $\tau$ \\
       \cmidrule{2-6}
TREC-3 &  .482 & .0143 && .487 (.471)  & .0113 (.0111)   \\
TREC-5 &  .571 & .0107 && .421 (.409)  & .0067 (.0043)   \\
TREC-6 &  .491 & .0131 && .458 (.452)  & .0052 (.0045)   \\
TREC-7 &  .423 & .0091 && .446 (.446)  & .0046 (.0046)   \\
TREC-8 &  .543 & .0102 && .533 (.538)  & .0043 (.0052)   \\
\bottomrule
\end{tabular}
\begin{tablenotes}
  \item[$\dagger$] The values in parentheses are those obtained with estimated $\mu$ and $\sigma$ (see Formula~\eqref{JDIQNorel:eq:estimate} and Table~\ref{JDIQNorel:tab:SNC_reproduced}).
  \item[$\ddagger$] Pool built with all the participating runs at depth 100, using the duplicates.
  \end{tablenotes}
  \end{threeparttable}
\end{table}
%
Table~\ref{JDIQNorel:tab:SNC_reproduced_cors} compares the SNC original values of mean $\tau$ 
and standard deviation of $\tau$ with those that we have obtain when reproducing such method. The left side of the table shows the values obtained considering the pool without duplicates and right side of the table shows the values when considering duplicates. Remember that we compute the mean $\tau$ score obtained over 20 repetitions.
As we can see from the table, apart from TREC-5, we obtain mean $\tau$ scores comparable to the ones of SNC, both when  duplicates are considered and when they are not.
We conjecture that the differences in the correlation values are caused by the possibility of selecting different runs to reproduce the original TREC pool.

\subsubsection{WUC}
\begin{table}[tbp]
\centering
\caption{
Average $r_S$ correlation over the topics of each collection: comparison for reproducibility. Original values from \citet[Table 1]{Wu:2003:MRI:952532.952693} (left) and our obtained values (right). 
}
\label{JDIQNorel:tab:WUC_reproduced_cors}
\begin{threeparttable}
\begin{adjustbox}{max width=\textwidth}
\begin{tabular}{l cccccc c cccccc}
\toprule
       & \multicolumn{6}{c}{Original WUC} & & \multicolumn{6}{c}{Our obtained values} \\
       \cmidrule{2-7} \cmidrule{9-14}
       & Basic & V1 & V2 & V3 & V4 & RS\tnote{$\dagger$}  &&  Basic & V1 & V2 & V3 & V4 & RS\tnote{$\dagger$}      \\
       \cmidrule{2-14}
TREC-3 & .246    & .248 & .548 & .567 & .587 & .628 && .513 & .522 & .512 & .504 & .283 & .629 \\
TREC-5 & .318    & .326 & .378 & .421 & .421 & .430 && .393 & .405 & .395 & .400 & .328 & .476 \\
TREC-6 & .309    & .316 & .371 & .383 & .384 & .436 && .442 & .451 & .485 & .497 & .498 & .522 \\
TREC-7 & .297    & .304 & .328 & .345 & .382 & .411 && .406 & .419 & .403 & .421 & .453 & .501 \\
TREC-01 & .279    & .288 & .377 & .401 & .413 & .463 && .449 & .460 & .448 & .459 & .443 & .571 \\
\bottomrule
\end{tabular}
\end{adjustbox}
\begin{tablenotes}
\item[$\dagger$] SNC, all runs, considering the duplicates and selecting \\randomly the 10\% of the documents as relevant.
  \end{tablenotes}
  \end{threeparttable}
\end{table}

Table~\ref{JDIQNorel:tab:WUC_reproduced_cors} shows a similar comparison for the WUC method. Here the comparison is based on  the average $r_S$ correlation values over the topics of each collection. This is of course different from SNC: we now focus on reproducing the results, thus we use the same correlation coefficients used in the original papers; we will provide a homogeneous comparison later. As we can see from the 
table, for almost all WUC variants (i.e., Basic, V1, V2, and V3) we obtain higher correlation values; concerning WUC V4, we obtain higher correlation values in three cases out of five.
In the original paper the highest correlation value is always obtained with the variant V4; instead, we find that the best variant depends on the collection.
Furthermore, in the original paper the four variants have increasing correlation values (i.e., for each collection, we always have Basic $<$ V1 $<$ V2 $<$ V3 $<$ V4); on the contrary, the correlation values that we obtain are usually comparable across the different WUC variants.

We conjecture that the differences between our values and the original ones might be due to:
\begin{itemize}
\item 
A different normalization score, especially for V4 (i.e., the most effective version in the original paper).
Regarding V4,
\citet[page~813]{Wu:2003:MRI:952532.952693} state: ``V4 consists of using each document's normalized score'', but a detailed explanation of the normalization process is not provided; for this reason we use each system ``retrieval status values'' (RSV) to compute each document normalized score.
We also consider the rank of the document with similar (or worse) outcomes.
Note that many different normalization formulas could have been used by original authors;
we tried different variants:
(i) normalize the RSV / rank in $[0,1]$, 
(ii) normalize the RSV / rank using the standard score 
$\frac{(x-\mu)}{\sigma}$), 
and
(iii) the above normalization using WUC Basic, V1, V2, or V3 in place of the RSV / rank.
We find that none of them leads to a successful reproducibility of original results.
Furthermore, we find surprising to not be able to reproduce WUC Basic version, since it does not require any normalization of the reference count score. 
%
\item 
A different approach used to compute correlations. 
\citet[page~813]{Wu:2003:MRI:952532.952693} state: ``In Table 1 we present the results when all runs of participant systems are taken. Each item in the table is the mean Spearman rank correlation coefficient over 50 topics of that year''.
If we have a collection with $n$ topics, we denote with $\AP(t_i)$ the vector of official $\AP$ for all the runs of topic $i$ (i.e., a column of Table~\ref{JDIQNorel:tab:AP}), and we use the subscripts $o$ and $w$ to differentiate respectively the official and the WUC measure, then
Table~\ref{JDIQNorel:tab:WUC_reproduced_cors} shows 
\begin{equation*}
\frac{1}{n} \sum_{i=1}^{n} r_{S}\left( \AP_{o}(t_i), \AP_{w}(t_i) \right). 
\end{equation*}
We try different variations, namely: 
\begin{equation*}
\begin{aligned}
&\mbox{(i) } \frac{1}{n} \sum_{i=1}^{n} r_{S}\left( \AP_{o}(t_i), \MAP_{w} \right) \mbox{, } 
\mbox{(ii) } \frac{1}{n} \sum_{i=1}^{n} r_{S}\left( \MAP_{o}, \AP_{w}(t_i) \right) \mbox{, and } \\ 
&\mbox{(iii) } r_{S}\left( \MAP_{o}, \MAP_{w} \right) \mbox{. }  
\end{aligned}
\end{equation*}
None of them lead to an effective reproducibility.
We also use our own implementation of Spearman rank $r_S$ correlation, which reflects the formula detailed by \citet[Eq.~4]{Wu:2003:MRI:952532.952693}; 
also in this case the results do not vary from the previous ones, obtained using the official Python 3 \emph{scipy.stats.spearmanr}\footnote{\url{https://docs.scipy.org/doc/scipy-0.14.0/reference/generated/scipy.stats.spearmanr.html}} implementation.
%
\item 
The number of runs used to compute the $r_S$ correlation score. Although \citet[Tab.~1]{Wu:2003:MRI:952532.952693} state: ``Mean correlation coefficient for average precision, all systems submitted to TREC'',
we try 
with a selection of systems. Also this attempt results in a failed reproducibility of the original scores.
\end{itemize}

We  remark that it is unlikely that by doing 
something wrong
we obtain higher correlation values than the original ones; one should expect exactly the opposite.
We release the code used to compute WUC scores (see Footnote~\ref{JDIQNorel:fn:URL}),
which could be useful for future reproducibility, and for a correct implementation of the method described by \citeauthor{Wu:2003:MRI:952532.952693}.

\citeauthor{Wu:2003:MRI:952532.952693} also reported, for comparison, the $r_S$ correlation values that they obtained when reimplementing SNC (in a slightly different version, i.e., considering the duplicates and selecting randomly the 10\% of the documents as relevant). Table~\ref{JDIQNorel:tab:WUC_reproduced_cors} shows, in the two ``RS'' columns, the $r_S$ correlation values obtained by \citeauthor{Wu:2003:MRI:952532.952693} and by ourselves when we reimplement SNC  following \citeauthor{Wu:2003:MRI:952532.952693} variation. 
We see that, also in this case, the values are in general different, apart from TREC-3 and TREC-5.
Referring to Table~\ref{JDIQNorel:tab:SNC_reproduced_cors}, recall that our $\tau$ correlation values are similar to SNC original ones. 
This might depend, again, on one of the justifications detailed above.

\subsubsection{SPO}

\begin{table}[tbp]
\centering
\caption{
Average $r_S$ correlation: comparison for reproducibility. Original values from from \citet[Table 1]{spoerri:2007} (first and third groups of three columns) and our obtained values (second and fourth groups of three columns). 
The  table shows on the left side the values obtained when using all the 1000 retrieved documents, and on  the right side when using the top 50 ones.
S\%, A\%, and S-A\% values computed using one short-run  per participant group (the one with highest MAP score). 
TREC-3, TREC-6, and TREC-7 correlations are estimated from \citet[Fig. 5]{spoerri:2007}.
}
\label{JDIQNorel:tab:SPO_reproduced_cors}
\begin{threeparttable}
\begin{adjustbox}{max width=\textwidth}
\begin{tabular}{l c@{\hspace{0.4cm}}cc@{} c c@{\hspace{0.5cm}}c@{}c@{} c c@{\hspace{0.4cm}}c@{}c@{} c c@{\hspace{0.5cm}}c@{}c@{}}
\toprule
 & \multicolumn{3}{c}{Orig. SPO} && \multicolumn{3}{c}{Our obtained  values} && \multicolumn{3}{c}{Orig. SPO top 50}  && \multicolumn{3}{c}{Our obtained top 50}  \\
       \cmidrule{2-4} \cmidrule{6-8} \cmidrule{10-12}  \cmidrule{14-16}
 & S\% & A\% &S-A\% && S\% & A\% & S-A\% && S\% & A\% &S-A\% && S\% & A\% & S-A\% \\
       \cmidrule{2-16}
TREC-3 & .67 & -- & .72  &&  .67 & .51 & .69  && .72 & -- & .72   &&   .46 & .25 & .44    \\
TREC-6 & .79 & -- & .78  &&  .69 & .40 & .68  && .86 & -- & .88   &&   .80 & .44 & .79    \\   
TREC-7 & .43 & -- & .47  &&  .59 & .38 & .57  && .70 & -- & .62   &&   .67 & .35 & .63    \\	  
TREC-8 & .89 & .88 & .95 &&  .79 & .48 & .76  && .92 & -- & .95   &&   .82 & .55 & .80    \\  
\bottomrule
\end{tabular}
\end{adjustbox}
  \end{threeparttable}
\end{table}

Table~\ref{JDIQNorel:tab:SPO_reproduced_cors} shows the  SPO S\%, A\%, and S-A\% scores
obtained by \citeauthor{spoerri:2007} and by ourselves
when considering one run (the short-run) per participant group;
the table shows 
correlation values (averaged over the $N$ random grouping of $N$ short-runs) when considering all the 1000 retrieved documents (left) and only the top 50  retrieved (right). 
%
For completeness, the table also contains all our A\% values, not always reported in the original paper. 
We see that our computed values, for both top 50 and top 1000, are comparable to \citeauthor{spoerri:2007}'s ones, even if they are usually lower; there are two exceptions, though: TREC-7 top 1000 and TREC-3 top 50.

%
The differences between the correlation values might depend on two different factors: 
(i) the division of systems into the sets containing five systems
is not trivial, and the implemented algorithm can affect the final result, and
(ii) the selection of which run to include from each participating group has a major impact on the correlation values, and a list of the used runs is not available in the original paper (we make available our list in the URL provided in Footnote~\ref{JDIQNorel:fn:URL}).

Another important remark is that in the original work the version with the highest correlation values is S-A\% for almost all collections, but according to our finding this is not always the case, especially when considering the top 50 rank positions.

\subsubsection{Comparison of SNC, WUC, and SPO} 
\begin{table}[tbp]
\centering
\caption{MAP comparison between all state-of-the-art used collections and all methods using $\tau$ correlation. We considered all the runs for each track, and the top 1000 retrieved documents. The highest values for each collection are in bold.
}
\label{JDIQNorel:tab:all_reproduced_cors}
\begin{threeparttable}
\begin{adjustbox}{max width=\textwidth}
\begin{tabular}{l cc c ccccc c ccc }
\toprule
     & \multicolumn{2}{c}{SNC} & & \multicolumn{5}{c}{WUC} & & \multicolumn{3}{c}{SPO} \\
       \cmidrule{2-3} \cmidrule{5-9} \cmidrule{11-13}
       & qrel & dups & & Basic & V1 & V2 & V3 & V4 & & S\% & A\% & S-A\% \\ 
       \cmidrule{2-13}
TREC-3  & .401 & \textbf{.486}  && .405 & .405 & .436 & .444 & .344  && .416 & .216 & .400\\
TREC-5  & .359 & \textbf{.422}  && .332 & .341 & .336 & .338 & .355  && .359 & .215 & .352 \\
TREC-6  & .391 & \textbf{.459}  && .422 & .427 & .451 & .448 & .453  && .421 & .288 & .421 \\
TREC-7  & .378 & \textbf{.445}  && .402 & .408 & .393 & .408 & .430  && .358 & .216 & .346 \\
TREC-8  & .460\tnote{$\dagger$} & \textbf{.532}  && .466 & .480 & .386 & .391\tnote{$\dagger$} & .322  && .490\tnote{$\dagger$} & .322 & .475\\
TREC-01 & .473 & \textbf{.625}  && .537 & .545 & .582 & .582 & .554  && .521 & .347 & .520 \\
\bottomrule
\end{tabular}
\end{adjustbox}
\begin{tablenotes}
\item[$\dagger$] shown in Figure~\ref{JDIQNorel:fig:reproduce} (see below).
  \end{tablenotes}
  \end{threeparttable}
\end{table}

As we have already noticed, and as it is clear from the last three tables, the correlation values reported in the original papers are not directly comparable.  To be able to compare in a more systematic and convenient way the three methods, in Table~\ref{JDIQNorel:tab:all_reproduced_cors} we show the $\tau$ correlation obtained by us when reproducing the methods considering all the runs participating in the track and the top 1000 documents retrieved for each topic.
As we can see from the table, SNC dups always achieves the highest correlation values. This might be surprising, since it is the most trivial method.

\begin{figure}[tbp]
  \centering
  \begin{tabular}{@{}c@{}c@{}c@{}}
    \includegraphics[width=.33\linewidth]{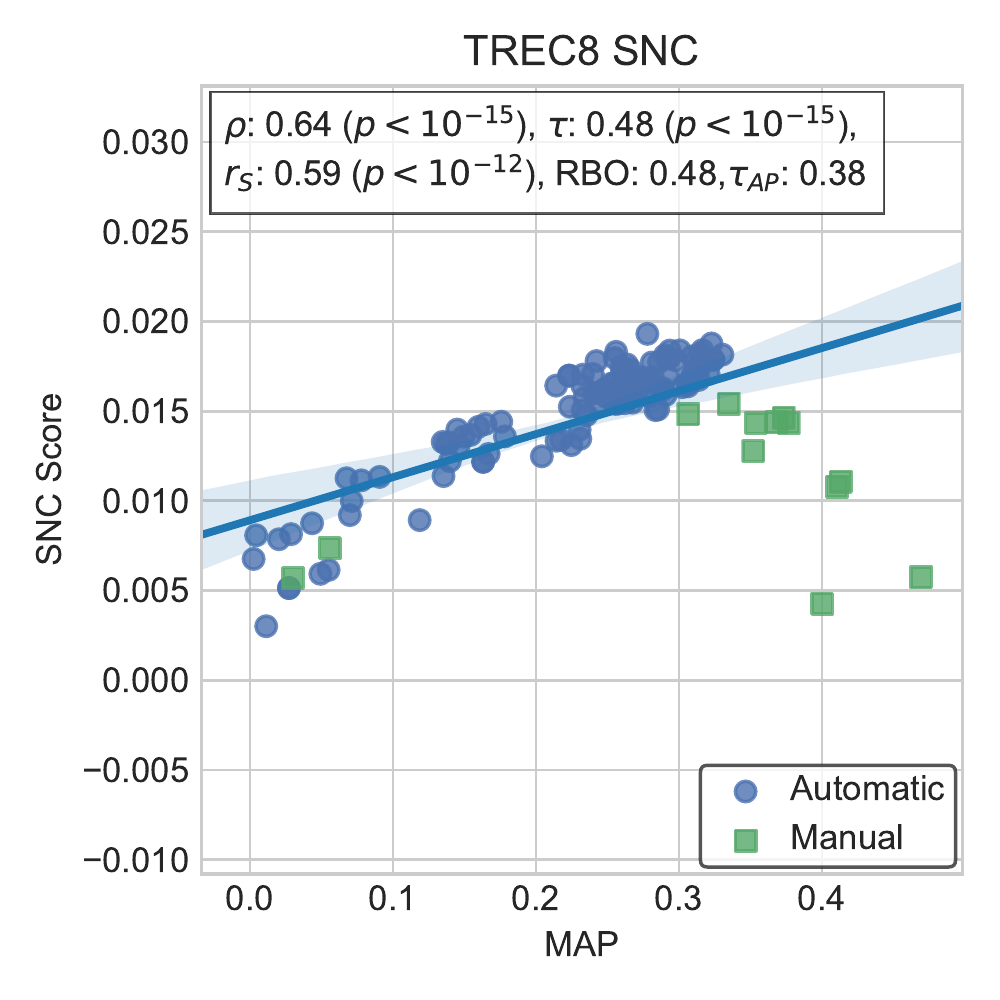}&
        \includegraphics[width=.33\linewidth]{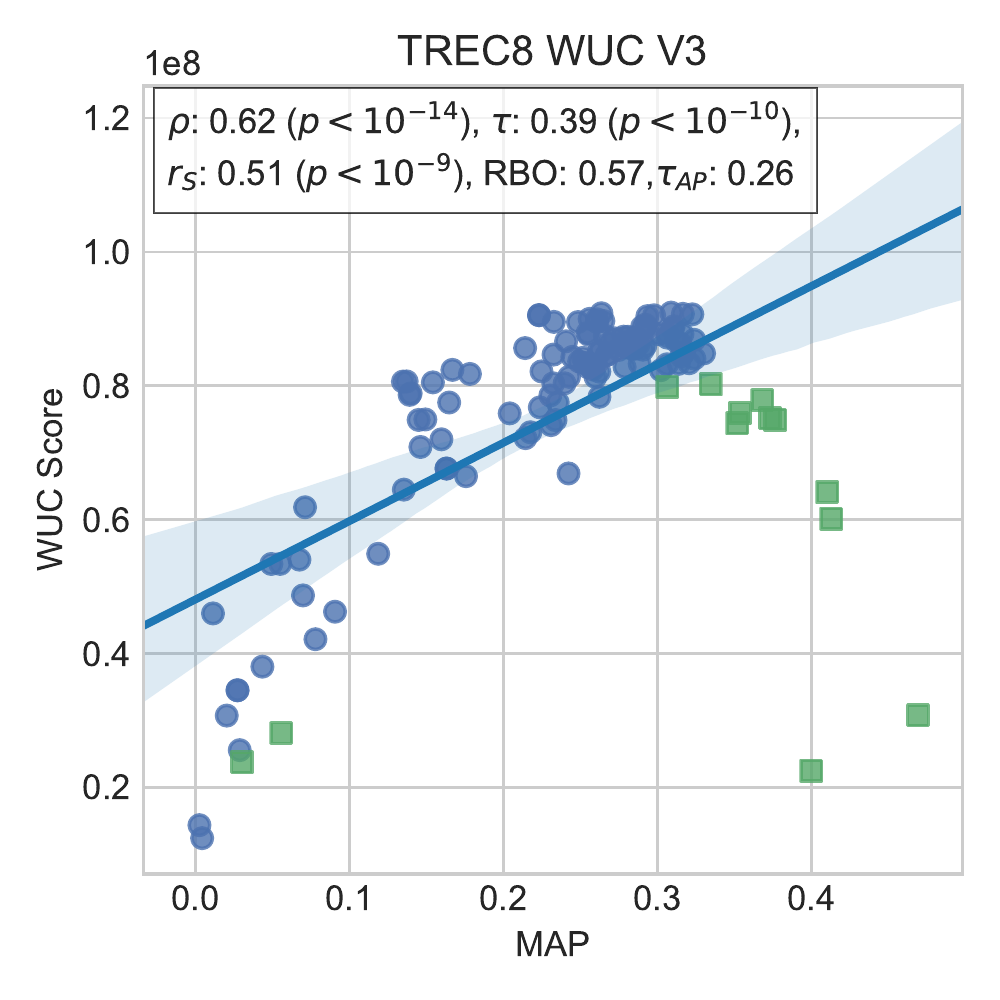}&
    \includegraphics[width=.33\linewidth]{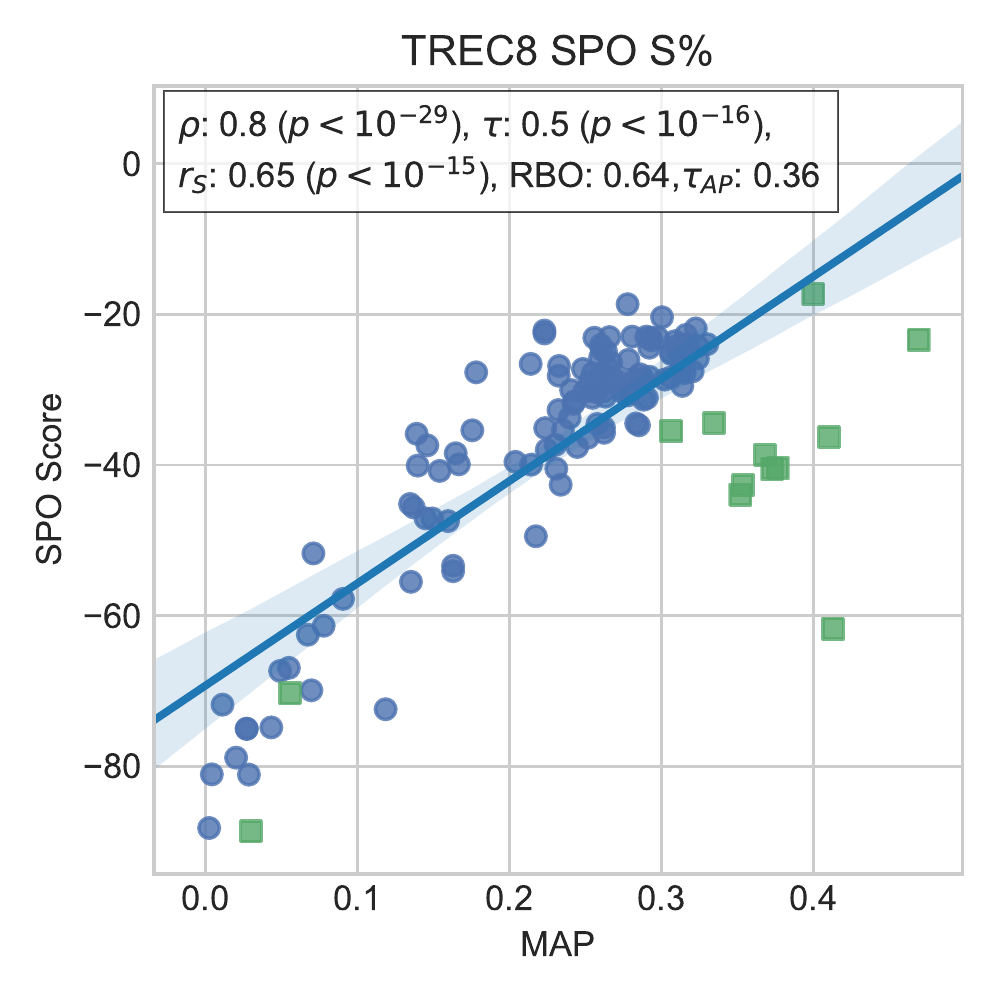}
  \end{tabular}
  \caption{Scatterplots for the three methods: 
  SNC
  (left), WUC  V3
  (center), and SPO S\%
  (right), on the TREC-8 collection.
  Each dot is a system / run. 
  The x-axis shows the real MAP value; 
  the y-axis shows the predicted MAP (for SNC) or a meaningful value of the specific method, i.e., a (normalized) count for  WUC  and a (negated) percentage for SPO. 
}
  \label{JDIQNorel:fig:reproduce}
\end{figure}

Figure~\ref{JDIQNorel:fig:reproduce} shows some selected results as scatter plots: the x-axis shows the real MAP value, while the y-axis shows the score of the SNC, SPO, and WUC methods; each dot is a system;  and automatic and manual runs are graphically different.
In the plots we also display the regression line as well as $\rho$, $\tau$, $r_S$, RBO, and $\tau_{AP}$ measures.
Note that the correlation values are minimally different from the ones in Table~\ref{JDIQNorel:tab:all_reproduced_cors}; the table shows the average correlation obtained over the 20 repetitions, the figure shows instead the plot and the corresponding correlation values where each (M)AP is the average (M)AP obtained over the 20 repetitions. 

%
For  WUC and SPO we selected the method variants featuring the best results, i.e., those having highest Kendall's correlation (V3 and S\%, respectively). 
The V4 version of WUC and S-A\% version of SPO have a very similar outcome.
In the following we report the WUC V3 and V4, and the SPO S\%: they are almost always the best ones; when they are not the absolute best, they are very close to the best or even indistinguishable.
An exhaustive appendix of our results can be found at the URL provided in Footnote~\ref{JDIQNorel:fn:URL}.

Figure~\ref{JDIQNorel:fig:reproduce} shows that all the three methods present an ``inverse U'' shape (less evident for SPO): all the methods fail in predicting the effectiveness of the most effective systems, which  in this case (for TREC-8) are  manual runs; we analyze the effect of the manual runs in \ref{JDIQNorel:sub:generalize_other_collections}; we can see that the methods are equally bad in predicting the top ranked systems by the values of $\tau_{AP}$ and RBO, that are very similar for the three methods. 
%


\subsection{Discussion}
Considering the results, we can make  two main remarks:
\begin{enumerate} [label=(\roman*)] 

\item  
On reproducing per-se: reproducing the work is never easy, with all the methods (i.e., SNC, WUC, and SPO), for many different reasons: specific choices are often not described in the original papers, like the choice of including a system instead of another to form a subset of systems/run (see Sections~\ref{JDIQNorel:sec:methods_config} and \ref{JDIQNorel:sec:repro_result}); 
when the method includes a non deterministic process, an exact reproducibility is not possible.
Furthermore, SNC is the only method which produces estimated AP values, the other ones produce a score, which can be of difficult interpretation.

\item On the results, the rather low correlation values, as well as the ``inverse U'' shape that can be seen in  Figure~\ref{JDIQNorel:fig:reproduce} seem to discourage the use and further development of these approaches.
\end{enumerate}

To investigate this latter remark, we generalize the three methods to other collections,  other evaluation metrics, and a shallow pool, which we discuss next.

\section{\ref{JDIQNorel:I:G}: Generalize} \label{JDIQNorel:sec:Generalize}
We now turn to the second aim \ref{JDIQNorel:I:G}; more precisely, we generalize the previous work results to
other test collections (Section~\ref{JDIQNorel:sub:generalize_other_collections}),  
other evaluation metrics (Section~\ref{JDIQNorel:sub:generalize_other_metrics}), 
and a shallow pool (Section~\ref{JDIQNorel:sub:generalize_shallow_pool}). We also briefly discuss
these results (Section~\ref{JDIQNorel:sec:generalize:discuss}).

Before detailing the results, 
we make some remarks on generalizing the three methods:
(i) SPO and WUC do not estimate AP values, thus we have to normalize the produced scores, which can be done in many different ways, as we discuss next;
(ii) all WUC versions are calibrated considering the top 1000 documents retrieved by each system; thus using other pool depths with WUC is not trivial;
(iii) SNC requires an estimation of the $\mu$ and $\sigma$ parameters:
in the case of a new collection, these parameters can be estimated using different techniques, which may condition the effectiveness of this method.
Furthermore, while SPO and WUC can in principle work for any relevance scale, SNC is restricted to the binary relevance scale.

\subsection{\ref{JDIQNorel:I:G:C}: Generalize to Other Collections} \label{JDIQNorel:sub:generalize_other_collections}
\begin{figure}[tbp]
  \centering
  \begin{tabular}{@{}c@{}c@{}c@{}}
    \includegraphics[width=.33\linewidth]{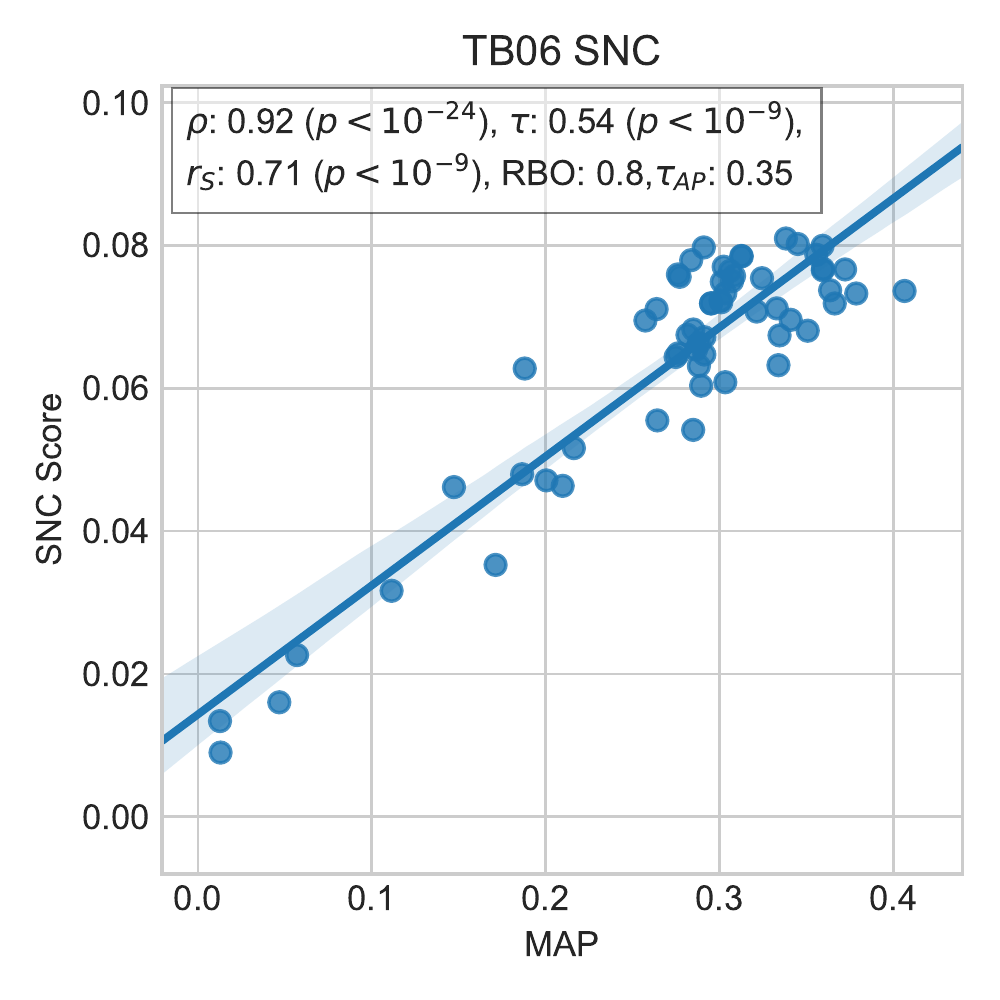}&
        \includegraphics[width=.33\linewidth]{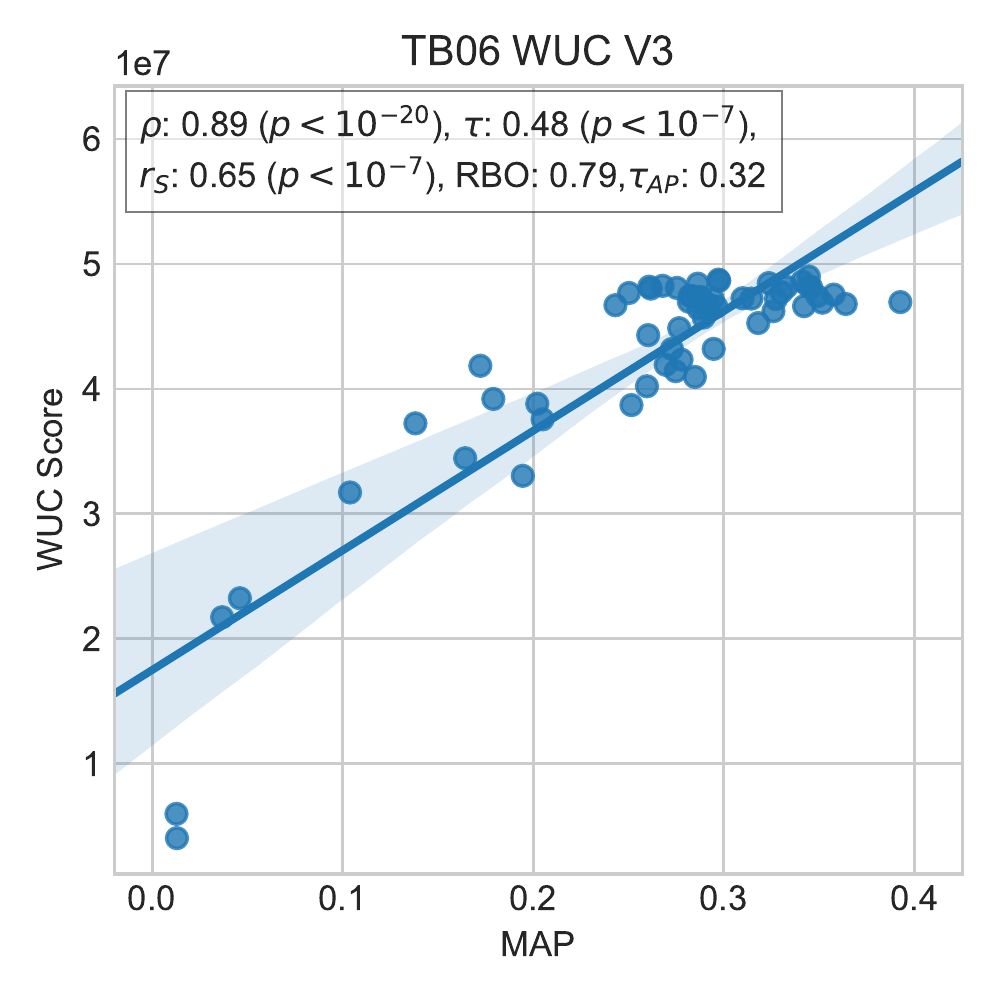}&
    \includegraphics[width=.33\linewidth]{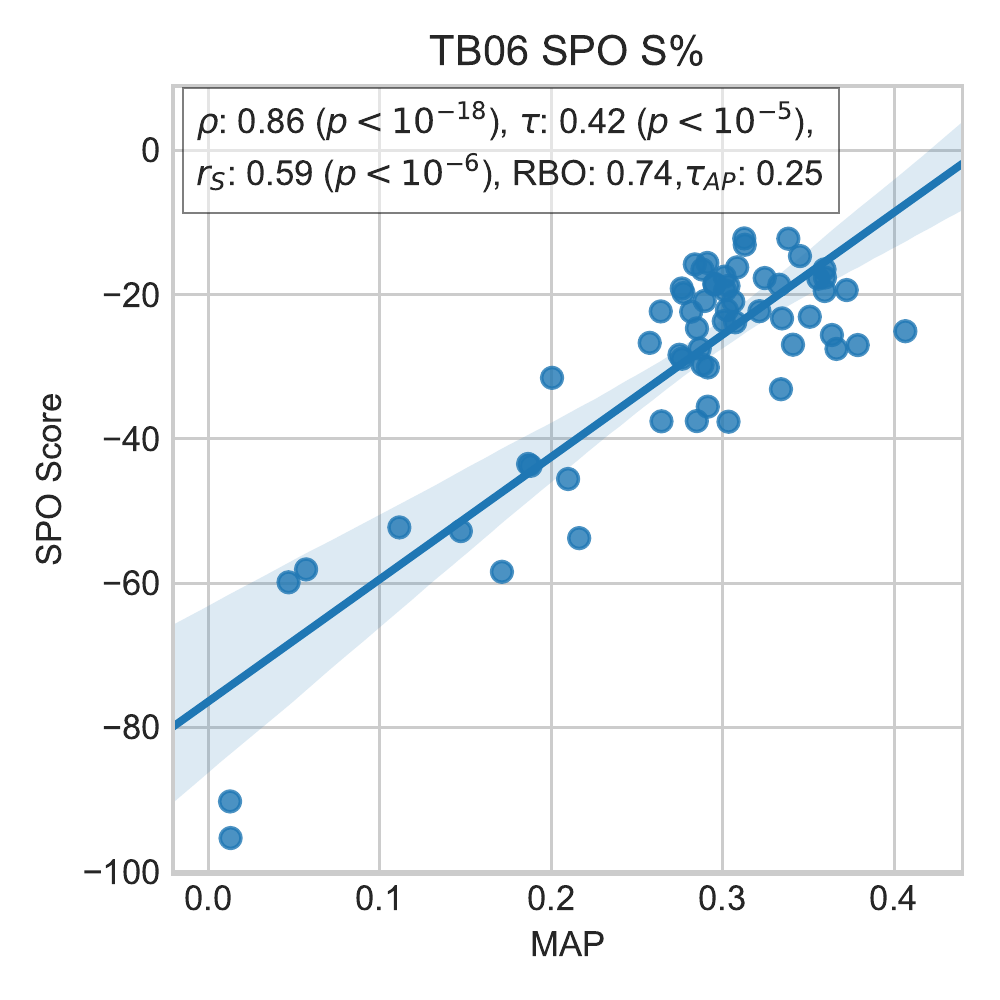}  \\
    \includegraphics[width=.33\linewidth]{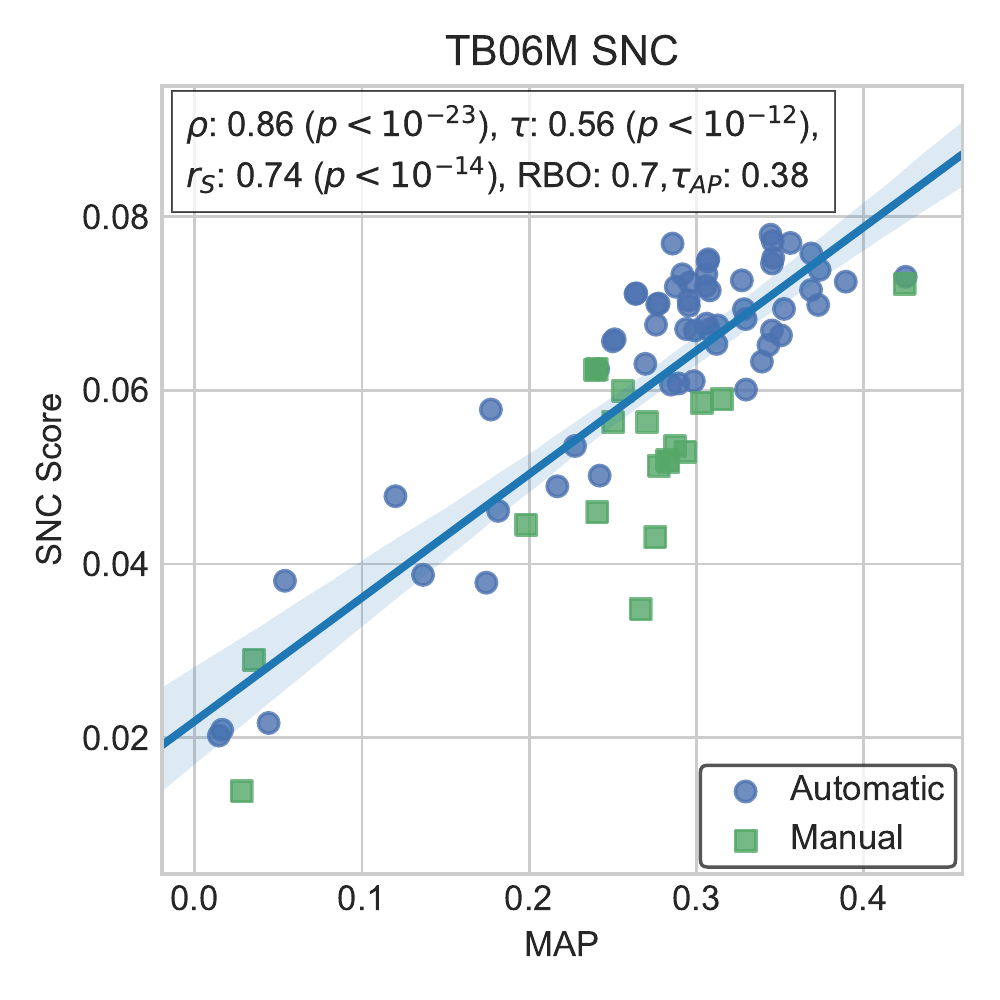}&
        \includegraphics[width=.33\linewidth]{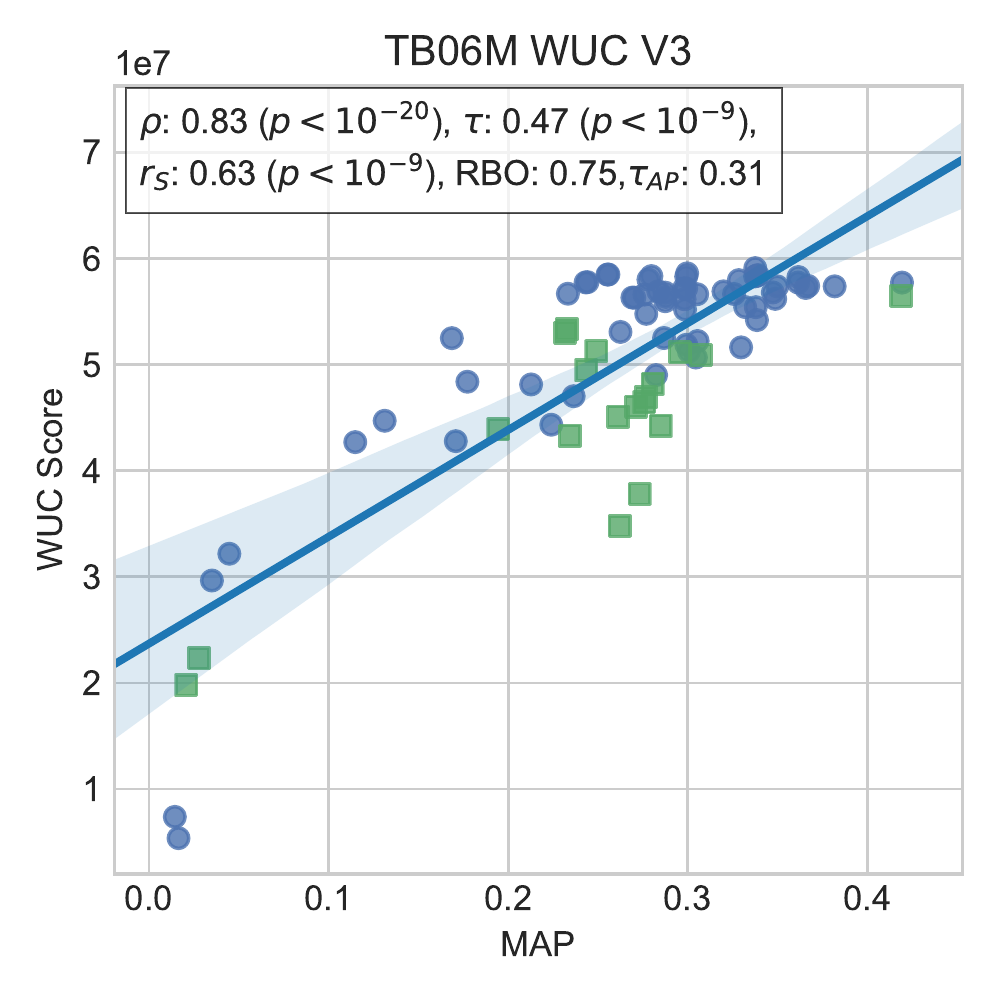}&
    \includegraphics[width=.33\linewidth]{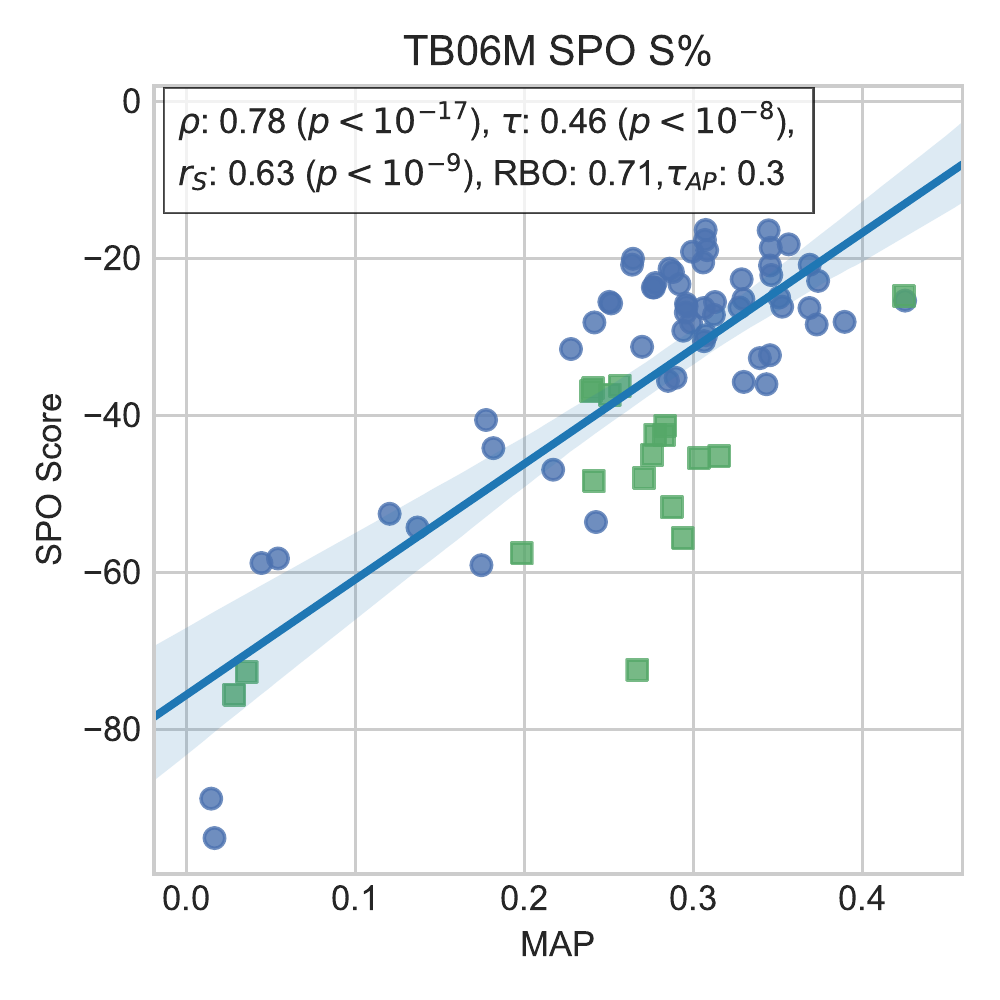}  \\
    \includegraphics[width=.33\linewidth]{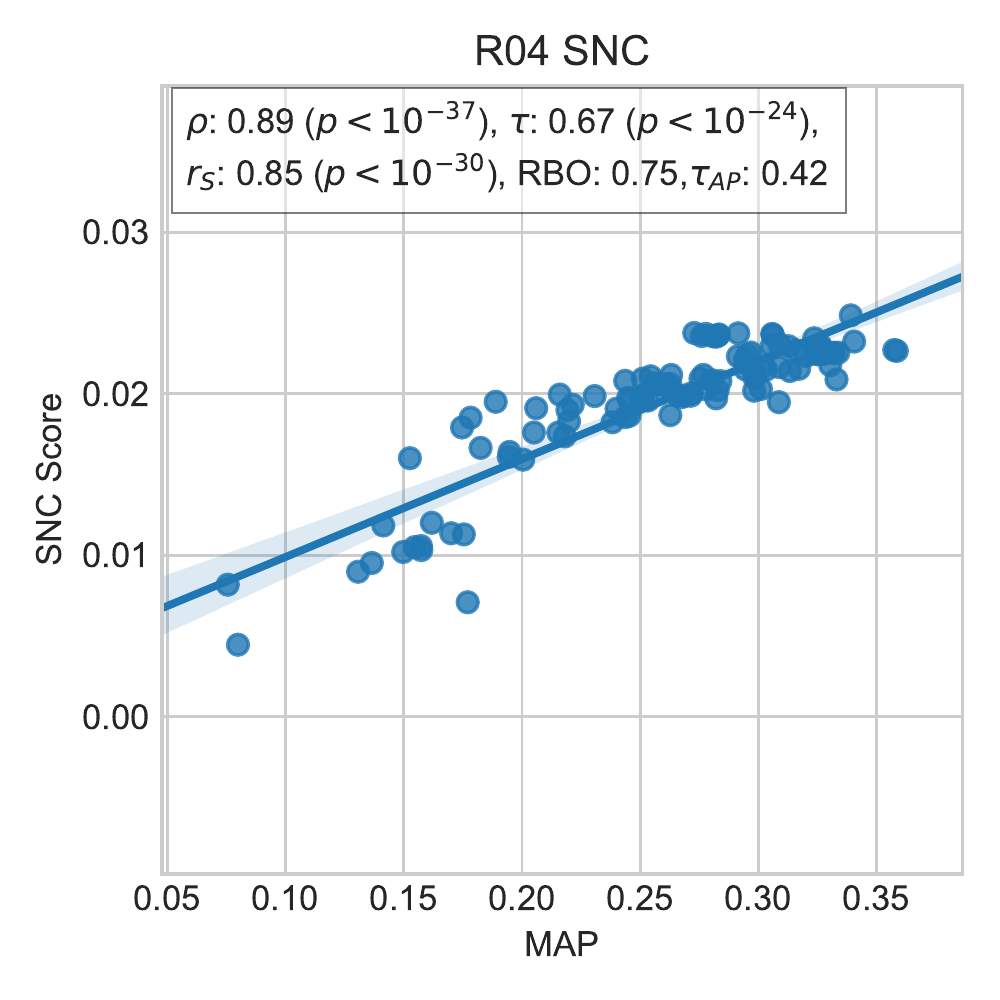}&
        \includegraphics[width=.33\linewidth]{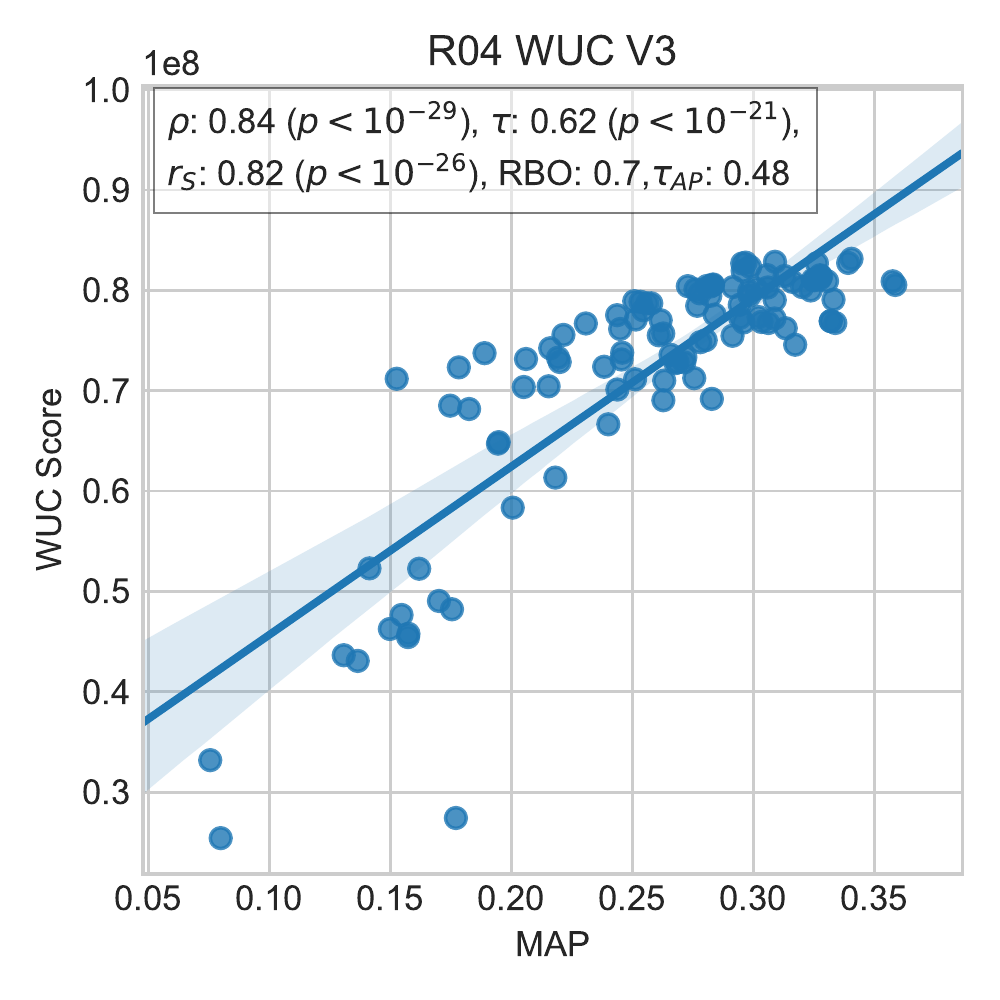}&
    \includegraphics[width=.33\linewidth]{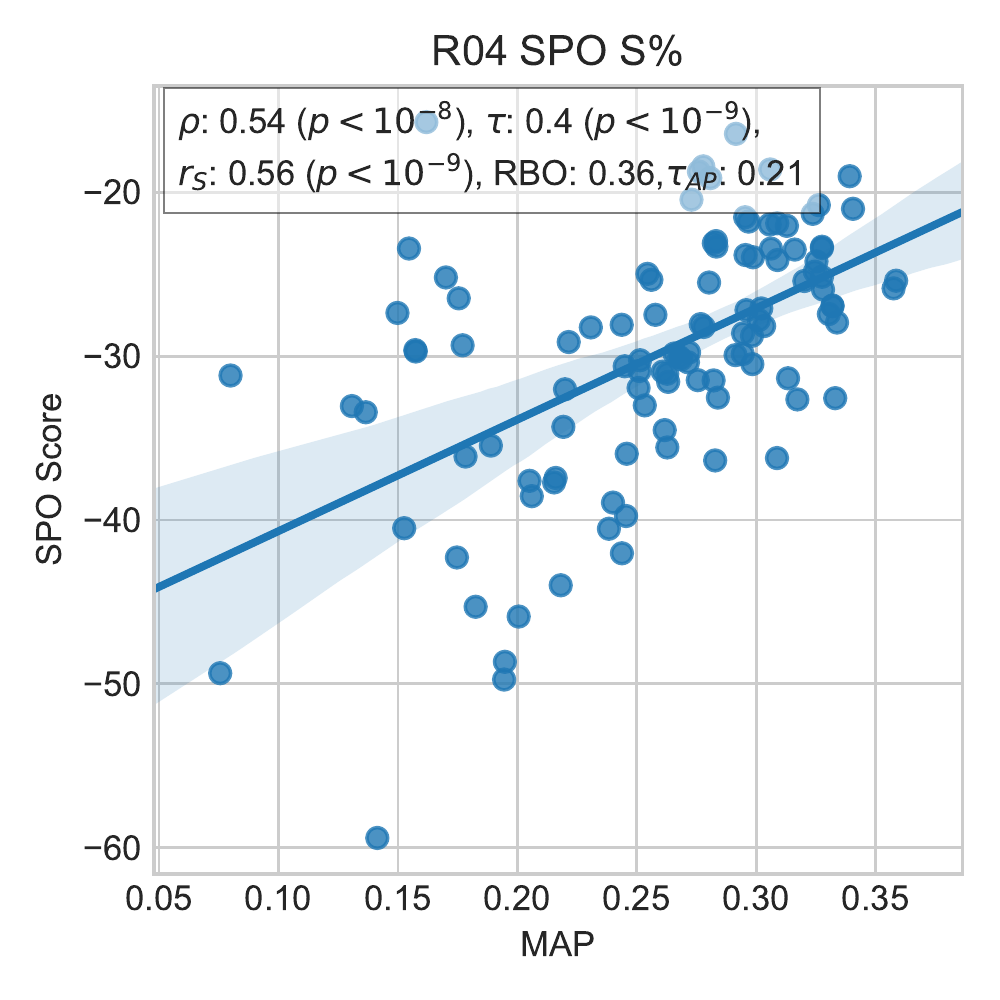} \\
     \includegraphics[width=.33\linewidth]{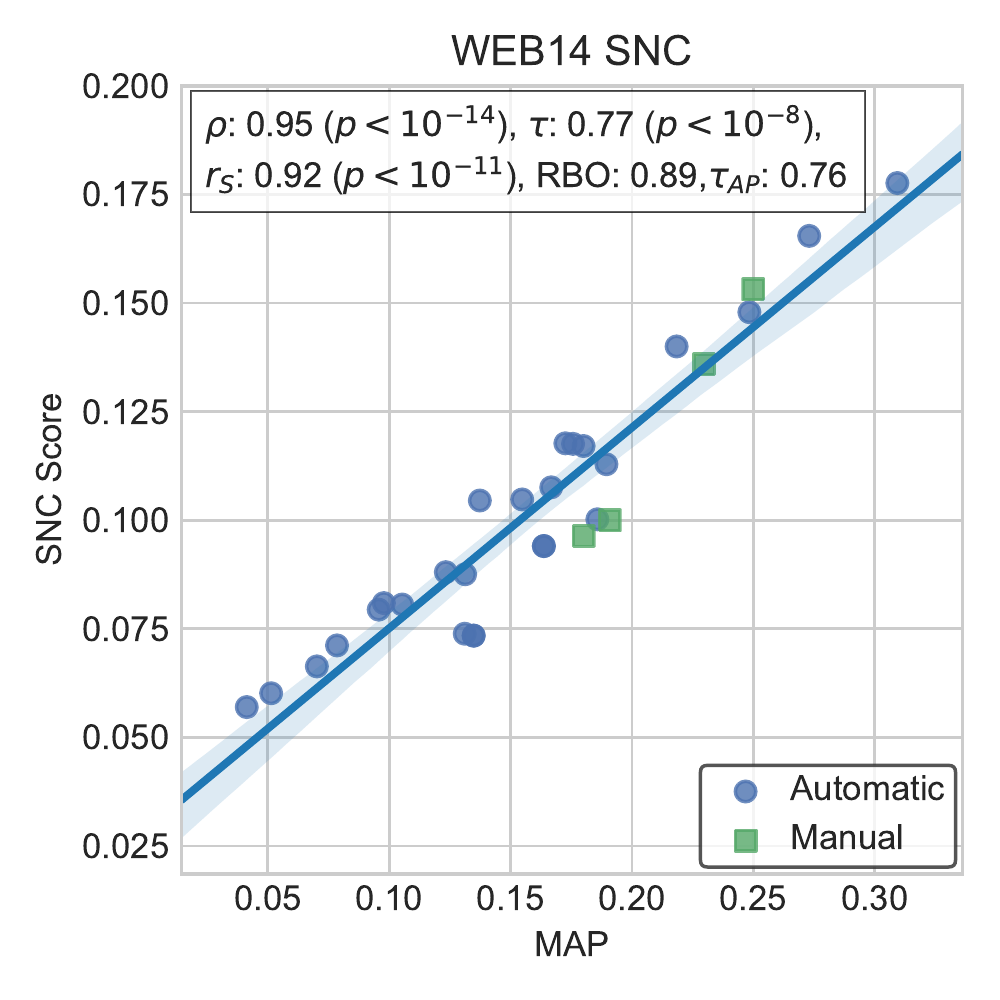}&
        \includegraphics[width=.33\linewidth]{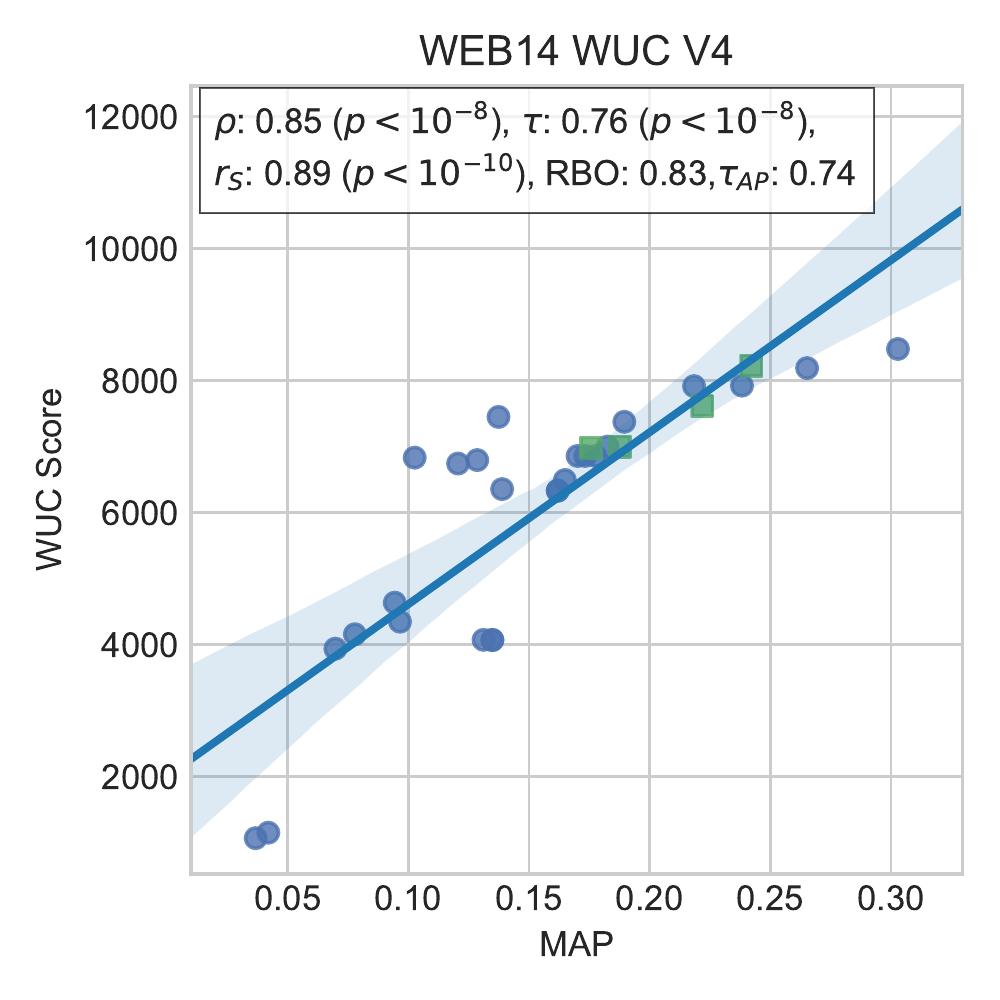}&
    \includegraphics[width=.33\linewidth]{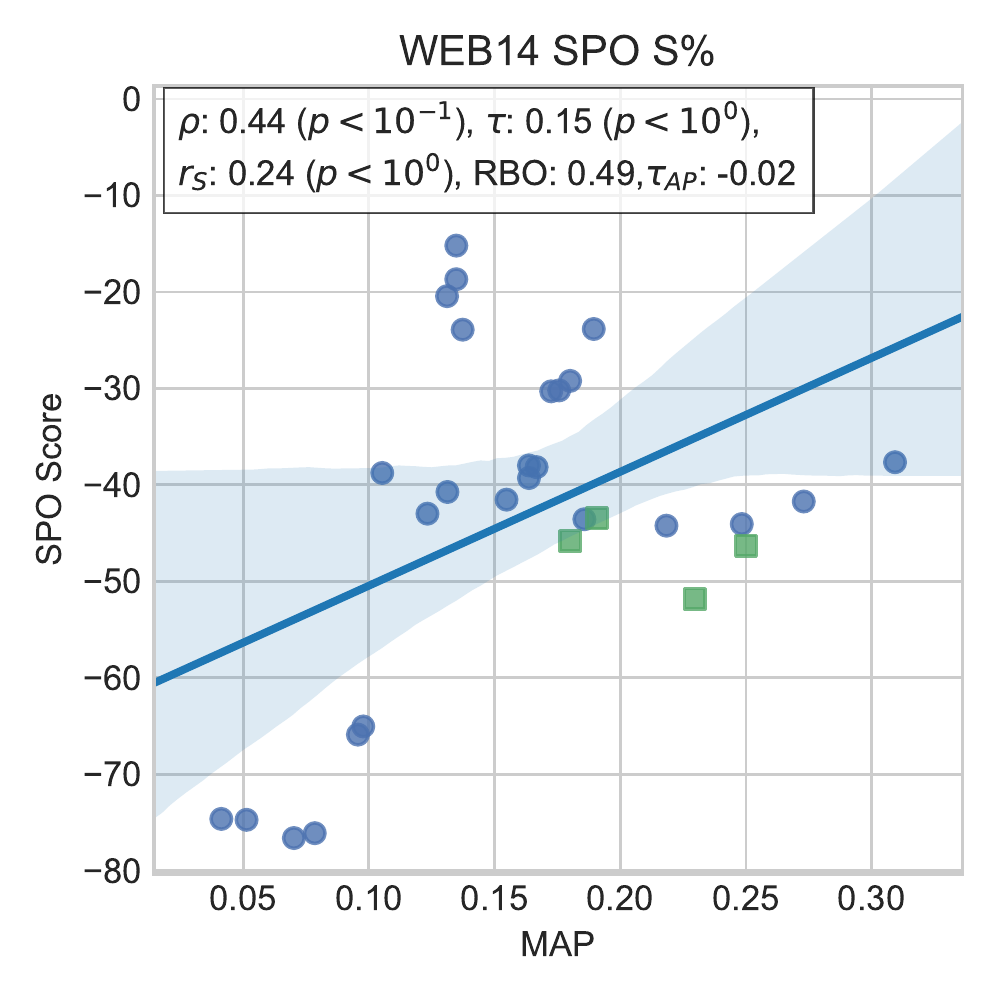}  \\
  \end{tabular}
  \caption{Generalization to other collections. Scatterplots for the three methods 
  SNC,
  WUC V3 (or V4), 
  and SPO S\%
  (on the first, second, and third columns, respectively)
  and four collections TB06, TB06M, R04, and WEB14 (on the first, second, third, and fourth rows, respectively). Compare with Figure~\ref{JDIQNorel:fig:reproduce}. 
}
  \label{JDIQNorel:fig:other-collections}
\end{figure}

To generalize the results to other collections, we choose  different TREC editions featuring a different number of systems/runs and  of topics (see Table~\ref{JDIQNorel:tab:coll}):  TB06, TB06M, R04, and WEB14.
Results on these other collections are shown in Figure~\ref{JDIQNorel:fig:other-collections}.
The charts in Figure~\ref{JDIQNorel:fig:other-collections} are organized by rows (each row represents a collection: TB06, TB06M, R04, and WEB14, respectively), and by columns (each column represents a 
different method: SNC, WUC, and SPO, respectively).

From the whole Figure~\ref{JDIQNorel:fig:other-collections} we notice that, on these collections, the correlations are usually higher for the  SNC and WUC methods, whereas they are always lower for SPO.
Furthermore,  the three methods do  not show a consistent  behavior across various datasets: by comparing 
Figure~\ref{JDIQNorel:fig:reproduce} with Figure~\ref{JDIQNorel:fig:other-collections}, and the rows of Figure~\ref{JDIQNorel:fig:other-collections} (i.e., considering different datasets but the same method), we notice that the performance of a single method is highly dependent on the particular dataset.
On the contrary, the three methods are consistent within each dataset.

From the figure
we also see that the most effective systems are not penalized as they were in Figure~\ref{JDIQNorel:fig:reproduce}: the ``inverse U'' shape of TREC-8 does not occur anymore. This is true for SNC and WUC, while SPO still penalizes the best systems, although in a less evident way.
This behavior can be caused by the fact that for TREC-8 the most effective systems are the manual runs, which are peculiar systems; in fact, the ``inverse U'' shape disappears if we imagine to remove all the manual runs from Figure~\ref{JDIQNorel:fig:reproduce}.
If we focus on the manual runs of TB06M and WEB14, their performance is still underestimated by the three methods, but, differently from TREC-8, their computed score for the three methods is not similar (i.e., they do not ``cluster'' as in TREC-8); this can be caused by the fact that for TB06M and WEB14 the manual runs are not the most effective.

This hints that the three methods for effectiveness evaluation without relevance judgements, rather than failing in predicting the most effective systems, fail in providing a correct prediction of the effectiveness of manual systems. Also taking into account the work by \citet{Aslam:Savell:2003}, this is probably due to the manual systems being ``unpopular'', both because they are intrinsically different from the automatic ones, and because there are many more automatic than manual systems in each TREC edition (see Table~\ref{JDIQNorel:tab:coll}).

\subsection{\ref{JDIQNorel:I:G:M}: Generalize to Other Evaluation  Metrics} \label{JDIQNorel:sub:generalize_other_metrics}
We now turn to Aim \ref{JDIQNorel:I:G:M}: generalize to other evaluation metrics; more precisely, we generalize previous results to the 
GMAP (Section~\ref{JDIQNorel:sub:generalize_gmap}),
logitAP (Section~\ref{JDIQNorel:sub:generalize_logitAP}), and 
NDCG (Section~\ref{JDIQNorel:sub:generalize_ndcg}) metrics.

\subsubsection{GMAP} \label{JDIQNorel:sub:generalize_gmap}
Geometric Mean Average Precision (GMAP), proposed by  \citet{Robertson:2006:GOT:1183614.1183630}, is the geometric mean of the AP values over the set of topics. Using the same notation as in Table~\ref{JDIQNorel:tab:AP} and in Formula~\eqref{JDIQNorel:eq:MAP}:
\begin{equation}\label{JDIQNorel:eq:GMAP}	
\GMAP(s_i) = \sqrt[n]{\prod_{j=1}^n \AP(s_i,t_j)} = \exp\left(\frac{1}{n} \sum_{j=1}^{n} \ln(\AP(s_i, t_j)) \right),
\end{equation}
where we use the natural logarithm ($\ln$) and its inverse function ($\exp$), and AP values of zero need to be replaced by a small $\epsilon$ value of $10^{-5}$ as done by \citeauthor{Robertson:2006:GOT:1183614.1183630}. Even though the GMAP measure averages the system effectiveness over all the topics as MAP does (see Formula~\eqref{JDIQNorel:eq:MAP}), GMAP gives emphasis on the low values of the effectiveness measure (i.e., the bottom of the scale for an effectiveness measure).
To better explain the concept we report an example taken from \citet[page~81]{Robertson:2006:GOT:1183614.1183630}:
\begin{displayquote}
$\GMAP$ treats a change in $\AP$ from 0.05 to 0.1 as having the same value as a change from 0.25 to 0.5. $\MAP$ would equate the former with a change from 0.25 to 0.3, regarding a change from 0.25 to 0.5 as five times larger.
\end{displayquote}
Furthermore, as discussed by \citet{Fuhr17},  when a researcher is interested in measuring relative changes, GMAP should be used; when absolute changes are studied, MAP should be used. 

To compute the GMAP values, we tried different normalizations of the WUC and SPO scores:
scores normalized in $\left[0,1\right]$,
scores normalized in $\left[0,0.5\right]$ (since the AP scores are often in that range),
scores normalized in\\ $\left[\min_{i,j}(\AP(s_i,t_j)), \max_{i,j}(\AP(s_i,t_j))  \right]$, and 
no normalization at all.
To normalize the scores, first we transform the AP values into log AP scores; 
then, we normalize those scores per-collection; e.g., to normalize in $[0,1]$ we consider the $min$ and $max$ value of the collection, (i.e., $max$ and $min$ values of Table~\ref{JDIQNorel:tab:AP}, all runs and topics together);
finally, we average the results over the set of topics.
Results are all very similar, and in the following we use the first normalization.

Results of the generalization to the GMAP measure are shown in Table~\ref{JDIQNorel:tab:GMAP}. By comparing the table to Figures~\ref{JDIQNorel:fig:reproduce} and~\ref{JDIQNorel:fig:other-collections}, no qualitative differences emerge. 

\begin{table}[tb]
\centering
\caption{Generalization to other metrics: GMAP. 
Correlations of
Pearson's $\rho$,
Kendall's $\tau$,
Spearman's $r_S$,
Rank Biased Overlap (RBO), and 
$\tau_{AP}$
for the three methods
SNC, WUC V3, and SPO S\% on the TREC-8, TB06, TB06M, R04, and WEB14 collections, for the GMAP metric.
Compare with Figures~\ref{JDIQNorel:fig:reproduce} and~\ref{JDIQNorel:fig:other-collections}.
}
\label{JDIQNorel:tab:GMAP}
\begin{small}
 \begin{threeparttable}
\begin{adjustbox}{max width=\textwidth}
\begin{tabular}{l c rrr@{\hskip 1mm}r@{\hskip 2mm}r c rrr@{\hskip 1mm}r@{\hskip 2mm}r c rrr@{\hskip 1mm}r@{\hskip 2mm}r}
\toprule
&& \multicolumn{5}{c}{\textbf{SNC}}           
&& \multicolumn{5}{c}{\textbf{WUC V3}}
&& \multicolumn{5}{c}{\textbf{SPO S\%}}
\\
\cmidrule{3-7} \cmidrule{9-13} \cmidrule{15-19}
&& \multicolumn{1}{c}{$\rho$} & \multicolumn{1}{c}{$\tau$} & \multicolumn{1}{c}{$r_S$} & \multicolumn{1}{c}{RBO} & \multicolumn{1}{c}{$\tau_{AP}$}
&& \multicolumn{1}{c}{$\rho$} & \multicolumn{1}{c}{$\tau$} & \multicolumn{1}{c}{$r_S$} & \multicolumn{1}{c}{RBO} & \multicolumn{1}{c}{$\tau_{AP}$}
&& \multicolumn{1}{c}{$\rho$} & \multicolumn{1}{c}{$\tau$} & \multicolumn{1}{c}{$r_S$} & \multicolumn{1}{c}{RBO} & \multicolumn{1}{c}{$\tau_{AP}$}
\\
\midrule
TREC-8 &&  .50 & .52 & .63 & .46 & .43 && .48 & .45 & .57 &  .68 & .32 && .69 & .55 & .68 & .73 & .41 \\
TB06&&  .88 & .51 & .67 & .76 & .37 && .87 & .46 & .62 &  .75 & .31 && .84 & .43 & .59 & .72 & .30  \\
TB06M&&  .79 & .49 & .66 & .77 & .34 && .78 & .42 & .58 &  .67 & .26 && .73 & .40 & .56 & .63 & .27  \\
R04&&   .88 & .71 & .88 & .85 & .53 && .81 & .65 & .83 & .83 & .52 && .63 & .48 & .66 & .43 & .34 \\
WEB14&& .93 & .74 & .89 & .66 & .70 && .73 & .54 & .75 & .51 & .35 && .35\tnote{\#} & .09\tnote{\#}& .19\tnote{\#} &  .49 & -.10 \\
\bottomrule
\end{tabular}
\end{adjustbox}
  \begin{tablenotes}
  \item[\#] $p>0.05$.
  \item All the other values have $p<0.01$.
  \end{tablenotes}
  \end{threeparttable}
  \end{small}
\end{table}

\subsubsection{logitAP} \label{JDIQNorel:sub:generalize_logitAP}

logitAP is similar to GMAP, but operates using the logistic transformation of AP values.
We compute logitAP as done by \citet{Robertson:2006:GOT:1183614.1183630}:
\begin{equation*}
\logitAP(s_i, t_j) = \ln \left( \frac{\AP(s_i, t_j)}{1-\AP(s_i, t_j)} \right) .
\end{equation*}
Here again we use the natural logarithm and AP values of zero (and one) need to be replaced by a small $\epsilon$ value of $10^{-5}$ (and $1-10^{-5}$), as done by \citeauthor{Robertson:2006:GOT:1183614.1183630}.
Using \citeauthor{Robertson:2006:GOT:1183614.1183630}'s words 
\citeyear[page~132]{Robertson:2006:GOT:1183614.1183630}:
\begin{displayquote}
Like the log transform, or equivalently like using the geometric mean GMAP, this pays attention to hard topics in a way that ordinary MAP does not.
\end{displayquote}

Results of the generalization to the logitAP measure are shown in  Table~\ref{JDIQNorel:tab:logitAP}. 
To compute the logitAP values, we normalized the WUC e SPO scores as done for the GMAP values: in this case the normalization makes a difference, and performing no normalization at all leads to lower correlation values.
The correlation values for 
SNC are almost identical to the ones obtained for GMAP (Table~\ref{JDIQNorel:tab:GMAP});
the correlation values for WUC are comparable, especially when considering top-heavy correlation measures (RBO, and $\tau_{AP}$);
the correlation values for SPO are more different, even thought the similarity is still high when considering top-heavy correlation measures.

When comparing GMAP (Table~\ref{JDIQNorel:tab:GMAP}), logitAP (Table~\ref{JDIQNorel:tab:logitAP}), and the correlation values of Figure~\ref{JDIQNorel:fig:other-collections} no significant differences emerge.

\begin{table}[tb]
\centering
\caption{Generalization to other metrics: logitAP. 
Table for the correlations of
Pearson's $\rho$,
Kendall's $\tau$,
Spearman's $r_S$, and 
Rank Biased Overlap (RBO), for the three methods
SNC, WUC V3, and SPO S\% on the TREC-8, TB06, TB06M, R04, and WEB14 collections, for the logitAP metric.
Compare with Figures~\ref{JDIQNorel:fig:reproduce}, ~\ref{JDIQNorel:fig:other-collections}, and with Table~\ref{JDIQNorel:tab:GMAP}.
}
\label{JDIQNorel:tab:logitAP}
\begin{small}
 \begin{threeparttable}
  \begin{adjustbox}{max width=\textwidth}
\begin{tabular}{l c rrr@{\hskip 1mm}r@{\hskip 2mm}r c rrr@{\hskip 1mm}r@{\hskip 2mm}r c rrr@{\hskip 1mm}r@{\hskip 2mm}r}
\toprule
&& \multicolumn{5}{c}{\textbf{SNC}}           
&& \multicolumn{5}{c}{\textbf{WUC V3}}
&& \multicolumn{5}{c}{\textbf{SPO S\%}}
\\
\cmidrule{3-7} \cmidrule{9-13} \cmidrule{15-19}
&& \multicolumn{1}{c}{$\rho$} & \multicolumn{1}{c}{$\tau$} & \multicolumn{1}{c}{$r_S$} & \multicolumn{1}{c}{RBO} & \multicolumn{1}{c}{$\tau_{AP}$}
&& \multicolumn{1}{c}{$\rho$} & \multicolumn{1}{c}{$\tau$} & \multicolumn{1}{c}{$r_S$} & \multicolumn{1}{c}{RBO} & \multicolumn{1}{c}{$\tau_{AP}$}
&& \multicolumn{1}{c}{$\rho$} & \multicolumn{1}{c}{$\tau$} & \multicolumn{1}{c}{$r_S$} & \multicolumn{1}{c}{RBO} & \multicolumn{1}{c}{$\tau_{AP}$}

\\
\midrule
TREC-8 &&  .34 & .52 & .63 & .46 & .42 && .33 & .42 & .55 & .69 & .30 && .58 & .55 & .69 & .71 & .43  \\
TB06&&  .83 & .51 & .67 & .76 & .33 && .84 & .49 & .65 & .75 & .33 && .61 & .42 & .59 & .69 & .27  \\
TB06M&&  .74 & .51 & .68 & .77 & .37 && .67 & .42 & .57 & .66 & .26 && .55 & .43 & .59 & .61 & .26  \\
R04&& .85 & .71 & .88 & .84 & .53 && .82 & .64 & .82 & .81 & .50 && .37 & .30 & .37 & .32 & .15  \\
WEB14&& .94 & .75 & .90 & .66 & .69 && .71 & .49 & .68 & .51 & .39 && .47 & .14\tnote{\#} & .21\tnote{\#} & .49 & -.05  \\
\bottomrule
\end{tabular}
\end{adjustbox}
  \begin{tablenotes}
  \item[\#] $p>0.05$.
  \item All the other values have $p<0.01$.
  \end{tablenotes}
  \end{threeparttable}
  \end{small}
\end{table}

\begin{figure}[tbp]
  \centering
  \begin{tabular}{@{}c@{}c@{}c@{}}
    \includegraphics[width=.33\linewidth]{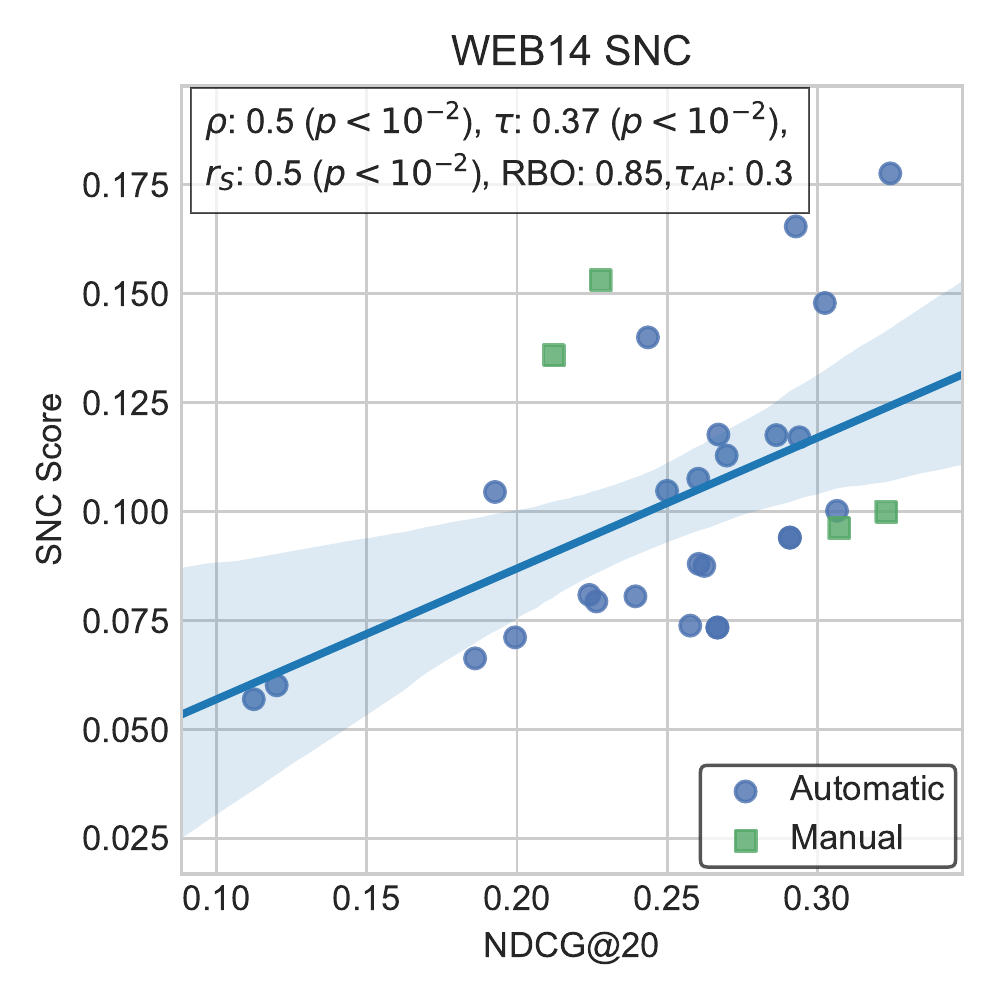}
    &
    \includegraphics[width=.33\linewidth]{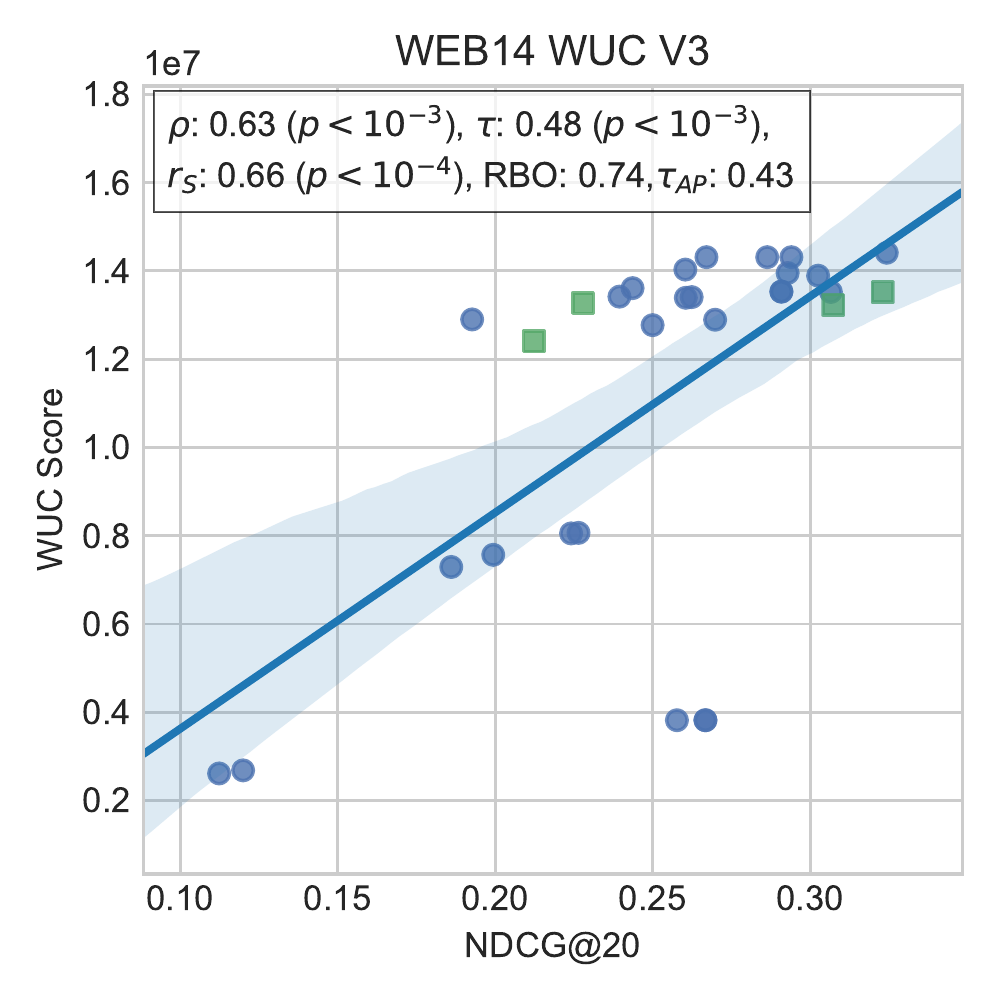}&
    \includegraphics[width=.33\linewidth]{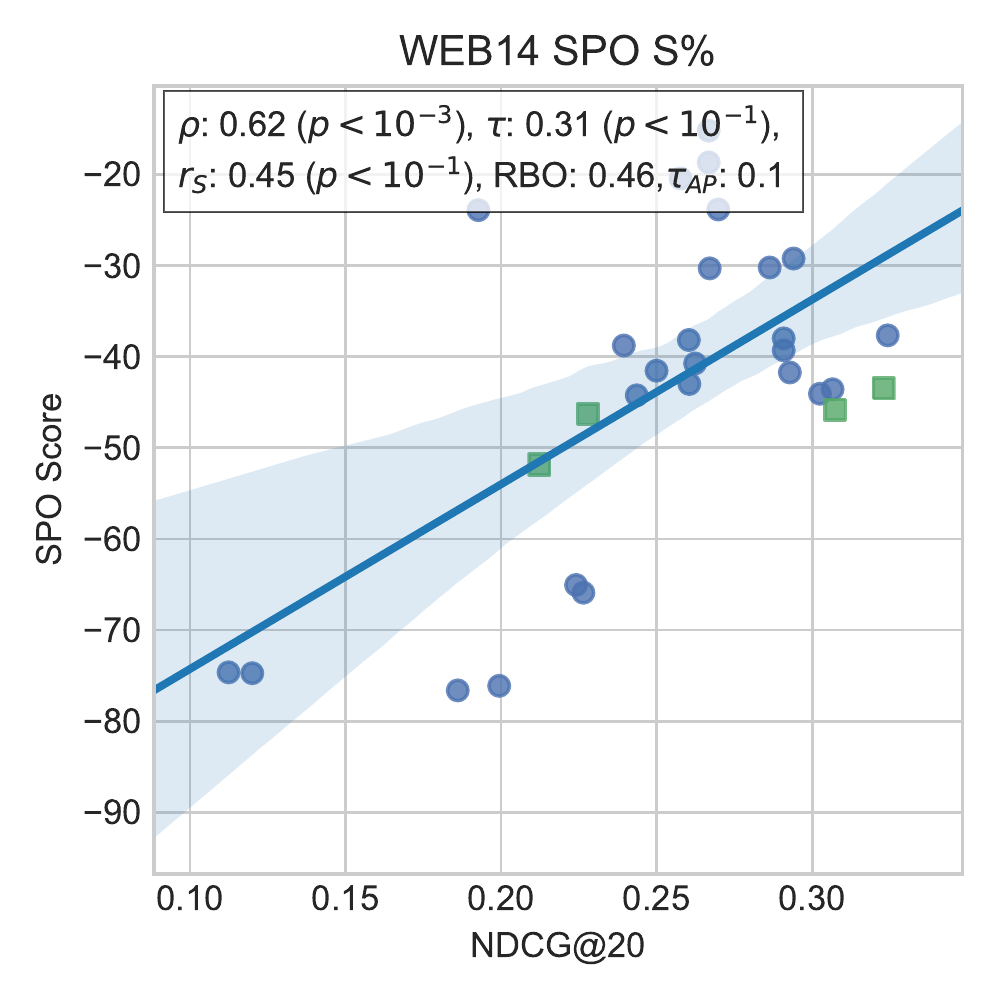}
  \end{tabular}
  \caption{Generalization to other metrics: NDCG. Scatterplots for the three methods SNC, WUC V3, SPO S\% from left to righton the WEB14 collection, for the NDCG metric. Compare with Figures~\ref{JDIQNorel:fig:reproduce} and~\ref{JDIQNorel:fig:other-collections}, and Tables~\ref{JDIQNorel:tab:GMAP} and~\ref{JDIQNorel:tab:logitAP}.  
  }
  \label{JDIQNorel:fig:NDCG}
\end{figure}

\subsubsection{NDCG}  \label{JDIQNorel:sub:generalize_ndcg}
To generalize previous results to the NDCG metric we used the official NDCG@20 measure for the WEB14 collection.
Results are shown in Figure~\ref{JDIQNorel:fig:NDCG}. 
When we compare Figure~\ref{JDIQNorel:fig:NDCG} to Tables~\ref{JDIQNorel:tab:GMAP} (GMAP) and~\ref{JDIQNorel:tab:logitAP} (logitAP) as well as Figures~\ref{JDIQNorel:fig:reproduce} and~\ref{JDIQNorel:fig:other-collections} (TREC-8, R04, TB06, and WEB14 AP),
we notice two visible differences: the correlation values are lower and the ``inverse U'' shape is not present.
The most effective systems are not penalized, but in this case the correlation is more scattered.
This behavior can be caused by the fact that having more than one relevance value (i.e., from 0 to 3) the error that can be made with a random assessment is much higher: for example, imagine a not relevant document (i.e., 0) assessed as highly relevant (i.e., 3) or vice-versa.

\subsection{\ref{JDIQNorel:I:G:S}: Generalize to a Shallow Pool} \label{JDIQNorel:sub:generalize_shallow_pool}

Since results from Section~\ref{JDIQNorel:sub:generalize_ndcg} are obtained  with both a new metric (NDCG) and a shallow pool (20), we computed the results when considering the standard MAP metric with a shallow pool at depth 20 (i.e., AP@20).
Table~\ref{JDIQNorel:tab:shallow} shows the results. 
Comparing these results with the ones obtained considering the top 1000 documents retrieved, we can see that the linear and rank correlations (i.e., $\rho$, $\tau$, and $r_S$) are generally lower for SNC, and higher or comparable for WUC (except for WEB14) as well as for SPO.
This can be caused by the fact that  a shallow pool results in low-quality qrels, which penalize SNC; on the contrary, WUC e SPO measures appear to be stable, probably because the top-ranked documents are more informative for providing an accurate final ranking of systems.
For SPO it is worth noticing that, for R04 and WEB14, the correlations obtained are significantly higher than the correlations obtained when considering all (i.e., 1000) documents retrieved.
When considering top-heavy correlations (i.e., RBO and $\tau_{AP}$), we notice that RBO values are generally comparable with the values obtained when considering the full set of documents, except, again, for WEB14 SNC and WUC V3 methods. The particular behavior of WEB14 can be explained by the fact that such dataset is known to be rather incomplete due to shallow pools and low number of participants \cite{Lu:2016:EPE:2975219.2975241}.
Thus, our results suggest that considering a shallow pool does not penalize the SNC ranking of the most effective systems, which we might have supposed at first when considering linear and rank correlation values.
When considering $\tau_{AP}$ correlations, we see that they are usually higher or comparable to the values obtained when considering the full set of documents, confirming previous findings.

\begin{table}[tb]
\centering
\caption{ Generalization to a shallow pool: AP@20. 
Table for the correlations of
Pearson's $\rho$,
Kendall's $\tau$,
Spearman's $r_S$,
Rank Biased Overlap (RBO), and 
$\tau_{AP}$
for the three methods
SNC, WUC V3, and SPO S\% on the TREC-8, TB06, TB06M, R04, and WEB14 collections. 
}
\label{JDIQNorel:tab:shallow}
\begin{small}
 \begin{threeparttable}
  \begin{adjustbox}{max width=\textwidth}
\begin{tabular}{l c rrr@{\hskip 1mm}r@{\hskip 2mm}r c rrr@{\hskip 1mm}r@{\hskip 2mm}r c rrr@{\hskip 1mm}r@{\hskip 2mm}r}
\toprule
&& \multicolumn{5}{c}{\textbf{SNC}\tnote{$\dagger$}}           
&& \multicolumn{5}{c}{\textbf{WUC V3}}
&& \multicolumn{5}{c}{\textbf{SPO S\%}}
\\
\cmidrule{3-7} \cmidrule{9-13} \cmidrule{15-19}
&& \multicolumn{1}{c}{$\rho$} & \multicolumn{1}{c}{$\tau$} & \multicolumn{1}{c}{$r_S$} & \multicolumn{1}{c}{RBO} & \multicolumn{1}{c}{$\tau_{AP}$}
&& \multicolumn{1}{c}{$\rho$} & \multicolumn{1}{c}{$\tau$} & \multicolumn{1}{c}{$r_S$} & \multicolumn{1}{c}{RBO} & \multicolumn{1}{c}{$\tau_{AP}$}
&& \multicolumn{1}{c}{$\rho$} & \multicolumn{1}{c}{$\tau$} & \multicolumn{1}{c}{$r_S$} & \multicolumn{1}{c}{RBO} & \multicolumn{1}{c}{$\tau_{AP}$}\\
\midrule
%
TREC-8 &&  .55  & .44 & .56  & .58  & .34  && .70 & .46 & .60 & .66 & .33 && .71 & .47 & .62 & .72 & .34 \\
TB06&&    .78  & .29 & .43  & .60  & .14  && .83 & .36 & .52 & .86 & .21 && .82 & .34 & .49 & .83 & .16 \\
R04&&     .78  & .51 & .70  & .74  & .29  && .90 & .65 & .82 & .79 & .45 && .90 & .65 & .82 & .80 & .42 \\
WEB14&&   .53  & .43 & .56  & .88  & .28  && .83 & .59 & .74 & .92 & .36 && .80 & .49 & .64 & .93 & .27 \\
\bottomrule
\end{tabular}
\end{adjustbox}
  \begin{tablenotes}
  \item[$\dagger$] all runs.
  \item All the 
  values have $p<0.01$.
  \end{tablenotes}
  \end{threeparttable}
  \end{small}
\end{table}

\subsection{Discussion}\label{JDIQNorel:sec:generalize:discuss}
Considering all generalization experiments, we can make two general remarks. First, with the other collections TB06, TB06M, R04, and WEB14, correlation values are higher than TREC-8, and the ``inverse U'' shape which was very clear in TREC-8 (see Figure~\ref{JDIQNorel:fig:reproduce}) disappears on the other collections:
the automatic evaluation process can obtain a reasonable rank of IR systems in a completely automatic way.
Second, although the choice of the metric can potentially impact the outcome of the methods for automatic evaluation of IR systems,  this impact is clear only when using NDCG, i.e., a metric based on a different relevance scale (from binary to four levels).

We can also make some more specific remarks.

One issue that worth considering in reproducibility is the choice of using or not the whole dataset \cite{ferro2016increasing,Ferro:2017:RCI:3035914.3020206}.
In TREC result analysis, the choice is whether to use all the dataset or only the top 75\% of most effective systems, as it is commonly done in the analysis of TREC data, see for example \citet{VoorheesBuckley02}.
The comparison between the results restricting or not to the top 75\% of most effective systems shows that there is
no effect for the SNC method, while in general to consider the top 75\% of runs leads to lower correlation values for the other two methods, i.e., SPO and WUC.


Another interesting issue  concerning reproducibility is whether to distinguish between systems that retrieve all the documents for all the topics (e.g., 1000 for each topic) or not. For all the analyzed collections there are systems that do not retrieve all the documents for all the topics: an example is system READWARE for TREC-8.
The result of including or not such systems depends on the method. For the SNC method, this corresponds simply to remove some points (i.e., systems) in the scatterplots. For the other two methods, i.e., SPO and WUC, the effect is twofold: some points are removed from the scatterplots, as well as the structure of overlap for SPO and the count value for WUC are recomputed; this results in lower correlation values for SPO and WUC. This is related to the well known fact that ``The quality of the pools is significantly enhanced by the presence of the recall oriented manual runs'' \cite[page~8]{Voorhees00overviewof} that remarks the peculiarity and the effect of some runs that may do not retrieve all the documents for all the topics.

Finally, SNC is the only method which gives AP values so is the only one which can be generalized naturally, whereas WUC and SPO require normalizations that introduce some arbitrariness in the process.

\section{\ref{JDIQNorel:I:E}: Extend} \label{JDIQNorel:sec:Expand}
 
We now turn to the last aim of this chapter, \ref{JDIQNorel:I:E}, and address two novel research questions that, although they arise in a quite natural way in the context of this research, have  been neglected so far.

\subsection{\ref{JDIQNorel:I:E:MIX}:  Semi-Automatic Approaches}

A research question that in our opinion is quite natural is what happens when a part of the evaluation is automatic and a part of it is manual, i.e., based on human relevance assessments (Aim~\ref{JDIQNorel:I:E:MIX}).
In other terms, what happens when some values in the AP matrix of  Table~\ref{JDIQNorel:tab:AP} are \emph{artificial}, i.e., obtained by one of the three methods, and some others are \emph{real}, i.e., obtained by means of human relevance assessments? In this section we focus on this issue.

\subsubsection{Injecting Columns}
We assume to work atomically on the topics: we do not work on individual cells of the AP matrix of Table~\ref{JDIQNorel:tab:AP} but on its columns, i.e. (since each column corresponds to a topic), on individual topics. 
We select some columns from the real matrix, and some others from the artificial one, i.e., we are downsampling the topics.
Note that besides downsampling the topics in this manner, other alternative approaches could have been used. One might for instance randomly sample the assessments.
On the one hand, this is a convenient working assumption (and in future work we do plan to discuss the possible alternative downsampling approaches);  on the other hand, however, this is also a reasonable strategy, since assessing the relevance of another document for the same topic costs less than assessing the relevance of a document for a new topic.

Therefore, instead of using all the $n$ columns of the artificial matrix only, we use $a \leq n$ columns from that matrix, and $b=n-a$ columns from the  original real matrix. The $b$ columns are ``injected'' into the artificial matrix, and the MAP value is computed accordingly. In other terms, injecting means to fully evaluate, with human relevance assessments, specific topics.

The $b$ columns, or topics, to be injected can be selected in many ways.
If the selection criterion of the $b$ columns does not depend on the real AP matrix, then this would correspond to a procedure that can be applied in practice: the artificial AP matrix is built completely automatically and, still in a completely automatic way (before any human relevance assessment takes place), some topics are selected to be injected (i.e., manually evaluated, by means of human relevance assessments). 
Conversely, if the selection criterion of the $b$ columns  depends on the real AP matrix, then this would be useless in a practical evaluation setting, since the full results of the evaluation would be needed to do the selection according to such a criterion. 
However, it might also be interesting to study this case, as  hidden properties of topics and evaluation in general might be revealed.
Of course it would be possible to imagine also mixed or approximated selection criteria, or even to add real columns (maybe approximated to a pool depth) rather than using them to replace the artificial ones; we leave that for future work.

We take into account the following theoretical selection criteria: 
\begin{enumerate}[label=(T\arabic*),ref=T\arabic*] 
	\item \label{JDIQNorel:i:t:real_art} Select the topics having higher (and lower as well) correlation between real AP and artificial AP. This means that we are injecting the real columns that are more (less) similar to the artificial ones. 
    Although it can be expected that injecting the most similar columns will have a smaller effect than the less similar ones, this needs to be verified experimentally.
	\item \label{JDIQNorel:i:t:real_real} Select the topics having higher (lower) correlation between real AP and real MAP. This means that we are selecting the topics whose real columns are individually most (least) similar to the real MAP values, i.e., those individual topics that somehow better (worse) resemble the overall real evaluation. 
    Those are the topics that provide a most (least) similar final ranking of IR systems, and that might be most (least) important to evaluate in an accurate way.
	\item \label{JDIQNorel:i:t:art_real} Select the topics having higher (lower) correlation between artificial AP and real MAP. This means that we are selecting the topics whose artificial columns  are individually most (least) similar to the real MAP values, i.e., those topics that somehow, in the artificial evaluation, better (worse) resemble the overall real evaluation. 
    Similarly to the first criterion T1, injecting a more similar columns is likely to have a smaller effect than a less similar one (replacing the latter would mean to ``correct the errors'' made by the automatic evaluation).
	\item \label{JDIQNorel:i:t:BestSubReal} Select the Best (and Worst as well) possible columns according to the BestSub method presented by \citet{Guiver:2009:FGT:1629096.1629099} on the real matrix. \citeauthor{Guiver:2009:FGT:1629096.1629099} described how to find ``a few good (bad) topics'', i.e., the topic subset of a given cardinality that evaluates the systems in the most (least) similar way to the full set of topics. This method would provide for any cardinality a subset of topics to be injected that, in the real matrix, better (worse) resemble the overall real evaluation. Differently from the previous criteria, this chapters on topic sets rather than individual topics, thus, it can be expected to work better than previous methods. Also this conjecture needs to be verified experimentally.
	\item \label{JDIQNorel:i:t:HITSReal} Select the Best (Worst) possible column according to the HITS method of \citet{ecir11} on the real matrix. This method computes for each topic its hubness, a measure of how much the topic is able to predict system effectiveness.
\end{enumerate}
Furthermore, we take into account the following practical selection criteria: 
\begin{enumerate}[label=(P\arabic*),ref=P\arabic*]
	\item \label{JDIQNorel:i:p:random} Select the topics randomly (repeating the experiment to avoid noise---we use 1,000 repetitions in the following).
	\item \label{JDIQNorel:i:p:art_art} Select the topics having higher (lower) correlation between artificial AP and artificial MAP. This means that we are injecting the artificial columns that are more (less) similar to  the artificial MAP values, i.e., those individual topics that somehow better (worse) resemble the overall artificial evaluation.
	\item \label{JDIQNorel:i:p:BestSubArtificial} Select the Best (Worst) possible columns according to the BestSub method on the artificial
    matrix. 
	\item \label{JDIQNorel:i:p:HITSArtificial} Select the Best (Worst) possible column according to the HITS method, i.e., computing its hubness, on the artificial matrix.
\end{enumerate}
Note that P2, P3, and P4 correspond to T3, T4, and T5 respectively, but the former ones can be used in practice. Furthermore, P1 (i.e., sample topics randomly) can be seen as a baseline both for theoretical and practical approaches.

Before turning to the results, we remark that with these experiments we are investigating two related but different things: 
\begin{itemize}
\item  It seems intuitive that by injecting real / correct values (i.e., substituting an SNC/SPO/WUC artificial column with a real one)  all the three methods will be improved. Besides verifying this, we also study which is the column selection method that provides the best results (i.e., increases most the correlation values with the ground truth, that is the real MAP value).
\item It is unclear whether any of the three methods can be exploited to improve the BestSub topic selection method by \citeauthor{Guiver:2009:FGT:1629096.1629099}. That is, by computing MAP values using not only the ``few good topics'' subset (i.e., just a few columns of the AP matrix), but using  a complete matrix with the real AP values for the best topic subsets and the artificial AP values for the other topics, do we get a system evaluation / ranking that correlates better to the real MAP value computed using the full real matrix?
\end{itemize}
In addition, with the practical experiments, we are investigating whether a practical semi-automatic evaluation is possible; and in the case of an affirmative answer, which is the best selection criterion that should be adopted. 

\subsubsection{Results}
\begin{figure}[tbp]
  \centering
  \begin{adjustbox}{max width=\textwidth}
  \begin{tabular}{cc}
    \includegraphics[width=.49\linewidth]{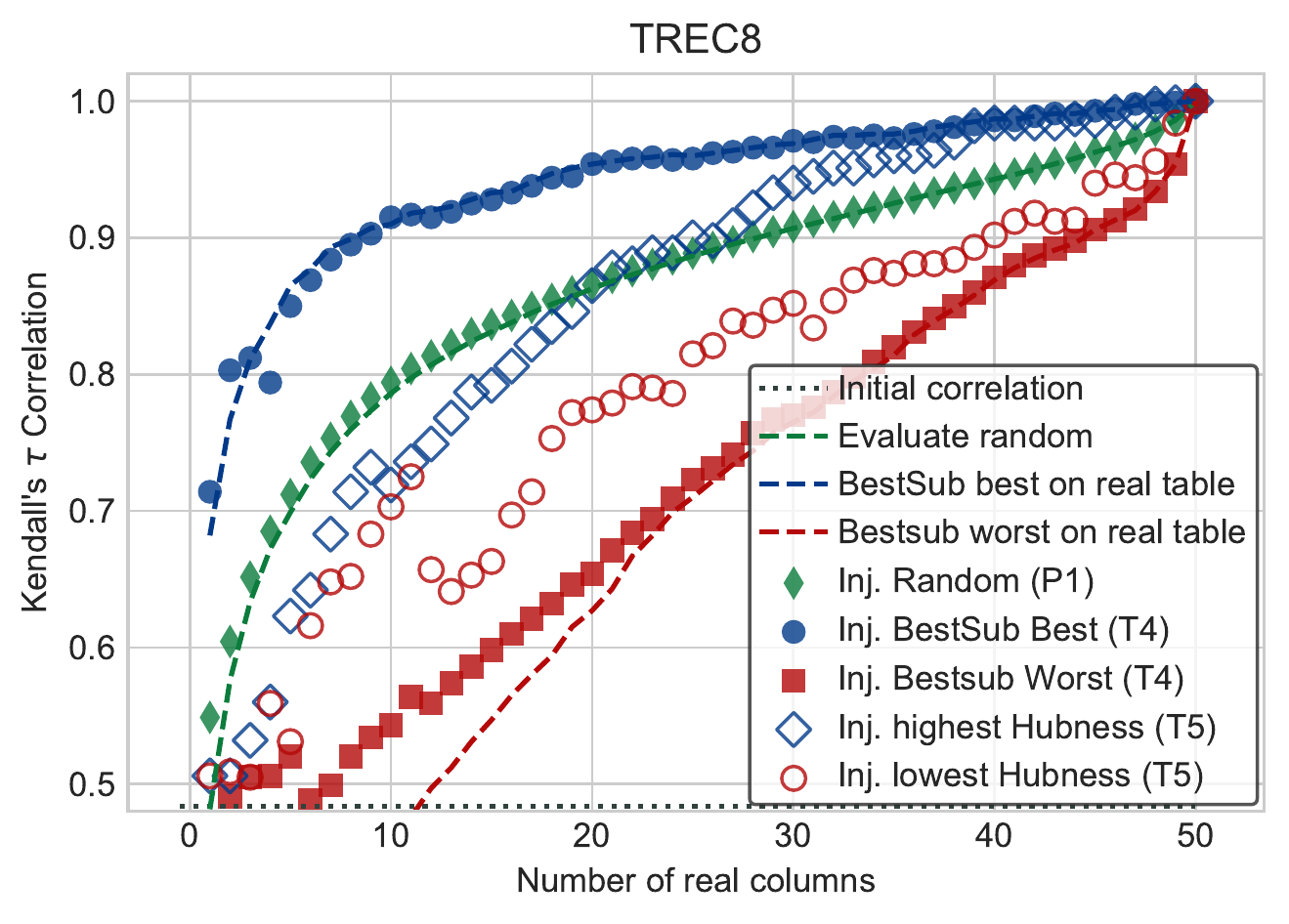}&
    \includegraphics[width=.49\linewidth]{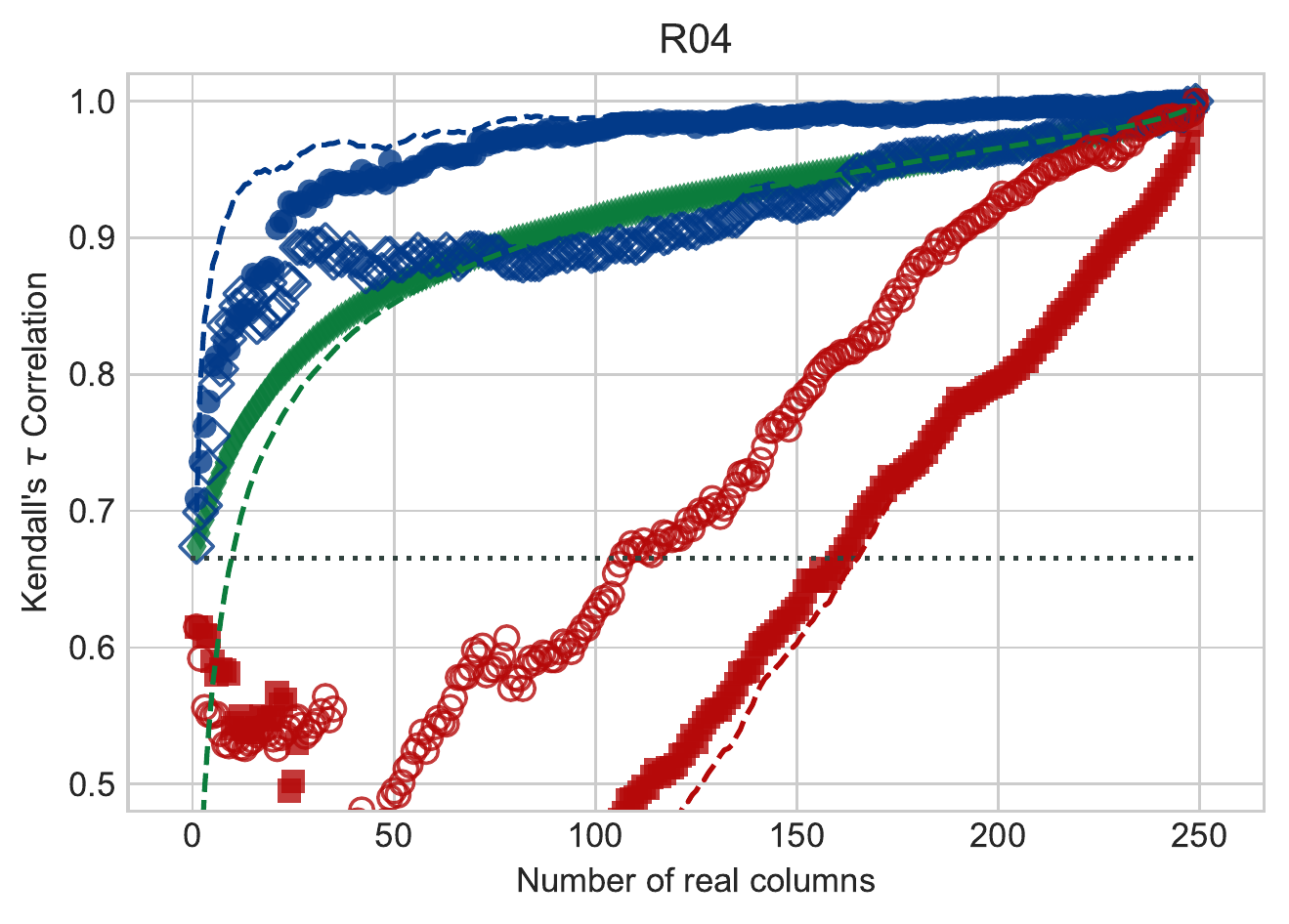}\\
    (a)&(b) \\
    \includegraphics[width=.49\linewidth]{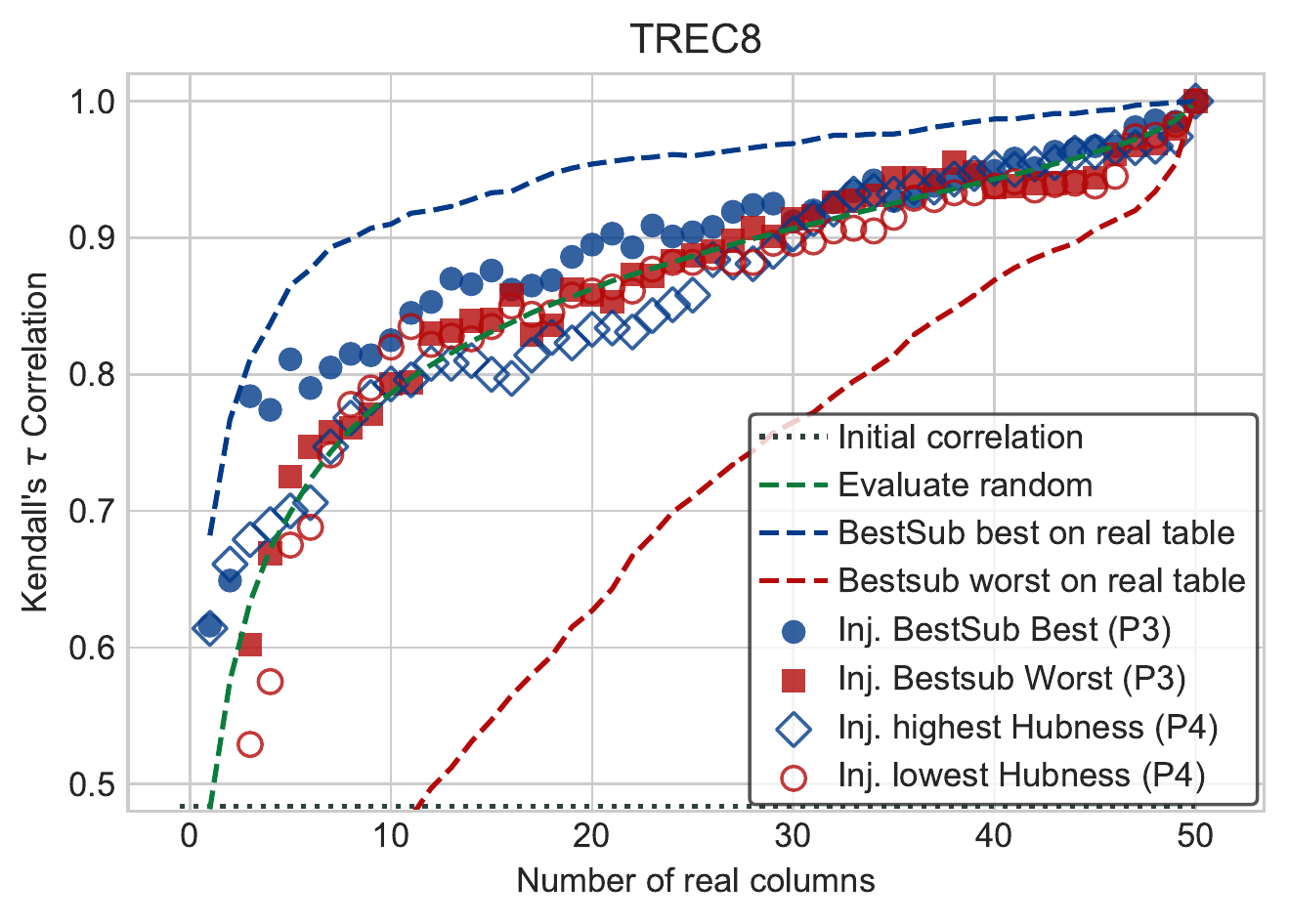}&
    \includegraphics[width=.49\linewidth]{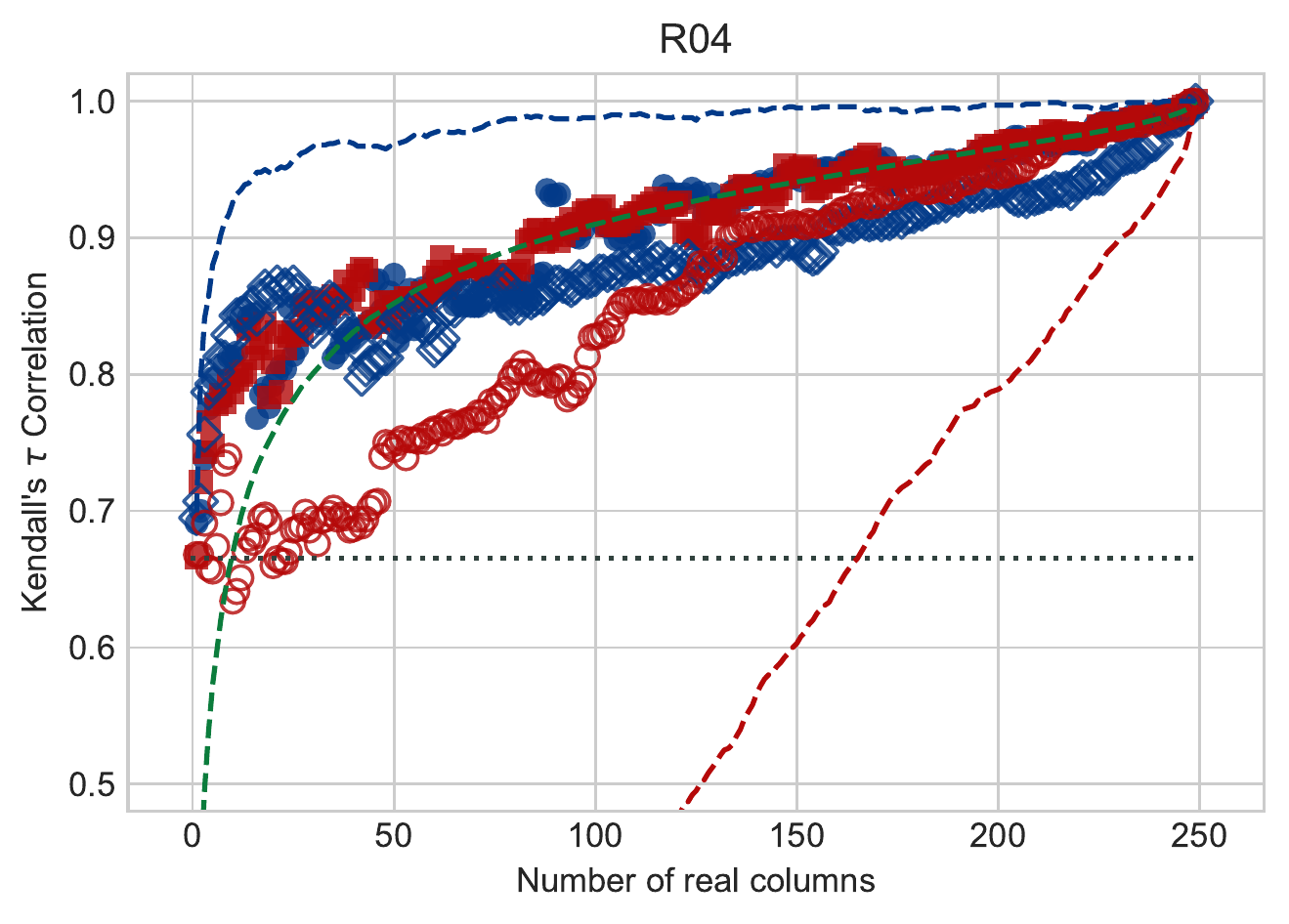}\\
    (c)&(d) \\
    \end{tabular}
    \end{adjustbox}
  \caption{
  Correlation curves, SNC for TREC-8 ((a) and (c)) and R04 ((b) and (d)), for the theoretical approaches T4 and T5  plus the random injection P1 ((a) and (b)), and the practical ones P3 and P4 ((c) and (d)); the horizontal black dotted line is the Kendall's correlation value for SNC method without injecting any column.
  }
  \label{JDIQNorel:fig:correlation}
\end{figure}

We report results for SNC only in this section; the results for the other two methods are generally worse (even despite the normalization attempts), probably because SNC is the only one that predicts the actual AP values, and during the injection process this is probably critical. We leave further analyses of the other methods to future work, as well as a complete analysis of the variability across datasets.
Figure~\ref{JDIQNorel:fig:correlation} displays the result for SNC for TREC-8 and R04.
We report the results by showing the same correlation charts used in the state-of-the-art work on topic set reduction, see for example \citet{Guiver:2009:FGT:1629096.1629099,Berto:2013:UFT:2499178.2499184,ecir11}):
the charts in figure show on the x-axis
the number of columns injected from the real table, and on the y-axis
the $\tau$ correlation values between the MAP obtained using the matrix composed by artificial and injected real topics, and the real MAP obtained using the full set of real topics.

The series represent:
\begin{itemize}
\item The Best/Worst series obtained by running \citeauthor{Guiver:2009:FGT:1629096.1629099}'s BestSub method on the original matrix, as well as the Average series, obtained by selecting topic subsets randomly (using 1,000 repetitions). These series are represented by the dashed lines in the plot. Note that in this case MAP is computed on a subset of topics, and not on a set composed of real and artificial topics, as it is for the other series.
\item The result of injecting topics/columns randomly, i.e., \ref{JDIQNorel:i:p:random}. This represents what happens when injecting topics without any strategy, thus it can be considered as a baseline.
\item The result of injecting the Best/Worst possible columns according to BestSub computed on the real matrix, i.e., \ref{JDIQNorel:i:t:BestSubReal} (in (a) and (b)), or on the artificial matrix, i.e., \ref{JDIQNorel:i:p:BestSubArtificial} (in (c) and (d)).
\item The result of injecting the Best/Worst possible column according to the HITS analysis computed on the on the real matrix,  i.e., \ref{JDIQNorel:i:t:HITSReal} (in (a) and (b)), or on the artificial matrix,  i.e., \ref{JDIQNorel:i:p:HITSArtificial} (in (c) and (d)).
\end{itemize}
The \ref{JDIQNorel:i:t:real_art}, \ref{JDIQNorel:i:t:real_real}, and \ref{JDIQNorel:i:t:art_real} selection methods are not shown in the charts, since they have a similar behavior to the random topic injection.

Results of Figure~\ref{JDIQNorel:fig:correlation}(a) and (b) concern  the theoretical selection criteria and show that:
\begin{itemize}
\item Perhaps surprisingly, the Best series are not improved by the topic injection, even when considering the theoretical Best possible columns (\ref{JDIQNorel:i:t:BestSubReal}). This is valid on both datasets, and more  evident in R04. Therefore, it is better to evaluate on a small subset of a few good topics rather than on a larger topic set obtained adding artificial SNC columns. Looking at injecting the Worst series, we remark that it does not decrease the correlation obtained by Worst BestSub.
\item Although injecting random topics is a practical selection criterion (\ref{JDIQNorel:i:p:random}), it is shown in the charts in Figure~\ref{JDIQNorel:fig:correlation}(a) and (b) (also for a clearer graphical representation). Let us remark here that it does improve slightly the average subset of topics: injecting randomly is better than using a random subset of topics.
\item Even though injecting the Best columns (\ref{JDIQNorel:i:t:BestSubReal}) does not improve BestSub, the series are always well above the average series. Conversely, the Worst stays well below.
\item The highest/lowest hubness selection criteria (\ref{JDIQNorel:i:t:HITSReal}), even if computed on the real topics, does not improve the random topic selection on TREC-8, at least not until the cardinality is higher than 20. Also for R04 the lowest hubness topic set is always well below the average, and up to cardinality 100 even below the horizontal dotted line: injecting low hubness topics is useless and, at least on R04, is even worse than not injecting topics at all. On the contrary, on R04 the highest hubness series is always above the average series for cardinalities up to 75, and it is even comparable to the BestSub selection criteria \ref{JDIQNorel:i:t:BestSubReal} for the low cardinalities. 
\end{itemize}

Figure~\ref{JDIQNorel:fig:correlation}(c) and (d) concern the corresponding practical selection criteria and show that:
\begin{itemize}
\item All the  practical selection methods improve the artificial only evaluation: there is almost no dot below the dotted line. This means that in the case of a semi-automatic evaluation, using practical approaches is always useful. 
\item As already mentioned, and shown in the (a) and (b) charts, SNC does improve the Average BestSub series, although to a small extent: injecting topics, even  randomly (\ref{JDIQNorel:i:p:random}), is still better than not injecting at all.
\item The injection method does matter. Concerning \ref{JDIQNorel:i:p:BestSubArtificial}, at low cardinalities the Best BestSub series tend to stay above random topic injection, and Worst BestSub series obtain almost always lower level of correlation than the random topic injection, for both datasets.
Concerning \ref{JDIQNorel:i:p:HITSArtificial}, low hubness topics obtain 
almost always lower correlation values of both the random series and the random topic injection; on the contrary this is not true for the highest hubness topic, especially for the TREC-8 dataset.
\item \ref{JDIQNorel:i:p:art_art}, not shown in the charts, has similar behavior to the random topic injection.
\end{itemize}

On a related issue, one might wonder what happens when combining automatic evaluation (we focus on SNC only in this analysis) 
with using fewer topics. In more detail, one could compare:
\begin{enumerate}[label=(\roman*)] 
\item the correlation between the real MAP and the SNC MAP, i.e., the MAP  obtained averaging all the artificial AP values obtained by SNC (this is the the horizontal dotted line in Figure~\ref{JDIQNorel:fig:correlation}); with 
\item the correlation between the real MAP and a ``reduced'' SNC MAP, i.e., a MAP obtained by considering only a limited number of topics and averaging only the corresponding  artificial AP values obtained by SNC. 
\end{enumerate}
In other terms, one would use all the topics to run SNC, but then selects a subset of them to compute the artificial MAP. 
One might expect that when using fewer topics, the obtained correlation is lower that using all the topics; however, of course the selection can be done in different ways, and an optimal selection of the best topics might lead to better results. 
We experiment with the selection criteria used above: random selection, the best and worst as found by BestSub, and high and low hubness.
\begin{figure}[tbp]
  \centering
  \begin{adjustbox}{max width=\textwidth}
  \begin{tabular}{cc}
    \includegraphics[width=.49\linewidth]{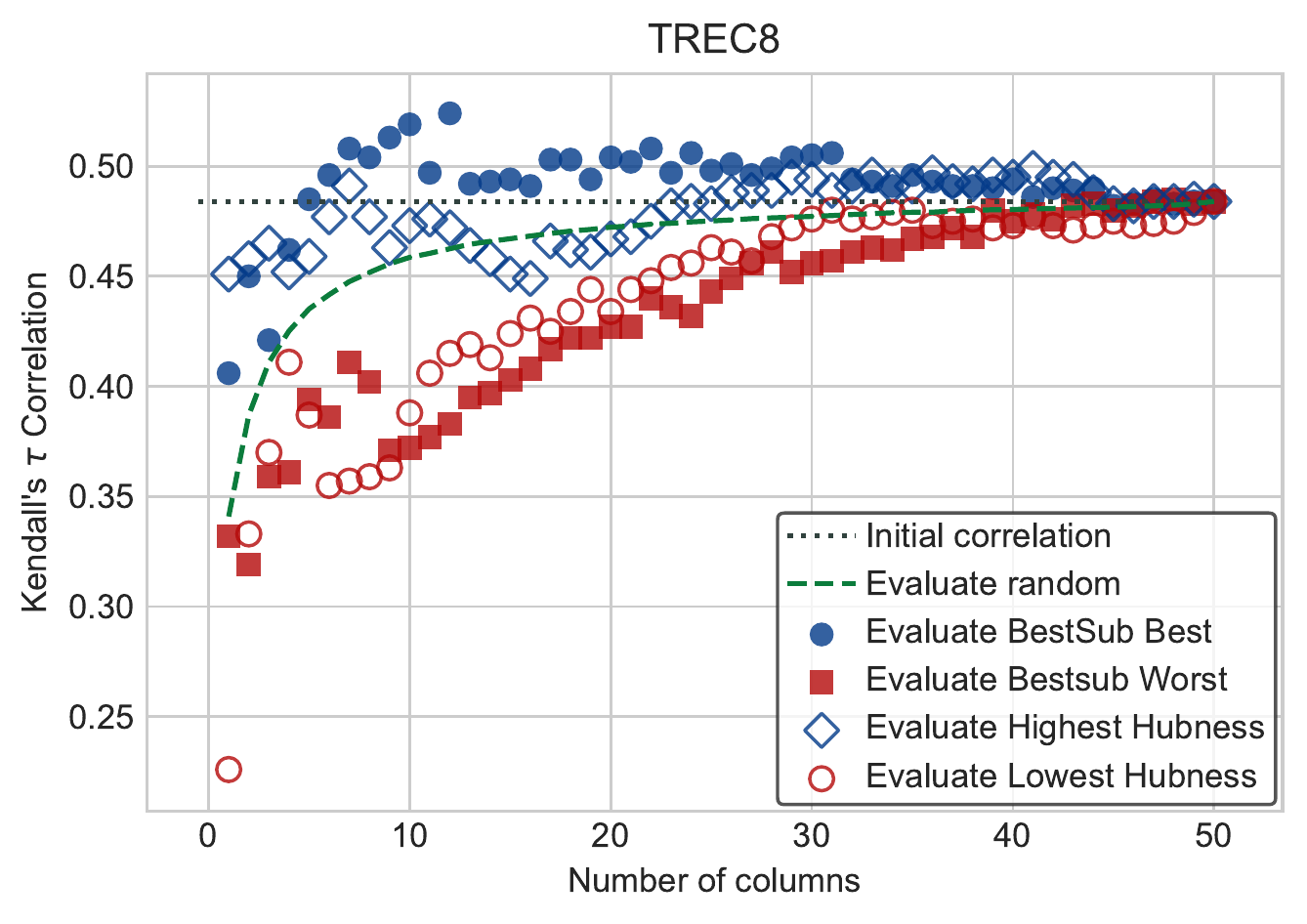}&
    \includegraphics[width=.49\linewidth]{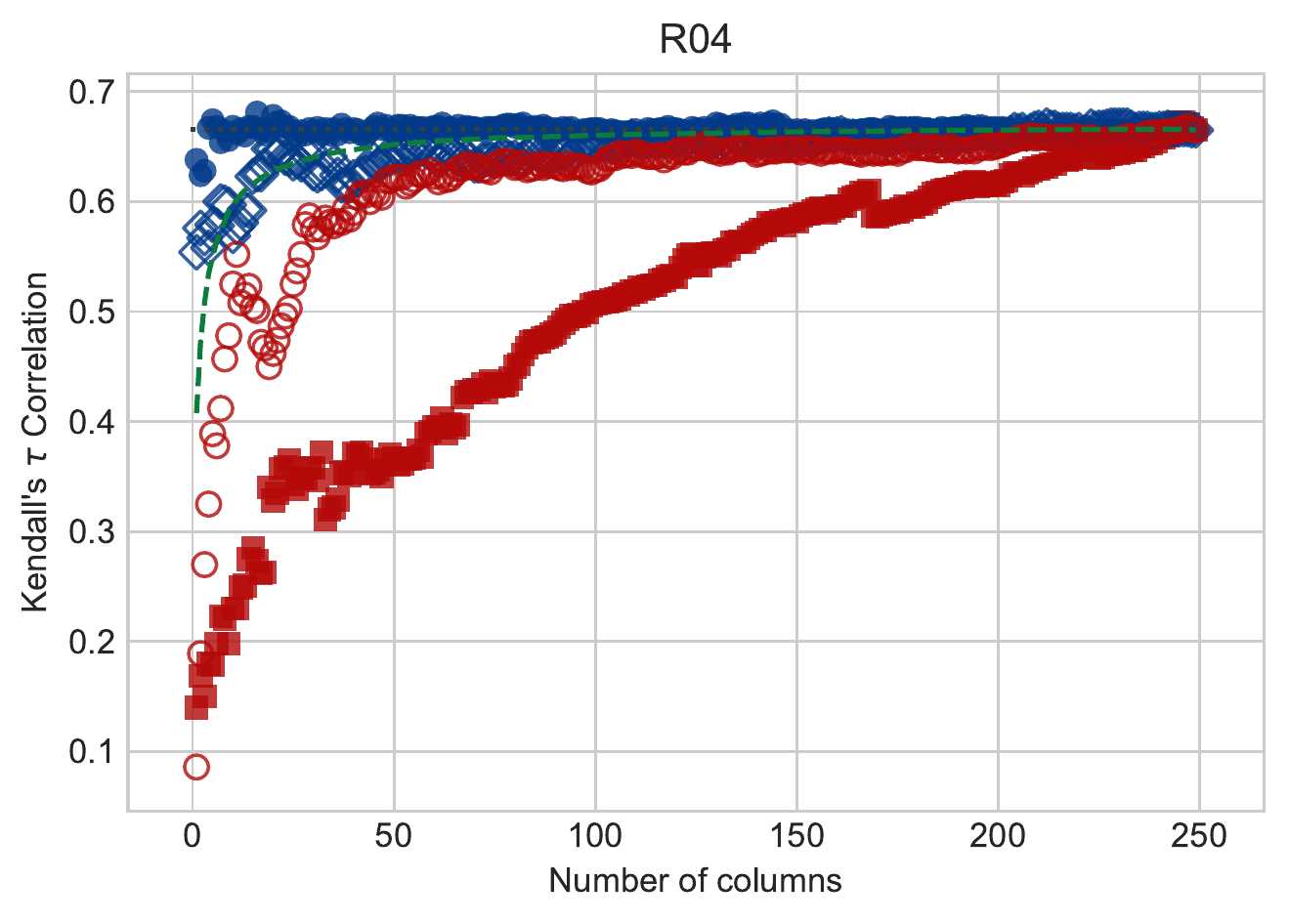}\\
    (a)&(b) \\
    \end{tabular}
    \end{adjustbox}
  \caption{
Correlation curves, SNC for TREC-8 (a) and R04 (b).   The horizontal black dotted line is the Kendall's correlation value for SNC method without injecting any column. The other  curves are obtained by using only a subset of topics (artificial AP values) to compute the correlation with the real MAP. }
  \label{JDIQNorel:fig:SNCplusfewer}
\end{figure}

Figure~\ref{JDIQNorel:fig:SNCplusfewer} shows the results, again on TREC-8 and R04: when using most selection methods, a subset of topics is always less effective than the whole set. However, when using the best subset of a few topics found by BestSub, the obtained correlation is higher than when using all topics. This is clearly the case for TREC-8 in the cardinality range $6$--$44$, and it is less clear cut for R04, where anyway the subset of best topics is never worse than the full set.

Summarizing, our results indicate that, when the ground truth (i.e., the real AP matrix) is available, injecting topics does not help to improve the few good topics approach suggested by  \citet{Guiver:2009:FGT:1629096.1629099}; this holds for any topic selection criterion we have tried.
On the contrary, in the case of a semi-automatic evaluation, the Best columns selected from running the BestSub method on SNC matrix (\ref{JDIQNorel:i:p:BestSubArtificial}) resemble the overall real evaluation better than not injecting topics at all, or  injecting topics randomly.

These results show that there are some patterns in semi-automatic evaluation that are worth studying. Not only semi-automatic evaluation clearly improves the automatic only evaluation; although we have presented a limited sample of the results, it is clear that there are also important practical consequences. For example, looking at the cardinalities 10 for TREC-8 and 50 for R04 in Figure~\ref{JDIQNorel:fig:correlation}(c) and (d), we see that a quite high Kendall's $\tau$ correlations 
in the $0.8$--$0.9$ range in system rankings can be obtained by using the SNC automatic evaluation method and ``augmenting'' it with the evaluation of only about 20\% of the topics. This would be a practical approach to decrease the costs and resources in a test collection evaluation exercise.  This needs to be further studied in future work.
The results shown in Figure~\ref{JDIQNorel:fig:SNCplusfewer}  hint at some promising future research directions as well.

\subsection{\ref{JDIQNorel:I:E:AAP}: Predicting Topic Difficulty}\label{JDIQNorel:sec:E:AAP}

Another, last, very natural research question to be asked in this scenario is whether it is possible to automatically predict not only system effectiveness but also topic difficulty (see Aim~\ref{JDIQNorel:I:E:AAP}). First we discuss the problem in more detail, then we  provide some background on the related issue of query difficulty prediction, then we present some more detailed motivations, and finally we describe our experiments and results.

\subsubsection{From System Effectiveness to the Dual Problem of Topic Ease}
In our context, automatically estimating topic ease corresponds to a sort of dual problem to estimating system effectiveness. To see why, let us  
go back to Table~\ref{JDIQNorel:tab:AP} and Formula~\ref{JDIQNorel:eq:MAP}. We can notice that while $\MAP$ represents a  measure of system effectiveness, a dual measure of topic ease can be defined as it has been proposed by  \citet{Mizzaro:2007:HHT:1277741.1277824}: 
\begin{equation}\label{JDIQNorel:eq:AAP}
\AAP(t_j)=\frac{1}{m}\sum_{i=1}^{m} \AP(s_i, t_j)
\end{equation}
(AAP stands for Average AP). Simply, in place of averaging the rows, one can average the columns.

Thus, in our context, we are now asking if the three methods can be used to predict not MAP of systems but rather AAP of topics. 
Although this seems a very natural research issue, it has not been addressed in the three studies that we have discussed at length so far, and by nobody else.
Furthermore, the importance of this problem can be better understood by considering the related problem of query difficulty prediction.



\subsubsection{Background on Predicting Query Difficulty} \label{JDIQNorel:sub:back_difficulty}
Predicting query difficulty is an important research issue. 
The knowledge that a query is going to be difficult might be exploited by a system, that could adopt appropriate countermeasures. 
The Reliable Information Access (RIA) Workshop \cite{Harman:2004:NRI:1008992.1009104,Harman:2009:ORI:1644394.1644419} has been the first large scale study aimed at understanding query variability and difficulty.
The many approaches, that have been developed, can be classified as pre-retrieval and post retrieval, depending on when the prediction takes place. 
The pre-retrieval approaches \cite{Carmel:2010:EQD:1855038,Hauff:2008:SPQ:1458082.1458311,SparckJones:1988:SIT:106765.106782} are more practical, but the correlation of the predicted query difficulty with the ground truth is rather weak. 
Pre-retrieval methods can be based on statistical or  linguistic approaches \cite{mothe2005linguistic};
some results on combining pre-retrieval techniques can be found in the work by \citet{Bashir:2014:CPQ:2592907.2592915}.

The post-retrieval approaches exploit the results of a retrieval phase, and  tend to provide slightly higher correlations with the ground truth, but they are less practical.
Some works using post-retrieval features can be found in the work by \citet{Shtok:2012:PQP:2180868.2180873}.
\citet{Carmel:2010:EQD:1855038} discuss post-retrieval prediction methods using various measures like clarity, robustness, and score distribution analysis. 
Pre- and post-retrieval approaches can be combined in various ways, as detailed by \citet{Carmel:2010:EQD:1855038}, that also provide a complete review of estimating topic difficulty as well as propose a general model for query difficulty together with some practical applications.

Other approaches have been tested. For example,  \citet{Chifu2017} and \citet{mizzaro2016you} use human prediction (crowdsourcing) in predicting topic difficulty. \citet{Buckley:2004:TPB:1008992.1009093} proposes a measure (called AnchorMap) to compute similarity between ranked document lists retrieved by systems; this measure allows a categorization of topics into easy and difficult ones.

It is important to remark that although query difficulty prediction is an interesting research issue, the state-of-the-art is such that no satisfying solution is available yet: when measuring the correlation between predicted and actual difficulty, the best methods reach Pearson correlation values around $0.5$ 
\cite{Carmel:2010:EQD:1855038,Zhao:2008:EPQ:1793274.1793285}.

\subsubsection{Topic Ease + Query Difficulty = Topic Difficulty}
%
%
The previous brief analysis of the literature on query difficulty highlights that the important issue is topic \emph{difficulty}, not \emph{ease}: it is important to understand which are the difficult topics (on which the current systems can be improved), rather than the easy topics (on which the state of the art is already satisfactory). 

We also need to understand  that topic and query difficulty, although related, are different. In query difficulty prediction, usually the AP of a single system is being predicted, rather than the AAP over a set of systems. 
However, the AAP measure is an interesting alternative \cite{Mizzaro:2007:HHT:1277741.1277824}. 
Studying the difficulty of a topic rather than a query makes sense also given the recent result by \citet{Thomas:2017:TQR:3166072.3166079} who find that ``task difficulty'' would be a more reliable notion  than ``query difficulty''.
Finally, although of course, the AAP of a topic will depend on the systems participating to the evaluation exercise, the measure seems quite stable. For example, there are  50 common topics between R04 and TREC-8; the AAP values computed using the two different systems populations feature high correlations (Pearson's $\rho$ correlation is $0.99$ and  Kendall's $\tau$ is $0.89$).

In other terms, what we are proposing is a post-retrieval topic difficulty prediction method that, at least in principle (i.e., not taking efficiency into account), could be used in practice.
Before turning to the results, let us notice that, given the unsatisfactory results obtained by state of the art query difficulty predictors (correlations higher than $\rho= 0.5$ are difficult to obtain), this seems an interesting and promising approach.


\subsubsection{Experimental Results on Predicting Topic Difficulty}\label{JDIQNorel:sec:dual}
\begin{figure}[tbp]
  \centering
  \begin{tabular}{@{}c@{}c@{}c@{}}
    \includegraphics[width=.33\linewidth]{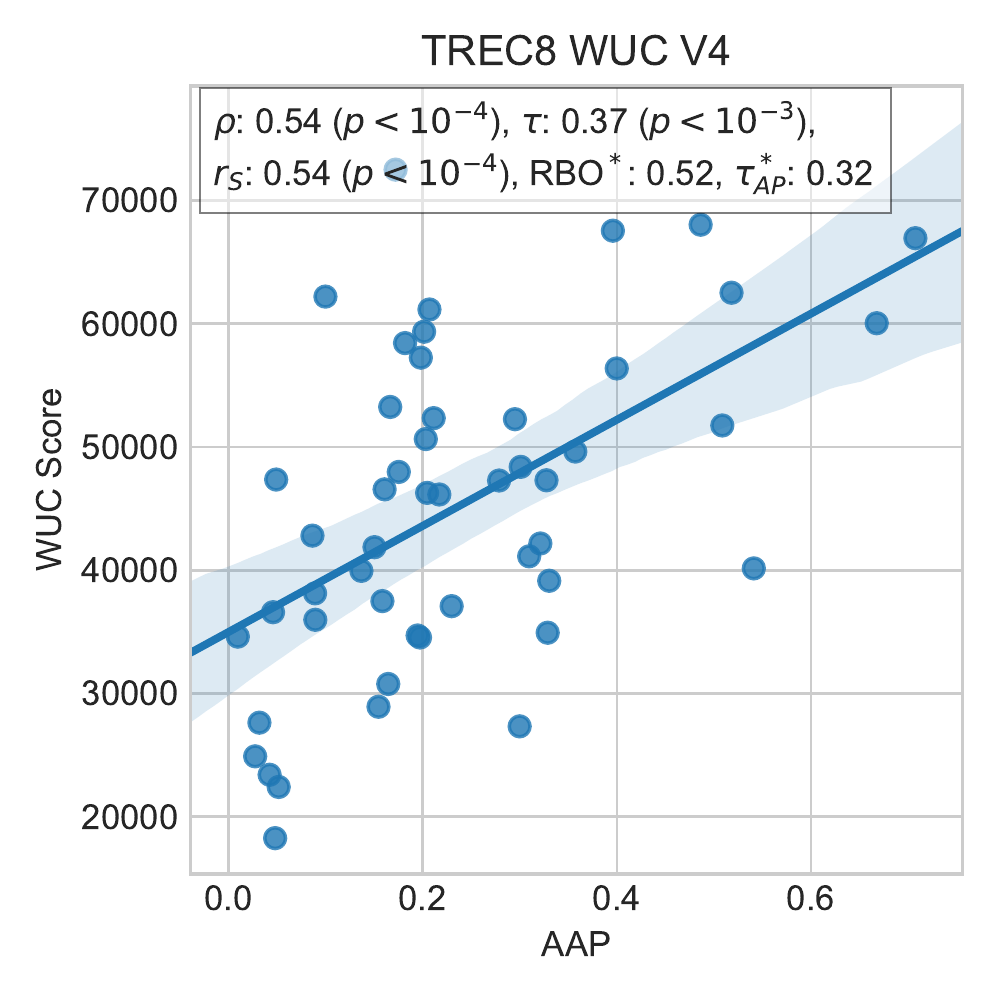}
    &
 \includegraphics[width=.33\linewidth]{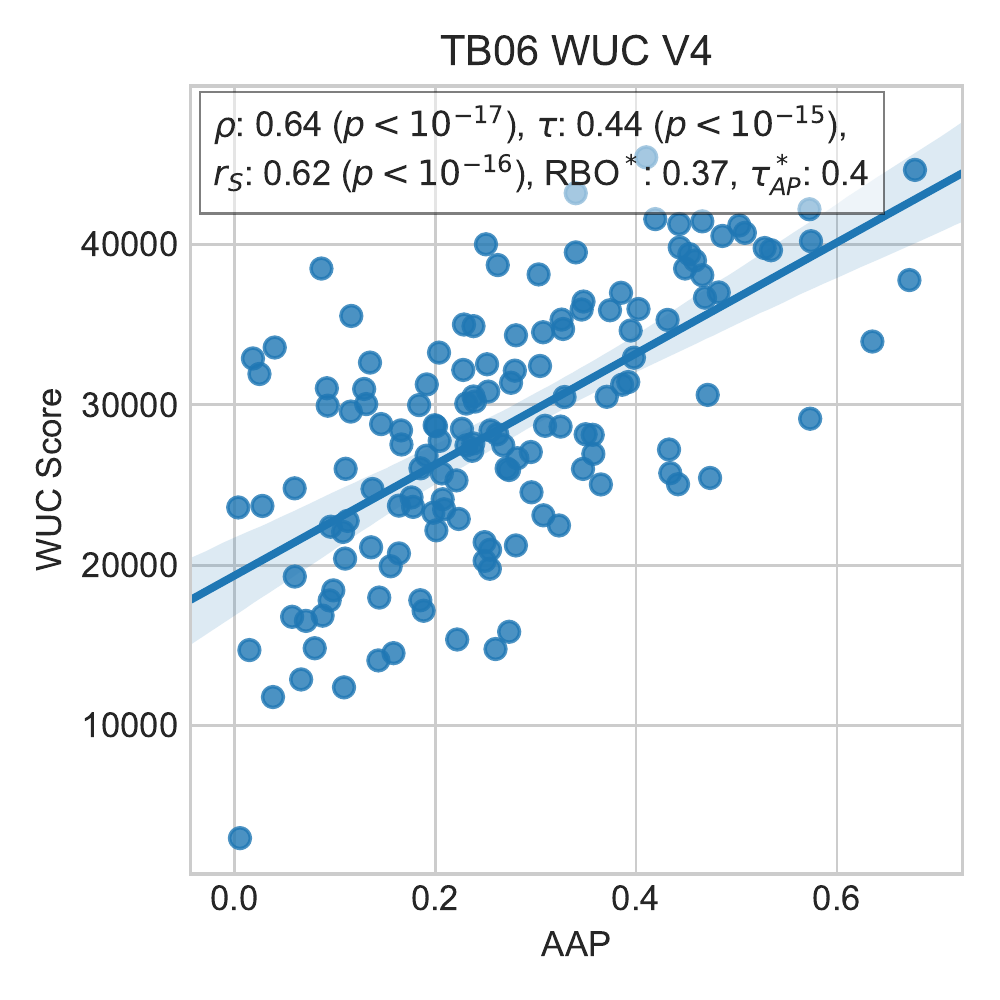}
 &
    \includegraphics[width=.33\linewidth]{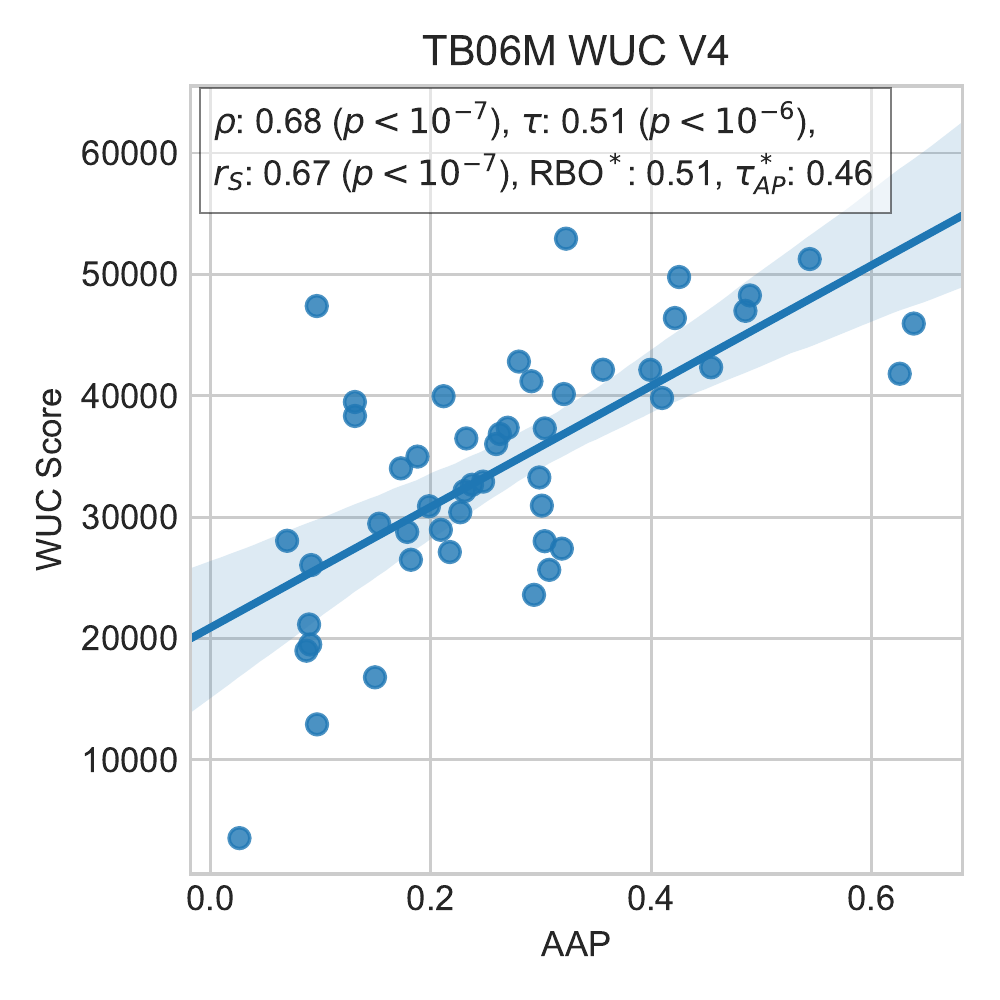}
    \\
 \includegraphics[width=.33\linewidth]{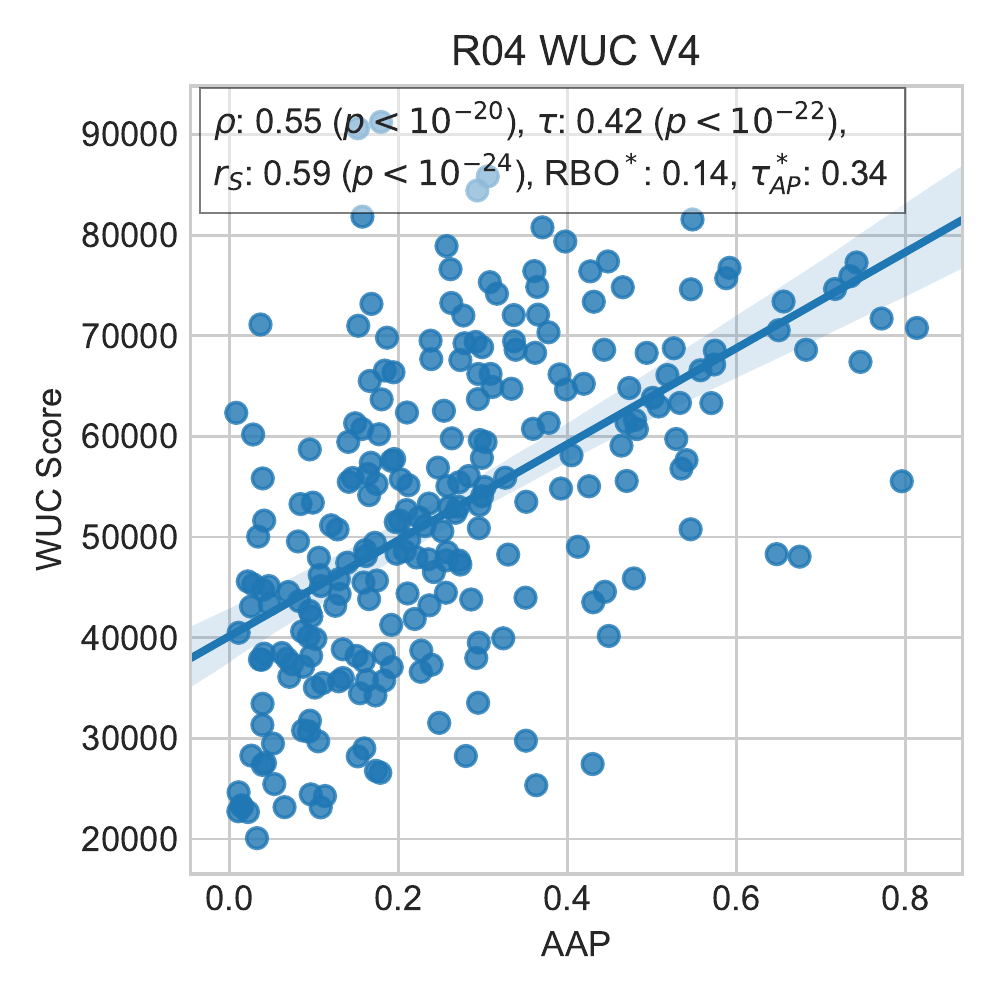} 
 &
  \includegraphics[width=.33\linewidth]{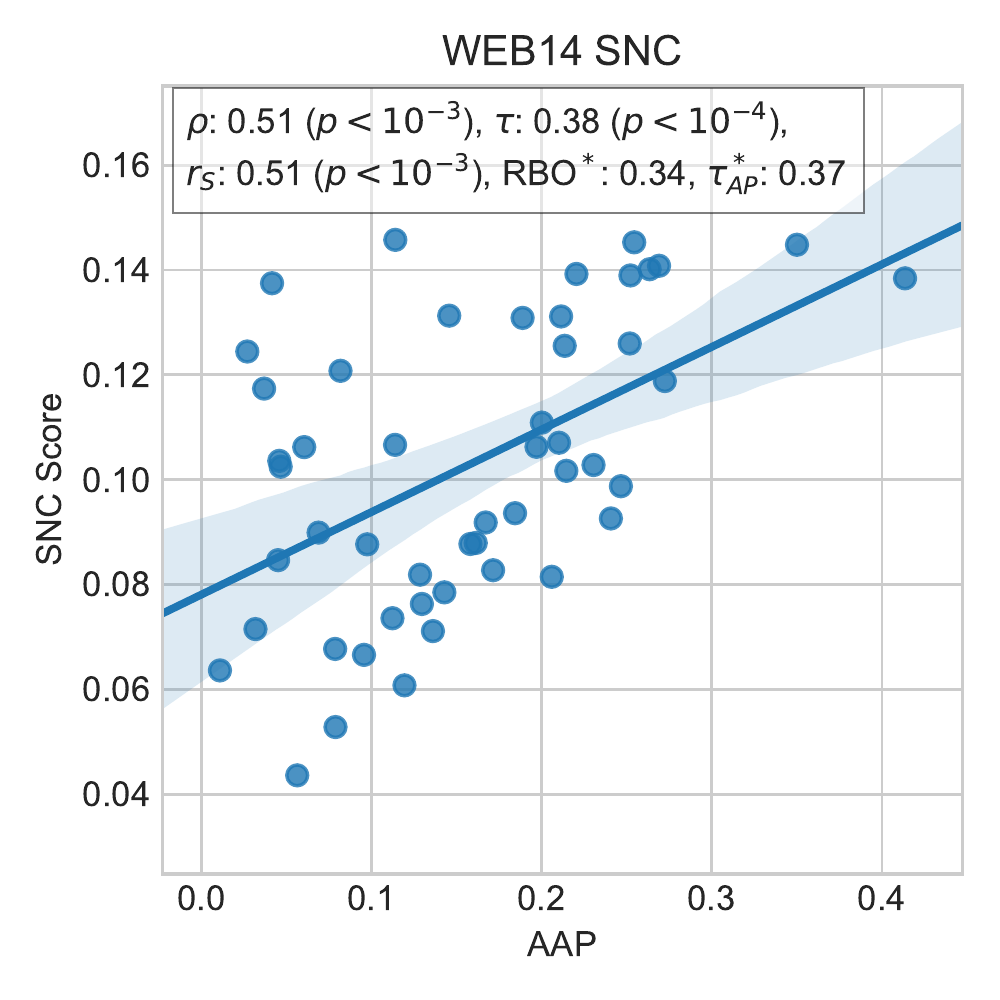}
 &
  \includegraphics[width=.33\linewidth]{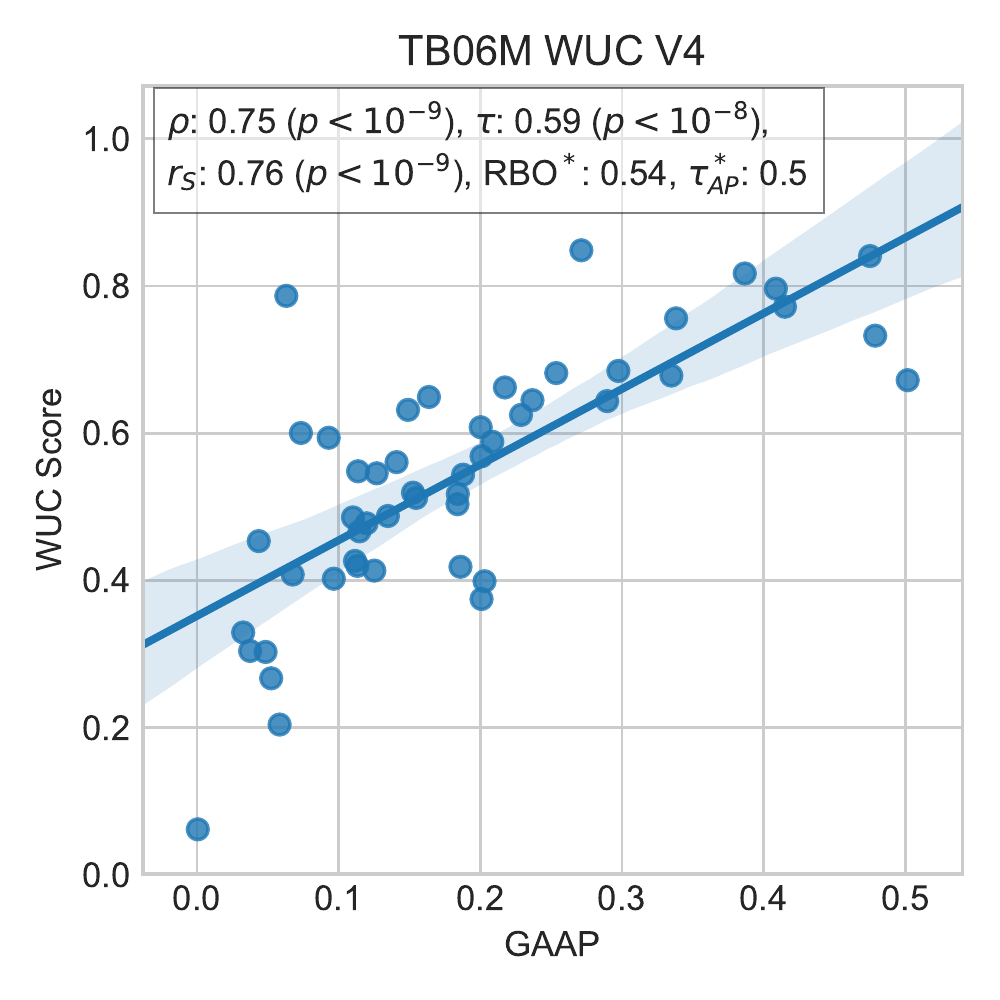}

  \end{tabular}    
  \caption{
  AAP scatterplots for some methods and collections as indicated. 
  Differently from previous scatterplots, here each dot is a topic.
  In the bottom right scatterplot, GAAP is used in place of AAP and, here and in the following, GAAP values have been normalized as it has been done for GMAP (see Section~\ref{JDIQNorel:sub:generalize_gmap}).
  \label{JDIQNorel:fig:AAP}
  }
\end{figure}

At first sight, results seem not exciting.
Figure~\ref{JDIQNorel:fig:AAP} shows the AAP scatterplots, for the five collections and some selected methods (we report the methods with the highest correlations: usually WUC 
V4, with the exception of the fifth chart --- on WEB14 SNC is slightly better). 
Differently from the  scatterplots presented so far in this chapter, here each dot is a topic, not a system, and the axes represent real and predicted AAP, not MAP. 
The last (bottom right) chart in figure uses, in place of AAP, a slightly different metric, Geometric AAP, that can be defined as:
\begin{equation*}
\GAAP(t_j) = \sqrt[m]{\prod_{i=1}^m \AP(s_i,t_j)} = \exp \left( {\frac{1}{m} \sum_{i=1}^{m} \ln(\AP(s_i, t_j))} \right)
\end{equation*}
(GAAP is to AAP as GMAP is to MAP, compare also to Formula~\eqref{JDIQNorel:eq:GMAP}). 
GAAP emphasizes more the difficult/low end of the topic difficulty scale, which seems the interesting one if one wants to work on difficult topics.\footnote{It is also possible to use logitAP in place of $\ln(\AP)$, but in our experiments we did not find any significant difference.} 
For the same reasons, in Figure~\ref{JDIQNorel:fig:AAP}  and in the following we report RBO$^*$ and $\tau^*_{AP}$, which are the bottom-heavy versions of the top-heavy RBO and $\tau_{AP}$, computed simply by reversing the order of the vectors to give more weight to the difficult topics.
Table~\ref{JDIQNorel:tab:AAP} shows  all the correlation values for the five collections, AAP (GAAP in one case), and the three methods (we selected the overall most effective method variants; V3 and V4 have similar correlation values).

\begin{table}[tb]
\centering
\small\addtolength{\tabcolsep}{-1pt}
\caption{AAP  correlations, and GAAP in the last row.
} 
\label{JDIQNorel:tab:AAP}
 \begin{threeparttable}
\begin{adjustbox}{max width=\textwidth}
\begin{tabular}{l c rrr@{~}r@{~~}r c rrr@{~}r@{~~}r c rrr@{~}r@{~~}r}
\toprule
&& \multicolumn{5}{c}{\textbf{SNC}}           
&& \multicolumn{5}{c}{\textbf{WUC V4}}
&& \multicolumn{5}{c}{\textbf{SPO S-A\%}}
\\
\cmidrule{3-7} \cmidrule{9-13} \cmidrule{15-19}
&& $\rho$ & $\tau$ & $r_S$ & $\mathrm{RBO}^{*}$ & $\tau^{*}_{AP}$
&& $\rho$ & $\tau$ & $r_S$ & $\mathrm{RBO}^{*}$ & $\tau^{*}_{AP}$
&& $\rho$ & $\tau$ & $r_S$ & $\mathrm{RBO}^{*}$ & $\tau^{*}_{AP}$
\\
\midrule
TREC-8 &&  .26\tnote{\#} & .24\tnote{+} & .35\tnote{+} & .44 & .18 && .54 & .37 & .54 & .52 & .32 && .33\tnote{+} & .30 & .42 & .46 & .25  \\
TB06&&  .51 & .38 & .53 & .14 & .35 && .64 & .44 & .62 & .37 & .40 && .53 & .37 & .51 & .24 & .36  \\
TB06M&&  .47 & .30 & .42 & .43 & .27 && .68 & .51 & .67 & .51 & .46 && .44 & .31 & .44 & .41 & .28  \\
R04&& .26 & .24 & .36 & .09 & .20 && .55 & .42 & .59 & .14 & .34 && .23 & .21 & .31 & .06 & .20  \\
WEB14&& .51 & .38 & .51 & .34 & .37 && .52 & .35 & .49 & .20 & .32 && .32\tnote{+} & .34 & .44 & .09 & .34  \\
\addlinespace
\multicolumn{5}{l}{\textbf{GAAP}}\\
TB06M && .58 & .44 & .58 & .47 & .41 && .75 & .59 & .76 & .54 & .50 && .54 & .44 & .59 & .41 & .42 \\
\bottomrule
\end{tabular}
\end{adjustbox}
  \begin{tablenotes}
  \item[+] $p<0.05$. 
  \item[\#] $p>0.05$.
  \item All the other values have $p<0.01$.
  \end{tablenotes}
  \end{threeparttable}
\end{table}

The best results in topic difficulty prediction are obtained for TB06M with WUC V4 (see the two right most charts in figure and the values in the table); these values are even higher when GAAP is used.
In these cases the correlation values are comparable to, if not even higher than, the state-of-the-art, which is around $0.5$. 
However, when comparing the scatterplots and the values with the previous ones, the most visible difference 
is that correlations are much lower: topic ease seems less predictable than system effectiveness. 

\begin{figure}[tbp]
  \centering
  \begin{tabular}{@{}c@{}c@{}c}
    \includegraphics[width=.33\linewidth]{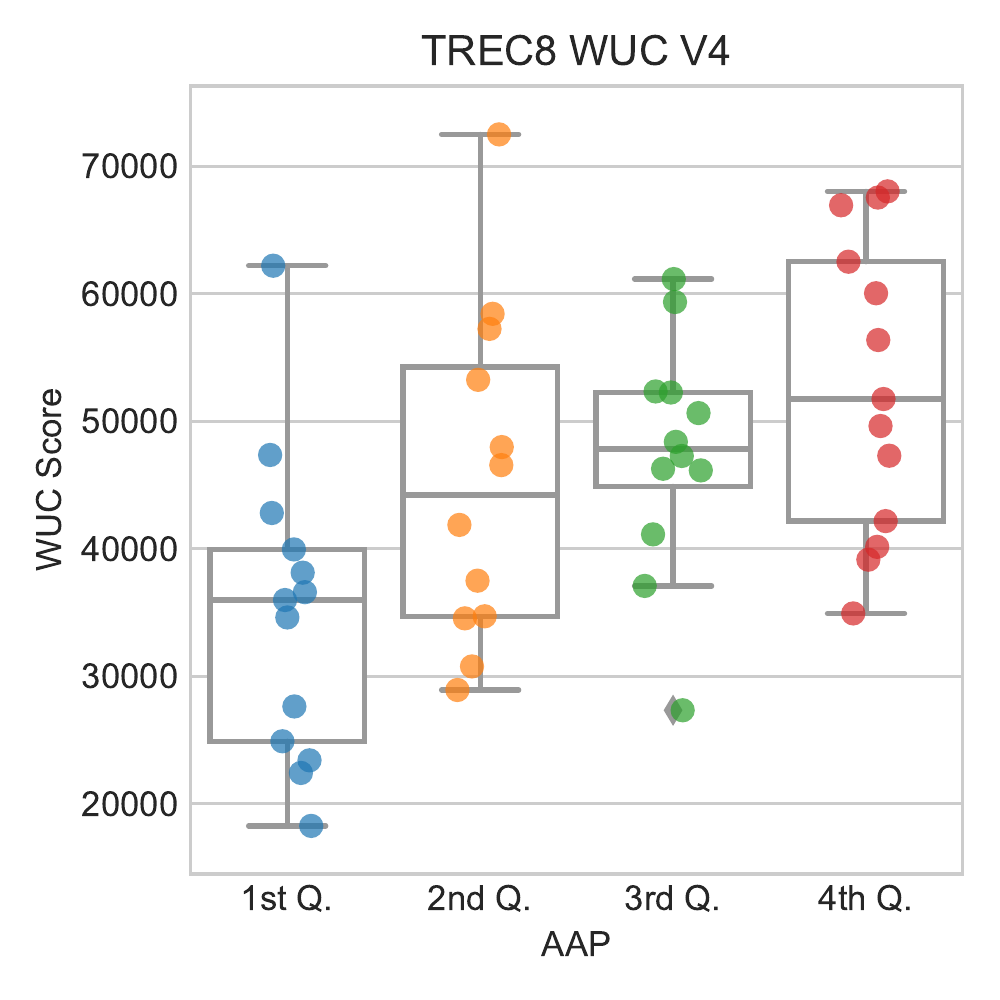}
    &
 \includegraphics[width=.33\linewidth]{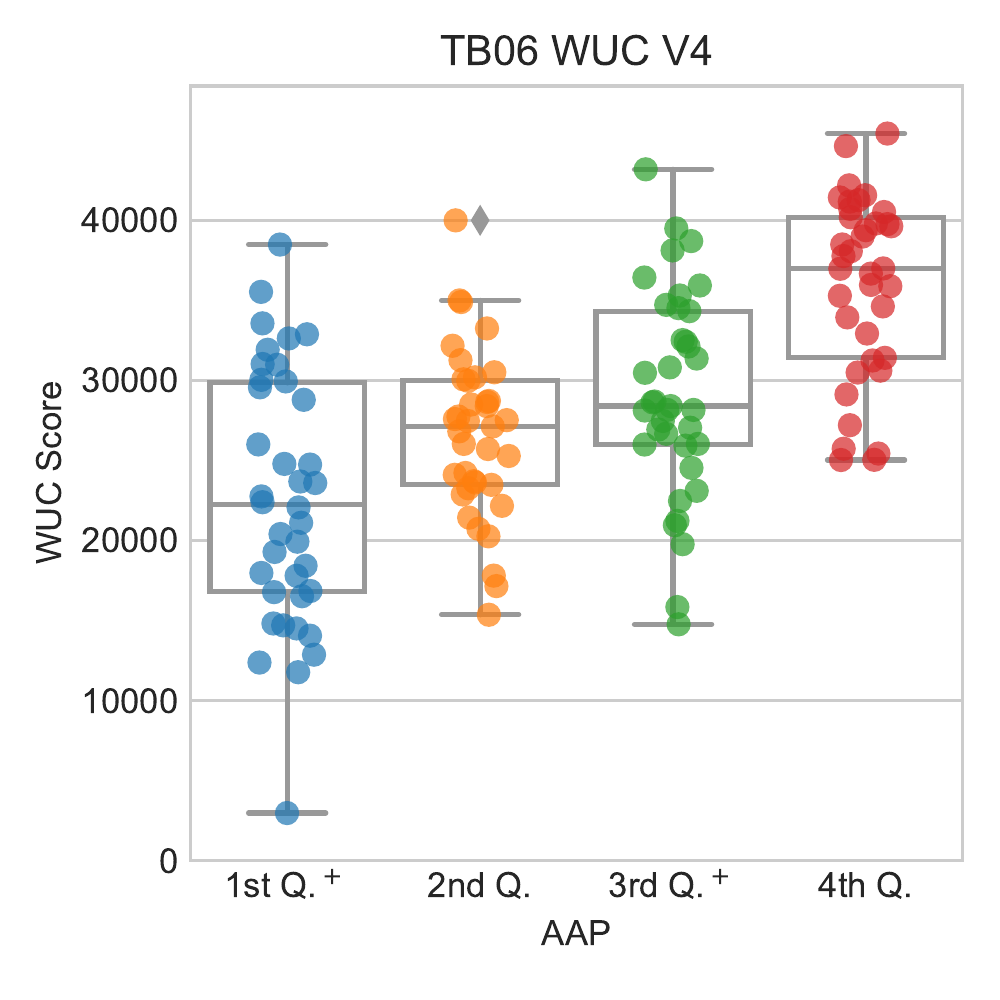}
    &
    \includegraphics[width=.33\linewidth]{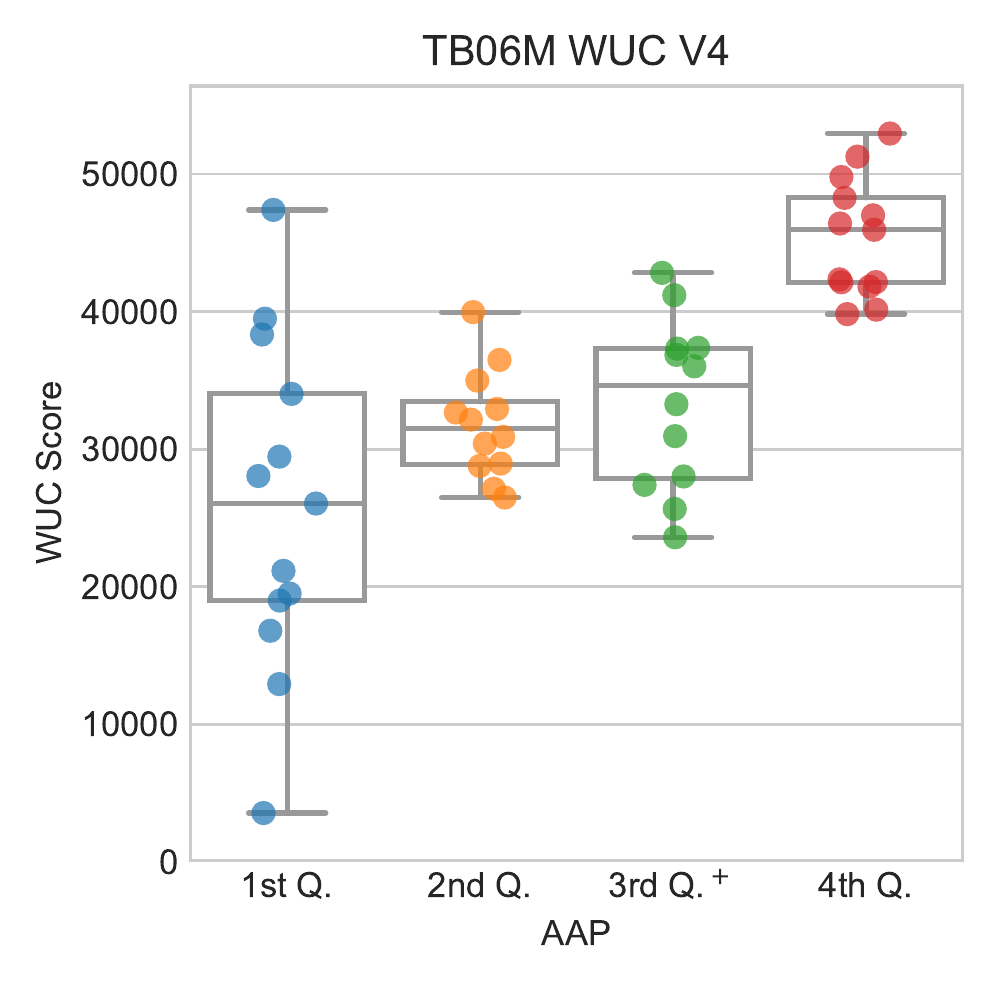}
    \\ 
    \includegraphics[width=.33\linewidth]{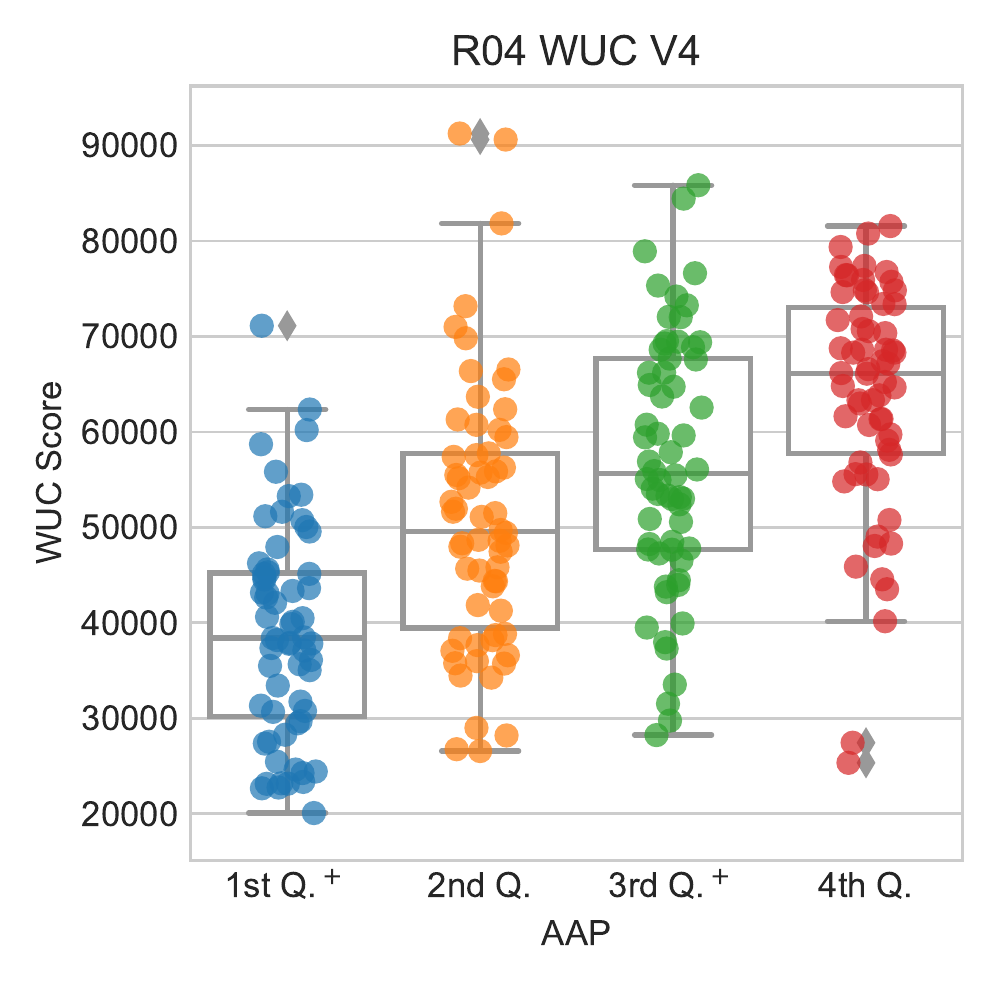}
    &
    \includegraphics[width=.33\linewidth]{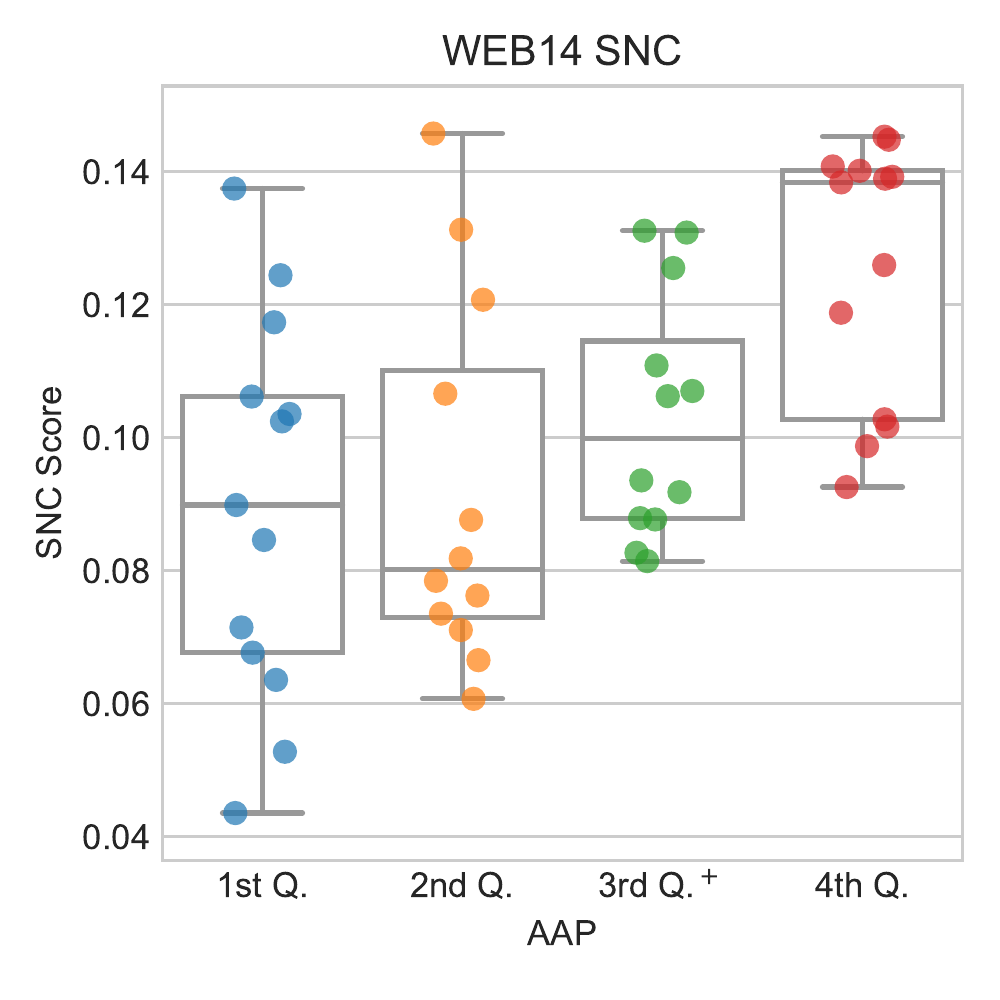}
    &
    \includegraphics[width=.33\linewidth]{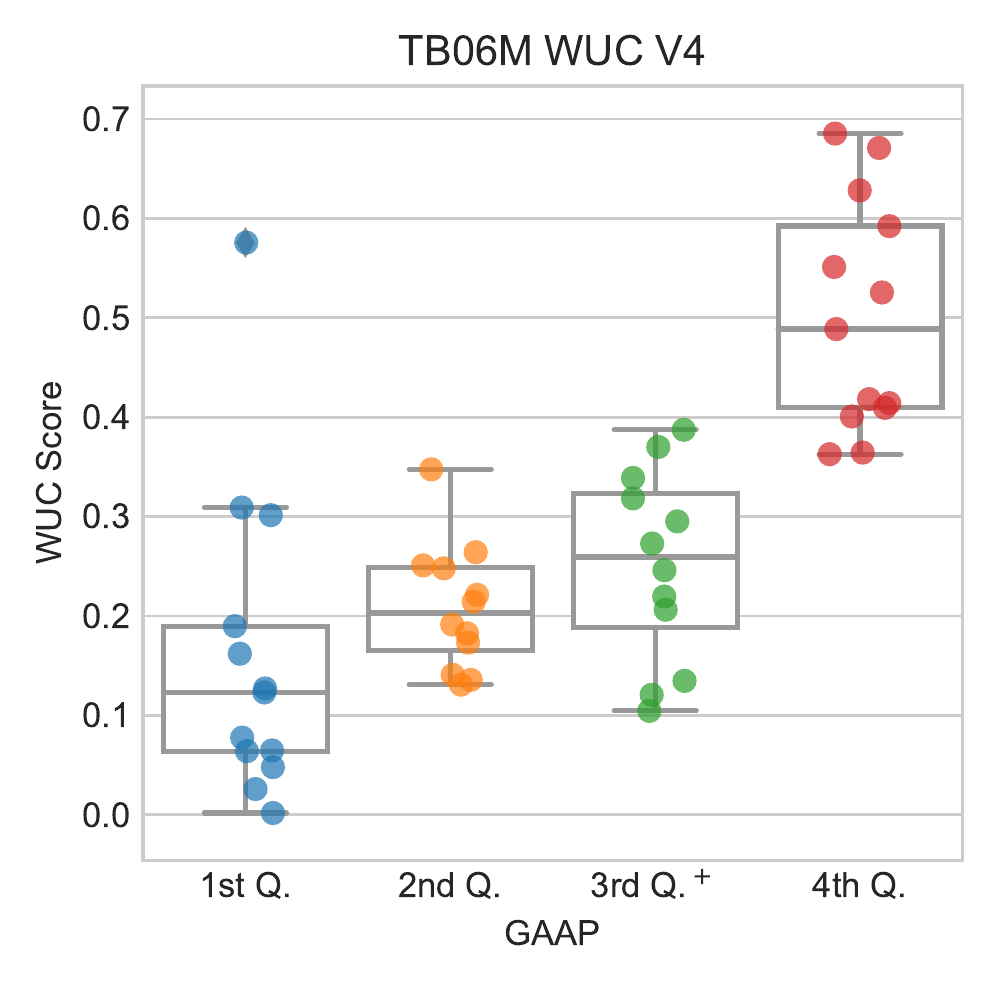}
  \end{tabular}    
  \caption{
  AAP boxplots for the scatterplots of Figure~\ref{JDIQNorel:fig:AAP} divided into quartiles according to AAP (or GAAP); statistical significance computed considering Bonferroni's correction.}
  \label{JDIQNorel:fig:AAP_BoxPlot}
\end{figure}

However, this is not the whole story.
While it is important to understand which system is the best  (the most effective one, having the higher effectiveness value), for the topics it is rather important to understand which are the difficult ones (the ones having the lowest effectiveness values), since on these alternative strategies can be used to improve effectiveness.
Some further analysis is shown in Figure~\ref{JDIQNorel:fig:AAP_BoxPlot}, that shows, for each scatterplot of Figure~\ref{JDIQNorel:fig:AAP}, one boxplot chart where the topics are grouped into quartiles according to their real AAP (GAAP) value. 
So, the x-axis represents the quartiles, the y-axis is still the topic ease predicted by the method, the dots are still topics (the small horizontal variations on the dots is just a random jitter for graphical reasons, to avoid overlapping dots), and the boxes summarize the distribution of the predicted values for each quartile, with the horizontal line corresponding to the median. For all the six charts, with just one exception, the median of each quartile is lower than the subsequent ones. This means that if the difficulty of topics is measured by categorizing them into the four difficulty categories, the three methods are reasonably good in predicting it 
(although, of course, the increasing median is not a sufficient condition to make the four different classes fully separable). 
Furthermore, we ran an unpaired t-test using the Bonferroni's correction \cite{dunn1961multiple}: 
whereas for the four charts about TREC-8, TB06M, and WEB14 the differences between adjacent quartiles are mostly not significant (also because only 50 topics occur in those charts, and of course even fewer in each quartile), the differences between the first and the third quartile, for the two charts about TB06 and R04, are significant.

\begin{figure}[tbp]
  \centering
  \begin{tabular}{@{}c@{}c@{}c}
    \includegraphics[width=.33\linewidth]{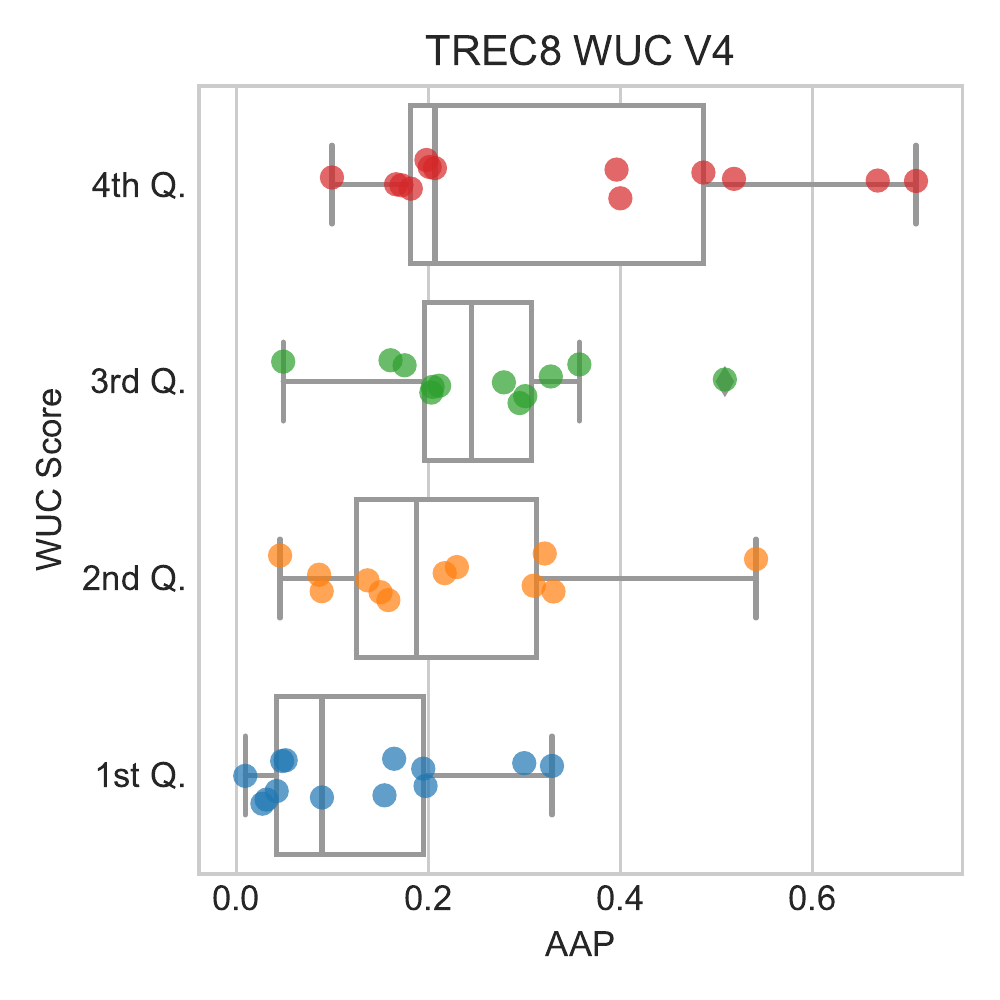}
    &
 \includegraphics[width=.33\linewidth]{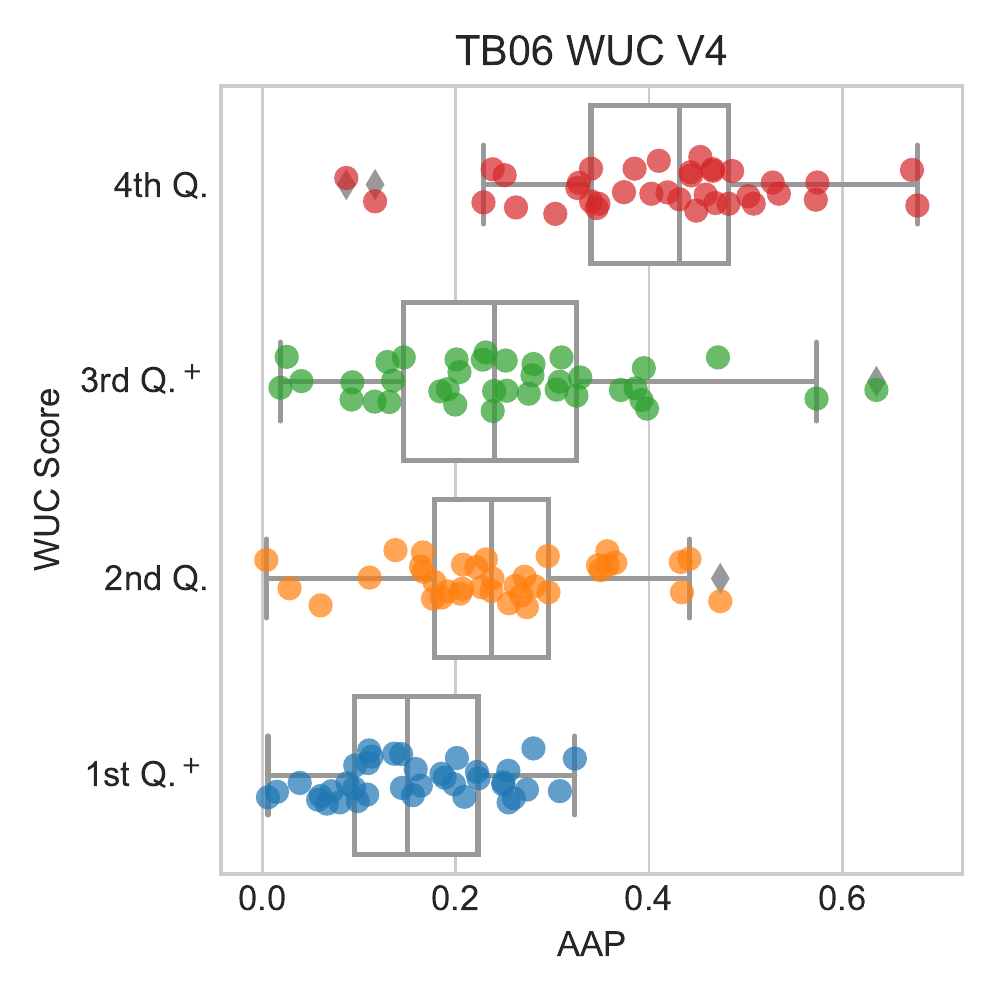}
    &
    \includegraphics[width=.33\linewidth]{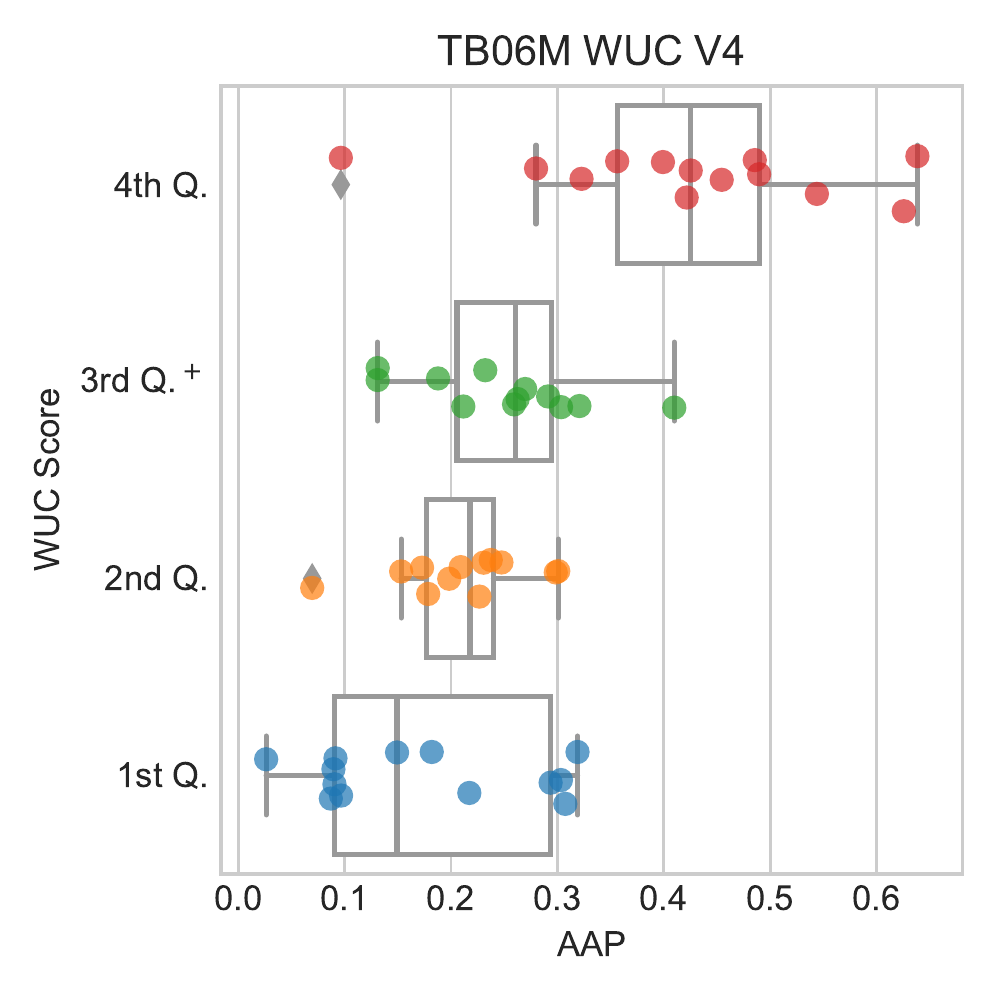}
    \\ 
    \includegraphics[width=.33\linewidth]{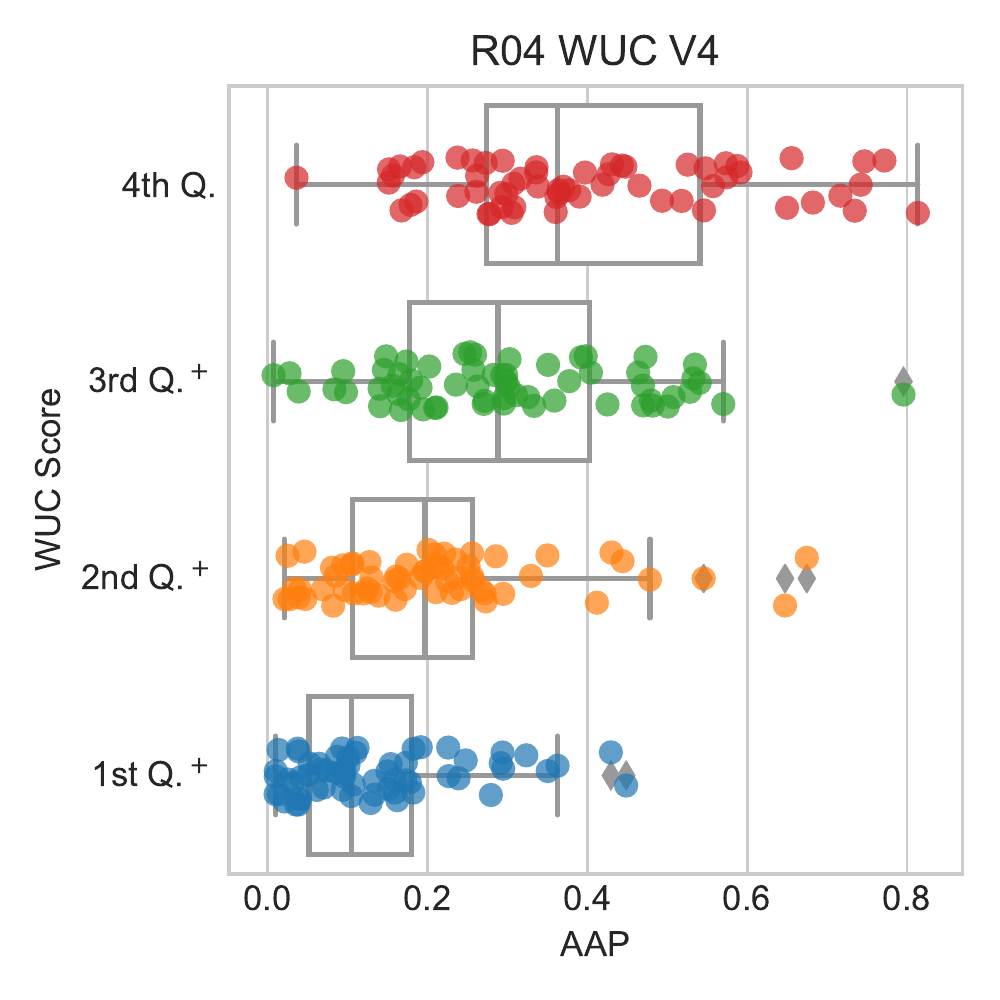}
    &
    \includegraphics[width=.33\linewidth]{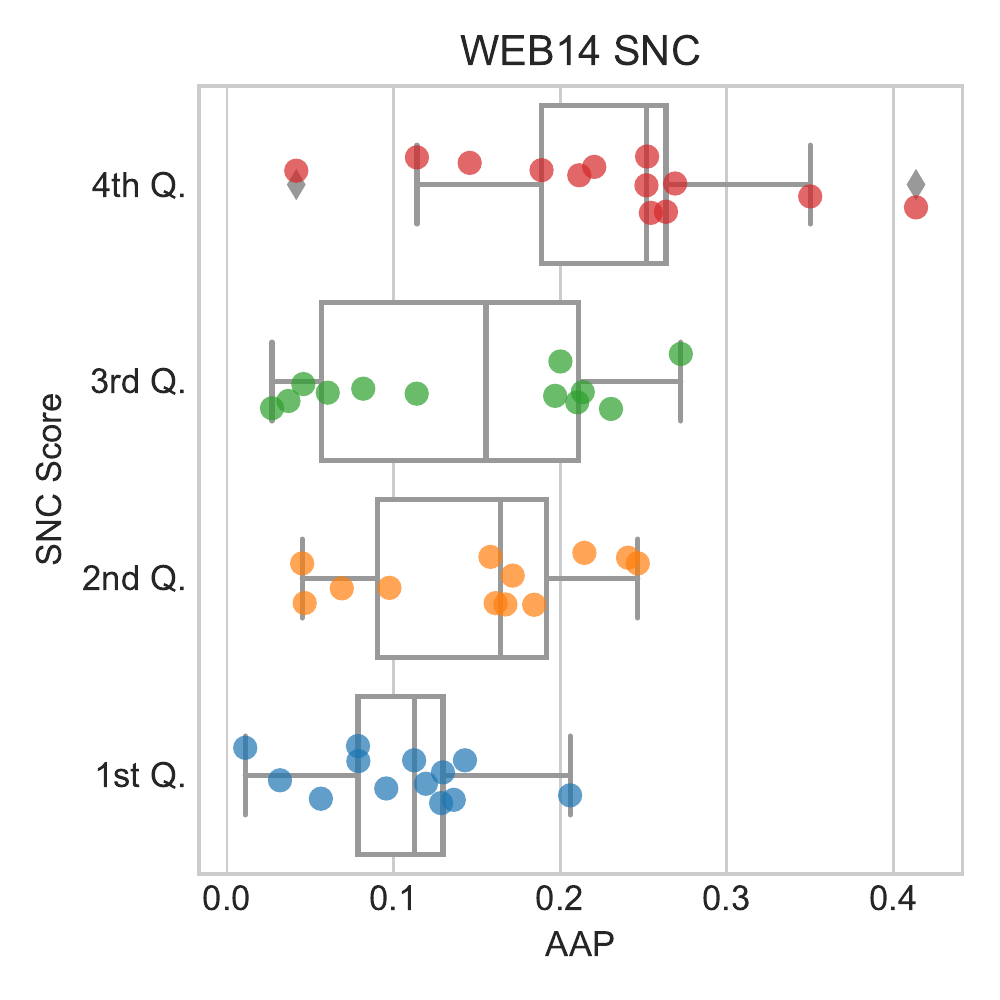}
    &
    \includegraphics[width=.33\linewidth]{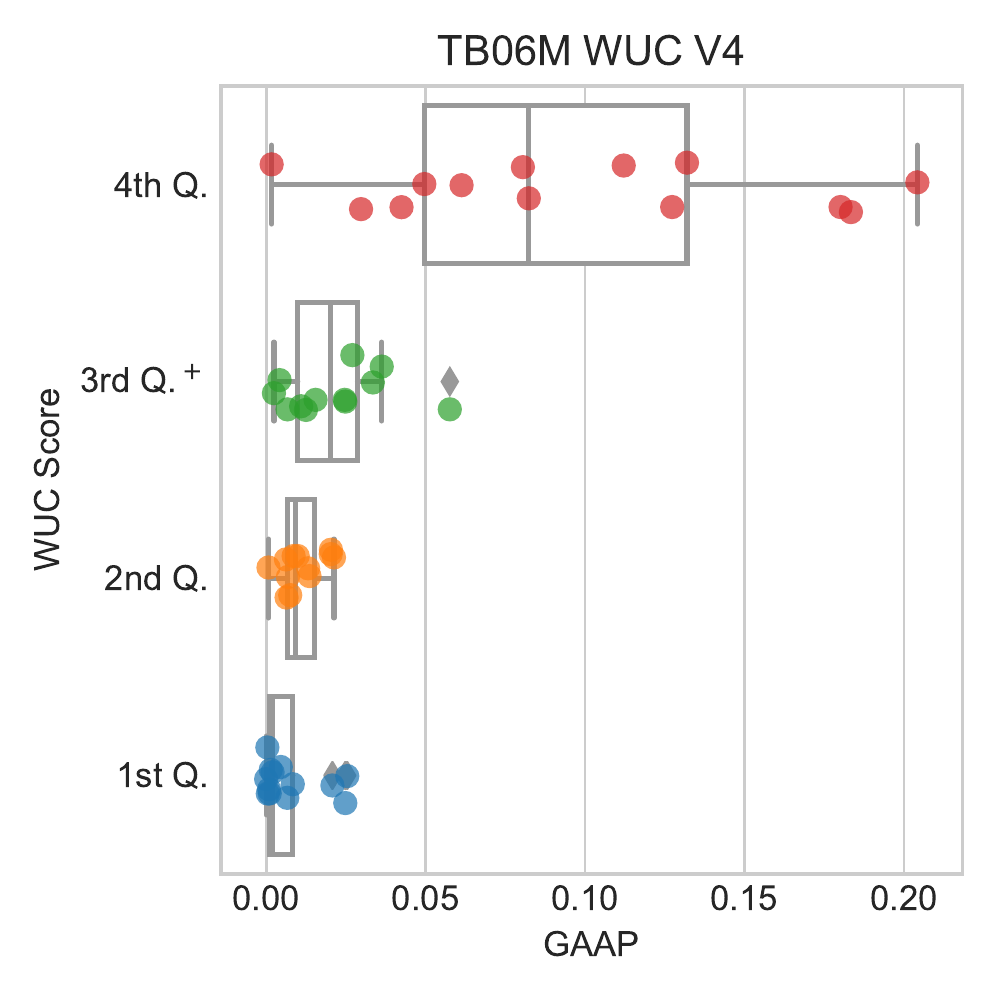}
  \end{tabular}    
  \caption{
  AAP boxplots for the scatterplots of Figure~\ref{JDIQNorel:fig:AAP} divided into quartiles according to SNC/SPO/WUC Score. Statistical significance computed considering Bonferroni's correction.}
  \label{JDIQNorel:fig:AAP_BoxPlot2}
\end{figure}

Figure~\ref{JDIQNorel:fig:AAP_BoxPlot2} presents yet another analysis. The six charts are again those corresponding to the charts in Figures~\ref{JDIQNorel:fig:AAP} and~\ref{JDIQNorel:fig:AAP_BoxPlot} but, in this figure, the topics are grouped into quartiles accordingly to the difficulty predicted by the SNC/WUC/SPO method, rather than their real difficulty. 
What is interesting here is that for each method the topics that are in the first quartile turn out to be indeed difficult ones also accordingly to their exact AAP value, with very few exceptions. 
%
Although this is also true for some topics in the 2nd, 3rd, and 4th quartile (especially for R04 and TB06, but note that the number of topics is much higher for those collections), this last result can have immediate practical applications. We could build an IR system that adopts some countermeasures (e.g., perform an automatic query reformulation, or ask further information to the user, and so on) for the difficult topics. And we can reliably predict which topics are difficult by selecting those in the first quartile according to each of the three methods. If the countermeasures are effective, this would lead to a more effective IR system. 

In other terms, for topic difficulty prediction, precision seems more important than recall: if a topic prediction system fails to recognize that a topic is difficult, no harm is done; conversely, if an IR system adopts some countermeasure on a topic that is easy but is wrongly predicted as false, this would likely decrease retrieval effectiveness. The three methods are indeed precision oriented ones. To state it in yet another way, going back to Figure~\ref{JDIQNorel:fig:AAP}, what is important is that no topics, or at least very few of them, are in the  bottom-right part
of the charts, which is indeed the case: the vast majority of dots are in the top left triangular part.


\section{Conclusions} \label{JDIQNorel:sec:con}

In this chapter we set to reproduce the most important work on automatic evaluation of IR effectiveness, i.e., evaluation without human relevance judgements. Instead of 
only reproducing the work, we provide a 
fourfold
contribution:
\begin{enumerate}[label=(\roman*)]
\item we succeeded in reproducing the main results of three previous similar studies, with only some minor caveats
, which we discussed in the respective sections;
we released all the code used to carry out the experiments; 
differently from the original works, we focused on future reproducibility, and we detailed all the parameters required to implement each method;
\item we presented the results in a uniform way; 
\item we generalized those results to other test collections, evaluation metrics, and a shallow pool; and
\item \label{JDIQNorel:i:res:exp} we expanded those results, obtaining two practical strategies that seem effective to, respectively, decrease the costs involved in test collection based evaluation, and improve retrieval effectiveness on difficult topics. 
\end{enumerate}
A general lesson learned of methodological nature is that we believe that this is the right attitude in a reproducibility setting: not only simply reproducing, but also providing a uniform representation; in our case this lead naturally to generalization, and was also inspiring to obtain the apparently very interesting results~\ref{JDIQNorel:i:res:exp}.



In the following chapter we will use the detailed methods and their combinations  to build a sort of meta-method that allows to evaluate e collection without the usage of relevance judgements.

\chapter{Evaluation Without Relevance judgements: Analysis of Existing Methods and of their Combinations} 
\chaptermark{Evaluation Without Relevance judgements}
\label{chapt:eewrj:combinations}

This chapter deals with the analysis of existing methods on evaluation without relevance judgements and their combinations.
Section~\ref{IPM:sec:introduction} introduces and frames research questions,
Section~\ref{IPM:sec:experimental-settings} describes the experimental setting,
Section~\ref{IPM:sec:accur-indiv-meth} investigates the accuracy of individual methods,
Section~\ref{IPM:sec:relat-among-meth} considers relationships between the methods,
Section~\ref{IPM:sec:comb-autom-eval} presents experiments on the combinations of the considered methods. Finally, Section~\ref{IPM:sec:concl-future-devel} concludes and sketches some directions for future work.

In our experiments, we will use the methods listed in Table~\ref{IPM:tab:methods}. 
We believe that we have included all the proposals from  the literature, with the only exception of \citeauthor{Diaz:2007}'s one \cite{Diaz:2007}: we leave it for future work since 
it uses the text of topics and documents, and we are interested in providing a complete and uniform account of  the  methods that do not use  the text of documents, nor the topic descriptors.

\section{Introduction and Research Questions}
\label{IPM:sec:introduction}
In Information Retrieval (IR), test-collection  based effectiveness evaluation is a well-known and quite standard method. The whole evaluation process has a cost, in terms of resources needed, effort made by the research community, and also money; thus it is not surprising that researchers tried and are still trying to reduce such costs, for example by using fewer topics,  more sensitive effectiveness metrics,  shallower pools, or cheaper (usually, crowdsourced) human relevance judgements. 
A more radical approach is to avoid human relevance judgements altogether, as it has been proposed by several researchers \cite{Soboroff:2001:RRS:383952.383961,Wu:2003:MRI:952532.952693,Aslam:Savell:2003,Nuray:2003,nuray:can:2006,spoerri:2007,Diaz:2007,sakai-lin2010}.
In this chapter, we set out to provide a detailed and complete analysis of the methods for effectiveness evaluation without human relevance judgements, as well as study if they can be fruitfully combined. 
%



When analyzing the literature on the methods for effectiveness evaluation without relevance judgements, one can notice that their accuracy is often evaluated using different measures, on different datasets, and on the basis of different effectiveness metrics (see the last three columns of Table~\ref{IPM:tab:methods}). 
This means that it is not clear what the relative accuracies are, and how these vary across the datasets. 
Therefore, our first researchquestion is aimed at establishing a solid baseline for these effectiveness evaluation prediction methods:
\begin{itemize}
\item \textbf{RQ1}: What is the  comparative accuracy of the various methods for effectiveness evaluation without relevance judgements when they are evaluated under the same conditions?
  What about different collections, and different measures? 
\end{itemize}
Some comparisons do exist, although they are made in a rather implicit and incomplete way. The most similar works to ours are those by  \citet{hauff:dejong:2010} and \citet{sakai-lin2010}. \citeauthor{hauff:dejong:2010} present a comparison of most of the methods, but their aim is to study the variations across topics, and to understand what happens when selecting the ``right'' topics subset. 
The work by \citeauthor{sakai-lin2010} is more related to our RQ1, but again it focuses on just 6 methods (while we analyze 17 of them) and uses 2 TREC and 3 NTCIR collections (instead, we test them on 14 TREC collections). Also, we report a more complete set of accuracy measures and, finally, we consider the actual accuracy of the methods as a means for the remaining two research questions, rather than as an end in itself.

Going beyond accuracy figures, one might wonder whether the methods are really different from each other, or rather whether they all measure, more or less, the same thing. Our second research question specifically addresses this issue:
\begin{itemize}
\item \textbf{RQ2}: What are the relationships among the methods? Do they tend to measure the same phenomenon with almost no differences, or is there any variability that can be exploited?
\end{itemize}
If the methods are indeed different, it is natural to ask whether this diversity can be exploited by combining them. Therefore, our third and last research question is:
\begin{itemize}
\item \textbf{RQ3}: Can the methods be combined in an effective way? What combination strategies lead to
the highest accuracy?
\end{itemize}

\section{Experimental Setting}
\label{IPM:sec:experimental-settings}

We describe the overall setting common to all the experiments. We present the basic definitions, the measures, and the datasets used.

\subsection{Notation, Background, and Terminology}
\label{IPM:sec:notation-background}

\begin{figure}[tbp]
\centering
  \begin{tabular}[t]{r|ccc| c |c|}
     \multicolumn{1}{c}{} &$t_1$ &  $\cdots$ & \multicolumn{1}{c}{$t_n$} &\multicolumn{1}{c}{}&\multicolumn{1}{c}{$\MAP$}\\
    \cline{2-4}\cline{6-6}
    $s_1$ & $\AP(s_1,t_1)$& $\cdots$ &$\AP(s_1,t_n)$&&$\MAP(s_1)$\\
    \vdots& \vdots &$\ddots$& \vdots  &&$\vdots$\\
    $s_m$& $\AP(s_m,t_1)$& $\cdots$ &$\AP(s_m,t_n)$&&$\MAP(s_m)$\\
    \cline{2-4} \cline{6-6}
    \multicolumn{6}{c}{}\\
    \cline{2-4} 
    $\AAP$ & $\AAP(t_1)$ & $\cdots$ & $\AAP(t_n)$ & \multicolumn{2}{c}{}\\
    \cline{2-4}
  \end{tabular}
\caption{AP ($\AP(s_i, t_j)$), MAP ($\MAP(s_i)$), and AAP ($\AAP(t_j)$) for $n$ topics and $m$ systems (adapted from \cite{Mizzaro:2007:HHT:1277741.1277824, Roitero2017}).
\label{IPM:fig:AP}}
\end{figure}

Figure~\ref{IPM:fig:AP} shows the basic outcome of a test collection evaluation exercise, represented as a matrix and two vectors. 
Each row of the matrix is a system $s_i$ (or run), each column is a topic $t_j$, and each cell $(i,j)$ is the effectiveness of system $s_i$ on topic $t_j$.     
Averaging each row on the $n$ topics one obtains a measure of system effectiveness (for all systems, this is the column vector on the right); averaging each column on the $m$ systems one obtains a measure of topic ease (the row vector on the bottom).

In this chapter we focus on Average Precision (AP) as the effectiveness measure. 
We use the following notation. 
$\AP(s_i,t_j)$ is the AP value of system $s_i$ on topic $t_j$, $\AP$ is the matrix of AP values, and $\MAP$ and $\AAP$ are the vectors of the MAP (Mean AP) and AAP (Average AP) values. 
Although we acknowledge that ``Mean Average Precision'' is a questionable term, we use it to distinguish from both average precision (the individual effectiveness value of a system on a topic) and Average Average Precision \cite{Mizzaro:2007:HHT:1277741.1277824, Roitero2017} (a measure of topic ease).
  
Turning to the effectiveness values predicted by a method, we denote with $\widehat{\AP}$ the matrix of predicted AP values, and with $\widehat{\MAP}$ and $\widehat{\AAP}$ the vectors of predicted MAP and AAP values, respectively.
We will first and mainly focus on MAP ($\MAP$ and $\widehat{\MAP}$) in this chapter, as others have done, but we will also study and exploit AAP and AP. 
Thus, the main question will be the accuracy of $\widehat{\MAP}$ as a prediction of the ground truth $\MAP$, but we will also study the accuracy of $\widehat{\AAP}$ and $\widehat{\AP}$ as predictions of the original $\AAP$ and $\AP$.


\subsection{Accuracy Measures}
\label{IPM:sec:accuracy-measures}

One can imagine several accuracy measures, and indeed many alternatives have been used in the past studies (see Table~\ref{IPM:tab:methods}). 
Kendall's or Spearman's rank correlations are reasonable choices when one is interested in the order of the values, an option that is quite common when a ranking of the systems according to their effectiveness is desired. Often, the top positions of a rank are the most important, and in such a case a top-heavy rank correlation like Tau-AP ($\tau_{ap}$) \cite{Yilmaz:2008:NRC:1390334.1390435} can be used.
Pearson's linear correlation can be used when one wants to understand if the predicted values have a linear relation with the original ones, i.e., when measuring the ability of methods in predicting the exact values, not just the ranks. 
Correlations are a natural measure when working on the vectors $\MAP$ and $\AAP$, but they can be used also on the matrix $\AP$ by converting it to a vector. 
Vectorization of a matrix is a linear operation that concatenates all the columns of the matrix into a column vector. 
However, for a matrix a similarity measure based on matrix difference is also meaningful. 
In the following we will use:
\begin{itemize}
\item Pearson's linear correlation (denoted with $\rho$, i.e., $\rho(\MAP,\widehat{\MAP})$,\\ $\rho(\AAP,\widehat{\AAP})$, $\rho(\AP,\widehat{\AP})$, the latter being the correlation between the vectorized AP matrices);
\item Kendall's rank correlation  ($\tau(\MAP,\widehat{\MAP})$, etc.);
\item Spearman's rank correlation ($r_s(\MAP,\widehat{\MAP})$, etc.);
\item Tau-AP, a top-heavy rank correlation \cite{Yilmaz:2008:NRC:1390334.1390435} ($\tau_{ap}(\MAP,\widehat{\MAP})$);
\item Matrix difference ($\delta(\AP,\widehat{\AP}) = \frac{1}{nm}\sum{|\widehat{(\AP}(i,j) - \AP(i,j)|}$). 

\end{itemize}

\subsection{Datasets}
\label{IPM:sec:datasets}


\begin{table}[tbp]
  \centering
  \caption{The 14 Datasets Used in this chapter
  \label{IPM:tab:coll}}
  \begin{threeparttable}
  \begin{tabular}{@{\makebox[1.5em][r]{\rrownumber\space \space}}  l@{ }l@{ }l @{ }r@{ }r l@{}}
  \toprule
    Acron. & Name  & Year & Topics &  Runs   & Used Topics 
        \gdef\rrownumber{\stepcounter{mmagicrownumbers}\arabic{mmagicrownumbers}} \\
        \midrule
    TREC3 &Ad Hoc&1994&50&40&151-200\\
    TREC5 &Ad Hoc&1996&50&61&251-300\\
    TREC6 &Ad Hoc&1997&50&74&301-350\\
    TREC7 &Ad Hoc&1998&50&103&351-400\\
    TREC8 & Ad Hoc &	1999 & 50  & 129&  401-450  \\ 
    TREC01 &Ad Hoc&2001&50&97&501-550\\
    R04 & 	Robust    &	2004 & 249 & 110  & 301-450, 601-700\tnote{1}  \\
    TB04 &TeraByte&2004&49&69&701-750\tnote{2}\\
    R05 &Robust&2005&50&74& See \cite[Figure 1]{voorhees2003overview} \\

    TB05 &TeraByte&2005&50&58&751-800\\
    TB06 & 	TeraByte  &	2006 & 149 & 61&  701-850\tnote{2} \\
    W11 &Web Track&2011&50&61&101-150\\
    W12 &Web Track&2012&50&48&151-200\\
    W13 &Web Track&2013&50&55&201-250\\
    \bottomrule
  \end{tabular}
  \begin{tablenotes}
 \item[1] 672 excluded.
 \item[2] 703 excluded.
 
  \end{tablenotes}
  \end{threeparttable}
\end{table}

Table~\ref{IPM:tab:coll} summarizes the 14 datasets considered in this chapter, showing an acronym (used in the following), a longer name, the year, the number of topics ($m$) and of systems ($n$), as well as the topic identifiers in the dataset.
We use several TREC collections, 
spanning 20 years,  selecting among those having large enough sets of topics and systems/runs. For each dataset we produced the corresponding table as in Figure~\ref{IPM:fig:AP}. The three Web track collections (last three rows) adopted a non-binary notion of relevance; we computed AP values collapsing relevance levels -2 and 0 into irrelevant and 1, 2, and 3 into relevant, and then running
trec\_eval.\footnote{See \url{https://trec.nist.gov/trec_eval/}}
%
%
The code to conduct the experiments can be found at \url{https://github.com/KevinRoitero/LeToE-Code}.

\section{RQ1:  Individual Methods Accuracy}
\label{IPM:sec:accur-indiv-meth}



\begin{figure}[tb]
\centering
\includegraphics[width=.99\linewidth]{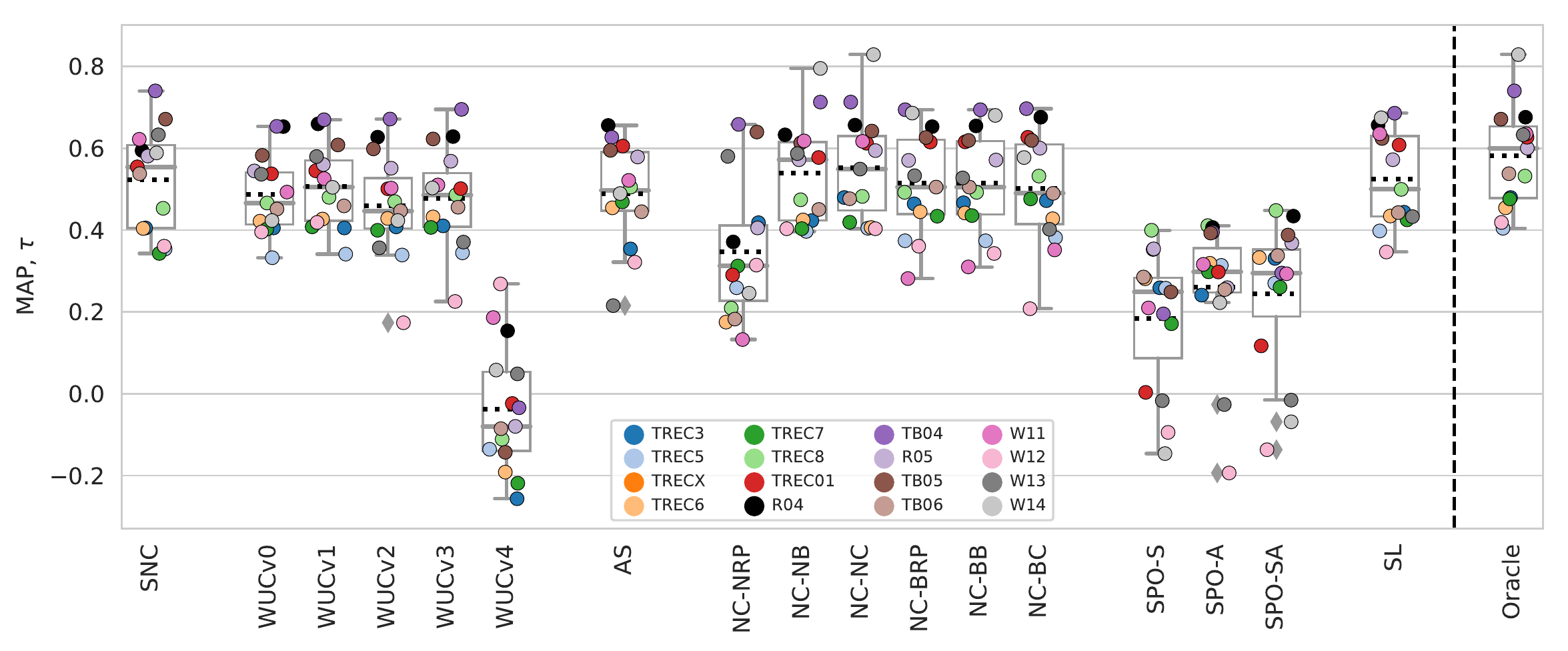}
\vspace*{-2mm}
\caption{Accuracy of the  methods: MAP $\tau$
  \label{IPM:fig:accuracy:MAP:tau}}
\end{figure}
\begin{figure}[tb]
  \centering
   \includegraphics[width=.99\linewidth]{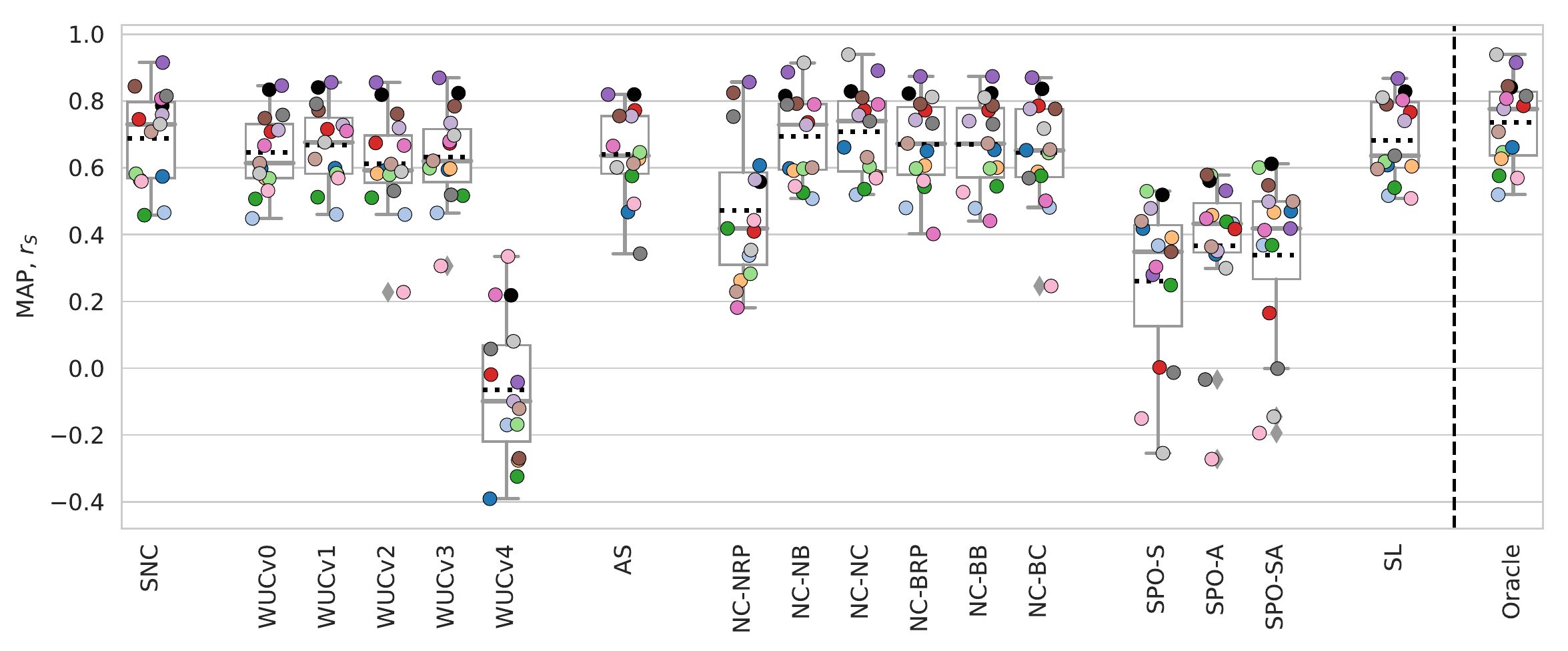}
\vspace*{-2mm}
 \caption{Accuracy of the  methods: MAP $r_s$
 \label{IPM:fig:accuracy:MAP:rs}}
\end{figure}
\begin{figure}[tb]
  \centering
   \includegraphics[width=.99\linewidth]{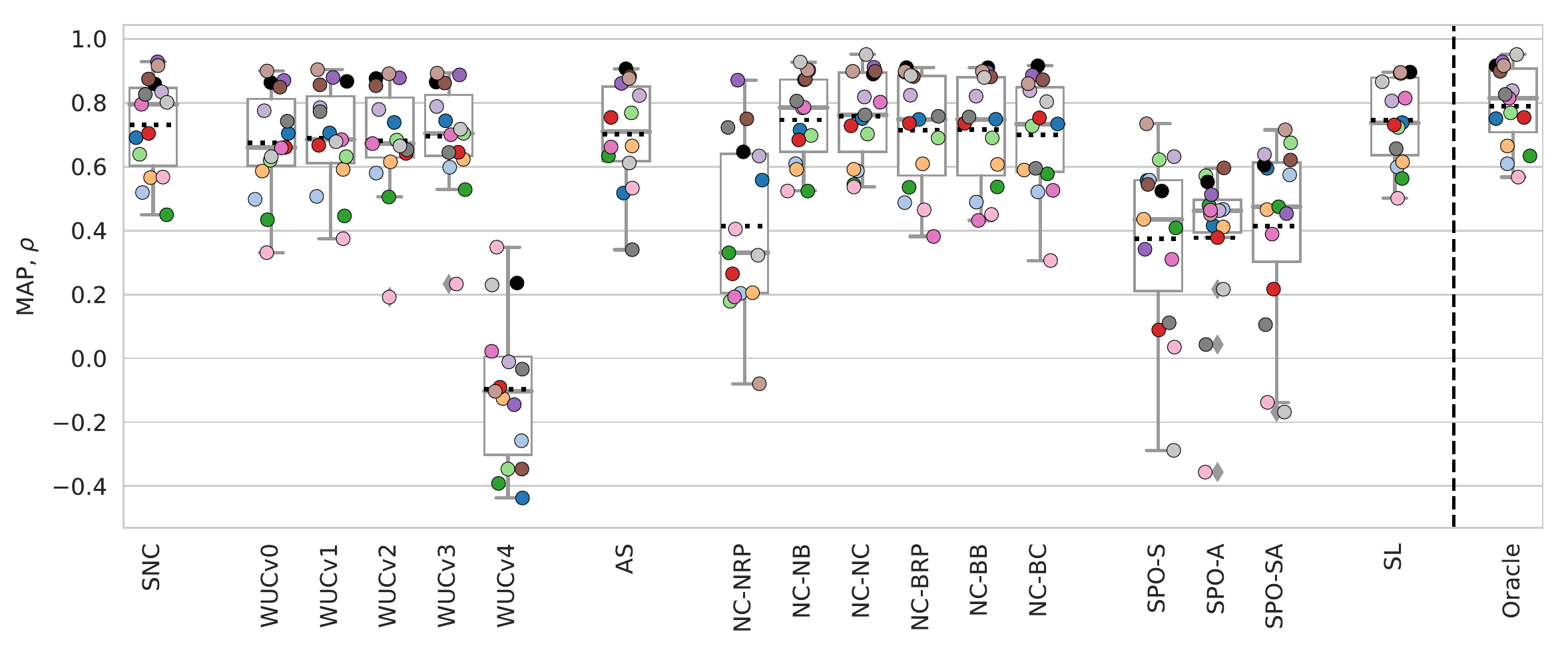}
\vspace*{-2mm}
 \caption{Accuracy of the  methods: MAP $\rho$
 \label{IPM:fig:accuracy:MAP:rho}}
\end{figure}
\begin{figure}[tb]
  \centering
   \includegraphics[width=.99\linewidth]{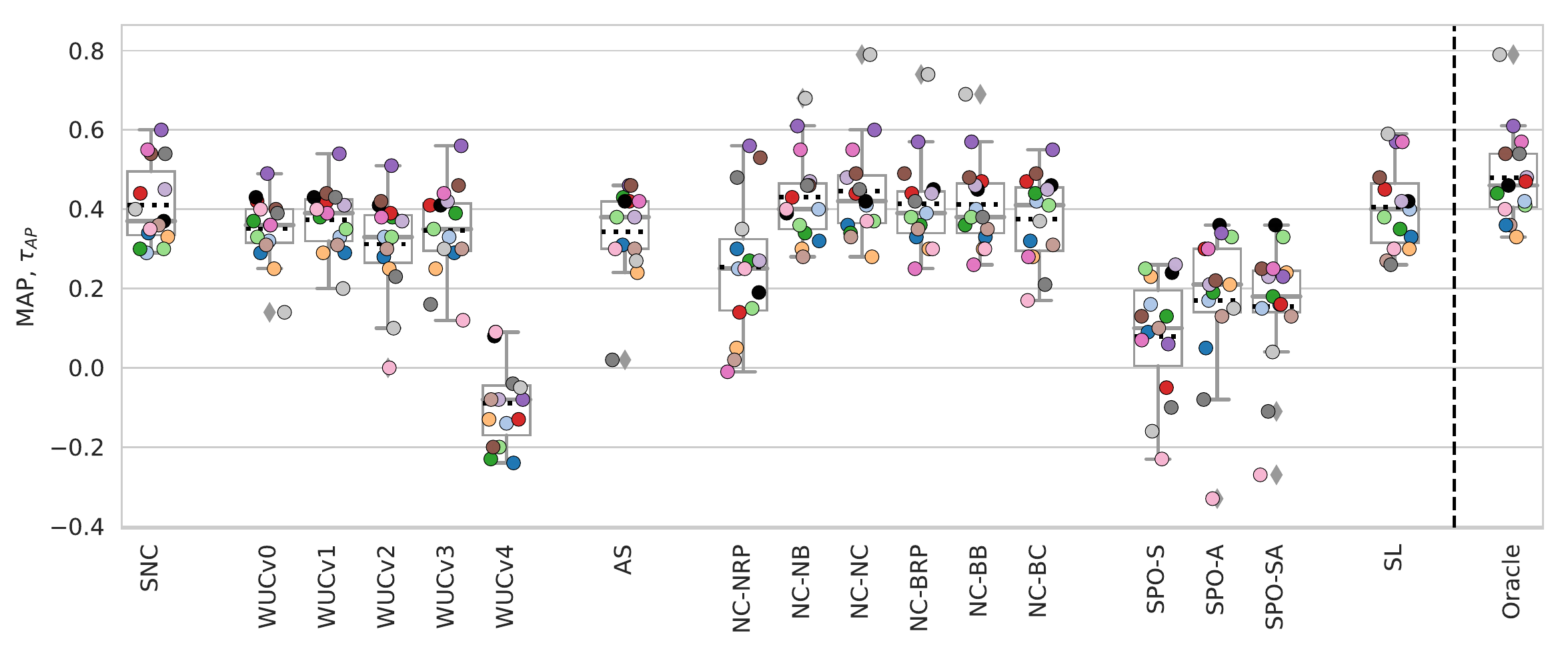}
\vspace*{-2mm}
 \caption{Accuracy of the  methods: MAP $\tau_{ap}$
 \label{IPM:fig:accuracy:MAP:tauAP}}
\end{figure}

\begin{figure}[tb]
  \centering
   \includegraphics[width=.99\linewidth]{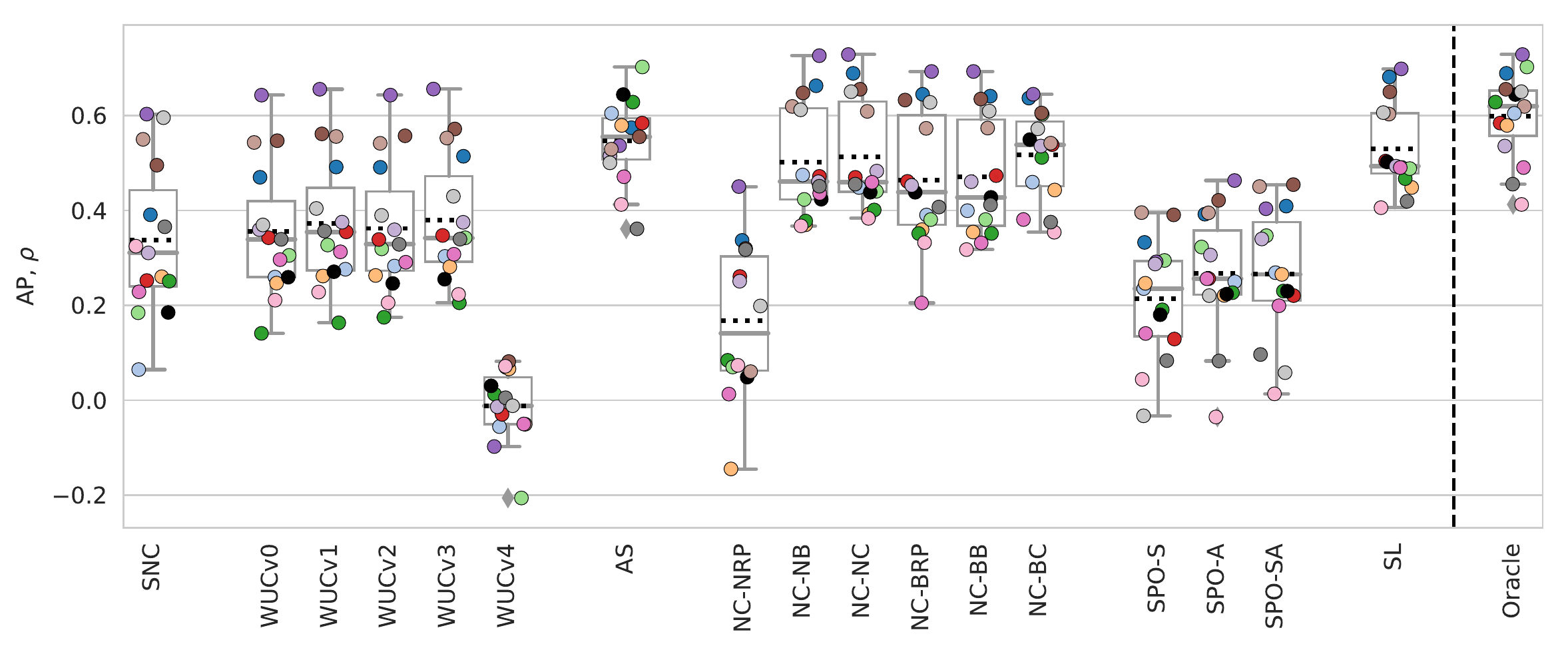}
\vspace*{-2mm}
 \caption{Accuracy of the  methods: AP $\rho$
 \label{IPM:fig:accuracy:AP:rho}}
\end{figure}
\begin{figure}[tb]
  \centering
   \includegraphics[width=.99\linewidth]{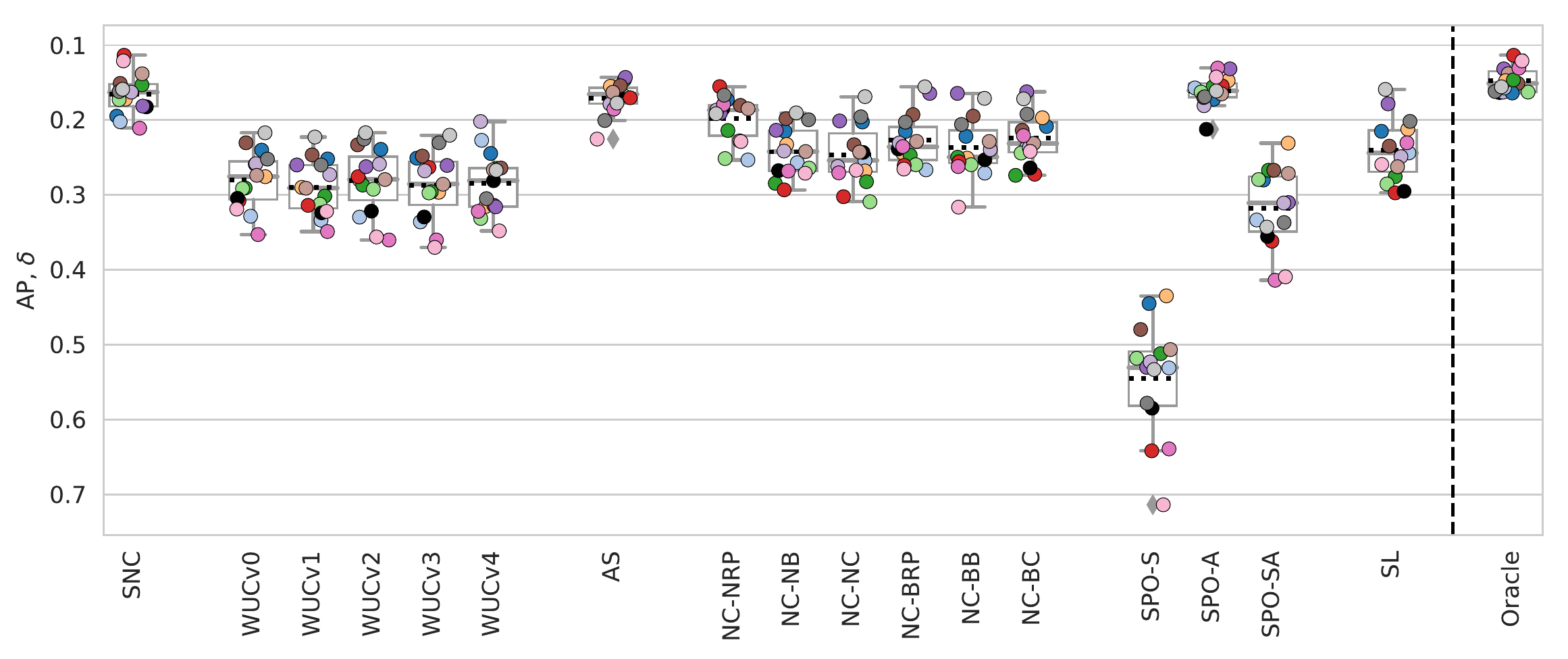}
\vspace*{-2mm}
 \caption{Accuracy of the  methods: AP $\delta$ 
 \label{IPM:fig:accuracy:AP:delta}}
\end{figure}
\begin{figure}[tb]
  \centering
   \includegraphics[width=.99\linewidth]{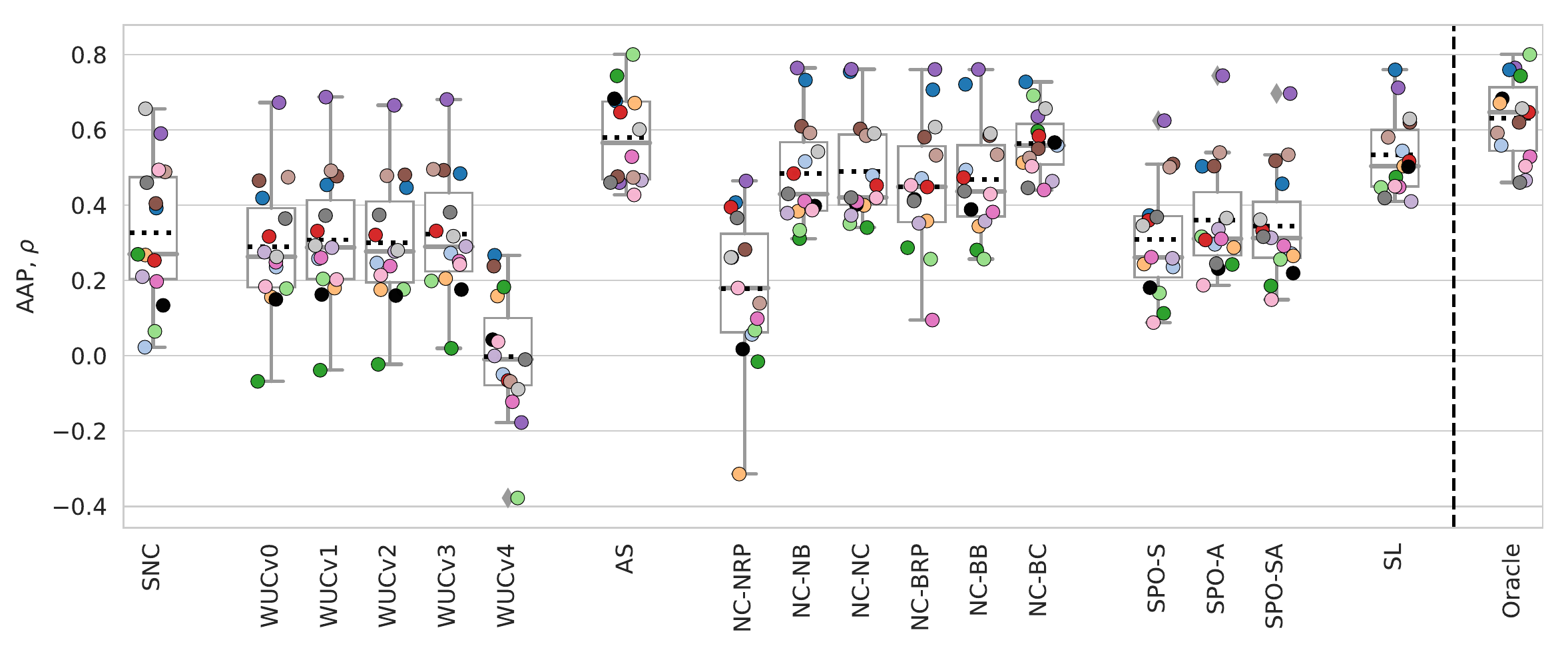}
\vspace*{-2mm}
 \caption{Accuracy of the  methods: AAP $\rho$
 \label{IPM:fig:accuracy:AAP:rho}}
\end{figure}
\begin{figure}[tb]
  \centering
   \includegraphics[width=.99\linewidth]{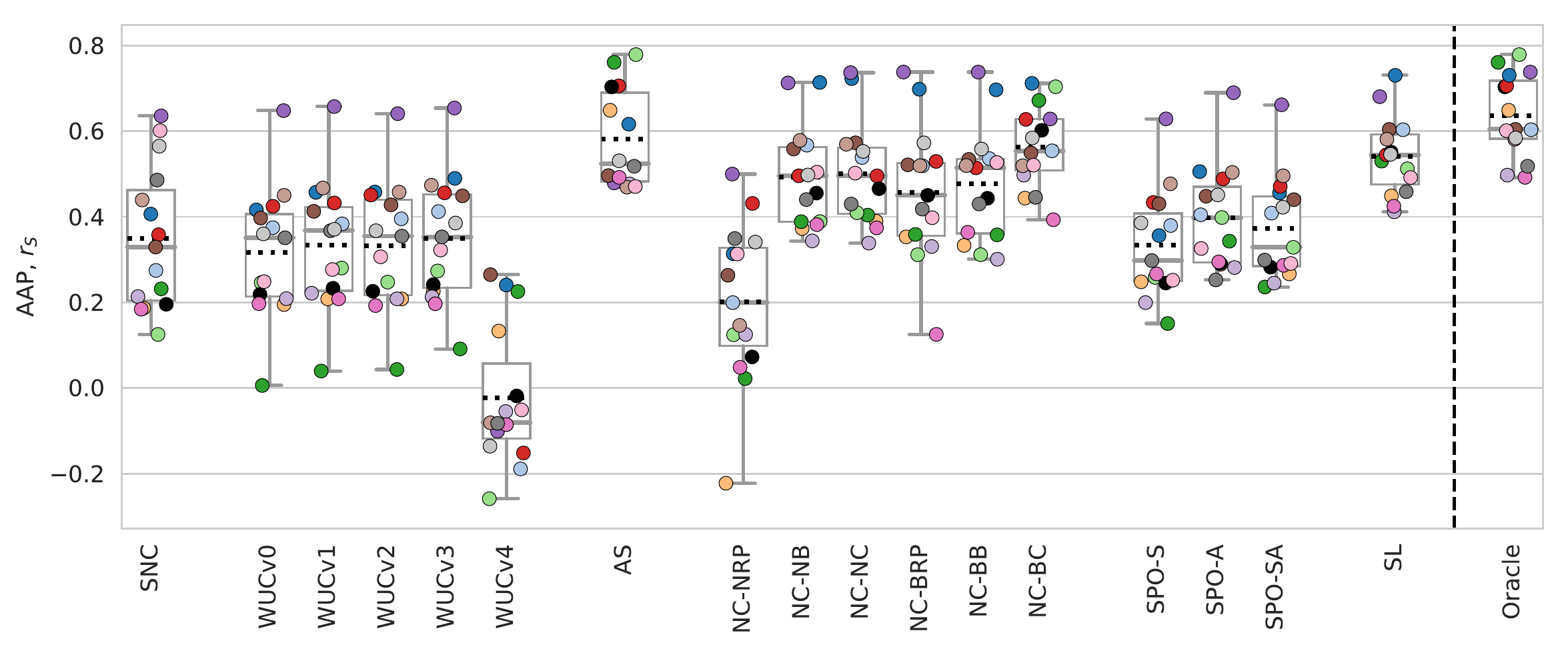}
\vspace*{-2mm}
 \caption{Accuracy of the  methods: AAP $r_s$
 \label{IPM:fig:accuracy:AAP:rs}}
\end{figure}



We now turn to presenting and discussing the results of our experiments. We start by focusing on RQ1, aimed at quantifying the accuracy of the individual methods.
Figures~\ref{IPM:fig:accuracy:MAP:tau} to \ref{IPM:fig:accuracy:AAP:rs} show
the accuracy of prediction as eight box-plot charts. 
These charts 
show the accuracy (Y-axis) of the individual methods (X-axis) on each dataset (legend). 
As indicated on the Y axes, the prediction is of MAP ($\MAP$) in the first four charts, of AP ($\AP$) in the following two, and of AAP ($\AAP$) in the last two; accuracy is measured by $\tau$, $r_s$, $\rho$, $\tau_{ap}$, and $\delta$. 
%
%
Each dot shows the accuracy of the prediction of a method on a dataset. 
So, for example, the  dot on the top-left of the first 
chart (Figure~\ref{IPM:fig:accuracy:MAP:tau}) is $\tau(\MAP,\widehat{\MAP})$, Kendall's $\tau$ correlation  of the actual MAP values in TB04 with the  MAP values predicted by SNC. 
The box-plots synthetically represent the distributions of the accuracy values (dots) for each method by showing the 95\% range, the 25th and 75th percentiles, and the median, as well as the mean (the dashed black horizontal line).
The rightmost panes will be discussed in Section~\ref{IPM:sec:oracle-combination}.

Analyzing each chart individually, we can make the following observations. 
Let us start by focusing on measuring accuracy of MAP prediction by $\tau$ (Figure~\ref{IPM:fig:accuracy:MAP:tau}). 
On average, i.e., looking at medians and means, the three most accurate individual methods seem to be SNC, NC-NB, and NC-NC. 
Other methods (WUCv1, NC-BRP, and NC-BB) look almost as accurate, other ones (WUCv0, WUCv2, WUCv3, NC-BC, AS, and SL) are not much less effective  and the last five (WUCv4, NC-NRP, CPO-S, SPO-A, and SPO-SA) are clearly outperformed, with WUCv4 showing a very low accuracy.

To better understand the differences in accuracy, we ran a paired Wilcoxon's significance test\footnote{See \url{https://docs.scipy.org/doc/scipy-0.14.0/reference/generated/scipy.stats.wilcoxon.html}.} \cite{wilcoxon1970critical}
between the methods, considering the series of $\tau$ values on the 14 datasets;
 we used the Bonferroni's method \cite{dunn1961multiple} to deal with multiple comparisons.
We found no statistical significant difference between the top six methods.


There is a somehow consistent behavior of datasets across methods: some of them have steady higher $\tau$ values (e.g., TB04) other ones have lower $\tau$ (e.g., TREC7). 
Clearly, MAP prediction is easier for some datasets, as others have already reported \cite{Hauff:2008:SPQ:1458082.1458311}. 
 
There is some variation over datasets, and this variation is quite similar across methods (i.e., the sizes of the box-plots, representing the inter-quartile range, are quite similar).  
When looking at individual collections, there are many exceptions to the average behavior: for example, 
SNC is slightly less accurate than SL for R04, TREC8, and TREC01. 
This means, when considered with the just noted consistent variation over datasets, that if a researcher wants to evaluate effectiveness on a new unseen dataset, it is not completely clear which method should be used, as well as which is the expected accuracy of the method.  

\begin{figure}[tb]
  \centering
\centering
 \includegraphics[width=.8\linewidth]{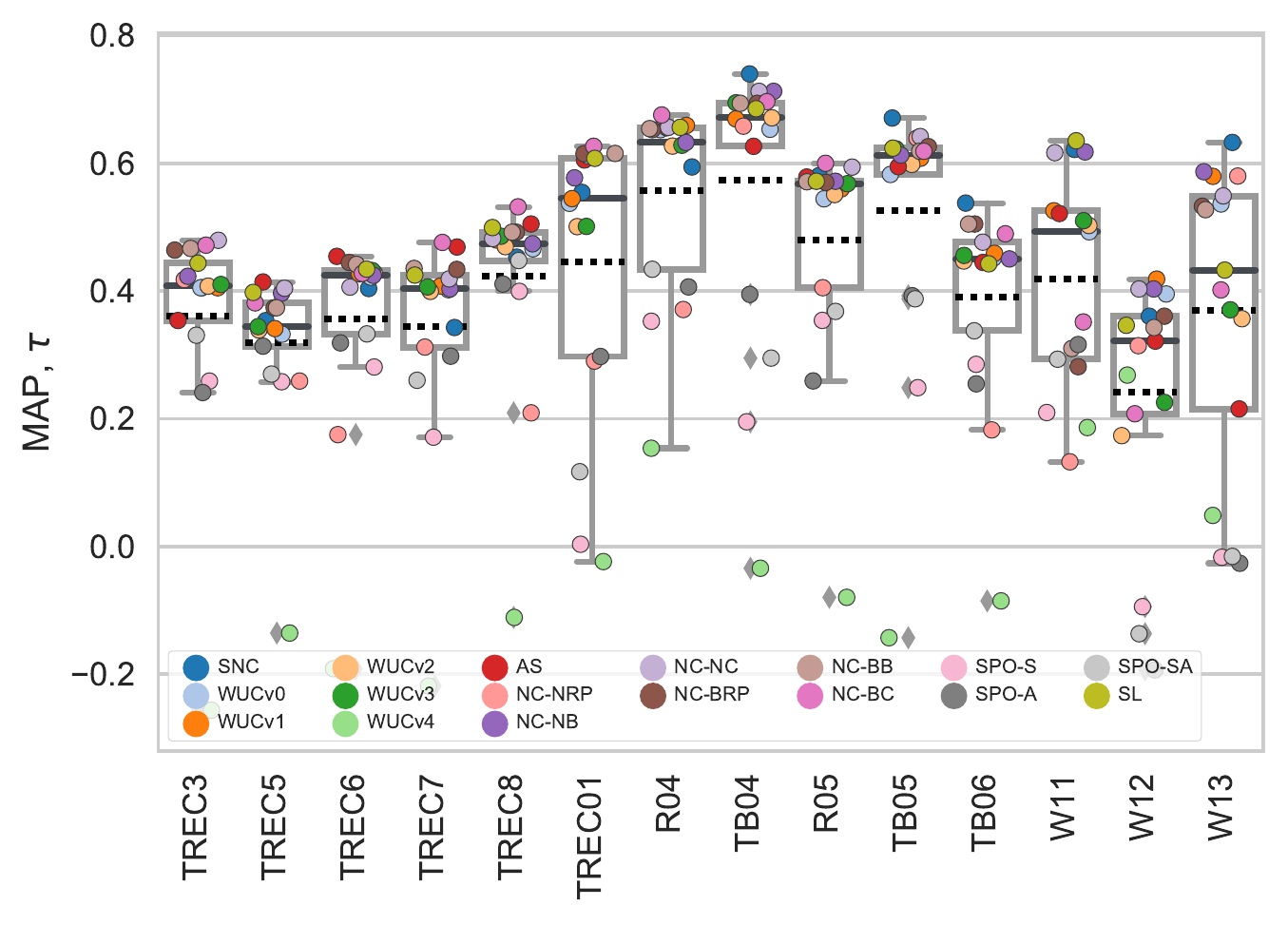}
  \caption{Accuracy (MAP $\tau$) of the  methods across  datasets.
  \label{IPM:fig:boxplot-dual}}
\end{figure}

Figure~\ref{IPM:fig:boxplot-dual} 
provides a  dual representation of the same data for MAP $\tau$ (the other accuracy measures show a similar behavior).
Here we group in each box-plot all the methods on a single dataset: 
Figures~\ref{IPM:fig:accuracy:MAP:tau}--\ref{IPM:fig:accuracy:AAP:rs} show the variability of the different methods when applied to different datasets and Figure~\ref{IPM:fig:boxplot-dual} shows the consistency of the different methods on each dataset.
Clearly, the average accuracy and variance of the methods depend on the specific dataset on which they are applied.

Going back to Figures~\ref{IPM:fig:accuracy:MAP:rs}--\ref{IPM:fig:accuracy:MAP:tauAP},
the observations on the basis of these three charts
are quite similar
Indeed, there is no significant statistical difference among the top accuracy methods (although the specific results are not reported here for brevity reasons).
The fact that $\tau_{ap}$ outcome is very similar 
means that method accuracies remain relatively stable even when weighting more the top ranks.
In the 5th chart
in Figure~\ref{IPM:fig:accuracy:AP:rho}
we see that the situation is slightly different for AP, with accuracy predicted by $\rho$.
Correlation values in the AP $\rho$ chart (usually below $0.6$) are clearly lower than in the previous MAP $\rho$ chart (usually above $0.6$): predicting AP
is more difficult than predicting MAP.
Also, here the most accurate method is now AS, which does not only outperform the other ones, but also shows a much smaller variation over the collections. 
Considering statistical significance, AS is indistinguishable from NC-NB, NC-NC,  NC-BC, and SL. 
These five methods are always statistically significantly more accurate than the other ones.

Accuracy of AP prediction measured by $\delta$ is shown in the 6th chart
in Figure~\ref{IPM:fig:accuracy:AP:delta}; 
since smaller differences are better, the best methods are those with the lower values, and the scale on the Y axis is inverted for consistency with the other charts. 
When using AP $\delta$ as the accuracy measure, the top three methods are SNC, AS, and SPO-A: a different set with respect to that in the previous AP $\rho$ chart. Also, they are statistically significant better than all the other ones at the $.01$ level.
Accuracy in predicting AP  seems to be neither necessary nor sufficient to accurately predict MAP, and the $\rho$ and $\delta$ measures do not agree much.  
Although AP $\delta$ measure seems quite a reasonable one in principle, it turns out that  its results show a different behaviour with respect to those exhibited by the other measures. This is probably due to normalization: since some methods try to predict AP values, while others are just interested in the rank, we normalized all the predicted AP values into $[0,1]$,\footnote{\label{IPM:fn:normalization}We used the standard normalization $\frac{\AP - \min(\AP)}{\max(\AP)-\min(\AP)}$, and we also tried another standard normalization ($\frac{\AP - \mbox{mean}(\AP)}{\mbox{std}(\AP)}$) but results looked very  similar.} to be able to compare the predicted AP values obtained from the methods with the real AP values that are within the same range. However,  the normalization that we used, although standard, might have harmed some methods more than others.
Because of the difference between the behaviour of $\delta$ and the other variables, in the following we do not report $\delta$ anymore. 

Accuracy in AP prediction might be considered an artificial measure, but this would be a mistake: it has a practical usefulness, since one might be interested in knowing the effectiveness of a specific system on a specific topic, or in comparing the effectiveness of all systems on specific topics. Moreover, a better AP prediction could be related to a better AAP prediction, as we now discuss.
The last 
two charts in Figures~\ref{IPM:fig:accuracy:AAP:rho} and~\ref{IPM:fig:accuracy:AAP:rs} show accuracy in AAP prediction. 
Whereas, when predicting MAP, rank-based correlations seem a better option as accuracy measure than linear correlation (usually one is interested to know which is the best system), for AAP the choice is less clear (ranking the topics by difficulty seems as interesting as knowing their difficulty values). 
Anyway, for both $\rho$ and $r_s$, AS is again (as for AP $\rho$) the best method on average, although its variation is not lower than the other methods as it was in the AP case. 
When considering statistical significance, however, AS is not more accurate than NC-NB, NC-NC, NC-BB, NC-BC, and SL.

Finally, we remark again the particularly low accuracy of WUCv4.   
This might be due to our failure in 
reproducing its normalization algorithm, which is not fully detailed in the original paper \cite{Wu:2003:MRI:952532.952693}. 
In the following we exclude this method from most of our analyses. 



\section{RQ2: Relations Between Methods}
\label{IPM:sec:relat-among-meth}

\begin{figure}[tb]
  \centering
  \includegraphics[width=.66\linewidth]{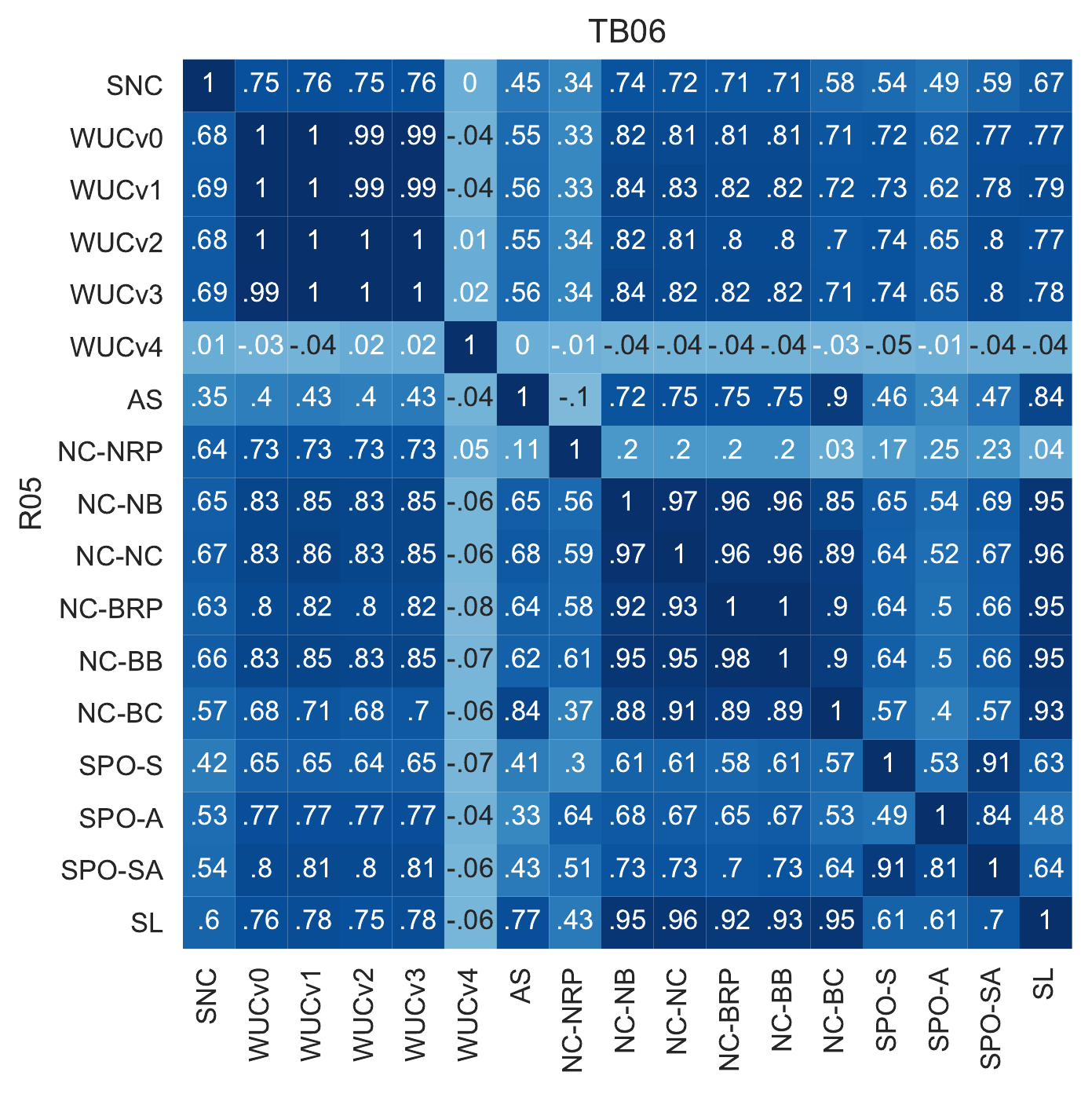}
  \caption{$\rho$ correlations between the AP values predicted by the  methods, on  both the R05 and TB06 datasets.
  \label{IPM:fig:heatmap-corr}}
\end{figure}

\begin{figure}[tb]
  \centering
  \begin{tabular}{@{}c@{}c@{}}
  \includegraphics[width=.5\linewidth]{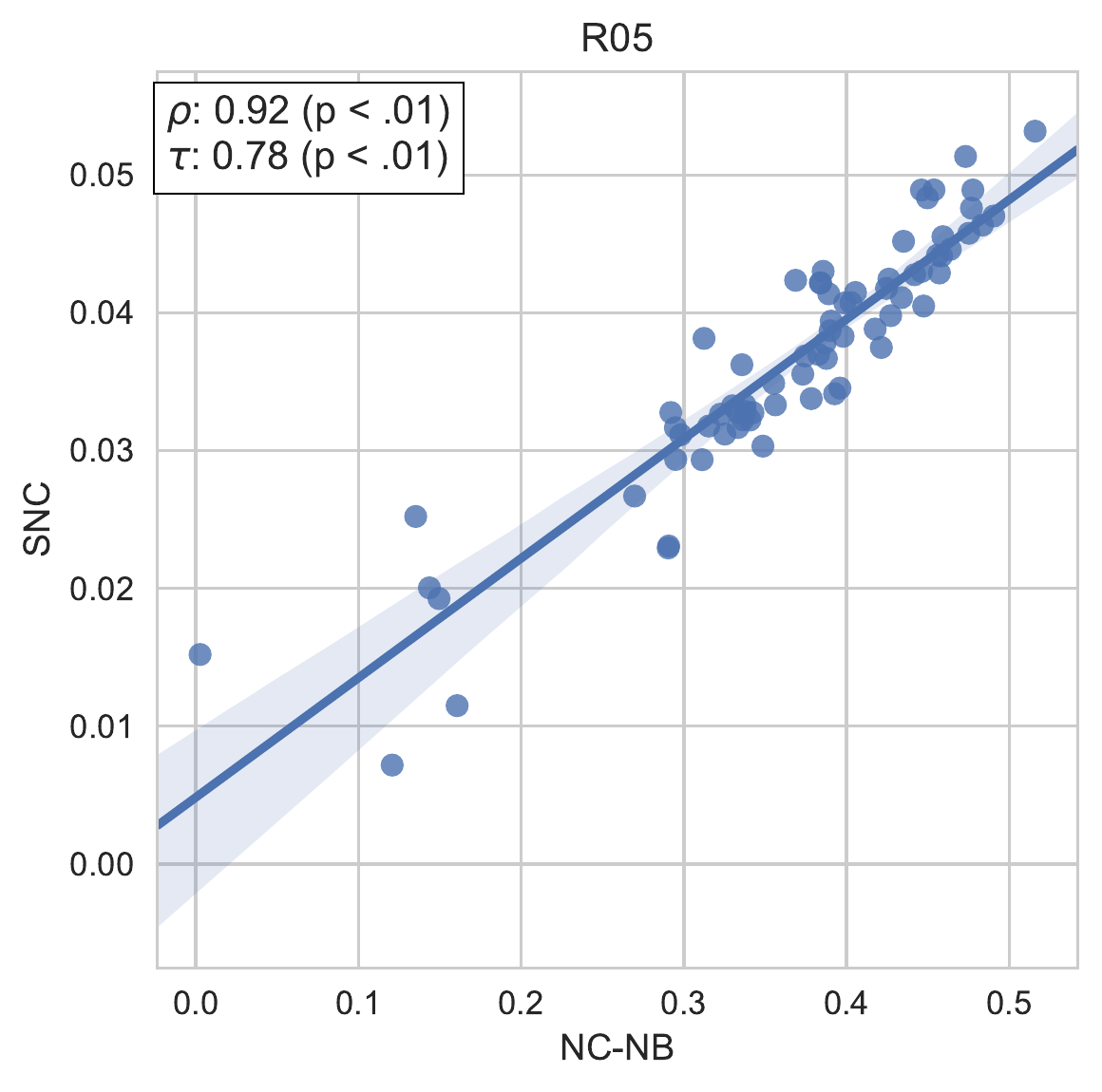}&
  \includegraphics[width=.5\linewidth]{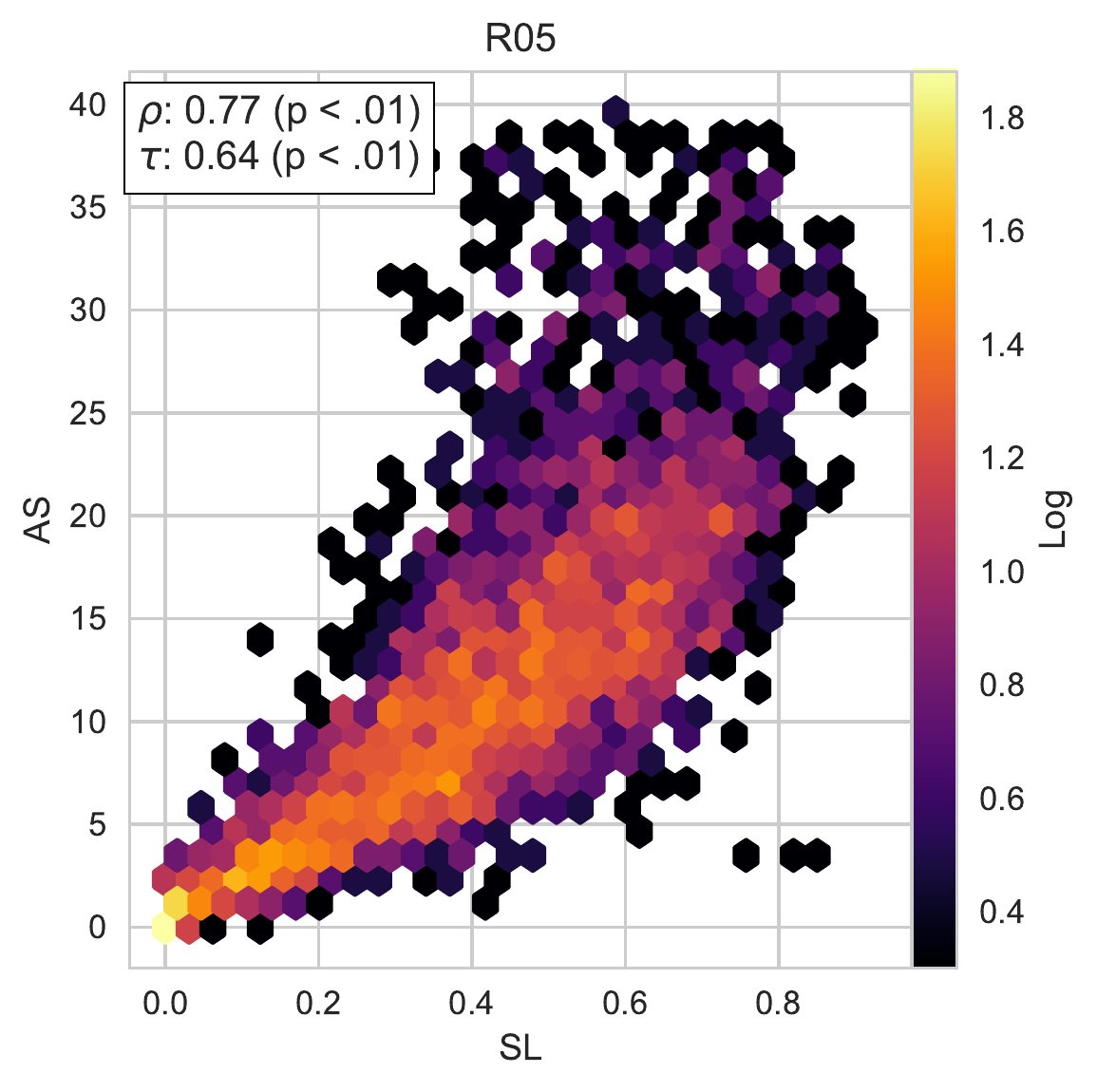}
  \end{tabular}
  \caption{
  Scatterplot of MAP values predicted by NC-NB and SNC; hexbin scatterplot of AP values predicted by SL and AS. 
  \label{IPM:fig:scatter}}
\end{figure}

Having established a common ground consistent with the previous literature, as well as some baselines to compare to, we now focus on RQ2, a question which is more central for our paper. 
Our analysis is aimed at understanding if the various individual methods measure different aspects or are very correlated. The heat-map in Figure~\ref{IPM:fig:heatmap-corr} shows $\rho$ correlations between the AP values predicted by the individual methods, for  each pair of the 17 individual methods, and  for the two datasets R05 (bottom left triangular part) and TB06 (upper right). 
In other terms, the heat-map contains, for each pair of methods $i$, $j$, $\rho(\widehat{\AP}_i,\widehat{\AP}_j)$.
Observe that, given the accuracy measures that we use, the heat-map is symmetric. Thus, we chose to report the results for two datasets into a single heat-map, in which the upper triangular part shows the outcomes on a dataset, and the lower triangular part  on the other one. Presenting two datasets in the same heat-map has also the advantage of clearly emphasizing  graphically that the values are similar across datasets.
We do not show the heat-maps for the other 12 datasets, that are anyway very similar.
%
A quick visual inspection immediately shows three properties:
\begin{itemize}
\item Some methods are highly correlated with each other (the darker cells and triangular/rectangular areas).
\item Conversely, some methods do not seem to correlate well (lighter areas): they are producing quite different predictions.  Some of these methods are not accurate (e.g., WUCv4), but the low correlations among SNC, AS,  SL, and NC-* methods are interesting: these are accurate methods that do not correlate well. This result is promising when considering methods combinations: the low-correlation methods might provide complementary information.
\item The correlations are quite consistent across the two datasets (the triangular areas usually become rectangular across the diagonal, i.e., when considering the two datasets together). This also happens for the other datasets.
\end{itemize}
The  two charts in Figure~\ref{IPM:fig:scatter} provide some further details. The MAP scatterplot (left) is an example of the rather high, though not perfect, correlation between SNC and NC-NB when predicting MAP ($\rho(\widehat{\MAP}_1,\widehat{\MAP}_{9})$ using the numbers in Table~\ref{IPM:tab:methods}). 
The AP hexbin scatterplot\footnote{A normal scatterplot would be too cluttered as it would have, in this case, 74x50=3700 points. 
The hexbin scatterplot bins the points  in hexagonal areas, and shows the density of points by a color gradient (logarithmic in our figure, as shown on the right).} (right) is a more detailed representation of the $.77$ $\rho$ value in the last row, 7th column of the heat-map, and shows that even two of the most accurate methods across measures (AS and SL) do not correlate much in terms of their AP prediction. 
It is also clear that the relation between AS and SL in this case is not linear.

In summary, it is clear that the methods do show some differences.  
Given the low correlations, occurring  even on accurate methods, it makes sense to try to combine them; 

\section{RQ3: Combining the Methods}
\label{IPM:sec:comb-autom-eval}

We now turn to our last research question RQ3, i.e., whether it is possible to combine in an effective way the individual prediction methods, and which is the best approach. 
We test two approaches to methods combination: a first one based on data fusion techniques, and a second one based on machine learning.

\subsection{Oracle Combination}
\label{IPM:sec:oracle-combination}

First, we compute an optimal result in which an oracle selects the best method.  In particular, for each collection, we select the method that achieves higher correlation (we call it the ``Oracle Method''). 
This is not the best that can be done, it is rather a sub-optimal best; indeed, it can be that combining together a subset of the methods will lead to achieve higher correlation values than the oracle. Nevertheless, the oracle method    sets a simple and reasonable upper bound to aim to with the combination of one or more methods.

Figures~\ref{IPM:fig:accuracy:MAP:tau}--\ref{IPM:fig:accuracy:AAP:rs} show, in the rightmost panes, the oracle accuracy of prediction for the eight box-plot charts. 
Analyzing the results, we can make the following observation:
concerning all of AP, MAP, and AAP, the correlation values the oracle achieves are always similar to any other method; furthermore, in cases where the oracle has a higher median correlation value, the differences from the oracle to any other method are not statistically significant.  
This means that any trivial combination of the methods can improve only partially the correlation values obtained by the best method. 
On the other hand, if we compare the oracle with the worst performing method, we can observe that the oracle has always higher correlation values, and this difference is statistically significant. 
Based on these observations we conclude it is  worth trying to combine the methods; indeed, this combination may be used when evaluating automatically a new collection, without having any prior knowledge of which of the methods will perform better.
More details are presented in the following sections.

\subsection{Data Fusion Approaches}
\label{IPM:sec:comb-first-attemps}
In the following subsections we detail the data fusion approaches used in this chapter: we define the setting (Section~\ref{IPM:sec:DFset}), list the algorithms that we use (Section~\ref{IPM:sec:DF:alg}), and present the results (Section~\ref{IPM:sec:DFres}).

\subsubsection{Data Fusion Setting}\label{IPM:sec:DFset}
The basic idea is to define a fusion operation that merges the results of the individual prediction methods. We can sketch the situation using the following three equations:
\begin{align}
\widehat{\MAP}_*&=\mathtt{DF}(\widehat{\MAP}_1,\ldots,\widehat{\MAP}_q) \label{IPM:eq:DF:MAP}\\
\widehat{\AAP}_*&=\mathtt{DF}(\widehat{\AAP}_1,\ldots,\widehat{\AAP}_q) \label{IPM:eq:DF:AAP}\\
\widehat{\AP}_*&=\mathtt{DF}(\widehat{\AP}_1,\ldots,\widehat{\AP}_q). \label{IPM:eq:DF:AP}
\end{align}
Focusing on MAP first (Equation~\eqref{IPM:eq:DF:MAP}), the  MAP vectors $\widehat{\MAP}_i$ predicted by the individual methods are combined into $\widehat{\MAP}_*$ by a data fusion function $\mathtt{DF}$. 
In our experimental setting we have $q=17$ individual methods (though we will use fewer as detailed below in Subsection \ref{IPM:sec:DF:alg}).
Besides working directly on MAP (and, symmetrically, on AAP, Equation~\eqref{IPM:eq:DF:AAP}), we also try the same techniques on AP values (Equation~\eqref{IPM:eq:DF:AP}). 
This makes sense in an attempt to avoid losing information: the predicted AP matrices $\widehat{\AP}_i$ are combined into $\widehat{\AP}_*$. 
The latter is the only possible approach when aiming at predicting AP; conversely when aiming at MAP (and AAP) prediction, two approaches can be used, as one can directly predict MAP and AAP, or predict AP and then average the obtained values.

\subsubsection{Data Fusion Algorithms}\label{IPM:sec:DF:alg}
We use four basic and well known data fusion approaches (some of which are also used 
by some individual methods:

\begin{itemize}
\item \emph{Average function}. Arithmetic average of predicted values $\widehat{\MAP}_i$,  $\widehat{\AAP}_i$ and $\widehat{\AP}_i$. We therefore obtain three data fusion functions: MAP-AVG, AAP-AVG, AP-AVG.

\item \emph{Rank}. Combination of $\widehat{\MAP}_i$,  $\widehat{\AAP}_i$ and $\widehat{\AP}_i$ according to the rank position of each systems: MAP-RP, AAP-RP, AP-RP. In summary, this approach assigns a score based on the rank in which the system occurs.
Let us consider the following  toy example, with two systems $s_i$, $s_j$ and three methods. Suppose the system $s_i$ occurs in the 1st, 2nd, and 3rd position in the ranked list of $\AP$ inferred from the respective methods, and the system $s_j$ occurs in the 2nd, 1st, and 1st position. The score for the system $s_i$ in the fusion list is
$1/\left( 1/1 + 1/2 + 1/3  \right) = 0.55$, and the score for the system  $s_j$ is 
$1/\left( 1/2 + 1/1 + 1/1  \right) = 0.4$.
Thus, in the fusion list, $s_j$ will be ranked before $s_i$ (the lower the score the better) and their respective scores will be 0.4 and 0.55.

\item \emph{Borda count} \cite{Emerson2013}. Predicted values $\widehat{\MAP}_i$,  $\widehat{\AAP}_i$ and $\widehat{\AP}_i$ are treated as expression of preferences, which are then combined based on the rank position of the systems:  MAP-B, AAP-B, AP-B. In summary, the  Borda count assigns a score to each so called candidate considering the reverse proportion of its ranking.
Referring to the previous example, the score for the system $s_i$ in the fusion list is
$ (3-1) + (3-2) + (3-3) = 3$,  and the score for the system  $s_j$ is
$ (3-2) + (3-1) + (3-1) = 5$.
Thus, in the fusion list, $s_j$ will be ranked before $s_i$ (the higher the score the better) and their respective scores will be of 5 and 3 (to be then normalized in [0,1]).

\item \emph{Condorcet} \cite{fishburn1977condorcet}. A majority method of pairwise comparisons between ranked retrieval systems: MAP-C, AAP-C, AP-C. In summary, in the Condorcet method the winner is the candidate that is preferred to any other candidate, when compared to the opponents one at a time according to a scoring system based on preferences.
The Condorcet method works as follows (see the example above): for each pair of systems, in this case just $(s_i, s_j)$, we build a table in which we count for the three methods how many times $s_i$ is preferred to $s_j$ (in this case we count a ``win'' for $s_i$), how many times it happens the opposite (in this case we count a ``lose'' for $s_i$), and how many times there is no preference (in this case we count a ``tie'' for  both systems). 
In our example we have that 
for method 1 $s_i$ is preferred to $s_j$,
for method 2 $s_j$ is preferred to $s_i$, and
for method 3 $s_j$ is preferred to $s_i$.
Thus, $s_i$ will have $\mbox{win}=1$, $\mbox{lose}=2$, $\mbox{tie}=0$, and
      $s_j$ will have $\mbox{win}=2$, $\mbox{lose}=1$, $\mbox{tie}=0$.
Then, we rank the systems according to their wins, lose, and tie values.
\end{itemize}
Since not all the individual methods aim to return AP values on the same scale of the original ones, we again apply the same standard normalization operation to map the predicted AP values into $[0,1]$ (see Footnote~\ref{IPM:fn:normalization}). 
We do not use WUCv4 in the data fusion approaches, given its low accuracy. 
For MAP prediction, we also try removing the four worst individual methods NC-NRP, SPO-S, SPO-A, SPO-SA as found in Section~\ref{IPM:sec:accur-indiv-meth}. 
The obtained data fusion functions are labeled with an ``s'' (for ``selected''): MAP-AVGs, MAP-RPs, MAP-Bs, and MAP-Cs.

\subsubsection{Results}\label{IPM:sec:DFres}


\begin{figure}[tb]
\centering
  \includegraphics[width=.99\linewidth]{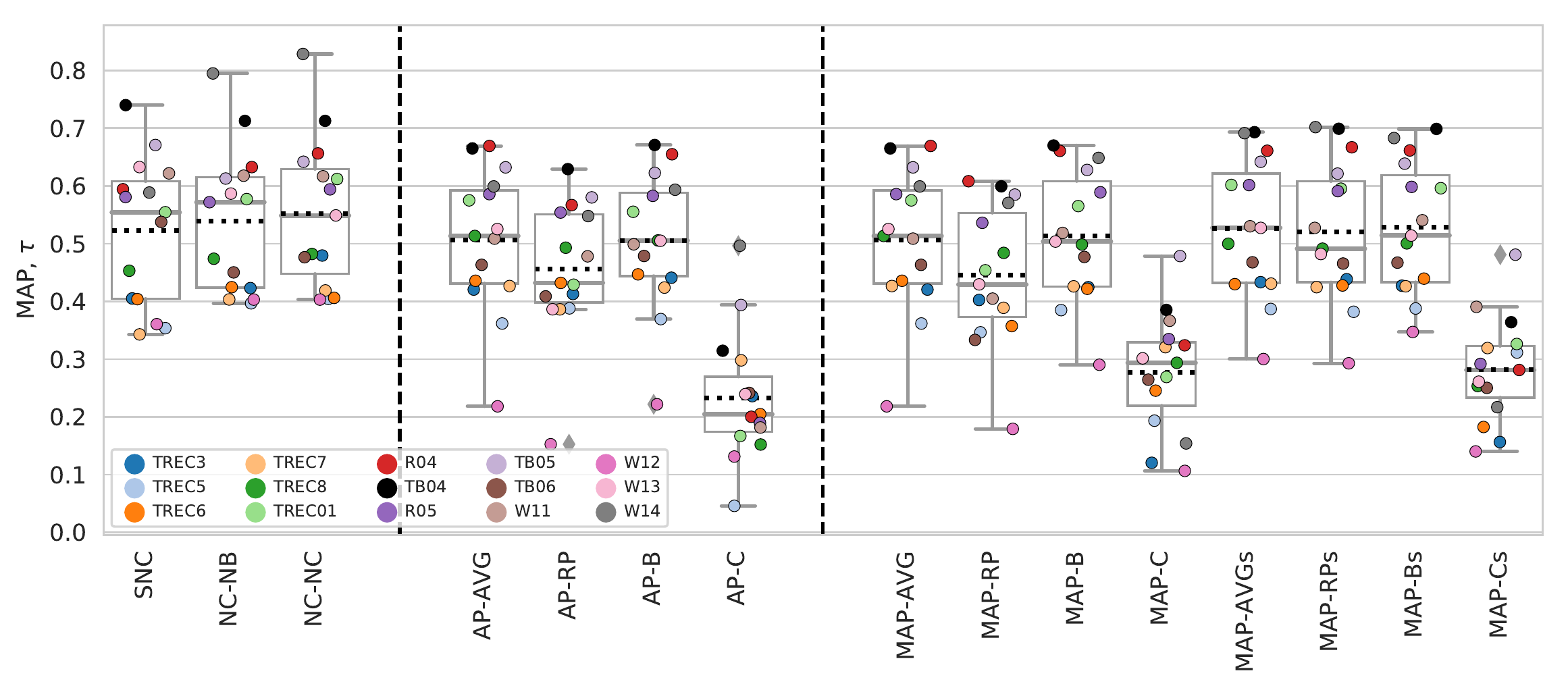}
\vspace*{-2mm}
 \caption{Accuracy of the data fusion approaches, compared with the three best individual methods from Figure~\ref{IPM:fig:accuracy:MAP:tau}:  MAP $\tau$.
 \label{IPM:fig:accuracy-DF-MAP-tau}}
\end{figure}




Figure~\ref{IPM:fig:accuracy-DF-MAP-tau} shows the results for MAP $\tau$ (as in Figure~\ref{IPM:fig:accuracy:MAP:tau}). The leftmost pane shows again the accuracy of the three top individual methods, according to the median  (i.e, these are selected from Figure~\ref{IPM:fig:accuracy:MAP:tau}); the other  two panes show the accuracy of using the data fusion approaches. 
Although combining methods seemed an interesting and promising idea, and despite the use of a spectrum of state of the art data fusion techniques, it is clear that no accuracy improvement is obtained with the data fusion approaches. Instead, usually the combinations by data fusion are less accurate in a statistically significant way than the best individual methods. Results do not change when using the other accuracy measures: all the corresponding  charts to  Figures~\ref{IPM:fig:accuracy:MAP:tau}--\ref{IPM:fig:accuracy:AAP:rs} look very similar to Figure~\ref{IPM:fig:accuracy-DF-MAP-tau}, and therefore we omit them for brevity. 
Anyway we remark that reporting the negative results and failed attempts is important, especially if they are obtained using non trivial techniques; this will prevent future researchers to waste time and resources trying the same ineffective approaches. This position is also supported by other authors: both in general \cite{knight2003negative,Fanelli2012}, and within IR  
 \cite{Ferro:2017:RCI:3035914.3020206, ferro2016increasing}.

 One possible reason for  the ineffectiveness of the data fusion approaches is that they somehow ``go towards the mean'' of the individual methods being combined,  i.e., they produce an outcome that is similar to the average of the individual methods, but they cannot improve the overall effectiveness: the best individual methods are somehow hampered by the other ones, and this negative effect remains also when removing the worst individual methods as we tried with our ``selected'' (``s'')  approaches  (see the end of Section~\ref{IPM:sec:DF:alg}). Indeed, the ``selected'' methods are more accurate than the all inclusive ones, but still not so effective as the individual methods.

Perhaps effectiveness could be improved with more tailored fusion approaches and/or more sophisticate normalization strategies, but these might depend on the method and on the dataset and it does not seem simple nor promising to follow this approach any further. We leave that for future work and we instead turn to a more general approach
, which can be more promising considering the results shown by the oracles in Figures~\ref{IPM:fig:accuracy:MAP:tau}--\ref{IPM:fig:accuracy:AAP:rs}.



\subsection{Machine Learning Approaches}
\label{IPM:sec:comb-second-attempt}
In the following subsections we detail the Machine Learning approaches we use in this chapter: first we discuss the setting (Section~\ref{IPM:sec:MLset}), the algorithms we use (Section~\ref{IPM:sec:MLalg}), and the the ML results (Section~\ref{IPM:sec:MLres}). 
Then, we report on a rather natural technique to be applied in our setting: Transfer Learning (Section~\ref{IPM:sec:TL}).

\subsubsection{Machine Learning Setting}\label{IPM:sec:MLset}


%

Instead of relying on  data fusion approaches, we now turn to the issue of automatically learning  the $\mathtt{DF}$ functions of Equations~\eqref{IPM:eq:DF:MAP}, \eqref{IPM:eq:DF:AAP}, and~\eqref{IPM:eq:DF:AP} relying on historical competitions data.
In such an experimental setting, we consider, in turn, each of the collections  as the test set, while the \lq\lq historical\rq\rq{} training set is composed of all  the instances belonging to the previously released collections, sorted according to their release year (see Table~\ref{IPM:tab:coll}).\footnote{If a collection is released in the same year as the test one, we choose not to consider it.}
To make an example, if we are considering as test collection TREC8, released in 1998, then the training set contains the collections released before 1998: TREC3, TREC5, TREC6, and TREC7.
Observe that, due to our setting, TREC3 can never be considered as test data: since it is the oldest collection, this implies that it would not have any older collection to be used as training data.
%
We generate features 
by running the individual methods (see Table~\ref{IPM:tab:methods}), thus obtaining their predicted values $\{\widehat{\AP}_1,\ldots,\widehat{\AP}_q\}$, while the labels are the actual MAP, AAP, and AP values, which for past data are also considered to be known.


Since predictors and labels are numeric continuous values, we focus on a subset of machine learning algorithms, namely regression algorithms. Although ranking (e.g., to rank systems according to their effectiveness) and  classification (e.g., into easy/difficult topics) are also possible, we leave those to future work. 

To train regression algorithms for estimating MAP the most intuitive choice is to consider a dataset with a row for each run, and a column for every distinct combination of individual metric and topic (plus a column for the label reporting the MAP value of the run).
However, such an approach has two criticalities: first of all, the resulting training set is often too small for machine learning algorithms to be trained effectively, given the number of features. For example, even considering the collection with the largest number of runs (TREC8, see Table \ref{IPM:tab:coll}), the samples in the training set would be just 129 (equal to the number of runs), while the number of columns would be 851 (50 topics $\times$ 17 methods, plus the MAP label).  Secondly, this kind of representation is strictly tied to the number, kind, and arrangement of topics. Therefore, training samples of a collection may only be combined with other collections sharing the same format of topics.
Equal considerations apply when predicting AAP. 

To overcome these limitations we focus on predicting AP values instead of MAP or AAP. 
By doing so, the dataset has a row for each distinct combination of run and topic, and a column for each individual method (plus a column for the label, which is the AP value of the run on the topic).
This kind of feature representation has three important characteristics: first of all, it is fine-grained, since it includes all the estimated values of each individual method; second, for each collection the training set is much larger than previous proposal (on TREC8 we will have 129 runs $\times$ 50 topics = 6450 rows, and just 18 columns); third, its dimensionality and column arrangement is totally independent of the format of topics,
therefore training samples of different collections can be combined together by simply stacking the rows.

%
By relying on the results of Figures~\ref{IPM:fig:accuracy:MAP:tau}--\ref{IPM:fig:boxplot-dual},
we removed feature WUCv4 given its consistently poor performance (as observed in Section \ref{IPM:sec:accur-indiv-meth}), 
therefore considering 16 methods, instead of the original 17.




\subsubsection{Machine Learning Algorithms}\label{IPM:sec:MLalg}

We tested several machine learning algorithms, all implemented 
using the following Python 3.5 libraries: Scikit-learn,\footnote{\url{https://scikit-learn.org/stable/}} and
Keras.\footnote{\url{https://keras.io/}}
We report the results for twelve of them:
\begin{itemize}
\item \emph{LinearRegression} \cite{witten2016data} (LR in the following), the standard linear regression technique.
\item \emph{RandomForest} \cite{Breiman2001} (RF in the following), an ensemble learning method that operates by constructing a set of decision trees during training, and outputting the average prediction of the trees when a new instance has to be predicted.

\item \emph{Ridge Regression} \cite{hoerl1970ridge} (RIDGE in the following), a regression algorithm that implements L2 regularization, and uses as objective function the minimization of the sum of square of coefficients.

\item \emph{Bayesian Ridge Regression} \cite{park2008bayesian} (BAYRIDGE in the following), a regression algorithm that uses Bayesian modeling and spherical Gaussian priors.

\item \emph{Lasso Regression} \cite{tibshirani1996regression} (LASSO in the following), a regression algorithm that implements L1 regularization, and uses as objective function the minimization of the sum of absolute value of coefficients.

%
%
\item{Neural Network} (NN-\emph{epochs}-\emph{loss} in the following): a neural network regression model with a Sequential architecture composed of two dense connected layers:
the first layer with 16 neurons, initialization function ``uniform'' and activation function ``ReLu''; the second one with dimension one, initialization function ``normal'', no activation function. We trained the model using ``Adam'' as optimizer, ``MSE'' and ``MAE'' as Loss functions (number of epoch set to 10 and 100).

\item 
%
%
\emph{Deeper Neural network} (DNN in the following), a neural network regression model with a sequential architecture composed of three dense connected layers:
the first layer with 32 neurons, initialization function ``uniform'' and activation function ``ReLu''; the second one with 16 neurons, initialization function ``uniform'' and activation function ``ReLu''; the last layer with
dimension one, initialization function ``normal'', no activation function. We trained the model using ``Adam'' as optimizer, ``MSE'' as Loss functions (number of epoch set to 10). We did some experimentation with 100 epochs, but results where worst than with 10.

\item \emph{SVM}, which is the Python implementation of the library for Support Vector Machines \cite{chang2011libsvm}, that are also capable of performing support vector regression. Specifically, we tested two nonlinear kernels: \emph{PolyKernel} (SVM-P in the following) and \emph{RBFKernel} (SVM-E in the following), both within the \emph{nu-SVR} SVM type and with the normalization step active.
\item \emph{Learning to Rank} (LtR in the following), 
which is typically used in information retrieval to predict the correct order of retrieved documents \cite{Liu:2009}. In this chapter, we use it to rank the systems of various competitions. Specifically, we rely on Python's \emph{XGBRegressor} package with a \emph{rank:pairwise} objective.
\end{itemize}

To avoid  over-fitting phenomena, as well as to ease reproducibility, we did not fine-tune the parameters of the algorithms, but instead relied on their default values, with the exception of \emph{XGBRegressor}, 
that typically requires a tuning phase to get the best results: specifically, to select the most appropriate choices for the model parameters, we performed a tuning phase, relying on \texttt{GridSearchCV} method from \texttt{Scikit-learn} library. As its name suggests, it performs a grid search in a given parameter space, returning their best combination, according to the performance exhibited by the trained model. Such score has been evaluated through 4-fold cross-validation on the training data. Table~\ref{IPM:tab:param} reports the tuned parameters, together with their search space and optimal values.

\begin{table}[tbp]
\caption{Search space and optimal values of the parameters used in \texttt{XGBRegressor}.\label{IPM:tab:param}}
\begin{adjustbox}{max width=\textwidth}
\begin{tabular}{llS[table-format=2.3]}
\toprule
Parameter  & Search    & \multicolumn{1}{r}{\mbox{Optimal}} \\
 Name      & Space     & \multicolumn{1}{r}{\mbox{Value}}  \\
\midrule
colsample\_bytree & $0.5, 0.7, 0.8, 1$    & 0.5        \\  
gamma & $0, 2, 5, 7, 10, 12, 15$  & 0        \\
learning\_rate & $0.001, 0.005, 0.01, 0.02, 0.04, 0.06, 0.08, 0.1, 0.2, 0.4$    & 0.005       \\  
max\_depth & $1, 2, 4, 8, 16, 32, 64, 128$    & 64        \\  
min\_child\_weight & $1, 2, 4, 8$    & 4       \\  
n\_estimators & $25, 50, 100, 200, 400$    & 50       \\  
subsample & $0.5, 0.7, 0.8, 1$    & 1        \\  
\bottomrule
\end{tabular}
\end{adjustbox}
\end{table}

\subsubsection{Results}\label{IPM:sec:MLres}

  

\begin{figure}[p]
\centering
  \includegraphics[width=.7\linewidth]{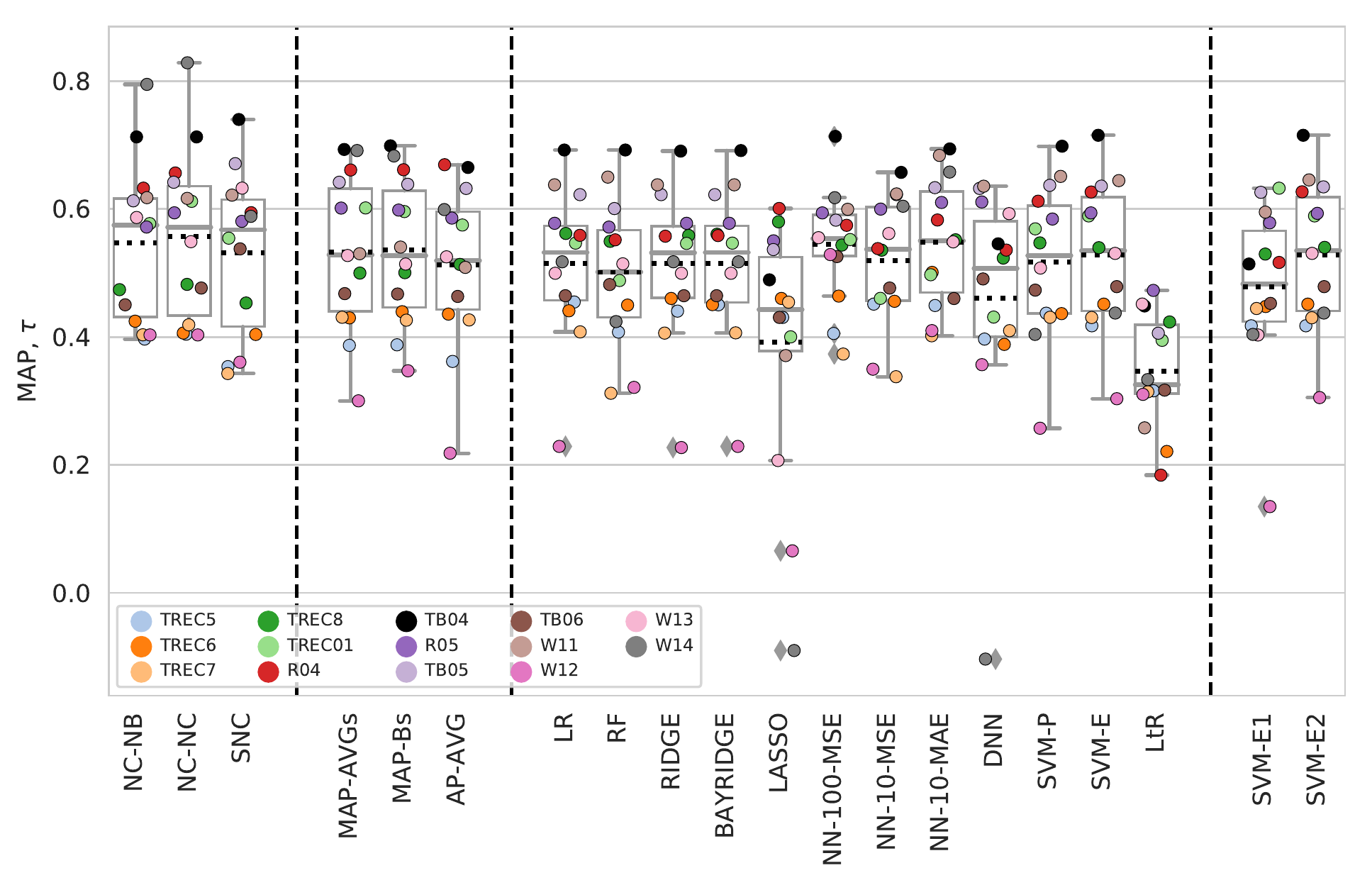}
 \caption{Accuracy of  machine learning approaches, 
 top three individual methods, and top three data fusion approaches: MAP $\tau$.
 \label{IPM:fig:accuracy-ML-MAP-tau}}
\end{figure}

\begin{figure}[p]
\centering
  \includegraphics[width=.7\linewidth]{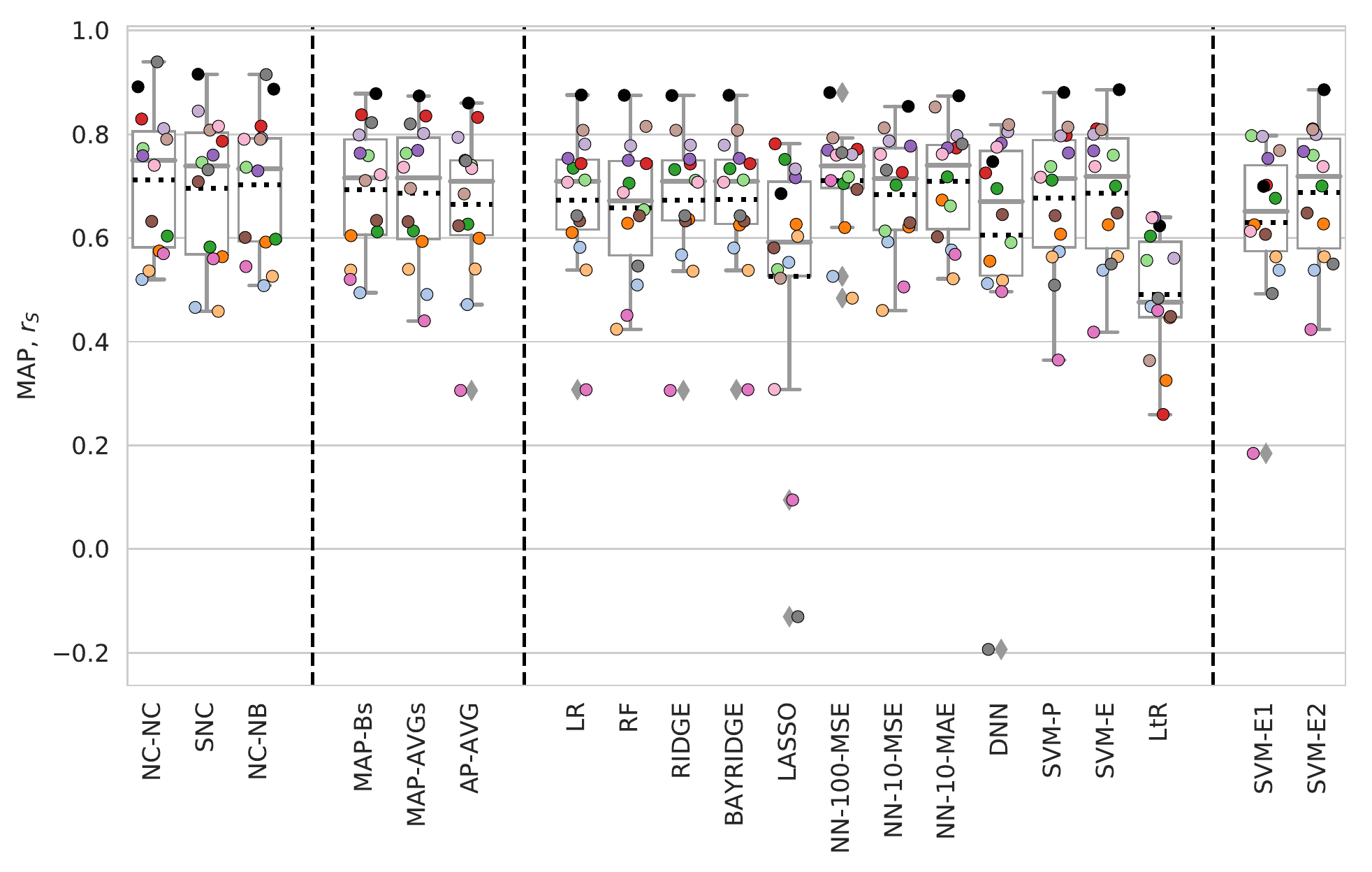} 
 \caption{Accuracy of  machine learning approaches, 
 top three individual methods, and top three data fusion approaches: MAP $r_s$.
 \label{IPM:fig:accuracy-ML-MAP-rs}}
\end{figure}

\begin{figure}[p]
\centering
  \includegraphics[width=.7\linewidth]{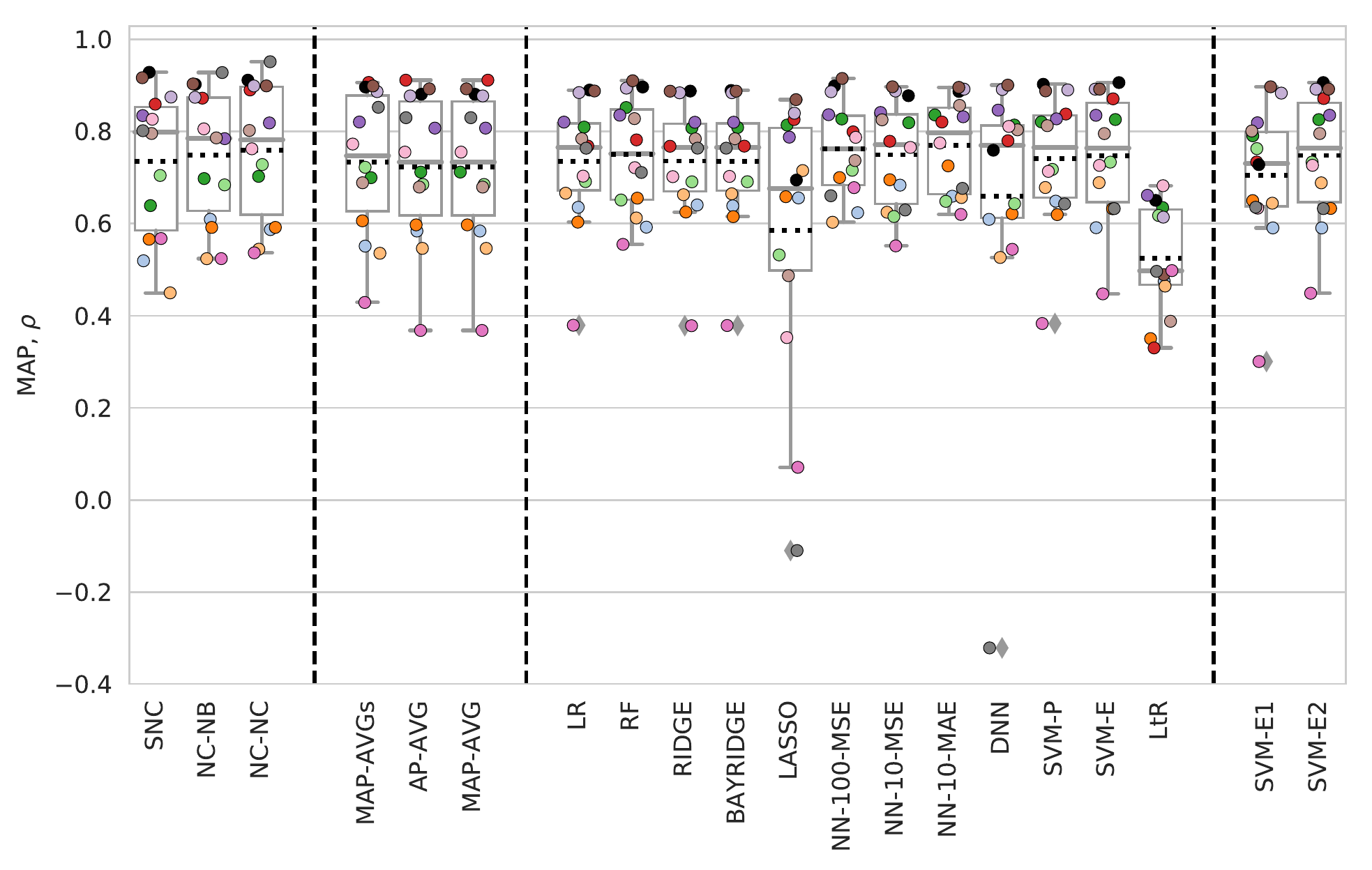}  
 \caption{Accuracy of  machine learning approaches, 
 top three individual methods, and top three data fusion approaches: MAP $\rho$.
 \label{IPM:fig:accuracy-ML-MAP-rho}}
\end{figure}

\begin{figure}[p]
\centering
  \includegraphics[width=.7\linewidth]{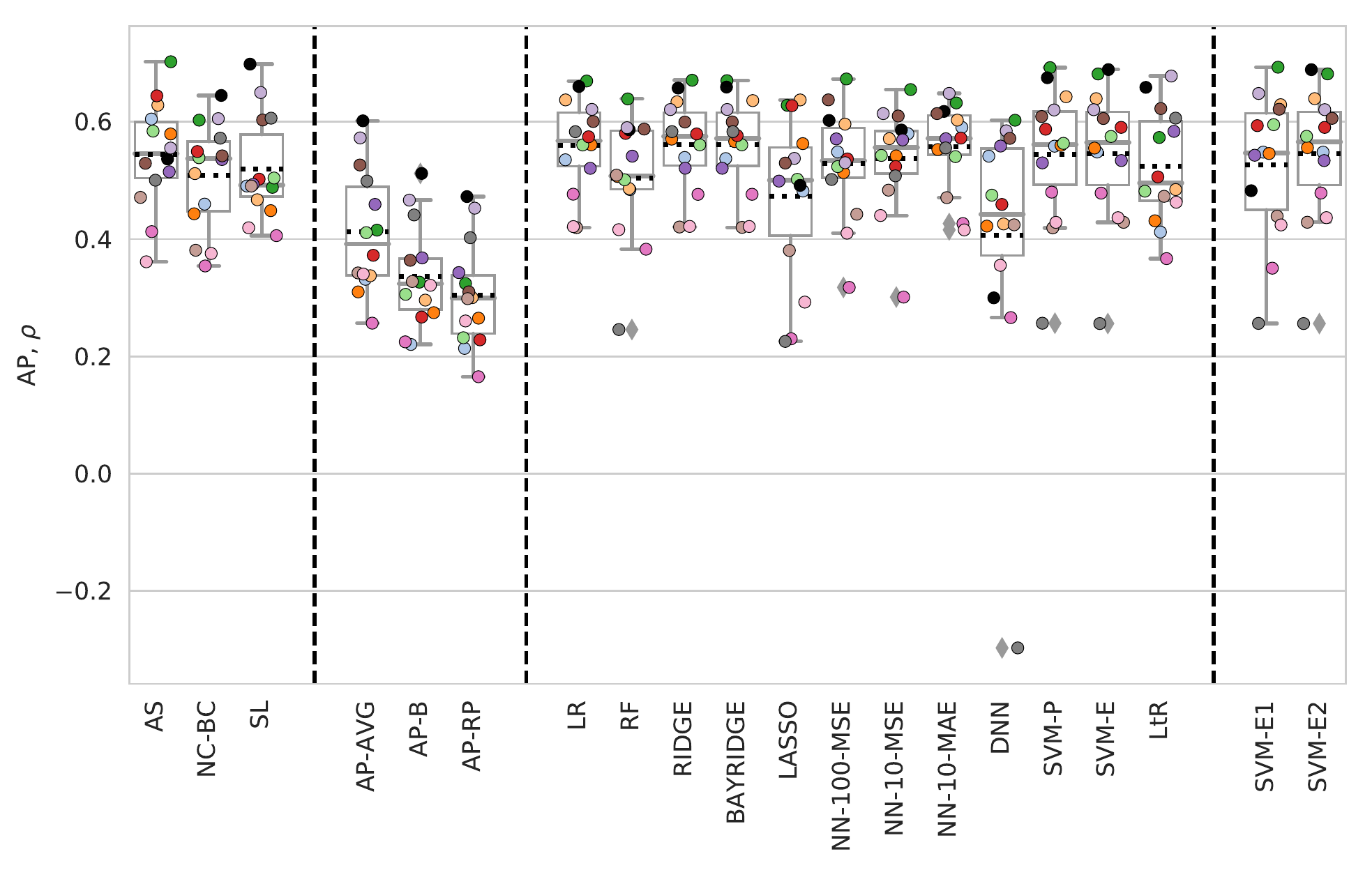}
 \caption{Accuracy of  machine learning approaches, 
 top three individual methods, and top three data fusion approaches: AP $\rho$.
 \label{IPM:fig:accuracy-ML-AP-rho}}
\end{figure}

\begin{figure}[p]
\centering
  \includegraphics[width=.7\linewidth]{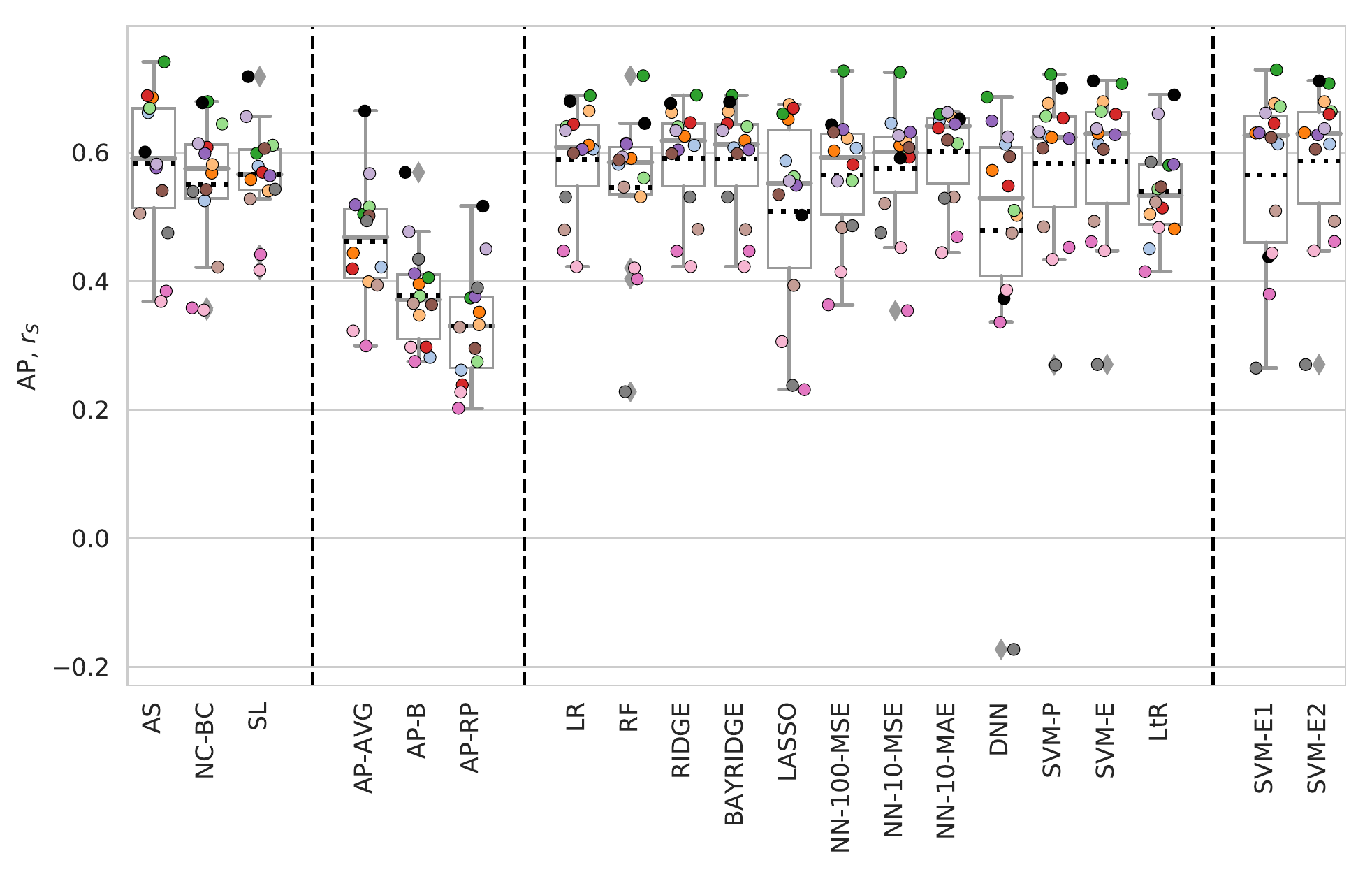}
 \caption{Accuracy of  machine learning approaches, 
 top three individual methods, and top three data fusion approaches: AP $r_s$.
 \label{IPM:fig:accuracy-ML-AP-rs}}
\end{figure}

\begin{figure}[tbp]
\centering
\includegraphics[width=.7\linewidth]{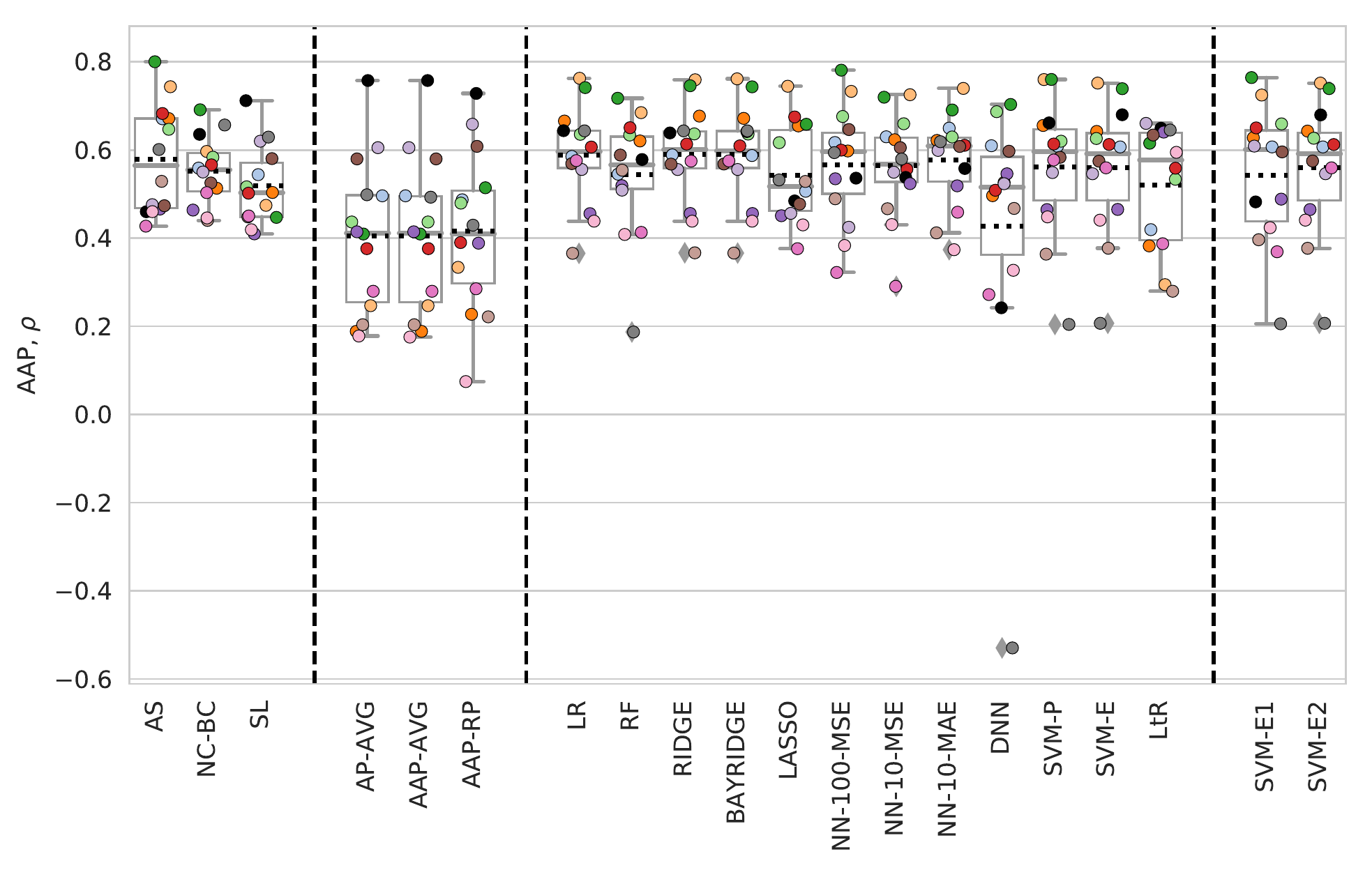}
 \caption{Accuracy of  machine learning approaches, 
 top three individual methods, and top three data fusion approaches: AAP $\rho$.
 \label{IPM:fig:accuracy-ML-AAP-rho}}
\end{figure}


Results are reported in Figures~\ref{IPM:fig:accuracy-ML-MAP-tau}--\ref{IPM:fig:accuracy-ML-AAP-rho}. The two leftmost panes show again the best individual methods (1st pane) and the best data fusion approaches (2nd pane). The accuracy of the  twelve machine learning approaches is presented next (3rd pane), as well as two variants discussed later (4th pane).
Differently from previous charts, those in this figure do not show the data point for the TREC3 dataset, 
the reason being that the machine learning techniques need at least a collection to be used as a training set.
Thus the box-plots in the first two panes are slightly different from those presented in the previous figures,
since there is one point less. 

We can draw several conclusions. 
Looking at the median values, SVM-E is consistently the most effective machine learning approach; SVM-E is never worse than the best data fusion technique. When using AP $\rho$ and $r_s$ (as well as for $\tau$, not shown here) as accuracy measures, SVM-E is the best possible option, as it outperforms the most effective individual methods, and the difference with NC-BC and SL is statistically significant at the $.05$ level (also note that data fusion methods are particularly ineffective in this case). 
SVM-E variation (measured as interquartile range) on AP $r_s$ is also much smaller than the variation on the best individual methods. 
Finally, SVM-E is also the best possible option for AAP $\rho$ ($r_s$ and $\tau$ are similar).
Moreover, 
the machine learning approaches are trained on AP values (the reason being the small amount of data available that makes it ineffective to work on MAP), whereas the individual methods are aimed at MAP prediction. This different objective reduces the effectiveness of learning algorithms in MAP prediction. While of course MAP prediction can be considered an interesting final aim, the fact that machine learning approaches outperform the best individual methods on AP is encouraging, also taking into account that the generality of the machine learning approach can allow to include same MAP tailoring as well. 

The variants shown in the rightmost panes of these charts 
are obtained by learning on the single most similar collection (called SVM-E1) and on the two most similar ones (called SVM-E2). Learning on the three most similar datasets (SVM-E3, not shown) is indistinguishable from SVM-E2.
As the similarity measure we use the average Kolmogorov-Smirnov distance \cite{massey1951kolmogorov} between the distributions of AP values predicted by the individual methods. 
The underlying hypothesis is that the more similar the training data to the object of prediction, the smaller the training set needed, and the higher the accuracy results.
Note that SVM-E1 and SVM-E2 are as effective as SVM-E. Moreover, SVM-E1 is more efficient than SVM-E as one does not need to train the regression SVM on many collections but can select just the most similar ones, thus decreasing computation times. 
Nevertheless, variation is usually larger on SVM-E2 than SVM-E: training on more datasets allows learning a more stable model. 

However, a more careful inspection of the charts 
reveals that both the data fusion and the machine learning approaches perform particularly badly on specific datasets, namely the Web track collections. This is even more manifest when looking at the AP box-plots, where W11, W12, and W13  are consistently among those with lowest accuracy. These collections, as remarked in Section~\ref{IPM:sec:datasets}, feature non-binary relevance judgements: it might be that the binarization that we performed to compute AP introduced too much noise (or, in general, due to other irregularities in the judgements of such collections).
We therefore performed the same analysis focusing on non-Web collections only, i.e.,  those with binary relevance. 
Results show (not reported here) that the top three individual methods are slightly different from Figures~\ref{IPM:fig:accuracy-ML-MAP-tau} and~\ref{IPM:fig:accuracy-ML-MAP-rho}, whereas the top three data fusion approaches do not change. When accuracy is measured with $\tau$, as well as $r_s$ (not shown), SVM-E  approach shows the same accuracy of the top individual methods. 

To conclude, we make two final remarks. When evaluating ML results, the well-known  \emph{cross-validation} technique is often used: a dataset is split into complementary  subsets, and the tested machine learning approach is learnt and evaluated multiple times using different partitions \cite{kohavi1995study}.
However, due to the intrinsic definition of our problem, we cannot rely on such a technique.
In fact, we cannot do cross-validation using each collection as a whole, since we select, as the test data, a collection that has been released on a specific year and then we use as training data all the collections that have been released over the previous years (see also Section~\ref{IPM:sec:MLset}). In other words, given a specific year we can only treat the collection of that year as testing set and past collections as training set.
Furthermore, we cannot perform cross-validation by  selecting/removing some individual AP scores (i.e., $\langle$system, topic$\rangle$ pairs) from the training and test set since, in order to test the effectiveness of our ML setting, we need all the AP scores for a given collection. Thus, we can only treat a collection as a monolithic item, which can not be split into sub parts.


The second remark is that a natural extension of this chapter would be to provide techniques and guidelines on which combination approach is the most effective giving some particular characteristics  of a dataset. We performed some preliminary analysis and tried to find patterns and correlations between the AP / MAP / AAP scores of a given collection and some of its most intuitive and straightforward features, such as the number of systems, the number of topics, the average scores of systems and topics, and so on. However, we failed to find any of such correlations.
We believe that a sound and complete analysis of the correlation between the collection features and the scores would require another paper to be investigated properly; thus, we leave such an analysis for future work.

\subsubsection{Transfer Learning}
\label{IPM:sec:TL}



In this setting, Transfer Learning (TL) seems a natural and promising direction to explore. 
In fact, TL is used when training and test data are not drawn from the same feature space and/or do not have the same distribution. Indeed, when the distribution changes, the results of a predictive learner can be degraded. 
Our research task falls in this context. In fact, samples of the past collections (training data) and those of the current collection (test data) are collected under different conditions, thus have different distribution. 
Moreover, TL has been proven to be an effective methodology in a somehow related setting: the vertical selection for web search \cite{Arguello:2010:VSP:1835449.1835564}.

Thus, as a final result of this chapter we attempt to study this idea and we report some results on six datasets: TREC3, TREC5,  TREC6, TREC7, TREC8, and TREC01.

We try TL on M5P, RF, SVM-P, SVM-E.
We investigate a specific transfer learning algorithm called ``Maximum Independence Domain Adaptation'' (MIDA) \cite{yan_2018TF}, which achieves state-of-the-art results in several contexts. 
%
%
We learn a model on a single dataset only, and transfer it to another one, for two different reasons:
first, the aggregation of multiple train collections into a sort of big training collection is not trivial, and might be wrong in our TL setting; the aim of TL algorithms is to transfer knowledge between different models/dataset, leveraging their differences; thus the fusion of different models/dataset should be avoided.
Second, all TL algorithms, including MIDA, present a high computational complexity, and algorithm convergence issues, that prevent them to run on a large amount of data.

We did some experiments with the TL algorithm TCA \cite{5640675}, but results where almost indistinguishable from the ones obtained with MIDA.
 We leave to future work experiments on learning on more than one dataset, and on using different TL algorithms.


 \begin{figure}[tb]
  \centering
  \begin{tabular}{@{}c@{}c@{}c@{}}
    \includegraphics[width=.33\linewidth]{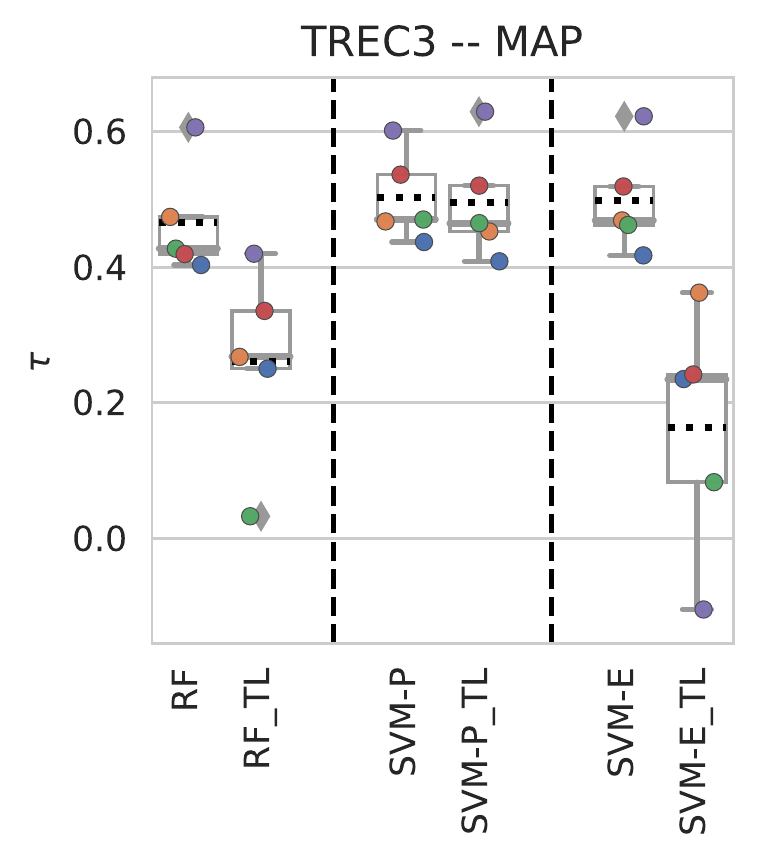}&
    \includegraphics[width=.33\linewidth]{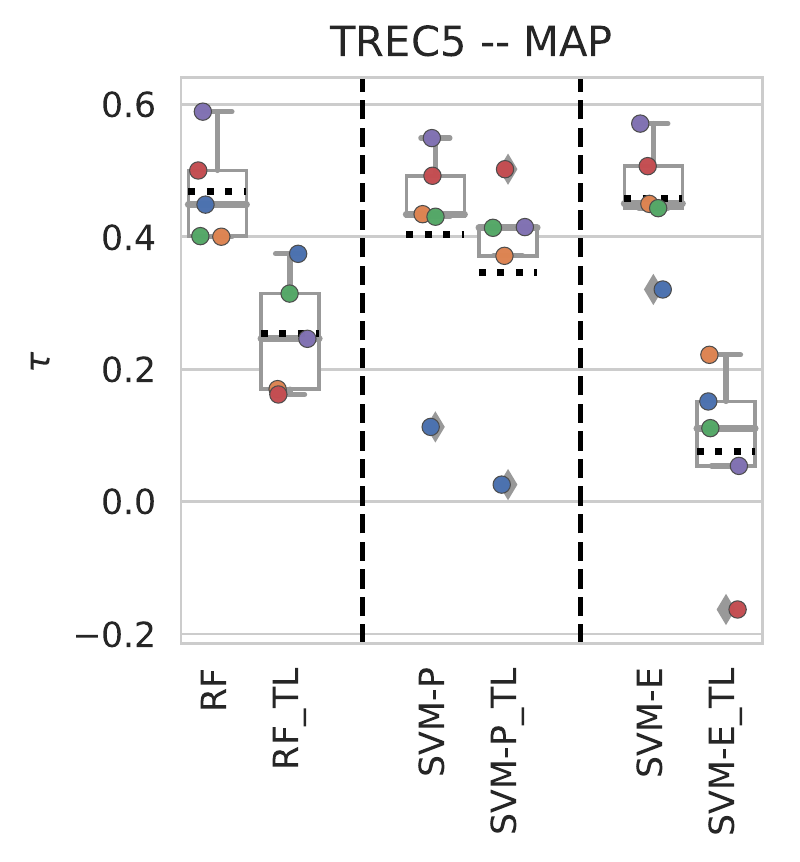}&
    \includegraphics[width=.33\linewidth]{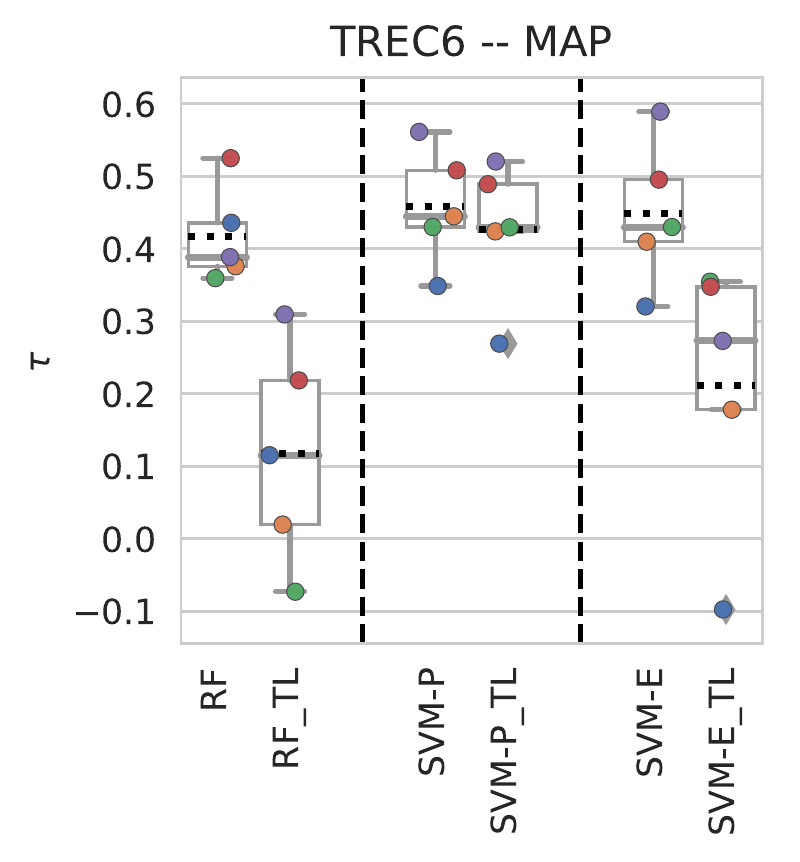}\\
    \includegraphics[width=.33\linewidth]{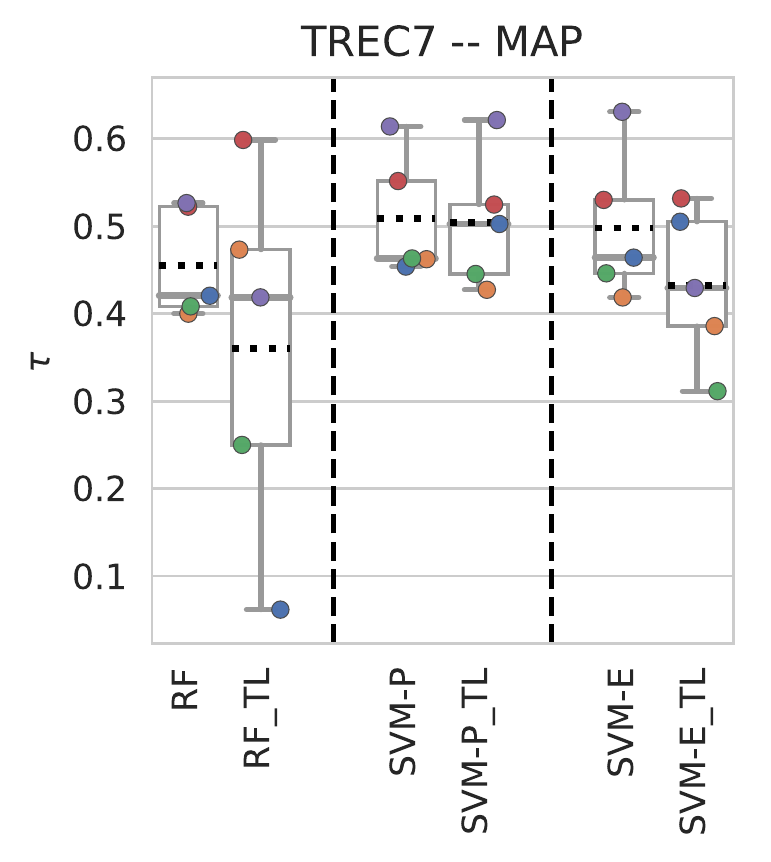}&
    \includegraphics[width=.33\linewidth]{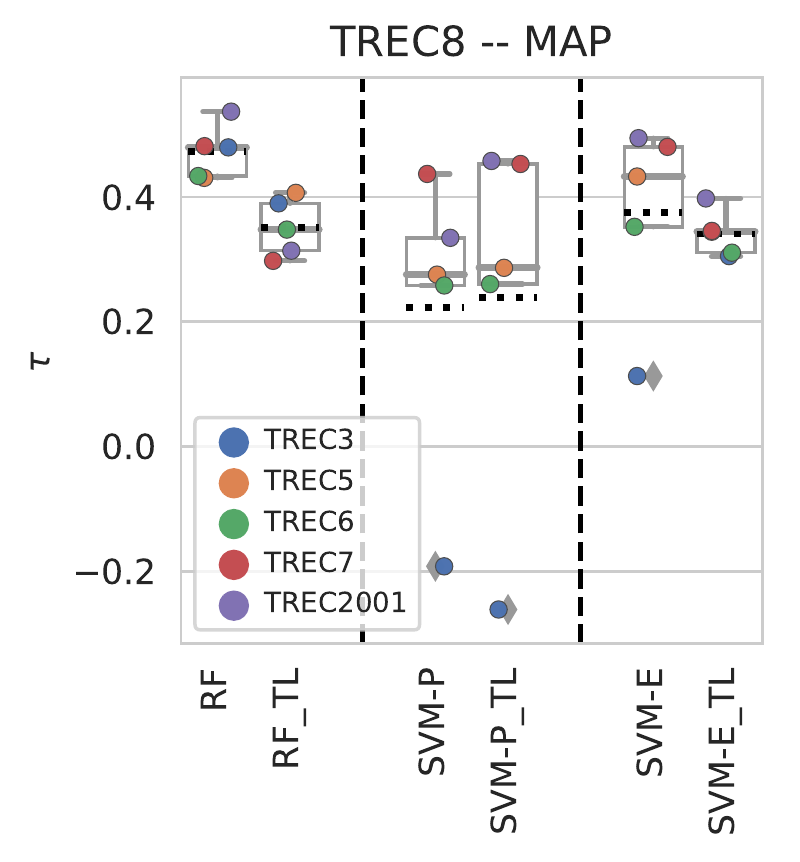}&
    \includegraphics[width=.33\linewidth]{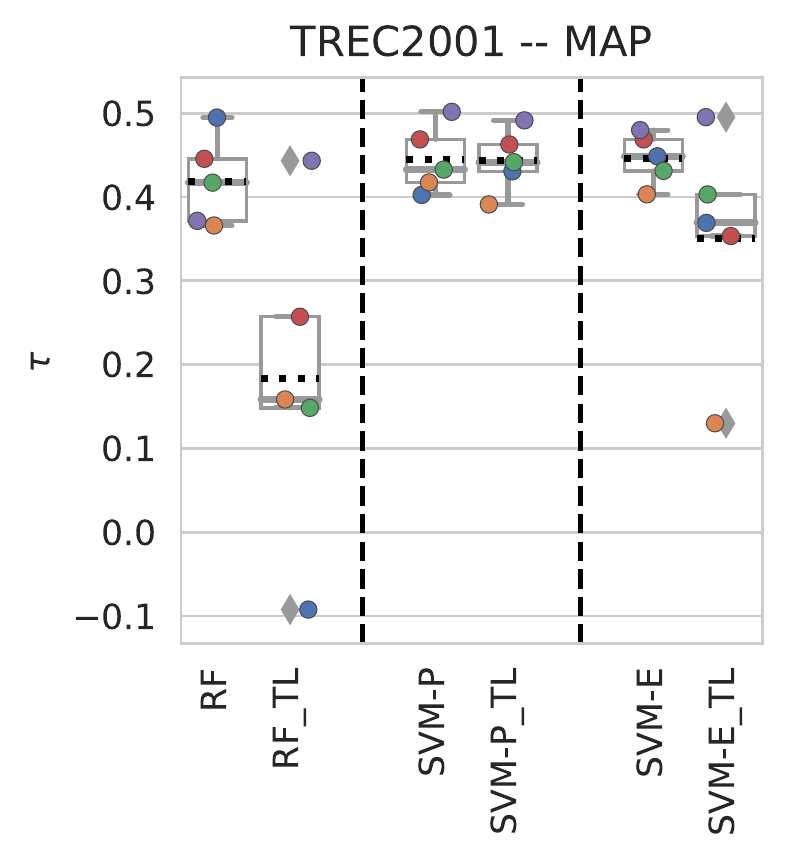}
    \end{tabular}
    \caption{ Accuracy of transfer learning approaches: MAP $\tau$. 
    }
  \label{IPM:fig:TL:MAP}
\end{figure}





Figure~\ref{IPM:fig:TL:MAP} compares TL to the classical learning methods on the six datasets. 
The comparison is for MAP ($\tau$) only, as the other measures show a similar behaviour and thus are not reported here. The charts show pairs of box-plots (one pair per panel): for each pair, the box-plot on the left shows some of the classical non-TL methods reported in previous figures, but when training on a single dataset; we report RF, SVM-P, and SVM-E: we include the former to use a tree based method, and the SVM variants because are the most effective in the non TL scenario. 
The box-plot on the right of each panel is the corresponding TL.  Perhaps surprisingly, in general TL is not effective; rather, it systematically and significantly harms RF, and SVM-E. The only case in which TL is competitive with the non-TL counterpart is SVM-P. 

\section{Conclusions}\label{IPM:sec:concl-future-devel} 

We presented the results of a battery of experiments and of a rather extensive analysis over 17 prediction methods, 14 TREC collections, 15 accuracy measures (obtained by combining the three MAP, AP, and AAP with the five $\rho$, $\tau$, $r_s$, $\delta$, and $\tau_{ap}$), four data fusion approaches (plus variants), and twelve machine learning algorithms (plus variants). 
We have provided a fourfold contribution:
(i) Figures~\ref{IPM:fig:accuracy:MAP:tau}--\ref{IPM:fig:accuracy:AAP:rs} are a solid account of individual method effectiveness across different collections; 
(ii) the analysis of Section~\ref{IPM:sec:relat-among-meth} highlights some interesting, potentially useful, and so far unnoticed relationships between the individual methods; 
(iii) the negative results on the two promising techniques of data fusion and transfer learning techniques, although not useful in practice, will avoid other researchers to perform the same attempts; 
%
and 
(iv) the results on method combinations by means of machine learning algorithms provide a practical methodology for the researcher that wants to run an effectiveness evaluation without human relevance assessments. 
Overall, our results show that the combination of  the methods for effectiveness evaluation without relevance assessments is a viable approach, is effective and robust when using off-the-shelf, state-of-the-art machine learning algorithms, and provides a useful framework for future improvements. In particular, despite being sometimes outperformed by the best single method, the combination of the methods for evaluation without relevance assessments via machine learning is more effective than a random selection of the individual methods, and less risky in the real case scenario, where neither the knowledge on the performance of the individual methods nor the performance of the participating runs is known a-priori.

\part{On Crowdsourcing Relevance judgements and The Effect of The judgement Scale}\label{part:cs}

\chapter{Introduction and Background}
This chapter is structured as follows:
Section~\ref{part:cs:intro} introduces,
Section~\ref{part:cs:binaryvsgraded} discusses binary and graded relevance scales,
Section~\ref{part:cs:continuous} discusses continuous relevance scales, and
Section~\ref{part:cs:relevancedimensions} discusses the dimensions of relevance and biases.

\section{Introduction} \label{part:cs:intro}
Over the last few years, the increasing size of document collections created the need to scale the gathering of relevance judgements. For this reason, crowdsourcing has become a consolidated methodology to create relevance labels for query-document pairs given a judgement pool. In order to produce crowdsourced relevance labels at a quality level comparable with that of expert assessors a number of techniques have been proposed and evaluated in literature.
A common approach is to collect relevance judgements for the same query-document pair from different crowd workers and to aggregate them together \cite{alonso,kazai,Venanzi:2014:CBA:2566486.2567989} thus allowing to remove noise in the labels. Past research also showed that asking for a justification for the judgements \cite{leasehcomp} and that limiting the time to judge \cite{time} can increase crowdsourced relevance judgement quality. 
In our work we leverage crowdsourcing to collect relevance judgements over different scales and build on top of existing crowdsourcing research in terms of quality checks and HIT design best practices.


\section{Binary vs Graded Relevance judgements}\label{part:cs:binaryvsgraded}
Relevance is a central concept in IR \cite{ASI:ASI20682} evaluation; IR systems are usually evaluated using test collections, which are composed of 
(i) a collection of documents,
(ii) a set of queries (called topics),
and (iii) a set of relevance assessment for each (topic, document) pair in a pooled set of documents; such assessments are made by human experts  according to an ordinal scale, which is usually binary.

Test collections can be created by means of a competition: participating systems return a ranked list of $n$ documents (usually 1000), which are then used to compute the judging pool (e.g., the top 100 documents returned by each system, for each topic).
The documents in the pool are the ones assessed by human experts.
The produced relevance judgements are used together with the ranked lists produced by the systems to compute an effectiveness metric for each (system, topic) pair; a commonly used metric is Average Precision (AP).
In order to provide a final rank of participant systems, the effectiveness scores are averaged over the set of topics; for example, the average of AP scores originates Mean AP (MAP).

Historically, relevance judgements were made by assessing whether a document is relevant or not to a topic; then, based on the observation that more than two levels might be needed, a set of novel metrics which incorporate multiple levels relevance scales were developed, such as, for example, Normalized Discounted Cumulative Gain (NDCG) \cite{Jarvelin:2002:CGE:582415.582418}, Expected Reciprocal Rank (ERR) \cite{Chapelle:2009:ERR:1645953.1646033}, and Q-Measure \cite{Sakai:2007:RIR:1232380.1232399}.

Concerning the ideal number of relevance levels to be used, over the years many proposal have been made: a three-level scale was used in TREC-Terabyte Track \cite{clarke2004overview},
a six-level scale was used in TREC-Web Track \cite{collins2015trec},
a seven-levels scale was proposed by \citet{tang1999towards} when studying evaluation of bibliographic records by students, using relevance scales with a range of levels from two to eleven.
Then, \citet{Maddalena:2017:CRM:3026478.3002172} proposed an 
unbounded scale based on Magnitude Estimation, which is described in the following. 
\citet{10.1007/978-3-540-78646-7_5,10.1145/2484028.2484094} discussed pairwise preference judgements.
Despite the many different approaches on relevance scales, the question of how many relevance levels should we use is far from answered. 
In our work we present a comprehensive study on the effects of relevance scales on IR evaluation proposing a fine-grained scale at 100 levels that incorporates the benefits of both bounded scales as well as the flexibility of an unbounded scale.

\section{Continuous Relevance and Magnitude Estimation}\label{part:cs:continuous}
We provide some more details on the use of ME since we compare against it in the following, and since our experiments rely on reassessing documents on a 100-level scale following the same experimental setting.
ME is psychophysical technique used to measure the intensity of sensations \cite{Maddalena:2017:CRM:3026478.3002172}. The ME technique asks a human subject to give as a first response a number in the range $\left(0; +\infty \right)$; the successive numbers are assigned to reflect their relative difference; the outcome of ME are a set of measurements in a ratio scale \cite{gescheider2013psychophysics}. 
\citeauthor{Maddalena:2017:CRM:3026478.3002172} \cite{Maddalena:2017:CRM:3026478.3002172} evaluated, using the CrowdFlower\footnote{https://www.crowdflower.com/} platform, 18 TREC-8 topics, for a total of 4,269 documents. The documents are the top 10 documents returned for IR systems competing in the ad-hoc track; some documents (i.e., 3,881) were previously evaluated by TREC assessors using a binary scale, and some of those documents (i.e., 805) have been reassessed in the study by \citeauthor{Sormunen:2002:LRC:564376.564433} \cite{Sormunen:2002:LRC:564376.564433} using a 4-level scale.

Results from \cite{Maddalena:2017:CRM:3026478.3002172} show that:
(i) ME aggregated judgements are closely aligned with the ordinal coarse-grained scale, both overall and across topics;
(ii) the gathered judgements have shown a high level of agreement with both TREC and \citeauthor{Sormunen:2002:LRC:564376.564433};
(iii) the impact on system evaluation, i.e., the correlation between the system ranking when using ME judgements, has a Kendall's $\tau$ correlation of 0.677 with the official TREC ranking using  NDCG@10.

In this chapter we look at the challenges and opportunities of using S100 as compared to ME, binary and 4-level scales by means of comparative experiments using crowdsourcing platforms to collect relevance judgements at scale. 

\section{Relevance Dimensions and Biases}\label{part:cs:relevancedimensions}
Recently, \citeauthor{Jiang:2017} \citep{Jiang:2017} looked at the use of a multi-dimensional relevance definition including novelty, understandability, reliability, and effort for contextual judgements that are performed by assessors when looking at the search engine result page. In our work we rather focus on the classic definition of relevance based on topicality and look at the effect of different scales on IR evaluation.

\citeauthor{eickhoff:2018} \cite{eickhoff:2018} looked at the effect of cognitive biases in crowdsourced relevance judgement tasks. He showed how crowd workers are affected by fellow workers' answer (Bandwagon effect) and by being presented with multiple options (Decoy effect). The existence of the Decoy effect proves that workers judgement is indeed affected by other documents they have seen before judging a given document, thus supporting even more the need for fine-grained relevance scales (as we propose in our work) that enable workers to express slight relevance differences across different documents.

\citeauthor{10.1145/3269206.3269261} \cite{10.1145/3269206.3269261} propose a framework to evaluate systems assuming the presence of multidimensional relevance, focusing on health search tasks. 
\citeauthor{10.1145/3121050.3121072} \cite{10.1145/3121050.3121072} consider both the effectiveness and credibility in ranked lists. 
\citeauthor{palotti2016assessors} \cite{palotti2016assessors} considered both disagreement and other factors in crowdsourcing when gathering labels and consider multiple relevance dimensions.
\citeauthor{10.1145/2911451.2914708} \cite{10.1145/2911451.2914708} consider different criteria and set of rankers that use a gain function based on a  multi dimensional relevance.
\citeauthor{zuccon2016understandability} \cite{zuccon2016understandability} proposes a methodology to integrate together multiple relevance dimensions.

\chapter{Crowdsourcing for IR Evaluation and Fine Grained Relevance Scales}
\chaptermark{Fine Grained Relevance Scales} 
\label{chapt:cs:S100}

This chapter deals with the usage of crowdsourcing for retrieval evaluation and experiments on the usage of fine grained scales.
Section~\ref{cs:s100:sect:intro} introduces and frames research questions,
Section~\ref{cs:s100:sect:dataset100} presents a 100 level relevance dataset,
Section~\ref{cs:s100:sec:agreement} investigates the comparison with other scales,
Section~\ref{cs:s100:sec:robust} the robustness to fewer judgements, 
Section~\ref{cs:s100:sec:runningout}  addresses the issue of running out of values,  and
Section~\ref{cs:s100:sect:time}  considers the time factor.
Finally, Section~\ref{cs:s100:sec:con} concludes the chapter


\section{Introduction and Research Questions}\label{cs:s100:sect:intro}
Relevance assessment is an integral part of Information Retrieval (IR) evaluation,
since the Cranfield experiments, through the TREC and TREC-like initiatives. In the recent years the collection of relevance judgements is being studied using crowdsourcing.
To gather relevance labels, several scales have been used in the past.
The most common are the classical binary scale, or ordered scales with a limited number of categories (usually ranging from 3 to 7).
It has recently been proposed \cite{Turpin:2015,Maddalena:2017:CRM:3026478.3002172} to use Magnitude Estimation (ME) to gather relevance assessments on a $]0 ,+\infty[$ scale  that has the following advantages:
\begin{enumerate}
\item It is more fine-grained than the above alternatives (and thus, at least potentially, allowing to capture relevance differences that would otherwise be lost);
\item It is able of always providing to the assessor a smaller or higher relevance value, and even a value in between other two, always allowing to assign to a new document a relevance value unforeseen in advance. This happens in particular at the extremes of the scale, but also for the values internal to the range; and  
\item It can adapt to different assessors' preferences (e.g., those who prefer to use a binary scale can do that and those who prefer to judge in a scale from 1 to 10 can also do that).
\end{enumerate}

ME is not free from disadvantages, though:
\begin{itemize}
\item It requires a normalization of the collected scores since each assessor is free to use a different ``internal'' relevance scale. This normalization is not simple, and it is not clear which is the best alternative, although some techniques seem to be reasonably effective \cite{Turpin:2015,Maddalena:2017:CRM:3026478.3002172};
\item It does not allow for a direct comparison of scores provided by different judges and/or on different topics as the score normalization is typically performed on a topic-by-topic basis;
\item It is somehow unnatural, or at least it requires some adaptation for the human assessor as compared to most common rating scales which are bounded; and
\item it leads to a log-normal distribution of relevance scores.
\end{itemize}

In this chapter we discuss and experimentally evaluate by means of large-scale crowdsourced relevance judgements the use a fine-grained scale on 100 levels (S100).
Using the proposed 100 levels scale, the human assessor judges the relevance of a document with respect to a query by means of a number in the $[0..100]$ range (extremes included, thus the levels are actually 101; we name it S100 anyway). 
Such a scale can be seen as a sort of compromise between the classical a-few-categories relevance scales and ME.
We run a large scale crowdsourcing experiment to collect more than 50 thousand labels on such a scale, we discuss its advantages and disadvantages with respect to the already proposed alternatives, and we experimentally compare our judgements with judgements on coarse-grained scales (i.e., binary and 4-levels) and with judgements using ME.

More specifically, our research questions are: 
\begin{itemize}
\item  \textbf{RQ1} Can relevance values be collected in a reliable way using a 100 levels scale in a crowdsourcing setting?
\begin{itemize}
\item     How do crowd workers choose to use the proposed scale?
\item     Are the collected labels consistent with standard ground truths?
\end{itemize}
\item  \textbf{RQ2} What are the differences between S100 and ME?
\begin{itemize}
\item What are the effects of the relevance scale on IR system evaluation/ranking?
\item Which scale is more robust to decreasing the number of judgements per topic/document pair? What happens when collecting fewer judgements per assessor?
\item ME requires some learning to be used effectively by crowd workers as it is not like rating scales they are already used to. What happens to judgement quality when the number of documents judged by each worker in each HIT\footnote{Human Intelligence Task, the task that each worker has to perform.} decreases?
\item     ME allows to go beyond the maximum and minimum judgement level previously used, and to always find a judgement in between two previously expressed judgements. Are these properties required and useful in practice when using S100? 
\item     Are S100 and ME different w.r.t. the time needed to express judgements? Does ME require more adaptation time when used for the first time (i.e., on the first documents)?
\end{itemize}
\end{itemize}

Our main findings are:
\begin{itemize}
\item w.r.t. binary and coarse-grained relevance scales, S100 gives assessors more flexibility in terms of preferential judgements over the documents they are presented during the judging task. Assessors using S100 also have the freedom to judge on a 10-level scale (or 4-level, etc.). It also better aligns with coarse-grained scales as compared to ME (see Section \ref{cs:s100:sec:agreement}).
\item w.r.t ME, S100 gives assessors a reference point by providing upper and lower scale boundaries (see Section \ref{cs:s100:sec:agreement}).
\item S100 is more robust than ME to both fewer assessors per document and fewer documents per assessor (see Section \ref{cs:s100:sec:robust}).
\item The theoretical problem of running out of values (at the extremes of the scale) does not occur often in practice, at least in our setting. Of course, with more document to judge for each worker, the problem might manifest  (see Section \ref{cs:s100:sec:runningout}).
\item If a fine-grained scale is preferred, using ME in a crowdsourcing setting can provide results faster while S100 enables direct comparison over topics and workers and does not require normalization (see Section \ref{cs:s100:sect:time}).
\item While ME shows a steeper learning curve with more time needed to judge the first few documents, it becomes faster for crowd workers to judge with ME compared to S100 in the long term. Considering that crowd work is long-tail distributed with most workers completing very few HITs, S100 may be a more efficient strategy for crowdsourced relevance judgements (see Section \ref{cs:s100:sect:time}). 
\end{itemize}



\section{S100: A 100-Level Relevance Dataset} \label{cs:s100:sect:dataset100}
In this section, we present the results of our crowdsourcing effort aimed at collecting  on a 100-level scale. To make our dataset comparable with others, we followed the experimental design defined by \cite{Maddalena:2017:CRM:3026478.3002172} and reassessed 4,269 documents from 18 topics of TREC-8 ad hoc collection, in a [0, 100] discrete scale. As done in \cite{Maddalena:2017:CRM:3026478.3002172}, we used  the CrowdFlower crowdsourcing platform and rewarded workers \$0.2 for each HIT performed (defined as a sequence of 8 documents which required to be judged in relation to one topic).

The main design difference as compared to that used for the ME collection by \cite{Maddalena:2017:CRM:3026478.3002172} is in the HIT graphical interface which, in our case, expects the relevance score to be given by using a [0,100] slider, instead than using a text field and an unbounded scale.
The adoption of the slider is motivated by the bounded and fine-grained scale and it commonly used for rating items on multi-level scales (see, for example, \cite{huynh2011study}).
In terms of quality checks, we performed the same checks 
as \cite{Maddalena:2017:CRM:3026478.3002172} (i.e., a test question on topic understanding; at least 20 seconds spent on at least 6 of the 8 documents in the HIT; consistency of judgements on two gold documents included in the 8 documents). Additionally, we required workers to move the slider (which was pre-set at 50) for at least 4 of the 8 documents.
When failing the quality checks, workers were allowed to restart the HIT and change their previous answers. Up to 3 attempts were allowed. We tracked the times spent by each worker on each document, and these were cumulated over different attempts. We observed that 85.3\% of workers completed the HIT after the first attempt, 11.2\% after the second, and 3.5\% after the third.
Workers could not work on a topic more than once, but they were given the chance to repeat the task on different topics. 

\subsection{judgement Distribution in S100}

\begin{figure*}[tbp]
  \centering
  \subfigure[]{\label{cs:s100:fig:a}\includegraphics[width=.31\linewidth]{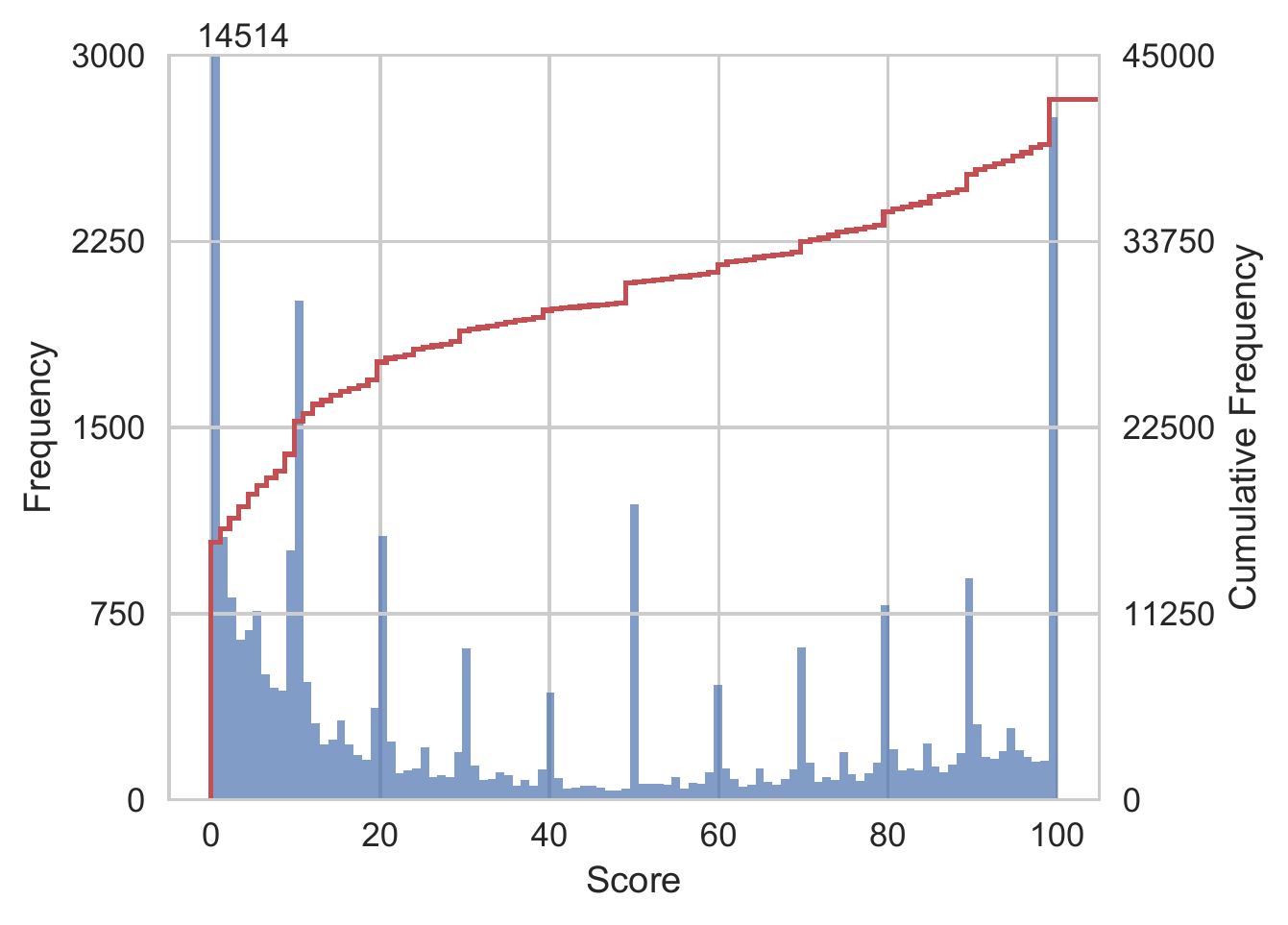}}
  \subfigure[]{\label{cs:s100:fig:b}\includegraphics[width=.31\linewidth]{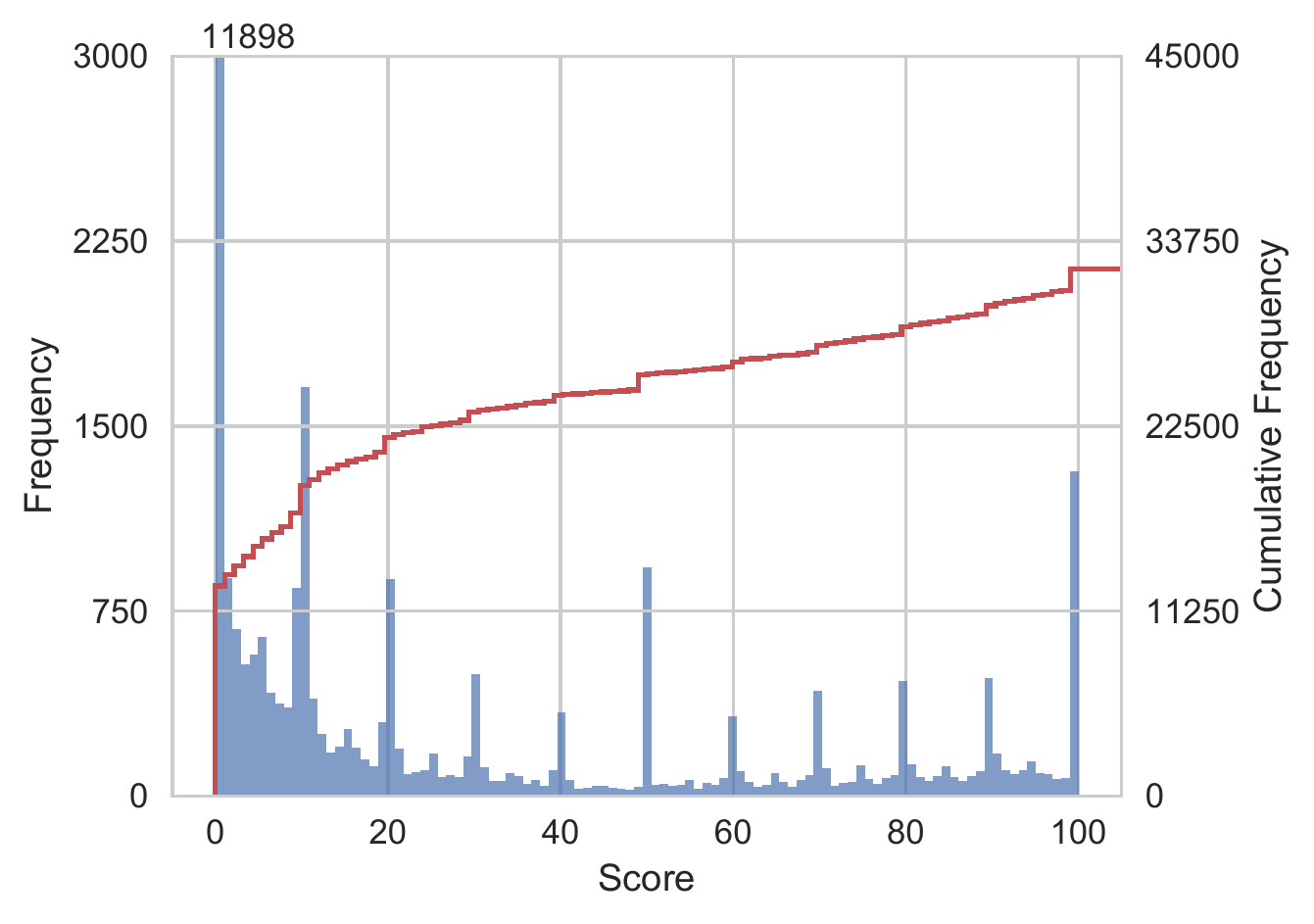}}
  \subfigure[]{\label{cs:s100:fig:c}\includegraphics[width=.31\linewidth]{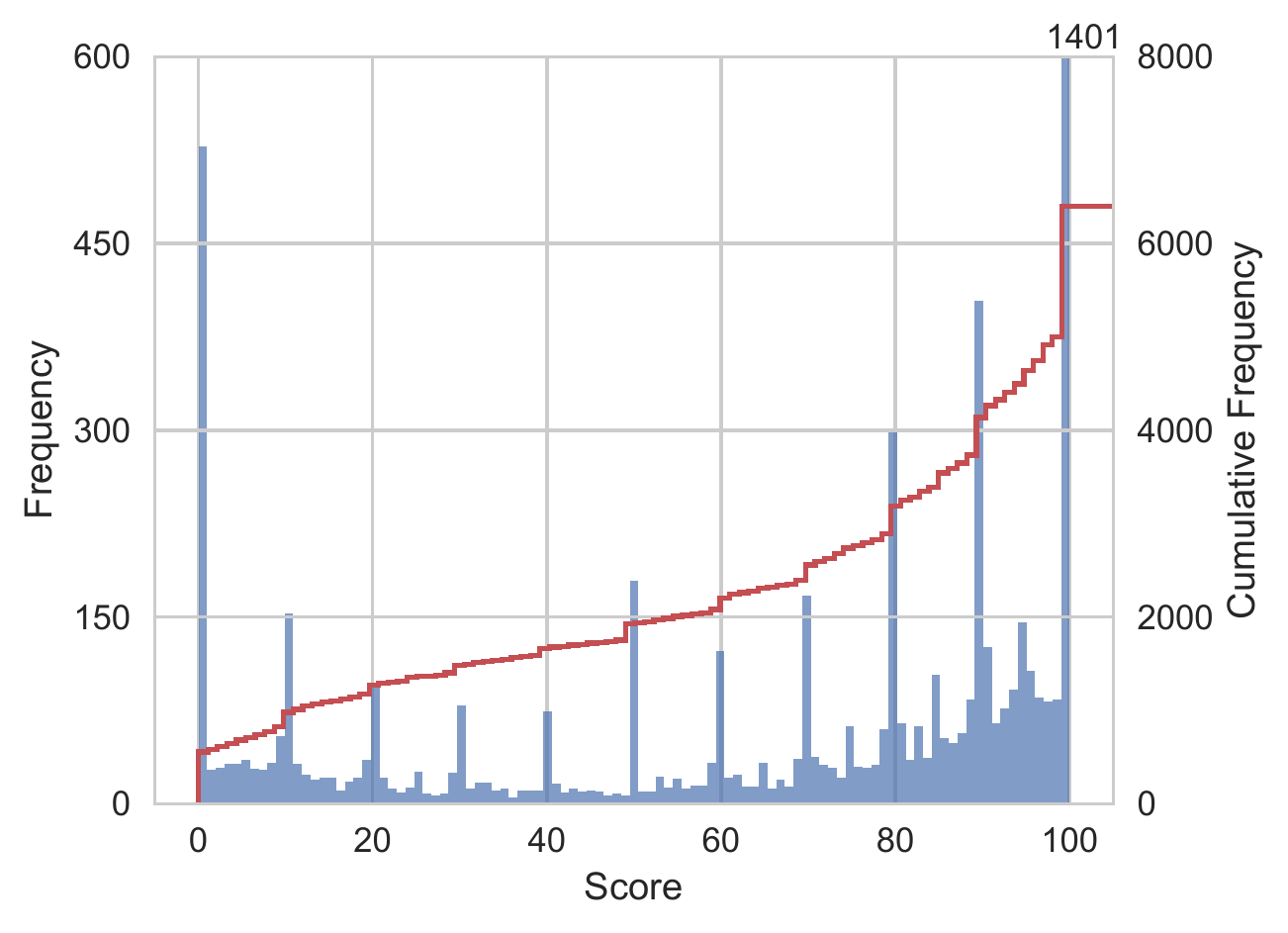}}\\
\vspace*{-2mm}
\subfigure[]			     
   {\label{cs:s100:fig:d}\includegraphics[width=.31\linewidth]{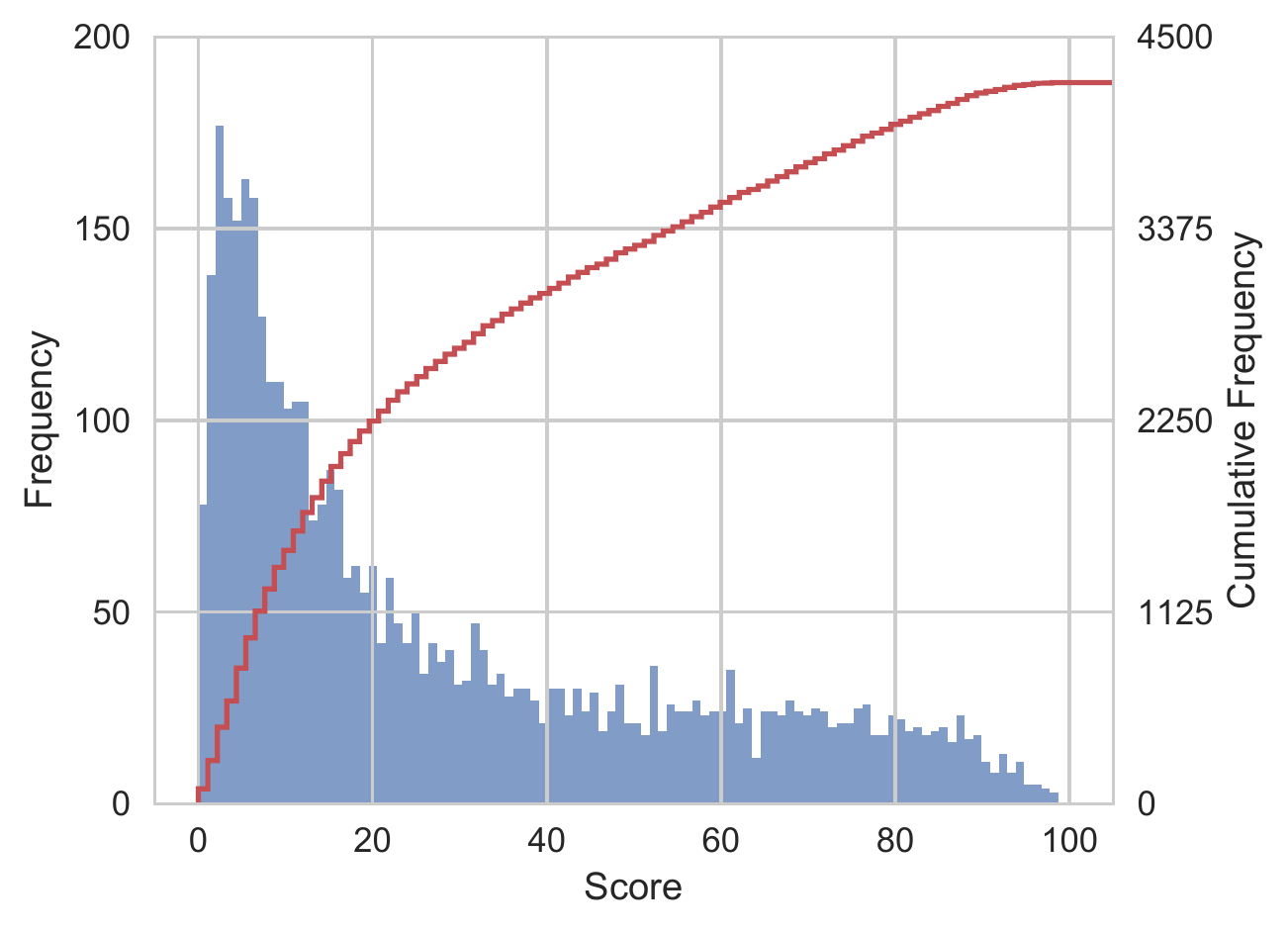}}
   \subfigure[]	
   {\label{cs:s100:fig:e}\includegraphics[width=.31\linewidth]{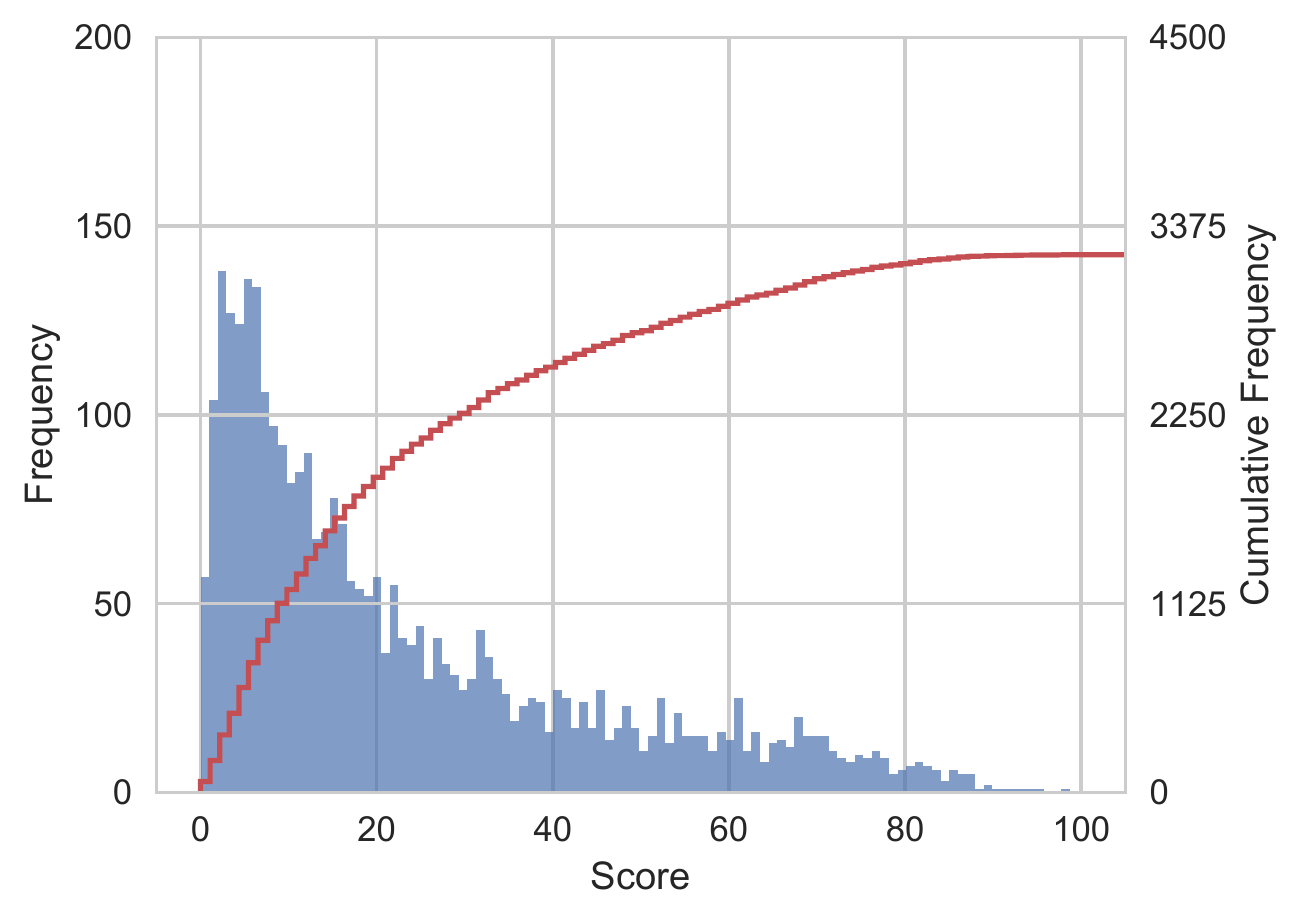}}
   \subfigure[]	
   {\label{cs:s100:fig:f}\includegraphics[width=.31\linewidth]{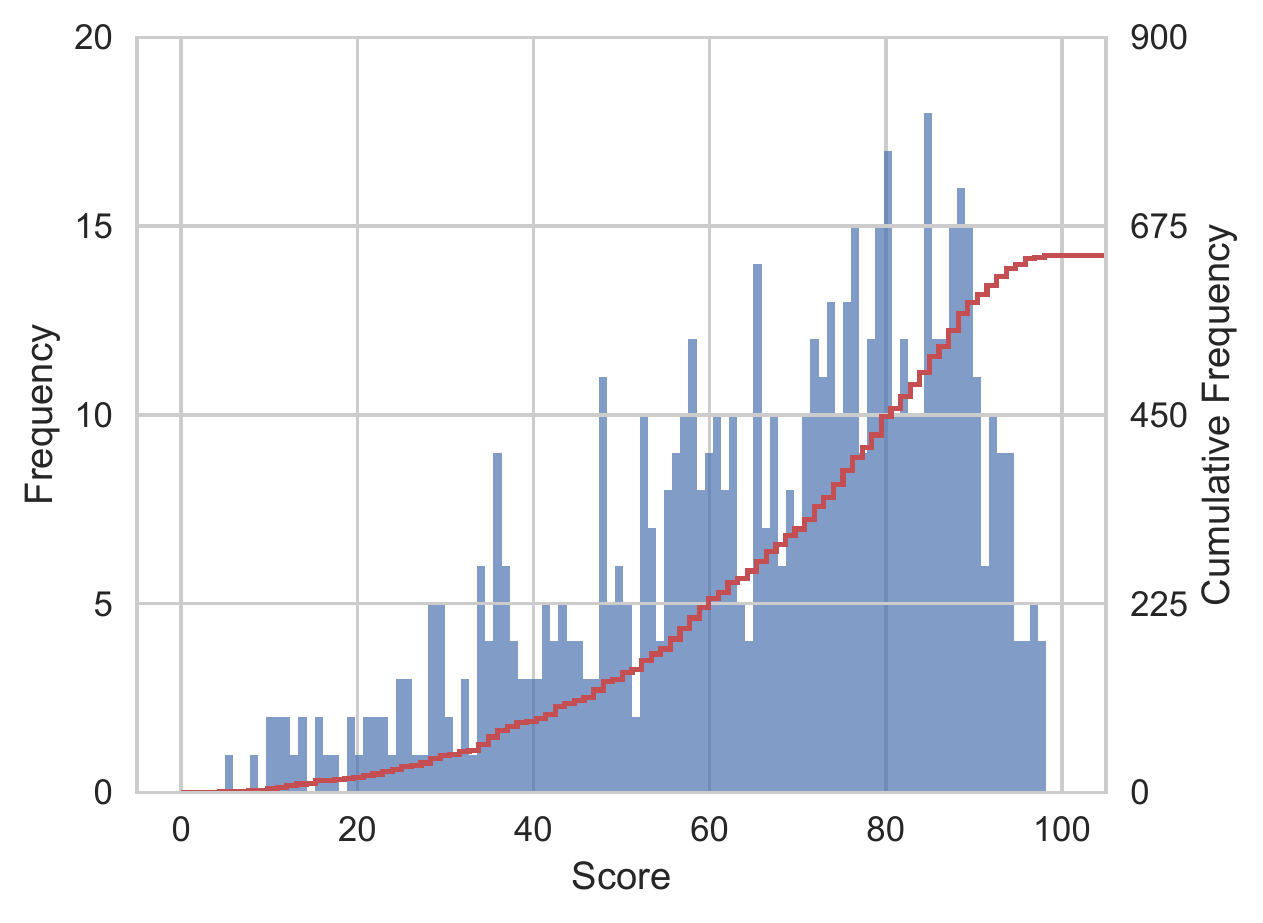}}\\
   \vspace*{-2mm}
 \caption{Individual score distribution in the S100 dataset for all (a), for non-relevant (b), and for relevant (c) documents according to TREC judgements. 
 Aggregated score distribution in the S100 dataset for all (d), for non-relevant (e), and for relevant (f) documents according to TREC judgements.  
 }
  \label{cs:s100:fig:scoreDistr_AllTopics_HueTREC}
\end{figure*}

Figure~\ref{cs:s100:fig:scoreDistr_AllTopics_HueTREC} (a) shows the distribution of the individual scores gathered for S100: the x-axis represent the score obtained by a document, the y-axis represent its frequency; the red line represent the cumulative distribution.  Figures~\ref{cs:s100:fig:scoreDistr_AllTopics_HueTREC} (b) and (c) show the distribution when doing a breakdown  on non-relevant documents  and on relevant documents, according to TREC assessors, respectively.
From the plot using all the judgements (Figure~\ref{cs:s100:fig:scoreDistr_AllTopics_HueTREC} (a)) we see that, as expected, the distribution is clearly skewed towards lower (less relevant) scores; furthermore, there is a clear tendency of giving scores which are a multiple of ten, and the two most frequent scores are 0 and 100. In fact, the scores which are divisible by 10 are 60\% of all the judgements in the dataset. The judgements on the scale boundaries (i.e., 0 and 100) are the 41\%. If we do not consider the scale boundaries, the number of judgements which are divisible by 10 are the 32\%. 
%
Due to the large presence of non relevant documents, the total distribution of scores is mainly influenced by and very similar to the distribution of the non-relevant documents according to TREC assessors (Figure~\ref{cs:s100:fig:scoreDistr_AllTopics_HueTREC} (b)), as we can see when comparing the cumulative distribution for the plots of all the documents with the one of non-relevant documents (i.e, the red lines in Figure~\ref{cs:s100:fig:scoreDistr_AllTopics_HueTREC} (a) and (b)).

When comparing Figure~\ref{cs:s100:fig:scoreDistr_AllTopics_HueTREC} (b) and (c), we see that  for non-relevant documents (Figure~\ref{cs:s100:fig:scoreDistr_AllTopics_HueTREC} (b)) the majority of S100 scores is in the lower part of the scale (left on the plot) and for relevant documents (Figure~\ref{cs:s100:fig:scoreDistr_AllTopics_HueTREC} (c)) the majority of S100 scores is in the higher part of the scale (right on the plot). Moreover, we observe that many non-relevant documents obtained the maximum possible score (i.e., 100), and many relevant documents obtained 0 as a score. This may depend on multiple factors: a misclassification by TREC experts, a document/topic ambiguity, or might even be an indicator of low quality crowd judgements, obtained despite the strict quality checks applied to the task. Furthermore, we notice that the ``decimal preference'' is still present both for relevant and non relevant documents.


\subsection{Aggregated judgements in S100}
Next, we proceed with aggregating the raw relevance judgements collected form the crowd for the same topic/document pair as commonly done to increase the quality of the collection.
Relevance scores in S100 are in the $\left[0,100 \right]$ range, thus a natural aggregation function is represented by the arithmetic mean of the individual scores, with no prior normalization of individual scores as done for ME.\footnote{We experimentally compared different aggregation functions and the use of score normalization functions but observed that the use of the arithmetic mean over non-normalized score lead to most accurate labels compared to the other datasets.} 
Figure~\ref{cs:s100:fig:scoreDistr_AllTopics_HueTREC} (d,e,f) show the distribution for the aggregated judgements. We can see that, as compared to the raw judgements given by individual workers (Figure \ref{cs:s100:fig:scoreDistr_AllTopics_HueTREC} (a,b,c)), the aggregated judgements follow the expected distribution of many non-relevant documents with a long-tail of more relevant documents.  The aggregation has also the effect of making the curves smoother, as well as making the tendency of scores to be a multiple of ten less prominent.

\begin{figure}[tbp]
  \centering
  \includegraphics[width=.6\linewidth]{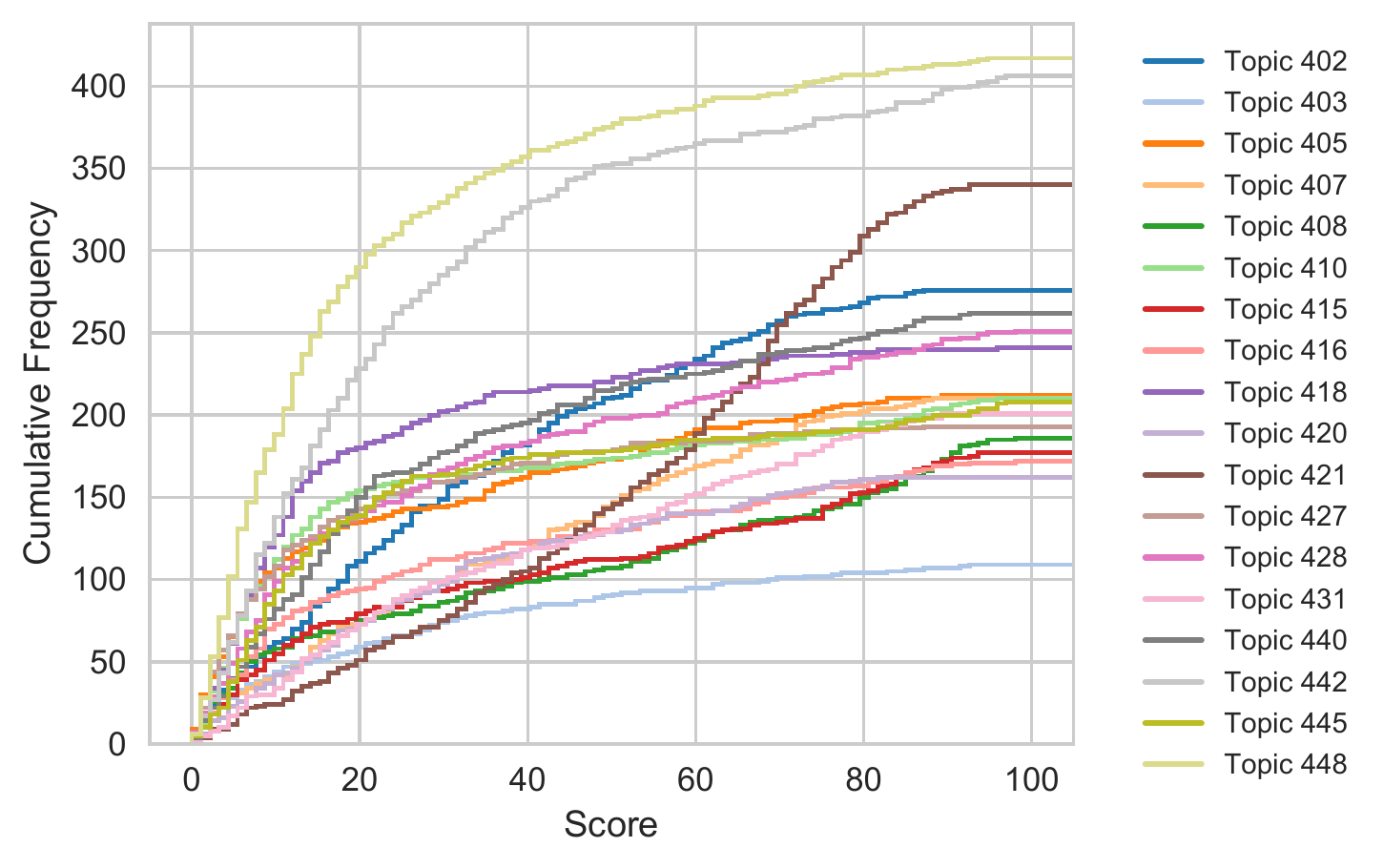}
\vspace*{-8mm}
\caption{Aggregated score cumulative distribution in the S100 dataset; breakdown on individual topics. 
}
  \label{cs:s100:fig:AGG_scoreDistr_HueTopic_cumulative}
\end{figure}

Figure~\ref{cs:s100:fig:AGG_scoreDistr_HueTopic_cumulative} shows the aggregated judgement cumulative distributions broken down by topic. We can observe similar trends over all topics with some topics (e.g., 448 and 442) having a cumulative curve growing faster (i.e., having many `not so relevant' documents) and others having a much slower growth (e.g., 403) showing a presence of more relevant documents. 
 A slightly different pattern is shown by topic 421 which grows towards the end of the relevance score interval. 
 This is explained by the fact that this topic has a small fraction of low relevance documents (Sormunen 0 and 1)  and a high fraction of high relevant documents (Sormunen 2 and 3).

\section{Comparison with other Scales}\label{cs:s100:sec:agreement}

In this section we compare the judgements collected for S100 with judgements performed on the same documents over different relevance scales. We  introduce an agreement measure that allows us to compute agreement across judgement scales and report agreement values for S100 with the TREC binary scale, the \citeauthor{Sormunen:2002:LRC:564376.564433} 4-level scale (S4), and ME.

\subsection{Score Distribution as Compared to Other Scales}
\begin{figure}[tbp]
  \centering
    \includegraphics[width=.6\linewidth]{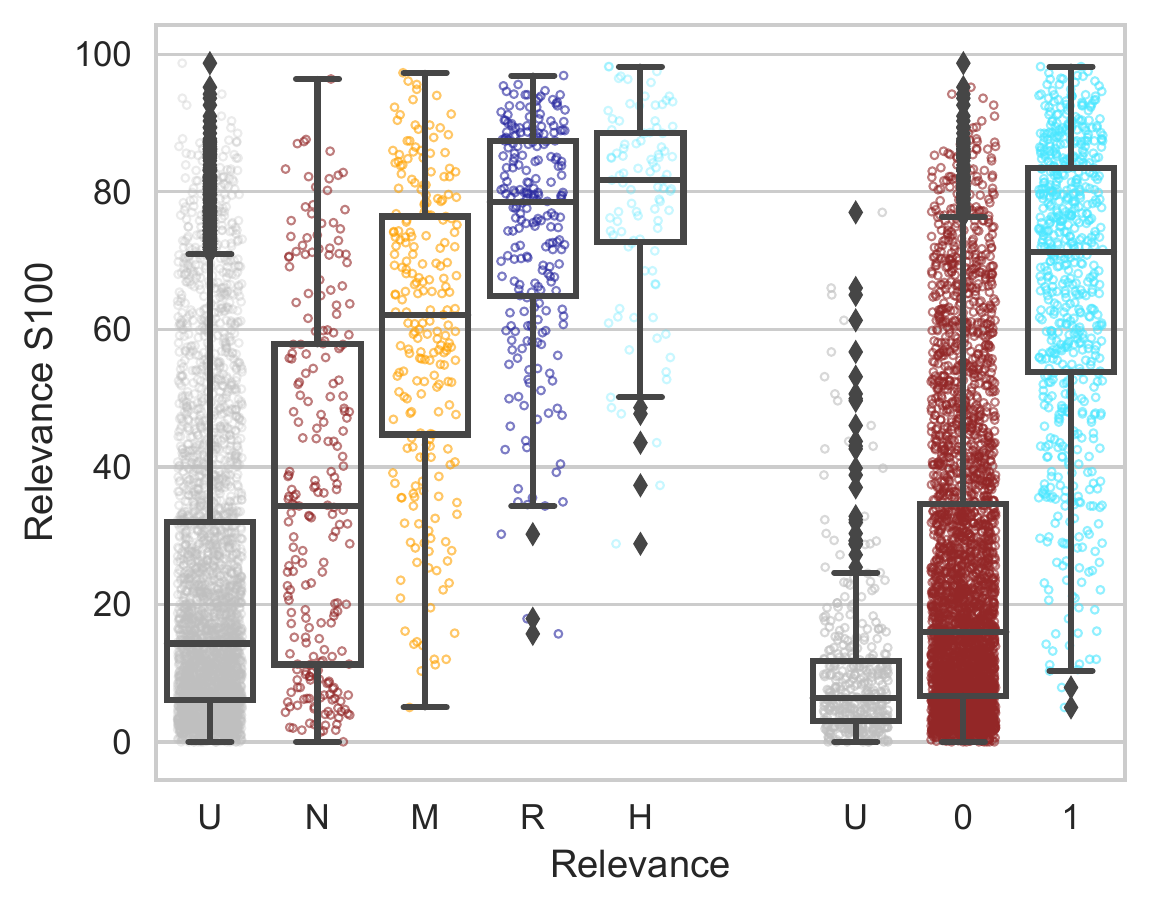}
\vspace*{-2mm}
\caption{Aggregated judgement scores collected for S100 vs TREC labels (right) and vs S4 (left). U indicates unjudged documents in the TREC and \citeauthor{Sormunen:2002:LRC:564376.564433} collections. 
}
  \label{cs:s100:fig:corr_BLS_TREC_Sorm}
\end{figure}

Figure \ref{cs:s100:fig:corr_BLS_TREC_Sorm} shows how the judgements performed on S100 compared with the binary labels collected by TREC and the 4-level judgements performed by \citeauthor{Sormunen:2002:LRC:564376.564433}.
We can observe that while the  median
value for documents judged as relevant and non-relevant by TREC is  different in S100, the distribution of S100 scores covers the entire score interval for both type of documents.
The distribution of S100 scores compared to  S4 labels by \citeauthor{Sormunen:2002:LRC:564376.564433} shows the non linearity of the 4-level labels that have been collected on the scale N-M-R-H (i.e., not relevant, marginally relevant, relevant, and highly relevant).
When comparing to the similar Figure~3 by \citet{Maddalena:2017:CRM:3026478.3002172} we can notice that in S100 the relevance scores are better distributed across the full scale with highest levels of relevance in S4 and the binary TREC scale having a median score closer to the upper bound of the scale as compared to ME. Such behavior is not observed for ME as the scale is unbounded at the top making scores for highly relevant documents having a wider distribution.

\begin{figure}[tbp]
  \centering
    \includegraphics[width=.6\linewidth]{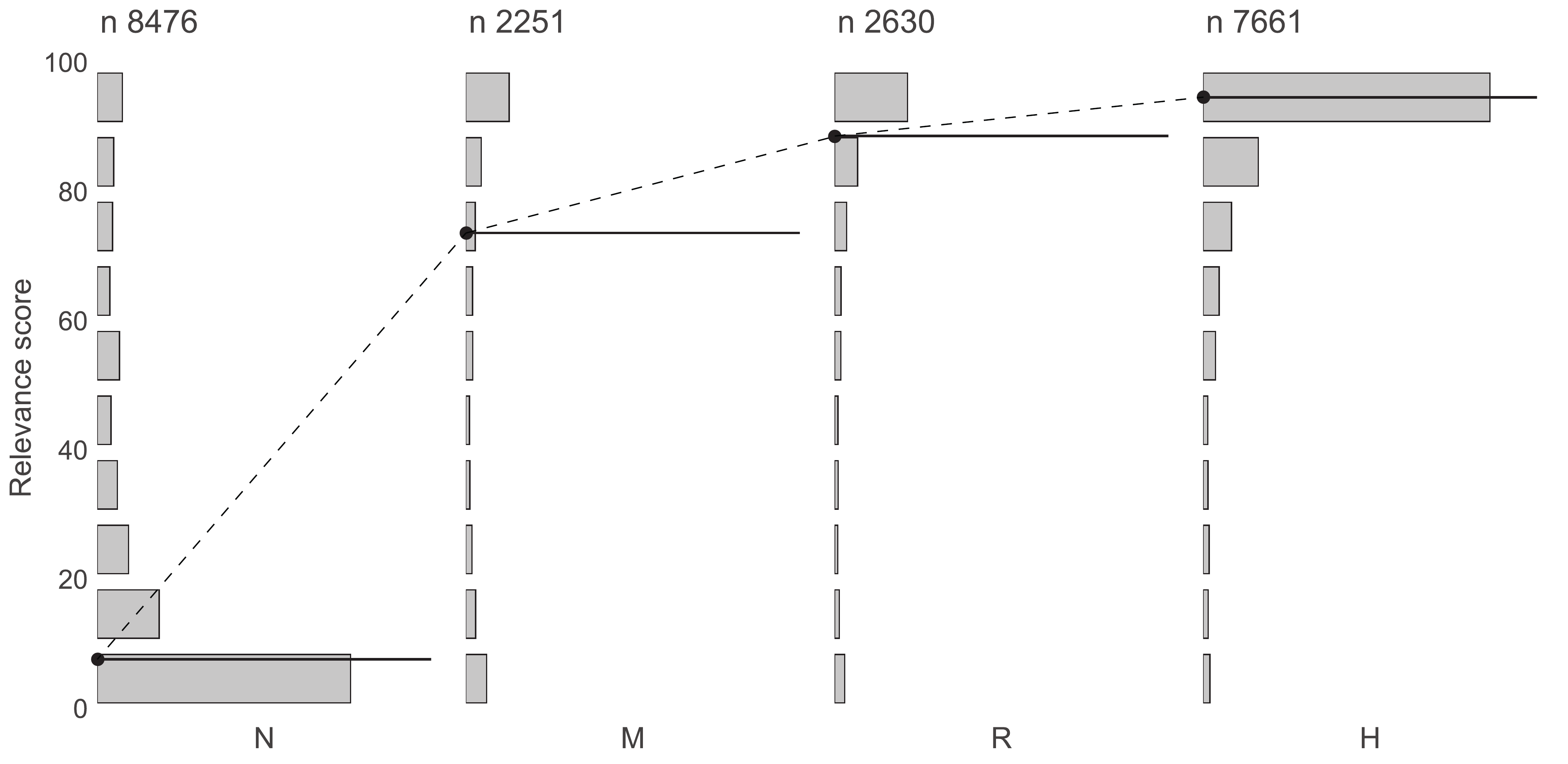}
\caption{Distribution of individual scores in S100 for each Sormunen level.
}
  \label{cs:s100:fig:distr_scores_sorm_levels}
\end{figure}

Figure~\ref{cs:s100:fig:distr_scores_sorm_levels}, similar to Figure~14 by \citet{Maddalena:2017:CRM:3026478.3002172}, presents a clearer evidence of the difference between ME and S100: whereas in ME the gain profiles seem to be exponential, or at least super-linear, in our case they are clearly sub-linear. We consider this a reason to prefer S100 to ME, since it better reflects the definition of relevance levels introduced for S4 which assumes that already marginally relevant documents should be substantially better than not relevant ones with a sub-linear step increase for the subsequent relevance levels R and H.
Such differences (Figures ~\ref{cs:s100:fig:corr_BLS_TREC_Sorm} and \ref{cs:s100:fig:distr_scores_sorm_levels}) between ME and S100 scores are likely due to the effect of the end of scale which is unbounded in ME thus making high-relevance scores disperse. This is not the case in S100, a bounded scale that allows assessors to implicitly map their judgements against the scale upper bound. 

\subsection{Agreement with TREC}
First we introduce a new measure that allows us to check assessor agreement across rating scales. We then adopt this measure to evaluate the quality of S100 scores as compared to other datasets.

\subsubsection{An agreement measure for ratings given over different scales.}\label{cs:s100:sec:agreement-def}
Given two rating vectors $X = \{ x_1, \ldots, x_n \}$ and $Y = \{ y_1, \ldots, y_n \}$ where $x_i$ ($y_i$ respectively) represents the  $i$-th document in a sequence of relevance judgements (e.g., a HIT), we define $X'$ as the sorted vector X and $Y'$ the re-ordering of $Y$ maintaining the relation to $X'$. That is,
$X' = \left\{ 
x_i\ |\ x_i \in X \land 
x_i\leq x_{i+1} ,\
i \in \{1, \ldots, n\}  
\right\}$
and
$Y' = \left\{
z_i\ |\  z_i = y_{index\_of(x_i)} \land 
y_i \in Y,\
i \in \{1, \ldots, n\} 
\right\}$.
Based on such two lists, we define the following agreement function\footnote{We consider $y_i < y_j$ rather than $x_i \le x_j$ as we assume $X$ to use a coarser-grained scale (e.g., binary) as compared to $Y$ (e.g., S100).}
$$ \mbox{pos\_agr}(A,B,i,j) = \begin{cases*}
			1 & if $x_i \neq x_j \land y_i < y_j \land x\in A \land y \in B $\\
    		0 & otherwise, 
    		\end{cases*} 
$$
that tells us whether the ordering of two documents is consistent across the two judging sets.
Thanks to this we can now define the agreement score as the ratio of consistent document pairs over all possible pairs:
$$ \left( \sum\limits_{i=1}^n\
\sum\limits_{\substack{j=i+1}}^n  
\mbox{pos\_agr}\left(X',Y',i,j\right) \right ) \cdot  {\binom{n}{2}}^{-1}. 
$$

Note that this is not a symmetric measure, but it rather computes agreement of $Y$ ratings as compared to $X$ considered the baseline judgements.
This measure computes the number of agreement pairs between the two datasets. That is, if a document $w$ has a higher relevance judgement score than a document $z$ according to judgements in $X$, we would like the same order $w \geq z$ to be maintained in $Y$. That is, the relevance judgement score of $w$ should be higher than $z$ according to $Y$ judgements. 

\subsubsection{Comparison with other scales}
\begin{figure}[tbp]
  \centering
  \includegraphics[width=.6\linewidth]{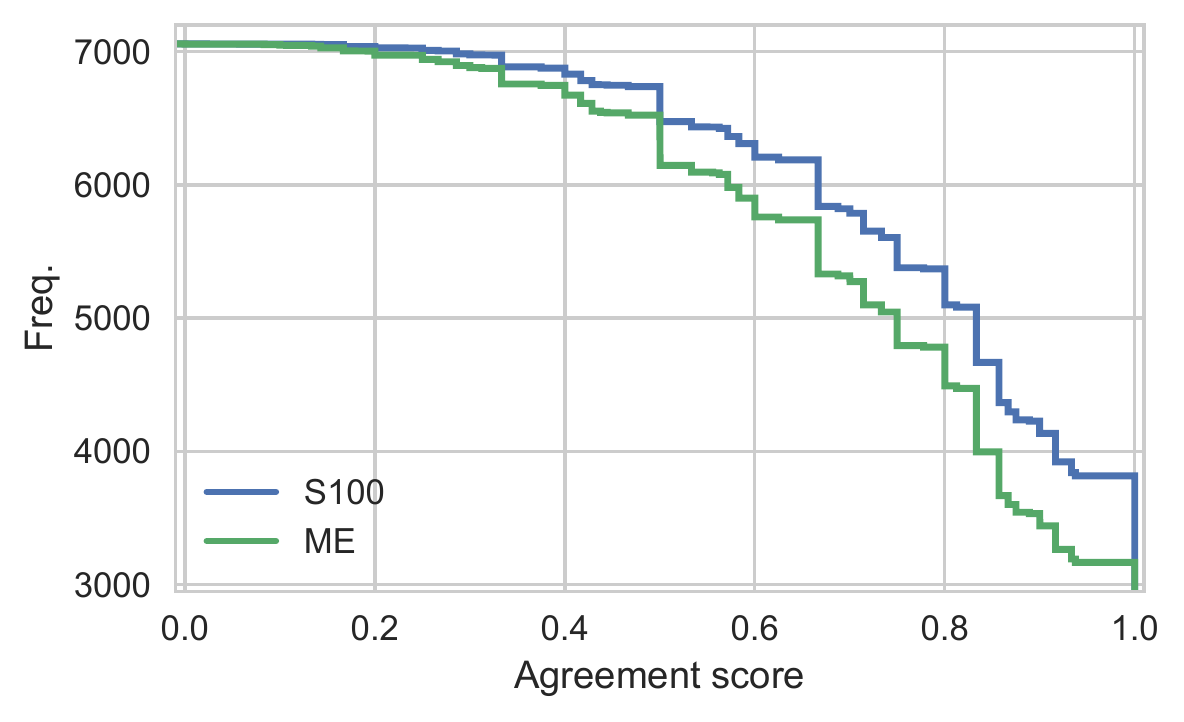}
  \vspace*{-2mm}
\caption{Complementary cumulative distribution function of pairwise agreement as defined in Section~\ref{cs:s100:sec:agreement-def} of S100 and ME with TREC. 
  \label{cs:s100:fig:Robust_FewDocPerUnit1}}
\end{figure}

Figure~\ref{cs:s100:fig:Robust_FewDocPerUnit1} shows the complementary 
cumulative distribution function (showing how often agreement is above a given value) of pairwise agreement for S100 and ME with respect to TREC binary judgements. The ME series is another representation of the data in \cite[Figure~6]{Maddalena:2017:CRM:3026478.3002172}. The comparison highlights that  agreement levels in S100 are 
higher than ME. 

\begin{figure}[tbp]
  \centering
  \subfigure{\label{cs:s100:fig:alpha:a}\includegraphics[width=.6\linewidth]{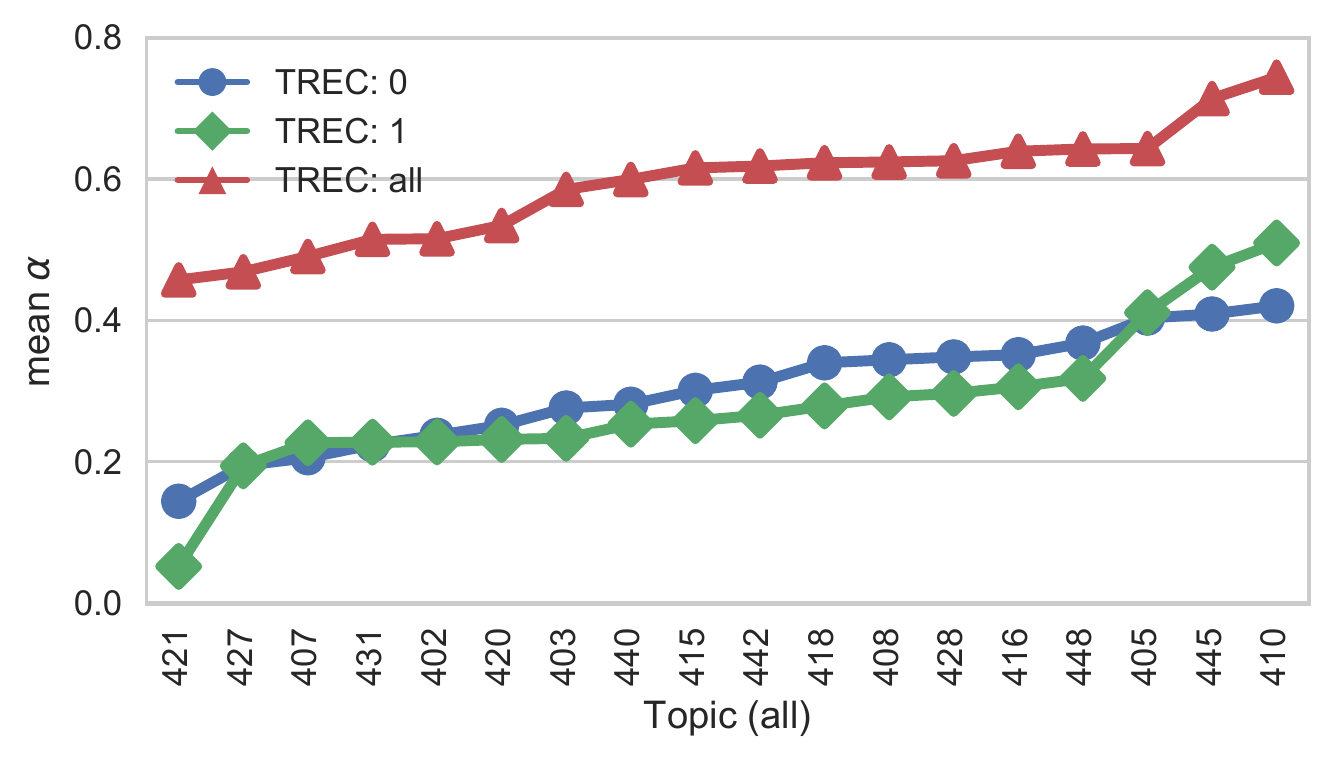}}\\
  \vspace*{-4mm}
  \subfigure{\label{cs:s100:fig:alpha:b}\includegraphics[width=.6\linewidth]{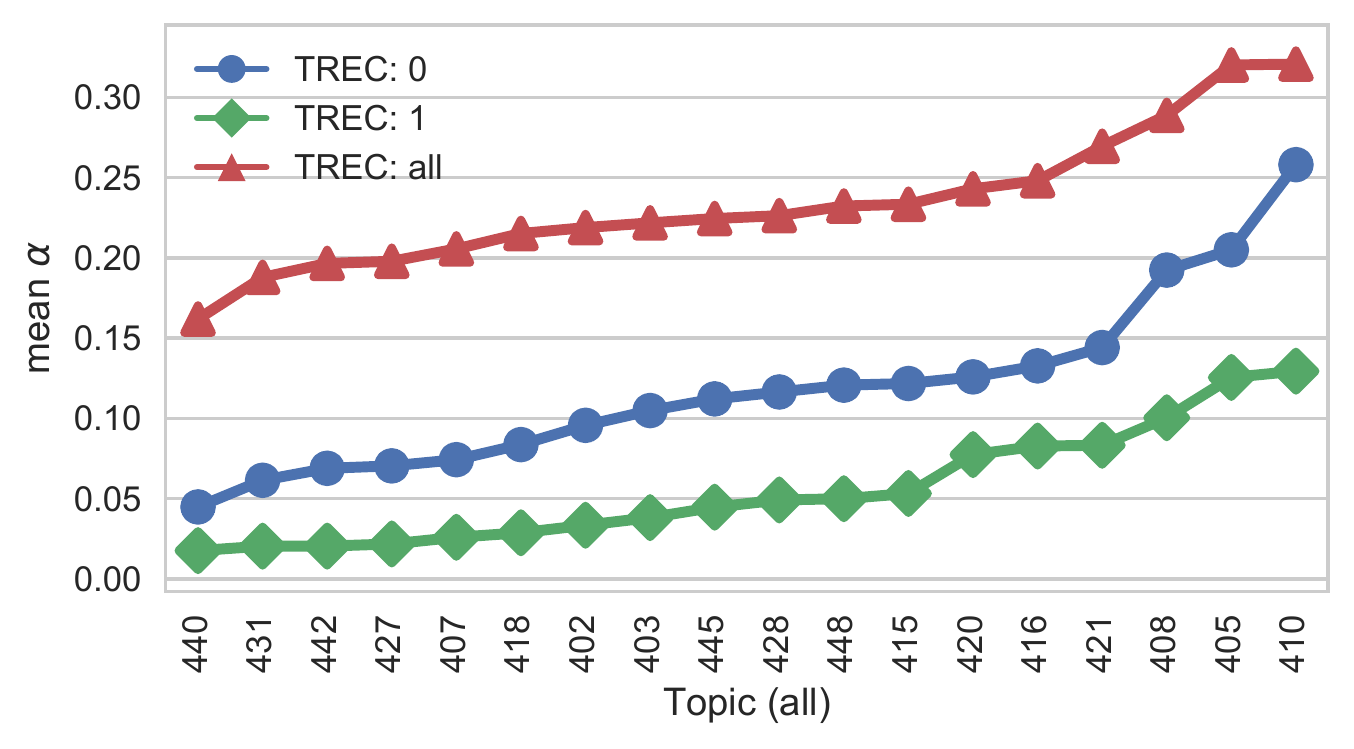}}\\
\vspace*{-2mm}
\caption{Agreement ($\alpha$) of the individual topics, in S100 (above) and ME (below). 
}
  \label{cs:s100:fig:AlphaScores}
\end{figure}

Figure~\ref{cs:s100:fig:AlphaScores} shows the topics ordered by another standard measure of agreement, \citeauthor{krippendorff2007computing}'s $\alpha$ \cite{krippendorff2007computing}.
We can make the following observations:
\begin{itemize}
\item Agreement scores for judgements collected with S100 are substantially higher than those collected with ME.
\item There is some consistency across S100 and ME in the sense that topics with high/low agreement tend to be the same.
\item Agreement over TREC non-relevant documents is higher as compared to relevant ones. 
\item 
Agreement on the non-relevant documents as compared to agreement on all the documents is similar in the two figures, whereas agreement on the relevant documents as compared to agreement on all the documents is higher in S100 than in ME (the green ``TREC: 1'' series is ``pulled up'' in the S100 chart). In other terms, S100 improves, w.r.t. ME, $\alpha$ agreement on the relevant documents.
\end{itemize}

\subsection{IR System Ranking Correlation}\label{cs:s100:sec:sysrank}

\begin{figure*}[tbp]
  \centering
  \subfigure[]{\includegraphics[width=.31\linewidth]{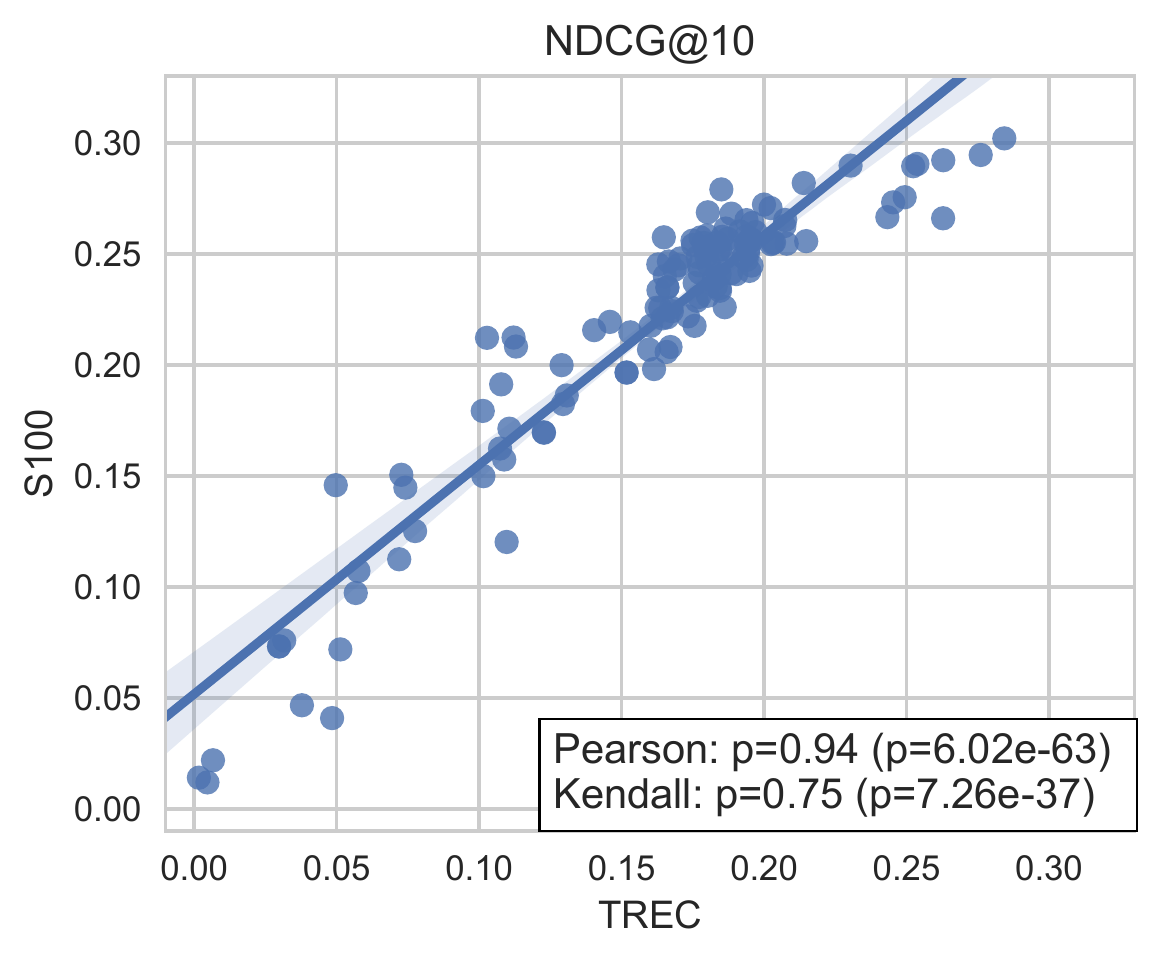}}
    \subfigure[]{\includegraphics[width=.31\linewidth]{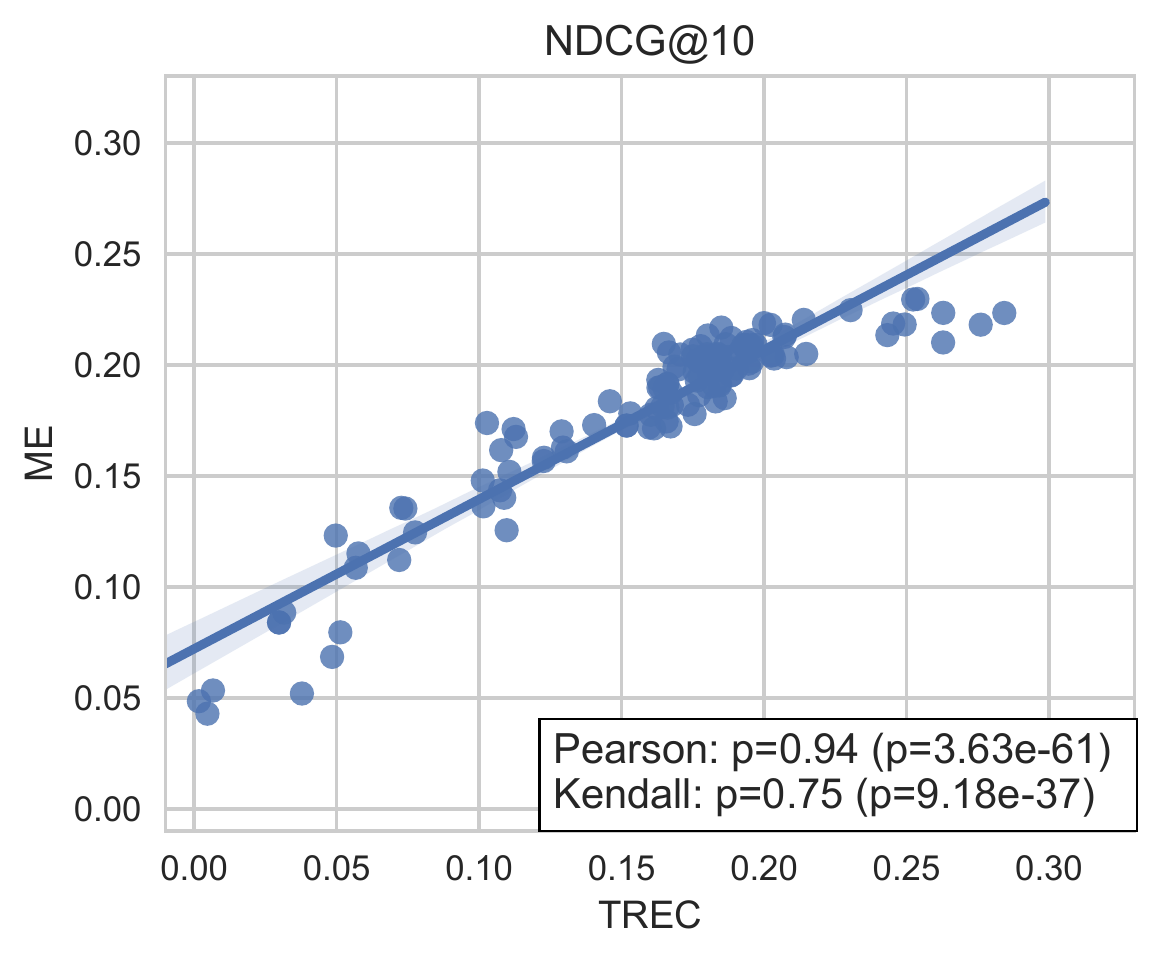}}
      \subfigure[]{\includegraphics[width=.31\linewidth]{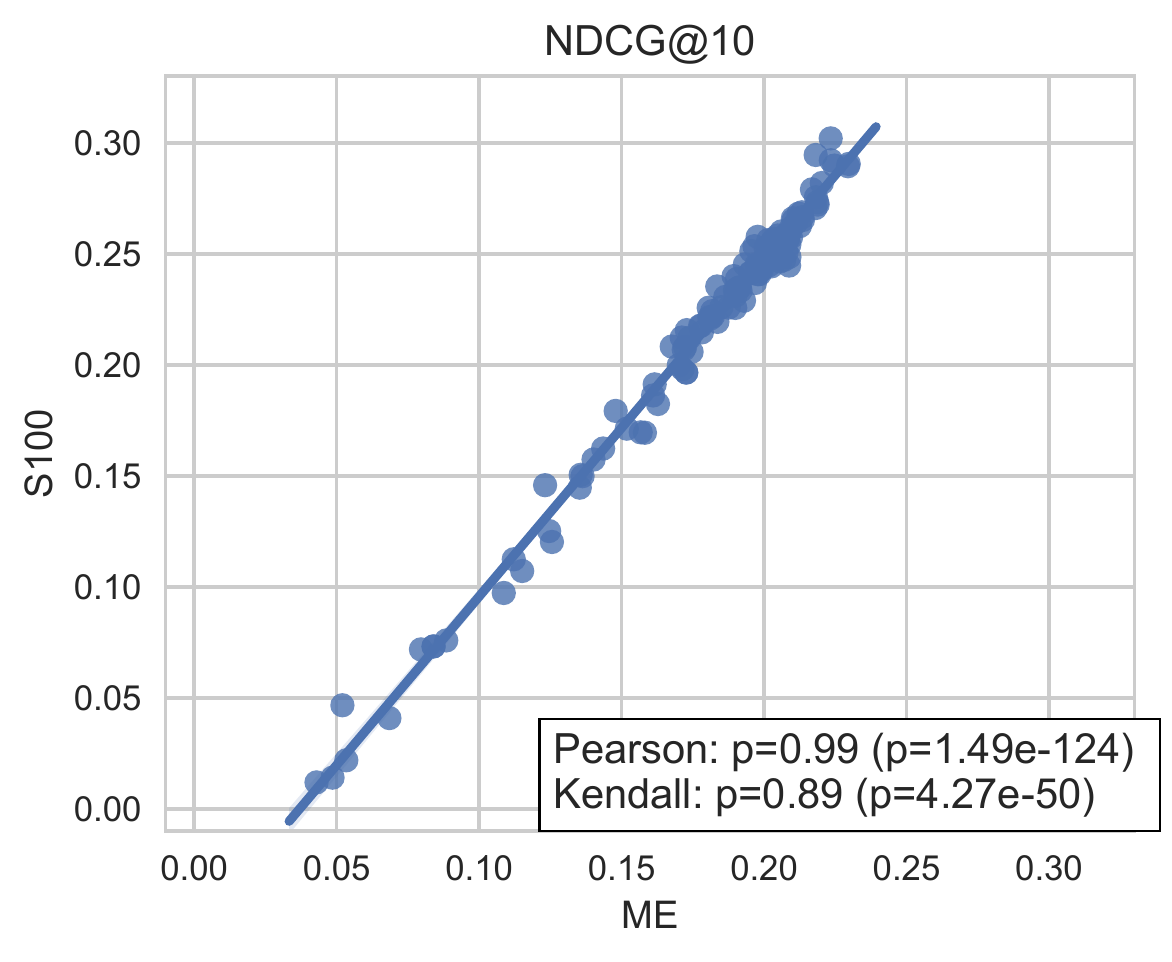}}\\
\vspace*{-2mm}
\caption{NDCG@10 scores for TREC-8 runs and judgements collected over different scales: TREC, ME and S100.}
  \label{cs:s100:fig:NDCG}
\end{figure*}

Finally, we computed Kendall $\tau$ correlation of IR systems ranked by effectiveness computed using judgements collected over different relevance scales.
Figure \ref{cs:s100:fig:NDCG} shows the IR system ranking correlation using NDCG@10 \citep{Jarvelin:2002:CGE:582415.582418} when using binary judgements as compared to S100 (a), using binary judgements as compared to ME (b); and using S100 as compared to ME (c).
Each dot is a system and the charts show its NDCG@10 values over two different scales.
We can observe that, while all judgements result in high system ranking correlation values, the best correlation is obtained when comparing S100 and ME.
This demonstrates how S100 lead to results similar to ME by providing assessors the flexibility to judge document relevance on a fine-grained basis. This is also explained by the fact that S100 and ME have been collected following the same crowdsourcing setup while the TREC and S4 did not use crowdsourcing.
Looking at how S100 and ME compare with TREC (Figure \ref{cs:s100:fig:NDCG} a and b) we can see that while correlation values are similar, NDCG@10 scores obtained using S100 are more consistent with those obtained with TREC labels whereas ME judgements tend to result in lower NDCG scores.



\section{Robustness to Fewer judgements}\label{cs:s100:sec:robust}
In this section we study how different relevance scales behave in terms of robustness to fewer judgements. That is, we look at how crowdsourced relevance label quality decreases as compared to editorial 
judgements by experts like TREC and Sormunen's S4 assessors. In detail, we study two kinds of robustness:
\begin{itemize}
\item Shorter HITs: including fewer documents to be judged in a HIT so that each worker has the option to do less if they wish to.
\item Fewer assignments per document: using fewer workers judging the same document, and averaging their judgements.
\end{itemize}
We measure robustness by observing how pairwise agreement with TREC and S4 decreases.

\subsection{Fewer Documents per HIT}

In the crowdsourcing setup used to create the S100 and ME collections each worker is required to judge 8 documents in one HIT. When using fewer documents per HIT, we assume we could lose on training effects (i.e., workers becoming proficient in the judging task) with the benefit of work flexibility.

\begin{figure*}[tbp]
  \centering
  \subfigure{\label{cs:s100:fig:r:2}\includegraphics[width=.31\linewidth]{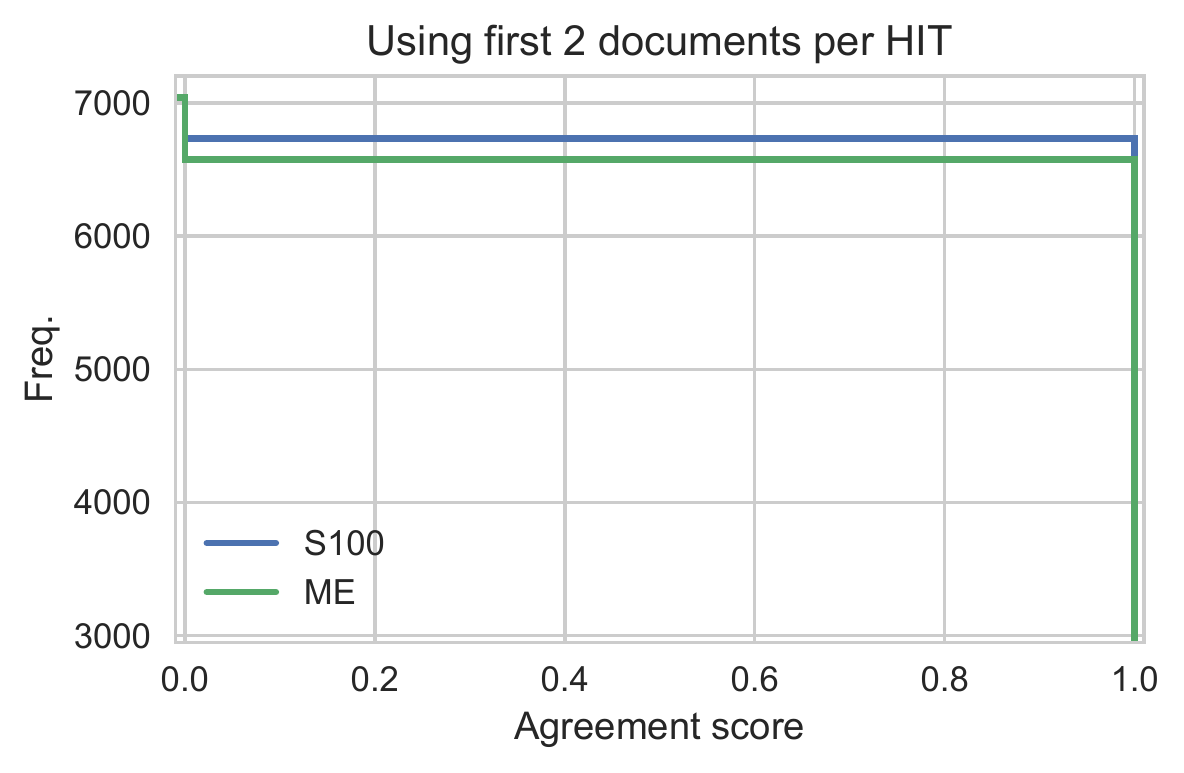}}
  \subfigure{\label{cs:s100:fig:r:3}\includegraphics[width=.31\linewidth]{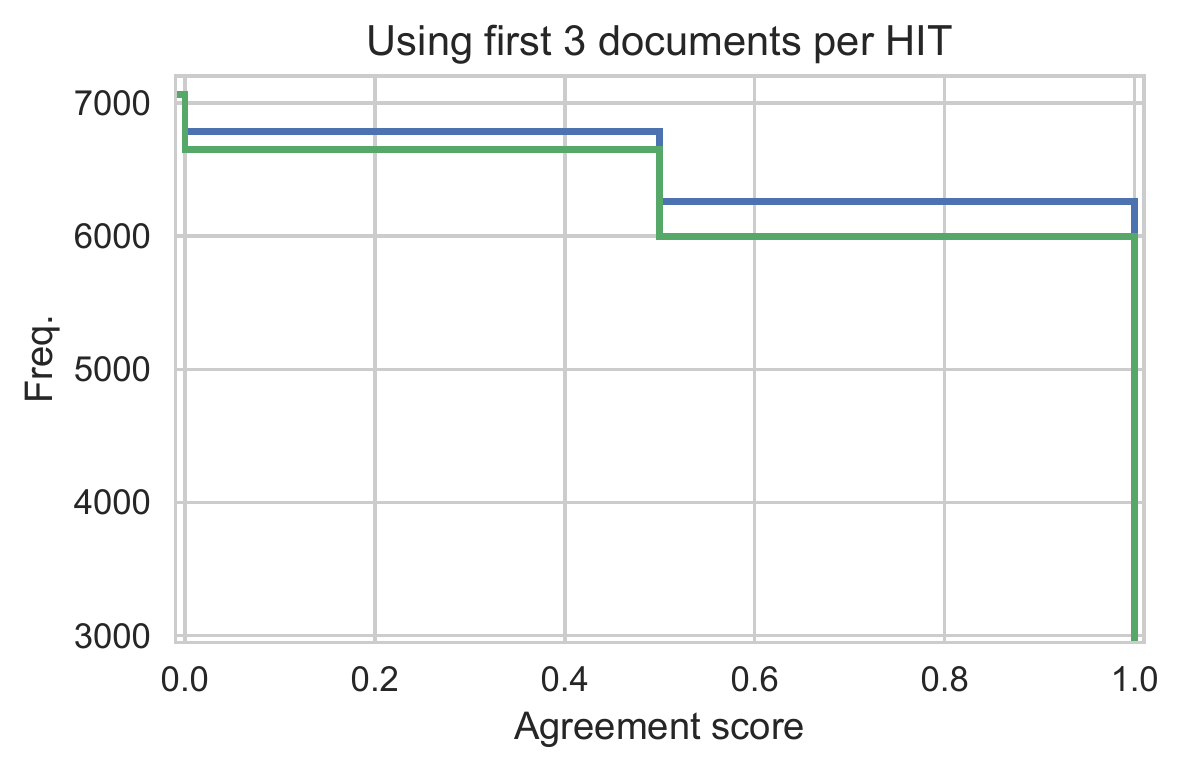}}
    \subfigure{\label{cs:s100:fig:r:4}\includegraphics[width=.31\linewidth]{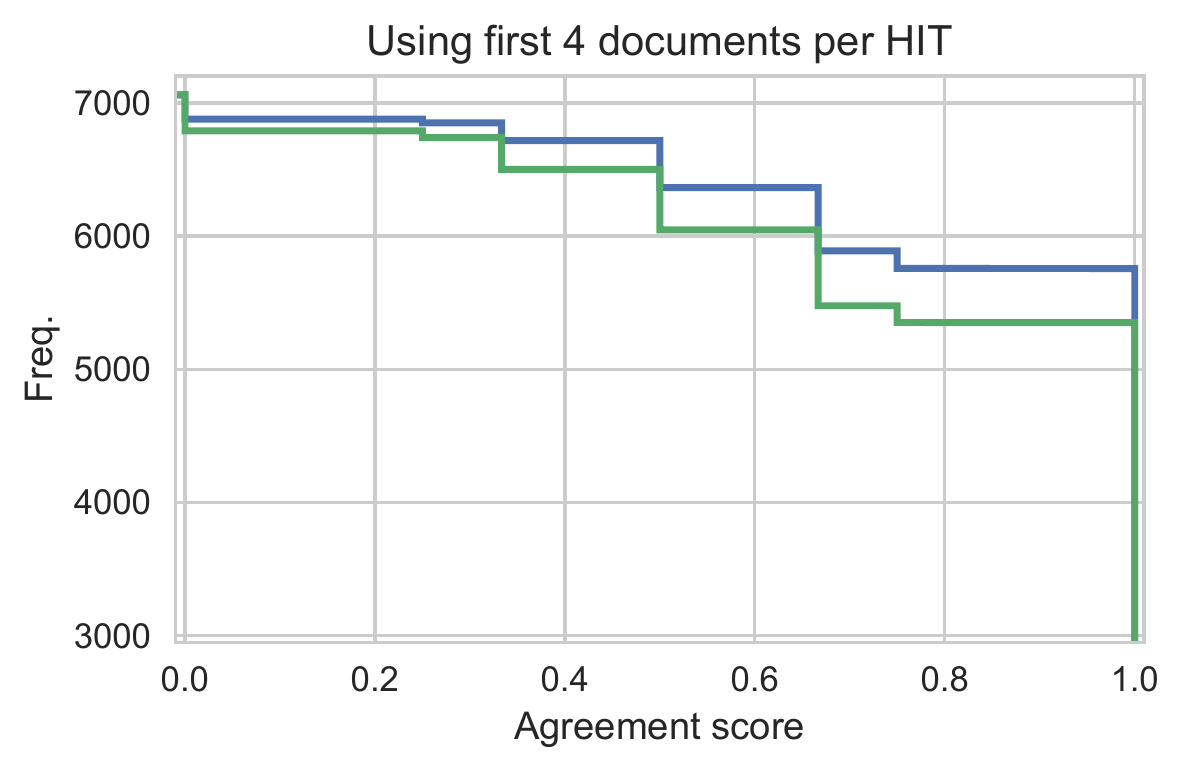}}\\
\vspace*{-2mm}
\subfigure{\label{cs:s100:fig:r:5}\includegraphics[width=.31\linewidth]{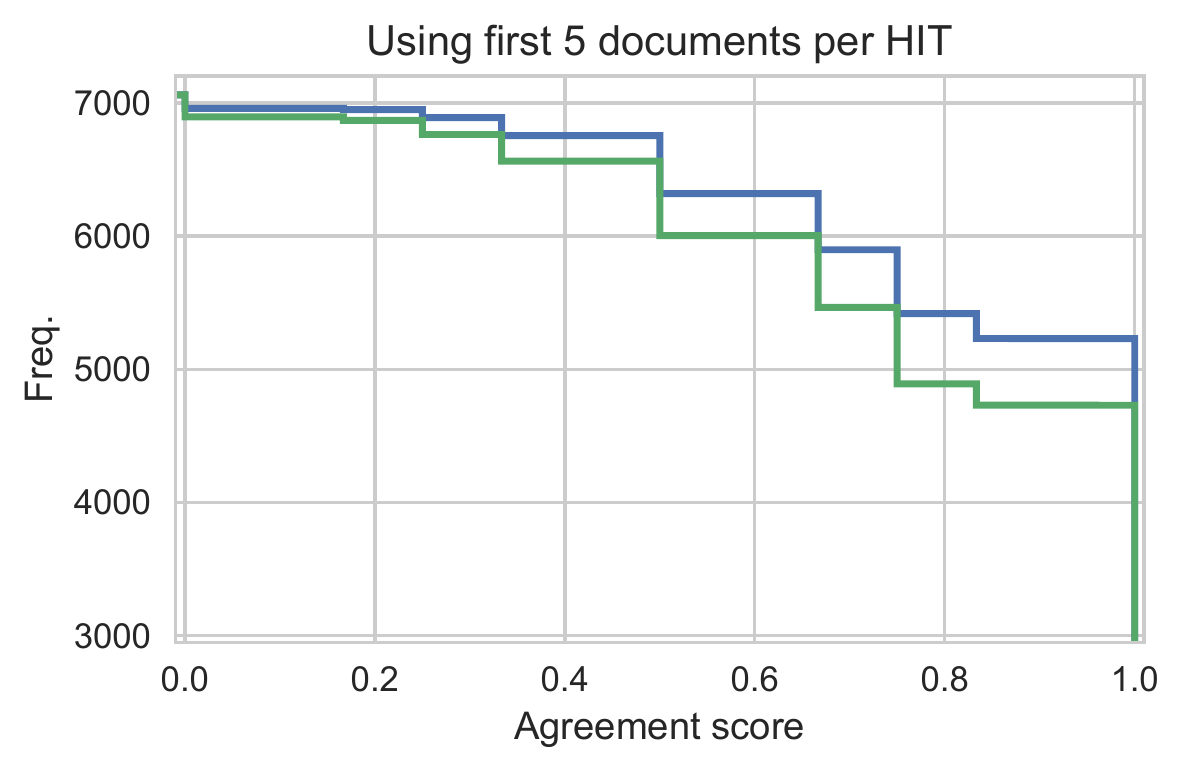}}
        \subfigure{\label{cs:s100:fig:r:6}\includegraphics[width=.31\linewidth]{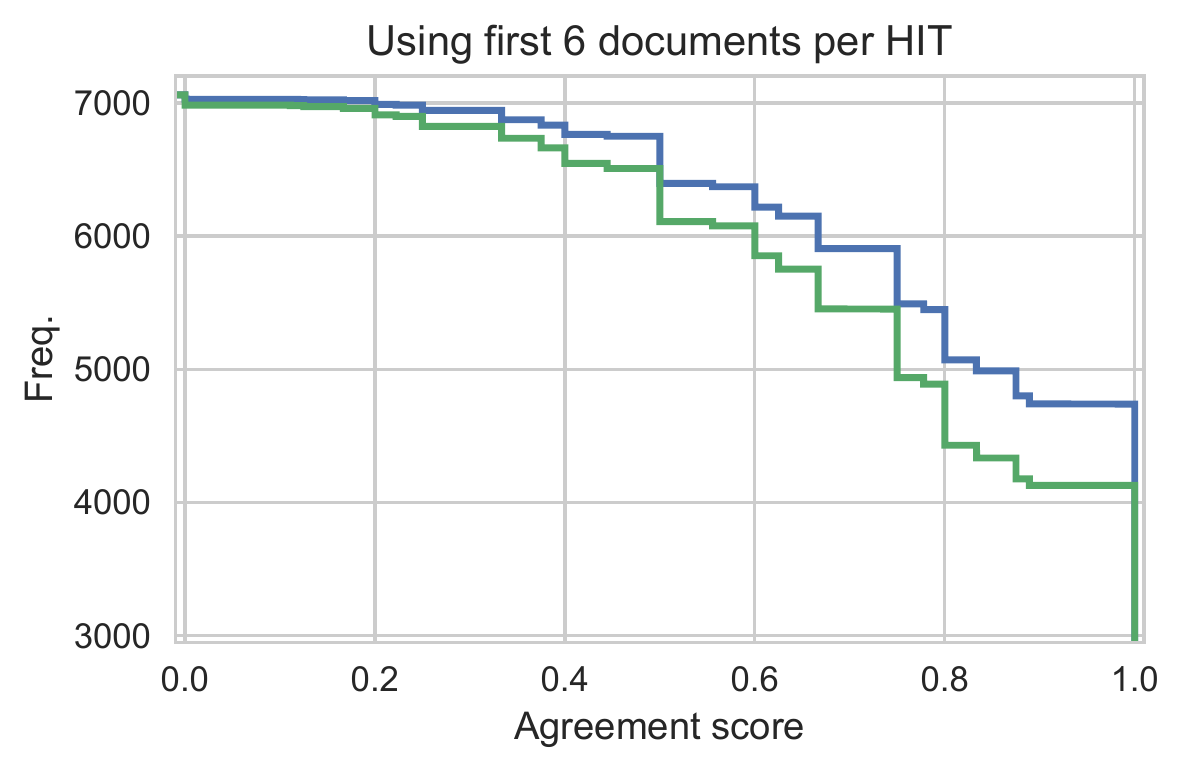}}
    \subfigure{\label{cs:s100:fig:r:7}\includegraphics[width=.31\linewidth]{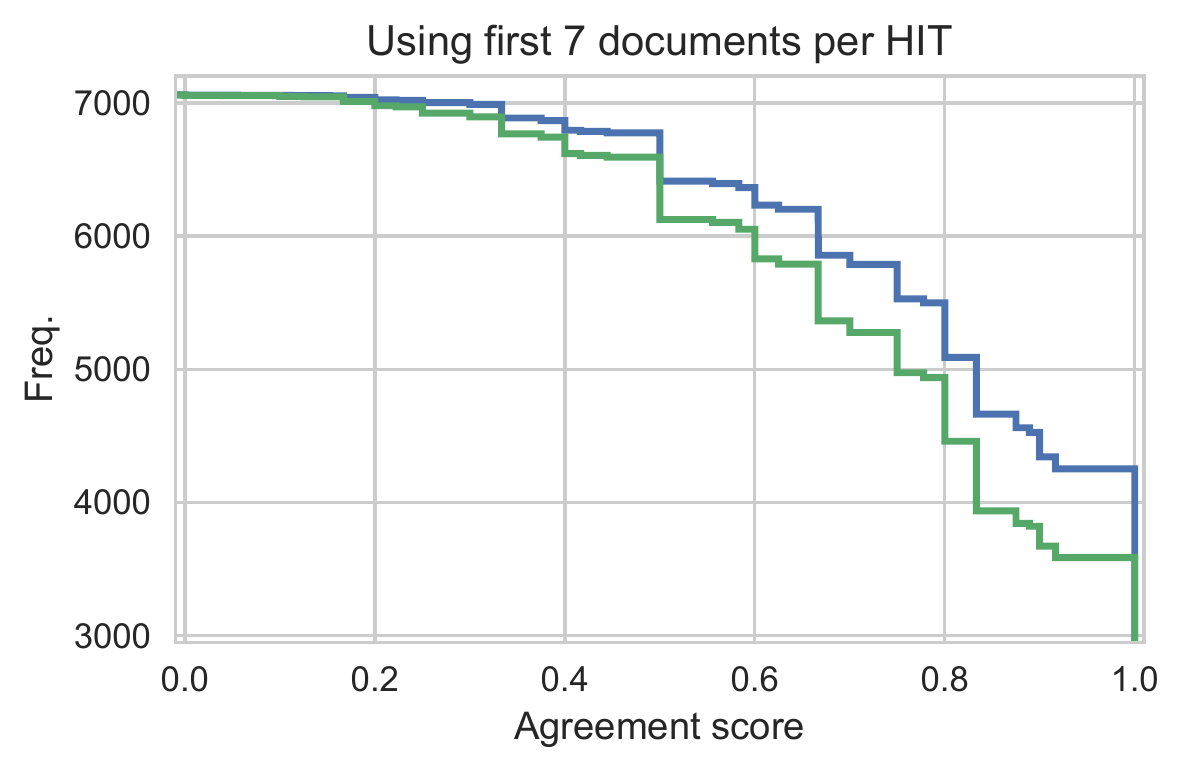}}
\vspace*{-5mm}
\caption{Complementary cumulative distribution function of pairwise agreement as defined in Section~\ref{cs:s100:sec:agreement-def} of S100 and ME with TREC, when using the first $i \in \{2, \ldots, 7\}$ documents for each HIT ($i=8$  shown in Figure~\ref{cs:s100:fig:Robust_FewDocPerUnit1}). 
}
  \label{cs:s100:fig:Robust_FewDocPerUnit}
\end{figure*}

Figure~\ref{cs:s100:fig:Robust_FewDocPerUnit} shows how pairwise agreement varies when using shorter HITs (i.e., looking at judgement quality based on the document position in the HIT). 
For any HIT length, the pairwise agreement of individual judgements is higher for S100 than ME with an increasing gain in agreement the longer the HIT.


\subsection{Fewer judgements per Document}

\begin{figure*}[tbp]
  \centering
  \includegraphics[width=0.75\linewidth]{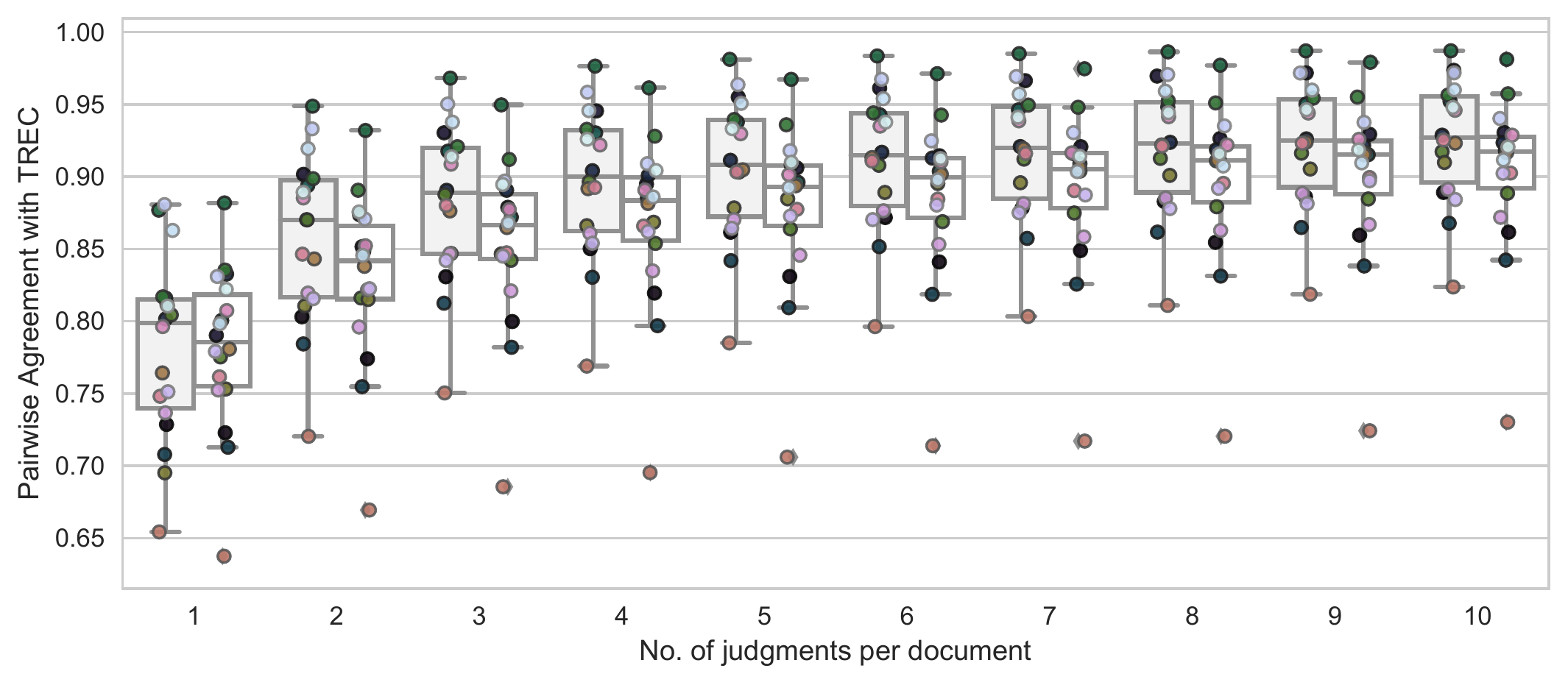}
  \vspace*{-2mm}
\caption{Pairwise agreement of S100 (boxplot on the left) and ME (boxplot on the right) with TREC, when sampling randomly (mean over 100 repetitions) $i \in \{1, \ldots, 10\}$ judgements for each document. Each dot is a topic. 
}
  \label{cs:s100:fig:Robust_FewJudgPerDoc}
\end{figure*}

In both S100 and ME, 10 judgements per document have been collected. When using fewer assignments, as it is expected, the quality of the aggregated judgements decreases. We analyze this by showing how aggregated pairwise agreement decreases when using a random subsample of the 10 judgements in Figure~\ref{cs:s100:fig:Robust_FewJudgPerDoc}: the figure reports the pairwise agreement average values, for each topic, over 100 random repetitions.
For all sizes of the subsample, S100 median pairwise agreement is higher for S100 than ME.\footnote{We observed the same result when computing pairwise agreement against Sormunen's S4 judgements but we did not include the figure for space limitations.} 
%
These results show a higher robustness to fewer assignments of S100 as compared to ME, making it a more economically viable scale to use to collect crowdsourced relevance judgements.

\section{Running out of Values}\label{cs:s100:sec:runningout}
As recalled above, the ME scale has the advantage that the assessor never ``runs out of values'', neither (i) at the scale extremes (which are unbounded) nor (ii) inside the scale (which is continuous). This advantage is lost when using limited scales as S100, and it is of course further compounded for scales with a lower amount of values like S4 and TREC. In this section we aim at understanding if these potential problems have actually been a practical constraint for the judges using the S100 scale.
An initial analysis can be performed comparing the number of ``back'' actions crowd workers performed which indicate a desire to change or look at their previously judged documents. This value is much higher in S100 than in ME. In ME the number of single ``back'' actions was 106, and the number of two or more ``back'' actions was 9 \cite[Table I]{Maddalena:2017:CRM:3026478.3002172}, whereas for S100 these two figures are 113 and 182 respectively: although the numbers are still very limited (more than 95\% of S100 workers did not use the ``back'' button at all), the differences are noticeable and might be ascribed to a higher difficulty in finding the ``right'' relevance score.
In the next two subsections we present a more detailed analysis, addressing the two issues (i) and (ii). 

\subsection{Reaching the Scale Boundaries}


\begin{table}
\centering
\caption{The number of cases with exactly $k$ 0s or 100s in the same HIT, and the corresponding number of potential run-out-of-values cases in the S100 dataset.  \label{cs:s100:tab:0sand100s}}
\vspace*{-2mm}
\begin{tabular}{l  r@{ }r r@{ }r@{ }r@{ }r@{ }r@{ }r r}
\toprule
$k$ & 0 & 1 & 2 & 3 & 4 & 5 & 6 & 7 &Tot\\
\addlinespace
0s &\emph{2355}&\emph{803}&694&689&736&786&641&355\\
100s &\emph{3796}&\emph{1808}&846&398&133&55&19&4\\
\midrule
$k-1$&  &  & 1&2&3&4&5&6\\
\addlinespace
0s*&0&0&694&1378&2208&3144&3205&2130&12759\\
100s*&0&0&846&796&399&220&95&24&2380\\
\bottomrule
\end{tabular}
\end{table}


Table~\ref{cs:s100:tab:0sand100s} shows in the first part the number of HITs with ``boundary judgements'', i.e., with exactly $k$ 0s or 100s in the S100 dataset. The HITs without or with only one boundary judgement (i.e., only one 0 or one 100, in italics in the table) do not create any potential problem; instead, the boundary judgements after another boundary judgement (i.e., in the same HIT, one or more 0s after a first 0, or one or more 100s after a first 100) might be cases in which the worker could have used a lower or higher value if available. 
So, the HITs with at least two boundary values (the following columns) are those in which, at least potentially, the worker ``ran out of values'' at each extreme of the scale. A lower (higher) value,  if available, could have been selected for each of the $k-1$ ``boundary judgements'' after the first one. The numbers of such ``boundary judgements following other boundary judgement(s)'' are quantified in the lower part of the table: these are obtained multiplying by $k-1$ the figures in the previous two rows (for example, the 2'208 value is obtained as 736x(4-1): in 736 HITs the workers used 0 for 4 times, and the last three in each HIT are candidates for ``run-out-of-value'' cases). 

To provide an understanding of the frequency of the problem, let us remember that we had a total of 7'059 HITs, each one containing 8 documents. Since the first expressed judgement in each unit can not be preceded by another (boundary) judgement, we have 7'059x7=49'413 judgements that could have manifested the problem. Of those, the problem manifests for a total of 12'759+2'380=15'139 cases (31\%). Of course this is not negligible: in almost one case out of three a worker might have been restricted in expressing the true intended judgement. 
However, this also means that in 69\% of the expressed judgement we can say that  the worker was not restricted by the boundaries of the S100 scale.
%
%
%
Moreover, these 31\% of cases are only potential problems, as it might well be that the worker intended to express exactly the same judgement and did not actually run out of values. Therefore, we further analyzed these potential run-out-of-values cases in two ways.

First, we looked in our S100 dataset what fraction of the boundary judgements 0 (or 100)  expressed by a worker in a HIT after the first boundary judgement corresponds to a document that has a strictly lower (higher) aggregated score. These are cases in which the worker ran out of values, assuming that the intended score corresponds to the aggregated one. This happened 7'523 cases, namely 50\% of the 15'139 potential problematic cases, or 15\% of the total 49'413 judgements expressed.

Second, we looked in the ME dataset how many of the 15'139 potential run-out-of-value cases received an ME score that was lower (for the 0s), or, respectively, higher (for the 100s) than the first corresponding boundary judgement in the unit. These are cases in which the worker ran out of values assuming that the judgement expressed by the ME worker in the corresponding unit was exact.
This happened for 4'309 cases, namely 28\% of the 15'139 potential problematic cases or 9\% of the total 49'413 judgements expressed.

So, the two analyses roughly agree that in only around one out of ten cases the bounded scale seems to have indeed limited the assessor, and therefore in about 90\% of the expressed judgements the S100 scale did not create any obstacle to judgement expression. 

\subsection{Discrete vs. Continuous Scale}

\begin{figure}[tbp]
  \centering
    \includegraphics[width=\linewidth]{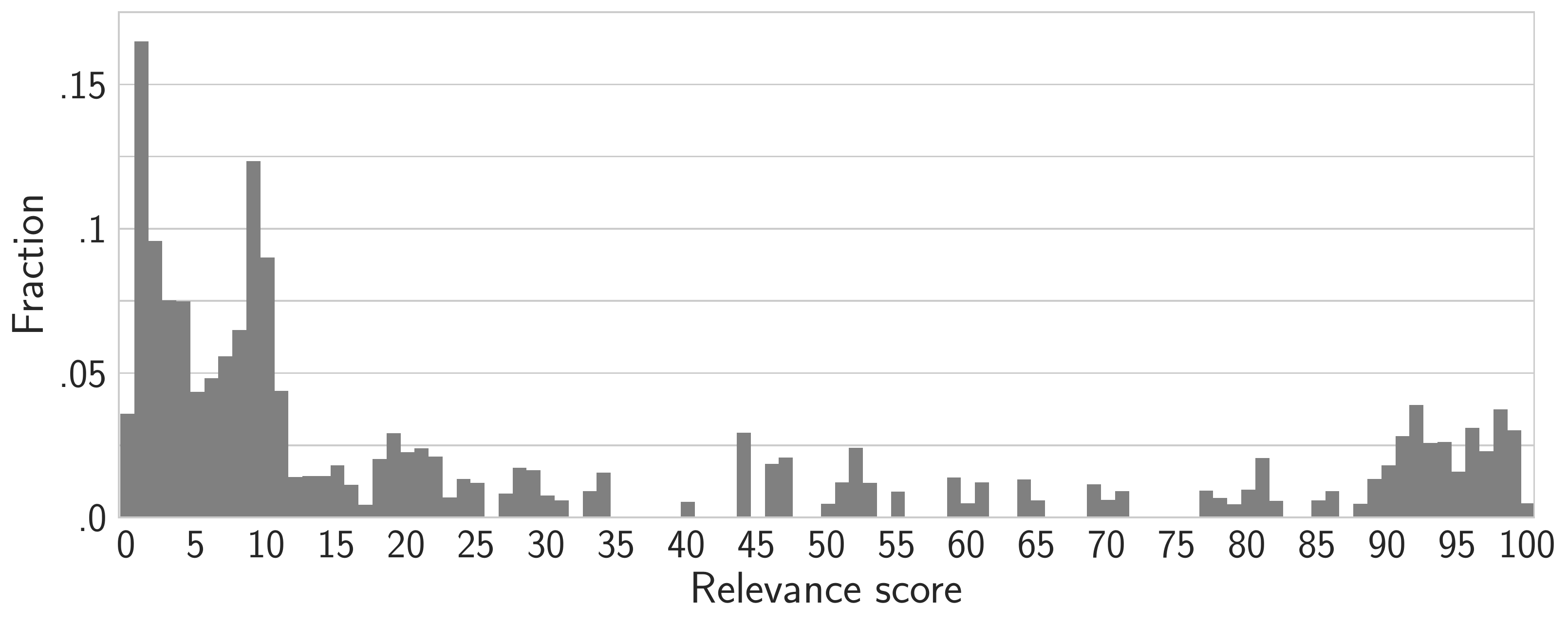}
\vspace*{-2mm}
\caption{Fraction of expressed judgements, for each value of the S100 scale, that might have been constrained by previously expressed judgements in the same HIT.   
  \label{cs:s100:fig:blocks}}
\end{figure}

The second situation in which the S100 scale could constrict judgement expression is when contiguous values are selected, thus making impossible for the worker to select another, different, value in between in the following judgements in the same HIT.
We counted in our S100 dataset how many scores $x$ were preceded  by both $x$ and $x+1$ (or by both $x-1$ and $x$) in the same HIT. These are the cases in which, potentially, the worker could not give a $y \in ]x-1,x[$ (or $y \in ]x,x+1[$) value because of the discrete scale. There were 1'911 such cases, out of the 7'059x6=42'354 possible ones (as we need to count starting from the third judgement in each HIT), which is less than 5\%. Notice again that this fraction is consistent with the number of `back' actions. Moreover, the vast majority of these cases (around 1'500) concern judgements between 0 and 10, which could be considered not critical (as it is probably more important to focus on the ``relevant'' end of the scale rather than on the ``not relevant'' end). 
This is confirmed by Figure~\ref{cs:s100:fig:blocks} that shows, for each value between 0 and 100, the fraction of judgements at a given level that may have been affected by previous judgements given at the same level. These values indicate the percentage of cases in which there may be a limitation because of the use of S100 as compared to ME which would allow to give a slightly higher or lower judgement score as compared to previous ones. Note that this is an upper bound of such expressiveness limitation as assessors may have assigned the same score multiple times on purpose.
From Figure~\ref{cs:s100:fig:blocks} we can also observe that most problematic cases are, as expected, at the boundaries of the scale but also that such cases are not prevalent (about 5\% of judgements are affected). More potentially constrained judgements are present at the lower end of the S100 scale. This should be less problematic than constrains at the upper end of the scale as we can expect more score ties for not relevant documents.

In summary, taking into account that these are only potential problems, since it is possible that the worker indeed intended to express the very same score again, we can conclude that these do not seem worrying problems in practice and that the theoretical constraints imposed by the S100 scale on judgements expression did not significantly obstacle the workers in practice.

\section{judgement Time}\label{cs:s100:sect:time}




We  compared the time required by crowd workers to assess documents using S100 and ME considering that the experimental design was consistent across the two studies with the only 
difference being the way relevance was expressed by assessors (i.e., a number versus a slider from 0 to 100).

\begin{figure}[tbp]
  \centering
    \includegraphics[width=.6\linewidth]{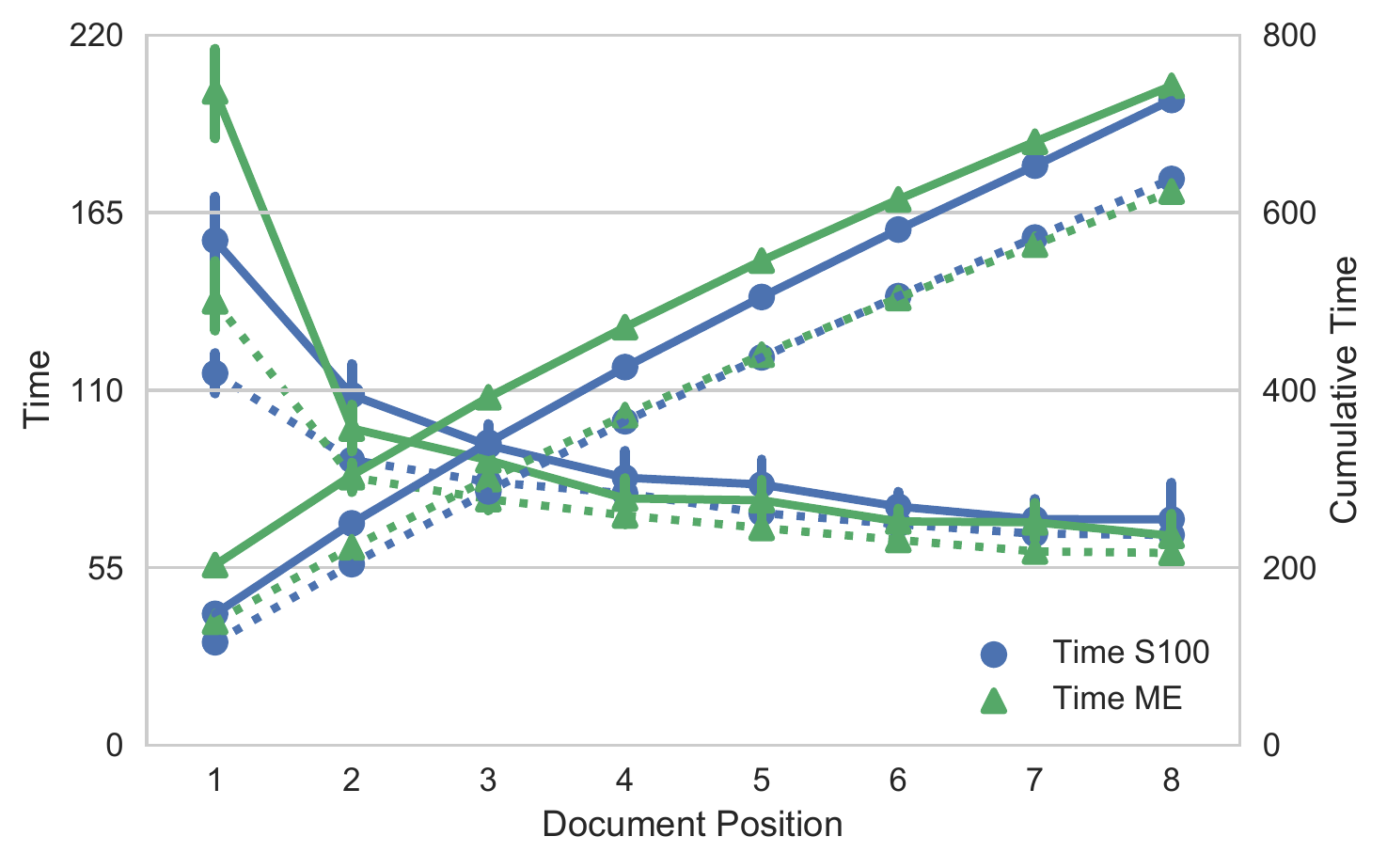}
\vspace*{-5mm}
\caption{Mean time (seconds) for expressing relevance judgements, over the 8 positions in each HIT, considering either only the first HIT for each worker (straight lines) or considering all HITs (dotted lines). All times difference are significant (Wilcoxon signed-rank test $p<0.01$). 
  \label{cs:s100:fig:times}}
\end{figure}

The dotted series in Figure \ref{cs:s100:fig:times} show the mean time taken by crowd workers to assess documents, based on the order in which they were presented. The cumulated time shows that S100 leads to slightly quicker judgements than ME for the first 5 documents. Starting from the 6th document in the HIT, the use of the ME scale leads to overall faster judgements.
This can be explained by the fact that workers are probably at first disoriented by the uncommon ME scale, but they become more efficient in using it as they progress completing more judgements. The time behavior is quite stable across workers and topics (as it can be seen by the small quartile bars in Figure \ref{cs:s100:fig:times}). Moreover, time differences are quite small, though always statistically significant, for each document position in the HIT: apart from the first position, the differences are around 5s for a total judgement time of 50-100s.

\begin{figure}[tbp]
  \centering
    \includegraphics[width=0.6\linewidth]{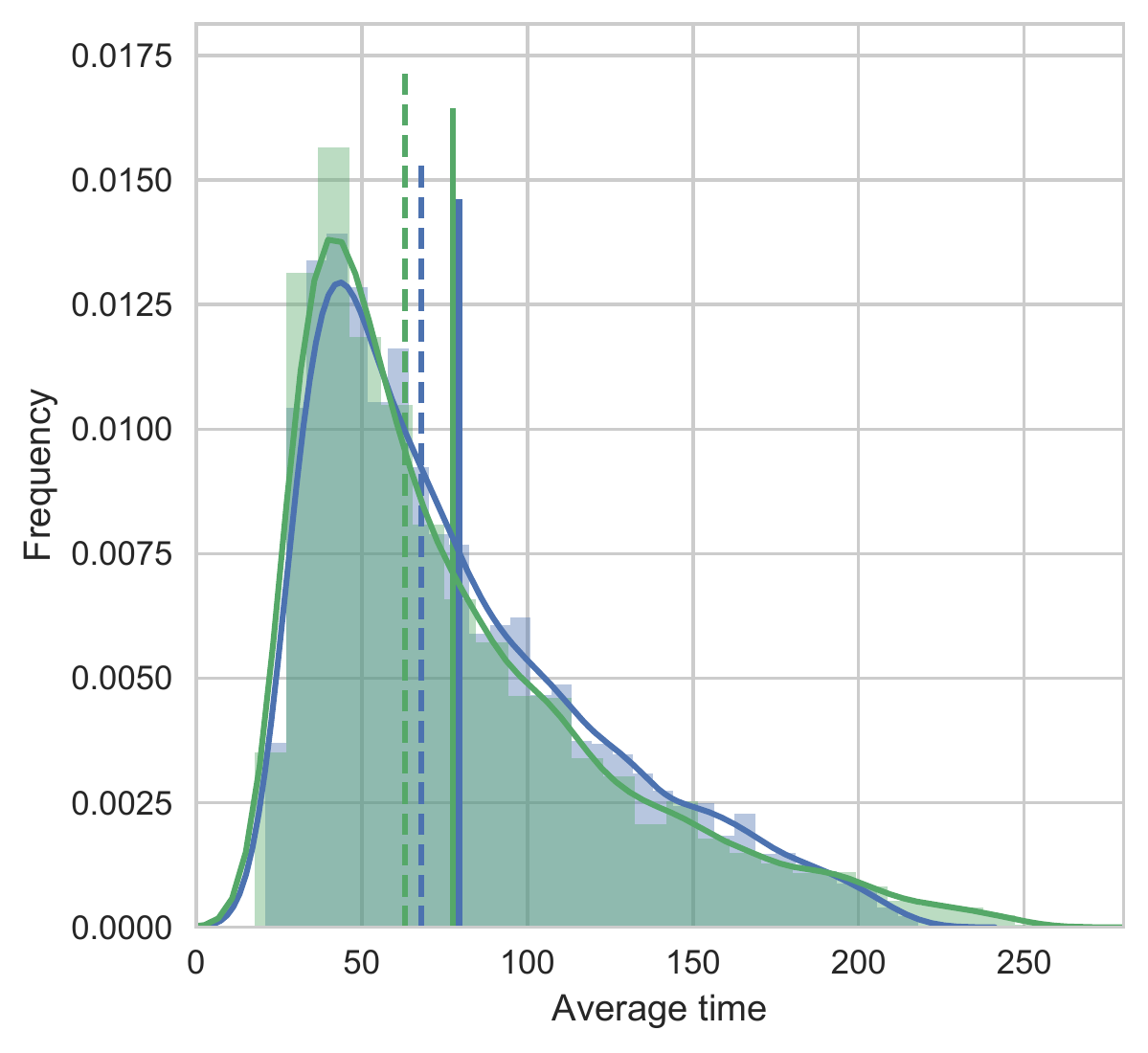}
    \vspace*{-2mm}
\caption{Judging time distributions, combined for all topics, for S100 and ME. Average (vertical lines) and median (dashed vertical lines) time is also shown.
\label{cs:s100:fig:time:alltopics:avg}}
\end{figure}

Figure~\ref{cs:s100:fig:time:alltopics:avg} shows in more detail the distributions of judging time for both S100 and ME: the two are rather similar, with ME having a slightly longer tail. The behavior does not significantly change across individual topics (not shown for brevity). Given the small and constant time difference, a possible explanation is the presence of the slider, which might require a small longer amount of time than inserting a number in a text box, especially on mobile devices as shown by \citet{modusoperandi}. Further analysis is needed to confirm this conjecture, and we leave that to future work. 

To further study the learning effects, we repeated the same analysis considering only the first HIT performed by each worker (remember that workers could redo the task, on a different topic). Going back to Figure~\ref{cs:s100:fig:times}, the straight, not dotted, lines represent only these first HITs, for both S100 and ME. We notice three main variations with respect to the dotted lines: (i) overall, average times are higher, indicating that indeed some learning effect is present and workers become more efficient after the first HIT; (ii) the time difference on the first expressed judgement is larger, thus confirming that starting with the ME scale is somehow more difficult than with the S100 one; and (iii) on this data, the ME cumulative curve stays above the S100 one, meaning that the disadvantage cumulated on the first document by ME can not be recovered even after 8 judgements are judged. 

Overall, we can conclude that time does not seem a critical factor when choosing between S100 and ME: differences are small and a longer time due to learning effects on ME tends to be compensated after some documents are judged.

\section{Conclusions}\label{cs:s100:sec:con}
In this chapter we presented a systematic study comparing the effects of different relevance scales on IR evaluation. We have shown many advantages of the S100 scale as compared to coarse-grained scales like binary and S4 and to unbounded scales like ME. 
S100 preserves many of the advantages of ME like, for example, allowing to gather  relevance judgements that are much more fine-grained than the usual binary or 4-value scales. Assessors use the full spectrum, although sometime with  a preference for scores that are a multiple of ten.
S100  has also demonstrated advantages over ME in terms of agreement with judgements collected on a binary and four level scales. This can be explained by the fact that ME requires a step of score normalization which makes judgements less comparable across assessors and topics. On the other hand, S100 leads to more similar judgements (i.e., higher agreement) to the classic binary and four level scales.
S100 has shown to be more robust than ME in terms of less assessors per documents (to be aggregated, as typically done for crowdsourced relevance judgements) and to less documents per assessor thus giving the freedom to crowd workers to perform few or many judging tasks.
Our results show that the potential constraints in judgement expression that the S100 scale might create with respect to the complete freedom of ME almost do not occur in practice since about 90\% of the judgements did not suffer from this problem.
The S100 scale also seems easy to learn for the workers and turns out to be faster than ME for short HITs with 5 or less documents to be judged, and of comparable speed for longer HITs.
Overall, our results show that S100 is an effective, robust, and usable scale to gather fine-grained relevance labels.


\chapter{Transforming Relevance Scales used for IR Evaluation} \label{chapt:cs:trans}

This chapter deals with the usage of crowdsourcing for retrieval evaluation and in particular discusses transofmrations between relevance scales.
Section~\ref{cs:trans:sec:intro} introduces, motivates the work, and frames research questions,
Section~\ref{cs:trans:sec:experiment} presents the experimental setting,
Section~\ref{cs:trans:sec:results} presents and discusses the results,
Finally, Section~\ref{cs:trans:sec:conc} concludes the chapter.


\section{Introduction and Research Questions}\label{cs:trans:sec:intro}

One of the design decision to make when creating an Information Retrieval (IR) evaluation collection is which \emph{relevance scale} to use for judgements. While, historically, binary relevance judgements have usually been collected, and metrics based on binary relevance like, e.g., precision, recall, and average precision, have been used, more recently multi-level relevance scales have become popular. There are several reasons for this. One reason is the availability of gain-based evaluation metrics like NDCG \cite{Jarvelin:2002:CGE:582415.582418}, which are defined also for non-binary relevance judgements.
 
Another reason can be found in the increasingly popular use of Crowdsourcing, which has become a standard methodology to create relevance labels for IR test collections.
%
%
One of the main challenges of the use of crowdsourcing to collect relevance judgements is quality. To address this issue a number of approaches have been proposed including task design methods \cite{time,priming} as well as multiple assignments of the same topic-document pair in order to aggregate judgements from different workers to improve the overall judgement quality \cite{Venanzi:2014:CBA:2566486.2567989,dawid1979maximum}. Some of the existing aggregation functions produce a value that is still in the original scale; some others like, e.g., arithmetic mean or median, can produce an aggregated value that does not belong to the original scale, thereby obtaining more fine-grained relevance values.
Following this trend, recent research has specifically and deliberately looked at the effect of unbounded and fine-grained relevance scales, thereby producing relevance labels on scales that have 101 values \cite{Roitero:2018:FRS:3209978.3210052} or are even unbounded \cite{Turpin:2015,Maddalena:2017:CRM:3026478.3002172}.

The existence of IR evaluation collections using different relevance scales does not allow for possible comparisons of evaluation results or merging of collections. Additionally, the evaluation of IR systems is limited to the relevance scale used by the specific test collection.
This leads to a natural research question: \textit{how to transform one scale into another?} Indeed, quite often in the past, IR researchers have transformed relevance judgements collected on a multi-level scale into a coarser-grained scale, e.g., from a 4-level (highly relevant, relevant, marginally relevant, and not relevant) scale to binary for the sake of using IR evaluation metrics that require binary judgements (e.g, Average Precision, Mean Reciprocal Rank, etc.) \cite{inex03,inex04,ntcir4,ntcir5,ntcir6}. More in general, whenever an evaluation metric requires a coarse-grained scale, the relevance judgements collected on a fine-grained scale need to be transformed.

In these examples, researchers have arbitrarily defined certain scale \textit{transformation thresholds} to map relevance judgements from a \emph{source scale} (e.g., 4-levels labelled as 0, 1, 2, and 3) into a \emph{target scale} (e.g., binary with levels labelled 0 and 1), for example, by transforming 0 and 1 judgements in the source scale as 0 in the target scale and 2 and 3 in the source scale  as 1 in the target scale, and doing so for all topics in the collection. Another solution has been to transform 0 into 0 and 1, 2, 3 into 1. These approaches were named ``rigid'' and ``relaxed'', respectively, in early NTCIR editions (e.g., NTCIR-4 CLIR task) \cite{ntcir4}.
Thus, the problem is defined as finding the best thresholds in the source scale (which we call \textit{\cuts} in this chapter) that allow us to then transform judgements from the source to the target scale.

We believe that a more principled approach is needed, and possible. The \cuts should not be selected arbitrarily but rather on the basis of more informed decisions. For example, the same \cut has been typically used uniformly across all topics, but different \cuts for different topics (for example, based on the number of relevant documents available for that topic) may be more accurate. Even more, different \cuts might be needed for different assessors.
Indeed, there seems to be some evidence that the ``one-cut-fits-all'' approaches are inadequate since different topics show different relevance profiles \cite{Turpin:2015,Maddalena:2017:CRM:3026478.3002172}.
To the best of our knowledge, such a systematic study of scale transformations has never been performed so far.

Besides aiming at better understanding scale transformations, we also mention that another reason for this research is to save resources: in the above cited large scale crowdsourcing experiments \cite{Turpin:2015,Maddalena:2017:CRM:3026478.3002172,Roitero:2018:FRS:3209978.3210052}, we estimate that gathering relevance judgements costed about \$100 for each topic considering 129 systems and a pool depth of 10. If fine-grained to coarse-grained scale transformations were available and reliable, an experimenter could collect the fine-grained relevance labels only, and rely on scale transformations when coarse-grained labels are needed. Equivalently, one could run all the experiments anyway but scale transformations would allow to gather more reliable data with the same resources (i.e., money, time to design and run the experiments, data cleaning, etc.). 

Finally, although in this chapter we focus on relevance judgements, the same issue might be found in other scenarios like fake news detection (where, for example, six levels have been used  \cite{fakenewsdataset} but it seems reasonable to use accuracy as an effectiveness measure, which requires binary values), 
sentiment classification where 5-star ratings are often transformed in three classes \cite{threesentimentclasses} or binarized in an ad-hoc manner (e.g., reviews with more than 3 stars are considered as positive and less than 3 stars as negative \cite{moraes2013document}), and many other classification-like problems. Our results might be useful also for those cases.
While we do not address the last two issues in this chapter, we rather present an in-depth analysis of the effects of relevance scale transformations on IR evaluation and present a comprehensive set of transformation approaches for crowdsourced relevance judgements systematically observing the effects on IR evaluation results. We also show that judgements collected for different topics are best transformed using different \cuts in order to maximize assessor agreement 
and IR system ranking correlation.
The main contribution of this chapter is an understanding of the effects of transforming relevance judgements into a more coarse-grained scale than that used to originally collect the judgements and a set of guidelines on how to best select \cuts to transform judgements into a different scale.


Thus, the research questions we focus on are the following:
\begin{enumerate}[label=RQ\arabic*,ref=RQ\arabic*]
    \item How to transform relevance scales with target scale data?
    \item How to transform scales without target scale data?
    \item Is there any difference when transforming expert or crowdsourced judgements?
    \item Should we transform scales differently for each topic or in the same way for an entire judgement collection?
    \item What is the effect of assuming unjudged documents as not-relevant on transforming scales?
    \item Which scale transformation method should we adopt to obtain IR evaluation results more similar to when using expert judgements?
\end{enumerate}

\section{Experimental Setting}\label{cs:trans:sec:experiment}
In this section we describe the datasets, the measures, and the transformation methods used in our experiments. We also briefly discuss the effect of unjudged documents.

\subsection{Relevance judgement Datasets}\label{cs:trans:sec:data}


Aiming at investigating the effect of scale transformations, we identified a set of 18 search topics from TREC-8~\cite{trec8} judged by NIST experts using a binary scale. Some of the documents retrieved for such topics were subsequently re-judged by Sormunen \cite{Sormunen:2002:LRC:564376.564433} on a 4-level ordinal relevance scale: \nn--not relevant (0); \mm--marginally relevant (1); \rr--relevant (2); \hh--highly relevant (3).
Then, Roitero et al. \cite{Roitero:2018:FRS:3209978.3210052} ran a crowdsourcing re-assessing exercise using a 101-level ordinal relevance scale.
Such a crowdsourced reassessment produced a total of \num{4269} topic-document judgements, of which 90.9\%  have binary TREC relevance judgements available, and 18.9\%
have Sormunen 4-level ordinal judgements available. 
The differences in the overlap of judged documents among these collections are due to the different sampling strategies adopted: the 4-level collection  \cite{Sormunen:2002:LRC:564376.564433}  contains only a sample of the documents judged by TREC, constructed by skipping many documents already judged as not relevant by TREC assessors; the 101-level collection contains judgements for all documents retrieved in the first 10 ranking positions by at least one system; and in TREC-8 some systems did not contributed to the pool (so they might have unjudged documents in the first ranking positions). 
Additionally to these datasets, in this chapter we also use new crowdsourced re-assessments of the same documents used by \citet{Roitero:2018:FRS:3209978.3210052}
using both a binary  and a 4-level scale. 

\begin{table}[tb]
    \centering
    \caption{The five datasets we use in this chapter.}
    \begin{tabular}{llll}
        \toprule
         & \textbf{2-levels}  & \textbf{4-levels}  & \textbf{101-levels} \\
         \cmidrule{2-4}
         \textbf{Expert}     & TR2 (TREC-8)   & So4 (Sormunen)  & \\
         \textbf{Crowd}      & S2        & S4        & S100 \\
         \bottomrule
    \end{tabular}
    \label{cs:trans:tab:datasets}
\end{table}

Thus, in summary (see Table~\ref{cs:trans:tab:datasets}), we use two expert-generated collections: one generated by NIST assessors for TREC-8 using a binary scale (in the following we refer to this data set with TR2, for TREC and binary) and one generated by assessors used for \cite{Sormunen:2002:LRC:564376.564433} using a 4-level scale (So4).
We also use three crowdsourced collections: one generated using a 101-level scale (S100) where workers assign values by a slider bar in the same way as in \cite{Roitero:2018:FRS:3209978.3210052}, one using a 4-level scale (S4), i.e., the same one used in \cite{Sormunen:2002:LRC:564376.564433}, where workers choose the values by radio buttons, and one using a binary scale (S2).
In the crowdsourced collections we use in this chapter, each crowd worker had to judge 8 documents for a single topic in a HIT (Human Intelligence Task, the basic unit of work to be performed by a crowd worker) and each document has been independently judged by 10 different crowd workers to be able to aggregate their judgements and improve the dataset quality. The three crowdsourced datasets contain exactly the same \num{4269} topic-document pairs and only differ for workers who completed the HITs and for the relevance scale used to collect the judgements.
The crowdsourced data collections have received ethics approval from the review board by the authors' institutions.


\subsection{Measuring the Similarity of Relevance judgement Sets}
\label{cs:trans:sec:measure}
When transforming a judgement set into a different target scale, we are able to use the transformed judgements to evaluate IR system effectiveness and to compare the evaluation results with those obtained using the original judgement set. We can do this by means of Kendall's $\tau$ correlation between the IR system rankings generated using the two evaluation sets.
Another method to compare the original relevance judgements with those transformed to the target scale is to rely on assessor agreement measures.
Using all judgements in a crowdsourced IR collection, we measure the agreement among different assessors who contribute to the judgement set, which we define as \emph{internal agreement}. When using this approach to compare original and transformed judgement sets, we can measure how much the  judgements transformed in the target scale agree among themselves as compared to the internal agreement of the judgements in the original scale. 
%
When another dataset collected in the target scale is available, we can also define \textit{external agreement} by measuring assessor agreement between the transformed judgements and the ones collected natively in the target scale, to check how the transformed judgements align with the (expert or crowd) judgements in the other available target-scale dataset.
Note that by using internal agreement we identify the best \cut that maximizes the agreement of the transformed judgements with themselves, while by using external agreement we select the best \cut where the transformed judgement set is the closest to the one obtained in the target scale w.r.t. judgement quality.
In the following section we report our experimental results comparing different transformation methods in terms of assessor agreement using Krippendorff's \cite{krippendorff2007computing} $\alpha$ and in terms of IR system effectiveness evaluation using $\tau$.

As shown in previous research \cite{Roitero:2018:FRS:3209978.3210052,Maddalena:2017:CAA:3121050.3121060}, there is large variance in assessor agreement values across different topics;
thus, we measure per-topic agreement to perform per-topic \cuts and transformations.
This approach is unconventional for the classical IR evaluation setting, where a certain relevance cutoff is chosen a priori to be the same over all the topics in the collection. 
Note, however, that a per-topic approach does not create issues in performing IR evaluation. For example,  
NDCG values for different topics would still be comparable despite the fact that they originate from topics with different \cuts
since the relevance scale and gain values across topics are the same, and due to the fact that the computed gains are normalized.
Different \cuts over different topics may affect the number of relevant documents in each topic, which is anyway something that varies considerably across topics in IR evaluation collections \cite{Moffat:2008:RPM:1416950.1416952}. Indeed, per-topic \cuts may help reducing the variance in the number of relevant documents across topics.

\subsection{Scale Transformation Methods}\label{cs:trans:sec:methods}

Given the datasets used in this chapter (see Table~\ref{cs:trans:tab:datasets}), we perform the following transformations of
relevance judgement datasets into a target scale: So4 into binary, S4 into binary, S100 into binary, and S100 into 4 relevance levels.
Note that we only perform transformations from fine-grained scales to coarse-grained scale as the opposite would require new information not available in the source dataset. 

We distinguish two main classes of approaches to transform the relevance scale used by a collection of judgements.
In the former only the judgement set to be transformed is available (in the source scale); in this case we use internal agreement (see Section~\ref{cs:trans:sec:measure}). In the latter scenario both the set to be transformed and a set of judgements created in the target scale (either by experts or crowdsourced) are available, and therefore we use external agreement (see Section~\ref{cs:trans:sec:measure}).

\subsubsection{Single Dataset Scale Transformation}\label{cs:trans:sec:singleDatasetMethod}
To transform a crowdsourced judgement set into a target scale we need to select one of the possible \cuts (e.g., to transform a 4-level judgement set into binary we have three possible choices). 
There are different possible approaches we can follow to decide on the best \textit{\cuts} for our crowdsourced judgement sets; for each of them we provide a long name, a short one (defined more in detail in Table~\ref{cs:trans:tab:transformations}), and a description:
\begin{description}
    \item [HIT-centric, transform then aggregate (H\_t+a$^1$).] Given all documents judged by an individual crowd worker, we can first transform each individual judgements into the target scale (using one of the possible \cuts) and then do the same for the other 9 judgements collected from the crowd for each document. We can then aggregate the 9 judgements for the same document (e.g., using majority vote or another aggregation function) thus obtaining two transformed judgements in the target scale: the individual worker judgement and the crowd aggregated judgement. This allows us to compute the Krippendorf's $\alpha$ agreement between an individual worker with respect to rest of the crowd.
    By computing this version of inter-annotator agreement for all possible \cuts, we are able to identify the \cut which
    maximizes the $\alpha$ value,
    in order to keep the highest judgement agreement.
    \item [HIT-centric, aggregate then transform (H\_a+t$^1$).] As variant of the previous approach, we first aggregate the 9 crowd judgements for the same document in the source scale and  then transform both the individual worker judgement and the crowd aggregated judgement into the target scale to compute $\alpha$ agreement for a specific \cut.
    \item [Topic-wide $\alpha$ (Tw\_$\alpha^1$).] A third  approach is to compute $\alpha$ on the entire worker-document matrix of judgements for a topic transformed in the target scale for each possible \cut to find the one that maximizes $\alpha$.
    Note that this method has a potential issue given by the fact that $\alpha$ is not a very reliable measure for sparse matrices~\cite{phi}.
\end{description}

\subsubsection{Double Dataset Scale Transformation}\label{cs:trans:sec:multiDatasetMethod}
In this context, we assume that both a judgement set in the source scale and one in the target scale are available. For example, besides a fine-grained crowdsourced judgement set, we could also have expert judgements collected in the target scale available (e.g., TR2 binary judgements) which we might leverage to better decide on the best \cut to be used on the source scale dataset (e.g., S4).
In this case, the possible scale transformation approaches are:
\begin{description}
    \item [HIT-centric (H$^2$).] All 8 documents judged by an individual crowd worker in a HIT are first transformed into the target scale using a certain \cut. Then, the transformed judgements are compared to the second dataset (that was created using the target scale) to compute the $\alpha$ agreement value for a given worker and \cut.
    We then make an average  of $\alpha$ values over all workers contributing judgements for a certain topic and, thus, are able to identify the best \cut using the highest $\alpha$ value.
    \item [Document-centric aggregate then transform (D\_a+t$^2$).] We first aggregate all relevance judgements collected for a document (e.g., 10 in our experimental design) in the source scale and then transform the aggregated document judgements into the target scale using a certain \cut and we then measure the $\alpha$ agreement score for the specific \cut. The best \cut is identified using the highest $\alpha$ value.
    \item [Document-centric transform then aggregate (D\_t+a$^2$).] A variant of the previous approach consists in first transforming judgements into the target scale using a certain \cut and then aggregating them (e.g., by majority vote) to compare them with judgements in the target scale (e.g., by experts) and measure $\alpha$ agreement for the specific \cut. The best \cut is identified using the highest $\alpha$ value.
\end{description}
 
We additionally distinguish two possible types of judgement datasets in the target scale: expert judgements (e.g., TR2 binary judgements) or crowd-generated aggregated judgements collected natively in the target scale (e.g, S2).

\begin{table}[tbp]
\small
\centering
\caption{The 27 possible scale transformation methods either using a single crowd dataset in the source scale or using two datasets.
We then apply these methods to perform three transformations: S4 into binary, S100 into binary and S100 into 4-level. Notation: Hit-centric (H), Document-centric (D), Topic-wide (Tw$_\alpha$), Aggregate (a), Transform (t), Single dataset (1), Double dataset (2) as superscript.}\label{cs:trans:tab:transformations}

\begin{tabular}{@{ }lll@{ }}
\toprule
     \multicolumn{3}{c}{\emph{Single dataset (by internal agreement)}}  \\
     \midrule
      H\_t+a$^1$(S4$\rightarrow$2) & H\_t+a$^1$(S100$\rightarrow$2) & H\_t+a$^1$(S100$\rightarrow$4)\\
      H\_a+t$^1$(S4$\rightarrow$2) & H\_a+t$^1$(S100$\rightarrow$2) & H\_a+t$^1$(S100$\rightarrow$4)\\
      Tw\_$\alpha^1$(S4$\rightarrow$2) & Tw\_$\alpha^1$(S100$\rightarrow$2) & Tw\_$\alpha^1$(S100$\rightarrow$4)\\
     \addlinespace
     \multicolumn{3}{c}{\emph{Double dataset (by external agreement)}}\\
     \midrule
    H$^2$(S4$\rightarrow$2, TR2) & H$^2$(S100$\rightarrow$2, TR2) & H$^2$(S100$\rightarrow$4, So4)\\
    H$^2$(S4$\rightarrow$2, S2) & H$^2$(S100$\rightarrow$2, S2) & H$^2$(S100$\rightarrow$4, S4)\\
    \addlinespace
    D\_a+t$^2$(S4$\rightarrow$2, TR2) & D\_a+t$^2$(S100$\rightarrow$2, TR2) & D\_a+t$^2$(S100$\rightarrow$4, So4)\\
    D\_a+t$^2$(S4$\rightarrow$2, S2) & D\_a+t$^2$(S100$\rightarrow$2, S2) & D\_a+t$^2$(S100$\rightarrow$4, S4)\\
\addlinespace
    D\_t+a$^2$(S4$\rightarrow$2, TR2) & D\_t+a$^2$(S100$\rightarrow$2, TR2) & D\_t+a$^2$(S100$\rightarrow$4, So4)\\
    D\_t+a$^2$(S4$\rightarrow$2, S2) & D\_t+a$^2$(S100$\rightarrow$2, S2) & D\_t+a$^2$(S100$\rightarrow$4, S4)\\
     \bottomrule
\end{tabular}
\end{table}

\subsubsection{Possible Transformations}
Given all the transformation methods presented so far and the datasets used in our experiments, we can generate 27 possible transformations (listed in Table~\ref{cs:trans:tab:transformations} and described below) which we will experimentally analyze and compare in the remaining of this chapter with the goal of drawing an understanding of how to best transform relevance judgements into a different relevance scale.

Using the single dataset scale transformation method defined in Section~\ref{cs:trans:sec:singleDatasetMethod} we can transform three crowd-generated judgement collections (i.e., S4 into binary, S100 into binary, and S100 into four levels) using the two HIT-centric methods or, alternatively, using topic-based $\alpha$ to select the best \cut for each topic.

This leads to nine possible transformations (see Table~\ref{cs:trans:tab:transformations}).
%
By using the double dataset scale transformation method (see Section~\ref{cs:trans:sec:multiDatasetMethod}), using either of the
two judgement sets (i.e., by experts like TR2 binary and So4 4-level judgements, or by the crowd such as S2 and S4) in the target scale we can transform judgements using the HIT-centric approach or the two possible document-centric approaches (i.e., performing first an aggregation and then a scale transformation or first a scale transformation and then a judgement aggregation). This leads to additional 18 possible transformations (double dataset by external agreement in Table~\ref{cs:trans:tab:transformations}).
%
In addition, using the document-centric approach we are able to perform the transformation of 4-level expert judgement set (i.e., So4) into binary, which allows us to compare with binary expert TR2 judgements. Note that in this transformation, the aggregation step is not needed.

We also notice that when transforming a 4-level scale into a binary scale the number of possible \cuts is 3: \emph{left} (0 into 0 and 1, 2, 3 into 1), \emph{middle} (0, 1 into 0 and  2, 3 into 1), and \emph{right} (0, 1, 2 into 0 and  3 into 1). When transforming a 101-level scale into binary the number of \cuts is 100, and when transforming a 101-level scale into four levels it is \num{161700}.

\subsection{The Effect of Unjudged Documents}\label{cs:trans:sec:assumption}

\begin{table}[tbp]
\centering
\caption{Agreement $\alpha$ between transformed Sormunen judgements and TREC with and without the assumption that unjudged documents are not relevant.}
\scalebox{.99}{
  \begin{tabular}{ll S[table-format=4.0] S[table-format=2.5] *{2}{S[table-format=0.3]}}
  \toprule
  \textbf{Source} & \textbf{Target} & \textbf{Available} & \multicolumn{3}{c}{\textbf{$\alpha$ values for transforming on}} \\
  \cline{4-6}
  \textbf{scale} & \textbf{scale} & \textbf{documents} & \textbf{left} & \textbf{middle} & \textbf{right} \\
  \midrule
 So4       & TR2       & 805  & 0.595 & 0.136 & -0.356 \\
 So4$_{a}$ & TR2       & 3881 & 0.882 & 0.620 & 0.212 \\
 So4$_{a}$ & TR2$_{a}$ & 4269 & 0.884 & 0.625 & 0.220 \\
  \bottomrule
  \end{tabular}
  }
\label{cs:trans:tab:sormunenVsTrec}
\end{table}

%
In both TR2 and So4 judgements, only a subset of documents have been judged by experts (see Section \ref{cs:trans:sec:data}). Thus, we make the common assumption of unjudged documents to be not relevant.
To verify this assumption, we add non-relevant labels to both TR2 and So4 judgements for unjudged documents that appear in our other experimental datasets (i.e., S2, S4, and S100). We thus obtain two additional datasets: TR2$_{a}$ and So4$_{a}$, where \textit{a} denotes adding the additional non-relevant labels based on this assumption.
Next, we perform the transformation of So4 4-level judgements into binary, and measure their agreement with TR2 judgements by means of $\alpha$ using the \textit{doc-centric} approach in two ways: i) only over judgements which are available in both datasets and ii) over all documents with the assumptions that unjudged documents are not relevant.

%
Table~\ref{cs:trans:tab:sormunenVsTrec} presents the number of judged documents in both So4 and TR2 datasets, with and without the assumption that unjudged documents imply non-relevance, and the results of the best \cuts selected to transform So4 judgements into binary using TR2 as target scale dataset. 
The results show that by adding more non-relevant labels to So4 judgements, the agreement between the two expert judgements becomes higher. This shows (also due to the selection strategy followed to build So4, see Section~\ref{cs:trans:sec:data}) that unjudged documents in So4 are often labelled as non-relevant by experts in TR2 judgements, which is in line with the assumption that unjudged documents are assumed as non-relevant ones. 
Note also that, in all cases, the best \cut selected using $\alpha$ values is the left one. As it was already observed by \citet{Sormunen:2002:LRC:564376.564433}, the relevance threshold is low in TREC judgements which is explained by the risk of possibly missing relevant documents \cite{zobel98}. 
%
Based on this observation, in the following we only focus on the results where we make this assumption.
We report  results both making the assumption and removing unjudged documents for S100 transformed into a 4-level scale using So4, as the results vary substantially in that case because of the many missing judgements in So4.


\subsection{Evaluation of Scale Transformations}\label{cs:trans:sec:evalmethod}
To understand which of the proposed scale transformation methods leads to better relevance judgement datasets in the target scale, we look at what results the generated judgements produce when used for the evaluation of IR system effectiveness. 
Specifically, we compare the evaluation results of the transformed judgements against expert judgements collected natively in the target scale.
By using NDCG \cite{Jarvelin:2002:CGE:582415.582418} as evaluation metric,
we compute Kendall's $\tau$ correlation between the system ranking generated with the transformed judgements and the ranking generated with expert judgements assuming that the desired outcome is to obtain a transformed judgement set that leads to IR evaluation results as similar as possible to expert judgements.
Thus, when comparing different alternative methods to transform a judgement set from a source scale to a target scale, we prefer the ones achieving higher $\tau$ values within the proposed evaluation approach.

\section{Results and Discussion}
\label{cs:trans:sec:results}

\begin{figure*}[t]
\includegraphics[width=0.99\textwidth]{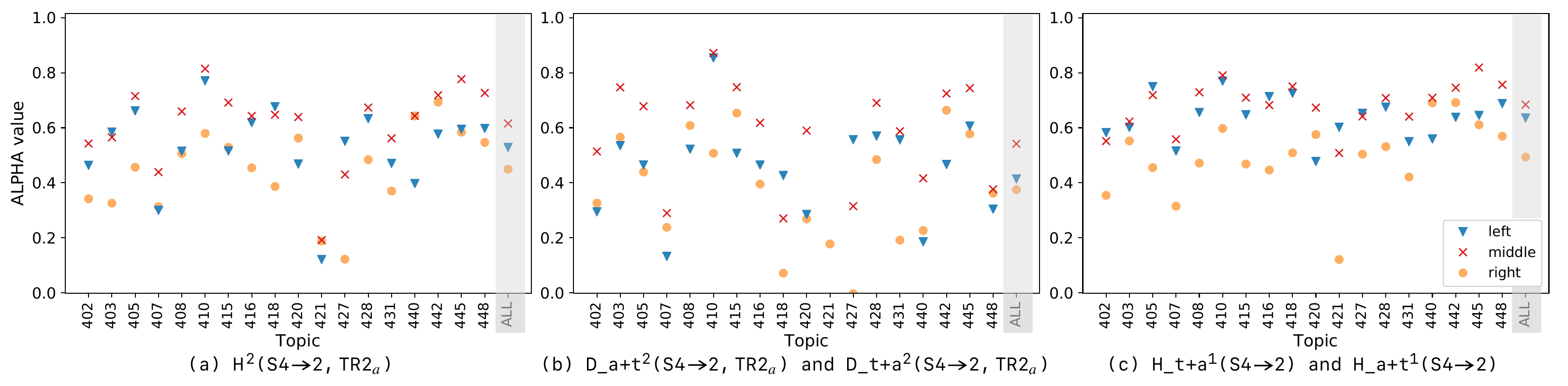}
\vspace*{-3mm}\caption{Transformation of S4 into 2 levels with TR2$_a$ (a) and (b) and HIT-centric single dataset transformation (c).}
\label{cs:trans:fig:S4vsTrec(noMissing)}
\end{figure*}
\begin{figure*}[t]
\includegraphics[width=0.99\textwidth]{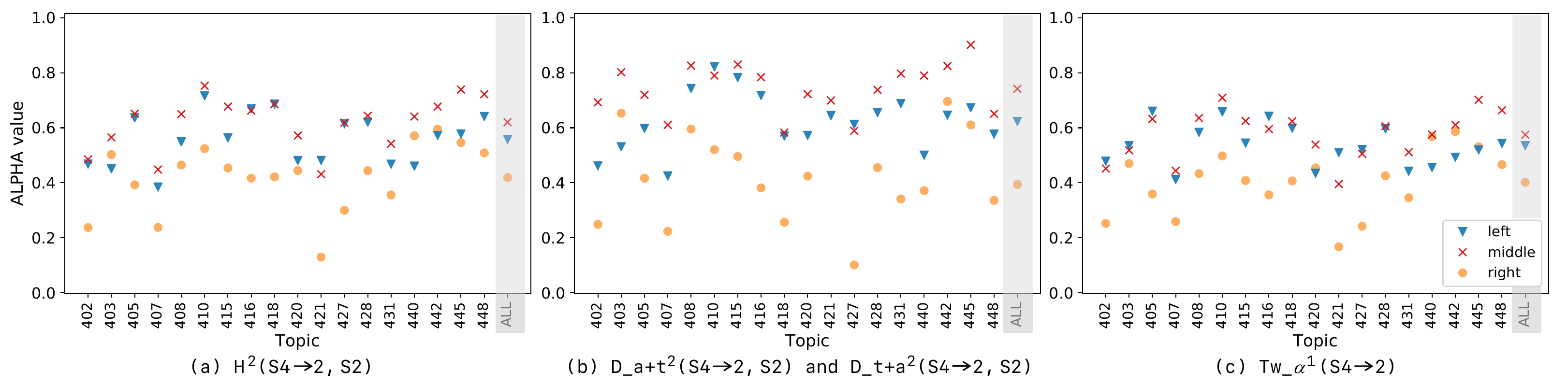}
\vspace*{-3mm}\caption{Transformation of S4 into 2 levels with S2 (a) and (b) and single dataset transformation with Tw\_$\alpha^1$ (c).}
\label{cs:trans:fig:S4vsS2}
\end{figure*}

In this section, we present our findings from applying the scale transformation approaches presented above to the considered expert and crowdsourced relevance judgements datasets (So4, S2, S4, S100). Additionally to the 27 transformations of crowdsourced datasets (see Table \ref{cs:trans:tab:transformations}), we report the transformation of the expert-generated dataset So4 into binary (So4$_a$ $\rightarrow$ 2, TR2$_a$).





\subsection{\texorpdfstring{So4$\rightarrow$2 and S4$\rightarrow$2}{So4->2 and S4->2}}

We begin with the simplest scale transformations, So4 and S4 to binary, where we have in total three possible choices to set \cuts between any two neighbor relevance levels.
%
\subsubsection{\texorpdfstring{So4$\rightarrow$2}{So4->2}}
Table~\ref{cs:trans:tab:sormunenVsTrec} shows agreement values between expert judgements collected on a 4-level scale (So4) which are transformed into a  binary scale with all possible \cut choices. The results indicate that setting \cuts just after 0 (left \cut) outperforms the other two choices (middle and right).
%
%
On a per-topic basis, the left \cut works best in 17 out of the 18 considered topics, with only one exception where the difference between left- and middle-\cut is not large.
%
This is consistent with the definition of relevance used by experts in both TR2 and So4 judgements, as discussed at the end of Section~\ref{cs:trans:sec:assumption}.


\subsubsection{\texorpdfstring{S4$\rightarrow$2}{S4->2}}
Next, we present the results involving judgements from crowd workers (i.e., S4).
Figures~\ref{cs:trans:fig:S4vsTrec(noMissing)}c and~\ref{cs:trans:fig:S4vsS2}c show the results of single dataset (i.e., S4) scale transformation with all three possible \cuts (left, middle, and right), and scale transformation methods described in Section~\ref{cs:trans:sec:singleDatasetMethod}: HIT-centric a+t and 
t+a (Fig.~\ref{cs:trans:fig:S4vsTrec(noMissing)}c) and Topic-wide $\alpha$ (Fig.~\ref{cs:trans:fig:S4vsS2}c). 
%
%
%
Note  that HIT-centric a+t and t+a produce exactly the same result,
just because we adopt majority vote (the same as median in the binary case) as our aggregation function for 2-level scales while take median values for 4-level scales. 
%
Actually, since we aggregate judgements from nine workers, the result of majority vote is always the same as the fifth value in a ranked list of 2-level judgements. On the other hand, the median value of a ranked list of nine 4-level judgements is the fifth value by definition.
Therefore, no matter whether we use t+a or a+t in the scale transformation method, the result of aggregating nine judgements exactly equals to transforming the fifth value in a ranked list, which means that both methods produce the same transformation.

From the charts it is evident that while the agreement level measured by means of $\alpha$ varies from topic to topic, the best \cut selected by each method is stable (i.e., middle \cut).
%
%
Figures~\ref{cs:trans:fig:S4vsTrec(noMissing)}ab and~\ref{cs:trans:fig:S4vsS2}ab show the agreement between transformed judgements from source scale and target scale judgements with all possible \cuts, by the double dataset methods defined in Section~\ref{cs:trans:sec:multiDatasetMethod} where we use both TR2 (Fig.~\ref{cs:trans:fig:S4vsTrec(noMissing)}ab) and S2 (Fig.~\ref{cs:trans:fig:S4vsS2}ab) as judgement dataset in the target scale.
The results show differences between using expert (i.e., TR2) and crowd (i.e., S2) judgements to measure the agreement used to select the best \cut. The best \cuts of Topics 421 and 427, for example, are identified as right- and left-transformation when compared to expert judgements, while both are picked up as middle-transformation as the best when using crowdsourced labels. This is because crowd workers have given different judgements as compared to experts (e.g., only 34 out of all 342 documents are labelled as relevant by experts in Topic 421, while crowd workers in S2 have judged 197 documents in this topic as relevant).

For each of these nine results on transforming S4, as well as for the result on So4, we show (on the right-hand side of each chart) the average of $\alpha$ values at each \cut across all topics, by which we are able to identify a single best \cut for the entire judgement collection (i.e., S4 or So4).\footnote{Note that an alternative approach to select a single \cut for the entire collection is to re-run a transformation method over all judgements regardless of their topic. This approach, which we do not report about for space limitations, leads to consistent results to those reported in this chapter.}
We observe that when transforming crowd judgements (i.e., S4) into binary, middle \cut  works best on average, regardless of what method is chosen to make the transformation. This is different from binarizing expert judgements (i.e., So4), where the best \cut is  the left one (Tab. \ref{cs:trans:tab:sormunenVsTrec}). 
Such result reveals how the relevance definition used by crowd workers differs from that of experts.





\subsection{\texorpdfstring{S100$\rightarrow$2}{S100->2}}



\begin{figure*}[t]
\includegraphics[width=0.99\textwidth]{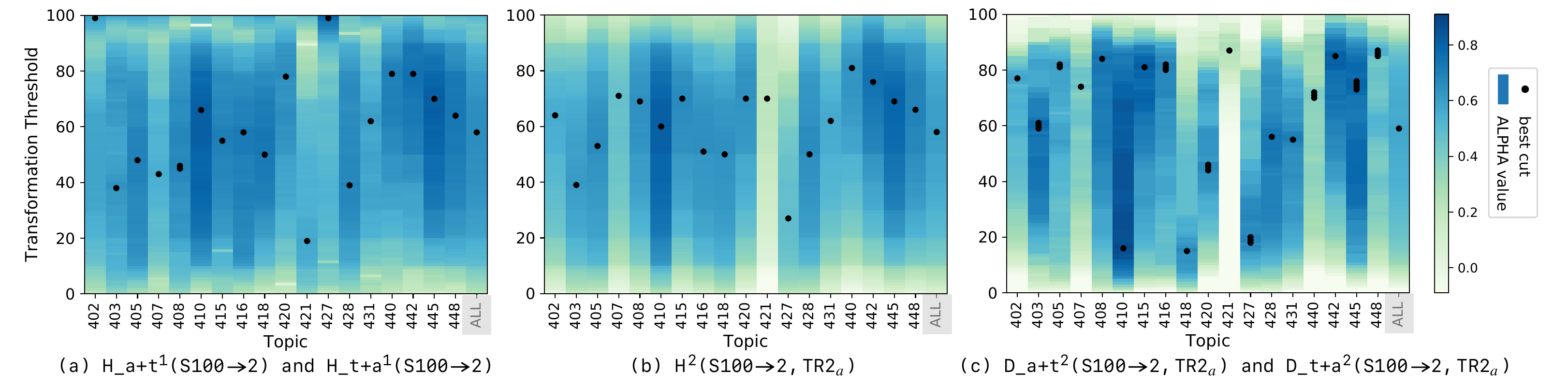}
\vspace*{-2mm}\caption{Transformation of S100 into 2 levels using a single dataset (a) and using TR2$_a$ (b) and (c).}
\label{cs:trans:fig:S100vsTrec(noMissing)}
\end{figure*}

\begin{figure*}[t]
\includegraphics[width=0.99\textwidth]{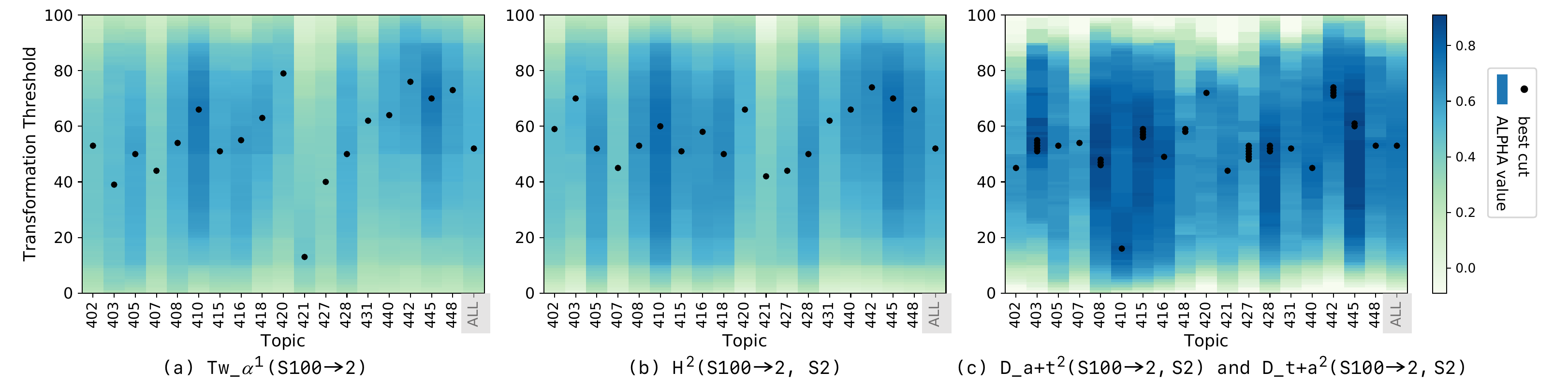}
\vspace*{-2mm}\caption{Transformation of S100 into 2 levels using a single dataset (a) and using S2 (b) and (c).}
\label{cs:trans:fig:S100vsS2}
\end{figure*}

\begin{figure}[t]
\centering
\includegraphics[width=0.4\textwidth]{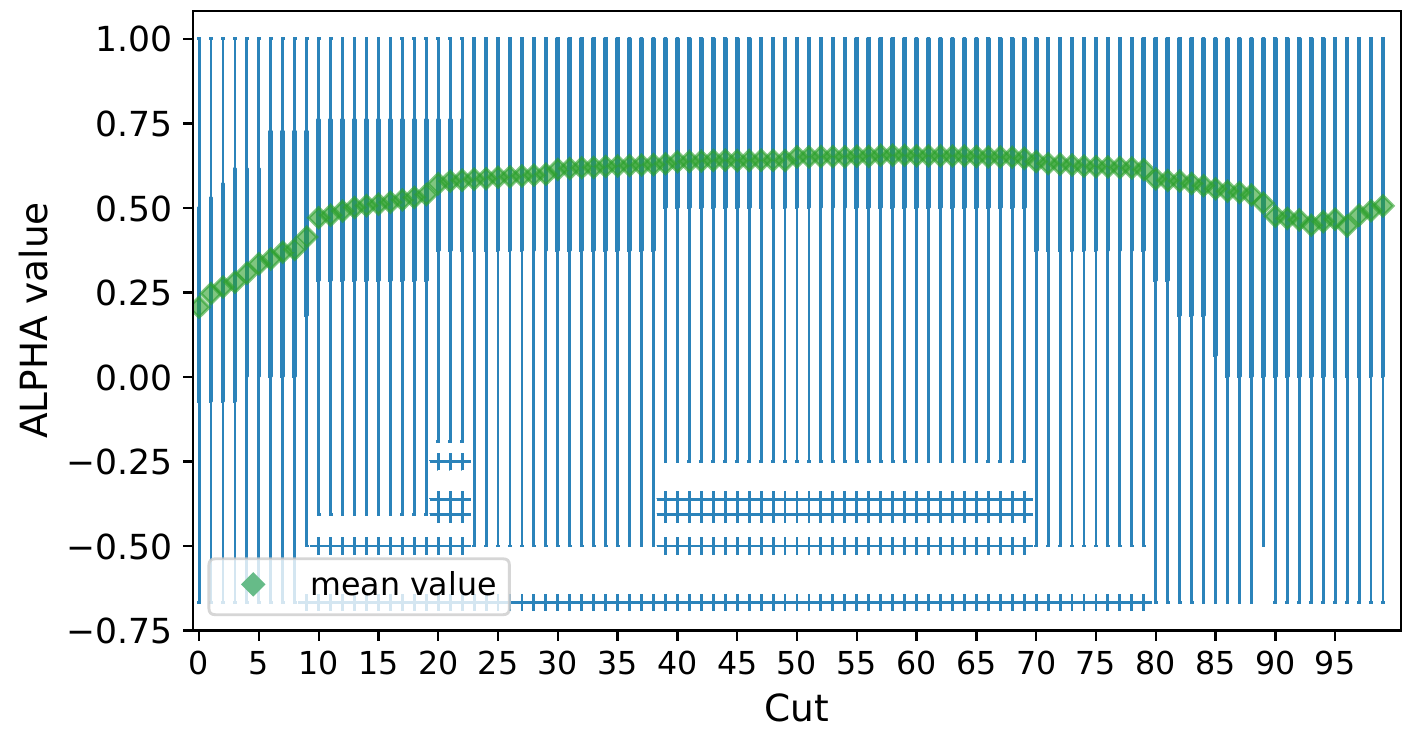}
\vspace*{-3mm}\caption{Inter-rater agreement over different cuts with H\_t+a$^1$(S100$\rightarrow$2) on the entire collection of S100.}
\label{cs:trans:fig:cut100to2}
\end{figure}

In this section, we focus on transformation of S100 into a binary scale. In this scenario there are a total of 100 possible choices among which to pick the best \cut.
Figures~\ref{cs:trans:fig:S100vsTrec(noMissing)} and \ref{cs:trans:fig:S100vsS2} present the results of both single and double dataset scale transformations defined in Section~\ref{cs:trans:sec:methods} for all possible \cut choices.
%
Note that in these charts (and in the following ones) the y-axis is not the $\alpha$ value like in the previous section for S4 transformation into binary, but rather the \cut in the $0..100$ range; the $\alpha$ value for each \cut is shown by the color density. The small black dot shows the best \cut. In case of multiple \cuts having the same best $\alpha$, there are multiple black dots shown for some topics, such as, for example, Topic 403 in Figure~\ref{cs:trans:fig:S100vsS2}c.

Unlike binarizing S4 scales, the selection of a method to transform S100 into 2 levels has an impact on picking up the best \cut for different topics. In Topic 402, for example, the best \cut is identified as 99 by HIT-centric methods in the single dataset transformation (see Figures~\ref{cs:trans:fig:S100vsTrec(noMissing)}a), while it is around 60 according to HIT-centric methods in the double dataset transformation (see Figures~\ref{cs:trans:fig:S100vsTrec(noMissing)}b and \ref{cs:trans:fig:S100vsS2}b) and below 50 in Doc-centric methods when comparing to S2 judgements (see Figure~\ref{cs:trans:fig:S100vsS2}c). 
This shows how different scale transformation methods may lead to very different judgements in the target scale and potentially impact the evaluation of IR systems.
There are two extreme cases in single dataset  HIT-centric method: Topic 402 and 427 (in Figures~\ref{cs:trans:fig:S100vsTrec(noMissing)}a). The best \cuts for both topics are selected as 99. This is because among all judgements in S100, no judgement at level 100 has been given by crowd workers to any document in these topics, and, therefore, by setting the \cut at 99 (i.e., mapping 100 to 1 and others to 0) every judgement is transformed into 0, which, in this case, maximises $\alpha$ agreement. 
We note, however, that alternative scale transformation methods select more appropriate \cuts to transform S100 judgements into binary.
%


From Figure~\ref{cs:trans:fig:cut100to2}, we can observe that the \cut that maximises $\alpha$ when applied on the entire collection rather than on a per-topic basis is 58. We can also notice that the lowest \cuts (i.e., 0-10) lead to lower $\alpha$ values ($\alpha=0.21$ when cut$=0$) than the highest possible \cuts (i.e., 90-100 with $\alpha=0.51$ when cut$=99$). This can be explained by the unbalanced distribution of relevant/non-relevant documents in the collection. That is, by selecting a \cut on the lower side of the scale we generate a binary judgement sets with many relevant documents. Contrary, by selecting a \cut on the higher extreme of the scale we generate many not-relevant judgements, which is more aligned with the natural distribution of judgements (with few relevant and many not-relevant judgements).

\subsection{\texorpdfstring{S100$\rightarrow$4}{S100->4}}
The most computationally challenging transformation is S100 into 4 levels as it implies selecting the best of 160K possible \cuts.


\subsubsection{Comparing Scale Transformation Methods}

\begin{figure*}[t]
\includegraphics[width=0.99\textwidth]{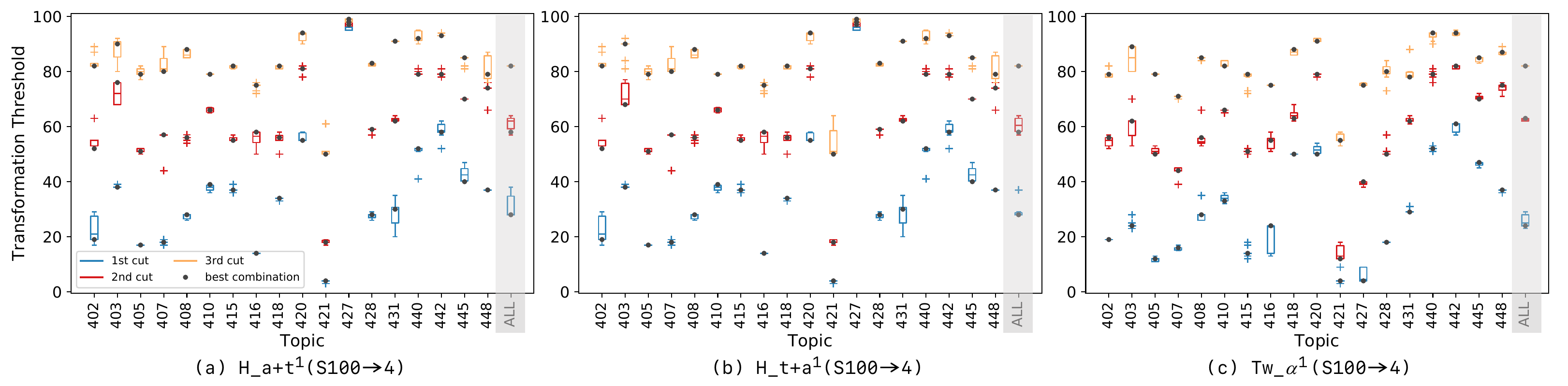}
\vspace*{-2.5mm}\caption{Single dataset transformation of S100 into 4 levels.}
\label{cs:trans:fig:S100singleDataset_100to4}
\end{figure*}

\begin{figure*}[t]
\includegraphics[width=0.99\textwidth]{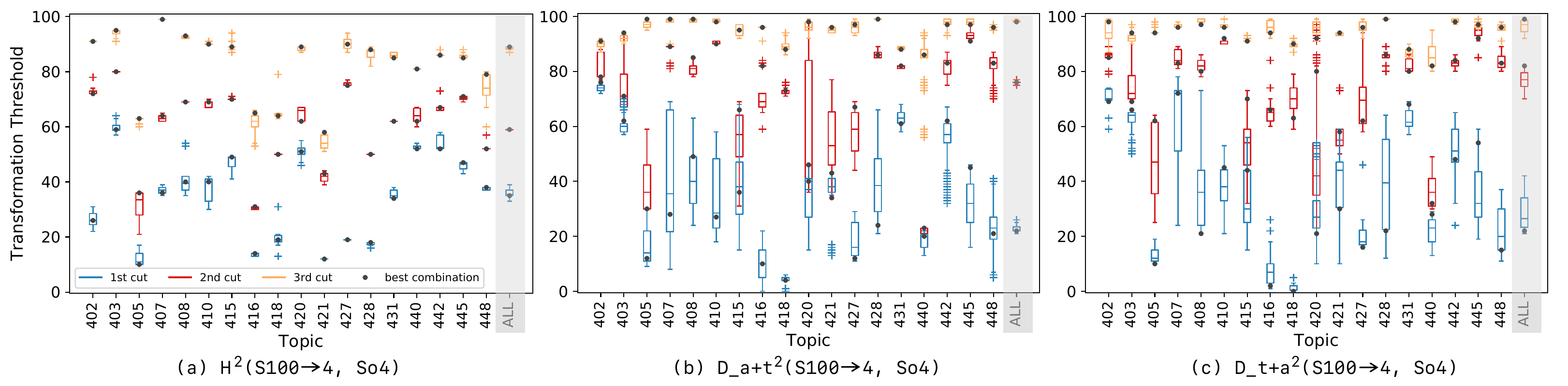}
\vspace*{-2.5mm}\caption{Transformation of S100 into 4 levels with So4.}
\label{cs:trans:fig:S100vsSormunen}
\end{figure*}

\begin{figure*}[t]
\includegraphics[width=0.99\textwidth]{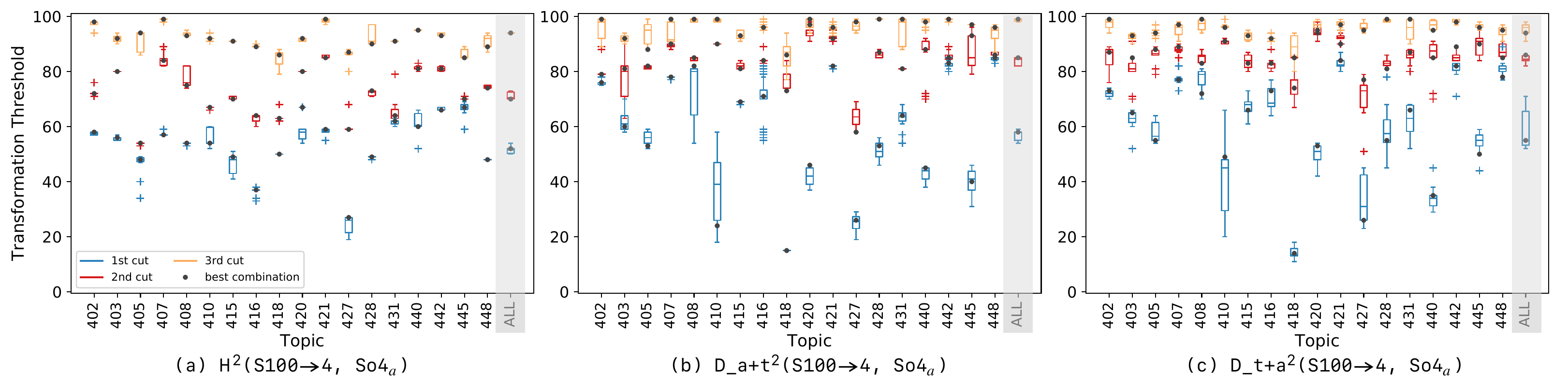}
\vspace*{-2.5mm}\caption{Transformation of S100 into 4 levels with So4$_a$.}
\label{cs:trans:fig:S100vsSormunen(noMissing)}
\end{figure*}

\begin{figure*}[t]
\includegraphics[width=0.99\textwidth]{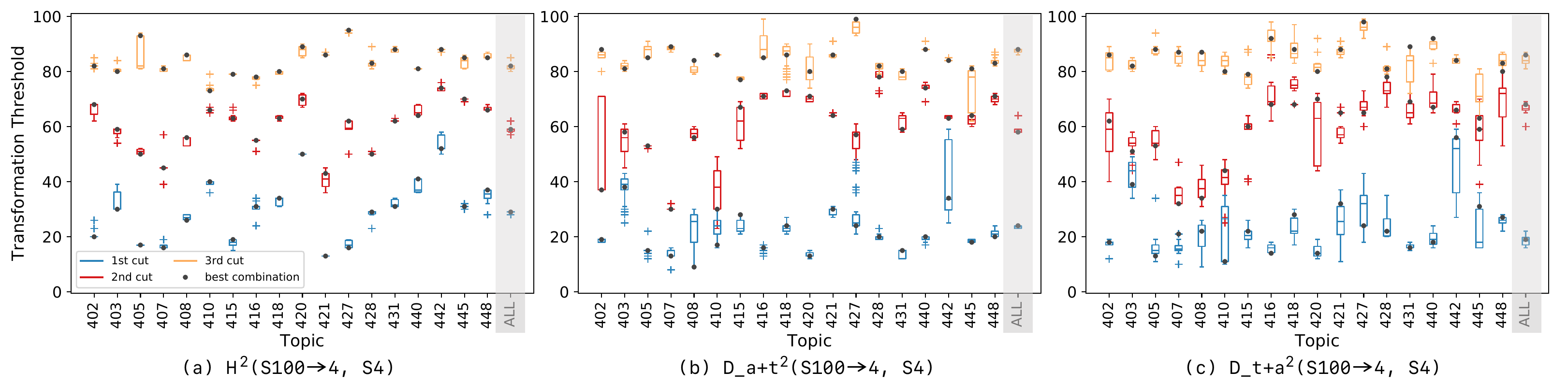}
\vspace*{-3.5mm}\caption{Transformation of S100 into 4 levels with S4.}
\label{cs:trans:fig:S100vsS4}
\end{figure*}

Figures~\ref{cs:trans:fig:S100singleDataset_100to4}, \ref{cs:trans:fig:S100vsSormunen}, \ref{cs:trans:fig:S100vsSormunen(noMissing)}, and \ref{cs:trans:fig:S100vsS4}
show for each transformation method a plot where for each topic we show the range of top 10 $\alpha$ values for each of  the three points of a scale transformation cut. The black dots indicate the best \cut for each topic.
Looking at these results we can make the following observations.
%
In HIT-centric single datasets methods, for topic 427 the methods select an extreme \cut (which also happened for this topic in the previous results on transforming S100 into binary. This issue gets however fixed by using double dataset methods that allows to select a more natural cut by measuring agreement against a judgement set in the target scale.
%
Because of So4 having a lot of unjudged documents, in this experiment we also compare the results of selecting the best \cut using either So4 or So4$_a$ as target scale dataset (Fig. \ref{cs:trans:fig:S100vsSormunen} and \ref{cs:trans:fig:S100vsSormunen(noMissing)}). 
We can observe that when using So4 without the assumption that unjudged are not relevant (and thus removing the unjudged from the agreement computation) we obtain a large interval of best cut points (shown by the large boxes in the plots of Fig.~\ref{cs:trans:fig:S100vsSormunen}). This is because 81\% of the documents have been removed from the calculation of $\alpha$ values. When we make the assumption that unjudged are not-relevant we have less variability in the selection of the best cut point (Fig. \ref{cs:trans:fig:S100vsSormunen(noMissing)}).
%
When using So4$_a$ we tend to get higher \cuts (i.e., to the right side of the scale) as compared to when we use S4 as target scale dataset. In detail, the top cut point selected by So4$_a$ methods tends to be 90 or higher and the difference from that selected by S4 methods is statistically significant (t-test $p<0.001$). This shows how using either expert or crowdsourced datasets has a strong impact on the selection of the best cut and on the judgement scale transformation results.
%
Finally, we also observe that double dataset doc-centric methods tend to display higher variability in the best \cuts as compared to HIT-centric methods. This is because in HIT-centric methods, there is no aggregation function used in the scale transformation process. When performing judgement aggregation,  with either D\_a+t$^2$ or D\_t+a$^2$, the distribution of the original relevance judgements shrinks into just one single value, which introduces \textit{gaps} (i.e., not all scale values are used) in the distribution of the crowdsourced judgements. As a result, whichever \cut is selected within the gap interval (i.e., the scale interval between two scale points used within aggregated relevance judgements), the transformed judgements remain the same, and consequently all these \cuts receive the same $\alpha$ and can be equally considered the best. 
%
In general, a comparison between the a+t and t+a approaches does not show significant differences.

\subsubsection{Analysis of the \cut distribution}
\begin{figure}[t]

\includegraphics[width=\linewidth]{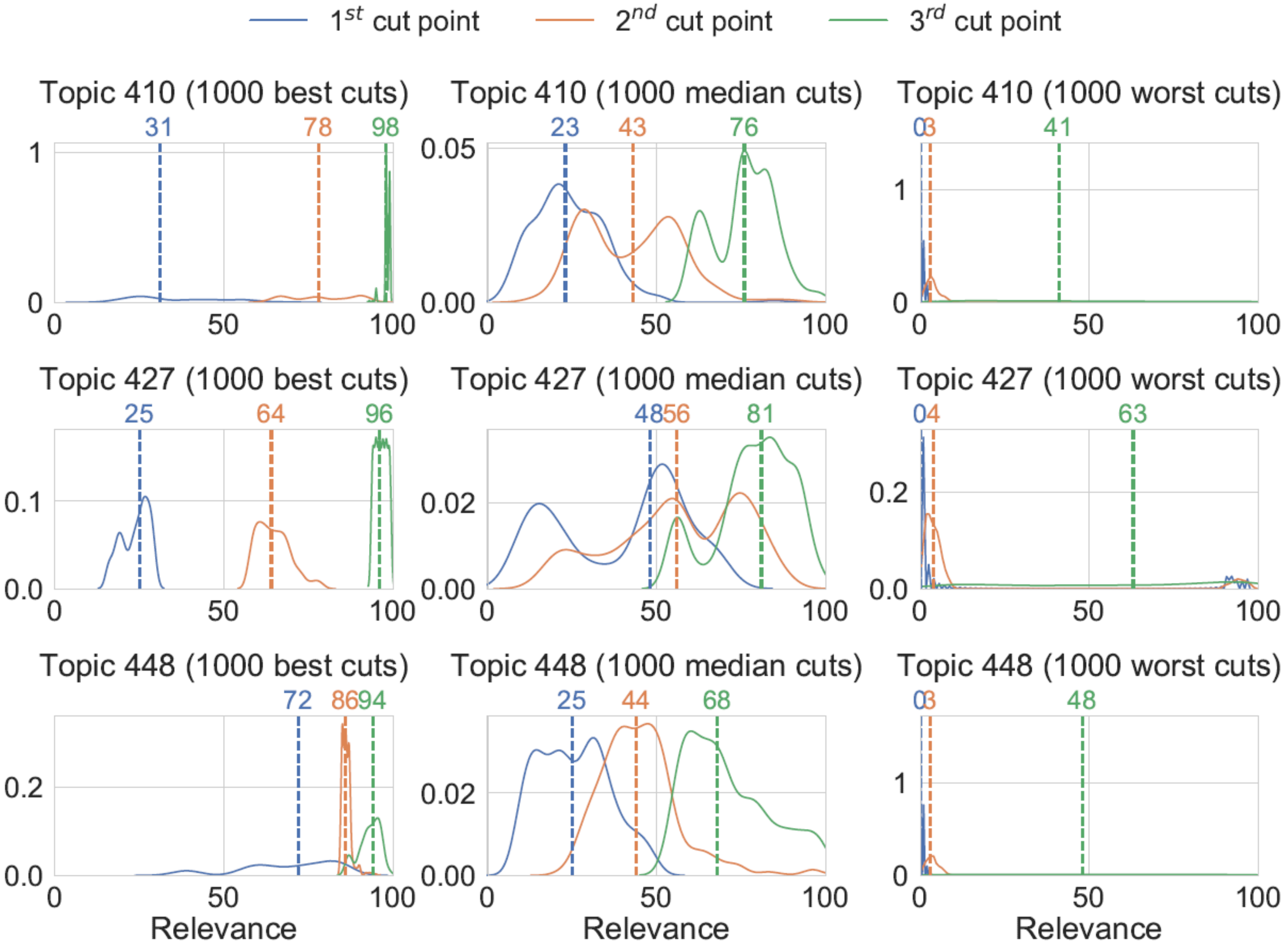}
\vspace*{-5mm}

\caption{Distributions of the three cut points for the 1000 best (left column), median (center column), and worst (right column) \cuts, over three topics in the collection. 
}
\label{cs:trans:fig:best_worst_cuts}
\vspace{-0.3cm}
\end{figure}

To investigate how all possible \cuts are distributed, we focus on D\_{a+t}$^2$(S100$\rightarrow$4, So4) which is the transformation method having the highest $\tau$ with
per-topic \cuts
(0.820, see 
Tab. \ref{cs:trans:tab:27_cuts_scores-mean-ndcg}).
We selected, out of the 160K possible \cuts, 3 subgroups having respectively the 1,000 \emph{best}, \emph{median}, and \emph{worst} \cuts.
Figure~\ref{cs:trans:fig:best_worst_cuts} shows that in such groups the \cut variability is quite limited for the best \cuts (left column) and the distance in terms of relevance is quite consistent among the three cut points. For median \cuts (central column), the noise increases but the three cut points are still uniformly distributed and distinguishable. 
The cut points of the worst \cuts (right columns) have values very close to scale boundaries. This happens in a consistent way across topics and different transformation methods.

We now look at relations between properties of \cuts and intrinsic characteristics of topics. First, we group the \emph{best}, \emph{median}, and \emph{worst} 1000 \cuts, then we compute the mean of topics of the widths of the 1000 \cuts ranges (intended as difference of relevance between the first and third cut point).
We find that the mean cut width of \emph{median} \cuts correlate positively (Pearson $0.54$, $p<0.05$) with the ratio of relevant documents in the topic, whereas such a correlation is negative when considering the group of the 1000 worst \cuts (Pearson $-0.5$, $p<0.05$).
This suggests that topics having a high ratio of relevant tend to have \emph{worst} cuts within a small range, and \emph{median} cuts within a wider range.

\subsection{Evaluating Scale Transformation Methods}
Table~\ref{cs:trans:tab:27_cuts_scores-mean-ndcg} shows the system ranking correlations computed when using  relevance judgements obtained with the selected best cut  and expert judgements.
This allows us to compare the effect of different scale transformation methods on IR evaluation results, as described in Section \ref{cs:trans:sec:evalmethod}.
We can observe that in the S4 transformation to binary all methods select the middle \cut as the best one, and thus the correlation values are very similar and only differ because of ties in the system rankings generated by the transformed judgements \cite{Carterette:2009:RCD:1571941.1572017}.
We thus cannot draw conclusions on which scale transformation method  is best for this experiment as they all pick the same best \cut.
When comparing single-dataset and double-dataset methods we see that  using a second dataset in the target scale consistently leads to more similar IR evaluation results. 
In the case of double-dataset methods, using expert generated judgements leads to higher correlation as compared to using crowd-generated judgements in the target scale.
These results are also due to the fact that the second dataset that we are using to  select the \cuts is  the same one we compare against when looking at IR system correlation.
We can also observe that, when transforming S100 into 4 levels, document-centric methods lead to higher evaluation results especially when using expert judgements (i.e., So4) as target scale dataset.
We  also see that for document-centric methods selecting different \cuts for each topic leads to more similar results as compared to selecting a single \cut for the entire collection.
However, on average, selecting the best \cut independently for each topic is not necessarily always better than selecting a single best \cut for the entire collection.
Comparing the use of So4 and S4 in double dataset methods, we see that So4 leads to significantly higher \cuts than S4. This confirms previous observations about the differences 
between experts and the crowd.

\begin{table}[tb]
    \centering
        \caption{Kendall's $\tau$ correlation between the IR system ranking generated using the transformed judgements and expert judgements for both selecting the best \cut on a per-topic basis or for the entire collection.
        Bold indicates the best values per experiment only considering expert judgements in the same scale as the target (not reported for the first experiment as all methods lead to the same best cut). So4 values are grayed-out when the transformation is into binary.
      }
\small
\begin{tabular}{@{}l rr c rr c@{}}
\toprule
{} & \multicolumn{2}{c}{Per-topic-cut} & & \multicolumn{2}{c}{Single-cut} & Single-Cut\\
\cmidrule{2-3} \cmidrule{5-6}
 &      \multicolumn{1}{c}{TR2} & \multicolumn{1}{c}{So4} &&  \multicolumn{1}{c}{TR2} & \multicolumn{1}{c}{So4} \\
\midrule
H\_t+a$^1$(S4$\rightarrow$2)         &  0.70  &  \cellcolor{shadowcells}0.71 &&  0.71 & \cellcolor{shadowcells}0.72 & middle\\
H\_a+t$^1$(S4$\rightarrow$2)         &  0.70  &  \cellcolor{shadowcells}0.71 &&  0.73 & \cellcolor{shadowcells}0.74 & middle\\
Tw\_$\alpha^1$(S4$\rightarrow$2)   &  0.69  &  \cellcolor{shadowcells}0.70 &&  0.72 & \cellcolor{shadowcells}0.74 & middle\\

\addlinespace

H$^2$(S4$\rightarrow$2, TR2)       	 &  0.72  &  \cellcolor{shadowcells}0.74 &&  0.72 & \cellcolor{shadowcells}0.73 & middle\\
H$^2$(S4$\rightarrow$2, S2)          &  0.72  &  \cellcolor{shadowcells}0.73 &&  0.71 & \cellcolor{shadowcells}0.72 & middle\\

D\_a+t$^2$(S4$\rightarrow$2, TR2)    &  0.73  &  \cellcolor{shadowcells}0.75 &&  0.73 & \cellcolor{shadowcells}0.74 & middle\\
D\_a+t$^2$(S4$\rightarrow$2, S2)     &  0.74  &  \cellcolor{shadowcells}0.75 &&  0.73 & \cellcolor{shadowcells}0.74 & middle\\

D\_t+a$^2$(S4$\rightarrow$2, TR2)    &  0.73  &  \cellcolor{shadowcells}0.75 &&  0.71 & \cellcolor{shadowcells}0.72 & middle\\
D\_t+a$^2$(S4$\rightarrow$2, S2)     &  0.73  &  \cellcolor{shadowcells}0.73 &&  0.71 & \cellcolor{shadowcells}0.72 & middle\\

\midrule

H\_t+a$^1$(S100$\rightarrow$2)       &  0.71  &  \cellcolor{shadowcells}0.72 &&  \textbf{0.77} & \cellcolor{shadowcells}0.81 & 58 \\
H\_a+t$^1$(S100$\rightarrow$2)       &  0.70  &  \cellcolor{shadowcells}0.71 &&  0.75 & \cellcolor{shadowcells}0.80 & 58 \\
Tw\_$\alpha^1$(S100$\rightarrow$2) &  0.72  &  \cellcolor{shadowcells}0.74 &&  0.76 & \cellcolor{shadowcells}0.79 & 52 \\

\addlinespace

H$^2$(S100$\rightarrow$2, TR2)       &  0.72  &  \cellcolor{shadowcells}0.74 &&  0.75 & \cellcolor{shadowcells}0.80 & 58 \\
H$^2$(S100$\rightarrow$2, S2)        &  0.76  &  \cellcolor{shadowcells}0.79 &&  0.76 & \cellcolor{shadowcells}0.79 & 52 \\

D\_a+t$^2$(S100$\rightarrow$2, TR2)  &  0.76  &  \cellcolor{shadowcells}0.78 &&  0.76 & \cellcolor{shadowcells}0.80 & 59 \\
D\_a+t$^2$(S100$\rightarrow$2, S2)   &  0.77  &  \cellcolor{shadowcells}0.80 &&  0.76 & \cellcolor{shadowcells}0.79 & 53 \\

D\_t+a$^2$(S100$\rightarrow$2, TR2)  &  \textbf{0.77}  &  \cellcolor{shadowcells}0.78 &&  0.75 & \cellcolor{shadowcells}0.80 & 55 \\
D\_t+a$^2$(S100$\rightarrow$2, S2)   &  0.77  &  \cellcolor{shadowcells}0.81 &&  0.75 & \cellcolor{shadowcells}0.81 & 56 \\

\midrule

H\_t+a$^1$(S100$\rightarrow$4)       &  0.75  &  0.76 &&  0.76 & 0.77 & $(28, 58, 82)$ \\
H\_a+t$^1$(S100$\rightarrow$4)       &  0.74  &  0.74 &&  0.74 & 0.76 & $(28, 58, 82)$ \\
Tw\_$\alpha^1$(S100$\rightarrow$4) &  0.74  &  0.76 &&  0.75 & 0.77 & $(24, 63, 82)$ \\

\addlinespace

H$^2$(S100$\rightarrow$4, So4)       &  0.77  &  0.77 &&  0.76 & \textbf{0.79} & $(52, 70, 94)$ \\
H$^2$(S100$\rightarrow$4, S4)        &  0.75  &  0.76 &&  0.76 & 0.77 & $(29, 59, 82)$ \\

D\_a+t$^2$(S100$\rightarrow$4, So4)  &  0.77  &  \textbf{0.82} &&  0.73 & 0.77 & $(58, 85, 99)$ \\
D\_a+t$^2$(S100$\rightarrow$4, S4)   &  0.74  &  0.76 &&  0.73 & 0.76 & $(24, 58, 88)$ \\

D\_t+a$^2$(S100$\rightarrow$4, So4)  &  0.76  &  0.80 &&  0.74 & 0.75 & $(55, 86, 94)$ \\
D\_t+a$^2$(S100$\rightarrow$4, S4)   &  0.73  &  0.76 &&  0.74 & 0.76 & $(19, 68, 86)$  \\
\bottomrule
\end{tabular}
    \label{cs:trans:tab:27_cuts_scores-mean-ndcg}
\vspace*{-0.3cm}
\end{table}

\section{Conclusions}\label{cs:trans:sec:conc}



Selecting the right relevance scale to be used within the creation of an IR evaluation collection is a key 
decision to make. When reusing existing collections, it may be necessary to transform judgements that have been originally collected  in a fine-grained scale into a different relevance scale.
In this chapter we presented an extensive study of relevance scale transformation methods over different datasets.
To the best of our knowledge, this is the first study of this kind.
We looked both at classic transformations previously adopted in the IR evaluation literature (i.e., the binarization of 4-level judgements) up to extreme cases in which we transformed a fine-grained (i.e., 101 relevance levels) scale into 4 levels thus considering 160K possible ways to transform it. 
Our results indicate that the method we select to transform judgements have strong implications on the results of IR evaluation experiments. 
We observed that:
\begin{itemize}
    \item Transforming the scale of a judgement collection is best done \textit{on a per-topic basis} rather than selecting the same \cut for the entire collection as all proposed methods tend to select quite different \cuts for different topics in our collection. Selecting \cuts per-topic or one single \cut for the entire collection appears, however, not to have a large impact on the IR evaluation results as compared to those obtained with expert judgements. 
    \item Transforming the scale of an expert-judged collection and a crowd-judged collection should not necessarily be done in the same way. In our experiments, when binarizing S4 and So4 we found that the best \cuts for these two dataset are different (i.e.,  middle and  left \cut, respectively, for S4 $\rightarrow$ 2 and So4 $\rightarrow$ 2).
    \item The classic assumption that unjudged documents are considered not-relevant may have strong implication on the way we transform judgements from one scale to another. 
    This results is not surprising as such an assumption is known to be invalid \cite{zobel98}.
    \item When comparing the IR evaluation outcomes obtained with transformed judgements with those obtained with native expert judgements, we observed that document-centric methods lead to more similar results and should thus be preferred.
\end{itemize}
%


\chapter{Conclusions and Future Directions}\label{chapt:thesisconclusion}
This chapter concludes the thesis.
Section~\ref{sect:concl:summary} summarizes the contributions, while
Section~\ref{sect:concl:fw} provides directions for future developments.

\section{Summary and Main Contributions}\label{sect:concl:summary}

In this thesis we presented works which deals with the reduction of the cost and the effort in the evaluation of Information Retrieval. 
More in detail, we considered three different approaches: 
the reduction of the topic set currently used (Part~\ref{part:fewtopics}),
the evaluation performed with no human intervention (Part~\ref{part:eewrj}), and 
the evaluation performed collecting crowdsourced relevance judgements (Part~\ref{part:cs}). 

The contributions of this thesis are as follows.
Part~\ref{part:fewtopics}, which deals with the usage of few topics in the effectiveness evaluation, provides the following contributions.

Chapter~\ref{chapt:few:evolutionary} presents a re-implementation of the BestSub software achieved using a novel approach based on evolutionary algorithms. Then, it presents a successful attempt of the reproduction of the the results by \citet{Guiver:2009:FGT:1629096.1629099}, \citet{ecir11}, and \citet{Berto:2013:UFT:2499178.2499184}. Finally, it provides the generalization of such results to other effectiveness metrics and other TREC collections.

Chapter~\ref{chapt:few:topicsubsets} presents a complete and exhaustive analysis on using fewer topics in the evaluation of retrieval systems.
We show that a larger ground truth topic set results in average and best subsets that are more highly correlated with the ground truth topic set than found in previous work. 
More in detail, as the cardinality of the ground
truth increases, the size of the subset (relative to ground truth) required to
obtain a high correlation also decreases.
Moreover, for large cardinalities, worst topic sets can be found that show  very low correlations with ground truth.
We also provide a detailed analysis on the role of statistically significant differences among runs considering different topic sets.
Finally, our analysis show that an effective clustering techniques can be exploited to find more representative topics.

Part~\ref{part:eewrj} deals with the evaluation of retrieval systems without relevance judgements, and provides the following contributions.

 Chapter~\ref{chapt:eewrj:reproduce} discusses the reproduction the most important work on evaluation of retrieval systems performed in absence of relevance judgements.
 Furthermore, we present many details useful for future reproducibility	, we present the results in a uniform way, and we generalize such results to other test collections, evaluation metrics, and a shallow pool. Finally, we expand those results, obtaining two practical strategies that seem effective to, respectively, decrease the costs involved in test collection based evaluation.
 
 Chapter~\ref{chapt:eewrj:combinations} presents an extensive analysis  over 17 prediction methods, 14 TREC collections, 15 accuracy measures, obtained by combining  three effectiveness measures (i.e., MAP, AP, and AAP) with five correlation measures (i.e., $\rho$, $\tau$, $r_s$, $\delta$, and $\tau_{ap}$), four data fusion approaches (plus variants), and twelve machine learning algorithms (plus variants) for the evaluation and combination of evaluation with out relevance judgement techniques. 
 We provide a systematic and uniform analysis on individual method effectiveness across different collections, and previously unnoticed relationships between the individual methods.
 Furthermore, we show that  practical results on method combinations by means of machine learning algorithms can be exploited to provide a practical methodology for the researcher that wants to run an effectiveness evaluation without human relevance assessments. 
 
 Part~\ref{part:cs} focuses on experiments on crowdsourcing relevance judgements and on the effect of the relevance scale adopted to collect relevance judgements.
 
 Chapter~\ref{chapt:cs:S100} presents a systematic study comparing the effects of different relevance scales on IR evaluation. The chapter shows many advantages of the fine grained (i.e., S100) scale as compared to coarse-grained scales like binary (i.e., S2) as well as S4, and to unbounded scales like ME. 
S100 preserves many of the advantages of ME like, for example, allowing to gather  relevance judgements that are much more fine-grained than the usual binary or 4-value scales. Assessors use the full spectrum, although sometimes with  a preference for scores that are a multiple of ten.
S100  has also demonstrated advantages over ME in terms of agreement with judgements collected on a binary and four level scales. 
Overall, our results show that S100 is an effective, robust, and usable scale to gather fine-grained relevance labels.

Chapter~\ref{chapt:cs:trans} addresses the issue of transforming relevance scales. The chapter shows that 
when reusing existing collections, it may be necessary to transform judgements that have been originally collected  in a fine-grained scale into a different relevance scale.
This chapter presents an extensive study of relevance scale transformation methods over different datasets.

\section{Future Developments}\label{sect:concl:fw}

This thesis leaves plenty of space for future work.

Concerning Part~\ref{part:fewtopics}, we could improve NewBestSub, in many ways.
The first improvement that we can study is a fine tuning of the algorithm parameters, such as: experimenting with other operators to perform crossover, like  XAND and  XOR;
study the relations between, and give an initial accurate estimate of, \emph{population number} and \emph{number of iterations};
in the case of more than one execution and the merge of the results, find the optimal number of executions.
We can also aim at finding the most (or a more) general topic set, i.e., a topic set which maximizes both the correlation with the ground truth and the ability to be a Best/Worst set in other collections.
Furthermore, with the consistent speedup of NewBestSub we aim at reproducing and extending the generalization experiments of \citet{Guiver:2009:FGT:1629096.1629099}, not by using a single split of the original AP matrix (both topic-wise and system-wise), but by performing many iterations of the process. With the old BestSub this experiment would be unfeasible. 
We aim also at developing an effective topic selection strategy, integrating 
 state-of-the-art methods to compute approximated matrix, such as the work discussed in Part~\ref{part:eewrj}. We believe that NewBestSub will be a useful tool to perform these experiments, as well as several other ones. 
Furthermore, we could employ NewBestSub to investigate the finding of few topics where multiple query variations are present, which has gathered increasing attention, as for example in \citet{10.1145/2911451.2914671} and \citet{jimmy2018overview}. 
We also plan to implement top heavy metrics such as $\tau_{\mathrm{AP}}$ and Rank Biased Overlap in NewBestSub, which will give us the possibility of studying and focusing on the most effective systems.

Concerning the few topics approach,
we plan to consider the correlation between topic subsets (rather than between a topic subset and the full topic set) as well as top-heavy measures of association such as Rank Biased Overlap or $\tau_{\mathrm{AP}}$, to give more importance to the most effective systems.
We believe we have only started to analyze how best and worst topic sets are formed. 
Considering the extreme nature of Best and Worst series, extreme value theory might be useful  to better understand and  model the stochastic behavior of Best / Worst series and topic subset distributions.
We also plan to deepen the analysis by finding more semantic features that characterize a good/bad topic set. 
Indeed, as in previous research, we have not attempted to devise methods to find good topic subsets before the evaluation exercise is performed, or while it is ongoing; the focus of our research so far has been on working to understand how different topic sets
interact. Future work studying more semantic topic features, combined with many runs, will
hopefully help to provide a set of guidelines for sound
topic set engineering.

Concerning Part~\ref{part:eewrj}, 
we plan to make further experiments on data fusion techniques and other similar methods. We will also experiment further on how to optimally combine the three methods. Injecting strategies to improve AAP estimation, instead of MAP, are less straightforward but can be devised, perhaps at the individual AP level. Combinations of injecting and shallow pool based (i.e., approximated metric computation) methods can also be devised. 
So far we have used only the individual prediction methods based on systems outcomes; \citeauthor{Diaz:2007}'s method \cite{Diaz:2007}, that requires the document collection as well, is an obvious candidate to be added.
We have not focused yet on the computational complexity and time needed to learn a model on the basis of the past datasets available, and we plan to do so.
Roughly, the training phase for learning a model even on all the past datasets is a matter of a few hours, and once the model has been learned its application on a new test set is very fast (a few seconds).
The machine learning approach suggests a more general framework that could include other features, also derived from completely different methods (for example, analyzing the text of topic descriptions and/or documents; properties of the systems; and so on).
This seems a promising approach, and we intend to pursue this research direction in the future. It would also be a way to address some limitations of the individual methods, that are quite rigid and difficult to extend.
Finally, 
Transfer Learning could
be exploited to adapt the models learned on past datasets to a new one with different properties \cite{Li:2016:EQW:2983323.2983852}. 
On more technical issues, in our approach we learn AP, not MAP, since as already discussed we do not have enough data to build a regressor on MAP values. This might be one reason for the better accuracy on AP (as well as AAP), than on MAP. Thus, we might refine our learning system to take into account MAP to some extent. Also, most individual methods generally aim at predicting MAP: it might be possible to tailor them as well for more accurate predictions of AP and AAP.
We also plan to test more sophisticate data fusion and ML techniques:
we plan to adapt to the setting of query performance prediction the learning-to-rank approach proposed by \citet{Raiber:2014:QPS:2600428.2609581}, as well as data fusion techniques 
\cite{Shtok:2016:QPP:2954381.2926790, Jayasinghe:2014:ITC:2682862.2682864}.

Concerning Part~\ref{part:cs}, 
we plan to extend the work on fine grained scales and do a systematic study in which we evaluate classical scales by means of crowdsourcing.
We also plan to develop and merge the fine grained scale approach with the scale transformations approach. We plan to differentiate the results from identifying the best \cut using the closest $\alpha$ of transformed judgement set by internal agreement to that of original judgement set, instead of picking up the best \cut by the highest $\alpha$. We will also investigate the use of system ranking correlation measures (e.g., $\tau$) as an alternative to assessor agreement for the selection of the best cut, perform in-depth comparative analysis of per-topic versus single-cut approaches, and design supervised methods to select \cuts for new datasets.

\appendix



\backmatter



\bibliographystyle{abbrvnat}
\bibliography{tesi}

\end{document}